\documentclass[a4paper,11pt]{book}

\pdfoutput=1

\usepackage[a4paper,vmargin={30mm,30mm},hmargin={30mm,30mm}]{geometry}

\usepackage{fancyhdr}
\usepackage{lmodern}
\usepackage[T1]{fontenc}
\usepackage[utf8]{inputenc}

\usepackage[inline]{enumitem}

\usepackage{amsfonts}
\usepackage{amstext}
\usepackage{amsmath}
\usepackage{amssymb}
\usepackage{mathtools}
\usepackage{mathdots}
\usepackage{mathrsfs}

\usepackage{arydshln}
\usepackage[margin=\parindent,labelfont=bf]{caption}

\usepackage{pbox}

\usepackage[usenames,dvipsnames]{xcolor}

\usepackage{tikz}
\usetikzlibrary{decorations.markings}
\usetikzlibrary{arrows}
\usetikzlibrary{matrix}
\usetikzlibrary{fadings}
\usetikzlibrary{decorations.pathreplacing}

\numberwithin{equation}{chapter}
\numberwithin{figure}{chapter}

\usepackage[numbers]{natbib}

\usepackage{bibentry}

\makeatletter
\renewcommand\bibentry[1]{\nocite{#1}{\frenchspacing
     \@nameuse{BR@r@#1\@extra@b@citeb}}}
\makeatother

\usepackage{hyperref}
\hypersetup{
pdfauthor={Nils Kanning}, 
pdftitle={On the Integrable Structure of Super Yang-Mills Scattering Amplitudes},
pdfkeywords={super Yang-Mills theory, scattering amplitudes,
  Graßmannian integral, Yangian invariance, oscillator representations,
  integrable spin chains, Bethe ansatz, vertex models, unitary matrix
  models, Bargmann transformation},
pdflang=en,
bookmarksnumbered=true,
colorlinks=true,
linkcolor=Red,
citecolor=Red,
urlcolor=Red,
unicode=true,
hypertexnames=false,
bookmarksdepth=4,
linktocpage=true
}

\usepackage{ulem}
\usepackage{microtype}

\usepackage{accents}

\newcommand*{\dt}[1]{%
  \accentset{\mbox{\large\bfseries .}}{#1}}

\DeclareMathOperator{\D}{d\!}
\DeclareMathOperator{\tr}{tr}
\DeclareMathOperator{\str}{str}
\DeclareMathOperator{\Min}{min}
\DeclareMathOperator{\Real}{Re}
\DeclareMathOperator{\Imag}{Im}
\DeclareMathOperator{\sgn}{sgn}

\usepackage[ddmmyyyy]{datetime}

\usepackage{currfile}

\usepackage[backgroundcolor=red!10!white,bordercolor=red,linecolor=red]{todonotes}

\newcommand{\indnm}{\mathcal}
\newcommand{\indssub}{\mathsf}

\newcommand{\mon}{M}
\newcommand{\bvac}{|\Omega\rangle}

\newcommand{\osca}{\mathbf{a}}

\newcommand{\s}{\mathbf{s}}

\newcommand{\rap}{\theta}
\newcommand{\spec}{u}
\newcommand{\specp}{u'}
\newcommand{\brt}{u}
\newcommand{\magn}{x}
\newcommand{\inhdiff}{z}
\newcommand{\inh}{v}
\newcommand{\inht}{w}

\newcommand{\epb}{j}
\newcommand{\epe}{i}

\newcommand{\graph}{\mathbf{G}}
\newcommand{\inhtset}{\mathbf{w}}
\newcommand{\brtset}{\mathbf{u}}
\newcommand{\magnset}{\mathbf{x}}

\newcommand{\rapset}{\boldsymbol{\theta}}

\newcommand{\sites}{N}
\newcommand{\lines}{L}
\newcommand{\brts}{P}
\newcommand{\dsites}{K}

\newcommand{\oscrep}{\mathcal{D}}

\newcommand{\elemm}{E}
\newcommand{\elemv}{e}

\begin{document} 

\frontmatter
\pagestyle{plain}

\begin{titlepage}
  \begin{center}
    {\Huge\textbf{
        On the Integrable Structure\\
        of Super Yang-Mills\\
        Scattering Amplitudes}\par}
    ~\\
    ~\\
    ~\\
    \textls[300]{Dissertation}\\
    ~\\
    (Überarbeitete Fassung)
    ~\\
    ~\\
    ~\\
    ~\\
    eingereicht an der\\
    ~\\
    Mathematisch-Naturwissenschaftlichen Fakultät\\
    ~\\
    der Humboldt-Universität zu Berlin\\
    ~\\
    ~\\
    ~\\
    von\\
    ~\\
    \textbf{Nils Kanning}\\
    ~\\
    \texttt{\href{mailto:kanning@physik.hu-berlin.de}{\color{black}{kanning@physik.hu-berlin.de}}}\\
    ~\\
    ~\\
    ~\\
    \textit{
      Institut für Mathematik und Institut für Physik,
      Humboldt-Universität zu Berlin,\\
      IRIS-Adlershof, Zum Großen Windkanal 6, 12489 Berlin, Germany\\
      ~\\
      Max-Planck-Institut für Gravitationsphysik, Albert-Einstein-Institut,\\
      Am Mühlenberg 1, 14476 Potsdam, Germany\\
      ~\\
      Arnold Sommerfeld Center for Theoretical Physics,
      Ludwig-Maximilians-Universität,\\
      Theresienstraße 37, 80333 München, Germany\\
    }
    ~\\
    ~\\
    ~\\
    Tag der mündlichen Prüfung: 09.12.2016
  \end{center}
\end{titlepage}

\chapter{Zusammenfassung}
\label{cha:zusammen}

Die maximal supersymmetrische Yang-Mills-Theorie im vierdimensionalen
Minkowski-Raum, $\mathcal{N}=4$ SYM, ist ein außergewöhnliches Modell
der mathematischen Physik. Dies gilt vor allem im planaren Limes, in
dem die Theorie integrabel zu sein scheint. Diese Integrabilität wurde
zunächst für das Spektrum der anomalen Dimensionen
entdeckt. Inzwischen ist sie auch bei anderen Observablen zu Tage
getreten. Insbesondere sind Streuamplituden auf Baumgraphenniveau
Invarianten einer Yangschen Algebra, die die superkonforme Algebra
$\mathfrak{psu}(2,2|4)$ beinhaltet. Diese unendlichdimensionale
Symmetrie ist ein Kennzeichen für Integrabilität. In dieser
Dissertation untersuchen wir Verbindungen zwischen solchen Amplituden
und integrablen Modellen. Wir verfolgen zweierlei Ziele. Zum einen
wollen wir Grundlagen für eine effiziente, auf der Integrabilität
basierende Berechnung von Amplituden legen. Zum anderen sind wir
bestrebt einen Zugang zu schaffen der neue Ideen zu Amplituden auch
für integrable Systeme im Allgemeinen anwendbar macht. Dazu
charakterisieren wir Yangsche Invarianten innerhalb der
Quanten-Inverse-Streumethode (QISM), die Werkzeuge zur Behandlung
integrabler Spinketten bereitstellt. Wir arbeiten mit einer Klasse von
Oszillatordarstellungen der Lie-Algebra $\mathfrak{u}(p,q|m)$, die
etwa vom Spektralproblem der $\mathcal{N}=4$ SYM bekannt ist. In
diesem Rahmen entwickeln wir Methoden zur Konstruktion von Yangschen
Invarianten. Wir zeigen, dass der algebraische Bethe-Ansatz, ein
Bestandteil der QISM, für die Erzeugung von Yangschen Invarianten für
$\mathfrak{u}(2)$ und im Prinzip auch für $\mathfrak{u}(n)$ verwendet
werden kann. Diese Invarianten sind spezielle Zustände inhomogener
Spinketten. Die zugehörigen Bethe-Gleichungen lassen sich leicht
lösen. Unser Zugang ermöglicht es zudem diese Invarianten als
Zustandssummen von Vertexmodellen zu interpretieren. Außerdem führen
wir ein unitäres Graßmannsches Matrixmodell zur Konstruktion Yangscher
Invarianten mit Oszillatordarstellungen von $\mathfrak{u}(p,q|m)$
ein. Es ist angeregt durch eine Formulierung von Amplituden als
mehrdimensionale Konturintegrale auf Graßmannschen
Mannigfaltigkeiten. Für einen Spezialfall reduziert sich unsere Formel
zu dem Brezin-Gross-Witten-Integral über unitäre Matrizen. Ferner
führt es zu einem $U(2)$-Integralausdruck für eine Invariante die
einer R-Matrix entspricht. Solche R-Matrizen bilden die Basis für
integrable Spinketten. Wir wenden eine auf Bargmann zurückgehende
Integraltransformation auf unser unitäres Graßmannsches Matrixmodell
an, welche die Oszillatoren in Spinor-Helizitäts-artige Variablen
überführt. Dadurch gelangen wir zu einer Weiterentwicklung des
bereits erwähnten Graßmannschen Integrals für bestimmte
Amplituden. Die maßgeblichen Unterschiede sind, dass wir in der
Minkowski-Signatur arbeiten und die Kontur auf die unitäre
Gruppenmannigfaltigkeit festgelegt ist. Wir vergleichen durch unser
Integral gegebene Yangsche Invarianten für $\mathfrak{u}(2,2|4)$ mit
bekannten Ausdrücken für Amplituden und kürzlich eingeführten
Deformationen derselben.
~\\
~\\
\noindent
\textbf{Schlagwörter:} Super-Yang-Mills-Theorie, Streuamplituden,
Graßmannsches Integral, Yangsche Invarianz, Oszillatordarstellungen,
integrable Spinketten, Bethe-Ansatz, Vertexmodelle, unitäre
Matrixmodelle, Bargmann-Transformation

\chapter{Abstract}
\label{cha:abstract}

The maximally supersymmetric Yang-Mills theory in four-dimensional
Minkowski space, $\mathcal{N}=4$ SYM, is an exceptional model of
mathematical physics. Even more so in the planar limit, where the
theory is believed to be integrable. This integrable structure was
first revealed for the spectrum of anomalous dimensions. By now it has
begun to surface also for further observables. In particular, the
tree-level scattering amplitudes were shown to be invariant under the
Yangian of the superconformal algebra $\mathfrak{psu}(2,2|4)$. This
infinite-dimensional symmetry is a hallmark of integrability. In this
dissertation we explore connections between these amplitudes and
integrable models. Our aim is twofold. First, we want to lay
foundations for an efficient integrability-based computation of
amplitudes. Second, we intend to create a formulation that makes new
ideas about amplitudes applicable to integrable systems in general. To
this end, we characterize Yangian invariants within the quantum
inverse scattering method (QISM), which is an extensive toolbox for
integrable spin chains. Throughout the thesis we work with a class of
oscillator representations of the Lie algebra $\mathfrak{u}(p,q|m)$,
that is known e.g.\ from the $\mathcal{N}=4$ SYM spectral
problem. Making use of this setup, we develop methods for the
construction of Yangian invariants. We show that the algebraic Bethe
ansatz from the QISM toolbox can be specialized to yield Yangian
invariants for $\mathfrak{u}(2)$ and in principle also for
$\mathfrak{u}(n)$. These invariants are special states of
inhomogeneous spin chains. The associated Bethe equations can be
solved easily. Our approach also allows to interpret these Yangian
invariants as partition functions of vertex models. What is more, we
establish a unitary Graßmannian matrix model for the construction of a
subset of $\mathfrak{u}(p,q|m)$ Yangian invariants with oscillator
representations. It is inspired by a formulation of amplitudes as
multi-dimensional contour integrals on Graßmannian manifolds. In a
special case our formula reduces to the Brezin-Gross-Witten integral
over unitary matrices. Furthermore, it yields a $U(2)$ integral
expression for an invariant corresponding to an R-matrix. Such
R-matrices generate integrable spin chain models. We apply an integral
transformation due to Bargmann to our unitary Graßmannian matrix
model, which turns the oscillators into spinor helicity-like
variables. Thereby we are led to a refined version of the
aforementioned Graßmannian integral for certain amplitudes. The most
decisive differences are that we work in Minkowski signature and that
the integration contour is fixed to be a unitary group manifold. We
compare $\mathfrak{u}(2,2|4)$ Yangian invariants defined by our
integral to known expressions for amplitudes and recently introduced
deformations thereof.
~\\
~\\
\noindent
\textbf{Keywords:} super Yang-Mills theory, scattering amplitudes,
Graßmannian integral, Yangian invariance, oscillator representations,
integrable spin chains, Bethe ansatz, vertex models, unitary matrix
models, Bargmann transformation

\chapter{List of Publications}
\label{cha:publications}

\nobibliography*

This dissertation is based on the following publications:
\begin{enumerate}
\item[\cite{Frassek:2013xza}] \bibentry{Frassek:2013xza}.
\item[\cite{Kanning:2014cca}] \bibentry{Kanning:2014cca}.
\end{enumerate}
In addition, it contains a considerable amount of unpublished
work. The author also contributed to a publication on a closely
related subject:
\begin{enumerate}
\item[\cite{Kanning:2014maa}] \bibentry{Kanning:2014maa}.
\end{enumerate}

\cleardoublepage
\phantomsection
\addcontentsline{toc}{chapter}{Contents}
\tableofcontents

\mainmatter
\pagestyle{fancy}

\addtocontents{toc}{\protect\setcounter{tocdepth}{2}}
\chapter{Introduction}
\label{cha:intro}

\section{Integrable Models}
\label{sec:int-mod}

The exact solution of the gravitational two-body problem dates back to
Newton's ``Principia'' published in 1687. It is a landmark in
mathematical physics that led to a most profound advance in our
physical understanding by establishing the sound theoretical
foundation of Kepler's empirical laws of planetary motion. Throughout
the centuries this problem has been revisited employing new concepts
and machinery developed in the field of classical mechanics. For
instance, the relative motion of the two bodies can be derived by
exploiting the conserved quantities of the problem, i.e. the energy,
the angular momentum vector and the Runge-Lenz vector. In the
Hamiltonian formulation this may be achieved by a specific canonical
transformation, as reviewed e.g.\ in
\cite{Cordani:2012,Arnold:2007}. This transformation in particular
replaces the three momenta by three of the conserved quantities, which
are independent and Poisson-commuting, just as the original
momenta. One can choose the energy and the third components of both
conserved vectors as these new momenta. The construction of this
transformation reduces to integrals and inversions of algebraic
equations. In the new coordinates the Hamiltonian equations of motion
are solved trivially because the momenta are conserved. The procedure
outlined here is a special case of a theorem by Liouville, cf.\
\cite{Babelon:2003,Arnold:2013}: The solution of a Hamiltonian system
in a $2N$-dimensional phase space with $N$ conserved quantities, which
are independent and Poisson-commuting, reduces to a number of
integrals and inversions. Therefore such a system is called
\emph{integrable} or sometimes synonymously \emph{exactly solvable}.

The gravitational two-body problem is a shining example for the
importance of integrable systems. They are mathematically much more
accessible than generic models. Their exact solutions can exhibit
features that are absent or easily overlooked in approximate
methods. Consequently, these solutions can even be critical to settle
conceptual questions arising in a new theory. Integrable systems can
be found in various branches of physics and there is an opulence of
mathematical approaches to investigate them. Naturally, our aim in
this introduction can in no way be to provide a comprehensive overview
of this fascinating field at the border between physics and
mathematics. Thus we concentrate on a selection of examples, which
illustrate important themes and are headed towards the subject of this
thesis. To compensate for this focus, most of the works we refer to
are books or review articles that embed the examples into a broader
context. In this section special \emph{emphasis} is put on topics that
reappear in later chapters.

The concept of integrability is not limited to the ordinary
differential equations of classical mechanics. Let us continue with an
example of an integrable partial differential equation appearing in
fluid dynamics. Water waves can be modeled by highly non-linear Euler
equations, whose general solution is not known. However, the situation
changes radically in a certain limiting case. Restricting to surface
waves in shallow water that propagate only in one dimension, one
arrives at the Korteweg-de~Vries equation. This equation is still
non-linear, yet it possesses remarkable hidden structures, as
explained e.g.\ in \cite{Drazin:1989,Ablowitz:1991}. For instance,
there are solitary wave solutions that can be superposed despite the
non-linearity. What is more, the Korteweg-de~Vries equation can be
formulated as a Hamiltonian system with infinitely many
Poisson-commuting conserved quantities. Even the initial value problem
for this equation has been solved by means of the so-called inverse
scattering method. This method can be understood as a generalization
of the Liouville theorem to infinitely many degrees of freedom because
in essence it is a canonical transformation to coordinates in which
the dynamics becomes linear and in this sense trivial. Thus the
Korteweg-de~Vries equation and many further partial differential
equations solvable by this method are said to be integrable. Let us
remark that the Korteweg-de~Vries equation also appears in fields that
seem to be far removed from fluid dynamics. In particular, classes of
solutions can be formulated as \emph{matrix models}, i.e.\ as
integrals over certain matrices, see e.g.\
\cite{DiFrancesco:1993cyw,Morozov:1994hh}.

The examples discussed up to this point are models of classical
physics. The exact solution of a prototypical quantum integrable model
was obtained by Bethe already shortly after the advent of quantum
mechanics in 1931. He studied a one-dimensional chain of electron
spins described by Pauli matrices, which Heisenberg had proposed
earlier as a model of a ferromagnet. Bethe made an ansatz for the wave
functions. He showed that for this ansatz to yield eigenfunctions of
the Hamilton operator, its parameters have to obey a set of algebraic
equations. This resulted in an efficient method for the
diagonalization of the Hamiltonian. Detailed accounts on this
\emph{Bethe ansatz} are provided e.g.\ in
\cite{Gaudin:2014,Sutherland:2004}.

Bethe's solution of the Heisenberg spin chain contains no direct
reference to the notion of integrability we encountered in the
previous paragraphs. This link was established only several decades
later by Faddeev and his coworkers in the context of what they termed
\emph{quantum inverse scattering method} (QISM)
\cite{Faddeev:1996iy,Korepin:1997}. At the core of this approach lies
the cubic \emph{Yang-Baxter equation}, see \cite{Jimbo:1990} for a
compilation of pioneering publications. The group around Faddeev
managed to express the Hamiltonian of the Heisenberg spin chain and
its eigenfunctions in terms of a particular solution of the
Yang-Baxter equation and thereby incorporated the Bethe ansatz into
their framework. This reformulation immediately led to a family of
commuting operators that contains the Hamiltonian. Therefore these
operators are analogous to the Poisson-commuting conserved quantities
in classically integrable models. In addition, the reparameterization
of the model from Pauli matrices to a solution of the Yang-Baxter
equation allows for an interpretation as the counterpart of the
canonical transformation that is at the heart of the Liouville theorem
and the inverse scattering method, cf.\ \cite{Faddeev:1995}. To this
effect, the QISM is a quantization of these classical results.

A significant feature of the QISM is that it can be extended to a
large class of \emph{integrable spin chains} by choosing different
solutions of the Yang-Baxter equation. Many of these solutions have a
representation theoretic origin. From this perspective an integrable
spin chain model is specified by the choice of a symmetry algebra and
a representation thereof. The Heisenberg model arises from the Lie
algebra $\mathfrak{su}(2)$ with the spin $\frac{1}{2}$ representation.
In fact, a closer look reveals that this Lie algebra is extended to an
infinite-dimensional symmetry algebra referred to as \emph{Yangian},
see e.g.\ \cite{Bernard:1992ya,MacKay:2004tc,Molev:2007}. Such hidden
extended symmetries are characteristic of integrable models. The QISM
has found applications beyond spin chain models. It is of utility for
two-dimensional \emph{vertex models} in statistical physics, where
solutions of the Yang-Baxter equation provide the Boltzmann weights at
the vertices, see also \cite{Baxter:2007}. Furthermore, it applies to
certain $1+1$-dimensional integrable quantum field theories, in which
the Yang-Baxter equation characterizes a factorization of the
scattering matrix into a succession of two-particle scattering
events. However, most applications of the QISM are restricted to such
low-dimensional theories.

The world we experience is $3+1$-dimensional. The standard model of
particle physics forms the foundation of our current physical
understanding of this world at a subatomic level. It incorporates the
electromagnetic interactions along with the weak and the strong
nuclear force. Its theoretical predictions are in impressive agreement
with experimental results. Nevertheless, most of our knowledge about
this model is based on perturbative calculations rather than exact
results. Especially in quantum chromodynamics (QCD), the theory of the
strong nuclear force, there is a strongly coupled regime that is
hardly accessible by such methods.

Can one use ideas from integrable models to gain deeper insights into
QCD? The situation is in some sense similar to that of water waves
discussed above. There are certain limiting cases of QCD that can be
mapped to integrable models. An example is provided by the scattering
of two hadrons in the multicolor limit, which is also called planar
limit for reasons that will be explained in the following section, and
Regge kinematics, i.e.\ at high energies and fixed momentum
transfer. This system can be described in terms of a spin chain that
is exactly solvable by a variant of the QISM
\cite{Lipatov:1993yb,Faddeev:1994zg}. It is closely related to the
Heisenberg model, the essential difference being the replacement of
the spin $\frac{1}{2}$ representation of $\mathfrak{su}(2)$ by an
infinite-dimensional representation of $\mathfrak{sl}(\mathbb{C}^2)$.
A readable discussion of this example of integrability in QCD may be
found in \cite{Korchemsky:1994um}, see also the extensive review
\cite{Belitsky:2004cz}.

Despite these achievements, an exact solution of the complete
Yang-Mills dynamics of QCD, even in the planar case, seems to be out
of reach, if it exists at all. In fact, to this day no
$3+1$-dimensional interacting quantum field theory has been solved
exactly. However, this unsatisfactory situation may change in the
foreseeable future as an integrable structure has began to surface in
a supersymmetric relative of QCD during the past 15 years. By now
there is overwhelming evidence that \emph{planar maximally
  supersymmetric Yang-Mills theory}, for short planar $\mathcal{N}=4$
SYM, is integrable \cite{Beisert:2010jr}. Like Newton's solution of
the two-body problem centuries ago, the integrability of this model
could prove to be a blueprint for aspects of current fundamental
physics. It might contribute to the mathematical underpinnings of
quantum field theories in general and advance our understanding of QCD
in particular. These prospects constitute our motivation for the
investigation of the quantum integrable structure of this planar
$\mathcal{N}=4$ model in the thesis at hand.

\section{Planar \texorpdfstring{$\mathcal{N}=4$}{N=4} Super Yang-Mills
  Theory}
\label{sec:spectrum}

The Lagrangian of pure Yang-Mills theory in four-dimensional Minkowski
space solely comprises bosonic gauge fields. To obtain a
supersymmetric theory, these have to be supplemented by further
fields, in particular fermionic ones. The number of supersymmetry
transformations is characterized by a positive integer
$\mathcal{N}$. It cannot be larger than four in order to be able to
avoid gravitational degrees of freedom, which makes $\mathcal{N}=4$
SYM the \emph{maximally supersymmetric Yang-Mills theory} in four
dimensions. This theory was introduced in
\cite{Brink:1976bc,Gliozzi:1976qd}, see also the reviews
\cite{Sohnius:1985qm,D'Hoker:2002aw} and some historical recollections
in \cite{Brink:2015ust}. It is essentially specified by only three
parameters: the coupling constant $g_{\text{YM}}$, the number of
colors $N_{\text{C}}$ of the gauge group $SU(N_{\text{C}})$ and the
instanton angle $\theta_{\text{I}}$. In this model the gauge bosons in
the Lagrangian are accompanied by scalar fields and fermions, which
are all $N_{\text{C}}\times N_{\text{C}}$ matrices in color space,
i.e.\ they transform in the adjoint representation of the gauge
group. The details of this field content are dictated by the
representation theory of the supersymmetry algebra. In fact, beyond
that the fields transform in a representation of the
\emph{superconformal algebra}
$\mathfrak{psu}(2,2|\mathcal{N}=4)$. This Lie superalgebra contains
the conformal algebra $\mathfrak{su}(2,2)\simeq\mathfrak{so}(4,2)$ as
well as the internal R-symmetry algebra
$\mathfrak{su}(\mathcal{N}=4)$. Let us interject here that bosonic
Yang-Mills theory is conformally invariant as a classical field
theory. However, this symmetry is broken at the quantum level.
Remarkably, this changes in the supersymmetric model. $\mathcal{N}=4$
SYM is a superconformal \emph{quantum} field theory. This suggests
that, in spite of the larger field content, the supersymmetric theory
is in effect simpler than its bosonic counterpart.

We already mentioned one motivation for the investigation of
$\mathcal{N}=4$ SYM in the previous section. Although it is not
realized in nature, it may serve as a mathematical toy model for more
realistic theories like QCD. Further impetus comes from the
\emph{AdS/CFT correspondence} proposed in 1997
\cite{Maldacena:1997re,Gubser:1998bc,Witten:1998qj}, see also the
rather recent review \cite{Polchinski:2010hw} and the references to
more comprehensive treatises therein. It conjectures the equivalence
of certain string theories on backgrounds containing an anti-de Sitter
(AdS) space and conventional quantum field theories with conformal
symmetry (CFT) that are basically defined on the boundary of that
space. The most thoroughly studied example of this correspondence
relates type IIB superstrings on the product of five-dimensional AdS
space and a five-sphere to $\mathcal{N}=4$ SYM on four-dimensional
Minkowski space. As a plausibility check, let us mention that the
isometries of the string background match the superconformal symmetry
of the quantum field theory as they are both described by the algebra
$\mathfrak{psu}(2,2|4)$. A feature which makes the correspondence
between the two different types of models particularly interesting,
and at the same time hard to prove, is its strong/weak type. Strongly
coupled regimes in $\mathcal{N}=4$ SYM are related to weakly coupled
string theory and therefore accessible by perturbative string
calculations, and vice versa. Initial evidence for the AdS/CFT
correspondence was found in the 't Hooft limit of $\mathcal{N}=4$ SYM,
where $N_{\text{C}}\to \infty$, the coupling
$g_{\text{YM}}^2N_{\text{C}}$ is kept at a fixed value and
$\theta_{\text{I}}$ is believed to be irrelevant. This is also
referred to as \emph{planar limit} because in this case only Feynman
graphs without intersections contribute in the perturbative expansion
of the SYM theory. On the string side of the correspondence this
translates into the limit of free strings. As both theories simplify
considerably in this limit, it is where most precision tests of the
AdS/CFT correspondence have been performed. In the past almost two
decades the correspondence has received the attention of many
scientists, which resulted in considerable progress. Despite these
efforts, a proof is still missing.

The simplicity of $\mathcal{N}=4$ SYM in the planar limit is closely
connected to an underlying \emph{integrable structure}, whose
emergence is detailed in the comprehensive reviews series
\cite{Beisert:2010jr}, see e.g.\ \cite{Serban:2010sr} for a more
concise presentation. These developments enabled crucial tests of the
AdS/CFT correspondence for this theory. One might even argue that a
complete understanding of the integrable structure is a prerequisite
for its proof in the planar limit. The evidence of integrability in
planar $\mathcal{N}=4$ SYM is rooted in the concepts discussed in the
previous section~\ref{sec:int-mod}. A fertile perspective is to view
integrability as an infinite-dimensional extension of the
superconformal symmetry algebra $\mathfrak{psu}(2,2|4)$ that should
determine all observables of the quantum field theory. For important
classes of observables, and often to a certain order in perturbation
theory, this point of view was worked out in detail and the
integrability is proven. In countless further cases the predictions of
assuming integrability were verified by laborious quantum field theory
calculations. These accumulated results point towards the
integrability of planar $\mathcal{N}=4$ SYM. Yet at present, the
origin of this extraordinary structure remains mostly a
mystery. Unraveling it will not only lead to the exact solution of
this one $3+1$-dimensional quantum field theory but likely go along
with new insights into integrable models in general.

The first class of observables that was related to an integrable
system are the anomalous dimensions of local gauge invariant
operators, which determine the two-point functions of these
operators. At one loop in the perturbative expansion it was proven
that they can be described by an integrable spin chain
\cite{Minahan:2002ve,Beisert:2003tq,Beisert:2003yb,Beisert:2003jj},
see also \cite{Minahan:2010js} of the review series mentioned above.
The aforesaid operators make up the states of this spin chain. They
consist of a color trace over a number of fields and derivatives
thereof at a single spacetime point. The fields in the operator
correspond to the sites of the spin chain. These sites transform in an
infinite-dimensional representation of the superconformal algebra
$\mathfrak{psu}(2,2|4)$, which replaces the spin $\frac{1}{2}$
representation of $\mathfrak{su}(2)$ from the Heisenberg spin chain of
section~\ref{sec:int-mod}. Scale transformations of planar
$\mathcal{N}=4$ SYM are generated by a dilatation operator that acts
on the spin chain states. This dilatation operator receives quantum
corrections. At one loop these can be identified with the Hamiltonian
of an integrable $\mathfrak{psu}(2,2|4)$ spin chain, which is a
straightforward generalization of the Heisenberg Hamiltonian and
contains only interactions of neighboring sites. Therefore the
spectrum of these quantum corrections, the so-called set of anomalous
dimensions, agrees with the energy spectrum of the integrable spin
chain. Consequently, a Bethe ansatz can be applied to compute the
anomalous dimensions. This method circumvents standard Feynman graph
calculations and thereby significantly simplifies the computation of
the anomalous dimensions, i.e.\ the solution of the \emph{spectral
  problem}. Lastly, let us note how this solution relates to the
perspective put forward in the previous paragraph. The spin chain of
the one-loop spectral problem is based on an infinite-dimensional
algebra, the Yangian of $\mathfrak{psu}(2,2|4)$. The role of this
Yangian was also emphasized in \cite{Dolan:2003uh,Dolan:2004ps}.

Strong evidence for the integrability of the spectral problem persists
beyond one loop, where the structure becomes much more intricate. The
range of the interactions in the dilatation operator increases with
the loop order. The explicit form of this operator beyond one loop is
only known for some subsectors of the full $\mathfrak{psu}(2,2|4)$
symmetric field content. In such subsectors it can be mapped to
long-range integrable spin chains. However, this description neglects
wrapping effects, which occur if the range of interaction equals or
exceeds to the length of the spin chain. What is more, neglecting
these effects it is even possible to formulate asymptotic all-loop
Bethe equations. They are based on a choice of a vacuum state which
reduces the $\mathfrak{psu}(2,2|4)$ symmetry to a residual
$\mathfrak{su}(2|2)\oplus\mathfrak{su}(2|2)$ algebra at one-loop
level. The all-loop result can then be obtained by encoding the
coupling constant into a central extension of this residual
algebra. These developments are summarized in
\cite{Sieg:2010jt,Rej:2010ju,Ahn:2010ka} of the review
series.\footnote{Here we merely sketched the developments on the
  $\mathcal{N}=4$ SYM side of the AdS/CFT correspondence. These were
  paralleled by the discovery of an integrable structure on the string
  side, namely at the classical level the string theory is described
  by an integrable sigma model, see once again the review series
  \cite{Beisert:2010jr}.} There are even proposals for the complete
all-loop spectrum of anomalous dimensions including wrapping. The most
elaborate one is a system of equations called ``quantum spectral
curve'' \cite{Gromov:2013pga,Gromov:2014caa}. Its predictions have
passed important tests. Notwithstanding these impressive achievements,
the all-loop dilatation operator itself, whose spectrum those
equations are believed to describe, remains unknown. Consequently,
also the infinite-dimensional symmetry algebra generalizing the
Yangian of $\mathfrak{psu}(2,2|4)$ at one loop to finite values of the
coupling has not yet been revealed.\footnote{At this point it is worth
  noting some recent work on a Yangian structure, or rather more
  generally a quantum group structure, at the level of the centrally
  extended residual symmetry algebra
  \cite{Beisert:2014hya,Beisert:2016qei}.}

If planar $\mathcal{N}=4$ SYM is integrable, this should manifest
itself not only in the spectrum of anomalous dimensions, which is
closely related to two-point functions, but also for further
observables. A natural step is to investigate the integrability of
three-point functions, see e.g.\ the recent approach
\cite{Basso:2015zoa} and the list of references therein. Another class
of observables are \emph{scattering amplitudes}, where in particular
at tree-level a Yangian symmetry has been exposed
\cite{Drummond:2009fd}. It is the very same Yangian of
$\mathfrak{psu}(2,2|4)$ that also features in the one-loop spectral
problem. This connection might shed light on a uniform integrable
structure underlying the whole of planar $\mathcal{N}=4$ SYM. For this
reason we focus on the integrability of scattering amplitudes in the
present thesis.

In the subsequent section~\ref{sec:amplitudes} we provide an
introduction to tree-level amplitudes of $\mathcal{N}=4$ SYM
emphasizing their integrable structure. The technical level is such
that it covers all the concepts and formulas needed later on. It also
enables us to formulate in detail the objectives and the outline of
this thesis in section~\ref{sec:objectives-outline}.

\section{Super Yang-Mills Scattering Amplitudes}
\label{sec:amplitudes}

In this section, we present a brief review of planar $\mathcal{N}=4$
SYM scattering amplitudes. After introducing significant fundamentals,
we focus on those results and techniques that form the basis for the
original research presented later in this thesis.  In particular, we
discuss the integrable structure of these amplitudes and recently
proposed deformations thereof, which preserve integrability. These
deformations are a step towards using powerful integrability
techniques for the efficient computation of scattering
amplitudes. Broader discussions of scattering amplitudes in gauge
theories can be found in the recent books
\cite{Henn:2014yza,Elvang:2015rqa} and e.g.\ in the concise review
article \cite{Dixon:2013uaa}. The intriguing features of amplitudes in
the planar $\mathcal{N}=4$ model are highlighted in the reviews
\cite{Roiban:2010kk,Drummond:2010km}. Except for the recently
introduced deformations, the topics covered in this section are
discussed in these texts. In addition, we provide references to the
original literature for key results and in case we find the exposition
particularly instructive.

\subsection{Color-Decomposition and Spinor Helicity Variables}
\label{sec:color-spin}

Before we can present formulas for $\mathcal{N}=4$ SYM scattering
amplitudes and discuss their properties, such as symmetries, we have
to organize the amplitudes in a way that makes these features most
accessible. This is achieved by stripping off the color structure from
the gauge theory amplitudes by means of a color decomposition. The
resulting partial amplitudes depend on the kinematics, i.e.\ the
particle momenta. They allow for a perturbative expansion and we
restrict our discussion for the most part to the tree-level
contribution. Furthermore, to obtain manageable formulas for these
tree-level partial amplitudes a clever parameterization of the momenta
is of importance. We choose to express the momenta using spinor
helicity variables. In what follows, we explain these techniques in
more detail.

Let us first discuss the \emph{color decomposition} of perturbative
gauge theory scattering amplitudes, see \cite{Bern:1990ux} and
references therein. The model of interest for us is four-dimensional
$\mathcal{N}=4$ SYM with gauge group $SU(N_{\text{C}})$. In a
scattering process in this theory each of the participating massless
particles $i=1,\ldots,N$ is associated with a momentum null vector
$p^i\in\mathbb{R}^{1,3}$, a helicity
$h^i=-1,-\frac{1}{2},0,+\frac{1}{2},+1$, and a color index
$a^i=1,\ldots, {N_{\text{C}}}^2-1$. Additional internal R-symmetry
quantum numbers of the particles are not important for the
color-decomposition and therefore suppressed here. The full gauge
theory scattering amplitude
$\mathscr{A}_N(p^1,h^1,a^1;\ldots;p^N,h^N,a^N)\equiv\mathscr{A}_N(\{p^i,h^i,a^i\})$
can be expressed in terms of \emph{partial amplitudes}
$A_N(\{p^i,h^i\})$, also called \emph{color-stripped amplitudes}, that
are independent of the color indices,
\begin{align}
  \label{eq:color-decomp}
    \mathscr{A}_N(\{p^i,h^i,a^i\})
    =
    g_{\text{YM}}^{N-2}\;\;\sum_{\mathclap{\sigma\in S_N/\mathbb{Z}_N}}\;\;
    \tr(T_{a^{\sigma(1)}}\cdots T_{a^{\sigma(N)}})
    A_N(\{p^{\sigma(i)},h^{\sigma(i)}\})+
    \parbox{2cm}{\centering multi-trace\\terms}.
\end{align}
Hence this procedure allows to separate the color structure of the
amplitude from the kinematics. Here $g_{\text{YM}}^{\phantom{N}}$
denotes the Yang-Mills coupling constant. Moreover, $T_a$ are
Hermitian traceless $N_{\text{C}}\times N_{\text{C}}$ matrices, which
are normalized such that $\tr(T_aT_b)=\delta_{ab}$. The summation in
this formula runs over all $(N-1)!$ non-cyclic permutations $\sigma$
of $N$ elements. We suppressed the explicit form of terms involving
products of multiple color traces. These terms do not contribute in
the planar limit of the theory, where $N_{\text{C}}\to \infty$ while
the 't Hooft coupling $g_{\text{YM}}^2N_{\text{C}}$ is kept constant.

The partial amplitudes have a perturbative expansion in the 't Hooft
coupling, see e.g.\ the exposition in
\cite{Bern:2005iz}. Schematically, we may write
\begin{align}
  \label{eq:color-decomp-expansion}
  A_N=A_N^{(\text{tree})}
  +\sum_{L=1}^\infty \bigg(\frac{g_{\text{YM}}^2N_{\text{C}}}{8\pi^2}\bigg)^LA_N^{(L)}\,.
\end{align}
The $L$-loop partial amplitude $A_N^{(L)}$ suffers from infrared
singularities that are commonly dealt with using dimensional
regularization. From now on we focus on the \emph{tree-level partial
  amplitude} $A_N^{(\text{tree})}$. The aforementioned review articles
\cite{Roiban:2010kk,Drummond:2010km} also cover loop amplitudes. Let
us remark that at tree-level the multi-trace terms in
\eqref{eq:color-decomp} are absent even in the non-planar theory. In
slight abuse of terminology, we sometimes refer to
$A_N^{(\text{tree})}$ simply as ``tree-level amplitude'' or even just
as ``amplitude''. As we will review shortly, already this tree-level
contribution displays some remarkable properties that are deeply
related to the integrable structure of planar $\mathcal{N}=4$ SYM.

After eliminating the color structure, we can focus on the kinematics
encoded in the partial amplitudes. For this purpose the choice of
\emph{spinor helicity variables}
\cite{Waerden:1928,Weyl:1929,DeCausmaecker:1981bg,Berends:1981uq,Kleiss:1985yh,Xu:1986xb}
for the particle momenta is most appropriate because it leads to
particularly simple expressions for the amplitudes. To introduce these
variables we use a bijection between Minkowski space
$\mathbb{R}^{1,3}$ and the space of Hermitian $2\times 2$ matrices. A
Minkowski vector $p=(p_\mu)$ is represented by the matrix
\begin{align}
  \label{eq:minkvec-hermmat}
  (p_{\alpha\dot{\beta}})=
  \begin{pmatrix}
    p_0+p_3&p_1-ip_2\\
    p_1+ip_2&p_0-p_3\\
  \end{pmatrix}\,,
\end{align}
where the indices take the values $\alpha,\dot\beta=1,2$. Using the
Minkowski inner product $p\cdot q=p_0\,q_0-\vec{p}\cdot\vec{q}$ one
verifies that $\det(p_{\alpha\dot{\beta}})=p^2$. For the scattering of
massless particles we are dealing with null momenta, $p^2=0$. Hence
the corresponding matrix is at most of rank $1$ and can thus be
expressed as
\begin{align}
  \label{eq:nullvec-spinors}
  p_{\alpha\dot{\beta}}=\lambda_\alpha\tilde{\lambda}_{\dot{\beta}}
\end{align}
with two spinors
$\lambda=(\lambda_\alpha),\tilde{\lambda}=(\tilde{\lambda}_{\dot{\beta}})\in\mathbb{C}^2$. Imposing
this matrix to be Hermitian restricts $\tilde{\lambda}$ to be a real
multiple of the complex conjugate spinor
$\overline{\lambda}$. Furthermore, \eqref{eq:nullvec-spinors} is
invariant under the rescaling $\lambda\mapsto z\lambda$ and
$\tilde\lambda\mapsto z^{-1}\tilde\lambda$ with $z\in\mathbb{C}$. This
allows us to restrict to spinors satisfying the reality condition
\begin{align}
  \label{eq:spinors-real}
  \tilde\lambda=\pm\overline{\lambda}\,.
\end{align}
The sign determines the sign of the energy $\sgn(p_0)$ of the null
momentum as $\pm 2 p_0=|\lambda_1|^2+|\lambda_2|^2$. In the field of
scattering amplitudes one often works with complexified momenta, i.e.\
independent spinors $\lambda$ and $\tilde\lambda$ that do \emph{not}
obey the reality condition \eqref{eq:spinors-real}. While we adapt to
this habit in some parts of the present introduction, the condition
will be of crucial importance later on in this thesis. Let us also
mention that the helicity $h^i$ of particle $i$ in a scattering
process can be measured by applying a differential operator in the
spinor variables,
\begin{align}
  \label{eq:spinors-hel-amp}
  \mathfrak{h}^iA_N^{(\text{tree})}(\{p^i,h^i\})=h^iA_N^{(\text{tree})}(\{p^i,h^i\})\,,
\end{align}
where
\begin{align}
  \label{eq:spinors-hel-op}
  \mathfrak{h}^i=
  \frac{1}{2}\Bigg(-\sum_{\alpha=1}^2\lambda_\alpha^i\partial_{\lambda_{\alpha}^i}
  +\sum_{\dot\alpha=1}^2\tilde{\lambda}_{\dot\alpha}^i\partial_{\tilde{\lambda}_{\dot\alpha}^i}\Bigg)\,.
\end{align}

The partial amplitudes $A_N^{(\text{tree})}$ are conveniently
expressed in terms of the \emph{angle} and \emph{square spinor
  brackets}
\begin{align}
  \label{eq:spinors-brackets}
  \langle i j\rangle=\lambda^i_1\lambda^j_2-\lambda^i_2\lambda^j_1\,,\quad
  [i j]=-\tilde\lambda^i_1\tilde\lambda^j_2+\tilde\lambda^i_2\tilde\lambda^j_1\,,
\end{align}
respectively. Up to a phase, these can be thought of as square roots
of the Mandelstam variable
\begin{align}
  \label{eq:spinors-innerprod}
  s_{ij}=(p^i+p^j)^2= \langle ij\rangle[ji]
\end{align}
for null momenta $p^i,p^j$. It is important to manipulate these
brackets efficiently. Therefore we state some of their
properties. Making use of \eqref{eq:spinors-real}, one obtains
\begin{align}
  \label{eq:spinors-brackets-conj}
  [ij]
  =
  -\sgn(p_0^i)\sgn(p_0^j)
  \overline{\langle ij\rangle}\,.
\end{align}
From \eqref{eq:spinors-brackets} the antisymmetry is immediate,
\begin{align}
  \label{eq:spinors-antisym}
  \langle ij\rangle=-\langle ji\rangle\,,\quad
  [ij]=-[ji]\,.
\end{align}
Furthermore, we have the \emph{Schouten identity}
\begin{align}
  \label{eq:spinors-schouten}
  \langle ij\rangle \langle kl\rangle
  -
  \langle ik\rangle\langle jl\rangle
  =
  \langle il\rangle\langle kj\rangle\,,
\end{align}
which also holds for square brackets. In the considered scattering
processes of $N$ massless particles with momenta $p^i$ the total
momentum $P$ is conserved. This condition,
\begin{align}
  \label{eq:spinors-momcons}
  P=\sum_{i=1}^Np^i=0\,,
  \quad
  \text{reads}
  \quad
  \sum_{i=1}^N\langle ki\rangle[il]=0\,
\end{align}
for all $k,l=1,\ldots,N$ when expressed in terms of spinor
brackets. We stress that for momentum conservation to hold both signs
in \eqref{eq:spinors-real} are needed, i.e. there have to be particles
with positive and negative energy.

Besides the spinor helicity variables discussed here, also twistors
\cite{Penrose:1967wn} or rather supertwistors are well suited for the
study of scattering amplitudes in $\mathcal{N}=4$ SYM. They have been
intensively investigated after featuring prominently in
\cite{Witten:2003nn}. Let us also mention the momentum twistor
variables introduced in \cite{Hodges:2009hk}. In this work we keep
hold of the spinor helicity variables mainly for two reasons. First,
it is easy to work with real momenta by imposing
\eqref{eq:spinors-real}. What is more, while these variables are
associated with the conformal algebra $\mathfrak{su}(2,2)$ of
four-dimensional Minkowski space, they naturally generalize to certain
oscillator representations of superalgebras
$\mathfrak{su}(p,q|m)$. These representations will play an important
role in this thesis.

\subsection{Gluon Amplitudes}
\label{sec:gluon-amp}

At this point everything is set up to present expressions for
tree-level partial amplitudes $A_N^{(\text{tree})}$. While in
$\mathcal{N}=4$ SYM each particle can have a helicity
$h^i=-1,-\frac{1}{2},0,+\frac{1}{2},+1$, let us for the moment
restrict to amplitudes with $h^i=\pm 1$, i.e.\ scattering of positive
and negative helicity gluons, which we denote by $g_{\pm}^i$. These
tree-level \emph{gluon amplitudes} of the $\mathcal{N}=4$ model are
identical to those in QCD. Hence they are even of phenomenological
importance and were studied intensively already in the 1980s. However,
the textbook approach to the computation of scattering amplitudes via
Feynman diagrams quickly reaches its limit. The number of diagrams
contributing grows very fast with the number of gluons $N$, see e.g.\
the discussion in \cite{Kleiss:1988ne}. Moreover, the computation of
individual Feynman diagrams completely obscures a remarkable
simplicity of the expressions for the complete amplitudes. Therefore
alternative techniques have been developed. Rather than explaining
these approaches in detail, in this section we merely state their
output: very handy expressions for amplitudes that are useful
numerically as well as analytically.

\begin{figure}[!t]
  \begin{center}
    \begin{tikzpicture}[scale=1.1]
      \node[align=center] (10) at (-1,1) {$A^{(\text{tree})}_{1,0}$};
      \node[align=center] (11) at (1,1) {$A^{(\text{tree})}_{1,1}$};
      \node[align=center] (20) at (-2,0) {$A^{(\text{tree})}_{2,0}$};
      \node[align=center] (21) at (0,0) {$A^{(\text{tree})}_{2,1}$};
      \node[align=center] (22) at (2,0) {$A^{(\text{tree})}_{2,2}$};
      \node[align=center] (30) at (-3,-1) {$A^{(\text{tree})}_{3,0}$};
      \node[align=center] (31) at (-1,-1) {$A^{(\text{tree})}_{3,1}$};
      \node[align=center] (32) at (1,-1) {$A^{(\text{tree})}_{3,2}$};
      \node[align=center] (33) at (3,-1) {$A^{(\text{tree})}_{3,3}$};
      \node[align=center] (40) at (-4,-2) {$A^{(\text{tree})}_{4,0}$};
      \node[align=center] (41) at (-2,-2) {$A^{(\text{tree})}_{4,1}$};
      \node[align=center] (42) at (0,-2) {$A^{(\text{tree})}_{4,2}$};
      \node[align=center] (43) at (2,-2) {$A^{(\text{tree})}_{4,3}$};
      \node[align=center] (44) at (4,-2) {$A^{(\text{tree})}_{4,4}$};
      \node[align=center] (50) at (-5,-3) {$A^{(\text{tree})}_{5,0}$};
      \node[align=center] (51) at (-3,-3) {$A^{(\text{tree})}_{5,1}$};
      \node[align=center] (52) at (-1,-3) {$A^{(\text{tree})}_{5,2}$};
      \node[align=center] (53) at (1,-3) {$A^{(\text{tree})}_{5,3}$};
      \node[align=center] (54) at (3,-3) {$A^{(\text{tree})}_{5,4}$};
      \node[align=center] (55) at (5,-3) {$A^{(\text{tree})}_{5,5}$};
      \node[align=center] (61) at (-4,-4) {$A^{(\text{tree})}_{6,1}$};
      \node[align=center] (62) at (-2,-4) {$A^{(\text{tree})}_{6,2}$};
      \node[align=center] (63) at (0,-4) {$A^{(\text{tree})}_{6,3}$};
      \node[align=center] (64) at (2,-4) {$A^{(\text{tree})}_{6,4}$};
      \node[align=center] (65) at (4,-4) {$A^{(\text{tree})}_{6,5}$};
      \node[align=center] (71) at (-5,-5) {$A^{(\text{tree})}_{7,1}$};
      \node[align=center] (72) at (-3,-5) {$A^{(\text{tree})}_{7,2}$};
      \node[align=center] (73) at (-1,-5) {$A^{(\text{tree})}_{7,3}$};
      \node[align=center] (74) at (1,-5) {$A^{(\text{tree})}_{7,4}$};
      \node[align=center] (75) at (3,-5) {$A^{(\text{tree})}_{7,5}$};
      \node[align=center] (76) at (5,-5) {$A^{(\text{tree})}_{7,6}$};
      \node[align=center] (MHV) at (-4,-6) {$\text{MHV}$};
      \node[align=center] (NMHV) at (-2,-6) {$\text{NMHV}$};
      \node[align=center] (MHVbar) at (2,-6) {$\overline{\text{NMHV}}$};
      \node[align=center] (NMHVbar) at (4,-6) {$\overline{\text{MHV}}$};
      \draw[densely dotted] (1,2) -- (6,-3);
      \draw[densely dotted] (1,0) -- (6,-5);\draw (0,1) -- (1,0);\draw[densely dotted] (0,1) -- (-1,2);
      \draw (-1,0) -- (5,-6);\draw[densely dotted] (-1,0) -- (-2,1);
      \draw (0,-3) -- (4,-7);\draw[densely dashed] (0,-3) -- (-1,-2);\draw[densely dotted] (-3,0) -- (-1,-2);
      \draw (0,-5) -- (2,-7);\draw[densely dashed] (0,-5) -- (-2,-3);\draw[densely dotted] (-2,-3) -- (-4,-1);
      \draw[densely dashed] (-1,-6) -- (-3,-4);\draw[densely dotted] (-3,-4) -- (-5,-2);
      \draw[densely dashed] (-3,-6) -- (-4,-5);\draw[densely dotted] (-4,-5) -- (-5,-4);
      \draw[densely dotted] (-6,-3) -- (-1,2);
      \draw[densely dotted] (-6,-5) -- (-1,0);\draw (-1,0) -- (0,1);\draw[densely dotted] (0,1) -- (1,2);
      \draw (-5,-6) -- (1,0);\draw[densely dotted] (1,0) -- (2,1);
      \draw (-4,-7) -- (0,-3);\draw[densely dashed] (0,-3) -- (1,-2);\draw[densely dotted] (1,-2) -- (3,0);
      \draw (-2,-7) -- (0,-5);\draw[densely dashed] (0,-5) -- (2,-3);\draw[densely dotted] (2,-3) -- (4,-1);
      \draw[densely dashed] (1,-6) -- (3,-4);\draw[densely dotted] (3,-4) -- (5,-2);
      \draw[densely dashed] (3,-6) -- (4,-5);\draw[densely dotted] (4,-5) -- (5,-4);
      \draw[-latex] (-5.75,3) -- (5.75,3) node[right] {$H$};
      \draw (0,3.1) -- (0,2.9) node[below] {$0$};
      \draw (-1,3.1) -- (-1,2.9) node[below] {$1$};
      \draw (-2,3.1) -- (-2,2.9) node[below] {$2$};
      \draw (-3,3.1) -- (-3,2.9) node[below] {$3$};
      \draw (-4,3.1) -- (-4,2.9) node[below] {$4$};
      \draw (-5,3.1) -- (-5,2.9) node[below] {$5$};
      \draw (1,3.1) -- (1,2.9) node[below] {$-1$};
      \draw (2,3.1) -- (2,2.9) node[below] {$-2$};
      \draw (3,3.1) -- (3,2.9) node[below] {$-3$};
      \draw (4,3.1) -- (4,2.9) node[below] {$-4$};
      \draw (5,3.1) -- (5,2.9) node[below] {$-5$};
      \draw[-latex] (-5.75,3) -- +(-90:1.41) node[below] {$N$};
      \draw[-latex] (-5.75,3) -- +(-45:1.41) node[below] {$K$};
    \end{tikzpicture}
    \caption{Tree-level partial amplitudes $A_{N,K}^{(\text{tree})}$
      with $N$ gluons out of which $K$ have negative helicity and with
      total helicity $H=\sum_ih^i=N-2K$ are classified by the degree
      of helicity violation, the so-called $\text{MHV}$ degree. This
      leads to two series of amplitudes: $\text{N}^{K-2}\text{MHV}$
      for $H\geq 0$ and $\overline{\text{N}^{N-K-2}\text{MHV}}$ for
      $H\leq 0$. Amplitudes enclosed by dotted lines vanish. The
      amplitude $A_{2,1}^{(\text{tree})}$ can be thought of as free
      propagation of a gluon. The classification carries over to
      superamplitudes $\mathcal{A}_{N,K}^{(\text{tree})}$ discussed in
      section~\ref{sec:superamplitudes}.}
    \label{fig:mhv-classification}
  \end{center}
\end{figure}
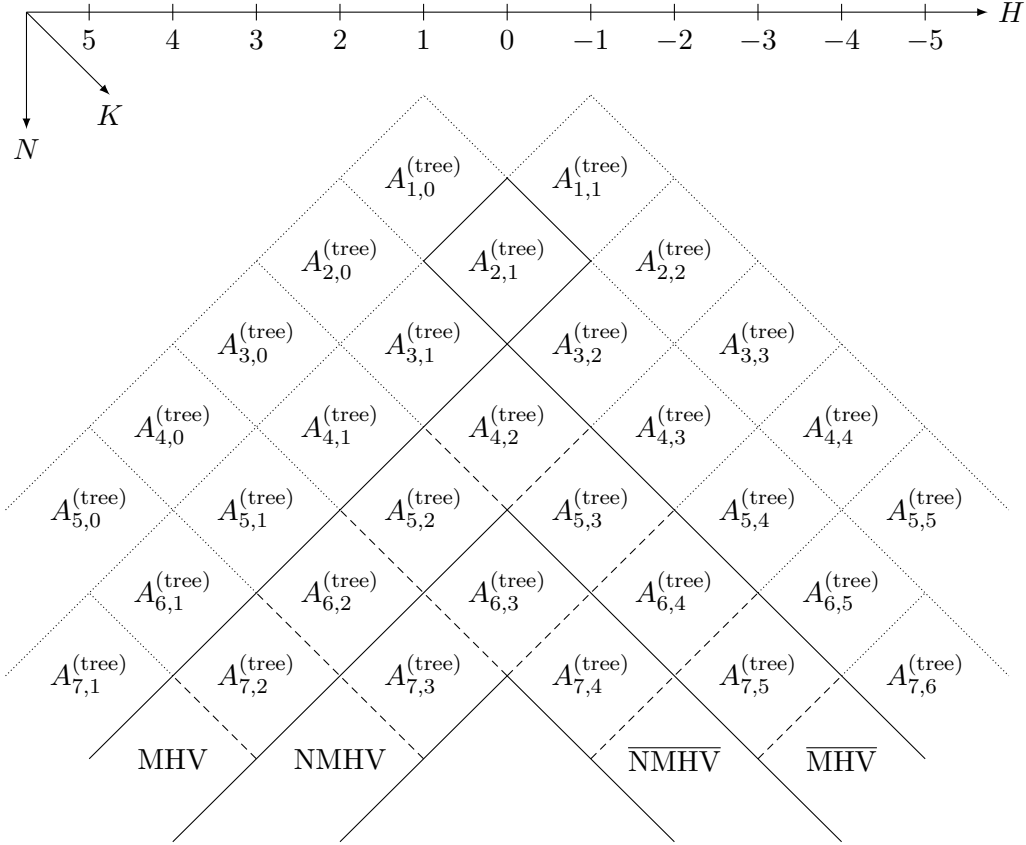

The total momentum in a scattering event is conserved,
$P=\sum_ip^i=0$. However, the total helicity $H=\sum_ih^i$ is not. If
$K$ is the number of negative helicity gluons, then $H=N-2K$.  It
turns out to be helpful to classify the gluon amplitudes by the degree
of helicity violation, see also
figure~\ref{fig:mhv-classification}. Therefore we denote them by
$A_{N,K}^{(\text{tree})}(\{p^i,\pm 1\})\equiv
A_{N,K}^{(\text{tree})}(\{g^i_\pm\})$,
where we added $K$ as a subscript. One finds that amplitudes with no
or one gluon of negative helicity vanish,
\begin{align}
  \label{eq:amp-bos-trivial}
  \begin{aligned}
    A_{N,0}^{(\text{tree})}(g^1_+,\ldots,g^N_+)&=0\,,\\
    A_{N,1}^{(\text{tree})}(g^1_+,\ldots,g^i_-,\ldots,g^N_+)&=0\,\quad\text{for}\quad N\geq 3\,.\\
  \end{aligned}
\end{align}
The same holds true for amplitudes with no or one positive helicity
gluon. The first non-trivial amplitudes are those with two negative
helicity gluons. Hence these are called \emph{maximally helicity
  violating} ($\text{MHV}$) \emph{amplitudes}. They are given by the
exceedingly simple formula\footnote{We do not keep track of overall
  numerical prefactors of amplitudes in this introductory chapter
  because we are primarily interested in symmetries that are
  determined by the functional dependence on the momenta, see
  section~\ref{sec:symmetries}.}
\begin{align}
  \label{eq:amp-bos-mhv}
  A_{N,2}^{(\text{tree})}(g_+^1,\ldots,g_-^i,\ldots,g_-^j,\ldots,g_+^N)
  =
  \frac{\delta^{4}(P)\langle ij\rangle^4}
  {\langle 12\rangle\langle 23\rangle\cdots\langle N-1\,N\rangle\langle N1\rangle}\,,
\end{align}
where $N\geq 4$ and the momentum conservation is implemented
by\footnote{Here we included the numerical prefactor $\frac{1}{2}$ in
  order to conform with our conventions in later chapters.}
\begin{align}
  \label{eq:amp-bos-delta}
  \delta^4(P)
  =
  \frac{1}{2}\prod_{\mu=0}^3\delta(P_\mu)
  =
  \delta(P_{11})\delta(P_{22})
  \delta(\Real P_{21})\delta(\Imag P_{21})
  \quad\text{with}\quad
  P_{\alpha\dot{\beta}}=\sum_{i=1}^N\lambda_\alpha^i\tilde{\lambda}_{\dot{\beta}}^i\,.
\end{align}
Amplitudes with exactly two gluons of positive helicity, so-called
$\overline{\text{MHV}}$ amplitudes, are given by the complex conjugate
of \eqref{eq:amp-bos-mhv}. The \emph{Parke-Taylor formula}
\eqref{eq:amp-bos-mhv} was conjectured in \cite{Parke:1986gb} and
proven in \cite{Berends:1987me} by recursion and employing gluons that
are off the mass shell. It is worth highlighting the simplicity of
this formula by bringing the earlier work \cite{Parke:1985ax} of Parke
and Taylor to our attention. This article was submitted only a few
months before \cite{Parke:1986gb} and contains the result of a clever
Feynman diagrammatic calculation for the $\text{MHV}$ amplitude with
$N=6$. However, in stark contrast to the single term in
\eqref{eq:amp-bos-mhv}, the result spreads over multiple
pages. Nevertheless, this result was considered useful for numerical
evaluation and therefore an experimentalist's delight. To quote
\cite{Parke:1985ax}: ``Furthermore, we hope to obtain a simple
analytic form for the answer, making our result not only an
experimentalist's, but also a theorist's delight.'' Indeed, this was
achieved with \eqref{eq:amp-bos-mhv}. By hindsight, one might be
tempted to attribute the simplicity of \eqref{eq:amp-bos-mhv} to some
underlying hidden symmetry like an integrable structure. Historically
however, the connection between amplitudes and integrability was
established differently, as we will discuss in
section~\ref{sec:symmetries}.

We continue the classification of amplitudes beyond the $\text{MHV}$
case. Amplitudes with $N\geq 6$ gluons out of which $K=3$ have
negative helicity are called \emph{Next-to-}$\text{MHV}$ ($\text{NMHV}$). An
example with a particular distribution of the negative helicity gluons
is
\begin{align}
  \label{eq:amp-bos-nmhv}
  \begin{aligned}
    &A_{6,3}^{(\text{tree})}(g^1_+,g^2_+,g^3_+,g^4_-,g^5_-,g^6_-)\\
    &\quad\quad\quad=
    \frac{\delta^{4}(P)}{[5|1+6|2\rangle}
    \Bigg(
    \frac{\langle 6|1+2|3]^3}
    {\langle 61\rangle\langle 12\rangle[34][45]s_{126}}
    +
    \frac{\langle 4|5+6|1]^3}
    {\langle 23\rangle\langle 34\rangle[16][65]s_{156}}
    \Bigg)\,.
  \end{aligned}
\end{align}
Here we used the shorthand notation $[5|1+6|2\rangle= [51]\langle 1
2\rangle+[56]\langle 6 2\rangle$ etc.\ and we defined the variable
\begin{align}
  \label{eq:amp-bos-mandel3}
  s_{ijk}
  =
  (p^i+p^j+p^k)^2
  =
  \langle ij\rangle[ji]+
  \langle ik\rangle[ki]+
  \langle jk\rangle[kj]
\end{align}
generalizing \eqref{eq:spinors-innerprod}. The two-term expression in
\eqref{eq:amp-bos-nmhv} for this amplitude was obtained in
\cite{Roiban:2004ix} as a limit of a seven-gluon amplitude. A slightly
more complicated expression containing three terms is known much longer
\cite{Mangano:1987xk}. 

To complete the classification, one refers to amplitudes with $N$
gluons out of which $K$ have negative helicity and which have total
helicity $H=N-2K\geq 0$ as \emph{Next-to-}$(K-2)$\emph{-}$\text{MHV}$
or, for short $\text{N}^{K-2}\text{MHV}$. For $H\leq 0$ these
amplitudes belong to the $\overline{\text{N}^{N-K-2}\text{MHV}}$
series. This classification of amplitudes is summarized in figure
\ref{fig:mhv-classification}. The complexity of the expressions for
the amplitudes tends to increase with an decreasing degree of helicity
violation. This is illustrated by comparing the single-term formula
\eqref{eq:amp-bos-mhv} for the $\text{MHV}$ amplitudes with the
two-term expression for the $\text{NMHV}$ amplitude in
\eqref{eq:amp-bos-nmhv}. Let us already mention that helicity
conserving amplitudes with $H=0$ will be of special interest later in
this thesis.

So far we in essence just \emph{presented} formulas for some sample
amplitudes. Let us also briefly mention the
\emph{Britto-Cachazo-Feng-Witten} (BCFW) \emph{on-shell recursion
  relations} \cite{Britto:2004ap,Britto:2005fq}, that can be used to
\emph{construct} tree-level amplitudes. This method makes use of the
analytic properties of the amplitudes as functions of the complexified
particle momenta. An amplitude is constructed iteratively from
multiple amplitudes with less particles. In particular, with the BCFW
recursion one recovers the sample amplitudes \eqref{eq:amp-bos-mhv}
and \eqref{eq:amp-bos-nmhv} presented in this section.

\subsection{Superamplitudes}
\label{sec:superamplitudes}

After concentrating on gluon scattering in the previous section, we
extend this discussion to the complete particle content of
$\mathcal{N}=4$ SYM. It can be deduced from the representation theory
of the superconformal algebra $\mathfrak{psu}(2,2|4)$ and is
summarized by:
\begin{align}
  \label{eq:amp-super-particles}
  \begin{tabular}{l l}
    particles&$\phantom{+}h$\\
    \hline
    1 positive helicity gluon $g_+$&$+1$\\
    4 gluinos $\tilde{g}_{\dot{a}}$&$+\frac{1}{2}$\\
    6 scalars $\varphi_{\dot{a}\dot{b}}=-\varphi_{\dot{b}\dot{a}}$&$\phantom{-}0$\\
    4 antigluinos $\bar{\tilde{g}}_{\dot{a}}$&$-\frac{1}{2}$\\
    1 negative helicity gluon $g_-$&$-1$\\
  \end{tabular}
\end{align}
Here the indices\footnote{A reader familiar with the field of
  $\mathcal{N}=4$ SYM scattering amplitudes would probably expect
  undotted indices $a,b$ here. We use $\dot{a},\dot{b}$ in order to
  comply with conventions that are natural in later chapters of this
  thesis. Throughout the present introductory chapter this is the main
  principle for selecting our notation. In particular, we also place
  the indices of $\bar{\tilde{g}}_{\dot{a}}$ and $\eta_{\dot{a}}$, see
  \eqref{eq:amp-superfield} below, downstairs for this reason.}
$\dot{a},\dot{b}=1,\ldots,4$ are associated with the internal
$\mathfrak{su}(4)$ R-symmetry of the model. Furthermore, we display
the helicity $h$ of each particle because as in the previous section
it is important to classify the amplitudes. Instead of discussing the
scattering amplitudes of the particles listed in
\eqref{eq:amp-super-particles} individually, it is convenient to
package them into \emph{superamplitudes}. For this the particles are
organized into a superfield \cite{Nair:1988bq},
\begin{align}
  \label{eq:amp-superfield}
  \begin{aligned}
    \Phi
    &=
    g_+
    +\sum_{\dot{a}}\eta_{\dot{a}}\,\tilde{g}_{\dot{a}}
    +\frac{1}{2!}\sum_{\dot{a},\dot{b}}\eta_{\dot{a}}\eta_{\dot{b}}\,\varphi_{\dot{a}\dot{b}}\\
    &\quad\quad\quad+\frac{1}{3!}\sum_{\dot{a},\dot{b},\dot{c},\dot{d}}\eta_{\dot{a}}\eta_{\dot{b}}\eta_{\dot{c}}\,\epsilon_{\dot{a}\dot{b}\dot{c}\dot{d}}\,\bar{\tilde{g}}_{\dot{d}}
    +\frac{1}{4!}\sum_{\dot{a},\dot{b},\dot{c},\dot{d}}\eta_{\dot{a}}\eta_{\dot{b}}\eta_{\dot{c}}\eta_{\dot{d}}\,\epsilon_{\dot{a}\dot{b}\dot{c}\dot{d}}\,g_-\,,
  \end{aligned}
\end{align}
where the completely antisymmetric symbol is fixed by
$\epsilon_{1234}=1$ and we introduced the Graßmann variables
$\eta_{\dot{a}}$ obeying
$\eta_{\dot{a}}\eta_{\dot{b}}=-\eta_{\dot{b}}\eta_{\dot{a}}$.
Extending the definition of the helicity operator in
\eqref{eq:spinors-hel-op}, we define a ``superhelicity'' operator that
has the eigenvalue $1$ when acting on the superfield
\eqref{eq:amp-superfield},
\begin{align}
  \label{eq:amp-super-hel}
  \frac{1}{2}\Bigg(-\sum_{\alpha=1}^2\lambda_\alpha\partial_{\lambda_{\alpha}}
  +\sum_{\dot\alpha=1}^2\tilde{\lambda}_{\dot\alpha}\partial_{\tilde{\lambda}_{\dot\alpha}}
  +\sum_{\dot{a}=1}^4\eta_{\dot{a}}\partial_{\eta_{\dot{a}}}\Bigg)\Phi=1\,\Phi\,.
\end{align}
Superamplitudes $\mathcal{A}_N^{(\text{tree})}$ can be understood as
scattering amplitudes of $N$ superfields $\Phi^i$. The individual
particle scattering amplitudes $A_N^{(\text{tree})}$ are then be
extracted as coefficients of an expansion in the Graßmann parameters
following \eqref{eq:amp-superfield}. The classification of amplitudes
in terms of helicity violation carries over to the superamplitudes. To
see this, we expand the superamplitude as
\begin{align}
  \label{eq:amp-super-expand}
  \begin{aligned}
  \mathcal{A}_N^{(\text{tree})}=
  \tikz[baseline]{
    \node[anchor=base] (amp) {$\mathcal{A}_{N,2}^{(\text{tree})}$};
    \node (label) [below=0.5cm of amp] {$\text{MHV}$};
    \draw[latex-] (amp) -- (label);
  }
  +
  \tikz[baseline]{
    \node[anchor=base] (amp) {$\mathcal{A}_{N,3}^{(\text{tree})}$};
    \node (label) [below=0.5cm of amp] {$\text{NMHV}$};
    \draw[latex-] (amp) -- (label);
  }
  +
  \tikz[baseline]{
    \node[anchor=base] (amp) {$\mathcal{A}_{N,4}^{(\text{tree})}$};
    \node (label) [below=0.5cm of amp] {$\text{N}^2\text{MHV}$};
    \draw[latex-] (amp) -- (label);
  }
  +\ldots+
  \tikz[baseline]{
    \node[anchor=base] (amp) {$\mathcal{A}_{N,N-3}^{(\text{tree})}$};
    \node (label) [below=0.5cm of amp] {$\overline{\text{NMHV}}$};
    \draw[latex-] (amp) -- (label);
  }
  +
  \tikz[baseline]{
    \node[anchor=base] (amp) {$\mathcal{A}_{N,N-2}^{(\text{tree})}$};
    \node (label) [below=0.5cm of amp] {$\overline{\text{MHV}}$};
    \draw[latex-] (amp) -- (label);
  }
  \,,\\
  \end{aligned}
\end{align}
where $\mathcal{A}_{N,K}^{(\text{tree})}$ contains products of $4 K$
Graßmann variables $\eta_{\dot a}^i$. Therefore the helicity
$H=\sum_ih^i=N-2K$ of all particle amplitudes packaged in the
superamplitude $\mathcal{A}_{N,K}^{(\text{tree})}$ is identical. In
particular, one expansion coefficient of
$\mathcal{A}_{N,K}^{(\text{tree})}$ is identical to the gluon
amplitude $A_{N,K}^{(\text{tree})}$ with helicity $H$. Hence the
classification of gluon amplitudes in terms of the degree of helicity
violation can be extended to the superamplitudes as indicated in
\eqref{eq:amp-super-expand}. See once again
figure~\ref{fig:mhv-classification}.

After setting up the formalism we can discuss actual
superamplitudes. We confine ourselves to the supersymmetric
generalizations of the $\text{MHV}$ gluon amplitudes
$A_{N,2}^{(\text{tree})}$ and the $\text{NMHV}$ amplitude
$A_{6,3}^{(\text{tree})}$, which also served as illustrative examples
in the previous section. The MHV superamplitude is \cite{Nair:1988bq}
\begin{align}
  \label{eq:amp-super-mhv}
  \mathcal{A}_{N,2}^{(\text{tree})}
  =  
  \frac{\delta^{4|0}(P)\delta^{0|8}(Q)}
  {\langle 12\rangle\langle 23\rangle\cdots\langle N-1\,N\rangle\langle N1\rangle}\,
\end{align}
for $N\geq 4$. The momentum conserving delta function
$\delta^{4|0}(P)\equiv\delta^{4}(P)$ is given by
\eqref{eq:amp-bos-delta}, where the notation here stresses that it is
purely bosonic. The other delta function implements what is called
supermomentum conservation,
\begin{align}
  \label{eq:amp-superdelta}
  \delta^{0|8}(Q)=
  \prod_{\alpha=1}^2\prod_{\dot{a}=1}^4Q_{\alpha\dot{a}}
  \quad\text{with}\quad
  Q_{\alpha\dot{a}}=\sum_{i=1}^Nq^i_{\alpha\dot a}
  \quad\text{and}\quad
  q^i_{\alpha\dot a}=
  \lambda^i_\alpha\eta^i_{\dot{a}}\,,
\end{align}
where in the product of anticommuting factors those with smaller
indices appear left. Let us discuss in more detail how to extract
particle amplitudes from this quantity. For this purpose we display
some parts of the Graßmann expansion of \eqref{eq:amp-super-mhv}
explicitly,
\begin{align}
  \label{eq:amp-recover-mhv}
  \begin{aligned}
    \mathcal{A}_{N,2}^{(\text{tree})}
    &={}\phantom{+}\ldots+
    (\eta^i_1\eta^i_2\eta^i_3\eta^i_4)(\eta^j_1\eta^j_2\eta^j_3\eta^j_4)\\
    &\quad\quad\quad\quad\quad\cdot 
    A_{N,2}^{(\text{tree})}(g_+^1,\ldots,g_-^i,\ldots,g_-^j,\ldots,g_+^N)
    \\
    &\phantom{{}={}}+\ldots+
    (\eta^k_1\eta^k_2\eta^k_3\eta^k_4)(\eta_2^l)(-\eta_2^m\eta_3^m\eta_4^m)\\
    &\quad\quad\quad\quad\quad\cdot 
    A_{N}^{(\text{tree})}(g_+^1,\ldots,g_-^k,\ldots,\tilde{g}_{2}^l,\ldots,\bar{\tilde{g}}_{1}^m,\ldots,g_+^N)\\
    &\phantom{{}={}}+\ldots\,.
  \end{aligned}
\end{align}
Comparing with the Graßmann expansion of the superfield
\eqref{eq:amp-superfield}, we identify the $\text{MHV}$ gluon
amplitude \eqref{eq:amp-bos-mhv}. As an illustration we displayed a
further term. According to \eqref{eq:amp-superfield} this has to be an
amplitude involving gluons $g_{\pm}$ as well as a gluino $\tilde{g}_2$
and an antigluino $\bar{\tilde{g}}_1$. Adding up the helicities of
these particles we obtain the total helicity $H=\sum_ih^i=N-4$, which
is the same as that of the $\text{MHV}$ gluon amplitude. Hence these
amplitudes are contained in the same superamplitude
$\mathcal{A}_{N,2}^{(\text{tree})}$.

We move on to the six-particle $\text{NMHV}$ superamplitude
\cite{Drummond:2008vq,Drummond:2008bq}
\begin{align}
  \label{eq:amp-super-nmhv6}
  \mathcal{A}^{(\text{tree})}_{6,3}
  =
  \frac{\delta^{4|0}(P)\delta^{0|8}(Q)}
  {\langle 12\rangle\langle 23\rangle\langle 34\rangle\langle 45\rangle\langle 56\rangle\langle 61\rangle}
  \big(\mathcal{R}^{1;46}+\mathcal{R}^{1;35}+\mathcal{R}^{1;36}\big)\,.
\end{align}
In order to define the quantities $\mathcal{R}^{r;st}$ in this
formula, we introduce so-called \emph{dual variables}
$x^i_{\alpha\dot{\beta}}$ and $\theta^i_{\alpha\dot{b}}$ by the
relations
\begin{align}
  \label{eq:amp-super-dualvars}
  p_{\alpha\dot\beta}^i=\lambda^i_\alpha\tilde{\lambda}^i_{\dot{\beta}}=x^i_{\alpha\dot{\beta}}-x^{i+1}_{\alpha\dot{\beta}}\,,\quad
  q_{\alpha\dot a}^i=\lambda^i_\alpha\eta^i_{\dot{a}}=\theta^i_{\alpha\dot{a}}-\theta^{i+1}_{\alpha\dot{a}}\,,
\end{align}
where $x^{N+1}=x^1$ and $\theta^{N+1}=\theta^1$. Furthermore, we
define the abbreviations $x^{ij}=x^i-x^j$ and
$\theta^{ij}=\theta^i-\theta^j$. Then
\begin{align}
  \label{eq:amp-super-r}
  \mathcal{R}^{r;st}=
  \frac{
    \langle s\,s-1\rangle\langle t\,t-1\rangle
    \delta^{0|4}\big(\langle r|x^{rs}x^{st}|\theta^{tr}\rangle+\langle r|x^{rt}x^{ts}|\theta^{sr}\rangle\big)
  }
  {
    (x^{st})^2
    \langle r|x^{rs}x^{st}|t\rangle
    \langle r|x^{rs}x^{st}|t-1\rangle
    \langle r|x^{rt}x^{ts}|s\rangle
    \langle r|x^{rt}x^{ts}|s-1\rangle
  }\,,
\end{align}
where the fermionic delta function is defined analogous to the one in
\eqref{eq:amp-superdelta}. The brackets occurring in this expression
can be reduced to ordinary spinor brackets, which we illustrate for
one example:
$\langle 6|x^{64}x^{42}|\theta^{26}\rangle=\langle
6\big(-4\rangle[4-5\rangle[5\big)\big(-2]\langle 2-3]\langle
3\big)\big(-1\rangle \eta^1-2\rangle \eta^2\big)$.
The gluon $\text{NMHV}$ amplitude \eqref{eq:amp-bos-nmhv} can be
recovered from \eqref{eq:amp-super-nmhv6} by means of an expansion in
the Graßmann parameters. We picked a certain order of gluon helicities
in \eqref{eq:amp-bos-nmhv} to obtain a particularly simple
formula. The superamplitude in \eqref{eq:amp-super-nmhv6} contains the
gluon amplitudes for any order of helicities. Let us also comment on
the dual variables $x^i$ and $\theta^i$ introduced in
\eqref{eq:amp-super-dualvars}. As we will discuss in the subsequent
section~\ref{sec:symmetries}, amplitudes have a superconformal
symmetry as functions of the momenta $p^i$ and the supermomenta
$q^i$. Remarkably, they exhibit a second so-called \emph{dual
  superconformal symmetry} in the variables $x^i$ and $\theta^i$. The
quantities $\mathcal{R}^{r;st}$ are superconformal as well as dual
superconformal invariants and they appear naturally as residues of
certain integrals that we will introduce in
section~\ref{sec:grassmannian-integral}.

All amplitudes $\mathcal{A}_{N,K}^{(\text{tree})}$ in $\mathcal{N}=4$
SYM were determined in \cite{Drummond:2008cr} by solving a
supersymmetric generalization of the BCFW recursion relations
\cite{Brandhuber:2008pf,ArkaniHamed:2008gz,Elvang:2008na}. As in the
bosonic version, these relations make use of complexified particle
momenta. The starting point of the recursion are the three-particle
scattering amplitudes\footnote{The amplitude
  $\mathcal{A}^{(\text{tree})}_{3,2}$ can be obtained from the formula
  \eqref{eq:amp-super-mhv} for $\text{MHV}$ superamplitudes for
  $N=3$.}
\begin{align}
  \label{eq:amp-3}
  \begin{aligned}
    \mathcal{A}^{(\text{tree})}_{3,2}
    &=  
    \frac{
      \delta^{4|0}(P)\delta^{0|8}(Q)
    }
    {
      \langle 12\rangle
      \langle 23\rangle
      \langle 31\rangle
    }\,,\\
    \mathcal{A}^{(\text{tree})}_{3,1}
    &=
    \frac{
      \delta^{4|0}(P)\delta^{0|4}([23]\eta^1+[31]\eta^2+[12]\eta^3)
    }
    {
      [23]
      [31]
      [12]
    }\,
  \end{aligned}
\end{align}
from which higher point amplitudes are constructed. For three real
null vectors, momentum conservation immediately implies that all
spinor brackets $\langle ij\rangle$ and $[ij]$ vanish. Hence, the
three-particle amplitudes are only meaningful in a complexified
setting where the spinors $\lambda$ and
$\tilde\lambda$ are treated as independent complex
variables not obeying \eqref{eq:spinors-real}. We present these
amplitudes at this point because they will reappear prominently in
sections~\ref{sec:grassmannian-integral} and \ref{sec:deform}.

\subsection{Symmetries and Integrability}
\label{sec:symmetries}

After introducing the tree-level partial gluon amplitudes
$A_{N,K}^{(\text{tree})}$ and their supersymmetric generalizations
$\mathcal{A}_{N,K}^{(\text{tree})}$ in the previous sections, we move
on to present their most important properties. From our perspective,
these are their symmetries because they severely constrain and
possibly even completely determine the amplitudes. We will discuss a
finite-dimensional Lie (super)algebra symmetry and a certain
infinite-dimensional extension thereof called \emph{Yangian} symmetry
\cite{Drinfeld:1985rx}. This Yangian symmetry algebra is closely tied
to integrability. Important integrable spin chain models, such as the
famous Heisenberg ferromagnet, and also integrable $1+1$-dimensional
quantum field theories are governed by it, see e.g.\
\cite{Bernard:1992ya,MacKay:2004tc} and
section~\ref{sec:int-mod}. Furthermore, the Yangian of the
superconformal algebra $\mathfrak{psu}(2,2|4)$ underlies the one-loop
spin chain of the planar $\mathcal{N}=4$ SYM spectral problem
\cite{Beisert:2003yb,Beisert:2003jj}, cf.\
section~\ref{sec:spectrum}. The very same Yangian algebra was also
found in the study of the scattering amplitudes
$\mathcal{A}_{N,K}^{(\text{tree})}$ of that theory
\cite{Drummond:2009fd}, see
\cite{Drummond:2010km,Beisert:2010jq,Ferro:2011ph} for reviews. This
discovery strengthened the believe that integrability is not just a
feature of the spectral problem but controls all observables in planar
$\mathcal{N}=4$ SYM.

Before becoming more technical, let us briefly sketch how the Yangian
symmetry of scattering amplitudes was discovered. The foundation is a
well established Lie (super)algebra symmetry. The gluon amplitudes
$A_{N,K}^{(\text{tree})}$ are annihilated by all generators of the
conformal algebra $\mathfrak{su}(2,2)$ acting on the particle momenta
$p^i$. Analogously, the superamplitudes
$\mathcal{A}_{N,K}^{(\text{tree})}$ are annihilated by the generators
of the superconformal algebra $\mathfrak{psu}(2,2|4)$ that act on
$p^i$ and on the supermomenta $q^i$. An illustrative calculation
showing this for the $\text{MHV}$ superamplitudes is contained in
\cite{Witten:2003nn}. Besides, a second copy of
$\mathfrak{psu}(2,2|4)$ termed \emph{dual superconformal symmetry}
that acts on the dual variables $x^i$ and $\theta^i$, cf.\
\eqref{eq:amp-super-dualvars}, was found
\cite{Drummond:2006rz,Drummond:2008vq,Brandhuber:2008pf}. Combining
these two superconformal algebras leads to the Yangian of
$\mathfrak{psu}(2,2|4)$ and consequently to the \emph{Yangian
  invariance} of the superamplitudes
$\mathcal{A}_{N,K}^{(\text{tree})}$ \cite{Drummond:2009fd}.

We continue by defining the Yangian invariance of amplitudes on a
technical level. To this end, we deviate from the ``historic'' route
of introducing the dual superconformal symmetry as just outlined. We
first discuss the Lie superalgebra symmetry, which will be extended to
a Yangian symmetry later. As $\mathcal{A}_{N,K}^{(\text{tree})}$ is
annihilated by all generators of the ordinary superconformal algebra
$\mathfrak{psu}(2,2|4)$, it is also annihilated by complex linear
combinations thereof and thus by the complexified algebra
$\mathfrak{psl}(\mathbb{C}^{4|4})\equiv\mathfrak{psl}(4|4)$. Hence, we
can work with complex algebras. Our aim is to state the generators
that annihilate the superamplitude. For this purpose, we start with a
set of generators $\mathfrak{J}_{\indnm{AB}}$ of the superalgebra
$\mathfrak{gl}(4|4)\supset \mathfrak{psl}(4|4)$, which are easily
realized in terms of spinor helicity variables. Arranged into a
supermatrix they read
\begin{align}
  \label{eq:sym-gen-gl44}
  (\mathfrak{J}_{\indnm{AB}})
  =
  \left(
    \begin{array}{c}    
      \lambda_{\alpha}\\[0.3em]
      \hdashline\\[-1.0em]
      -\partial_{\tilde{\lambda}_{\dot\alpha}}\\[0.3em]
      \hdashline\\[-1.0em]
      \partial_{\eta_{\dot a}}\\
    \end{array}
  \right)
  \left(
    \begin{array}{c:c:c}
    \partial_{\lambda_\beta}&
    \tilde{\lambda}_{\dot\beta}&
    \eta_{\dot b}\\
  \end{array}
  \right)
  =
  \left(
    \begin{array}{c:c:c}
    \lambda_\alpha\partial_{\lambda_\beta}&
    \lambda_\alpha\tilde{\lambda}_{\dot\beta}&
    \lambda_\alpha\eta_{\dot b}\\[0.3em]
    \hdashline\\[-1.0em]
    -\partial_{\tilde{\lambda}_{\dot\alpha}}\partial_{\lambda_\beta}&
    -\partial_{\tilde{\lambda}_{\dot\alpha}}\tilde{\lambda}_{\dot\beta}&
    -\partial_{\tilde{\lambda}_{\dot\alpha}}\eta_{\dot b}\\[0.3em]
    \hdashline\\[-1.0em]
    \partial_{\eta_{\dot a}}\partial_{\lambda_\beta}&
    \partial_{\eta_{\dot a}}\tilde{\lambda}_{\dot\beta}&
    \partial_{\eta_{\dot a}}\eta_{\dot b}\\
    \end{array}
  \right)\,.
\end{align}
Here we split the superindex $\indnm{A}=1,\ldots,8$ into bosonic
indices $\alpha,\dot\alpha=1,2$ with degree $|\alpha|=|\dot\alpha|=0$
and a fermionic index $\dot a=1,2,3,4$ with degree $|\dot a|=1$. The
generators obey the commutation relations
\begin{align}
  \label{eq:sym-alg-gl44}
  [\mathfrak{J}_{\indnm{AB}},\mathfrak{J}_{\indnm{CD}}\}
  =\delta_{\indnm{CB}}\mathfrak{J}_{\indnm{AD}}
  -(-1)^{(|\indnm{A}|+|\indnm{B}|)(|\indnm{C}|+|\indnm{D}|)}
  \delta_{\indnm{AD}}\mathfrak{J}_{\indnm{CB}}\,,
\end{align}
where the left hand side denotes the graded commutator, see
section~\ref{sec:yangian} for details on superalgebras. To obtain the
algebra $\mathfrak{psl}(4|4)$ we have to study the center of
$\mathfrak{gl}(4|4)$, i.e.\ those elements whose graded commutator
with all others vanishes. It is spanned by two generators,
\begin{align}
  \label{eq:sym-alg-central-c}
  \mathfrak{C}&=\tr(\mathfrak{J}_{\indnm{AB}})=\sum_{\indnm{A}}\mathfrak{J}_{\indnm{AA}}=
  \sum_{\alpha=1}^2\lambda_\alpha\partial_{\lambda_\alpha}
  -
  \sum_{\dot\alpha=1}^2\tilde{\lambda}_{\dot\alpha}\partial_{\tilde{\lambda}_{\dot\alpha}}
  -
  \sum_{\dot a=1}^4\eta_{\dot a}\partial_{\eta_{\dot a}}
  +
  2
\end{align}
and
$\mathfrak{B}=\str(\mathfrak{J}_{\indnm{AB}})=\sum_{\indnm{A}}(-1)^{|\indnm{A}|}\mathfrak{J}_{\indnm{AA}}$. The
subalgebra $\mathfrak{sl}(4|4)\subset\mathfrak{gl}(4|4)$ is defined by
imposing the relation $\mathfrak{B}=0$. A set of $\mathfrak{sl}(4|4)$
generators is
\begin{align}
  \label{eq:sym-alg-gen-sl}
  \mathfrak{J}'_{\indnm{AB}}
  =
  \mathfrak{J}_{\indnm{AB}}-\frac{1}{8}(-1)^{|A|}\delta_{\indnm{AB}}\mathfrak{B}\,
\end{align}
satisfying $\mathfrak{B}'=0$ and $\mathfrak{C}'=\mathfrak{C}$. One
obtains the simple algebra
$\mathfrak{psl}(4|4)\subset\mathfrak{sl}(4|4)$ by demanding that also
$\mathfrak{C}=0$. Generators $\mathfrak{J}_{\indnm{AB}}^i$ of the form
\eqref{eq:sym-gen-gl44} act on each particle of the superamplitude
$\mathcal{A}_{N,K}^{(\text{tree})}$. Thus an action of
$\mathfrak{gl}(4|4)$ on the whole amplitude is
\begin{align}
  \label{eq:sym-alg-level-1}
  M^{[1]}_{\indnm{AB}}=\sum_{i=1}^N\mathfrak{J}_{\indnm{BA}}^i\,.
\end{align}
We already argued that superamplitudes are invariant under
$\mathfrak{psl}(4|4)$. Noting that $\mathfrak{C}^i=0$ in
\eqref{eq:sym-alg-central-c} agrees with the condition on the
superhelicity in \eqref{eq:amp-super-hel}, the superamplitudes are
even invariant under $\mathfrak{sl}(4|4)$, 
\begin{align}
  \label{eq:sym-alg-annih-l1}
  M'^{[1]}_{\indnm{AB}}\mathcal{A}^{(\text{tree})}_{N,K}=0\,,
\end{align}
where the prime indicates that the generators $\mathfrak{J}_{\indnm{AB}}^i$
are to be replaced by $\mathfrak{J}'^i_{\indnm{AB}}$.  This concludes the
discussion of the Lie superalgebra invariance.

The Yangian extension of $\mathfrak{gl}(4|4)$ is obtained by appending
to the Lie superalgebra generators $M^{[1]}_{\indnm{AB}}$ an infinite
set of further generators $M^{[l]}_{\indnm{AB}}$ indexed by an integer
$l>1$. Often the generators $M^{[l]}_{\indnm{AB}}$ are said to be of
level $l-1$. There exists an elegant construction of these generators
within the so-called quantum inverse scattering framework, which we
will elaborate on in section~\ref{sec:yangian}. At this point,
however, we choose a rather pedestrian approach and just state the
explicit form of the generators with $l=2$,
\begin{align}
  \label{eq:sym-alg-level-2}
  \begin{aligned}
    M_{\indnm{AB}}^{[2]}
    &=\frac{1}{2}\sum_{\substack{i,j=1\\i<j}}^N\sum_{\indnm{C}} (-1)^{|\indnm{C}|}\Big(\mathfrak{J}_{\indnm{BC}}^j\mathfrak{J}_{\indnm{CA}}^i-\mathfrak{J}_{\indnm{BC}}^i\mathfrak{J}_{\indnm{CA}}^j\Big)\\
    &\quad+\sum_{i=1}^N \Big(v_i\mathfrak{J}_{\indnm{BA}}^i-\frac{1}{2}\sum_{\indnm{C}}(-1)^{|\indnm{A}||\indnm{B}|+|\indnm{A}||\indnm{C}|+|\indnm{B}||\indnm{C}|}\mathfrak{J}_{\indnm{CA}}^i\mathfrak{J}_{\indnm{BC}}^i\Big)\,,\\
  \end{aligned}
\end{align}
where $v_i$ are arbitrary complex parameters. This suffices because
all generators with larger $l$ can constructed from these. The graded
commutator between generators with $l=1,2$ evaluates to
\begin{align}
  \label{eq:sym-alg-comm-l1l2}
  [M^{[1]}_{\indnm{AB}},M^{[2]}_{\indnm{CD}}\}=
  \delta_{\indnm{AD}}M^{[2]}_{\indnm{CB}}
  -(-1)^{(|\indnm{A}|+|\indnm{B}|)(|\indnm{C}|+|\indnm{D}|)}\delta_{\indnm{CB}}M^{[2]}_{\indnm{AD}}\,.
\end{align}
We move on to the application of this Yangian algebra to scattering
amplitudes. For this we need a special case of the generators
\eqref{eq:sym-alg-level-2}. Form \eqref{eq:sym-gen-gl44} one derives
\begin{align}
  \label{eq:sym-alg-gen-rect}
  \sum_{\indnm{C}}(-1)^{|\indnm{A}||\indnm{B}|+|\indnm{A}||\indnm{C}|+|\indnm{B}||\indnm{C}|}
  \mathfrak{J}_{\indnm{CA}}^i\mathfrak{J}_{\indnm{BC}}^i
  =(\mathfrak{C}^i-1)\mathfrak{J}_{\indnm{BA}}^i+(-1)^{|\indnm{A}|}\delta_{\indnm{AB}}\mathfrak{C}^i\,.
\end{align}
Using this identity, the fact that $\mathfrak{C}^i=0$ on the amplitudes
$\mathcal{A}^{(\text{tree})}_{N,K}$ and setting $v_i=-\frac{1}{2}$, we
rewrite \eqref{eq:sym-alg-level-2} as
\begin{align}
  \label{eq:sym-alg-level-2-amp}
  \begin{aligned}
    M_{\indnm{AB}}^{[2]}
    &=\frac{1}{2}\sum_{\substack{i,j=1\\i<j}}^N\sum_{\indnm{C}} (-1)^{|\indnm{C}|}
    \Big(\mathfrak{J}_{\indnm{BC}}^j\mathfrak{J}_{\indnm{CA}}^i-\mathfrak{J}_{\indnm{BC}}^i\mathfrak{J}_{\indnm{CA}}^j\Big)\,.\\
  \end{aligned}
\end{align}
This is form of the Yangian generators is often found in the
literature on scattering amplitudes, cf. \cite{Drummond:2009fd}.  The
superamplitudes are annihilated by these generators,
\begin{align}
  \label{eq:sym-alg-annih-l2}
  M'^{[2]}_{AB}\mathcal{A}^{(\text{tree})}_{N,K}=0\,,
\end{align}
where again the prime indicates that $\mathfrak{gl}(4|4)$ generators
$\mathfrak{J}_{\indnm{AB}}^i$ are replaced by the $\mathfrak{sl}(4|4)$
generators $\mathfrak{J}'^i_{\indnm{AB}}$.\footnote{A short
  calculation shows that the Yangian generators in
  \eqref{eq:sym-alg-level-2-amp} are not affected by this replacement,
  $M'^{[2]}_{\indnm{AB}}=M^{[2]}_{\indnm{AB}}$.} The proof of this
condition constitutes the main result of
\cite{Drummond:2009fd}. Equations \eqref{eq:sym-alg-annih-l1} and
\eqref{eq:sym-alg-annih-l2} combined imply the invariance of the
superamplitudes under the Yangian of $\mathfrak{sl}(4|4)$. The
exploration of Yangian invariants of this and other algebras will
be the main subject of this thesis.

We finish this section with some remarks. Apart from the continuous
symmetries addressed here, amplitudes have a number of discrete
symmetries. Some of these follow directly from the definition of the
partial amplitudes \cite{Bern:1990ux}. As one example let us mention
the invariance of $\mathcal{A}^{(\text{tree})}_{N,K}$ under a cyclic
shift of the particles $i\mapsto i+1$. This cyclic invariance can also
be seen in the Yangian generators \cite{Drummond:2009fd} when acting
on an amplitude, manifestly in \eqref{eq:sym-alg-level-1} and
non-manifestly in \eqref{eq:sym-alg-level-2}. Further relations
between the $(N-1)!$ partial amplitudes, that enter the full
scattering amplitude in \eqref{eq:color-decomp}, reduce the number of
independent ones to $(N-3)!$ as shown in \cite{BjerrumBohr:2009rd}.

Another remark concerns the superconformal \eqref{eq:sym-alg-annih-l1}
and Yangian invariance \eqref{eq:sym-alg-annih-l2} of the amplitudes
$\mathcal{A}^{(\text{tree})}_{N,K}$. A careful analysis taking into
account the reality conditions \eqref{eq:spinors-real} of the spinor
variables reveals that these symmetries can be violated if two
particle momenta become collinear and thus a spinor bracket in the
denominator of the amplitude vanishes. This phenomenon can in
principle be overcome by introducing correction terms to the
generators of the symmetry algebra
\cite{Bargheer:2009qu}. Furthermore, the symmetries may break down at
multi-particle poles of the amplitude \cite{Sever:2009aa}.  All of
these issues are reviewed in \cite{Bargheer:2011mm}. Because they do
not occur for generic particle momenta, they are often neglected in
the discussion of tree-level amplitudes. This is also how we proceed
in most of this thesis.

It is worth emphasizing that although the amplitudes
$\mathcal{A}^{(\text{tree})}_{N,K}$ are annihilated by the Yangian of
the complex algebra $\mathfrak{sl}(4|4)$, the real form
$\mathfrak{su}(2,2|4)$ is of importance in the discussion of
scattering amplitudes. As a drastic illustration let us mention that
below in sections~\ref{sec:ncomp-susy-4-site} and
\ref{sec:sample-inv42} we will construct invariants associated with
the compact real form $\mathfrak{su}(4|4)$. Also these invariants are
annihilated by the Yangian of $\mathfrak{sl}(4|4)$, however they are
just polynomials and not distributions like the amplitudes.

\subsection{Graßmannian Integral}
\label{sec:grassmannian-integral}

So far we presented explicit formulas only for a few
superamplitudes. In \eqref{eq:amp-super-mhv} we introduced the
$\text{MHV}$ superamplitudes $\mathcal{A}^{(\text{tree})}_{N,2}$ and
in \eqref{eq:amp-super-nmhv6} we displayed the simplest $\text{NMHV}$
superamplitude $\mathcal{A}^{(\text{tree})}_{6,3}$. Comparing these
two equations, we notice that the complexity of the expression
increases significantly from $\text{MHV}$ to $\text{NMHV}$
superamplitudes. As already mentioned in the context of these
equations, an explicit formula for all
$\mathcal{A}^{(\text{tree})}_{N,K}$ is known
\cite{Drummond:2008cr}. However, for an decreasing degree of helicity
violation the representation of the superamplitude that this formula
produces gets more and more involved. In this section we discuss an
alternative and very compact formulation of these superamplitudes in
terms of certain multi-dimensional contour integrals, so-called
\emph{Graßmannian integrals} \cite{ArkaniHamed:2009dn}, see also
\cite{Mason:2009qx}. The matrix-valued integration variable can be
interpreted as a point in a Graßmannian manifold, hence the name. This
approach to superamplitudes especially suites our interests because
the compact Graßmannian integral formula allows for an easy
investigation of symmetries. While the superconformal symmetry is
manifest in this formula, also the Yangian symmetry can be verified
\cite{Drummond:2010qh,Drummond:2010uq}.

Before defining the Graßmannian integral, we first have to discuss
some bare essentials about Graßmannian manifolds. Many more details on
this topic can be found in books on algebraic geometry like e.g.\
\cite{Griffiths:2011}. The \emph{Graßmannian} $\text{Gr}(N,K)$
is the space of all $K$-dimensional linear subspaces of
$\mathbb{C}^N$. The complex entries of a $K\times N$ matrix $C$
provide ``homogeneous'' coordinates on this space. The transformation
$C\mapsto VC$ with $V\in GL(\mathbb{C}^K)$ corresponds to a change of
basis within a given subspace, and thus it does not change the point
in the Graßmannian. This allows us to describe a generic point in
$\text{Gr}(N,K)$ by the ``gauge fixed'' matrix
\begin{align}
  \label{eq:grassint-matrix}
  C=
  \left(
  \begin{array}{c:c}
    1_{K}&\mathcal{C}\\
  \end{array}
  \right)
  \quad
  \text{with}
  \quad
  \mathcal{C}=
  \begin{pmatrix}
    C_{1 K+1}&\cdots&C_{1 N}\\
    \vdots&&\vdots\\
    C_{K K+1}&\cdots&C_{K N}\\
  \end{pmatrix},
\end{align}
where $1_K$ denotes the $K\times K$ unit matrix. In what follows we
will also encounter the $(N-K)\times N$ matrix
\begin{align}
  \label{eq:grassint-matrix-perp}
  C^\perp=\left(
    \begin{array}{c:c}
      -\mathcal{C}^t&1_{N-K}
    \end{array}
\right)
\end{align}
that obeys $C(C^\perp)^t=0$. It may be considered as an element of
$\text{Gr}(N,N-K)$. These ingredients are sufficient to present the
\emph{Graßmannian integral} formulation of $\mathcal{N}=4$ SYM
superamplitudes \cite{ArkaniHamed:2009dn},
\begin{align}
  \label{eq:grassint-amp}
  \mathcal{A}_{N,K}^{(\text{tree})}
  =
  \int \D \mathcal{C}
  \frac{
    \delta^{2(N-K)|0}_\ast(C^\perp\boldsymbol{\lambda})
    \delta^{2K|0}_\ast(C\boldsymbol{\tilde{\lambda}})
    \delta^{0|4K}(C\boldsymbol{\eta})
  }
  {(1,\ldots,K)\cdots(N,\ldots,K-1)}\,
\end{align}
with the holomorphic $K(N-K)$-form
$\D\mathcal{C}=\bigwedge_{k,l}\D C_{k l}$. In this formula
$(i,\ldots,i+K-1)$ denotes the minor of the matrix $C$ consisting of
the consecutive columns $i,\ldots,i+K-1$. These are counted modulo $N$
such that they are in the range $1,\ldots,N$. The external data is
encoded in the $N\times 2$ matrices
$\boldsymbol{\lambda}=(\lambda^i_\alpha)$ and
$\boldsymbol{\tilde\lambda}=(\tilde\lambda^i_\alpha)$ as well as the
$N\times 4$ matrix $\boldsymbol{\eta}=(\eta_{\dot{a}}^i)$. The symbol
$\delta_\ast$ denotes a formal bosonic delta function whose argument
may be complex. It can be understood as a calculation rule to set the
argument to zero. 

Let us interject that in this thesis we encounter different types of
bosonic delta functions besides $\delta_\ast$. An ordinary delta
function of a real argument is always simply denoted by $\delta$, see
e.g.\ \eqref{eq:amp-bos-delta}. Below in
chapter~\ref{cha:grassmann-amp} we will encounter a complex delta
function $\delta_\mathbb{C}$ which is defined in terms of real ones.

We now discuss the evaluation of the Graßmannian integral
\eqref{eq:grassint-amp}. For this we have to specify the contour of
integration. Before doing so, it is helpful to note that the equations
imposed by the bosonic and fermionic delta functions in
\eqref{eq:grassint-amp} imply momentum conservation
\eqref{eq:amp-bos-delta} and supermomentum conservation
\eqref{eq:amp-superdelta},
\begin{align}
  \label{eq:grassint-consv}
  \begin{aligned}
  \boldsymbol{\lambda}^t\boldsymbol{\tilde\lambda}=0\quad
  \Leftrightarrow\quad
  P_{\alpha\dot{\beta}}=0\,,\\
  \boldsymbol{\lambda}^t\boldsymbol{\eta}=0\quad
  \Leftrightarrow\quad
  Q_{\alpha\dot{b}}=0\,.
  \end{aligned}
\end{align}
Of course, these constraints are understood to hold only in the
presence of those delta function. This allows us to assess the number
of integration variables remaining of the $K(N-K)$ variables contained
in $\mathcal{C}$ after solving for the $2N$ bosonic delta
functions. Taking into account that due to momentum conservation there
are four bosonic delta functions remaining, we are left with
\begin{align}
  \label{eq:grassint-numvars}
  K(N-K)-2N+4
\end{align}
complex integration variables. Thus a contour has to be specified only
for these variables. Let us first focus on those sample amplitudes
with which we are already familiar with from the previous
sections. For the $\text{MHV}$ amplitudes
$\mathcal{A}^{(\text{tree})}_{N,2}$ there is no integration remaining
according to \eqref{eq:grassint-numvars}. Thus the Graßmannian
integral \eqref{eq:grassint-amp} directly evaluates to the
Parke-Taylor-like formula \eqref{eq:amp-super-mhv} after solving the
bosonic delta functions. The next example is once again the
$\text{NMHV}$ amplitude $\mathcal{A}^{(\text{tree})}_{6,3}$, where we
are left with one complex integration variable. The integrand has six
poles in this variable, that are related to the points where the
minors in \eqref{eq:grassint-amp} vanish. The expression
\eqref{eq:amp-super-nmhv6} for $\mathcal{A}^{(\text{tree})}_{6,3}$ is
then obtained by picking a closed contour which encircles three of
those poles and applying Cauchy's residue theorem. Each of the three
terms in \eqref{eq:amp-super-nmhv6} corresponds to one residue. At
this point we can illustrate the situation for a general amplitude
$\mathcal{A}^{(\text{tree})}_{N,K}$. According to
\eqref{eq:grassint-numvars} we are left with a multi-dimensional
complex contour integral. In order to obtain a Yangian invariant
expression one should select a closed contour because the integrand of
\eqref{eq:grassint-amp} is only Yangian invariant up to an exact term
\cite{Drummond:2010qh,Drummond:2010uq}. This type of integrals can be
evaluated by means of a multi-dimensional generalization of Cauchy's
residue theorem, the so-called ``global residue theorem'', see the
discussion in \cite{ArkaniHamed:2009dn} and the references given
there. An explicit contour for all amplitudes
$\mathcal{A}^{(\text{tree})}_{N,K}$ was proposed in
\cite{Bourjaily:2010kw}, see also
\cite{Bullimore:2009cb,Kaplan:2009mh}.

Let us also mention some open challenges of the Graßmannian integral
approach in the form presented in \eqref{eq:grassint-amp}. The first
point concerns the spacetime signature. In case of the physical
Minkowski signature $(1,3)$, the spinors obey the reality conditions
\eqref{eq:spinors-real}. The spinors contained in
$\boldsymbol{\tilde\lambda}$ depend on those in $\boldsymbol{\lambda}$
and both variables are in general complex. Hence, the bosonic delta
functions $\delta_\ast$ in \eqref{eq:grassint-amp} have complex
arguments and therefore can strictly speaking not be treated as
ordinary real delta functions $\delta$. Moreover, the counting used to
obtain the number of remaining integrations in
\eqref{eq:grassint-numvars} is invalid because it assumes the
independence of $\boldsymbol{\lambda}$ and
$\boldsymbol{\tilde\lambda}$. Typically these issues are avoided by
either working in split signature $(2,2)$ or in a complexified
momentum space, where $\boldsymbol{\lambda}$ and
$\boldsymbol{\tilde\lambda}$ are treated as independent real or
complex variables, respectively. In the arguments presented in this
section we implicitly worked with the latter choice. The second point
to be discussed is the contour of integration. As just mentioned, a
contour for all amplitudes $\mathcal{A}^{(\text{tree})}_{N,K}$ was
given in \cite{Bourjaily:2010kw}. However, the explicit form of this
contour is quite intricate. One might argue that the complexity of
explicit formulas for general amplitudes
$\mathcal{A}^{(\text{tree})}_{N,K}$, which served as a motivation for
the Graßmannian integral formulation in the first paragraph, actually
persists in this approach. While the integrand of
\eqref{eq:grassint-amp} is simple, the complexity is contained in the
contour. In this thesis we will address both points. We will argue
that in Minkowski signature the reality conditions of the spinors and
the choice of the integration contour are tightly interrelated.

The Graßmannian integral formulation of scattering amplitudes as
introduced in \cite{ArkaniHamed:2009dn} presents merely the initial
step in a plethora of in part still ongoing developments. Let us
briefly mention some of the major advances of these
investigations. While in this section we confined ourselves to the
Graßmannian integral \eqref{eq:grassint-amp} for tree-level
superamplitudes $\mathcal{A}^{(\text{tree})}_{N,K}$, this integral
also contains certain data of loop amplitudes, cf.\
\eqref{eq:color-decomp-expansion}. Already in
\cite{ArkaniHamed:2009dn} it was argued that with a suitable contour
the integral computes leading singularities of loop
amplitudes. Further major steps were performed in
\cite{ArkaniHamed:2012nw}. In this work tree-level amplitudes and even
all-loop integrands were formulated in terms of \emph{on-shell
  diagrams}. These are networks composed out of the two types of
three-particle amplitudes in \eqref{eq:amp-3}, which are represented
graphically by trivalent vertices. It was realized that the on-shell
diagrams can be described utilizing the geometry of Graßmannian
manifolds and each diagram can be labeled by a permutation.  The
Graßmannian integral of \cite{ArkaniHamed:2009dn} is understood in
this setting as a means of computing on-shell diagrams. To obtain
amplitudes a number of on-shell diagrams have to be added up. Another
development is the introduction of a structure named
\emph{Amplituhedron} \cite{Arkani-Hamed:2013jha}, which aims at
interpreting amplitudes in geometric terms as certain
``volumes''. Finally, we want to emphasize that while the material
presented in the main part of this thesis has certainly interesting
connections to most of the developments sketched here, we will mostly
confine ourselves to relating it to the original Graßmannian integral
as introduced in \cite{ArkaniHamed:2009dn} and presented around
\eqref{eq:grassint-amp}.

\subsection{Integrable Deformations}
\label{sec:deform}

We already encountered one connection between scattering amplitudes
and integrability in section \ref{sec:symmetries}, namely the Yangian
invariance discovered in \cite{Drummond:2009fd}. A different link was
observed in \cite{Zwiebel:2011bx}, where in particular the amplitude
$\mathcal{A}_{4,2}^{(\text{tree})}$ was related to the one-loop
dilatation operator of the planar $\mathcal{N}=4$ SYM spectral
problem. This operator is the Hamiltonian of an integrable spin chain
and it can be constructed employing an R-matrix, which is a solution
of the Yang-Baxter equation \cite{Beisert:2003yb,Beisert:2003jj}. This
R-matrix depends on an arbitrary free complex parameter, a so-called
spectral parameter. The authors of \cite{Ferro:2012xw,Ferro:2013dga}
set out to expose this parameter also in the context of scattering
amplitudes, thus leading to \emph{deformed amplitudes}. The reasons
for pursuing this route are manifold. First, these deformations are
crucial to identify structures that are necessary to apply powerful
integrability-based methods like the quantum inverse scattering
framework to the study of amplitudes. In analogy to the situation for
the spectral problem almost 15 years ago, such an understanding of
tree-level amplitudes should also provide important clues of how to
proceed to loop-level. Second, as put forward in
\cite{Ferro:2012xw,Ferro:2013dga}, the deformation parameters could
even be of direct relevance for loop amplitudes as novel symmetry
preserving regulators of divergent loop integrals. Finally, the
deformations are of interest because they reveal exciting connections
between amplitudes and various areas of mathematics, ranging from
representation theory to hypergeometric functions. In the main part of
this thesis we will detail this enumeration and add some further
items. Let us mention that the developments reviewed in this section
happened in parallel to the author's research presented later in this
thesis.

The key principle in constructing deformed amplitudes
$\mathcal{A}^{(\text{def.})}_{N,K}$ is that they remain Yangian
invariant. We already know that the Yangian generators $M_{\indnm{AB}}^{[2]}$
in \eqref{eq:sym-alg-level-2-amp}, which appear for the undeformed
amplitudes $\mathcal{A}^{(\text{tree})}_{N,K}$, are a very particular
case of \eqref{eq:sym-alg-level-2}. The latter equation can be
rephrased as
\begin{align}
  \label{eq:sym-alg-level-2-amp-def}
  \begin{aligned}
    M_{\indnm{AB}}^{[2]}
    &=\frac{1}{2}\sum_{\substack{i,j=1\\i<j}}^N
    \sum_{\indnm{C}} (-1)^{|\indnm{C}|}\Big(\mathfrak{J}_{\indnm{BC}}^j\mathfrak{J}_{\indnm{CA}}^i
    -\mathfrak{J}_{\indnm{BC}}^i\mathfrak{J}_{\indnm{CA}}^j\Big)
    +\sum_{i=1}^N\hat{v}_i\mathfrak{J}^i_{\indnm{BA}}\,,\\
  \end{aligned}
\end{align}
where we used \eqref{eq:sym-alg-gen-rect}, assumed
$\sum_{i=1}^N\mathfrak{C}^i=0$ and introduced
$\hat{v}_i=v_i-\frac{c_i}{2}+\frac{1}{2}$ with $c_i$ being the eigenvalue of
$\mathfrak{C}^i$. Deformed amplitudes are then characterized by the
Yangian invariance condition
\begin{align}
  \label{eq:yi-def}
  {M'}_{\indnm{AB}}^{[1]}\mathcal{A}^{(\text{def.})}_{N,K}=0\,,\quad {M'}_{\indnm{AB}}^{[2]}\mathcal{A}^{(\text{def.})}_{N,K}=0\,,
\end{align}
where $M_{\indnm{AB}}^{[1]}$ is still given by
\eqref{eq:sym-alg-level-1} and $M_{\indnm{AB}}^{[2]}$ by
\eqref{eq:sym-alg-level-2-amp-def}. Furthermore, the prime signifies
that the $\mathfrak{gl}(4|4)$ generators $\mathfrak{J}_{\indnm{AB}}^i$
are to be replaced by the $\mathfrak{sl}(4|4)$ generators
$\mathfrak{J'}_{\indnm{AB}}^i$ in \eqref{eq:sym-alg-gen-sl}. These
deformations comprise the complex parameters $\hat{v}_i$ and
$c_i$. The parameters $\hat{v}_i$ appear directly in the Yangian
generators \eqref{eq:sym-alg-level-2-amp-def}. The solutions
$\mathcal{A}^{(\text{def.})}_{N,K}$ to the Yangian invariance
condition also contain the $c_i$, which can be understood as
deformations of the superhelicities, cf.\ \eqref{eq:amp-super-hel} and
\eqref{eq:sym-alg-central-c}.

As a first example we state the deformation of the four-particle
$\text{MHV}$ amplitude \cite{Ferro:2012xw,Ferro:2013dga},
\begin{align}
  \label{eq:amp-4-2-def}
  \mathcal{A}^{(\text{def.})}_{4,2}
  =
  \frac{
    \delta^{4|0}(P)\delta^{0|8}(Q)
  }{
    \langle 12\rangle
    \langle 23\rangle
    \langle 34\rangle
    \langle 41\rangle
  }
  \left(\frac{\langle 41\rangle}{\langle 34\rangle}\right)^{c_1}
  \left(\frac{\langle 12\rangle}{\langle 41\rangle}\right)^{c_2}
  \left(
    \frac{\langle 12\rangle\langle 34\rangle}
    {\langle 23\rangle\langle 41\rangle}
  \right)^{z}
\end{align}
with $z=\hat{v}_1+\frac{c_1}{2}-\hat{v}_2-\frac{c_2}{2}$. The
eigenvalues of the central elements and the parameters in the Yangian
generators obey
\begin{align}
  \label{eq:amp-4-2-def-para}
   c_3=-c_1\,,\quad c_4=-c_2\,,\quad
   \hat{v}_3=\hat{v}_1\,,\quad \hat{v}_4=\hat{v}_2\,.
\end{align}
The Yangian invariance condition \eqref{eq:yi-def} for this deformed
amplitude can be shown to be equivalent to the Yang-Baxter
equation. Hence, $\mathcal{A}^{(\text{def.})}_{4,2}$ is essentially an
R-matrix with spectral parameter $z$ and $c_1,c_2$ may be interpreted
as representation labels. In this language the undeformed amplitude
$\mathcal{A}^{(\text{tree})}_{4,2}$ is understood as an R-matrix with
the representations $c_1=c_2=0$ that is evaluated at a special value
$z=0$ of the spectral parameter. Let us interject that this
interpretation proposed in \cite{Ferro:2012xw,Ferro:2013dga} differs
slightly from the one in \cite{Zwiebel:2011bx}, where the amplitude is
related to a Hamiltonian and not to an R-matrix. Note that typically
such a Hamiltonian of an integrable spin chain is not Yangian
invariant, see the discussion at the end of
section~\ref{sec:yangian-inv} below. We believe that the conceptual
differences between the two interpretations deserve further
attention. However, in this thesis we do not dwell on this point and
stick to that of \cite{Ferro:2012xw,Ferro:2013dga}.

We move on to the deformations of the two complexified three-particle
amplitudes \eqref{eq:amp-3} that were also introduced in
\cite{Ferro:2012xw,Ferro:2013dga}. The first one reads
\begin{align}
  \label{eq:amp-3-2-def}
  \begin{aligned}
    \mathcal{A}^{(\text{def.})}_{3,2}
    &=  
    \frac{
      \delta^{4|0}(P)\delta^{0|8}(Q)
    }
    {
      \langle 12\rangle^{1+c_3}
      \langle 23\rangle^{1+c_1}
      \langle 31\rangle^{1+c_2}
    }
  \end{aligned}
\end{align}
with 
\begin{align}
  \label{eq:eq:amp-3-2-def-para}
  \hat{v}_3-\frac{c_3}{2}=\hat{v}_1+\frac{c_1}{2}\,,\quad
  \hat{v}_1-\frac{c_1}{2}=\hat{v}_2+\frac{c_2}{2}\,,\quad
  \hat{v}_2-\frac{c_2}{2}=\hat{v}_3+\frac{c_3}{2}\,.\quad
\end{align}
The second one becomes
\begin{align}
  \label{eq:amp-3-1-def}
  \begin{aligned}
    \mathcal{A}^{(\text{def.})}_{3,1}
    &=
    \frac{
      \delta^{4|0}(P)\delta^{0|4}([23]\eta^1+[31]\eta^2+[12]\eta^3)
    }
    {
      [23]^{1+c_1}
      [31]^{1+c_2}
      [12]^{1+c_3}
    }
  \end{aligned}
\end{align}
with 
\begin{align}
  \label{eq:amp-3-1-def-para}
  \hat{v}_2-\frac{c_2}{2}=\hat{v}_1+\frac{c_1}{2}\,,\quad
  \hat{v}_3-\frac{c_3}{2}=\hat{v}_2+\frac{c_2}{2}\,,\quad
  \hat{v}_1-\frac{c_1}{2}=\hat{v}_3+\frac{c_3}{2}\,.\quad
\end{align}
For these deformed three-particle amplitudes the Yangian invariance
condition \eqref{eq:yi-def} turns out to be equivalent to a so-called
bootstrap equation. This equation is known from the description of
bound states in $1+1$-dimensional integrable models, see
\cite{Zamolodchikov:1989zs} and e.g.\ \cite{Dorey:1997,Samaj:2013}.

Having these deformed three-point amplitudes at hand, one can start
deforming the on-shell diagrams of \cite{ArkaniHamed:2012nw}, which
are constructed by gluing together these amplitudes. In particular,
the deformation of an on-shell diagram for the one-loop amplitude
$\mathcal{A}_{4,2}^{(1)}$ was investigated in
\cite{Ferro:2012xw,Ferro:2013dga}. In the undeformed case this
on-shell diagram yields a divergent integral. It is usually addressed
by dimensional regularization, which breaks the superconformal
symmetry. The authors found that a particular choice of the
deformation parameters regularizes this integral. This observation
initiated the hope that the deformations might serve as symmetry
preserving regulators. However, while this deformed on-shell diagram
is indeed superconformally invariant, it violates the Yangian
symmetry. Essentially, the reason is that only the parameters $c_i$
but not the $\hat{v}_i$ of the three-particle amplitudes were matched
when gluing them together. Notice that the $\hat{v}_i$ do not enter in
the amplitudes \eqref{eq:amp-3-2-def} and \eqref{eq:amp-3-1-def} but
only in the Yangian generators. This is not satisfactory because to
have a chance of exploiting the integrable structure for the
computation of loop amplitudes, a regulator that preserves Yangian
symmetry should be essential.

The study of deformed on-shell diagrams was continued in
\cite{Beisert:2014qba}. Undeformed on-shell diagrams can be labeled by
permutations \cite{ArkaniHamed:2012nw}, as briefly mentioned at the
end of the previous section. The authors of \cite{Beisert:2014qba}
found that the Yangian invariance of the deformed diagrams can be
maintained by imposing a clever constraint on the deformation
parameters that is formulated in terms of the associated
permutation. With this technology they reexamined deformations of an
on-shell diagram for the one-loop amplitude
$\mathcal{A}_{4,2}^{(1)}$. They found that the Yangian invariant
deformation vanishes for generic deformation parameters. There are
only highly singular contributions for vanishing deformation
parameters. Further studies of deformed on-shell diagrams for one-loop
amplitudes were performed in \cite{Broedel:2014hca}. Apart from this,
\cite{Beisert:2014qba} contains an investigation of deformations of
the $\text{NMHV}$ amplitude $\mathcal{A}_{6,3}^{(\text{tree})}$. These
are of interest because $\mathcal{A}_{4,2}^{(1)}$ can be obtained from
this amplitude as a so-called forward limit, where two legs are
identified. Hence, they might provide an alternative approach to
regulate the one-loop amplitude $\mathcal{A}_{4,2}^{(1)}$. We already
know from the previous section that the three terms in the formula
\eqref{eq:amp-super-nmhv6} for $\mathcal{A}_{6,3}^{(\text{tree})}$ can
be understood as residues. Furthermore, each term is associated with
an on-shell diagram which can be deformed individually. Demanding that
these three deformed diagrams are compatible, i.e.\ that the
parameters $c_i$ and $\hat{v}_i$ for all diagrams agree, and that all
superhelicities take the physical value $c_i=0$ results in constraints
which are only satisfied by the undeformed amplitude.

Does this result imply that there is no deformation of
$\mathcal{A}_{6,3}^{(\text{tree})}$? In \cite{Ferro:2014gca} it was
argued that it does not suffice to deform the three on-shell diagrams
contributing to the amplitude individually. Instead, one should take
seriously that these diagrams are residues originating from a
one-dimensional contour integral, which arises upon the evaluation of
the Graßmannian integral \eqref{eq:grassint-amp}. This line of thought
led to a \emph{deformed Graßmannian
  integral} \cite{Ferro:2014gca,Bargheer:2014mxa}\footnote{In
  \cite{Ferro:2014gca} the result is presented in supertwistor
  variables. The spinor helicity formula \eqref{eq:grassint-amp-def}
  can formally, which essentially means for the signature $(2,2)$, be
  obtained by applying Witten's half Fourier transform
  \cite{Witten:2003nn}. However, compared to equation (13) of
  \cite{Ferro:2014gca} the variables $\hat{v}^+_i$ and $\hat{v}^-_i$
  are exchanged in \eqref{eq:grassint-amp-def}. The reason is a small
  typo in that $-c_i$ and not $c_i$ is the eigenvalue of the
  superhelicity operator given in (11) of that reference. This can be
  verified by evaluating (16) therein for the $\text{MHV}$ case.}
\begin{align}
  \label{eq:grassint-amp-def}
  \mathcal{A}_{N,K}^{(\text{def.})}
  =
  \int \D \mathcal{C}
  \frac{
    \delta^{2(N-K)|0}_\ast(C^\perp\boldsymbol{\lambda})
    \delta^{2K|0}_\ast(C\boldsymbol{\tilde{\lambda}})
    \delta^{0|4K}(C\boldsymbol{\eta})
  }
  {(1,\ldots,K)^{1+\hat{v}_K^--\hat{v}_1^+}
    \cdots 
    (N,\ldots,K-1)^{1+\hat{v}^-_{K-1}-\hat{v}_N^+}}\,.
\end{align}
Here the exponents are defined by \cite{Beisert:2014qba}
\begin{align}
  \label{eq:amp-wpm}
  \hat{v}_i^\pm=\hat{v}_i\pm\frac{c_i}{2}\,.
\end{align}
To ensure Yangian invariance they have to satisfy
\begin{align}
  \label{eq:amp-wperm}
  \hat{v}^-_{i+K}=\hat{v}^+_i\,
\end{align}
for $i=1,\ldots,N$, where we count modulo $N$. Actually,
\eqref{eq:amp-wperm} is the clever constraint on the deformation
parameters found in \cite{Beisert:2014qba}, which we referred to in
the previous paragraph. Writing the subscript on the left hand site as
$\sigma(i)=i+K$, we see a permutation $\sigma$ emerging. This
particular permutation is a cyclic shift, which is associated with a
special class of on-shell diagrams called \emph{top-cells} in the
language of \cite{ArkaniHamed:2012nw}.

The next step is to evaluate the deformed Graßmannian integral
\eqref{eq:grassint-amp-def}. The sample invariants
$\mathcal{A}^{(\text{def.})}_{3,1}$,
$\mathcal{A}^{(\text{def.})}_{3,2}$ and
$\mathcal{A}^{(\text{def.})}_{4,2}$ discussed above can easily be
obtained from \eqref{eq:grassint-amp-def} by solving for the bosonic
delta functions. In this way, we also obtain the deformed $\text{MHV}$
amplitudes \cite{Ferro:2014gca,Bargheer:2014mxa}
\begin{align}
  \label{eq:amp-def-mhv}
  \mathcal{A}_{N,2}^{(\text{def.})}
  =
  \frac{\delta^{4|0}(P)\delta^{0|8}(Q)}{
    \langle 12\rangle^{1+\hat{v}_2^--\hat{v}_1^+}
    \cdots
    \langle N1\rangle^{1+\hat{v}_1^--\hat{v}_N^+}
  }\,.
\end{align}
Let us note a subtlety at this point. Because the deformations
parameters $\hat{v}^\pm_i$ are complex, these deformed amplitudes are
multi-valued functions in the spinors
$\lambda^i,\tilde{\lambda}^i$. This multi-valuedness does not seem to
cause problems in the $\text{MHV}$ case. However, for deformed
$\text{NMHV}$ amplitudes, and in particular for
$\mathcal{A}_{6,3}^{(\text{def.})}$, solving for the bosonic delta
functions in \eqref{eq:grassint-amp-def} leaves us with a
one-dimensional contour integral. Due to the complex exponents in
\eqref{eq:grassint-amp-def}, the residue theorem does not apply any
longer for the evaluation of this integral. This may be viewed as an
explanation why deforming the three individual residues in
\cite{Beisert:2014qba} did not succeed. Instead, one should find an
appropriate closed contour taking into account the intricate structure
of the multi-valued integrand. Partial results in this direction were
obtained in \cite{Ferro:2014gca}. Yet ultimately the problem of
finding an appropriate contour that yields a closed form expression of
the deformed six-particle $\text{NMHV}$ amplitude
$\mathcal{A}_{6,3}^{(\text{def.})}$ remained open in this
reference. It is one of the issues addressed in this thesis.

We conclude this section with some remarks. First, in
\cite{Ferro:2014gca} close relations between the deformed Graßmannian
integral \eqref{eq:grassint-amp-def} and the theory of multivariate
hypergeometric functions \cite{Gelfand:1986} were observed, see also
the books \cite{Kita:2011} and \cite{Vilenkin1993}.

A further remark concerns an interesting construction of deformed
amplitudes that was introduced in
\cite{Chicherin:2013sqa,Chicherin:2013ora}. Simple sample amplitudes
were obtained by acting with a number of special solutions of a
Yang-Baxter equation on a vacuum state. This method was explored
systematically in \cite{Kanning:2014maa,Broedel:2014pia}. It led to a
construction of all on-shell diagrams that are relevant for tree-level
amplitudes. What is more, in \cite{Ferro:2014gca} it is argued that it
can be used to derive the deformed Graßmannian integral
\eqref{eq:grassint-amp-def}. We chose not to present this method here
in detail mainly for two reasons. First, although the central object
is a solution of a Yang-Baxter equation, its proper interpretation
remains somewhat unclear. In particular, this solution does not
commute with the central elements of the symmetry algebra
$\mathfrak{gl}(4|4)$, see equation (102) in
\cite{Kanning:2014maa}. Thus it is not a usual $\mathfrak{gl}(4|4)$
invariant R-matrix. Second, this solution is typically represented as
a formal integral operator. The proper contour of integration is
unknown at present. We believe that finding the contour for this
operator might be even more involved than finding directly the proper
contour for the deformed Graßmannian integral formula
\eqref{eq:grassint-amp-def}. That is because to obtain this formula
the integral operator has to be applied multiple times, probably each
time with a different contour. Addressing these points would clearly
be desirable.

As a final comment, an integrability-related approach to planar
$\mathcal{N}=4$ SYM scattering amplitudes was also initiated in
\cite{Basso:2013vsa}. It is distinct from the Yangian invariant
deformations outlined in this section and aims at computing amplitudes
at finite coupling, see e.g.\ the recent work \cite{Basso:2015rta} for
a comprehensive list of references.

\section{Objectives and Outline}
\label{sec:objectives-outline}

The long-term objective of the line of research presented in this
thesis is to make the powerful methods of integrable models available
for the computation of scattering amplitudes in planar $\mathcal{N}=4$
SYM, even at all-loop level. Thereby it may be possible to repeat
the tremendous success story of integrability in the spectral
problem. Moreover, we believe that such a development would help to
clarify the integrable structure that seems to underlie the planar
$\mathcal{N}=4$ model as a whole.

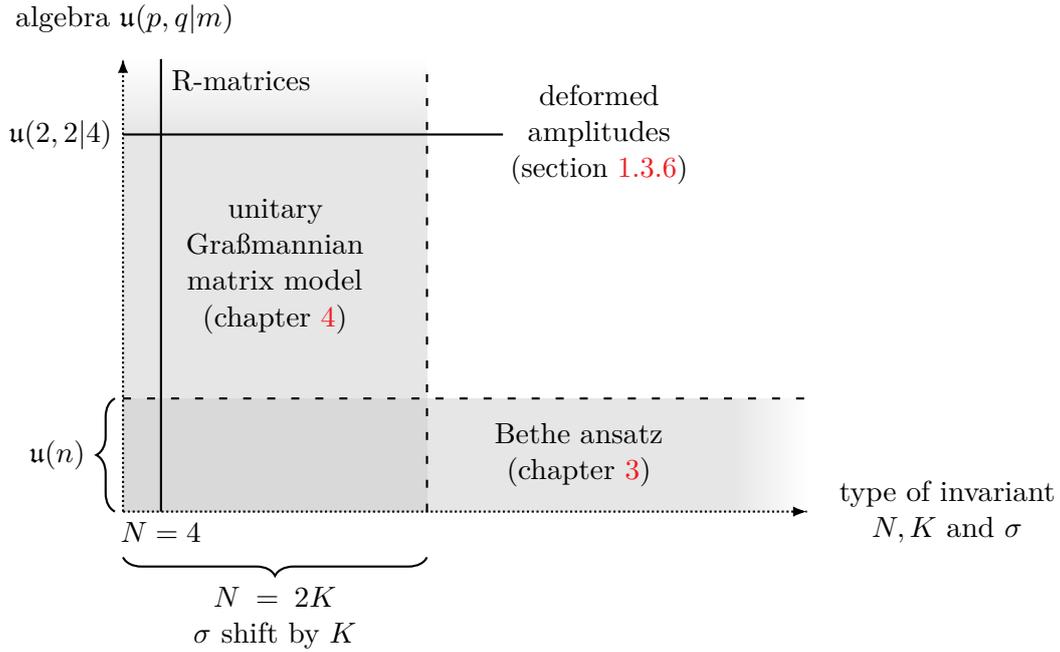
\begin{figure}[!t]
  \begin{center}
    \begin{tikzpicture}
      \fill [color=black!10] (0,0) rectangle (8,1.5);
      \fill [color=black!10,path fading=east] (8,0) rectangle (9,1.5);
      \fill [color=black!10] (0,0) rectangle (4,5);
      \fill [color=black!10,path fading=north] (0,5) rectangle (4,6);
      \fill [color=black!15] (0,0) rectangle (4,1.5);
      \node at (6,.75) [align=center,text width=3cm] 
      {Bethe ansatz\\ (chapter~\ref{cha:bethe-vertex})};
      \node at (2,3.25) [align=center,text width=3cm] 
      {unitary\\ Graßmannian\\ matrix model\\ (chapter~\ref{cha:grassmann-amp})};
      \draw[thick,loosely dashed] (0,1.5) -- (9,1.5);
      \draw[thick,loosely dashed] (4,0) -- (4,6);
      \draw[thick,densely dotted,
      decoration={markings, mark=at position 1.0 with {\arrow{latex}}},
      postaction={decorate}]
      (0,0) -- (9,0) node[right=.2,align=center,text width=3cm]
      {type of invariant\\ $N,K$ and $\sigma$};
      \draw[thick,densely dotted,
      decoration={markings, mark=at position 1.0 with {\arrow{latex}}},
      postaction={decorate}]
      (0,0) -- (0,6) node[above=.2,align=center,text width=3cm] 
      {algebra $\mathfrak{u}(p,q|m)$};
      \draw[thick,decorate,decoration={brace,amplitude=7pt},xshift=-3pt]
      (0,0) -- (0,1.5) node [midway,left=0.25]{$\mathfrak{u}(n)$};
      \draw[thick,decorate,decoration={brace,amplitude=7pt},yshift=-3pt]
      (4,-.5) -- (0,-.5) node [midway,below=0.25,align=center,text
      width=3cm] {$N=2K$\\$\sigma$ shift by
        $K$};
      \draw[thick] (0,5) node[left]{$\mathfrak{u}(2,2|4)$}
      -- (5,5) node[right,align=center,text width=2.25cm]{deformed\\
        amplitudes\\ (section~\ref{sec:deform})}; \draw[thick] (.5,0)
      node[below]{$N=4$} -- (.5,6) node[below right] {R-matrices};
    \end{tikzpicture}
    \caption{Schematic view of the ``landscape'' of Yangian
      invariants. The Bethe ansatz is applicable for the construction
      of Yangian invariants for compact algebras
      $\mathfrak{u}(n)$,
      the details being worked out for $\mathfrak{u}(2)$.
      In this setting it leads to a classification of all
      invariants. The unitary Graßmannian matrix model can be used for
      the large class of non-compact superalgebras
      $\mathfrak{u}(p,q|m)$.
      However, at the moment it is restricted to a certain type of
      invariants. In this spirit both methods are complementary.}
    \label{fig:map-inv}
  \end{center}
\end{figure}

In this thesis we undertake steps in this direction by building a
bridge between tree-level scattering amplitudes on one side and
integrable models, as they appear e.g.\ in the one-loop spectral
problem, on the other side. The reader might argue that the relation
of these amplitudes to integrability was already clarified by showing
their Yangian invariance as discussed in the preceding
section. However, this pioneering connection alone did not immediately
enable the application of the techniques associated with integrable
systems to the calculation of amplitudes. In the present thesis we
supplement this work by an effort for a robust construction whose
algebraic and representation theoretic foundations we lay in
chapter~\ref{cha:yang-rep}. Based on this groundwork we start erecting
our bridge on both riverbanks. In chapter~\ref{cha:bethe-vertex} we
work on the integrable systems side employing the mindset of
sections~\ref{sec:int-mod} and \ref{sec:spectrum}. We show that a
Bethe ansatz can be applied for the computation of certain Yangian
invariants. Our work continues in chapter~\ref{cha:grassmann-amp} on
the amplitudes side, where we make use of the ideas presented in
section~\ref{sec:amplitudes}. Here we develop a second method for the
computation of Yangian invariants which is based on the Graßmannian
integral approach. These two methods are in a sense complementary but
they also have an overlap, see figure~\ref{fig:map-inv} and the
explanations in the following paragraphs. Interestingly, this might
make it possible to use our bridge also in the reverse direction and
thus apply ideas from amplitudes to gain new insights into integrable
models as such. Finally, in chapter~\ref{cha:concl} we assess the
status of our constructions and plan future steps.

Let us outline our work on a more technical level. The key principle
on which our investigations are based on is the Yangian invariance of
tree-level amplitudes. In \textbf{chapter~\ref{cha:yang-rep}} we start
by reviewing the algebraic foundation of the Yangian of the Lie
superalgebra $\mathfrak{gl}(n|m)$
in the QISM language, which we encountered in
section~\ref{sec:int-mod} in connection with integrable spin
chains. What is more, we formulate the \emph{Yangian invariance
  condition within the QISM}. This will later allow us to use the QISM
framework for the construction of Yangian invariants. We move on by
discussing a class of unitary representations of the non-compact
superalgebra $\mathfrak{u}(p,q|m)\subset\mathfrak{gl}(n=p+q|m)$
that are built in terms of harmonic oscillator algebras. For
$\mathfrak{u}(2,2|4)$
essentially these representations are frequently used in the
$\mathcal{N}=4$
SYM spectral problem. Moreover, they are equivalent to the spinor
helicity variables employed for amplitudes in section
\ref{sec:amplitudes}. The compact $\mathfrak{u}(n)$
case is of interest to make contact with the common literature on spin
chains. Thus this class of representations provides a great
versatility. Working with $\mathfrak{u}(p,q|m)$
instead of directly focusing on the amplitude case
$\mathfrak{u}(2,2|4)$
also satisfies our mathematical curiosity to systematically explore
the notion of Yangian invariance. We conclude by discussing some
sample Yangian invariants to gain more familiarity with the
representations. In doing so we notice the necessity of using a
further family of oscillator representations, which are dual to the
ordinary ones. The Yangian invariants that we study are associated
with a spin chain of $N$
sites out of which $K$
carry a dual representation. We observe that for $(N,K)=(4,2)$
the Yangian invariance condition is equivalent to the Yang-Baxter
equation, whose solutions are called R-matrices. The foundations laid
in this chapter are used throughout the thesis.

In \textbf{chapter~\ref{cha:bethe-vertex}} we make use of the QISM for
the construction of Yangian invariants. To this end, we first review
the Bethe ansatz for inhomogeneous $\mathfrak{u}(2)$ spin chains in
the QISM setting. This class of spin chains includes the Heisenberg
model, which served as an example in section~\ref{sec:int-mod}. These
preparations allow us to develop a \emph{Bethe ansatz for Yangian
  invariants} with $\mathfrak{u}(2)$ representations by identifying
these invariants with specific eigenstates of particular spin
chains. These eigenstates are characterized by a special case of the
usual Bethe equations. Unlike the full equations, this special case
can easily be solved explicitly. Doing so we recover some of the
sample invariants from chapter~\ref{cha:yang-rep}. Furthermore, all
solutions of the equations can be classified in terms of permutations
$\sigma$. These are analogous to the permutations appearing for
deformed amplitudes in section~\ref{sec:deform}. Hence the compact
$\mathfrak{u}(2)$ Yangian invariants constructed in this chapter can
be considered as toy models for amplitudes. Their representation
labels and inhomogeneities can be identified with the deformation
parameters of those amplitudes. Some additional material on the Bethe
ansatz for Yangian invariants, in particular concerning the extension
to $\mathfrak{u}(n)$, is deferred to
\textbf{appendix~\ref{cha:app-bethe}}. On a different note, we show
that Yangian invariants for $\mathfrak{u}(n)$ can be interpreted as
partition functions of certain vertex models on in general
non-rectangular lattices. This allows us to view our Bethe ansatz
construction of Yangian invariants as a vast generalization of
Baxter's little known perimeter Bethe ansatz.

\textbf{Chapter~\ref{cha:grassmann-amp}} is devoted to an alternative
construction of Yangian invariants. Taking inspiration from the
Graßmannian integral for deformed amplitudes of
section~\ref{sec:deform}, we derive a Graßmannian integral formula for
a family of Yangian invariants with $N=2K$ and oscillator
representations of the non-compact superalgebra
$\mathfrak{u}(p,q|m)$. A key difference to the original formula is
that for these representations the formal delta function in the
integrand gets replaced by an exponential function of
oscillators. Furthermore, we are able to fix the contour of
integration to be a unitary group manifold. For special values of the
deformation parameters this integral reduces to a well-known unitary
matrix model due to Brezin, Gross and Witten. Consequently, we term
our formula \emph{unitary Graßmannian matrix model}. Evaluating it for
special cases, we once again rederive sample invariants from
chapter~\ref{cha:yang-rep}. To relate the invariants obtained from
this integral to deformed amplitudes, we apply a change of basis
mapping the oscillator variables to spinor helicity-like variables of
the algebra $\mathfrak{u}(p,p|m)$. In particular this provides us with
a proposal for a refined Graßmannian integral formula for deformed
amplitudes when specializing to $\mathfrak{u}(2,2|4)$. This formula
features several improvements over the one in
section~\ref{sec:deform}. Importantly, it is formulated in the
physical Minkowski signature. In addition, we find that the unitary
contour circumvents all issues with branch cuts pointed out in the
aforementioned section. Finally, we put our refined formula to the
test. We are able to rederive the already known deformed amplitude
$\mathcal{A}_{4,2}^{(\text{def.})}$. What is more, we obtain a natural
candidate for the presently unknown
$\mathcal{A}_{6,3}^{(\text{def.})}$. Curiously, there is a caveat
because in the undeformed limit the tree-level amplitude
$\mathcal{A}_{6,3}^{(\text{tree})}$ emerges only in a certain region
of the whole momentum space. Let us add that supplementary material on
the unitary Graßmannian integral is presented in
\textbf{appendix~\ref{cha:add-grass-int}}.

The final \textbf{chapter~\ref{cha:concl}} contains our
conclusions. Furthermore, we provide an outlook on possibilities to
extend our work. In particular, we ponder about the implications of
the just mentioned caveat on the symmetries of tree-level amplitudes.

\chapter{Yangians and Representations}
\label{cha:yang-rep}

The purpose of this chapter is to set up a common framework within
which we will develop methods for the construction of Yangian
invariants in the following chapters~\ref{cha:bethe-vertex} and
\ref{cha:grassmann-amp}.

We begin by reviewing the Yangian of the Lie superalgebra
$\mathfrak{gl}(n|m)$ in section~\ref{sec:yangian}. In contrast to the
brief discussion of the Yangian algebra in the introductory
section~\ref{sec:amplitudes}, we choose a QISM formulation that
highlights the connections with integrable systems, in particular with
the Yang-Baxter equation and spin chains. In section
\ref{sec:yangian-inv} we translate the important Yangian invariance
condition into this language. Before we are able to explore solutions
of this equation, we have to select representations of the
$\mathfrak{gl}(n|m)$ algebra. Therefore, in section~\ref{sec:osc-rep}
we present a class of unitary representations of
$\mathfrak{u}(p,q|m)\subset\mathfrak{gl}(n|m)$, that are constructed
from bosonic and fermionic oscillators. For $\mathfrak{u}(2,2|4)$
these turn out to be equivalent to the spinor helicity variables of
section~\ref{sec:amplitudes}. The oscillator formalism allows us to
work with more general non-compact superalgebras.

We continue in section~\ref{sec:sample-inv} by ``manually''
constructing sample solutions of the Yangian invariance condition that
are associated with spin chains consisting of two, three and four
sites. Initially we restrict to sample invariants with representations
of the compact bosonic algebra $\mathfrak{u}(n)$. These will be
recovered systematically by means of a Bethe ansatz in
chapter~\ref{cha:bethe-vertex}. Then we attempt to generalize the
sample invariants to the full non-compact supersymmetric
$\mathfrak{u}(p,q|m)$ setting. A Graßmannian integral construction for
such Yangian invariants will be developed in
chapter~\ref{cha:grassmann-amp}. Furthermore, we explain in
section~\ref{sec:sample-inv} that for the four-site sample invariant
the Yangian invariance condition is equivalent to a Yang-Baxter
equation. Hence for the algebra $\mathfrak{u}(2,2|4)$ this invariant
is essentially the R-matrix of the integrable spin chain governing the
planar $\mathcal{N}=4$ SYM one-loop spectral problem. Later in
chapter~\ref{cha:grassmann-amp} we will apply a change of basis from
oscillators to spinor helicity variables that turns this invariant
into the amplitude $\mathcal{A}_{4,2}^{(\text{tree})}$.

\section{Yangian Algebra}
\label{sec:yangian}

We encountered the Yangian of $\mathfrak{sl}(4|4)$ as an
infinite-dimensional extension of the complexified superconformal
algebra in the review of $\mathcal{N}=4$ SYM amplitudes in
section~\ref{sec:symmetries}. Here we provide a systematic account on
the \emph{Yangian of the Lie superalgebra}
$\mathfrak{gl}(n|m)$. Yangian algebras that are based on bosonic Lie
algebras, such as $\mathfrak{gl}(n)$, were introduced by Drinfeld
\cite{Drinfeld:1985rx}. In physics they occur in the study of
integrable models, cf.\ \cite{MacKay:2004tc,Bernard:1992ya}. Moreover,
they are of interest mathematically as prominent examples of a special
class of Hopf algebras called ``quantum groups''
\cite{Drinfeld:1986in,Jimbo:1985zk}, see also \cite{Chari:1995}. There
exist different formulations of Yangian algebras, one of which we saw
in section~\ref{sec:symmetries}. Here we pursue a more elaborate
approach that has its origins in the study of integrable models, in
particular spin chains. This is the \emph{quantum inverse scattering
  method} (QISM) mentioned already in section~\ref{sec:int-mod}. It
goes back to work of the Leningrad school around Faddeev in the 1970s
and 80s. It explores the algebraic and representation theoretic
consequences of the Yang-Baxter equation and provides a toolbox to
``solve'' integrable models. Authoritative reviews of the QISM can be
found in \cite{Faddeev:1996iy,Korepin:1997}. See also
\cite{Sklyanin:1991ss} for an alternative selection of topics and
\cite{Faddeev:1995} for a historic perspective. The formulation of
Yangians within the QISM and its representation theory are discussed
extensively in \cite{Molev:2007}. The presentation in this section is
to a large extent influenced by this reference. The works cited so far
concentrate on bosonic algebras. A $\mathfrak{gl}(n|m)$ version of the
QISM can be found in \cite{Kulish:1985bj}. The Yangian of this
superalgebra was defined in \cite{Nazarov:1991}, see also
\cite{Gow:2007}. In this section we immediately present the formulas
for the $\mathfrak{gl}(n|m)$ case. The $\mathfrak{gl}(n)$ analogues
can be easily obtained by dropping the grading factors $(-1)^{|\cdots|}$.

To begin with, we recapitulate some basic notions of the super vector
space $\mathbb{C}^{n|m}$ and introduce the general linear Lie
superalgebra
$\mathfrak{gl}(\mathbb{C}^{n|m})\equiv\mathfrak{gl}(n|m)$. Most of
these basics are nicely summarized in \cite{Kulish:1985bj}, see the
references therein for more details. An extensive discussion of Lie
superalgebras can be found e.g.\ in \cite{Frappat:1996pb}. Let the
vectors $\elemv_{\indnm{A}}$ with the superindex
$\indnm{A}=1,\ldots,n+m$ form a basis of the super vector space
$\mathbb{C}^{n|m}$. We allow for a non-standard grading by choosing
the integer partitions $n=p+q$ and $m=r+s$. Based thereon we assign
the degree
\begin{align}
  \label{eq:grading}
  |\elemv_{\indnm{A}}|=|\indnm{A}|=
  \begin{cases}
    0\quad\text{for}\quad \indnm{A}=\phantom{p+r+q+{}}1,\ldots,p\,,\\
    1\quad\text{for}\quad \indnm{A}=\phantom{r+q+{}}p+1,\ldots,p+r\,,\\
    0\quad\text{for}\quad \indnm{A}=\phantom{q+{}}p+r+1,\ldots,p+r+q\,,\\
    1\quad\text{for}\quad \indnm{A}=p+r+q+1,\ldots,p+r+q+s\,.\\
  \end{cases}
\end{align}
This freedom in the choice of the grading is already adapted to the
representations of $\mathfrak{u}(p,q|r+s)\subset\mathfrak{gl}(n|m)$,
which we will study in section~\ref{sec:osc-rep} below. We define
supermatrices $\elemm_{\indnm{AB}}$ by
$\elemm_{\indnm{AB}}\,\elemv_{\indnm{C}}=\delta_{\indnm{BC}}\,\elemv_{\indnm{A}}$. They
satisfy
\begin{align}
  \label{eq:elemsupermat-prop}
  \elemm_{\indnm{AB}}\elemm_{\indnm{CD}}=\delta_{\indnm{BC}}\elemm_{\indnm{AD}}
\end{align}
and are of degree $|\elemm_{\indnm{AB}}|=|\indnm{A}|+|\indnm{B}|$. The
$\elemm_{\indnm{AB}}$ can be considered as generators of the defining
representation of the \emph{general linear Lie superalgebra}
$\mathfrak{gl}(n|m)$. In this context the super vector space these
generators act on is denoted by $\square=\mathbb{C}^{n|m}$. The
$\mathfrak{gl}(n|m)$ algebra is defined by the relations
\begin{align}
  \label{eq:gl-superalg}
  [J_{\indnm{AB}},J_{\indnm{CD}}\} =\delta_{\indnm{CB}}J_{\indnm{AD}}-
  (-1)^{(|\indnm{A}|+|\indnm{B}|)(|\indnm{C}|+|\indnm{D}|)}\delta_{\indnm{AD}}J_{\indnm{CB}}\,
\end{align}
for the generators $J_{\indnm{AB}}$ of degree
$|\indnm{A}|+|\indnm{B}|$. Here we employed the graded commutator
\begin{align}
  \label{eq:gl-supercomm}
  [U,V\}=UV-(-1)^{|U||V|}VU
\end{align}
for homogeneous elements $U,V\in\mathfrak{gl}(n|m)$. We consider a
representation of this algebra where the generators $J_{\indnm{AB}}$
act as operators on some super vector space $\mathcal{V}$. In slight
abuse of notation we often refer to $\mathcal{V}$ itself as
representation. Operators acting on the tensor product
$\mathcal{V}\otimes\mathcal{V}'$ of two super vectors spaces obey
\begin{align}
  (U\otimes V)(U'\otimes V')=(-1)^{|V||U'|}UU'\otimes VV'\,.
\end{align}

At this point, we discus some automorphisms of the superalgebra
$\mathfrak{gl}(n|m)$ given in \eqref{eq:gl-superalg}, which we will
make use of later. One readily verifies that
\begin{align}
  \label{eq:auto-dual}
  J_{\indnm{AB}}\mapsto -(-1)^{|\indnm{A}|+|\indnm{A}||\indnm{B}|}J_{\indnm{AB}}^\dagger\,
\end{align}
is an automorphism. Here the conjugation satisfies
$(UV)^\dagger=V^\dagger U^\dagger$ and
$(U^\dagger)^\dagger=U$.\footnote{The superalgebra
  $\mathfrak{gl}(n|m)$ may also be equipped with a graded conjugation
  $\ddagger$ obeying
  $(UV)^\ddagger=(-1)^{|U||V|}V^\ddagger U^\ddagger$ and
  $(U^\ddagger)^\ddagger=(-1)^{|U|}U$, cf.\
  \cite{Scheunert:1976wi}. Then
  $J_{\indnm{AB}}\mapsto -J_{\indnm{AB}}^\ddagger$ defines an
  automorphism.} Notice that for bosonic algebras the map
\eqref{eq:auto-dual} is an antiinvolution. A similar but distinct
automorphism of the $\mathfrak{gl}(n|m)$ algebra
\eqref{eq:gl-superalg} is
\begin{align}
  \label{eq:auto-other}
  J_{\indnm{AB}}\mapsto -(-1)^{|\indnm{A}|+|\indnm{A}||\indnm{B}|}J_{\indnm{BA}}\,.
\end{align}
We will encounter \eqref{eq:auto-dual} and \eqref{eq:auto-other} in
section~\ref{sec:osc-rep} in the context of so-called dual
representations. Yet another automorphism of \eqref{eq:gl-superalg} is
\begin{align}
  \label{eq:auto-shift}
  J_{\indnm{AB}}\mapsto
  J_{\indnm{AB}}+v\delta_{\indnm{AB}}(-1)^{|\indnm{B}|}\,
\end{align}
with an arbitrary parameter $v\in\mathbb{C}$.

After introducing $\mathfrak{gl}(n|m)$, we move on to discuss the
\emph{Yangian} \cite{Drinfeld:1985rx} of this Lie superalgebra. It is
defined by the relation
\begin{align}
  \label{eq:yangian-def}
  R_{\square\,\square'}(\spec-\specp)
  (M(\spec)\otimes 1)(1\otimes M(\specp))
  =(1\otimes M(\specp))(M(\spec)\otimes 1)
  R_{\square\,\square'}(\spec-\specp)\,.
\end{align}
In what follows we explain this definition. The \emph{R-matrix}
\begin{align}
  \label{eq:yangian-def-r}
  \begin{aligned}
    R_{\square\,\square'}(\spec-\specp)
    &=
    1+(\spec-\specp)^{-1}\sum_{\indnm{A},\indnm{B}}\elemm_{\indnm{AB}}\otimes \elemm'_{\indnm{BA}}(-1)^{|\indnm{B}|}
    =
    \,\,\,\\\phantom{}
  \end{aligned}
  \begin{aligned}
    \begin{tikzpicture}
      \draw[thick,densely dashed,
      decoration={
        markings, mark=at position 0.85 with {\arrow{latex reversed}}},
      postaction={decorate}
      ] 
      (0,0) 
      node[left] {$\square,\spec$} -- 
      (1,0);
      \draw[thick,densely dashed,
      decoration={
        markings, mark=at position 0.85 with {\arrow{latex reversed}}},
      postaction={decorate}
      ] 
      (0.5,-0.5) 
      node[below] {$\square',\specp$} -- 
      (0.5,0.5);
    \end{tikzpicture}
  \end{aligned}
  \begin{aligned}
    ,\\\phantom{}
  \end{aligned}
\end{align}
acts on two copies of the defining representation
$\square\otimes\square'=\mathbb{C}^{n|m}\otimes\mathbb{C}^{n|m}$, see
e.g.\ \cite{Kulish:1985bj}. In the bosonic case it is attributed to
Yang. The prime at the defining generator $\elemm'_{\indnm{BA}}$
merely emphasizes that it acts on the space $\square'$. The two spaces
also enter the definition of the Yangian in
\eqref{eq:yangian-def}. They are called \emph{auxiliary spaces} and in
the graphical notation established in \eqref{eq:yangian-def-r} we
associate a dashed line with each of them. The R-matrix depends on the
complex \emph{spectral parameters} $\spec$ and $\specp$, each one
belonging to one of the representation spaces. They are also included
in the graphical representation. Note that the identity operator on
$\mathbb{C}^{n|m}$ may be written as
$\sum_{\indnm{A}} \elemm_{\indnm{AA}}$. The $1$ in
\eqref{eq:yangian-def-r} stands for the appropriate identity operator
on the tensor product. The R-matrix \eqref{eq:yangian-def-r} is a
solution of the \emph{Yang-Baxter equation}
\begin{align}
  \label{eq:yangian-ybe-def}
  \begin{aligned}
    &R_{\square\,\square'}(\spec-\spec')
    R_{\square\,\square''}(\spec-\spec'')
    R_{\square'\,\square''}(\spec'-\spec'')\\
    &=
    R_{\square'\,\square''}(\spec'-\spec'')
    R_{\square\,\square''}(\spec-\spec'')
    R_{\square\,\square'}(\spec-\spec')\,,
  \end{aligned}
\end{align}
which acts in the tensor product $\square\otimes \square'\otimes
\square''$. Graphically it reads
\begin{align}
  \label{eq:yangian-ybe-def-graphical}
  \begin{aligned}
    \begin{tikzpicture}
      \draw[thick,densely dashed,
      decoration={
        markings, mark=at position 0.9 with {\arrow{latex reversed}}},
      postaction={decorate}
      ]
      (-0.25,0.5) 
      node[left] {$\square,\spec$} -- 
      (1.75,0.5);
      \draw[thick,densely dashed,
      decoration={
        markings, mark=at position 0.85 with {\arrow{latex reversed}}},
      postaction={decorate}
      ]
      (0,0) 
      node[below] {$\square',\spec'$} -- 
      (1.5,1.5);
      \draw[thick,densely dashed,
      decoration={
        markings, mark=at position 0.85 with {\arrow{latex reversed}}},
      postaction={decorate}
      ]
      (1.5,0)
      node[below] {$\square'',\spec''$}  -- 
      (0,1.5);
    \end{tikzpicture}
  \end{aligned}
  \begin{aligned}
    \,\,\,=\,\,\,\\\phantom{}
  \end{aligned}
  \begin{aligned}
    \begin{tikzpicture}
      \draw[thick,densely dashed,
      decoration={
        markings, mark=at position 0.9 with {\arrow{latex reversed}}},
      postaction={decorate}
      ]
      (-0.25,1) 
      node[left] {$\square,\spec$} -- 
      (1.75,1);
      \draw[thick,densely dashed,
      decoration={
        markings, mark=at position 0.85 with {\arrow{latex reversed}}},
      postaction={decorate}
      ]
      (0,0) 
      node[below] {$\square',\spec'$} -- 
      (1.5,1.5);
      \draw[thick,densely dashed,
      decoration={
        markings, mark=at position 0.85 with {\arrow{latex reversed}}},
      postaction={decorate}
      ]
      (1.5,0) 
      node[below] {$\square'',\spec''$} -- 
      (0,1.5);
    \end{tikzpicture}
  \end{aligned}
  \begin{aligned}
    .\\\phantom{}
  \end{aligned}
\end{align}
The arrows of the lines define an orientation which translates into
the order in which the R-matrices act. R-matrices ``earlier'' on the
line are right of ``later'' ones in the corresponding formula.
Comparing with \eqref{eq:yangian-ybe-def}, we may interpret the
definition of the Yangian in \eqref{eq:yangian-def} as a Yang-Baxter
equation where the third space is left unspecified. The R-matrices
that would act on this space are replaced by the operator valued
\emph{monodromy matrix} $M(u)$. This matrix contains the infinitely
many generators $M_{\indnm{AB}}^{(l)}$ with $l=1,2,3,\ldots$ of the
Yangian. They are obtained from an expansion in the spectral parameter
\begin{align}
  \label{eq:yangian-mono}
  M(u)=\sum_{\indnm{A},\indnm{B}}\elemm_{\indnm{AB}}M_{\indnm{AB}}(u)(-1)^{|\indnm{B}|}\,,\quad
  M_{\indnm{AB}}(u)=M^{(0)}_{\indnm{AB}}+u^{-1}M^{(1)}_{\indnm{AB}}+u^{-2}M^{(2)}_{\indnm{AB}}+\ldots
\end{align}
with the normalization
\begin{align}
  \label{eq:yangian-norm}
  M^{(0)}_{\indnm{AB}}=\delta_{\indnm{AB}}(-1)^{|\indnm{B}|}.
\end{align}
The degree of $M_{\indnm{AB}}(u)$ is $|\indnm{A}|+|\indnm{B}|$. With
\eqref{eq:yangian-mono} the defining relation \eqref{eq:yangian-def}
is equivalent to
\begin{align}
  \label{eq:yangian-def-mono-elem}
  \begin{aligned}
  &(u'-u)[M_{\indnm{AB}}(u),M_{\indnm{CD}}(u')\}\\
  &\quad=
  (M_{\indnm{CB}}(u)M_{\indnm{AD}}(u')-M_{\indnm{CB}}(u')M_{\indnm{AD}}(u))
  (-1)^{|\indnm{A}||\indnm{D}|+|\indnm{A}||\indnm{C}|+|\indnm{C}||\indnm{D}|}\,.
  \end{aligned}
\end{align}
After expanding the monodromy elements this reads
\begin{align}
  \label{eq:yangian-def-gen}
  \begin{aligned}
    &[M_{\indnm{AB}}^{(k)},M_{\indnm{CD}}^{(l)}\}\\
    &\quad=\sum_{q=1}^{\text{min}(k,l)}
    (M_{\indnm{CB}}^{(k+l-q)}M_{\indnm{AD}}^{(q-1)}-M_{\indnm{CB}}^{(q-1)}M_{\indnm{AD}}^{(k+l-q)})
    (-1)^{|\indnm{A}||\indnm{D}|+|\indnm{A}||\indnm{C}|+|\indnm{D}||\indnm{C}|}\,.
  \end{aligned}
\end{align}
These quadratic relations provide an explicit definition of the
Yangian algebra \eqref{eq:yangian-def} in terms of its generators
$M_{\indnm{AB}}^{(l)}$.

We remark that \eqref{eq:yangian-def-gen} manifestly displays a
filtration of the Yangian algebra, with regard to which the generators
$M_{\indnm{AB}}^{(l)}$ are of level $l$, cf.\
\cite{Molev:2007}.\footnote{This notion of ``level'' differs from the
  one mentioned in the context of scattering amplitudes right before
  \eqref{eq:sym-alg-level-2}.} In addition, one easily deduces from
\eqref{eq:yangian-def-gen} that all generators $M_{\indnm{AB}}^{(l)}$
with $l>2$ can be expressed via $M_{\indnm{AB}}^{(1)}$ and
$M_{\indnm{AB}}^{(2)}$. By setting $r=s=1$ we see that the
$M_{\indnm{BA}}^{(1)}$ satisfy the $\mathfrak{gl}(n|m)$ algebra
\eqref{eq:gl-superalg}.  Furthermore, expanding
\eqref{eq:yangian-def-mono-elem} only in the spectral parameter $u$
leads to
\begin{align}
  \label{eq:yangian-gen-adjoint}
  [M^{(1)}_{\indnm{AB}},M_{\indnm{CD}}(u')\}
  =
  \delta_{\indnm{AD}}M_{\indnm{CB}}(u')
  -
  (-1)^{(|\indnm{A}|+|\indnm{C}|)(|\indnm{B}|+|\indnm{D}|)}
  \delta_{\indnm{CB}}M_{\indnm{AD}}(u')\,.
\end{align}
Thus the monodromy elements transform in the adjoint representation of
$\mathfrak{gl}(n|m)$.

Having introduced the Yangian, we study realizations of its
defining relation \eqref{eq:yangian-def} where the generators
$M_{\indnm{AB}}(u)$ act on the tensor product
$\mathcal{V}_1\otimes\cdots\otimes \mathcal{V}_N$ of
$\mathfrak{gl}(n|m)$ representations. This space is referred to as
\emph{quantum space}. Let us introduce the \emph{Lax operator}
\begin{align}
  \label{eq:yangian-def-lax}
  \begin{aligned}
    R_{\square\,\mathcal{V}_i}(\spec-\inh_i)
    &=
    f_{\mathcal{V}_i}(\spec-\inh_i)
    \Bigg(1+(\spec-\inh_i)^{-1}\sum_{\indnm{A},\indnm{B}}\elemm_{\indnm{AB}} J_{\indnm{BA}}^i(-1)^{|\indnm{B}|}\Bigg)
    =
    \,\,\,\\\phantom{}
  \end{aligned}
  \begin{aligned}
    \begin{tikzpicture}
      \draw[thick,densely dashed,
      decoration={
        markings, mark=at position 0.85 with {\arrow{latex reversed}}},
      postaction={decorate}
      ] 
      (0,0) 
      node[left] {$\square,\spec$} -- 
      (1,0);
      \draw[thick,
      decoration={
        markings, mark=at position 0.85 with {\arrow{latex reversed}}},
      postaction={decorate}
      ] 
      (0.5,-0.5) 
      node[below] {$\mathcal{V}_i,\inh_i$} -- 
      (0.5,0.5);
    \end{tikzpicture}
  \end{aligned}
  \begin{aligned}
    \\\phantom{}
  \end{aligned}
\end{align}
acting on the tensor product $\square\otimes\mathcal{V}_i$ of an
auxiliary and a local quantum space, which we indicate graphically by
a solid line. This Lax operator is characterized as the solution of a
Yang-Baxter equation like \eqref{eq:yangian-ybe-def-graphical}, where
the third space is replaced by $\mathcal{V}_i$. Hence it is already a
solution of \eqref{eq:yangian-def}. Clearly this equation does not
determine the scalar normalization $f_{\mathcal{V}_i}$. Note that up
to a change in this normalization the complex inhomogeneity parameter
$v_i$ can be altered by applying the $\mathfrak{gl}(n|m)$ automorphism
\eqref{eq:auto-shift} to the generators $J_{\indnm{BA}}^i$. Further
solutions of \eqref{eq:yangian-def} are obtained by multiplying
multiple Lax operators acting in different quantum spaces,
\begin{align}
  \label{eq:yangian-mono-spinchain}
  \begin{aligned}
    \mon(\spec)
    =
    R_{\square\,\mathcal{V}_1}(\spec-\inh_1)
    \cdots R_{\square\,\mathcal{V}_\sites}(\spec-\inh_\sites)
    =
    \,\,\,\\\phantom{}
  \end{aligned}
  \begin{aligned}
    \begin{tikzpicture}
      \draw[thick,densely dashed,
      decoration={
        markings, mark=at position 0.95 with {\arrow{latex reversed}}},
      postaction={decorate}
      ] 
      (0,0) 
      node[left] {$\square,\spec$} -- 
      (3,0);
      \draw[thick,
      decoration={
        markings, mark=at position 0.85 with {\arrow{latex reversed}}},
      postaction={decorate}
      ] 
      (0.5,-0.5) 
      node[below] {$\mathcal{V}_1,\inh_1$} -- 
      (0.5,0.5);
      \node at (1.5,-0.25) {$\ldots$};
      \draw[thick,
      decoration={
        markings, mark=at position 0.85 with {\arrow{latex reversed}}},
      postaction={decorate}
      ] 
      (2.5,-0.5) 
      node[below] {$\mathcal{V}_\sites,\inh_\sites$} -- 
      (2.5,0.5);
    \end{tikzpicture}
  \end{aligned}
  \begin{aligned}
    .\\\phantom{}
  \end{aligned}
\end{align}
This can be proved using the so-called ``train argument'' which makes
use of the Yang-Baxter equations for the individual Lax
operators. Notice that \eqref{eq:yangian-norm} imposes a condition on
the normalizations $f_{\mathcal{V}_i}$ in the Lax operators. It is
readily satisfied by setting $f_{\mathcal{V}_i}=1$. However, at times
we will use non-trivial normalization factors. The monodromy matrix in
\eqref{eq:yangian-mono-spinchain} is that of an \emph{inhomogeneous
  spin chain} with $N$ sites. Here the meaning of the word
``inhomogeneous'' is twofold. First, we associate a inhomogeneity
$\inh_i$ with each site. Second, each site carries a different
representation $\mathcal{V}_i$. The quantum space is the state space
of the spin chain. In section \ref{sec:bethe-ansatze} we will discuss
how the Heisenberg spin chain can be obtained from such a monodromy
matrix. From now on we focus on realizations of the Yangian with a
monodromy of the form \eqref{eq:yangian-mono-spinchain}.

Expanding the monodromy \eqref{eq:yangian-mono-spinchain} according to
\eqref{eq:yangian-mono} leads to expressions for the Yangian
generators in terms of $\mathfrak{gl}(n|m)$ generators. With
$f_{\mathcal{V}_i}=1$ this results in
\begin{align}
  \label{eq:yangian-mono-coeff-glnm}
  M_{\indnm{AB}}^{(1)}=\sum_{i=1}^N J_{\indnm{BA}}^i\,,\quad
  M_{\indnm{AB}}^{(2)}=\sum_{i=1}^N v_iJ_{\indnm{BA}}^i+\sum_{\substack{i,j=1\\i<j}}^N\sum_{\indnm{C}} (-1)^{|\indnm{C}|}J_{\indnm{BC}}^jJ_{\indnm{CA}}^i\,,\quad
  \ldots\,.
\end{align}
Let us also discuss a different way of expanding the monodromy
elements,
\begin{align}
  \label{eq:yangian-exp-alternative}
  \begin{aligned}
    M(u)
    &=1+u^{-1}M^{(1)}+u^{-2}M^{(2)}+\ldots\\
    &=\exp{\Big(u^{-1}M^{[1]}+u^{-2}M^{[2]}+\ldots\Big)}\,.
  \end{aligned}
\end{align}
The matrix elements $M_{\indnm{AB}}^{[l]}$ of the new expansion
coefficients $M^{[l]}$ are defined analogously to those of the
original coefficients $M^{(l)}$ in \eqref{eq:yangian-mono}. We obtain
\begin{align}
  \label{eq:yangian-diff-gen-rel}
  M^{[1]}=M^{(1)}\,,\quad
  M^{[2]}=M^{(2)}-\frac{1}{2}M^{(1)}M^{(1)}\,,\quad\ldots\,.
\end{align}
Using \eqref{eq:yangian-mono-coeff-glnm} this allows us to compute the
explicit form of the generators also for this expansion,
\begin{align}
  \label{eq:yangian-diff-gen-expl}
  \begin{aligned}
    M_{\indnm{AB}}^{[1]}&=\sum_{i=1}^N J_{\indnm{BA}}^i\,,\\
    M_{\indnm{AB}}^{[2]}&=\sum_{i=1}^N \Big(v_iJ_{\indnm{BA}}^i-\frac{1}{2}\sum_{\indnm{C}}(-1)^{|\indnm{A}||\indnm{B}|+|\indnm{A}||\indnm{C}|+|\indnm{B}||\indnm{C}|}J_{\indnm{CA}}^iJ_{\indnm{BC}}^i\Big)\\
    &\quad+\frac{1}{2}\sum_{\substack{i,j=1\\i<j}}^N\sum_{\indnm{C}} (-1)^{|\indnm{C}|}\Big(J_{\indnm{BC}}^jJ_{\indnm{CA}}^i-J_{\indnm{BC}}^iJ_{\indnm{CA}}^j\Big)\,,\\
    \ldots\,.
  \end{aligned}
\end{align}
This is precisely the form of the Yangian generators which we
encountered in \eqref{eq:sym-alg-level-1} and
\eqref{eq:sym-alg-level-2} in the discussion of $\mathcal{N}=4$ SYM
scattering amplitudes.

The definition \eqref{eq:yangian-def} of the Yangian in terms of a
monodromy matrix $M(u)$ is often called RTT-relation.\footnote{The
  name has its origin in the frequent use of the symbol ``$T(u)$'' for
  the monodromy $M(u)$ in the literature.} The generators
$M_{\indnm{AB}}^{[l]}$ that we obtained from the expansion in
\eqref{eq:yangian-exp-alternative} are essentially what is known as
Drinfeld's first realization of the Yangian \cite{Drinfeld:1985rx}. We
did not discuss Drinfeld's second realization
\cite{Drinfeld:1987sy}. It is based on a Gauß decomposition of the
monodromy matrix $M(u)$ into the product of an upper triangular, a
diagonal and a lower triangular matrix. The generators of the second
realization are obtained from an expansion of the elements of these
matrices in the spectral parameter, c.f.\ \cite{Brundan2005}. Let us
add that in this section we did not emphasize the quantum group and in
particular the Hopf algebra structure of the Yangian. It can be nicely
phrased in the QISM language employed here, see e.g.\
\cite{Chari:1995}.

\section{Yangian Invariance}
\label{sec:yangian-inv}

Recall from sections~\ref{sec:symmetries} and \ref{sec:deform} that
tree-level amplitudes $\mathcal{A}^{(\text{tree})}_{N,K}$ of
$\mathcal{N}=4$ SYM and deformations
$\mathcal{A}^{(\text{def.})}_{N,K}$ thereof are Yangian invariant. How
does this translate into the QISM language? A state $|\Psi\rangle$ in
the quantum space $\mathcal{V}_1\otimes\cdots\otimes \mathcal{V}_N$ is
\emph{Yangian invariant} iff
\begin{align}
  \label{eq:yi}
  M(u)|\Psi\rangle=|\Psi\rangle\,.
\end{align}
On the level of matrix elements this translates
into\footnote{Analogous equations were shown to be satisfied by the
  physical vacuum state of integrable two-dimensional quantum field
  theories in \cite{deVega:1984wk,Destri:1993qh}.}
\begin{align}
  \label{eq:yi-components}
  M_{\indnm{AB}}(u)|\Psi\rangle=(-1)^{|\indnm{B}|}\delta_{\indnm{AB}}|\Psi\rangle\,,
\end{align}
where we used \eqref{eq:yangian-mono}. With the help of
\eqref{eq:yangian-mono-spinchain} we represent \eqref{eq:yi}
graphically as
\begin{align}
  \label{eq:yi-inv-rmm-pic}
  \begin{aligned}
    \begin{tikzpicture}
      \draw[thick,densely dashed,
      decoration={
        markings, mark=at position 0.95 with {\arrow{latex reversed}}},
      postaction={decorate}
      ] 
      (0,0) 
      node[left=0.4cm] {$\square,\spec$}
      node[left] {} -- 
      (3,0)
      node[right] {};
      \draw[thick,
      decoration={
        markings, mark=at position 0.85 with {\arrow{latex reversed}}},
      postaction={decorate}
      ] 
      (0.5,-0.5) 
      node[below] {$\mathcal{V}_1,\inh_1$} -- 
      (0.5,0.5)
      node[above] {\phantom{$\mathcal{V}_1,\inh_1$}};
      \node at (1.5,-0.25) {$\ldots$};
      \draw[thick,
      decoration={
        markings, mark=at position 0.85 with {\arrow{latex reversed}}},
      postaction={decorate}
      ] 
      (2.5,-0.5) 
      node[below] {$\mathcal{V}_\sites,\inh_\sites$} -- 
      (2.5,0.5);
      \draw (1.5,1) 
      node[minimum height=1cm,minimum width=2.5cm,draw,
      thick,rounded corners=8pt,densely dotted] 
      {$|\Psi\rangle$};
      \path
      (0,2) 
      node[left] {\phantom{$\square,\spec$}} -- (3,2);
    \end{tikzpicture}
  \end{aligned}
  \,\,\,=
  \begin{aligned}
    \begin{tikzpicture}
      \draw[thick,densely dashed,
      decoration={
        markings, mark=at position 0.95 with {\arrow{latex reversed}}},
      postaction={decorate}
      ] 
      (0,2) 
      node[left=0.4cm] {$\square,\spec$}
      node[left] {} -- 
      (3,2)
      node[right] {};
      \draw[thick,
      decoration={
        markings, mark=at position 0.85 with {\arrow{latex reversed}}},
      postaction={decorate}
      ] 
      (0.5,-0.5) 
      node[below] {$\mathcal{V}_1,\inh_1$} -- 
      (0.5,0.5)
      node[above] {\phantom{$\mathcal{V}_1,\inh_1$}};
      \node at (1.5,-0.25) {\ldots};
      \draw[thick,
      decoration={
        markings, mark=at position 0.85 with {\arrow{latex reversed}}},
      postaction={decorate}
      ] 
      (2.5,-0.5) 
      node[below] {$\mathcal{V}_\sites,\inh_\sites$} -- 
      (2.5,0.5);
      \draw (1.5,1) 
      node[minimum height=1cm,minimum width=2.5cm,draw,
      thick,rounded corners=8pt,densely dotted] 
      {$|\Psi\rangle$};
    \end{tikzpicture}
  \end{aligned}\,.
\end{align}
Here the dashed line on the right hand side corresponds to an identity
operators acting on the auxiliary space $\square$. The invariant
$|\Psi\rangle$ itself is symbolized by a dotted ``black box'' without
specifying the interior. During the course of this thesis we will
construct solutions of the Yangian invariance condition and so to
speak learn about the structure of this interior. In order to make
contact with the notion of Yangian invariance used in
sections~\ref{sec:symmetries} and \ref{sec:deform}, we expand
\eqref{eq:yi} in the spectral parameter. Using the expansion in the
first line of \eqref{eq:yangian-exp-alternative} we obtain
\begin{align}
  \label{eq:yi-exp-1}
  M_{\indnm{AB}}^{(l)}|\Psi\rangle=0
\end{align}
for all $l=1,2,3,\ldots\,$. The expansion in the second line of
\eqref{eq:yangian-exp-alternative} yields
\begin{align}
  \label{eq:yi-exp-2}
  M_{\indnm{AB}}^{[l]}|\Psi\rangle=0
\end{align}
for all $l=1,2,3,\ldots\,$. Each set of conditions,
\eqref{eq:yi-exp-1} as well as \eqref{eq:yi-exp-2}, provides a
definition of Yangian invariance that is equivalent to
\eqref{eq:yi}. They show that $|\Psi\rangle$ is a one-dimensional
representation of the Yangian as it is annihilated by all its
generators. A further simplification is possible. Arguing as in the
paragraph after \eqref{eq:yangian-def-gen}, one shows that if
\eqref{eq:yi-exp-1} holds for $l=1$ and $l=2$, it is satisfied for all
$l$. The same is true for \eqref{eq:yi-exp-2}. This is essentially the
definition of Yangian invariance used for the amplitudes
$\mathcal{A}^{(\text{tree})}_{N,K}$ in \eqref{eq:sym-alg-annih-l1} and
\eqref{eq:sym-alg-annih-l2}, and for the deformed amplitudes
$\mathcal{A}^{(\text{def.})}_{N,K}$ in \eqref{eq:yi-def}. Thus we know
already something about the ``black box'' $|\Psi\rangle$ in
\eqref{eq:yi-inv-rmm-pic}. For the representation
\eqref{eq:sym-gen-gl44} of $\mathfrak{sl}(4|4)$ in terms of spinor
helicity variables, it may be identified with the amplitudes
$\mathcal{A}^{(\text{tree})}_{N,K}$ or the deformed amplitudes
$\mathcal{A}^{(\text{def.})}_{N,K}$. While the definition of Yangian
invariance in \eqref{eq:yi-exp-2} allows to make contact with the
literature on scattering amplitudes, let us emphasize the advantage of
\eqref{eq:yi}. It represents the Yangian invariance condition as an
eigenvalue problem for monodromy matrix elements and thus makes the
powerful QISM toolbox applicable. In particular, the formulation
\eqref{eq:yi} will be exploited in section~\ref{sec:bethe-yangian},
where this equation is solved using an algebraic Bethe ansatz.

We conclude with a comment on ``Yangian invariance'' or ``Yangian
symmetry'' of spin chain models. In this context both terms are used
synonymously and signify that the Hamiltonian of a model commutes with
the generators of a Yangian algebra. The Hamiltonians of the original
Heisenberg model with spin~$\frac{1}{2}$ $\mathfrak{su}(2)$~symmetry
and $\mathfrak{su}(n)$ generalizations thereof contain only nearest
neighbor interactions. They can be derived from a monodromy of the
form \eqref{eq:yangian-mono-spinchain}, cf.\
section~\ref{sec:bethe-ansatze} below. Nonetheless, the Yangian
invariance is spoiled for a finite number of sites $N$ by boundary
terms, cf.\ \cite{Bernard:1992ya}. For an infinite number of sites a
Yangian symmetry was also discovered in the Hubbard model
\cite{Uglov:1993jy}, see also \cite{Essler:2010}. Furthermore, for an
infinite number of sites there are classes of integrable spin chains
with long-range interactions that exhibit Yangian symmetry, see
e.g. \cite{Beisert:2007jv}. The Hubbard model as well as such
long-range spin chains are of relevance for describing the multi-loop
dilatation operator in planar $\mathcal{N}=4$ SYM, see the review
\cite{Rej:2010ju} and recall section~\ref{sec:spectrum}. The
Haldane-Shastry chain \cite{Haldane1988,Shastry1988} is a long-range
model with Yangian symmetry even for a finite number of sites. A class
of further models with Yangian invariance at finite length was
investigated recently in \cite{Finkel:2015}, see also the references
therein.

\section{Oscillator Representations}
\label{sec:osc-rep}

In the previous sections we discussed the Yangian of
$\mathfrak{gl}(n|m)$ and the Yangian invariance condition on an
algebraic level. Here we introduce those representations of the
$\mathfrak{gl}(n|m)$ algebra that we will employ at the sites of the
spin chain monodromy \eqref{eq:yangian-mono-spinchain} defining the
Yangian. We work with certain classes of unitary representations of
the non-compact algebra
$\mathfrak{u}(p,q|m)\subset\mathfrak{gl}(p+q|m)$ that are constructed
in terms of bosonic and fermionic oscillator algebras. Such
representations have a long history in the physics literature and are
sometimes referred to as ``ladder representations'', see e.g.\
\cite{Todorov:1966zz} for the bosonic case. We follow the presentation
of \cite{Bars:1982ep} which includes the generalization to
superalgebras. Our primary interest for the study of Yangian
invariants comes from planar $\mathcal{N}=4$ SYM scattering
amplitudes. As already mentioned above, for the algebra
$\mathfrak{u}(2,2|4)$ these oscillator representations are unitarily
equivalent to the realization in terms of spinor helicity variables in
\eqref{eq:sym-gen-gl44}. This equivalence will be shown below in
section~\ref{sec:osc-spinor}. For now, however, we discuss the
oscillator representations in the general setting of the algebra
$\mathfrak{u}(p,q|m)$.

The basic ingredient is a family of
\emph{superoscillators} obeying
\begin{align}
  \label{eq:super-osc}
  \begin{aligned}
    [\mathbf{A}_{\indnm{A}},\bar{\mathbf{A}}_{\indnm{B}}\}=\delta_{\indnm{AB}}\,,\quad
    [\mathbf{A}_{\indnm{A}},\mathbf{A}_{\indnm{B}}\}=0\,,\quad
    [\bar{\mathbf{A}}_{\indnm{A}},\bar{\mathbf{A}}_{\indnm{B}}\}=0\,,\quad
  \end{aligned}  
\end{align}
where the indices of the annihilation operators
$\mathbf{A}_{\indnm{A}}$ and creation operators
$\bar{\mathbf{A}}_{\indnm{A}}$ take the values
$\indnm{A}=1,\ldots,n+m$. It is equipped with a
conjugation $\dagger$ and acts on a Fock space $\mathcal{F}$ that is
spanned by monomials in $\bar{\mathbf{A}}_{\indnm{A}}$ acting on a
vacuum state $|0\rangle$,
\begin{align}
  \label{eq:super-osc-fock-conj}
    \mathbf{A}_{\indnm{A}}^\dagger=\bar{\mathbf{A}}_{\indnm{A}}\,,\quad
    \mathbf{A}_{\indnm{A}}|0\rangle=0\,.
\end{align}
The $\mathfrak{gl}(n|m)$ algebra \eqref{eq:gl-superalg} can be
realized as
\begin{align}
  \label{eq:gen-comp}
  \mathbf{J}_{\indnm{AB}}=\bar{\mathbf{A}}_{\indnm{A}}\mathbf{A}_{\indnm{B}}\,.
\end{align}
From now on, we mark $\mathfrak{gl}(n|m)$ generators which are
realized in terms of oscillators by bold letters. This construction
yields unitary representations of the \emph{compact} algebra
$\mathfrak{u}(n|m)\subset\mathfrak{gl}(n|m)$, cf.\ \cite{Bars:1982ep}.

We proceed to oscillator representations of the \emph{non-compact}
algebra $\mathfrak{u}(p,q|m)$.  For this we split the family of
superoscillators with $\mathfrak{gl}(n|m)$ index $\indnm{A}$ into two
parts. One carries a $\mathfrak{gl}(p|r)$ index
$\indssub{A}=1,\ldots,p+r$ and the other one a $\mathfrak{gl}(q|s)$
index $\dot{\indssub{A}}=p+r+1,\ldots,p+r+q+s$ with $p+q=n$ and
$r+s=m$. The degrees $|\indssub{A}|$ and $|\dot{\indssub{A}}|$ of
these indices can be inferred from $|\indnm{A}|$ specified in
\eqref{eq:grading}. For the annihilation operators this reads
\begin{align}
  \label{eq:osc-split}
  \left(\mathbf{A}_{\indnm{A}}\right)=
  \left(
    \begin{array}{c}    
      \vcenter{\vspace{1.2cm}}\mathbf{A}_{\indssub{A}}\\[0.3em]
      \hdashline\\[-1.0em]
      \vcenter{\vspace{1.2cm}}\mathbf{A}_{\dot{\indssub{A}}}\\
    \end{array}
  \right)=
    \left(
    \begin{array}{c}    
      \mathbf{a}_{\alpha}\\[0.3em]
      \hdashline\\[-1.0em]
      \mathbf{c}_{a}\\[0.3em]
      \hdashline\\[-1.0em]
      \mathbf{b}_{\dot{\alpha}}\\[0.3em]
      \hdashline\\[-1.0em]
      \mathbf{d}_{\dot{a}}\\
    \end{array}
  \right)\,.
\end{align}
Here we introduced an additional piece of notation that will be used
at times. We spelled out the superoscillators
$\mathbf{A}_{\indssub{A}}$ in terms of bosonic oscillators
$\mathbf{a}_\alpha$ and fermionic $\mathbf{c}_a$ with
$\alpha=1,\ldots,p$ and $a=1,\ldots,r$. In the same way
$\mathbf{A}_{\dot{\indssub{A}}}$ is written using bosonic
$\mathbf{b}_{\dot{\alpha}}$ and fermionic $\mathbf{d}_{\dot{a}}$ with
$\dot{\alpha}=1,\ldots,q$ and $\dot{a}=1,\ldots,s$. Analogous notation
applies for the creation operators
$\bar{\mathbf{A}}_{\indnm{A}}$. This terminology is inspired by
\cite{Ferro:2013dga} and \cite{Beisert:2003jj}. Next, one verifies
that the ``particle-hole'' transformation
\begin{align}
    \left(
    \begin{array}{c}    
      \mathbf{A}_{\indssub{A}}\\[0.3em]
      \hdashline\\[-1.0em]
      \mathbf{A}_{\dot{\indssub{A}}}\\
    \end{array}
  \right)
  \mapsto
    \left(
    \begin{array}{c}    
      \mathbf{A}_{\indssub{A}}\\[0.3em]
      \hdashline\\[-1.0em]
      \bar{\mathbf{A}}_{\dot{\indssub{A}}}\\
    \end{array}
  \right)\,,\quad
    \left(
    \begin{array}{c}    
      \bar{\mathbf{A}}_{\indssub{A}}\\[0.3em]
      \hdashline\\[-1.0em]
      \bar{\mathbf{A}}_{\dot{\indssub{A}}}\\
    \end{array}
  \right)
  \mapsto
    \left(
    \begin{array}{c}    
      \bar{\mathbf{A}}_{\indssub{A}}\\[0.3em]
      \hdashline\\[-1.0em]
      -(-1)^{|\dot{\indssub{A}}|}\mathbf{A}_{\dot{\indssub{A}}}\\
    \end{array}
  \right)\,
\end{align}
is an automorphism of the superoscillator algebra
\eqref{eq:super-osc}. Note, however, that it breaks
\eqref{eq:super-osc-fock-conj}. This is essential for the transition
from a compact to a non-compact algebra, because otherwise the reality
conditions of the $\mathfrak{gl}(n|m)$ generators
$\mathbf{J}_{\indnm{AB}}$ would not be altered. Applying this
automorphism to the generators \eqref{eq:gen-comp}
yields\footnote{Compared to \cite{Kanning:2014cca} we changed the
  position of the minus signs even in the purely bosonic case.}
\begin{align}
  \label{eq:gen-ordinary}
  (\mathbf{J}_{\indnm{AB}})=
  \left(
  \begin{array}{c:c}
    \mathbf{J}_{\indssub{A}\indssub{B}}&\mathbf{J}_{\indssub{A}\dot{\indssub{B}}}\\[0.3em]
    \hdashline\\[-1.0em]
    \mathbf{J}_{\dot{\indssub{A}}\indssub{B}}&\mathbf{J}_{\dot{\indssub{A}}\dot{\indssub{B}}}\\
  \end{array}
  \right)=
  \left(
  \begin{array}{c:c}
    \bar{\mathbf{A}}_{\indssub{A}} \mathbf{A}_{\indssub{B}}&
    \bar{\mathbf{A}}_{\indssub{A}} \bar{\mathbf{A}}_{\dot{\indssub{B}}}\\[0.3em]
    \hdashline\\[-1.0em]
    -(-1)^{|\dot{\indssub{A}}|}\mathbf{A}_{\dot{\indssub{A}}}\mathbf{A}_{\indssub{B}}&
    -(-1)^{|\dot{\indssub{A}}|}\mathbf{A}_{\dot{\indssub{A}}}\bar{\mathbf{A}}_{\dot{\indssub{B}}}\\
  \end{array}
  \right)\,.
\end{align}
Notice that the blocks $\mathbf{J}_{\indssub{AB}}$ and
$\mathbf{J}_{\dot{\indssub{A}}\dot{\indssub{B}}}$ realize the
subalgebras $\mathfrak{gl}(p|r)$ and $\mathfrak{gl}(q|s)$,
respectively. Let $\oscrep_c\subset\mathcal{F}$ be the eigenspace
of the central element 
\begin{align}
  \label{eq:central-ord}
  \mathbf{C}=\tr(\mathbf{J}_{\indnm{AB}})
  =\sum_{\indssub{A}}\bar{\mathbf{A}}_{\indssub{A}}\mathbf{A}_{\indssub{A}}
  -\sum_{\dot{\indssub{A}}}(-1)^{|\dot{\indssub{A}}|}\mathbf{A}_{\dot{\indssub{A}}}\bar{\mathbf{A}}_{\dot{\indssub{A}}}
\end{align}
with eigenvalue $c$. For each $c\in\mathbb{Z}$ this
infinite-dimensional space forms a unitary representation of the
algebra $\mathfrak{u}(p,q|r+s)$, see \cite{Bars:1982ep}. Hence we may
interpret $c$ as a representation label. The space $\oscrep_c$
contains a lowest weight state, which by definition is annihilated by
all $\mathbf{J}_{\indnm{AB}}$ with $\indnm{A}>\indnm{B}$, i.e.\ by the
strictly lower triangular entries of the matrix
\eqref{eq:gen-ordinary}. Notice that in the special case $q=0$ or
$p=0$ the space $\oscrep_c$ is finite-dimensional.

According to \eqref{eq:yi-exp-1}, Yangian invariants are in particular
$\mathfrak{gl}(n|m)$ singlet states. For such states to exist, we need
also spin chain sites with representations that are dual to the class
of representations $\oscrep_c$. We define such \emph{dual
  representations} with generators
$\bar{\mathbf{J}}_{\indnm{AB}}=-(-1)^{|\indnm{A}|+|\indnm{A}||\indnm{B}|}\mathbf{J}_{\indnm{AB}}^\dagger$
that are obtained from \eqref{eq:gen-ordinary} employing the
automorphism \eqref{eq:auto-dual},
\begin{align}
  \label{eq:gen-dual}
  (\bar{\mathbf{J}}_{\indnm{AB}})=
  \left(
  \begin{array}{c:c}
    \bar{\mathbf{J}}_{\indssub{AB}}&\bar{\mathbf{J}}_{\indssub{A}\dot{\indssub{B}}}\\[0.3em]
    \hdashline\\[-1.0em]
    \bar{\mathbf{J}}_{\dot{\indssub{A}}\indssub{B}}&\bar{\mathbf{J}}_{\dot{\indssub{A}}\dot{\indssub{B}}}\\
  \end{array}
  \right)=
  \left(
  \begin{array}{c:c}
    -(-1)^{|\indssub{A}|+|\indssub{A}||\indssub{B}|}\bar{\mathbf{A}}_{\indssub{B}} \mathbf{A}_{\indssub{A}}&
    -(-1)^{|\indssub{A}|+|\dot{\indssub{B}}|+|\indssub{A}||\dot{\indssub{B}}|}\mathbf{A}_{\dot{\indssub{B}}}\mathbf{A}_{\indssub{A}}\\[0.3em]
    \hdashline\\[-1.0em]
    (-1)^{|\dot{\indssub{A}}|+|\dot{\indssub{A}}||\indssub{B}|}\bar{\mathbf{A}}_{\indssub{B}} \bar{\mathbf{A}}_{\dot{\indssub{A}}}&
    (-1)^{|\dot{\indssub{A}}|+|\dot{\indssub{B}}|+|\dot{\indssub{A}}||\dot{\indssub{B}}|}\mathbf{A}_{\dot{\indssub{B}}}\bar{\mathbf{A}}_{\dot{\indssub{A}}}\\
  \end{array}
  \right)\,.
\end{align}
We denote by $\bar{\oscrep}_c\subset\mathcal{F}$ the eigenspace of
the central element
\begin{align}
  \label{eq:central-dual}
  \bar{\mathbf{C}}=
  \tr{(\bar{\mathbf{J}}_{\indnm{AB}})}=
  -\sum_{\indssub{A}}\bar{\mathbf{A}}_{\indssub{A}}\mathbf{A}_{\indssub{A}}
  +\sum_{\dot{\indssub{A}}}(-1)^{|\dot{\indssub{A}}|}\mathbf{A}_{\dot{\indssub{A}}}\bar{\mathbf{A}}_{\dot{\indssub{A}}}\,
\end{align}
with eigenvalue $c$. For each $c\in\mathbb{Z}$ this space carries a
unitary representation of $\mathfrak{u}(p,q|r+s)$. The representation
$\bar{\oscrep}_c$ is dual to $\oscrep_{-c}$. It contains a
highest weight state, which is annihilated by all
$\bar{\mathbf{J}}_{\indnm{AB}}$ with $\indnm{A}<\indnm{B}$. In case of
$q=0$ or $p=0$ the space $\bar{\oscrep}_c$ is finite
dimensional. Notice that for the compact generators
\eqref{eq:gen-comp} the automorphism \eqref{eq:auto-dual} does agree
with \eqref{eq:auto-other}. However, for the non-compact generators
\eqref{eq:gen-ordinary} it does not. 

Having defined the two classes of oscillator representations allows us
to use them at the sites of the monodromy $M(u)$ in
\eqref{eq:yangian-mono-spinchain}. At each site we chose either an
``ordinary'' representation $\oscrep_{c_i}$ with generators
$J_{\indnm{AB}}^i=\mathbf{J}_{\indnm{AB}}^i$ or a ``dual''
representation $\bar{\oscrep}_{c_i}$ with
$J_{\indnm{AB}}^i=\bar{\mathbf{J}}_{\indnm{AB}}^i$. The monodromy
$M(u)$, and hence the representation of the Yangian, is completely
specified by $2N$ parameters, i.e.\ $N$ inhomogeneities
$v_i\in\mathbb{C}$ and $N$ representation labels
$c_i\in\mathbb{Z}$. We remark that the tensor product decomposition of
the oscillator representations employed at the spin chain sites was
studied in \cite{Kashiwara:1978} for the $\mathfrak{u}(p,q)$ case, see
also e.g.\ \cite{Anderson:1968,Anderson:1968a} for exemplary results.

Let us add a comment on the necessity of the two classes of
representations $\oscrep_c$ and $\bar{\oscrep}_c$. In
section~\ref{sec:symmetries} the $\mathfrak{gl}(4|4)$ generators
\eqref{eq:sym-gen-gl44}, which are expressed in terms of the spinors
$\lambda$ and $\tilde{\lambda}$, seem to look alike at all $N$ sites
of the amplitude. However, taking into account the reality condition
\eqref{eq:spinors-real} for the spinors,
$\tilde{\lambda}=\pm\overline{\lambda}$, there are two different kinds
of generators. In section~\ref{sec:osc-spinor} we will show that these
correspond to the two classes of representations introduced here.

\section{Sample Invariants}
\label{sec:sample-inv}

\subsection{Compact Bosonic Invariants}
\label{sec:comp-boson-invar}

We have the Yangian algebra and oscillator representations of the
non-compact superalgebra $\mathfrak{u}(p,q|m)$ at our
disposal. Therefore we could in principle start looking for solutions
$|\Psi\rangle$ of the Yangian invariance condition \eqref{eq:yi} for
these representations. However, to gain a better understanding of the
oscillator formalism, we choose to concentrate on the special case of
the compact bosonic algebra $\mathfrak{u}(n)\subset\mathfrak{gl}(n)$
for the time being. We present some details on the oscillator
representations and the Lax operators in this case in
section~\ref{sec:deta-oscill-repr}. This is followed in
section~\ref{sec:yang-invar-as} by a reformulation of the Yangian
invariance condition as an intertwining relation and thereby
emphasizing its interpretation as a Yang-Baxter-like equation. The
remaining sections contain sample Yangian invariants. The structure of
these invariants is relatively simple. They are polynomials in the
creation operators acting on the Fock vacuum because the
representations we are dealing with are finite-dimensional. We will
encounter these compact bosonic sample invariants again in
section~\ref{sec:bethe-yangian} where we construct them using a Bethe
ansatz. 

Let us remark that these sample invariants with finite-dimensional
$\mathfrak{u}(n)$ representations can be brought into a form which
makes them look very much akin to tree-level $\mathcal{N}=4$ SYM
amplitudes with infinite-dimensional $\mathfrak{psu}(2,2|4)$
representations, see \cite{Frassek:2013xza}. This reformulation
involves the Graßmannian integral formulation of the amplitudes in
terms of supertwistors, which is similar to that with spinor helicity
variables reviewed in section~\ref{sec:grassmannian-integral}. By the
nature of the differing representations involved the argument is
necessarily somewhat formal. Nevertheless, it provides further
motivation for the study of compact bosonic sample invariants.

\subsubsection{Details on Representations and Lax Operators}
\label{sec:deta-oscill-repr}

After introducing the classes of oscillator representations
$\oscrep_c$ and $\bar{\oscrep}_{c}$ of the non-compact
superalgebra $\mathfrak{u}(p,q|r+s)\subset\mathfrak{gl}(n|m)$ in
section~\ref{sec:osc-rep}, we present additional details on these
representations for the special case $q=r=s=0$. Thus we are
concentrating on unitary representations of
$\mathfrak{u}(n)\subset\mathfrak{gl}(n)$. In this case the
$\mathfrak{gl}(n)$ generators~\eqref{eq:gen-ordinary} associated with
the class $\oscrep_c$ and \eqref{eq:gen-dual} of
$\bar{\oscrep}_{c}$ reduce to, respectively,
\begin{align}
  \label{eq:osc-gen-s-bs}
    \mathbf{J}_{\alpha\beta}=\bar\osca_{\alpha}\osca_{\beta}\,,\quad
    \bar{\mathbf{J}}_{\alpha\beta}=-\bar{\osca}_{\beta}\osca_{\alpha}
\end{align}
with $\alpha,\beta=1,\ldots,n$. Here we used
\eqref{eq:osc-split} to express the generators in terms of
\emph{bosonic oscillators},
\begin{align}
  \label{eq:osc-alg-bos}
    [\mathbf{a}_{\alpha},\bar{\mathbf{a}}_{\beta}]=\delta_{\alpha\beta}\,,\quad
    [\mathbf{a}_{\alpha},\mathbf{a}_{\beta}]=0\,,\quad
    [\bar{\mathbf{a}}_{\alpha},\bar{\mathbf{a}}_{\beta}]=0\,,\quad
    \mathbf{a}_{\alpha}^\dagger=\bar{\mathbf{a}}_{\alpha}\,,\quad
    \mathbf{a}_{\alpha}|0\rangle=0\,,
\end{align}
where the brackets denote the commutator. A review of these
realizations of the $\mathfrak{gl}(n)$ algebra, which are sometimes
said to be of Jordan-Schwinger-type, may be found e.g.\ in
\cite{Biedenharn:1981}. The central elements \eqref{eq:central-ord}
and \eqref{eq:central-dual} become, up to a sign, number operators,
\begin{align}
  \label{eq:central-comp-bos}
    \mathbf{C}=
    \sum_{\alpha=1}^n\bar{\mathbf{a}}_{\alpha}\mathbf{a}_{\alpha}\,,\quad
    \bar{\mathbf{C}}=
    -\sum_{\alpha=1}^n\bar{\mathbf{a}}_{\alpha}\mathbf{a}_{\alpha}\,.
\end{align}
In the non-compact case their eigenvalues $c$ can be arbitrary
integers. From \eqref{eq:central-comp-bos} we conclude that in the
compact case for the class $\oscrep_c$ we have $c\in \mathbb{N}$,
while for $\bar{\oscrep}_c$ one needs $c\in-\mathbb{N}$. Hence the
representation spaces $\oscrep_c$ and $\bar{\oscrep}_{-c}$ with
$c\in\mathbb{N}$ are the finite-dimensional subspace of the Fock space
$\mathcal{F}$ consisting of polynomials of degree $c$ in the creation
operators $\bar{\mathbf{a}}_\alpha$ acting on $|0\rangle$. Both
representations posses a \emph{highest weight state},
\begin{align}
  \label{eq:osc-hws}
  \begin{aligned}
    |\sigma\rangle=(\bar\osca_1)^c|0\rangle\in\oscrep_c\,,\quad
    |\bar\sigma\rangle=(\bar\osca_n)^c|0\rangle\in\bar{\oscrep}_{-c}\,.
  \end{aligned}
\end{align}
These states are characterized by
\begin{align}
  \label{eq:osc-hws-char}
  \begin{aligned}
    \mathbf{J}_{\alpha\beta}|\sigma\rangle&=0\quad\text{for}\quad \alpha<\beta\,,&
    \bar{\mathbf{J}}_{\alpha\beta}|\bar\sigma\rangle&=0\quad\text{for}\quad \alpha<\beta\,,\\
    \mathbf{J}_{\alpha\alpha}|\sigma\rangle&=c\,\delta_{1\,\alpha}|\sigma\rangle\,,&
    \bar{\mathbf{J}}_{\alpha\alpha}|\bar\sigma\rangle&=-c\,\delta_{n\,\alpha}|\bar\sigma\rangle\,.\quad
  \end{aligned}
\end{align}
Because both representations are finite-dimensional, they also contain
a lowest weight state, that we do not state here. The
$\mathfrak{gl}(n)$ weight $\Xi=(\xi^{(1)},\ldots,\xi^{(n)})$ of a
state $|\phi\rangle$ is defined by
$J_{\alpha\alpha}|\phi\rangle=\xi^{(\alpha)}|\phi\rangle$. Therefore,
from \eqref{eq:osc-hws-char} we read off the highest weights of the
two classes of oscillator representations,
\begin{align}
  \label{eq:comp-osc-weights}
  \Xi=(c,0,\ldots,0)\quad\text{for}\quad\oscrep_c\,,\quad
  \Xi=(0,\ldots,0,-c)\quad\text{for}\quad\bar{\oscrep}_{-c}\,.
\end{align}
From its highest weight we may identify $\oscrep_c$ with the class of
totally symmetric representations of $\mathfrak{gl}(n)$. Notice that
in the non-compact setting only $\bar{\oscrep}_{-c}$ has a highest
weight state, while $\oscrep_c$ contains merely a lowest weight state.
Knowing that in the compact case both types of oscillators
representations have a highest weight will be of great importance in
section~\ref{sec:bethe-yangian}. The Bethe ansatz employed there for
the construction of Yangian invariants makes crucial use of highest
weight states. For completeness, we recall from
section~\ref{sec:osc-rep} that $\bar{\oscrep}_{-c}$ is the \emph{dual
  representation} to $\oscrep_{c}$ is the sense that the generators
\eqref{eq:osc-gen-s-bs} satisfy
\begin{align}
  \label{eq:osc-gen-conj}
  \bar{\mathbf{J}}_{\alpha\beta}
  =-\mathbf{J}_{\alpha\beta}^\dagger\,.
\end{align}

Next, we provide some details on the Lax
operators~\eqref{eq:yangian-def-lax} for the compact bosonic
oscillator representations. With \eqref{eq:osc-gen-s-bs} they read
\begin{align}
  \label{eq:osc-lax-fund-s}
  &\begin{aligned}
    R_{\square\,\oscrep_c}(\spec-\inh)
    &=
    f_{\oscrep_c}(\spec-\inh)
    \Bigg(1+(\spec-\inh)^{-1}\sum_{\alpha,\beta=1}^n\elemm_{\alpha\beta}\bar\osca_\beta\osca_\alpha\Bigg)
    =
    \,\,\,\\\phantom{}
  \end{aligned}
  \begin{aligned}
    \begin{tikzpicture}
      \draw[thick,densely dashed,
      decoration={
        markings, mark=at position 0.85 with {\arrow{latex reversed}}},
      postaction={decorate}
      ] 
      (0,0) 
      node[left] {$\square,\spec$} -- 
      (1,0);
      \draw[thick,
      decoration={
        markings, mark=at position 0.85 with {\arrow{latex reversed}}},
      postaction={decorate}
      ] 
      (0.5,-0.5) 
      node[below] {$\oscrep_c,\inh$} -- 
      (0.5,0.5);
    \end{tikzpicture}
  \end{aligned}
  \begin{aligned}
    ,\\\phantom{}
  \end{aligned}
  \\
  \label{eq:osc-lax-fund-bs}
  &\begin{aligned}
    R_{\square\,\bar{\oscrep}_{-c}}(\spec-\inh)
    &=
    f_{\bar{\oscrep}_{-c}}(\spec-\inh)
    \Bigg(1-(\spec-\inh)^{-1}\sum_{\alpha,\beta=1}^n\elemm_{\alpha\beta}\bar\osca_\alpha\osca_\beta\Bigg)
    =
    \,\,\,\\\phantom{}
  \end{aligned}
  \begin{aligned}
    \begin{tikzpicture}
      \draw[thick,densely dashed,
      decoration={
        markings, mark=at position 0.85 with {\arrow{latex reversed}}},
      postaction={decorate}
      ] 
      (0,0) 
      node[left] {$\square,\spec$} -- 
      (1,0);
      \draw[thick,
      decoration={
        markings, mark=at position 0.85 with {\arrow{latex reversed}}},
      postaction={decorate}
      ] 
      (0.5,-0.5) 
      node[below] {$\bar{\oscrep}_{-c},\inh$} -- 
      (0.5,0.5);
    \end{tikzpicture}
  \end{aligned}
  \begin{aligned}
    .\\\phantom{}
  \end{aligned}
\end{align}
We severely constrain the normalizations
$f_{\bar{\oscrep}_{-c}}(u)$ and $f_{\oscrep_{c}}(u)$ by
demanding that each of the Lax operators satisfies two additional
equations, a unitarity and a crossing relation. The \emph{unitarity
  relations} read
\begin{align}
  \label{eq:osc-unitarity}
  R_{\square\,\oscrep_{c}}(\spec-\inh)R_{\oscrep_{c}\,\square}(\inh-\spec)=1\,,
  \quad
  R_{\square\,\bar{\oscrep}_{-c}}(\spec-\inh)R_{\bar{\oscrep}_{-c}\,\square}(\inh-\spec)=1\,.
\end{align}
They contain the two further Lax operators
\begin{align}
  \label{eq:osc-lax-sbs-fund}
  \begin{aligned}
    R_{\oscrep_{c}\,\square}(\inh-\spec)
    =
    \,\,\,\\\phantom{}
  \end{aligned}
  \begin{aligned}
    \begin{tikzpicture}
      \draw[thick,
      decoration={
        markings, mark=at position 0.85 with {\arrow{latex reversed}}},
      postaction={decorate}
      ] 
      (0,0) 
      node[left] {$\oscrep_{c},\inh$} -- 
      (1,0);
      \draw[thick,densely dashed,
      decoration={
        markings, mark=at position 0.85 with {\arrow{latex reversed}}},
      postaction={decorate}
      ] 
      (0.5,-0.5) 
      node[below] {$\square,\spec$} -- 
      (0.5,0.5);
    \end{tikzpicture}
  \end{aligned}
  \begin{aligned}
    ,\\\phantom{}
  \end{aligned}
  \quad
  \begin{aligned}
    R_{\bar{\oscrep}_{-c}\,\square}(\inh-\spec)
    =
    \,\,\,\\\phantom{}
  \end{aligned}
  \begin{aligned}
    \begin{tikzpicture}
      \draw[thick,
      decoration={
        markings, mark=at position 0.85 with {\arrow{latex reversed}}},
      postaction={decorate}
      ] 
      (0,0) 
      node[left] {$\bar{\oscrep}_{-c},\inh$} -- 
      (1,0);
      \draw[thick,densely dashed,
      decoration={
        markings, mark=at position 0.85 with {\arrow{latex reversed}}},
      postaction={decorate}
      ] 
      (0.5,-0.5) 
      node[below] {$\square,\spec$} -- 
      (0.5,0.5);
    \end{tikzpicture}
  \end{aligned}
  \begin{aligned}
    ,\\\phantom{}
  \end{aligned}
\end{align}
where the order of auxiliary and quantum spaces is exchanged. These
Lax operators solve the Yang-Baxter equation in
$\square\otimes \oscrep_{c}\otimes \square$ and
$\square\otimes \bar{\oscrep}_{-c}\otimes \square$, respectively. In
these equations they are the only unknowns, cf.\
\eqref{eq:yangian-ybe-def}. The Lax operators
\eqref{eq:osc-lax-sbs-fund} can be written in terms of those in
\eqref{eq:osc-lax-fund-s} and \eqref{eq:osc-lax-fund-bs},
\begin{align}
  \label{eq:osc-lax-symm}
  \begin{aligned}
    R_{\oscrep_{c}\,\square}(\inh-\spec)
    &=
    R_{\square\,\oscrep_{c}}(\inh-\spec-c+1)\,,\\
    R_{\bar{\oscrep}_{-c}\,\square}(\inh-\spec)
    &=
    R_{\square\,\bar{\oscrep}_{-c}}(\inh-\spec+n+c-1)\,.
  \end{aligned}
\end{align}
In this sense they are symmetric up to a shift of the spectral
parameter. Form the unitarity conditions in \eqref{eq:osc-unitarity}
we obtain constraints on the normalization of the Lax operators,
\begin{gather}
  \label{eq:osc-norm-unitarity}
  f_{\oscrep_{c}}(\spec)f_{\oscrep_{c}}(-\spec-c+1)
  =
  \frac{\spec(\spec+c-1)}{\spec(\spec+c-1)-c}\,,\quad
  f_{\bar{\oscrep}_{-c}}(\spec)f_{\bar{\oscrep}_{-c}}(-\spec+c-1+n)=1\,.
\end{gather}
The \emph{crossing relations} that we demand are
\begin{align}
  \label{eq:osc-crossing}
  R_{\square\,\bar{\oscrep}_{-c}}(\spec+\kappa_{\oscrep_{c}})
  =
  R_{\oscrep_{c}\,\square}(-\spec)^\dagger\,,\quad
  R_{\square\,\oscrep_{c}}(\spec+\kappa_{\bar{\oscrep}_{-c}})
  =
  R_{\bar{\oscrep}_{-c}\,\square}(-\spec)^\dagger\,,
\end{align}
where $\kappa_{\oscrep_{c}}$ and $\kappa_{\bar{\oscrep}_{-c}}$
are called crossing parameters and the conjugation only acts on the
oscillators. These equations impose
\begin{align}
  \label{eq:osc-crossing-norm}
  \kappa_{\oscrep_{c}}=c-1\,,
  \quad
  \kappa_{\bar{\oscrep}_{-c}}=-c+1-n\,,
  \quad
  f_{\bar{\oscrep}_{-c}}(\spec)=f_{\oscrep_c}(-\spec)\,.
\end{align}
Note that the two equations in \eqref{eq:osc-crossing} lead only to
one condition on the normalizations. A solution of
\eqref{eq:osc-norm-unitarity} and \eqref{eq:osc-crossing-norm} is
given by the well-known formula
\begin{align}
  \label{eq:osc-norm-solution}
  f_{\oscrep_c}(\spec)
  =
  \frac{
    \Gamma\big(\frac{1-\spec}{n}\big)
    \Gamma\big(\frac{n+\spec}{n}\big)
  }
  {
    \Gamma\big(\frac{1-c-\spec}{n}\big)
    \Gamma\big(\frac{n+c+\spec}{n}\big)
  }\,.
\end{align}
In case of $c=1$ it was obtained in \cite{Berg:1977dp}. For higher
integer values of $c$ it can be constructed by means of the recursion
relation
\begin{align}
  \label{eq:osc-norm-add-eq}
  f_{\oscrep_c}(\spec) f_{\oscrep_{c'}}(\spec+c)=f_{\oscrep_{c+c'}}(\spec)\,.
\end{align}
Let us remark that the solution \eqref{eq:osc-norm-solution} is not
unique, see once again \cite{Berg:1977dp}.

\subsubsection{Yangian Invariance as Yang-Baxter Equation}
\label{sec:yang-invar-as}

Here we present a reformulation of the Yangian invariance condition
\eqref{eq:yi} as a Yang-Baxter-like equation. By way of example, we
consider a monodromy \eqref{eq:yangian-mono} where the first $K$ sites
of the total quantum space carry a dual representation of the class
$\bar{\oscrep}_c$ and the remaining $N-K$ sites carry an ordinary
one of the type $\oscrep_c$. We denote such a monodromy by
$M_{N,K}(u)$ and an associated solution of \eqref{eq:yi} will be
labeled $|\Psi_{N,K}\rangle$. This is motivated by the notation
employed for the amplitudes $\mathcal{A}_{N,K}^{(\text{tree})}$ in
section~\ref{sec:amplitudes}. The monodromy reads
\begin{align}
  \label{eq:yi-mono-split}
  \begin{aligned}
    \mon_{N,K}(\spec)
    =
    &R_{\square\,\bar{\oscrep}_{c_1}}(\spec-\inh_1)\cdots R_{\square\,\bar{\oscrep}_{c_K}}(\spec-\inh_K)\\
    &\cdot R_{\square\,\oscrep_{c_{K+1}}}(\spec-\inh_{K+1})\cdots R_{\square\,\oscrep_{c_{N}}}(\spec-\inh_N)\,.
  \end{aligned}
\end{align}
Because we are in the compact case the representation labels satisfy
$c_1,\ldots,c_K\in-\mathbb{N}$ and
$c_{K+1},\ldots,c_{N}\in\mathbb{N}$. The Yangian invariance condition
\eqref{eq:yi} can be interpreted as an intertwining relation of the
tensor product of the first $K$ with the remaining $N-K$ spaces of the
total quantum space. To obtain this relation we conjugate
\eqref{eq:yi} in the first $K$ spaces and use \eqref{eq:osc-unitarity}
and \eqref{eq:osc-crossing} for these spaces. This yields
\begin{align}
  \label{eq:yi-intertwiner}
  \begin{aligned}
    &R_{\square\,\oscrep_{c_{K+1}}}(\spec-\inh_{K+1})\cdots R_{\square\,\oscrep_{c_{N}}}(\spec-\inh_N)
    \mathcal{O}_{\Psi_{N,K}}\\
    &=
    \mathcal{O}_{\Psi_{N,K}}
    R_{\square\,\oscrep_{-c_K}}(\spec-\inh_K+\kappa_{\bar{\oscrep}_{c_{K}}})\cdots
    R_{\square\,\oscrep_{-c_1}}(\spec-\inh_1+\kappa_{\bar{\oscrep}_{c_{1}}})\,,
  \end{aligned}
\end{align}
where
$\mathcal{O}_{\Psi_{N,K}}:=|\Psi_{N,K}\rangle^{\dagger_1\cdots\dagger_\dsites}$.
It is represented graphically as
\begin{align}
  \label{eq:yi-intertwiner-pic}
  \begin{aligned}
    \begin{tikzpicture}
      \draw[thick,densely dashed,
      decoration={
        markings, mark=at position 0.95 with {\arrow{latex reversed}}},
      postaction={decorate}
      ] 
      (0,0) 
      node[left] {$\square,\spec$} -- 
      (2.5,0);
      \node at (1.25,-0.25) {$\ldots$};
      \node at (1.25,2.25) {$\ldots$};
      \draw[thick,
      decoration={
        markings, mark=at position 0.85 with {\arrow{latex reversed}}},
      postaction={decorate}
      ] 
      (0.5,-0.5) 
      node[left=0.25cm,below] {$\oscrep_{c_{K+1}},\inh_{K+1}$} -- 
      (0.5,0.5);
      \draw[thick,
      decoration={
        markings, mark=at position 0.85 with {\arrow{latex reversed}}},
      postaction={decorate}
      ] 
      (2,-0.5) 
      node[right=0.25cm,below] {$\oscrep_{c_N},\inh_N$} -- 
      (2,0.5);
      \draw[thick,
      decoration={
        markings, mark=at position 0.85 with {\arrow{latex reversed}}},
      postaction={decorate}
      ] 
      (2,1.5) -- 
      (2,2.5)
      node[right=1.0cm,above] {$\oscrep_{-c_1},\inh_1-\kappa_{\bar{\oscrep}_{c_1}}$};
      \draw[thick,
      decoration={
        markings, mark=at position 0.85 with {\arrow{latex reversed}}},
      postaction={decorate}
      ] 
      (0.5,1.5) -- 
      (0.5,2.5)
      node[left=0.5cm,above] {$\oscrep_{-c_K},\inh_\dsites-\kappa_{\bar{\oscrep}_{c_K}}$};
      \draw (1.25,1) 
      node[minimum height=1cm,minimum width=2cm,draw,
      thick,rounded corners=8pt,densely dotted] 
      {$\mathcal{O}_{\Psi_{N,K}}$};
    \end{tikzpicture}
  \end{aligned}
  =
  \begin{aligned}
    \begin{tikzpicture}
      \draw[thick,densely dashed,
      decoration={
        markings, mark=at position 0.95 with {\arrow{latex reversed}}},
      postaction={decorate}
      ] 
      (0,2) 
      node[left] {$\square,\spec$} -- 
      (2.5,2);
      \node at (1.25,-0.25) {$\ldots$};
      \node at (1.25,2.25) {$\ldots$};
      \draw[thick,
      decoration={
        markings, mark=at position 0.85 with {\arrow{latex reversed}}},
      postaction={decorate}
      ] 
      (0.5,-0.5) 
      node[left=0.25cm,below] {$\oscrep_{c_{K+1}},\inh_{K+1}$} -- 
      (0.5,0.5);
      \draw[thick,
      decoration={
        markings, mark=at position 0.85 with {\arrow{latex reversed}}},
      postaction={decorate}
      ] 
      (2,-0.5) 
      node[right=0.25cm,below] {$\oscrep_{c_N},\inh_N$} -- 
      (2,0.5);
      \draw[thick,
      decoration={
        markings, mark=at position 0.85 with {\arrow{latex reversed}}},
      postaction={decorate}
      ] 
      (2,1.5) -- 
      (2,2.5)
      node[right=1.0cm,above] {$\oscrep_{-c_1},\inh_1-\kappa_{\bar{\oscrep}_{c_1}}$};
      \draw[thick,
      decoration={
        markings, mark=at position 0.85 with {\arrow{latex reversed}}},
      postaction={decorate}
      ] 
      (0.5,1.5) -- 
      (0.5,2.5)
      node[left=0.5cm,above] {$\oscrep_{-c_K},\inh_K-\kappa_{\bar{\oscrep}_{c_K}}$};
      \draw (1.25,1) 
      node[minimum height=1cm,minimum width=2cm,draw,
      thick,rounded corners=8pt,densely dotted] 
      {$\mathcal{O}_{\Psi_{N,K}}$};
    \end{tikzpicture}
  \end{aligned}.
\end{align}
The analogy of this equation with the Yang-Baxter equation will allow
us to give a graphical interpretation of some sample intertwiners
$\mathcal{O}_{\Psi_{N,K}}$ similar to the one for the R-matrix in
\eqref{eq:yangian-def-r}. An equation like \eqref{eq:yi-intertwiner}
was formulated in \cite{Ferro:2013dga} in the context of deformed
amplitudes of $\mathcal{N}=4$ SYM.

\subsubsection{Two-Site Invariant and Identity Operator}
\label{sec:comp-bos-2-site}

\begin{figure}[!t]
  \begin{center}  
    \begin{align*}
      \begin{aligned}
        \mathcal{O}_{\Psi_{2,1}}=\hspace{-.5cm}
      \end{aligned}
      \begin{aligned}
        \begin{tikzpicture}
          \draw[thick,rounded corners=10pt,
          decoration={
            markings, 
            mark=at position 0.95 with {\arrow{latex reversed}},
            mark=at position 0.08 with {\arrow{latex reversed}}
          },
          postaction={decorate}]
          (1.25,0) node[below] {$\oscrep_{c_2},v_2$}
          -- (1.25,3.5) node[above] {$\oscrep_{-c_1},v_2$};
        \end{tikzpicture}
      \end{aligned}
      \quad\quad
      \begin{aligned}
        \mon_{2,1}(\spec)=\,\,\,\\\vphantom{}
      \end{aligned}
      \begin{aligned}
        \begin{tikzpicture}
          \draw[densely dashed,thick,
          decoration={
            markings, mark=at position 0.95 with {\arrow{latex reversed}}},
          postaction={decorate}]
          (0,0)
          node[left] {$\square,\spec$} -- 
          (2.5,0);
          \draw[thick,
          decoration={
            markings, mark=at position 0.85 with {\arrow{latex reversed}}},
          postaction={decorate}] 
          (0.5,-0.5)
          node[below] {$\bar{\oscrep}_{c_1},\inh_1$} -- 
          (0.5,0.5); 
          \draw[thick,
          decoration={
            markings, mark=at position 0.85 with {\arrow{latex reversed}}},
          postaction={decorate}] 
          (2.0,-0.5) 
          node[below] {$\vphantom{\bar{\oscrep}_{c_1}}\oscrep_{c_2},\inh_2$} -- 
          (2.0,0.5); 
        \end{tikzpicture}
      \end{aligned}
    \end{align*}
    \caption{The left side shows a graphical depiction of the
      intertwiner version $\mathcal{O}_{\Psi_{2,1}}$ found in
      \eqref{eq:osc-opsi21} of the Yangian invariant vector
      $|\Psi_{2,1}\rangle$. This intertwiner is essentially an
      identity operator and is thus represented by a line. On the
      right we display the monodromy $M_{2,1}(u)$ belonging to the
      Yangian invariant. The representation labels and inhomogeneities
      of the intertwiner as well as the monodromy obey
      \eqref{eq:osc-m21-vs} to ensure Yangian invariance.}
    \label{fig:osc-psi21}
  \end{center} 
\end{figure} 

At this point everything is set up to construct the first and simplest
sample solution of the Yangian invariance condition~\eqref{eq:yi}. We
consider a monodromy of the form \eqref{eq:yi-mono-split} with $N=2$
sites out of which $K=1$ are dual,
\begin{align}
  \label{eq:osc-m21}
  \mon_{2,1}(\spec)
  =
  R_{\square\,\bar{\oscrep}_{c_1}}(\spec-\inh_1)
  R_{\square\,\oscrep_{c_2}}(\spec-\inh_2)\,,
\end{align}
see also figure~\ref{fig:osc-psi21}. The Lax operators of this
monodromy are given explicitly in \eqref{eq:osc-lax-fund-s} and
\eqref{eq:osc-lax-fund-bs}. Recall from \eqref{eq:yi-exp-1} that the
Yangian invariant $|\Psi_{2,1}\rangle$ we want to construct is in
particular a $\mathfrak{gl}(n)$ singlet. Hence we pick a monodromy
where one site carries an ordinary representation $\oscrep_{c_2}$ and
the other one a dual representation $\bar{\oscrep}_{c_1}$. In trying
to solve \eqref{eq:yi} with this monodromy we find that a solution
only exists if the inhomogeneities and representation labels obey
\begin{align}
  \label{eq:osc-m21-vs}
    \inh_1=\inh_2-n-c_2+1\,,
    \quad
    c_1+c_2=0\,.
\end{align}
The latter equation was to be expected because it guarantees that the
two representations are in fact dual to each other, cf.\ the
discussion around \eqref{eq:osc-gen-conj}. The normalization of the
monodromy \eqref{eq:osc-m21} derives from those of the Lax operators
in \eqref{eq:osc-lax-fund-s} and \eqref{eq:osc-lax-fund-bs}. It turns
out to be trivial,
\begin{align}
  \label{eq:osc-m21-norm}
  f_{\bar{\oscrep}_{c_1}}(\spec-\inh_1)f_{\oscrep_{c_2}}(\spec-\inh_2)=1\,.
\end{align}
This is shown using \eqref{eq:osc-m21-vs}, the unitarity condition for
$f_{\bar{\oscrep}_{-c}}(\spec)$ in \eqref{eq:osc-norm-unitarity}
and the relation between the normalizations
$f_{\oscrep_{c}}(\spec)$ and $f_{\bar{\oscrep}_{-c}}(\spec)$
in \eqref{eq:osc-crossing-norm}. At this stage the solution of
\eqref{eq:yi} is easily shown to be
\begin{align}
  \label{eq:osc-psi21}
  |\Psi_{2,1}\rangle
  =
  (1\bullet 2)^{c_2}|0\rangle
  \quad
  \text{with}
  \quad
  (l\bullet k)=\sum_{\alpha=1}^n\bar{\mathbf{a}}_\alpha^k\bar{\mathbf{a}}_\alpha^l\,.
\end{align} 
Here the upper indices on the oscillators indicate the site of the
monodromy. The invariant $|\Psi_{2,1}\rangle$ is unique up to a scalar
prefactor, which evidently drops out of \eqref{eq:yi}. 

The intertwiner belonging to the invariant $|\Psi_{2,1}\rangle$ is
obtained from \eqref{eq:yi-intertwiner} with $\dsites=1$ and using
$\kappa_{\bar{\oscrep}_{c_1}}$ given in
\eqref{eq:osc-crossing-norm}. We are led to
\begin{align}
  \label{eq:osc-int-psi21}  
  R_{\square\,\oscrep_{c_2}}(\spec-\inh_2)\mathcal{O}_{\Psi_{2,1}}
  =
  \mathcal{O}_{\Psi_{2,1}}R_{\square\,\oscrep_{-c_1}}(\spec-\inh_2)
\end{align}
with
\begin{align}
  \label{eq:osc-opsi21}
  \mathcal{O}_{\Psi_{2,1}}:=|\Psi_{2,1}\rangle^{\dagger_1}
  =
  \sum_{\alpha_1,\ldots,\alpha_{c_2}=1}^n
  \!\!\!
  \bar\osca^2_{\alpha_1}\cdots\bar\osca^2_{\alpha_{c_2}}
  |0\rangle\langle 0|
  \osca^1_{\alpha_1}\cdots\osca^1_{\alpha_{c_2}}\,.
\end{align} 
The intertwiner $\mathcal{O}_{\Psi_{2,1}}$ may be understood as $c_2!$
times the identity operator because with $c_1=-c_2$ in
\eqref{eq:osc-m21-vs} the representation spaces $\oscrep_{-c_1}$
and $\oscrep_{c_2}$ can be identified. Hence we represent
$\mathcal{O}_{\Psi_{2,1}}$ graphically in figure~\ref{fig:osc-psi21}
by a line. In analogy we may also think of the ``black box'' for the
invariant $|\Psi_{2,1}\rangle$ in the graphical version
\eqref{eq:yi-inv-rmm-pic} of the Yangian invariance condition as a
line.

\subsubsection{Three-Site Invariants and Bootstrap Equations}
\label{sec:comp-bos-3-site}

\begin{figure}[!t]
  \begin{center}
    \begin{align*}
      \begin{aligned}
        \mathcal{O}_{\Psi_{3,1}}=\hspace{-.5cm}
      \end{aligned}
      \begin{aligned}
        \begin{tikzpicture}
          \draw[thick,rounded corners=10pt,
          decoration={
            markings, 
            mark=at position 0.14 with {\arrow{latex reversed}}
          },
          postaction={decorate}] 
          (0.5,0) node[below] {$\oscrep_{c_2},v_2$\phantom{aa}}
          -- (0.5,1) 
          -- (1.25,1.75);
          \draw[thick,rounded corners=10pt,
          decoration={
            markings, 
            mark=at position 0.14 with {\arrow{latex reversed}}
          },
          postaction={decorate}] 
          (2,0) node[below] {\phantom{aa}$\oscrep_{c_3},v_2-c_2$}
          -- (2,1) 
          -- (1.25,1.75);
          \draw[thick,rounded corners=10pt,
          decoration={
            markings, 
            mark=at position 0.9 with {\arrow{latex reversed}}
          },
          postaction={decorate}]
          (1.25,1.75)
          -- (1.25,3.5) node[above] {$\oscrep_{-c_1},v_2$};
        \end{tikzpicture}
      \end{aligned}
      \quad\quad
      \begin{aligned}
        \mon_{3,1}(\spec)=\,\,\,\\\vphantom{}
      \end{aligned}
      \begin{aligned}
        \begin{tikzpicture}
          \draw[densely dashed,thick,
          decoration={
            markings, mark=at position 0.97 with {\arrow{latex reversed}}},
          postaction={decorate}]
          (0,0)
          node[left] {$\square,\spec$} --
          (4,0);
          \draw[thick,
          decoration={markings,
            mark=at position 0.85 with {\arrow{latex reversed}}},
          postaction={decorate}] 
          (0.5,-0.5)
          node[below] {$\vphantom{\bar{\oscrep}_{c_i}}\bar{\oscrep}_{c_1},\inh_1$} -- 
          (0.5,0.5);
          \draw[thick,
          decoration={markings,
            mark=at position 0.85 with {\arrow{latex reversed}}},
          postaction={decorate}] 
          (2.0,-0.5)
          node[below] {$\vphantom{\bar{\oscrep}_{c_i}}\oscrep_{c_2},\inh_2$} -- 
          (2.0,0.5);
          \draw[thick,
          decoration={markings,
            mark=at position 0.85 with {\arrow{latex reversed}}},
          postaction={decorate}] 
          (3.5,-0.5) 
          node[below] {$\vphantom{\bar{\oscrep}_{c_i}}\oscrep_{c_3},\inh_3$} -- 
          (3.5,0.5);
        \end{tikzpicture}
      \end{aligned}
    \end{align*}
    \caption{The left part represents the intertwiner
      $\mathcal{O}_{\Psi_{3,1}}$ that we computed in
      \eqref{eq:osc-opsi31}. It is represented by a trivalent
      vertex. To its right we depict the monodromy $M_{3,1}(u)$
      associated with the Yangian invariant $|\Psi_{3,1}\rangle$. The
      necessary constraints on the representation labels and
      inhomogeneities may be found in \eqref{eq:osc-m31-vs}.}
    \label{fig:osc-psi31}
  \end{center} 
\end{figure}
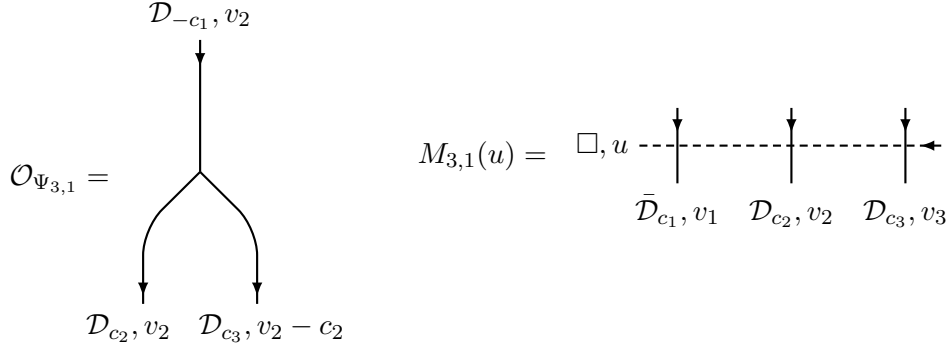

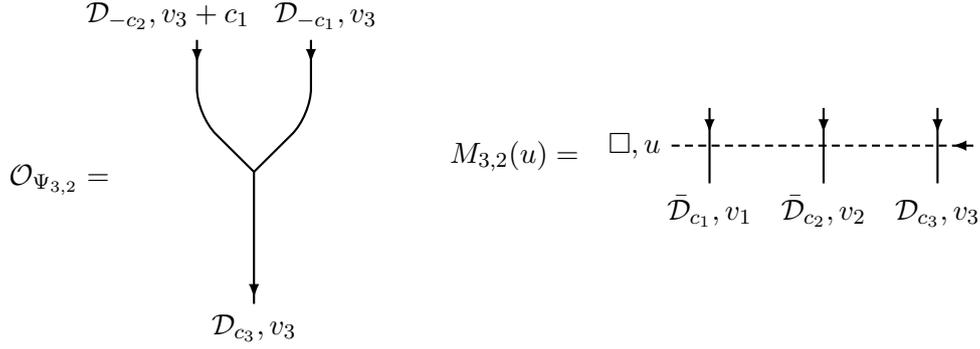
\begin{figure}[!t]
  \begin{center}
    \begin{align*}
      \begin{aligned}
        \mathcal{O}_{\Psi_{3,2}}=\hspace{-.5cm}
      \end{aligned}
      \begin{aligned}
        \begin{tikzpicture}
          \draw[thick,rounded corners=10pt,
          decoration={
            markings, 
            mark=at position 0.94 with {\arrow{latex reversed}}            
          },
          postaction={decorate}] 
          (1.25,1.75) 
          -- (2,2.5) 
          -- (2,3.5) node[above] {\phantom{aa}$\oscrep_{-c_1},v_3$};
          \draw[thick,rounded corners=10pt,
          decoration={
            markings, 
            mark=at position 0.94 with {\arrow{latex reversed}}
          },
          postaction={decorate}] 
          (1.25,1.75) 
          -- (0.5,2.5) 
          -- (0.5,3.5) node[above] {$\oscrep_{-c_2},v_3+c_1$\phantom{aaaa}};
          \draw[thick,rounded corners=10pt,          
          decoration={
            markings, 
            mark=at position 0.15 with {\arrow{latex reversed}}
          },
          postaction={decorate}]
          (1.25,0) node[below] {$\oscrep_{c_3},v_3$}
          -- (1.25,1.75);
        \end{tikzpicture}
      \end{aligned}
      \quad\quad
      \begin{aligned}
        \mon_{3,2}(\spec)=\,\,\,\\\vphantom{}
      \end{aligned}
      \begin{aligned}
        \begin{tikzpicture}
          \draw[densely dashed,thick,
          decoration={
            markings, mark=at position 0.97 with {\arrow{latex reversed}}},
          postaction={decorate}]
          (0,0)
          node[left] {$\square,\spec$} --
          (4,0);
          \draw[thick,
          decoration={markings,
            mark=at position 0.85 with {\arrow{latex reversed}}},
          postaction={decorate}] 
          (0.5,-0.5)
          node[below] {$\vphantom{\bar{\oscrep}_{c_i}}\bar{\oscrep}_{c_1},\inh_1$} -- 
          (0.5,0.5);
          \draw[thick,
          decoration={markings,
            mark=at position 0.85 with {\arrow{latex reversed}}},
          postaction={decorate}] 
          (2.0,-0.5)
          node[below] {$\vphantom{\bar{\oscrep}_{c_i}}\bar{\oscrep}_{c_2},\inh_2$} -- 
          (2.0,0.5);
          \draw[thick,
          decoration={markings,
            mark=at position 0.85 with {\arrow{latex reversed}}},
          postaction={decorate}] 
          (3.5,-0.5) 
          node[below] {$\vphantom{\bar{\oscrep}_{c_i}}\oscrep_{c_3},\inh_3$} -- 
          (3.5,0.5);
        \end{tikzpicture}
      \end{aligned}
    \end{align*}
    \caption{The intertwiner $\mathcal{O}_{\Psi_{3,2}}$ from
      \eqref{eq:osc-opsi32} is visualized on the left side as a
      trivalent vertex. The monodromy $M_{3,2}(u)$ of the
      corresponding Yangian invariant $|\Psi_{3,2}\rangle$ can be
      found on the right side. The parameters of the intertwiner and
      the monodromy obey the constraints \eqref{eq:osc-m32-vs}.}
    \label{fig:osc-psi32}
  \end{center} 
\end{figure}

Apart from the two-site Yangian invariant $|\Psi_{2,1}\rangle$, the
next simplest invariants are $|\Psi_{3,1}\rangle$ or
$|\Psi_{3,2}\rangle$ with three sites. These are characterized by
three-site monodromies, where also here we place the sites carrying a
dual representation left of those with an ordinary one. A graphical
interpretation of the three-site invariants $|\Psi_{3,1}\rangle$ and
$|\Psi_{3,2}\rangle$ as trivalent vertices is provided in figures
\ref{fig:osc-psi31} and \ref{fig:osc-psi32}, respectively.

The said monodromy with one dual site characterizing
$|\Psi_{3,1}\rangle$ is
\begin{align}
  \label{eq:osc-m31}
  \mon_{3,1}(\spec)
  =
  R_{\square\,\bar{\oscrep}_{c_1}}(\spec-\inh_1)
  R_{\square\,\oscrep_{c_2}}(\spec-\inh_2)
  R_{\square\,\oscrep_{c_3}}(\spec-\inh_3)\,,
\end{align}
cf.\ the left site of figure~\ref{fig:osc-psi31}. In order to find a
solution of the Yangian invariance condition \eqref{eq:yi}, we choose
the parameters
\begin{equation}
  \label{eq:osc-m31-vs}
  \inh_2=\inh_1+n+c_2+c_3-1\,,
  \quad
  \inh_3=\inh_1+n+c_3-1\,,
  \quad
  c_1+c_2+c_3=0\,.
\end{equation}
Taking into account this choice the normalization of the monodromy
\eqref{eq:osc-m31} with Lax operators of the types
\eqref{eq:osc-lax-fund-s} and \eqref{eq:osc-lax-fund-bs} trivializes,
\begin{align}
  \label{eq:osc-m31-norm}
  f_{\bar{\oscrep}_{c_1}}(\spec-\inh_1)f_{\oscrep_{c_2}}(\spec-\inh_2)f_{\oscrep_{c_3}}(\spec-\inh_3)=1\,.
\end{align}
To show this we used \eqref{eq:osc-norm-add-eq} for
$f_{\oscrep_c}(\spec)$, the unitarity condition for
$f_{\bar{\oscrep}_{-c}}(\spec)$ and we expressed
$f_{\bar{\oscrep}_{-c}}(\spec)$ in terms of
$f_{\oscrep_{c}}(\spec)$ by means of \eqref{eq:osc-crossing-norm}.
Then by a straightforward computation one verifies that the Yangian
invariant solving \eqref{eq:yi} is
\begin{equation}
  \label{eq:osc-psi31}
  |\Psi_{3,1}\rangle
  =
  (1\bullet 2)^{c_2}
  (1\bullet 3)^{c_3}
  |0\rangle\,,
\end{equation}
where we fixed a scalar prefactor. 

We continue with the discussion of the corresponding intertwining
relation. It is obtained from the general relation
\eqref{eq:yi-intertwiner} by restricting to $K=1$ and using
$\kappa_{\bar{\oscrep}_{c_1}}$ from \eqref{eq:osc-crossing-norm},
\begin{equation}
  \label{eq:osc-int-psi31}
  R_{\square\,\oscrep_{c_2}}(\spec-\inh_2)
  R_{\square\,\oscrep_{c_3}}(\spec-\inh_2+c_2)
  \mathcal{O}_{\Psi_{3,1}}
  =
  \mathcal{O}_{\Psi_{3,1}}
  R_{\square\,\oscrep_{-c_1}}(\spec-\inh_2)
\end{equation} 
with
\begin{equation}
  \label{eq:osc-opsi31}
  \mathcal{O}_{\Psi_{3,1}}
  :=
  |\Psi_{3,1}\rangle^{\dagger_1}
  =
  \sum_{\substack{\alpha_1,\ldots,\alpha_{c_2}\\\beta_1,\ldots,\beta_{c_3}}}
  \!\!\!
  \bar\osca_{\alpha_1}^2\cdots\bar\osca_{\alpha_{c_2}}^2
  \bar\osca_{\beta_1}^3\cdots\bar\osca_{\beta_{c_3}}^3
  |0\rangle
  \langle 0|
  \osca_{\alpha_1}^1\cdots\osca_{\alpha_{c_2}}^1
  \osca_{\beta_1}^1\cdots\osca_{\beta_{c_3}}^1\,.
\end{equation}
Such an intertwining relation is known as bootstrap equation
\cite{Zamolodchikov:1989zs}, see also e.g.\
\cite{Dorey:1997,Samaj:2013}.

Next we investigate the monodromy
\begin{equation}
  \label{eq:osc-m32}  
  \mon_{3,2}(\spec)
  =
  R_{\square\,\bar{\oscrep}_{c_1}}(\spec-\inh_1)
  R_{\square\,\bar{\oscrep}_{c_2}}(\spec-\inh_2)
  R_{\square\,\oscrep_{c_3}}(\spec-\inh_3)\,
\end{equation}
with two dual sites on the left. It is depicted on the right side
of figure~\ref{fig:osc-psi32}. In order to find the solution
$|\Psi_{3,2}\rangle$ of \eqref{eq:yi} with this monodromy, we demand
\begin{equation}
  \label{eq:osc-m32-vs}
  \inh_1=\inh_3-n+c_1+1\,,
  \quad
  \inh_2=\inh_3-n-c_3+1\,,
  \quad
  c_1+c_2+c_3=0\,.
\end{equation}
Analogous to the case of the other three-site invariant, these
conditions ensure that the normalization of the monodromy
\eqref{eq:osc-m32} trivializes,
\begin{align}
  \label{eq:osc-m32-norm}
  f_{\bar{\oscrep}_{c_1}}(\spec-\inh_1)
  f_{\bar{\oscrep}_{c_2}}(\spec-\inh_2)
  f_{\oscrep_{c_3}}(\spec-\inh_3)=1\,.
\end{align}
We solve \eqref{eq:yi} by an explicit calculation that yields
\begin{align}
  \label{eq:osc-psi32}
  |\Psi_{3,2}\rangle
  =
  (1\bullet 3)^{-c_1}
  (2\bullet 3)^{-c_2}
  |0\rangle\,,
\end{align}
where once more we picked a scalar prefactor. Recall that
$c_1,c_2\in-\mathbb{N}$ for the compact representations under
consideration here.

The intertwiner version of the Yangian invariance condition is
obtained from \eqref{eq:yi-intertwiner} by setting $K=2$ and using the
values of $\kappa_{\bar{\oscrep}_{c_1}}$,
$\kappa_{\bar{\oscrep}_{c_2}}$ from \eqref{eq:osc-crossing-norm}. This
results in the bootstrap equation
\begin{equation}
  \label{eq:osc-int-psi32}
  R_{\square\,\oscrep_{c_3}}(\spec-\inh_3)
  \mathcal{O}_{\Psi_{3,2}}
  =
  \mathcal{O}_{\Psi_{3,2}}
  R_{\square\,\oscrep_{-c_2}}(\spec-\inh_3-c_1)
  R_{\square\,\oscrep_{-c_1}}(\spec-\inh_3)
\end{equation}
with the solution
\begin{equation}
  \label{eq:osc-opsi32}
  \mathcal{O}_{\Psi_{3,2}}
  :=
  |\Psi_{3,2}\rangle^{\dagger_1\dagger_2}
  =
  \sum_{\substack{\alpha_1,\ldots,\alpha_{-c_1}\\\beta_1,\ldots,\beta_{-c_2}}}
  \!\!\!
  \bar\osca_{\alpha_1}^3\cdots\bar\osca_{\alpha_{-c_1}}^3
  \bar\osca_{\beta_1}^3\cdots\bar\osca_{\beta_{-c_2}}^3
  |0\rangle \langle 0|
  \osca_{\alpha_1}^1\cdots\osca_{\alpha_{-c_1}}^1
  \osca_{\beta_1}^2\cdots\osca_{\beta_{-c_2}}^2\,.
\end{equation}

\subsubsection{Four-Site Invariant and Yang-Baxter Equation}
\label{sec:comp-bos-4-site}

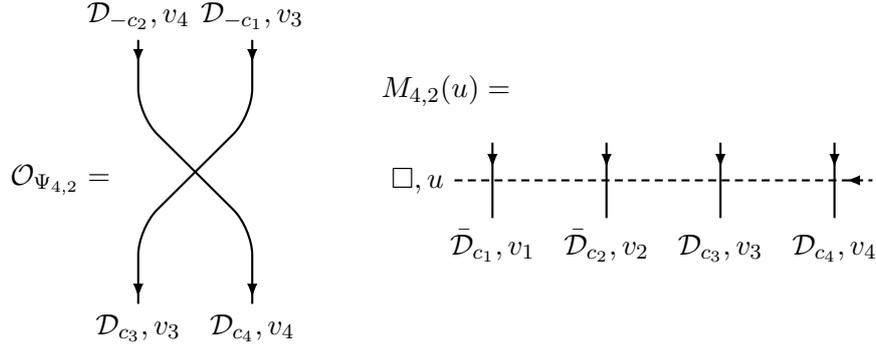
\begin{figure}[!t]
  \begin{center}
    \begin{align*}
      \begin{aligned}
        \mathcal{O}_{\Psi_{4,2}}=\hspace{-.5cm}
      \end{aligned}
      \begin{aligned}
        \begin{tikzpicture}
          \draw[thick,rounded corners=10pt,
          decoration={
            markings, 
            mark=at position 0.97 with {\arrow{latex reversed}},
            mark=at position 0.2 with {\arrow{latex reversed}}
          },
          postaction={decorate}] 
          (0.5,0) node[below] {$\oscrep_{c_3},v_3$}
          -- (0.5,1) 
          -- (2,2.5) 
          -- (2,3.5) node[above] {$\oscrep_{-c_1},v_3$};
          \draw[thick,rounded corners=10pt,
          decoration={
            markings, 
            mark=at position 0.97 with {\arrow{latex reversed}},
            mark=at position 0.2 with {\arrow{latex reversed}}
          },
          postaction={decorate}] 
          (2,0) node[below] {$\oscrep_{c_4},v_4$}
          -- (2,1) 
          -- (0.5,2.5) 
          -- (0.5,3.5) node[above] {$\oscrep_{-c_2},v_4$};
        \end{tikzpicture}
      \end{aligned}
      \quad\quad
      \begin{aligned}
        &
        \begin{aligned}
          \mon_{4,2}(\spec)=\\[-8pt]\phantom{}
        \end{aligned}\\
        &
        \begin{aligned}
          \begin{tikzpicture}
            \draw[densely dashed,thick,
            decoration={
              markings, mark=at position 0.97 with {\arrow{latex reversed}}},
            postaction={decorate}]
            (0,0)
            node[left] {$\square,\spec$} --
            (5.5,0);
            \draw[thick,
            decoration={markings,
              mark=at position 0.85 with {\arrow{latex reversed}}},
            postaction={decorate}] 
            (0.5,-0.5)
            node[below] {$\vphantom{\bar{\oscrep}_{c_i}}\bar{\oscrep}_{c_1},\inh_1$} -- 
            (0.5,0.5);
            \draw[thick,
            decoration={markings,
              mark=at position 0.85 with {\arrow{latex reversed}}},
            postaction={decorate}] 
            (2.0,-0.5)
            node[below] {$\vphantom{\bar{\oscrep}_{c_i}}\bar{\oscrep}_{c_2},\inh_2$} -- 
            (2.0,0.5);
            \draw[thick,
            decoration={markings,
              mark=at position 0.85 with {\arrow{latex reversed}}},
            postaction={decorate}] 
            (3.5,-0.5) 
            node[below] {$\vphantom{\bar{\oscrep}_{c_i}}\oscrep_{c_3},\inh_3$} -- 
            (3.5,0.5);
            \draw[thick,
            decoration={markings,
              mark=at position 0.85 with {\arrow{latex reversed}}},
            postaction={decorate}] 
            (5.0,-0.5) 
            node[below] {$\vphantom{\bar{\oscrep}_{c_i}}\oscrep_{c_4},\inh_4$} -- 
            (5.0,0.5);
          \end{tikzpicture}
        \end{aligned}
      \end{aligned}
    \end{align*}
  \end{center}
  \caption{The left side shows the intertwiner
    $\mathcal{O}_{\Psi_{4,2}}$ given in \eqref{eq:osc-opsi42}. We
    represent it by the intersection of two lines because it can be
    understood as an R-matrix. The monodromy $M_{4,2}(u)$ of the
    corresponding Yangian invariant $|\Psi_{4,2}\rangle$ is displayed
    on the right side. The necessary identifications of the
    representation labels and the inhomogeneities are written in
    \eqref{eq:osc-m42-vs}.}
  \label{fig:osc-psi42}
\end{figure}

In this section we address Yangian invariants with four
sites. Importantly, we will show that a particular four-site invariant
is nothing but the well-known $\mathfrak{gl}(n)$ invariant
R-matrix~\cite{Kulish:1981gi} ``in disguise''. This R-matrix is a
solution of the Yang-Baxter equation. For $K=2$ dual sites in the
monodromy \eqref{eq:yi-mono-split}, the intertwining relation
\eqref{eq:yi-intertwiner} reduces to this Yang-Baxter
equation. Therefore we study the monodromy
\begin{equation}
  \label{eq:osc-m42}
  \mon_{4,2}(\spec)
  =
  R_{\square\,\bar{\oscrep}_{c_1}}(\spec-\inh_1)R_{\square\,\bar{\oscrep}_{c_2}}(\spec-\inh_2)
  R_{\square\,\oscrep_{c_3}}(\spec-\inh_3)R_{\square\,\oscrep_{c_4}}(\spec-\inh_4)\,,
\end{equation}
see also figure~\ref{fig:osc-psi42}. For the inhomogeneities and
representation labels we choose
\begin{equation}
  \label{eq:osc-m42-vs}
  \inh_1=\inh_3-n-c_3+1\,,
  \quad
  \inh_2=\inh_4-n-c_4+1\,,
  \quad
  c_1+c_3=0\,,
  \quad
  c_2+c_4=0\,.
\end{equation}
The constraints on the representation labels are needed to indeed
obtain the Yang-Baxter equation from \eqref{eq:yi-intertwiner}. Notice
that \eqref{eq:osc-m42-vs} consists of two sets of the conditions we
used for the two-site invariant in \eqref{eq:osc-m21-vs}. Hence the
normalization trivializes similar as in
section~\ref{sec:comp-bos-2-site},
\begin{align}
  \label{eq:osc-m42-norm}
  f_{\bar{\oscrep}_{c_1}}(\spec-\inh_1)f_{\bar{\oscrep}_{c_2}}(\spec-\inh_2)
  f_{\oscrep_{c_3}}(\spec-\inh_3)f_{\oscrep_{c_4}}(\spec-\inh_4)=1\,.
\end{align}
To solve the Yangian invariance condition \eqref{eq:yi} we start from
the ansatz
\begin{equation}
  \label{eq:osc-psi42}
  |\Psi_{4,2}(\inh_3-\inh_4)\rangle
  :=
  |\Psi_{4,2}\rangle
  =
  \sum_{k=0}^{\Min(c_3,c_4)}
  \!\!\!
  d_k(\inh_3-\inh_4)
  |\Upsilon_k\rangle
\end{equation}
with the $\mathfrak{gl}(n)$ invariant vectors
\begin{align}
  \label{eq:osc-phi}
  |\Upsilon_k\rangle
  =
  (1\bullet 3)^{c_3-k}
  (2\bullet 4)^{c_4-k}
  (2\bullet 3)^{k}
  (1\bullet 4)^{k}
  |0\rangle\,.
\end{align}
It turns out that the four-site invariant depends on a free complex
spectral parameter which is the difference of two inhomogeneities,
\begin{align}
  \label{eq:osc-diffinh}
  \inhdiff:=\inh_3-\inh_4.
\end{align}
This dependence is made explicit by the notation
$|\Psi_{4,2}(\inhdiff)\rangle$. The coefficients $d_k$ obey a
recursion relation that is obtained from inserting the ansatz
\eqref{eq:osc-psi42} into \eqref{eq:yi},
\begin{equation}
  \label{eq:osc-recrel}
  \frac{d_k(\inhdiff)}{d_{k+1}(\inhdiff)}
  =
  \frac{(k+1)(\inhdiff-c_3+k+1)}{(c_3-k)(c_4-k)}\,.
\end{equation}
Its solution is, up to multiplication by a function of $k$ with period
$1$,
\begin{equation}
  \label{eq:osc-coeff}
 d_k(\inhdiff)
 =
 \frac{1}{(c_3-k)!(c_4-k)!k!^2}\,
 \frac{k!}{\Gamma(\inhdiff-c_3+k+1)}\,.
\end{equation}

The intertwiner corresponding to $|\Psi_{4,2}(\inhdiff)\rangle$ solves
\eqref{eq:yi-intertwiner} with $\dsites=2$ and
$\kappa_{\bar{\oscrep}_{c_1}}$, $\kappa_{\bar{\oscrep}_{c_2}}$
from \eqref{eq:osc-crossing-norm}. This equation has the form of a
Yang-Baxter equation,
\begin{equation}
  \label{eq:osc-int-psi42}
 R_{\square\,\oscrep_{c_3}}(\spec-\inh_3)
 R_{\square\,\oscrep_{c_4}}(\spec-\inh_4)
 \mathcal{O}_{\Psi_{4,2}(\inhdiff)}
 =
 \mathcal{O}_{\Psi_{4,2}(\inhdiff)}
 R_{\square\,\oscrep_{-c_2}}(\spec-\inh_4)
 R_{\square\,\oscrep_{-c_1}}(\spec-\inh_3)\,,
\end{equation} 
where
\begin{equation}
  \label{eq:osc-opsi42}
  \mathcal{O}_{\Psi_{4,2}(\inhdiff)}
  :=
  |\Psi_{4,2}(\inhdiff)\rangle^{\dagger_1\dagger_2}
  =
  \sum_{k=0}^{\Min(s_3,s_4)}d_k(\inhdiff)
  \mathcal{O}_{\Upsilon_k}\,,
\end{equation}
and
\begin{align}
  \label{eq:osc-ophi}
  \begin{aligned}
    \mathcal{O}_{\Upsilon_k}
    :=
    |\Upsilon_k\rangle^{\dagger_1\dagger_2}
    =
    \smash{\sum_{\substack{\alpha_1,\ldots,\alpha_{c_3}\\\beta_1,\ldots,\beta_{c_4}}}}
    &\bar\osca^3_{\alpha_1}\cdots\bar\osca^3_{\alpha_{c_3}}
    \bar\osca^4_{\beta_1}\cdots\bar\osca^4_{\beta_{c_4}}
    |0\rangle\\
    &\quad\cdot
    \langle 0| \osca^1_{\alpha_1}\cdots\osca^1_{\alpha_{c_3-k}}
    \osca^1_{\beta_{c_4-k+1}}\cdots\osca^1_{\beta_{c_4}}\\
    &\quad\quad\cdot
    \osca^2_{\beta_1}\cdots\osca^2_{\beta_{c_4-k}}
    \osca^2_{\alpha_{c_3-k+1}}\cdots\osca^2_{\alpha_{c_3}}\,.
  \end{aligned}
\end{align}
To obtain a more standard formulation of the Yang-Baxter equation, we
identify space the $\oscrep_{-c_1}$ with $\oscrep_{c_3}$ and
$\oscrep_{-c_2}$ with $\oscrep_{c_4}$ and rename
$\mathcal{O}_{\Psi_{4,2}(\inhdiff)}$ as
$R_{\oscrep_{c_3}\!\oscrep_{c_4}}(\inhdiff)$. Then \eqref{eq:osc-ophi}
translates into
\begin{equation}
  \label{eq:osc-psi42-ybe}
  \begin{aligned}
    R_{\square\,\oscrep_{c_3}}(\spec-\inh_3)
    R_{\square\,\oscrep_{c_4}}(\spec-\inh_4)
    R_{\oscrep_{c_3}\!\oscrep_{c_4}}(\inhdiff)
    =
    R_{\oscrep_{c_3}\!\oscrep_{c_4}}(\inhdiff)
    R_{\square\,\oscrep_{c_4}}(\spec-\inh_4)
    R_{\square\,\oscrep_{c_3}}(\spec-\inh_3)\,.
  \end{aligned}
\end{equation}
Consequently $R_{\oscrep_{c_3}\!\oscrep_{c_4}}(\inhdiff)$ is the
$\mathfrak{gl}(n)$ invariant R-matrix of \cite{Kulish:1981gi} for
symmetric representations, which are realized in terms of oscillators
in our approach.

Let us remark that the invariant given in \eqref{eq:osc-psi42} with
the coefficients \eqref{eq:osc-coeff} can be expressed as a Gauß
hypergeometric function ${}_2F_1(a,b;c;x)$,
\begin{align}
  \label{eq:inv42-bos-hyper}
  \begin{aligned}
    |\Psi_{4,2}\rangle
    =\;&\frac{1}{c_3!c_4!\Gamma(1-c_3+z)}\\
    &\cdot{}_2F_1\left(-c_3,-c_4,1-c_3+z,\frac{(2\bullet 3)(1\bullet 4)}{(1\bullet 3)(2\bullet 4)}\right)
    (1\bullet 3)^{c_3}(2\bullet 4)^{c_4}|0\rangle\,.
  \end{aligned}
\end{align}

Besides the invariant \eqref{eq:osc-psi42} that we identified with an
R-matrix, there are further Yangian invariants that can be obtained
from the four-site monodromy \eqref{eq:osc-m42}. For these solutions
the conditions on the representation labels in \eqref{eq:osc-m42-vs}
are relaxed to $c_1+c_2+c_3+c_4=0$. However, they to not depend on a
complex spectral parameter.

\subsection{Non-Compact Supersymmetric Invariants}
\label{sec:non-comp-supersymm}

The aim of this section is to generalize the compact bosonic sample
invariants to the non-compact supersymmetric case. Hence we use the
oscillator representations of $\mathfrak{u}(p,q|m=r+s)$ of
section~\ref{sec:osc-rep} in full generality. Again we work with a
monodromy that has $K$ dual sites left of $N-K$ ordinary ones,
\begin{align}
  \label{eq:yi-mono-nc}
  \begin{aligned}
    \mon_{N,K}(\spec)
    =
    &R_{\square\,\bar{\oscrep}_{c_1}}(\spec-\inh_1)\cdots R_{\square\,\bar{\oscrep}_{c_K}}(\spec-\inh_K)\\
    &\cdot R_{\square\,\oscrep_{c_{K+1}}}(\spec-\inh_{K+1})\cdots R_{\square\,\oscrep_{c_{N}}}(\spec-\inh_N)\,.
  \end{aligned}
\end{align}
The Lax operators are given in \eqref{eq:yangian-def-lax}. They
contain the $\mathfrak{gl}(n|m)$ generators
$J_{\indnm{AB}}=\bar{\mathbf{J}}_{\indnm{AB}}$ from
\eqref{eq:gen-dual} at the sites with a dual representation of the
type $\bar{\oscrep}_{c}$ and $J_{\indnm{AB}}=\mathbf{J}_{\indnm{AB}}$
from \eqref{eq:gen-ordinary} at sites with an ordinary representation
$\oscrep_{c}$. Recall that in the non-compact case the representation
label $c$ can be any integer. In all the examples discussed in the
previous section the overall normalization of the monodromy reduced to
unity. Therefore we choose the normalization
$f_{\oscrep_{c}}=f_{\bar{\oscrep}_{c}}=1$ directly at the level of the
Lax operators. The solutions $|\Psi_{N,K}\rangle$ to the Yangian
invariance condition \eqref{eq:yi}, that we will present momentarily,
will be expressed in terms of the contractions of oscillators
\begin{align}
  \label{eq:bullets-nc}
  \begin{aligned}
    (k\bullet l)&=
    \sum_{\indssub{A}}\bar{\mathbf{A}}^l_{\indssub{A}}\bar{\mathbf{A}}^k_{\indssub{A}}
    =\sum_{\alpha=1}^p\bar{\mathbf{a}}^l_\alpha\bar{\mathbf{a}}^k_\alpha
    +\sum_{a=1}^r\bar{\mathbf{c}}^l_a\bar{\mathbf{c}}^k_a\,,\\
    (k\circ l)&=
    \sum_{\dot{\indssub{A}}}\bar{\mathbf{A}}^l_{\dot{\indssub{A}}}\bar{\mathbf{A}}^k_{\dot{\indssub{A}}}
    =\sum_{\dot{\alpha}=1}^q\bar{\mathbf{b}}^l_{\dot\alpha}\bar{\mathbf{b}}^k_{\dot\alpha}
    +\sum_{\dot{a}=1}^s\bar{\mathbf{d}}^l_{\dot a}\bar{\mathbf{d}}^k_{\dot a}\,.
  \end{aligned}
\end{align}
Here site $l$ carries an ordinary representation $\oscrep_{c_l}$ and
site $k$ a dual representation $\bar{\oscrep}_{c_k}$.  Notice that the
contractions are bosonic because they contain fermionic oscillators
only in quadratic terms. Furthermore, they are $\mathfrak{gl}(p|r)$
and $\mathfrak{gl}(q|s)$ invariant, respectively,
\begin{align}
  \label{eq:bullets-nc-symm}
  [\bar{\mathbf{J}}_{\indssub{AB}}^k+\mathbf{J}_{\indssub{AB}}^l,(k\bullet l)\}=0\,,\quad
  [\bar{\mathbf{J}}_{\dot{\indssub{A}}\dot{\indssub{B}}}^k+\mathbf{J}_{\dot{\indssub{A}}\dot{\indssub{B}}}^l,(k\circ l)\}=0\,.
\end{align}
They generalize the oscillator contraction given in
\eqref{eq:osc-psi21} that we used to build the compact bosonic
invariants.

\subsubsection{Two-Site Invariant}
\label{sec:ncomp-susy-2-site}

We consider the monodromy $M_{2,1}(u)$. To proceed we make an ansatz
for the Yangian invariant state $|\Psi_{2,1}\rangle$ as a power series
in $(1\bullet 2)$ and $(1\circ 2)$ acting on the Fock vacuum
$|0\rangle$. Next, we demand Yangian invariance \eqref{eq:yi} of this
ansatz. Furthermore, we impose that each site carries an irreducible
representation of $\mathfrak{u}(p,q|r+s)$, i.e.\
$\bar{\mathbf{C}}^1|\Psi_{2,1}\rangle=c_1|\Psi_{2,1}\rangle$ and
$\mathbf{C}^2|\Psi_{2,1}\rangle=c_2|\Psi_{2,1}\rangle$. This fixes the
invariant, up to a normalization constant, to be
\begin{align}
  \label{eq:sample-inv21}
  \begin{aligned}
    |\Psi_{2,1}\rangle
    &=\;\;\sum_{\mathclap{\substack{g_{12},h_{12}=0\\g_{12}-h_{12}=q-s-c_1}}}^\infty\;\;
    \frac{(1\bullet 2)^{g_{12}}}{g_{12}!}\frac{(1\circ 2)^{h_{12}}}{h_{12}!}|0\rangle\\
    &=
    \frac{I_{q-s-c_1}\big(2\sqrt{(1\bullet 2)(1\circ 2)}\big)}
    {\sqrt{(1\bullet 2)(1\circ 2)}^{\,q-s-c_1}}
    (1\bullet 2)^{q-s-c_1}|0\rangle\,,\\
  \end{aligned}
\end{align}
where we identified the sum with the series expansion of the modified
Bessel function of the first kind $I_\nu(x)$.\footnote{In the double
  sum in \eqref{eq:sample-inv21} the expression $q-s-c_1$ can also
  take negative values. The validity of the Bessel function
  formulation in this case is easily verified using the series
  expansion.} The parameters of the monodromy have to obey
\begin{align}
  \label{eq:nc21-para}
  v_1-v_2=1-n+m-c_2\,,\quad c_1+c_2=0\,.
\end{align}
We observe that the invariant \eqref{eq:sample-inv21} can be expressed
as the complex contour integral
\begin{align}
  \label{eq:nc21-int}
  |\Psi_{2,1}\rangle
  =\frac{1}{2\pi i}\oint\D C_{1 2}\frac{e^{C_{1 2}(1\bullet 2)+C^{-1}_{1 2}(1\circ 2)}|0\rangle}
  {C_{1 2}^{1+q-s-c_1}}\,.
\end{align}
Here the contour is a counterclockwise unit circle around the
essential singularity at $C_{12}=0$. It can be interpreted as group
manifold of the unitary group $U(1)$. The integral is easily evaluated
using the residue theorem. This yields the series representation in
\eqref{eq:sample-inv21}. As we will see in
section~\ref{sec:grassmann-osc}, \eqref{eq:nc21-int} can be considered
as a simple Graßmannian integral. In that section we will generalize
this simple formula to a Graßmannian integral formulation of the
Yangian invariants $|\Psi_{N=2K,K}\rangle$ with oscillator
representations.

We finish this section with some remarks. We note that recently a
two-site Yangian invariant for oscillator representations of
$\mathfrak{psu}(2,2|4)$ was used in \cite{Jiang:2014cya} based on a
construction in \cite{Alday:2005kq}. It takes the form of an
exponential function instead of a Bessel function as in
\eqref{eq:sample-inv21}. This difference occurs because the sites of
that invariant of \cite{Jiang:2014cya} do not transform in irreducible
representations of the symmetry algebra, i.e.\ in our terminology this
invariant would not be an eigenstate of $\bar{\mathbf{C}}^1$ and
$\mathbf{C}^2$.\footnote{We thank Ivan Kostov and Didina Serban for
  clarifying this point.}  Furthermore, we remark that employing the
identity
\begin{align}
  \label{eq:hyper-bessel}
  \frac{I_\nu(2\sqrt{x})}{\sqrt{x}^{\,\nu}}
  =\frac{_0F_1(\nu+1;x)}{\Gamma(\nu+1)}\,,
\end{align}
cf.\ \cite{AbramowitzStegun:1964}, the invariant
\eqref{eq:sample-inv21} can alternatively be expressed in terms of a
generalized hypergeometric function $_0F_1(a;x)$. Sometimes this form
is more convenient because it avoids the ``spurious'' square roots,
which are absent in the series expansion. Additionally, the invariant
in \eqref{eq:sample-inv21} has infinite norm and thus is technically
speaking not an element of $\bar{\oscrep}_{c_1}\otimes\oscrep_{c_2}$
considered as a Hilbert space. As a last aside, let us consider the
special case of the compact bosonic algebra $\mathfrak{u}(n=p,0|0)$,
i.e.\ we set $q=r=s=0$. Then the sum in \eqref{eq:sample-inv21}
simplifies to a single term
\begin{align}
  \label{eq:nc21-comp}
  |\Psi_{2,1}\rangle=
  \frac{(1\bullet 2)^{c_2}}{c_2!}|0\rangle
\end{align}
with $c_2\geq 0$, where we used $(1\circ 2)^{h}=\delta_{0 h}$. Up to a
normalization factor, this is the compact bosonic two-site Yangian
invariant known from \eqref{eq:osc-psi21}.

\subsubsection{Three-Site Invariants}
\label{sec:ncomp-susy-3-site}

The generalization of the compact bosonic three-site invariants
$|\Psi_{3,1}\rangle$ and $|\Psi_{3,2}\rangle$ to the non-compact
supersymmetric setting is not as straightforward as for two
sites. Therefore we start by focusing on a particular case. For the
non-compact bosonic algebra $\mathfrak{u}(p,1)$ we verify by an
explicit computation that the vector
\begin{align}
  \label{eq:inv31up1}
  |\Psi_{3,1}\rangle=
  \bar{\mathbf{A}}_{p+1}^1(1\circ 2)^{-(c_2+1)}(1\circ 3)^{-(c_3+1)}
  \sum_{k=0}^\infty\frac{\big((1\bullet 2)(1\circ 2)+(1\bullet 3)(1\circ 3)\big)^k}
  {k!\big(-(c_2+1)-(c_3+1)+1+k\big)!}
  |0\rangle
\end{align}
solves the Yangian invariance condition \eqref{eq:yi} with the
monodromy $M_{3,1}(u)$ in \eqref{eq:yi-mono-nc}. For this the
parameters of the monodromy have to obey
\begin{gather}
  \label{eq:inv31up1-cond}
  \begin{gathered}
    v_1-v_3=1-n-c_3\,,\quad
    v_1-v_2=1-n-c_2-c_3\,,\\
    c_1+c_2+c_3=0\,,\quad
    c_2+1\leq 0\,,\quad
    c_3+1\leq 0\,.
  \end{gathered}
\end{gather}
Notice that this invariant cannot be expressed entirely in terms of
$(k\bullet l)$ and $(k\circ l)$ but contains the individual oscillator
$\bar{\mathbf{A}}_{p+1}^1$. Furthermore, note that one can easily
identify \eqref{eq:inv31up1} with the series expansion of the
hypergeometric function ${}_0F_1(a;x)$.

The invariant in \eqref{eq:inv31up1} immediately raises the question
about the generalization to other algebras. The $\mathfrak{u}(2,2)$
case is of special interest because of its relevance for
amplitudes. We did not manage to find solutions $|\Psi_{3,1}\rangle$
of the Yangian invariance condition for this algebra. Bearing in mind
the discussion of three-particle scattering amplitudes around
\eqref{eq:amp-3}, this was to be expected. We argued that these
amplitudes do not exist for real particle momenta. Although
\eqref{eq:amp-3} are superamplitudes, the same holds true for the
purely bosonic three-particle amplitudes. The $\mathfrak{u}(2,2)$
representations in terms of spinor helicity variables are equivalent
to those in terms of oscillators employed here, see
\cite{Stoyanov:1968tn} and also \cite{Mack:1969dg} as well as
section~\ref{sec:transf-boson-non} below. This means there should be
no $|\Psi_{3,1}\rangle$. Let us briefly sketch an alternative argument
supporting this conclusion. It is based on the decomposition of the
tensor product of two oscillator representations
$\oscrep_{c_2}\otimes \oscrep_{c_3}$ into a sum of irreducible
representations of $\mathfrak{u}(2,2)$. For the invariant
$|\Psi_{3,1}\rangle\in\bar{\oscrep}_{c_1}\otimes\oscrep_{c_2}\otimes
\oscrep_{c_3}$
to exist, this sum must contain an oscillator representation, namely
$\oscrep_{-c_1}$ that is dual to $\bar{\oscrep}_{c_1}$. However, from
(4.12) and (4.13) of \cite{Williams:1982} one concludes that the
tensor product decomposition of two $\mathfrak{u}(2,2)$
representations that are each built from one family of oscillators
does not contain the same type of oscillator representation. This
explains why we were not able to solve the Yangian invariance
condition.

Let us remark that the formulas of \cite{Williams:1982} that we used
for our argument are just a special case of the decomposition of the
multi-fold tensor product for $\mathfrak{u}(p,q)$ oscillator
representations in \cite{Kashiwara:1978}. With this reference it
should be straightforward to analyze for which bosonic algebras and
representations labels the three-site invariants $|\Psi_{3,1}\rangle$
and $|\Psi_{3,2}\rangle$ as well as further invariants
$|\Psi_{N,K}\rangle$ can exist. The tensor product decomposition for
superalgebras including $\mathfrak{u}(p,q|m)$ was worked out in
\cite{Cheng:2004}.

\subsubsection{Four-Site Invariant}
\label{sec:ncomp-susy-4-site}

In contrast to the situation for three sites, in case of the four-site
invariant $|\Psi_{4,2}\rangle$ the generalization to the non-compact
superalgebra $\mathfrak{u}(p,q|m)$ is possible again. It is given by
the unwieldy formula
\begin{align}
  \label{eq:sample-inv42}
  \begin{aligned}
    |\Psi_{4,2}\rangle=
    \;\;\smash{\sum_{\mathclap{\substack{g_{13},\ldots,g_{24}=0\\h_{13},\ldots,h_{24}=0\\\text{with \eqref{eq:sample-inv42-constraints}}}}}^\infty}\,\quad\quad
    &\frac{(1\bullet 3)^{g_{13}}}{g_{13}!}\frac{(1\bullet 4)^{g_{14}}}{g_{14}!}
    \frac{(2\bullet 3)^{g_{23}}}{g_{23}!}\frac{(2\bullet 4)^{g_{24}}}{g_{24}!}\\
    \cdot\,&\frac{(1\circ 3)^{h_{13}}}{h_{13}!}\frac{(1\circ 4)^{h_{14}}}{h_{14}!}
    \frac{(2\circ 3)^{h_{23}}}{h_{23}!}\frac{(2\circ 4)^{h_{24}}}{h_{24}!}
    |0\rangle\\
    \cdot\,(-1)^{g_{14}+h_{14}}&\text{B}(g_{14}+h_{23}+1,h_{13}+g_{24}-v_1+v_2)\,.
  \end{aligned}
\end{align}
In this expression the summation range is constrained by
\begin{align}
  \label{eq:sample-inv42-constraints}
  \begin{aligned}
    g_{13}-h_{13}+g_{14}-h_{14}&=-c_1+q-s\,,&\quad
    g_{23}-h_{23}+g_{24}-h_{24}&=-c_2+q-s\,,\\
    g_{13}-h_{13}+g_{23}-h_{23}&=\phantom{-}c_3+q-s\,,&\quad
    g_{14}-h_{14}+g_{24}-h_{24}&=\phantom{-}c_4+q-s\,.\\
  \end{aligned}
\end{align}
These constraints assure that the eigenvalues of
$\bar{\mathbf{C}}^1,\bar{\mathbf{C}}^2,\mathbf{C}^3,\mathbf{C}^4$
acting on the invariant are, respectively, $c_1,c_2,c_3,c_4$.
Furthermore, we made use of the Euler beta function
\begin{align}
  \label{eq:beta}
  \text{B}(x,y)
  =\frac{\Gamma(x)\Gamma(y)}{\Gamma(x+y)}\,.
\end{align}
To ensure the Yangian invariance of \eqref{eq:sample-inv42}, the
parameters of the monodromy \eqref{eq:yi-mono-nc} have to obey
\begin{equation}
  \label{eq:osc-m42-vs-nc}
  \inh_1-\inh_3=1-n+m-c_3\,,
  \quad
  \inh_2-\inh_4=1-n+m-c_4\,,
  \quad
  c_1+c_3=0\,,
  \quad
  c_2+c_4=0\,.
\end{equation}
For all compact and non-compact sample invariants discussed until here
the Yangian invariance condition \eqref{eq:yi} can be verified by a
straightforward explicit calculation. Of course, this is also possible
for the invariant $|\Psi_{4,2}\rangle$ in
\eqref{eq:sample-inv42}. However, the complexity of the formula
already foreshadows that such a computation is somewhat laborious in
this case. Therefore we do not display it here. Instead we refer the
reader to section~\ref{sec:sample-invariants} where
\eqref{eq:sample-inv42} is derived from a Yangian invariant
Graßmannian integral.

It is worth noting that in the compact bosonic case
$\mathfrak{u}(n=p,0|0)$ the invariant \eqref{eq:sample-inv42}
simplifies to
\begin{align}
  \label{eq:nc42-compact}
  \begin{aligned}
    |\Psi_{4,2}\rangle=
    \sum_{g_{14}=0}^\infty
    &\frac{(1\bullet 3)^{c_3-g_{14}}}{(c_3-g_{14})!}\frac{(1\bullet 4)^{g_{14}}}{g_{14}!}
    \frac{(2\bullet 3)^{g_{14}}}{g_{14}!}\frac{(2\bullet 4)^{c_4-g_{14}}}{(c_4-g_{14})!}|0\rangle\\
    \cdot\,&(-1)^{g_{14}}\text{B}(g_{14}+1,-v_3+v_4+c_3-g_{14})\,.
  \end{aligned}
\end{align}
This agrees with the compact invariant in \eqref{eq:osc-psi42} up to a
normalization factor. An interesting question is whether the
formulation of this compact invariant as a hypergeometric function in
\eqref{eq:inv42-bos-hyper} generalizes to the non-compact invariant
\eqref{eq:sample-inv42}.

Finally, let us mention that one can also work out an intertwiner
version $\mathcal{O}_{\Psi_{4,2}}$ of the non-compact supersymmetric
invariant $|\Psi_{4,2}\rangle$ in \eqref{eq:sample-inv42}. As shown in
detail for the compact bosonic case in \eqref{eq:osc-psi42-ybe}, this
intertwiner corresponds to an R-matrix satisfying a Yang-Baxter
equation. This R-matrix was worked out explicitly in
\cite{Ferro:2013dga} employing the same oscillator representations of
$\mathfrak{u}(p,q|m)$ that we use here. For the algebra
$\mathfrak{u}(2,2|4)$ it is essentially the R-matrix of the spin chain
governing the planar $\mathcal{N}=4$ SYM one-loop spectral problem
\cite{Beisert:2003yb,Beisert:2003jj}. A word of caution is in order
here. The oscillator representations used in this thesis and in
\cite{Ferro:2013dga} are those of \cite{Bars:1982ep}, see also
\cite{Gunaydin:1984fk} for the $\mathfrak{u}(2,2|4)$ case. The bosonic
generators $\bar{\mathbf{a}}_\alpha\mathbf{a}_\beta$ and
$\bar{\mathbf{b}}_{\dot{\alpha}}\mathbf{b}_{\dot{\beta}}$, cf.\
\eqref{eq:osc-split}, are associated with the compact subalgebra
$\mathfrak{su}(2)\oplus\mathfrak{su}(2)$ of the conformal algebra
$\mathfrak{su}(2,2)$. In contrast, the generators named
$\bar{\mathbf{a}}_\alpha\mathbf{a}_\beta$ and
$\bar{\mathbf{b}}_{\dot{\alpha}}\mathbf{b}_{\dot{\beta}}$ in
\cite{Beisert:2003jj} are associated with the non-compact Lorentz
subalgebra $\mathfrak{sl}(2)\equiv\mathfrak{sl}(\mathbb{C}^2)$ of
$\mathfrak{su}(2,2)$. Thus the ``oscillators'' in
\cite{Beisert:2003jj} satisfy non-standard reality conditions, see
e.g.\ \cite{Beisert:2010kp}. Some clarifications on related issues can
be found in \cite{Gunaydin:1998jc}.  The difference between the
oscillator representation used in this thesis and in
\cite{Ferro:2013dga} versus that in \cite{Beisert:2003jj} does not
seem to affect the R-matrix. Probably this is because of its
$\mathfrak{gl}(4|4)$ invariance. Leaving aside this subtlety, the
connection between Yangian invariance and the integrability discovered
in the $\mathcal{N}=4$ SYM spectral problem is a good point to finish
this chapter.

\chapter{Bethe Ansätze and Vertex Models}
\label{cha:bethe-vertex}

In the present chapter we employ the technology introduced in
chapter~\ref{cha:yang-rep} for a systematic construction of Yangian
invariants with finite-dimensional $\mathfrak{gl}(n)$
representations. In particular, the QISM formulation of the Yangian
algebra proves to be well suited for the construction of such
invariants by means of Bethe ansatz methods. Furthermore, we explain
that the partition functions of a rather general class of vertex
models can be interpreted as Yangian invariants. 

Let us outline the content of this chapter in more detail. The Bethe
ansatz is a powerful technique to solve integrable spin chain
models. The arguably simplest model to which it applies is the
Heisenberg spin chain, that is based on the Yangian of
$\mathfrak{gl}(2)$, cf.\ section~\ref{sec:int-mod}. In
section~\ref{sec:bethe-ansatze} we review its solution using an
algebraic Bethe ansatz, which makes crucial use of the QISM. The main
results of this chapter are contained in
section~\ref{sec:bethe-yangian}. We argue that Yangian invariants for
$\mathfrak{gl}(n)$ are particular eigenstates of special spin chains
and therefore in principle accessible by a Bethe ansatz. We detail our
argument in the $\mathfrak{gl}(2)$ case by using the algebraic Bethe
ansatz to construct Yangian invariants. This leads to a
characterization of those Yangian invariants in terms of certain
functional relations. In addition, this Bethe ansatz reproduces the
compact bosonic sample invariants of
section~\ref{sec:comp-boson-invar}. We also explain a classification
of the solutions to the functional relations. This classification is
of relevance more generally for non-compact supersymmetric Yangian
invariants and tree-level amplitudes of planar $\mathcal{N}=4$ SYM.
Further results on the Bethe ansatz for Yangian invariants, in
particular concerning its extension to the $\mathfrak{gl}(n)$ case,
have been shifted to appendix~\ref{cha:app-bethe}.

A different avenue is explored in the remainder of the chapter. In
section~\ref{sec:six-vertex} we review the rational six-vertex
model. The algebraic structure of this two-dimensional classical
statistical model is closely related to that of the quantum Heisenberg
spin chain. To be precise, both models are based on the same
$\mathfrak{gl}(2)$ symmetric solution of the Yang-Baxter equation. The
partition function of this vertex model, even on a non-rectangular
lattice, can be computed by the perhaps little known perimeter Bethe
ansatz. In section~\ref{sec:vertex-yangian} we show that the partition
functions of even more general vertex models with $\mathfrak{gl}(n)$
symmetry correspond to a certain class of Yangian invariants. Combined
with the results of the previous sections, this allows us to
understand our Bethe ansatz for Yangian invariants as a generalization
of the perimeter Bethe ansatz.

\section{Bethe Ansatz for Heisenberg Spin Chain}
\label{sec:bethe-ansatze}

In 1928 Heisenberg introduced a mathematical model of ferromagnetism
\cite{Heisenberg:1928mqa}. It consists of electron spins on a lattice
that interact with their nearest neighbors. In the one-dimensional
case it degenerates into a linear chain of spins, the so-called
\emph{Heisenberg spin chain}. Its Hamiltonian
\begin{align}
  \label{eq:heisenberg-hamiltonian}
  \mathscr{H}=-\sum_{i=1}^N\vec{\sigma}^{\,i}\cdot\vec{\sigma}^{\,i+1}\,
\end{align}
acts on the Hilbert space $(\mathbb{C}^2)^{\otimes N}$ and we assume
periodicity, $\vec{\sigma}^{\,i+N}=\vec{\sigma}^{\,i}$. The spins are
modeled by the Pauli matrices
\begin{align}
  \sigma_1=
  \begin{pmatrix}
    0&1\\
    1&0\\
  \end{pmatrix}\,,\quad
  \sigma_2=
  \begin{pmatrix}
    0&-i\\
    i&0\\
  \end{pmatrix}\,,\quad
  \sigma_3=
  \begin{pmatrix}
    1&0\\
    0&-1\\
  \end{pmatrix}\,,
\end{align}
that generate the Lie algebra $\mathfrak{su}(2)$. Remarkably, this
system can be \emph{solved exactly}, which here means that there exist
efficient analytical methods to diagonalize the Hamiltonian. This was
achieved by Bethe in 1931 using a technique nowadays called
\emph{coordinate Bethe ansatz} \cite{Bethe:1931hc}. Let us emphasize
that although the seemingly simple Heisenberg spin chain has been
under investigation for the better part of a century, it is still an
active field of research. Some more recent developments are reviewed
in \cite{Maillet:2007pda}. Theoretically, the exact solvability
extends beyond the energy spectrum, shedding light also on the
structure of correlation functions. The model can even be probed
experimentally as one-dimensional magnetic chains are realized in
certain crystals.

In this section we discuss the diagonalization of the Heisenberg
Hamiltonian \eqref{eq:heisenberg-hamiltonian} by means of an
\emph{algebraic Bethe ansatz}, see the authoritative reviews
\cite{Faddeev:1996iy,Korepin:1997}. It provides an alternative to
Bethe's original technique. We choose this flavor of the Bethe ansatz
because it is deeply rooted in the QISM, that we used to discuss the
Yangian algebra in section~\ref{sec:yangian}. Furthermore, it
straightforwardly generalizes to a large class of more elaborate spin
chain models.

In order to apply the algebraic Bethe ansatz, we have to rephrase the
Hamiltonian \eqref{eq:heisenberg-hamiltonian} in the QISM
language. This may seem like a detour at first but it will pay off
eventually. We start with a monodromy matrix
\eqref{eq:yangian-mono-spinchain} associated with the Yangian of
$\mathfrak{gl}(2)$,
\begin{align}
  \label{eq:heisenberg-monodromy}
  M(u)=R_{\square\,\square_1}(u)\cdots R_{\square\,\square_N}(u)\,.
\end{align}
Here we specialized to the case of a homogeneous spin chain with
$v_i=0$. In addition, we chose the defining representation of
$\mathfrak{gl}(2)$ for each local quantum space
$\mathcal{V}_i=\square_i=\mathbb{C}^2$. The generators
$J^i_{\alpha\beta}=\elemm^i_{\alpha\beta}$ are specified in
\eqref{eq:elemsupermat-prop}. In accordance with
section~\ref{sec:osc-rep} we use Greek letters for the bosonic
indices. The total quantum space
$\mathcal{V}_1\otimes\cdots\otimes\mathcal{V}_N$ already matches that
of the Heisenberg chain. Next, we introduce a \emph{transfer matrix}
by taking the trace over the auxiliary space $\square$,
\begin{align}
  T(u)=\tr_{\square}M(u)\,.
\end{align}
This operator acts solely on the total quantum space and it comprises
the Hamiltonian \eqref{eq:heisenberg-hamiltonian}. To show this, we
fix the normalization of the Lax operators \eqref{eq:yangian-def-lax}
entering the monodromy matrix to be
$f_{\square_i}(u)=u$.\footnote{This choice does not respect the
  condition \eqref{eq:yangian-norm} but is most convenient for this
  section.} At a special value of the spectral parameter these
operators reduce to permutation operators on
$\square\otimes\square_i$,
\begin{align}
  R_{\square\,\square_i}(u)\Big|_{u=0}=\sum_{\alpha,\beta=1}^2\elemm_{\alpha\beta}\elemm^{i}_{\beta\alpha}=: P_{\,\square\,\square_i}\,.
\end{align}
Such permutation operators satisfy
\begin{align}
  P_{\,\square_i\,\square_j}P_{\,\square_i\,\square_k}=P_{\,\square_j\,\square_k}P_{\,\square_i\,\square_j}\,,\quad
  P_{\,\square_i\,\square_j}=P_{\,\square_j\,\square_i}\,,\quad
  P_{\,\square_i\,\square_j}=P_{\,\square_i\,\square_j}^{-1}\,.
\end{align}
These properties allow us to extract a shift operator from the
transfer matrix,
\begin{align}
  T(u)\Big|_{u=0}=P_{\,\square_{N}\,\square_{N-1}}\cdots P_{\,\square_3\,\square_2}P_{\,\square_2\,\square_1}=:e^{\mathscr{P}}\,,
\end{align}
which we express in terms of its generator $\mathscr{P}$. The
Heisenberg Hamiltonian \eqref{eq:heisenberg-hamiltonian} is obtained
by taking the logarithmic derivative,
\begin{align}
  \label{eq:transfer-expansion}
  T(u)^{-1}\frac{\D}{\D u}T(u)\Big|_{u=0}
  =
  \sum_{i=1}^NP_{\,\square_i\,\square_{i+1}}
  =
  \frac{1}{2}(-\mathscr{H}+N)\,
\end{align}
with the identification $\square_{i+N}=\square_i$. Thus in the expansion of
the transfer matrix with respect to the spectral parameter $u$,
\begin{align}
  T(u)=\exp\Big(\mathscr{Q}^{[0]}+u\mathscr{Q}^{[1]}+u^2\mathscr{Q}^{[2]}+\ldots\Big)\,,
\end{align}
we can identify the first two coefficients with the generator of the
shift and the Hamiltonian, respectively,
\begin{align}
  \mathscr{Q}^{[0]}=\mathscr{P}\,,\quad
  \mathscr{Q}^{[1]}=\frac{1}{2}(-\mathscr{H}+N)\,.
\end{align}
The algebraic Bethe ansatz is a method to diagonalize the transfer
matrix. This matrix commutes for different values of the spectral
parameter,
\begin{align}
  [T(u),T(u')]=0\,.
\end{align}
This follows from the Yang-Baxter-like defining relation
\eqref{eq:yangian-def} of the Yangian after taking the trace over the
auxiliary spaces $\square$ and $\square'$. It implies that all the
coefficients of the expansion in \eqref{eq:transfer-expansion}
commute,
\begin{align}
  [\mathscr{Q}^{[r]},\mathscr{Q}^{[s]}]=0\,.
\end{align}
Therefore the algebraic Bethe ansatz in particular diagonalizes the
Hamiltonian \eqref{eq:heisenberg-hamiltonian} of the Heisenberg spin
chain. Let us remark that the expansion coefficients
$\mathscr{Q}^{[r]}$ are denoted as ``conserved quantities'' because
the Hamiltonian is among them. Furthermore, we may say that all these
charges are ``in involution'' because they commute. A sufficient
number of conserved quantities in involution, i.e.\ that
Poisson-commute, is the key ingredient of the Liouville theorem, that
defines integrable models with finitely many degrees of freedom in
classical mechanics, cf.\ \cite{Babelon:2003,Arnold:2013}. In this
sense, the QISM can be understood as a quantization of that classical
theorem, as already pointed out in section~\ref{sec:int-mod}.

After embedding the one-dimensional Heisenberg model in the QISM, we
discuss its solution via the algebraic Bethe ansatz. In fact, we
present this method for a more general class of $\mathfrak{gl}(2)$
spin chains. Our presentation highlights the essential features. More
detailed expositions can be found in the aforementioned reviews
\cite{Faddeev:1996iy,Korepin:1997}. We employ a $\mathfrak{gl}(2)$
monodromy matrix \eqref{eq:yangian-mono-spinchain} with general
finite-dimensional representations $\mathcal{V}_i$, inhomogeneity
parameters $v_i$ and normalizations $f_{\mathcal{V}_i}$ at the
sites. The monodromy \eqref{eq:heisenberg-monodromy} of the Heisenberg
model is a special case thereof. We express the general
$\mathfrak{gl}(2)$ monodromy as a matrix in the auxiliary space
$\square=\mathbb{C}^2$,
\begin{align}
  \label{eq:bethe-gl2-monodromy}
  \mon(\spec)=
  \begin{pmatrix}
    A(\spec)&B(\spec)\\
    C(\spec)&D(\spec)\\
  \end{pmatrix}.
\end{align}
The entries of this matrix are operators on the total quantum space
$\mathcal{V}_1\otimes\cdots\otimes\mathcal{V}_N$ of the spin
chain. This notation leads to the transfer matrix 
\begin{align}
  \label{eq:bethe-gl2-trans}
  T(\spec)=\tr_{\square}M(u)=A(\spec)+D(\spec)\,.
\end{align}
In what follows, we use the Yang-Baxter equation
\eqref{eq:yangian-def} obeyed by the monodromy $M(u)$ to diagonalize
this transfer matrix. We assume the existence of a reference state
$\bvac$ that is characterized by
\begin{align}
  \label{eq:bethe-gl2-vacuum}
  A(\spec)\bvac=\alpha(\spec)\bvac\,,
  \quad
  D(\spec)\bvac=\delta(\spec)\bvac\,,
  \quad
  C(\spec)\bvac=0\,
\end{align}
with some scalar functions $\alpha(u)$ and $\delta(u)$. These
conditions are fulfilled if we choose finite-dimensional
$\mathfrak{gl}(2)$ representations $\mathcal{V}_i$ for the generators
$J_{\alpha\beta}^i$ in the Lax operators \eqref{eq:yangian-def-lax} at
the spin chain sites. Such representations contain a highest weight
state $|\sigma_i\rangle$ that obeys
\begin{align}
  \label{eq:bethe-gl2-hws-loc}
  J^i_{11}|\sigma_i\rangle=\xi_i^{(1)}|\sigma_i\rangle\,,
  \quad  
  J^i_{22}|\sigma_i\rangle=\xi_i^{(2)}|\sigma_i\rangle\,,
  \quad  
  J^i_{12}|\sigma_i\rangle=0\,.
\end{align}
The scalar coefficients in these equations may be arranged into a
highest weight vector $\Xi_i=(\xi_i^{(1)},\xi_i^{(2)})$ that
characterizes the representation. In case of the Heisenberg model with
the defining representation $\mathcal{V}_i=\square_i$, we have
$\Xi_i=(1,0)$ and $|\sigma_i\rangle=\bigl(
\begin{smallmatrix}
  0\\
  1\\
\end{smallmatrix}\bigr)
$.
For a monodromy built from representations characterized by
\eqref{eq:bethe-gl2-hws-loc} the reference state in
\eqref{eq:bethe-gl2-vacuum} becomes the tensor product of the highest
weight states at the individual sites,
\begin{align}
  \label{eq:bethe-gl2-hws-tot}
  \bvac=|\sigma_1\rangle\otimes\cdots\otimes|\sigma_\sites\rangle\,.
\end{align} 
Moreover, the scalar functions in \eqref{eq:bethe-gl2-vacuum} can be
computed,
\begin{align}
  \label{eq:bethe-gl2-alphadelta}
  \begin{aligned}
    \alpha(\spec)
    =
    \prod_{i=1}^\sites f_{\mathcal{V}_i}(\spec-\inh_i)
    \frac{\spec-\inh_i+\xi_i^{(1)}}{\spec-\inh_i}\,,
    \quad
    \delta(\spec)
    =
    \prod_{i=1}^\sites f_{\mathcal{V}_i}(\spec-\inh_i)
    \frac{\spec-\inh_i+\xi_i^{(2)}}{\spec-\inh_i}\,.
  \end{aligned}
\end{align} 
However, in the following we do not use this explicit form of the
functions. It is sufficient to demand \eqref{eq:bethe-gl2-vacuum}. We
continue by making an ansatz for the eigenstates of the transfer
matrix \eqref{eq:bethe-gl2-trans},
\begin{align}
  \label{eq:bethe-gl2-eigenvector}
  |\Psi\rangle=B(\brt_1)\cdots B(\brt_\brts)\bvac\,.
\end{align}
It depends on $P$ complex parameters $u_k$, the so-called \emph{Bethe
  roots}. For generic values of these parameters the ansatz is not an
eigenstate of the transfer matrix. However, it turns into one if the
parameters satisfy a set of algebraic equations referred to as
\emph{Bethe equations}. To derive these equations we need some
commutation relations among the elements of the monodromy
\eqref{eq:bethe-gl2-monodromy}, which follow from the defining
relation \eqref{eq:yangian-def},
\begin{align}
  \label{eq:bethe-gl2-commrel-abd}
  \begin{aligned}
    A(\spec)B(\specp)
    &=
    \frac{\spec-\specp-1}{\spec-\specp}
    B(\specp)A(\spec)+\frac{1}{\spec-\specp}B(\spec)A(\specp)\,,\\
    D(\spec)B(\specp)
    &=
    \frac{\spec-\specp+1}{\spec-\specp}
    B(\specp)D(\spec)-\frac{1}{\spec-\specp}B(\spec)D(\specp)\,,\\
    B(\spec)B(\specp)
    &=
    B(\specp)B(\spec)\,.
  \end{aligned}
\end{align}
We continue by acting with the operators $A(u)$ and $D(u)$ on the
ansatz \eqref{eq:bethe-gl2-eigenvector}. Using
\eqref{eq:bethe-gl2-commrel-abd} we commute these operator to the
right and once they hit the reference state we employ
\eqref{eq:bethe-gl2-vacuum}. After a laborious calculation this
yields, see e.g.\ \cite{Faddeev:1996iy},
\begin{align}
  \label{eq:bethe-gl2-apsi-dpsi}
  \begin{aligned}
    A(\spec)|\Psi\rangle
    &=
    \alpha(\spec)\frac{Q(\spec-1)}{Q(\spec)}|\Psi\rangle-
    \sum_{k=1}^\brts\frac{\alpha(\brt_k)Q(\brt_k-1)}{\spec-\brt_k}B(\spec)
    \prod_{\substack{i=1\\i\neq k}}^\brts\frac{B(\brt_i)}{\brt_k-\brt_i}\bvac\,,\\
    D(\spec)|\Psi\rangle
    &=
    \delta(\spec)\frac{Q(\spec+1)}{Q(\spec)}|\Psi\rangle-
    \sum_{k=1}^\brts\frac{\delta(\brt_k)Q(\brt_k+1)}{\spec-\brt_k}B(\spec)
    \prod_{\substack{i=1\\i\neq k}}^\brts\frac{B(\brt_i)}{\brt_k-\brt_i}\bvac\,.\\
  \end{aligned}
\end{align}
We expressed the result using Baxter's Q-function, that is a
polynomial of degree $P$ whose roots are the Bethe roots $u_k$,
\begin{align}
  \label{eq:bethe-gl2-qfunct}
  Q(\spec)=\prod_{k=1}^\brts(\spec-\brt_k)\,.
\end{align}
To render $|\Psi\rangle$ into an eigenstate of the transfer matrix
\eqref{eq:bethe-gl2-trans}, the Bethe equations
\begin{align}
  \label{eq:bethe-gl2-betheeq}
  \alpha(\brt_k)Q(\brt_k-1)+\delta(\brt_k)Q(\brt_k+1)=0
\end{align}
for $k=1,\ldots,P$ have to be obeyed. They guarantee that after
summing up the two lines in \eqref{eq:bethe-gl2-apsi-dpsi} the
``unwanted terms'', which are not proportional to $|\Psi\rangle$,
disappear. A more common form of the Bethe equations is obtained by
solving for the fraction of Q-functions and using
\eqref{eq:bethe-gl2-alphadelta} as well as
\eqref{eq:bethe-gl2-qfunct},
\begin{align}
  \label{eq:bethe-gl2-betheeq-ordinary}
  \prod_{i=1}^\sites\frac{\brt_k-\inh_i+\xi_i^{(1)}}{\brt_k-\inh_i+\xi_i^{(2)}}
  =
  -\prod_{j=1}^\brts\frac{\brt_k-\brt_j+1}{\brt_k-\brt_j-1}\,.
\end{align}
However, we will mostly work with the Bethe equations in the form
\eqref{eq:bethe-gl2-betheeq} because for the special solutions that we
will examine in section~\ref{sec:bethe-yangian}, we would divide by
zero in \eqref{eq:bethe-gl2-betheeq-ordinary}. From
\eqref{eq:bethe-gl2-apsi-dpsi} it follows that the eigenvalue
$\tau(u)$ of $T(u)$ for the eigenstate $|\Psi\rangle$ is given by the
\emph{Baxter equation}
\begin{align}
  \label{eq:bethe-gl2-baxtereq}
  \tau(\spec)
  =
  \alpha(\spec)\frac{Q(\spec-1)}{Q(\spec)}
  +\delta(\spec)\frac{Q(\spec+1)}{Q(\spec)}\,.
\end{align}
The Bethe equations are a consequence of this equation alone if one
assumes regularity of the functions $\tau(u)$, $\alpha(u)$ and
$\delta(u)$ at the values of Bethe roots $u=u_k$ and furthermore
demands $Q(u)$ to be of the form \eqref{eq:bethe-gl2-qfunct}. With
these assumptions taking the residues of \eqref{eq:bethe-gl2-baxtereq}
at $u=u_k$ yields the Bethe equations \eqref{eq:bethe-gl2-betheeq}.

In this section we demonstrated how the transfer matrix $T(u)$ in
\eqref{eq:bethe-gl2-trans} of a finite-dimensional $\mathfrak{gl}(2)$
spin chain, and in particular of the Heisenberg model, is diagonalized
employing the algebraic Bethe ansatz. This method led to the formula
\eqref{eq:bethe-gl2-eigenvector} for the eigenvectors $|\Psi\rangle$
of the transfer matrix and to \eqref{eq:bethe-gl2-baxtereq} for its
eigenvalues $\tau(u)$. Both formulas are parameterized in terms of
Bethe roots $u_k$, which have to obey the Bethe equations
\eqref{eq:bethe-gl2-betheeq}. Obtaining solutions of these algebraic
equations is an important and difficult part of the analysis of the
integrable model at hand. For this one often has to resort to
numerical approximations. However, in certain cases exact analytical
solutions can be obtained. In the next section we will study in detail
a situation where this is possible. Let us also add that while we
focused on $\mathfrak{gl}(2)$ spin chains, the algebraic Bethe ansatz
can be extended to $\mathfrak{gl}(n)$. This higher rank case is
referred to as \emph{nested} algebraic Bethe ansatz and it is
technically considerably more involved, see e.g.\
\cite{Kulish:1983rd}.

\section{Bethe Ansatz for Yangian Invariants}
\label{sec:bethe-yangian}

We concluded chapter~\ref{cha:yang-rep} by presenting some sample
Yangian invariants. These were constructed ``by hand'' in the sense
that we explicitly inspected the Yangian invariance condition. The aim
of this section is to put forward a systematic construction of such
Yangian invariants. To explain the basic idea, let us focus on the
bosonic case of the Yangian of $\mathfrak{gl}(n)$. The crucial
starting point of this method is the Yangian invariance condition
\eqref{eq:yi} in terms of a spin chain monodromy $M(u)$ rather than
the expanded versions \eqref{eq:yi-exp-1} or \eqref{eq:yi-exp-2}
thereof. Taking the trace over the auxiliary space
$\square=\mathbb{C}^n$ in this formulation of the Yangian invariance
condition yields
\begin{align}
  \label{eq:bethe-inv-eigen}
  T(\spec)|\Psi\rangle=n|\Psi\rangle\,
\end{align} 
with the transfer matrix 
\begin{align}
  \label{eq:bethe-trans}
  T(\spec)=\tr_{\square}\mon(\spec)\,.
\end{align}
Therefore a Yangian invariant $|\Psi\rangle$ is a special eigenstate
of a transfer matrix. This transfer matrix can be diagonalized using a
Bethe ansatz, at least for finite-dimensional representations in the
quantum space. This renders the invariant $|\Psi\rangle$ into a
special Bethe vector and thereby makes it amenable to a Bethe ansatz
construction.

Here we will implement this general idea for simplicity in case of
compact invariants of the Yangian of $\mathfrak{gl}(2)$. In
section~\ref{sec:bethe-gl2} we specialize the Bethe ansatz of
section~\ref{sec:bethe-ansatze} to the case of Yangian invariant Bethe
vectors. We find that these are characterized by functional relations,
which emerge as a special case of the Baxter equation
\eqref{eq:bethe-gl2-baxtereq}. These functional relations determine
the Bethe roots and also severely constrain the inhomogeneities and
representation labels of the monodromy matrix. Remarkably, we find
that these relations can easily be solved analytically. In
section~\ref{sec:bethe-gl2-sol} we exemplify this observation by
discussing sample invariants that include in particular those of
section~\ref{sec:comp-boson-invar}. These sample invariants illustrate
the general structure of the solutions. The Bethe roots form strings
in the complex plane. The position of these strings is determined by
the inhomogeneities and their length by the representation labels.
Section~\ref{sec:class-solut} contains a classification of all
solutions to the functional equations and therefore basically of all
Yangian invariants within the studied class of representations. Each
Yangian invariant is associated with a permutation. These are
essentially those permutations that we already encountered in the
introductory section~\ref{sec:deform} on deformed SYM scattering
amplitudes. Supplementary material on the Bethe ansatz for Yangian
invariants is provided in appendix~\ref{cha:app-bethe}. It contains
results on the extension to $\mathfrak{gl}(n)$ and on the evaluation
of Yangian invariant Bethe vectors.

\subsection{Derivation of Functional Relations}
\label{sec:bethe-gl2}

To begin with, we write out the definition \eqref{eq:yi} of Yangian
invariants for $\mathfrak{gl}(2)$ using the notation
\eqref{eq:bethe-gl2-monodromy} for the monodromy elements,
\begin{align}
  \label{eq:bethe-gl2-diag} A(\spec)|\Psi\rangle&=|\Psi\rangle\,,&
D(\spec)|\Psi\rangle&=|\Psi\rangle\,,\\
  \label{eq:bethe-gl2-offdiag} B(\spec)|\Psi\rangle&=0\,,&
C(\spec)|\Psi\rangle&=0\,.
\end{align} 
Here we grouped the equations for diagonal elements of the monodromy
in \eqref{eq:bethe-gl2-diag} and those for off-diagonal elements in
\eqref{eq:bethe-gl2-offdiag}. For the construction of solutions to
these equations we proceed in two steps. First, we solve
\eqref{eq:bethe-gl2-diag} by specializing the algebraic Bethe ansatz
reviewed in section~\ref{sec:bethe-ansatze}. Second, we show that for
finite-dimensional representations \eqref{eq:bethe-gl2-diag} implies
\eqref{eq:bethe-gl2-offdiag}. These steps lead to a characterization
of Yangian invariants in terms of functional relations, which we
summarize at the end of this section.

Let us address the equations in \eqref{eq:bethe-gl2-diag} for the
diagonal monodromy elements. In the Bethe ansatz we seek eigenvectors
$|\Psi\rangle$ of the transfer matrix $T(u)=A(u)+D(u)$. Here we demand
in addition that $|\Psi\rangle$ is a common eigenvector of $A(u)$ and
$D(u)$ individually. We proceed as for usual the Bethe ansatz by
making the ansatz \eqref{eq:bethe-gl2-eigenvector} for
$|\Psi\rangle$. Next, and still in full analogy to the Bethe ansatz,
we derive the action \eqref{eq:bethe-gl2-apsi-dpsi} of $A(u)$ and
$D(u)$ on this ansatz. The following step differs from the usual Bethe
ansatz in that we demand the ``unwanted terms'' in both lines of
\eqref{eq:bethe-gl2-apsi-dpsi} to vanish individually. This leads to a
special case of the Bethe equations \eqref{eq:bethe-gl2-betheeq},
\begin{align}
  \label{eq:bethe-gl2-specbethe} \alpha(\brt_k)Q(\brt_k-1)=0\,, \quad
\delta(\brt_k)Q(\brt_k+1)=0\,.
\end{align}
To obtain the correct eigenvalues of $A(u)$ and $D(u)$ in
\eqref{eq:bethe-gl2-diag}, we further have to demand in
\eqref{eq:bethe-gl2-apsi-dpsi} that
\begin{align}
  \label{eq:bethe-gl2-specbaxter}
1=\alpha(\spec)\frac{Q(\spec-1)}{Q(\spec)}\,, \quad
1=\delta(\spec)\frac{Q(\spec+1)}{Q(\spec)}\,.
\end{align}
These equations are a degenerate case of the Baxter equation
\eqref{eq:bethe-gl2-baxtereq}. Each of the terms on the right hand
site of \eqref{eq:bethe-gl2-baxtereq} is fixed to $1$ and thus we
obtain the transfer matrix eigenvalue $\tau(u)=2$ in accordance with
\eqref{eq:bethe-inv-eigen}. Assuming regularity of $\alpha(u)$ and
$\delta(u)$ at the Bethe roots $u=u_k$, the residues of the functional
relations \eqref{eq:bethe-gl2-specbaxter} at these points yield the
special case of the Bethe equations in
\eqref{eq:bethe-gl2-specbethe}. Consequently, we reduced the solution
of the diagonal part \eqref{eq:bethe-gl2-diag} of the Yangian
invariance condition to the functional relations
\eqref{eq:bethe-gl2-specbaxter}.

We move on to discuss the equations in \eqref{eq:bethe-gl2-offdiag}
for the off-diagonal monodromy elements. Our argument makes use of
\eqref{eq:bethe-gl2-diag} for the diagonal elements, which we just
solved.  Expanding these equations in the spectral parameter $u$ as in
\eqref{eq:yangian-mono} yields
\begin{align}
  \label{eq:bethe-gl2-gl2-weight} \mon_{11}^{(1)}|\Psi\rangle=0\,,
\quad \mon_{22}^{(1)}|\Psi\rangle=0\,.
\end{align}
We mentioned after \eqref{eq:yangian-def-gen} that
$M^{(1)}_{\beta\alpha}$ are generators of a $\mathfrak{gl}(2)$
algebra. Therefore, \eqref{eq:bethe-gl2-gl2-weight} is equivalent to
saying that $|\Psi\rangle$ has $\mathfrak{gl}(2)$ weight
$(0,0)$. Next, we expand $C(u)|\Omega\rangle=0$ in
\eqref{eq:bethe-gl2-vacuum} to obtain
$M^{(1)}_{21}|\Omega\rangle=0$. Employing this condition, the
commutation relations \eqref{eq:yangian-gen-adjoint} and
\eqref{eq:bethe-gl2-apsi-dpsi} results in
\begin{align}
  \label{eq:bethe-gl2-m21psi} \mon^{(1)}_{21}|\Psi\rangle =
  -\sum_{k=1}^\brts(\alpha(\brt_k)Q(\brt_k-1)+\delta(\brt_k)Q(\brt_k+1))
  \prod_{\substack{i=1\\i\neq
  k}}^\brts\frac{B(\brt_i)}{\brt_k-\brt_i}\bvac = 0\,.
\end{align}
Here we used \eqref{eq:bethe-gl2-specbethe} for the last equality. In
case of finite-dimensional representations, which we are dealing with,
\eqref{eq:bethe-gl2-gl2-weight} and \eqref{eq:bethe-gl2-m21psi} are
sufficient to guarantee that $|\Psi\rangle$ is a $\mathfrak{gl}(2)$
singlet. This in turn implies
\begin{align}
  \label{eq:bethe-gl2-m12psi} \mon^{(1)}_{12}|\Psi\rangle=0\,.
\end{align} 
Lastly, we use \eqref{eq:yangian-gen-adjoint} once more to show
\begin{align}
  \label{eq:bethe-gl2-comm}
[\mon_{12}^{(1)},A(\spec)-D(\spec)]=2B(\spec)\,, \quad
[\mon_{21}^{(1)},D(\spec)-A(\spec)]=2C(\spec)\,.
\end{align} 
Acting with these equations on $|\Psi\rangle$ and employing
\eqref{eq:bethe-gl2-diag}, \eqref{eq:bethe-gl2-m21psi} as well as
\eqref{eq:bethe-gl2-m12psi} implies \eqref{eq:bethe-gl2-offdiag}, the
other part of the Yangian invariance condition for the off-diagonal
monodromy elements.

To summarize, we reduced the construction of solutions $|\Psi\rangle$
to the Yangian invariance condition \eqref{eq:yi} for
$\mathfrak{gl}(2)$ to the functional relations
\eqref{eq:bethe-gl2-specbaxter}. Given a solution
$(\alpha(u),\delta(u),Q(u))$ of these relations, where $\alpha(u)$ and
$\delta(u)$ are regular at $u=u_k$ and $Q(u)$ is of the form
\eqref{eq:bethe-gl2-qfunct}, the Yangian invariant $|\Psi\rangle$ is
the Bethe vector \eqref{eq:bethe-gl2-eigenvector}. The Bethe roots in
this vector are the zeros of $Q(u)$. There exists are remarkable
reformulation of the two functional relations in
\eqref{eq:bethe-gl2-specbaxter}. They decouple into one equation that
only contains the eigenvalues of the monodromy,
\begin{align}
  \label{eq:bethe-gl2-ad} 1=\alpha(\spec)\delta(\spec-1)\,,
\end{align}
and another equation in which also the Bethe roots enter,
\begin{align}
  \label{eq:bethe-gl2-qaq}
\frac{Q(\spec)}{Q(\spec+1)}=\delta(\spec)\,.
\end{align} 
The crucial step in analyzing this system is to find solutions of
\eqref{eq:bethe-gl2-ad}. The eigenvalues of the monodromy depend on
the inhomogeneities and the representation labels, cf.\
\eqref{eq:bethe-gl2-alphadelta}. Consequently, this equation selects
those monodromies which admit solutions of the Yangian invariance
condition \eqref{eq:yi}. Provided a solution of
\eqref{eq:bethe-gl2-ad}, the difference equation
\eqref{eq:bethe-gl2-qaq} can typically be solved with ease for the
Bethe roots $u_k$ contained in $Q(u)$. This is in stark contrast to
the usual application of the Bethe ansatz to a spin chain spectral
problem, where the Bethe equations are very hard to solve. We refer to
the construction of Yangian invariants described in this section as
\emph{Bethe ansatz for Yangian invariants}.

\subsection{Sample Solutions}
\label{sec:bethe-gl2-sol}

To get acquainted with the Bethe ansatz construction of Yangian
invariants, we use this method to rederive the compact bosonic sample
invariants of section~\ref{sec:comp-boson-invar} in the
$\mathfrak{gl}(2)$ case. The oscillator representations $\oscrep_c$
and $\bar{\oscrep}_{-c}$ with $c\in\mathbb{N}$, which we used there at
the sites of the monodromy, possess respectively the highest weights
$\Xi=(c,0)$ and $\Xi=(0,-c)$, cf.\
\eqref{eq:comp-osc-weights}. Therefore these invariants fall within
the reach of the Bethe ansatz. We discuss the two-site invariant
$|\Psi_{2,1}\rangle$, the three-site invariants $|\Psi_{3,1}\rangle$
and $|\Psi_{3,2}\rangle$ as well as the four-site invariant
$|\Psi_{4,2}(z)\rangle$ corresponding to the R-matrix. For each of
these sample invariants we present the solution of the functional
relations \eqref{eq:bethe-gl2-ad} and \eqref{eq:bethe-gl2-qaq}. In
particular, we find that the Bethe roots arrange into strings in the
complex plane. What is more, we present a superposition principle for
solutions of the functional relations.

\subsubsection{Two-Site Invariant}
\label{sec:bethe-gl2-sol-line}

\begin{figure}[!t]
  \begin{center}
    \begin{tikzpicture} \draw[thick] (0,0) circle (3pt)
node[above=0.1cm]{$\inh_2$}; \filldraw[thick] (-1,0) circle (1pt)
node[below]{$\brt_1$}; \filldraw[thick] (-2,0) circle (1pt)
node[below]{$\brt_2$}; \node at (-3,0) {\ldots}; \filldraw[thick]
(-4,0) circle (1pt) node[below]{$\brt_{c_2-1}$}; \filldraw[thick]
(-5,0) circle (1pt) node[below]{$\brt_{c_2}$}; \draw[thick] (-6,0)
circle (3pt) node[above=0.1cm]{$\inh_1$}; \draw[thick,|-|] (-6,1) --
node[midway,above] {$c_2+1$} (0,1);
    \end{tikzpicture}
    \caption{The Yangian invariant $|\Psi_{2,1}\rangle$ of
      section~\ref{sec:bethe-gl2-sol-line} is constructed from Bethe
      roots $u_k$ that form a string in the complex plane between the
      inhomogeneities $\inh_1$ and $\inh_2$, cf.\
      \eqref{eq:bethe-gl2-sol-line-roots}. The string contains $c_2$
      roots with a uniform real spacing of $1$.}
    \label{fig:bethe-gl2-sol-line}
  \end{center}
\end{figure}
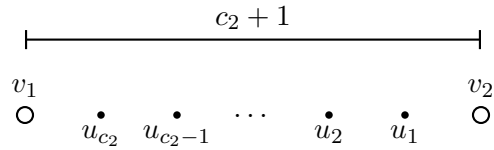

Here we discuss the $\mathfrak{gl}(2)$ case of the invariant
$|\Psi_{2,1}\rangle$ that was constructed ``by hand'' in
section~\ref{sec:comp-bos-2-site}. Let us recall the representations
and inhomogeneities of the monodromy $\mon_{2,1}(\spec)$ associated
with that invariant, cf.\ \eqref{eq:osc-m21} and
\eqref{eq:osc-m21-vs},
\begin{align}
  \label{eq:bethe-gl2-sol-line-constr}
  \begin{gathered} \mathcal{V}_1=\bar{\oscrep}_{c_1}\,, \quad \mathcal{V}_2=\oscrep_{c_2}\,,\\
    \inh_1=\inh_2-1-c_2\,, \quad c_1+c_2=0\,.
  \end{gathered}
\end{align} 
We use this data to compute the monodromy eigenvalues in
\eqref{eq:bethe-gl2-alphadelta},
\begin{align}
  \label{eq:bethe-gl2-sol-line-ad-eval}
\alpha(\spec)=\frac{\spec-\inh_2+c_2}{\spec-\inh_2}\,, \quad
\delta(\spec)=\frac{\spec-\inh_2+1}{\spec-\inh_2+1+c_2}\,,
\end{align} 
where we also inserted the trivial normalization
\eqref{eq:osc-m21-norm} and the highest weights
\eqref{eq:comp-osc-weights}. One readily verifies that these
eigenvalues obey the functional relation \eqref{eq:bethe-gl2-ad}.  The
solution of the remaining relation \eqref{eq:bethe-gl2-qaq} is
\begin{align}
  \label{eq:bethe-gl2-sol-line-q} Q(\spec) =
\frac{\Gamma(\spec-\inh_2+c_2+1)}{\Gamma(\spec-\inh_2+1)} =
\prod_{k=1}^{c_2}(\spec-\inh_2+k)\,.
\end{align}
Demanding the Q-function to be of the polynomial form
\eqref{eq:bethe-gl2-qfunct} eliminates the freedom to multiply a
solution of \eqref{eq:bethe-gl2-qaq} by any function of period $1$ in
$u$. Noting that $c_2$ is a positive integer, the gamma functions
reduce to a polynomial. We extract the Bethe roots as its zeros,
\begin{align}
  \label{eq:bethe-gl2-sol-line-roots}
\brt_k=\inh_2-k\quad\text{for}\quad k=1,\ldots,c_2\,.
\end{align}
These roots form a string in the complex plane, see
figure~\ref{fig:bethe-gl2-sol-line}. As always for the
$\mathfrak{gl}(2)$ Bethe ansatz, we may permute the labels of the
Bethe roots because the operators $B(u)$ entering the Bethe vector
\eqref{eq:bethe-gl2-eigenvector} commute for different values of the
spectral parameter $u$, cf.\
\eqref{eq:bethe-gl2-commrel-abd}. Ultimately, we want to obtain the
Yangian invariant Bethe vector \eqref{eq:bethe-gl2-eigenvector}
associated with the solution of the functional relations presented
here. First of all, this requires the reference state
\eqref{eq:bethe-gl2-hws-tot}. For the representations listed in
\eqref{eq:bethe-gl2-sol-line-constr} it is the tensor product of the
highest weight states \eqref{eq:osc-hws},
\begin{align}
  \label{eq:bethe-gl2-sol-line-vac} \bvac = (\bar\osca_2^1)^{c_2}
(\bar\osca_1^2)^{c_2} |0\rangle\,.
\end{align} 
Next, we evaluate \eqref{eq:bethe-gl2-eigenvector} employing
\eqref{eq:bethe-gl2-sol-line-constr},
\eqref{eq:bethe-gl2-sol-line-roots} and
\eqref{eq:bethe-gl2-sol-line-vac}. Details of this computation for
general $c_2\in\mathbb{N}$ are presented in
appendix~\ref{sec:derivation-two-site}. It is worth noting that the
normalization of the operators $B(\brt_k)$ trivializes due to
\eqref{eq:osc-m21-norm}. We obtain
\begin{align}
  \label{eq:bethe-gl2-sol-line-inv} 
  |\Psi \rangle = B(\brt_1)\cdots
  B(\brt_{c_2})\bvac = (-1)^{c_2} (1\bullet 2)^{c_2}|0\rangle 
  \propto |\Psi_{2,1}\rangle\,.
\end{align} 
Thus the Bethe ansatz for Yangian invariants indeed reproduces
$|\Psi_{2,1}\rangle$ as computed in \eqref{eq:osc-psi21}.

\subsubsection{Three-Site Invariants}
\label{sec:bethe-gl2-sol-three}

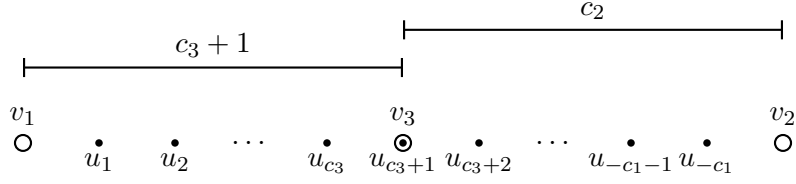
\begin{figure}[!t]
  \begin{center}
    \begin{tikzpicture} 
      \draw[thick] (0,0) circle (3pt)
      node[above=0.1cm]{$\inh_1$}; \filldraw[thick] (1,0) circle (1pt)
      node[below]{$\brt_1$}; \filldraw[thick] (2,0) circle (1pt)
      node[below]{$\brt_2$}; \node at (3,0) {\ldots}; \filldraw[thick] (4,0)
      circle (1pt) node[below]{$\brt_{c_3}$}; \filldraw[thick] (5,0) circle
      (1pt) node[below]{$\brt_{c_3+1}$}; \draw[thick] (5,0) circle (3pt)
      node[above=0.1cm]{$\inh_3$}; \filldraw[thick] (6,0) circle (1pt)
      node[below]{$\brt_{c_3+2}$}; \node at (7,0) {\ldots}; \filldraw[thick]
      (8,0) circle (1pt) node[below]{$\brt_{-c_1-1}$}; \filldraw[thick] (9,0)
      circle (1pt) node[below]{$\brt_{-c_1}$}; \draw[thick] (10,0) circle
      (3pt) node[above=0.1cm]{$\inh_2$}; \draw[thick,|-|] (0,1) --
      node[midway,above] {$c_3+1$} (5,1); \draw[thick,|-|] (5,1.5) --
      node[midway,above] {$c_2$} (10,1.5);
    \end{tikzpicture}
    \caption{The invariant $|\Psi_{3,1}\rangle$ is constructed from a
      real string of $-c_1=c_2+c_3\in\mathbb{N}$ uniformly spaced
      Bethe roots $\brt_k$ in the complex plane, cf.\
      \eqref{eq:bethe-gl2-sol-three1-roots}. The roots lie between the
      inhomogeneities $\inh_1$, $\inh_2$ and one of them coincides
      with $\inh_3$.}
    \label{fig:bethe-gl2-three1-line}
  \end{center}
\end{figure}

Two different tree-site sample invariants were introduced in
section~\ref{sec:comp-bos-3-site}. The monodromy $M_{3,1}(u)$
associated to the first invariant $|\Psi_{3,1}\rangle$ is defined by
the representations and inhomogeneities, cf.\ \eqref{eq:osc-m31} and
\eqref{eq:osc-m31-vs},
\begin{align}
  \label{eq:bethe-gl2-sol-three1-constr}
  \begin{gathered} 
    \mathcal{V}_1=\bar{\oscrep}_{c_1}\,, \quad 
    \mathcal{V}_2=\oscrep_{c_2}\,, \quad
    \mathcal{V}_3=\oscrep_{c_3}\,,\\
    \inh_2=\inh_1+1+c_2+c_3\,, \quad
    \inh_3=\inh_1+1+c_3\,, \quad 
    c_1+c_2+c_3=0\,,
  \end{gathered}
\end{align} 
where we specialized to the $\mathfrak{gl}(2)$ case and we recall that
$c_2,c_3\in\mathbb{N}$. With these parameters, the trivial
normalization of the monodromy \eqref{eq:osc-m31-norm} and the highest
weights \eqref{eq:comp-osc-weights}, we evaluate the monodromy
eigenvalues \eqref{eq:bethe-gl2-alphadelta} on the reference state,
\begin{align}
  \label{eq:bethe-gl2-sol-three1-ad-eval}
\alpha(\spec)=\frac{\spec-\inh_1-1}{\spec-\inh_1+c_1-1}\,, \quad
\delta(\spec)=\frac{\spec-\inh_1+c_1}{\spec-\inh_1}\,.
\end{align}
These satisfy the functional relation
\eqref{eq:bethe-gl2-ad}. Assuming the Q-function to be of the form
\eqref{eq:bethe-gl2-qfunct}, the other functional relation
\eqref{eq:bethe-gl2-qaq} has the unique solution
\begin{align}
  \label{eq:bethe-gl2-sol-three1-q} Q(\spec) =
\frac{\Gamma(\spec-\inh_1)}{\Gamma(\spec-\inh_1-c_2-c_3)} =
\prod_{k=1}^{c_2+c_3}(\spec-\inh_1-k)\,.
\end{align} 
Its zeros define the Bethe roots
\begin{align}
  \label{eq:bethe-gl2-sol-three1-roots}
  \brt_k=\inh_1+k\quad\text{for}\quad k=1,\ldots,c_2+c_3\,,
\end{align} 
which again form a string in the complex plane, see
figure~\ref{fig:bethe-gl2-three1-line}. We continue to calculate the
associated Bethe vector. With the representations given in
\eqref{eq:bethe-gl2-sol-three1-constr} the reference state
\eqref{eq:bethe-gl2-hws-tot} for this vector reads
\begin{align}
  \label{eq:bethe-gl2-sol-three1-vac} \bvac =
  (\bar\osca_2^1)^{c_2+c_3} (\bar\osca_1^2)^{c_2}
  (\bar\osca_1^3)^{c_3}|0\rangle\,.
\end{align} 
Note that the Bethe root $\brt_{c_3+1}=\inh_3$ is identical with an
inhomogeneity. As a consequence the Lax operator
$R_{\square\,{\oscrep_{c_3}}}(\brt_{c_3+1}-\inh_3)$, which enters
$B(\brt_{c_3+1})$ in the Bethe vector
\eqref{eq:bethe-gl2-eigenvector}, diverges, cf.\
\eqref{eq:osc-lax-fund-s}. To nevertheless obtain a finite Bethe
vector  we rely on an ad hoc prescription, which we checked for small
values of $c_2$ and $c_3$: In a first step, the parameter
$\brt_{c_3+1}$ is kept at a generic value, while all other Bethe roots
are inserted into \eqref{eq:bethe-gl2-eigenvector}. This renders the
resulting expression finite at $\brt_{c_3+1}=\inh_3$. After then also
inserting this last root, we are left with
\begin{align}
  \label{eq:bethe-gl2-sol-three1-inv} 
  |\Psi\rangle=B(\brt_1)\cdots
  B(\brt_{c_2+c_3})\bvac = (-1)^{c_2+c_3}
  (1\bullet 2)^{c_2}
  (1\bullet 3)^{c_3} |0\rangle \propto
  |\Psi_{3,1}\rangle\,.
\end{align} 
Hence, we constructed the three-site Yangian invariant
$|\Psi_{3,1}\rangle$ from \eqref{eq:osc-psi31} by means of a Bethe
ansatz. It would clearly be desirable to obtain a better understanding
of the divergence and to derive \eqref{eq:bethe-gl2-sol-three1-inv}
for general $c_2,c_3\in\mathbb{N}$. A promising avenue to address
these points might be a generalization of
appendix~\ref{sec:derivation-two-site} to the three-site case.

So-called ``singular solutions'' of the Bethe equations, which
superficially lead to divergent Bethe vectors, are well-known for the
Heisenberg spin chain and certain generalizations thereof. Recent
discussions of this phenomenon can be found in
\cite{Nepomechie:2013mua} and \cite{Baxter:2001sx}, see also the
references therein. The problem was even known to Bethe
\cite{Bethe:1931hc}. It also surfaced in the early days of the planar
$\mathcal{N}=4$ SYM spectral problem \cite{Beisert:2003xu}. There
exist different approaches to treat theses solutions properly, cf.\
\cite{Nepomechie:2013mua}. Some of them might also be applicable for
the inhomogeneous spin chain with mixed representations that is
associated with the Yangian invariant $|\Psi_{3,1}\rangle$.

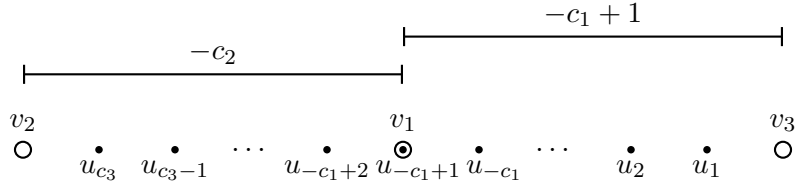
\begin{figure}[!t]
  \begin{center}
    \begin{tikzpicture} 
      \draw[thick] (0,0) circle (3pt)
      node[above=0.1cm]{$\inh_2$}; \filldraw[thick] (1,0) circle (1pt)
      node[below]{$\brt_{c_3}$}; \filldraw[thick] (2,0) circle (1pt)
      node[below]{$\brt_{c_3-1}$}; \node at (3,0) {\ldots}; \filldraw[thick]
      (4,0) circle (1pt) node[below]{$\brt_{-c_1+2}$}; \filldraw[thick] (5,0)
      circle (1pt) node[below]{$\hphantom{aa}\brt_{-c_1+1}$}; \draw[thick] (5,0) circle
      (3pt) node[above=0.1cm]{$\inh_1$}; \filldraw[thick] (6,0) circle (1pt)
      node[below]{$\hphantom{aa}\brt_{-c_1}$}; \node at (7,0) {\ldots}; \filldraw[thick]
      (8,0) circle (1pt) node[below]{$\brt_2$}; \filldraw[thick] (9,0)
      circle (1pt) node[below]{$\brt_1$}; \draw[thick] (10,0) circle (3pt)
      node[above=0.1cm]{$\inh_3$}; \draw[thick,|-|] (0,1) --
      node[midway,above] {$-c_2$} (5,1); \draw[thick,|-|] (5,1.5) --
      node[midway,above] {$-c_1+1$} (10,1.5);
    \end{tikzpicture}
    \caption{The $c_3=-c_1-c_2\in\mathbb{N}$ Bethe roots $\brt_k$
      associated with the invariant $|\Psi_{3,2}\rangle$ form a
      string. They lie between the inhomogeneities $\inh_2$ and
      $\inh_3$. One of the roots coincides with $\inh_1$.}
    \label{fig:bethe-gl2-three2-line}
  \end{center}
\end{figure}

The second three-site invariant discussed in
section~\ref{sec:comp-bos-3-site} is $|\Psi_{3,2}\rangle$. Its
monodromy is characterized by, cf.\ \eqref{eq:osc-m32} and
\eqref{eq:osc-m32-vs},
\begin{align}
  \label{eq:bethe-gl2-sol-three2-constr}
  \begin{gathered} 
    \mathcal{V}_1=\bar{\oscrep}_{c_1}\,, \quad
    \mathcal{V}_2=\bar{\oscrep}_{c_2}\,, \quad
    \mathcal{V}_3=\oscrep_{c_3}\,,\\
    \inh_1=\inh_3-1+c_1\,, \quad 
    \inh_2=\inh_3-1-c_3\,, \quad
    c_1+c_2+c_3=0\,,
  \end{gathered}
\end{align}
where $c_1,c_2\in -\mathbb{N}$. With the trivial normalization
\eqref{eq:osc-m32-norm} of this monodromy and the form of the highest
weights in \eqref{eq:comp-osc-weights}, this turns
\eqref{eq:bethe-gl2-alphadelta} into
\begin{align}
  \label{eq:bethe-gl2-sol-three2-ad-eval}
\alpha(\spec)=\frac{\spec-\inh_3+c_3}{\spec-\inh_3}\,, \quad
\delta(\spec)=\frac{\spec-\inh_3+1}{\spec-\inh_3+1+c_3}\,.
\end{align} 
Evidently, these functions solve the relation
\eqref{eq:bethe-gl2-ad}. The other relation \eqref{eq:bethe-gl2-qaq}
is then uniquely solved by
\begin{align}
  \label{eq:bethe-gl2-sol-three2-q} Q(\spec) =
\frac{\Gamma(\spec-\inh_3+c_3+1)}{\Gamma(\spec-\inh_3+1)} =
\prod_{k=1}^{c_3}(\spec-\inh_3+k)\,,
\end{align}
assuming we require the solution to be of the form
\eqref{eq:bethe-gl2-qfunct}. This yields the Bethe roots
\begin{align}
  \label{eq:bethe-gl2-sol-three2-roots}
  \brt_k=\inh_3-k\quad\text{for}\quad k=1,\ldots,c_3\,.
\end{align} 
Also for this sample solution they form a string, see
figure~\ref{fig:bethe-gl2-three2-line}. From the representations in
\eqref{eq:bethe-gl2-sol-three2-constr} we work out the reference state
\eqref{eq:bethe-gl2-hws-tot} of the corresponding Bethe vector,
\begin{align}
  \label{eq:bethe-gl2-sol-three2-vac} \bvac = (\bar\osca_2^1)^{-c_1}
(\bar\osca_2^2)^{-c_2} (\bar\osca_1^3)^{c_3}|0\rangle\,.
\end{align} 
The operator $B(\brt_{-c_1+1})$ diverges analogously to the one in the
first three-site invariant because $\brt_{-c_1+1}=\inh_1$. Using the
prescription from above we are led to a finite Bethe vector. By means
of an explicit calculation for small absolute values of the
representation labels we obtain
\begin{align}
  \label{eq:bethe-gl2-sol-three2-inv} |\Psi\rangle=B(\brt_1)\cdots
  B(\brt_{c_3})\bvac =(-1)^{c_3} 
  (1\bullet 3)^{-c_1}
  (2\bullet 3)^{-c_2} |0\rangle \propto
|\Psi_{3,2}\rangle\,.
\end{align} 
This agrees with the three-site invariant $|\Psi_{3,2}\rangle$ from
\eqref{eq:osc-psi32}.

\subsubsection{Four-Site Invariant and Superposition}
\label{sec:bethe-gl2-sol-four}

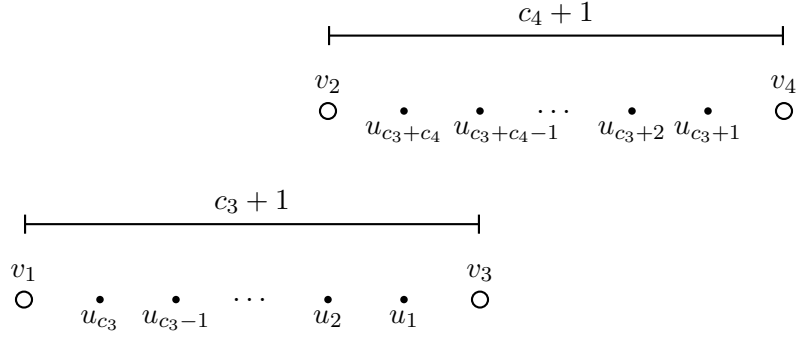
\begin{figure}[!t]
  \begin{center}
    \begin{tikzpicture}
      \begin{scope} 
        \draw[thick] (0,0) circle (3pt)
        node[above=0.1cm]{$\inh_3$}; \filldraw[thick] (-1,0) circle (1pt)
        node[below]{$\brt_1$}; \filldraw[thick] (-2,0) circle (1pt)
        node[below]{$\brt_2$}; \node at (-3,0) {\ldots}; \filldraw[thick]
        (-4,0) circle (1pt) node[below]{$\brt_{c_3-1}$}; \filldraw[thick]
        (-5,0) circle (1pt) node[below]{$\brt_{c_3}$}; \draw[thick] (-6,0)
        circle (3pt) node[above=0.1cm]{$\inh_1$}; \draw[thick,|-|] (-6,1) --
        node[midway,above] {$c_3+1$} (0,1);
      \end{scope}
      \begin{scope}[shift={(4,2.5)}] 
        \draw[thick] (0,0) circle (3pt)
        node[above=0.1cm]{$\inh_4$}; \filldraw[thick] (-1,0) circle (1pt)
        node[below]{$\brt_{c_3+1}$}; \filldraw[thick] (-2,0) circle (1pt)
        node[below]{$\brt_{c_3+2}$}; \node at (-3,0) {\ldots};
        \filldraw[thick] (-4,0) circle (1pt)
        node[below]{\hspace{20pt}$\brt_{c_3+c_4-1}$}; \filldraw[thick] (-5,0)
        circle (1pt) node[below]{$\brt_{c_3+c_4}$}; \draw[thick] (-6,0) circle
        (3pt) node[above=0.1cm]{$\inh_2$}; \draw[thick,|-|] (-6,1) --
        node[midway,above] {$c_4+1$} (0,1);
      \end{scope}
    \end{tikzpicture}
    \caption{The Bethe roots utilized for the construction of the
      four-site Yangian invariant $|\Psi_{4,2}(\inhdiff)\rangle$,
      i.e.\ to the R-matrix
      $R_{\oscrep_{c_3}\oscrep_{c_4}}(\inhdiff)$, form two real
      strings in the complex plane. The difference between the
      endpoints of the two strings is the spectral parameter
      $\inhdiff:=\inh_3-\inh_4$ of the R-matrix, cf.\
      \eqref{eq:osc-diffinh}. The number of Bethe roots per string is
      given by the representation labels $c_3$ and $c_4$.}
    \label{fig:bethe-gl2-four-line}
  \end{center}
\end{figure}

In section~\ref{sec:comp-bos-4-site} we investigated the four-site
sample invariant $|\Psi_{4,2}(\inh_3-\inh_4)\rangle$, which
corresponds to an R-matrix. The associated monodromy matrix
$\mon_{4,2}(\spec)$ is determined by, cf.\ \eqref{eq:osc-m42} and
\eqref{eq:osc-m42-vs},
\begin{align}
  \label{eq:bethe-gl2-sol-four-constr}
  \begin{gathered} 
    \mathcal{V}_1=\bar{\oscrep}_{c_1}\,, \quad 
    \mathcal{V}_2=\bar{\oscrep}_{c_2}\,, \quad
    \mathcal{V}_3=\oscrep_{c_3}\,, \quad 
    \mathcal{V}_4=\oscrep_{c_4}\,,\\ 
    \inh_1=\inh_3-1-c_3\,, \quad
    \inh_2=\inh_4-1-c_4\,, \quad 
    c_1+c_3=0\,, \quad c_2+c_4=0\,.
  \end{gathered}
\end{align} 
With the trivial normalization \eqref{eq:osc-m42-norm} of the
monodromy and the highest weights \eqref{eq:comp-osc-weights} of the
representations, the eigenvalues \eqref{eq:bethe-gl2-alphadelta}
become
\begin{align}
  \label{eq:bethe-gl2-sol-four-ad-eval} \alpha(\spec) =
  \frac{\spec-\inh_3+c_3}{\spec-\inh_3}\,
  \frac{\spec-\inh_4+c_4}{\spec-\inh_4}\,, \quad \delta(\spec) =
  \frac{\spec-\inh_3+1}{\spec-\inh_3+1+c_3}\,
  \frac{\spec-\inh_4+1}{\spec-\inh_4+1+c_4}\,.
\end{align} 
They satisfy the functional relation \eqref{eq:bethe-gl2-ad}.
The other relation \eqref{eq:bethe-gl2-qaq} is solved by
\begin{align}
  \label{eq:bethe-gl2-sol-four-q}
  \begin{aligned} Q(\spec) =
    \frac{\Gamma(\spec-\inh_3+c_3+1)}{\Gamma(\spec-\inh_3+1)}\,
    \frac{\Gamma(\spec-\inh_4+c_4+1)}{\Gamma(\spec-\inh_4+1)} =
    \prod_{k=1}^{c_3}(\spec-\inh_3+k)\prod_{k=1}^{c_4}(\spec-\inh_4+k)\,.
  \end{aligned}
\end{align} 
Assuming this Q-function to be of the form \eqref{eq:bethe-gl2-qfunct}
assures the uniqueness of this solution. The zeros of this Q-function
yield the Bethe roots
\begin{align}
  \label{eq:bethe-gl2-sol-four-roots}
  \begin{aligned} 
    \brt_k&=\inh_3-k\quad\text{for}\quad
    k=1,\ldots,c_3\,,\\ \brt_{k+c_3}&=\inh_4-k\quad\text{for}\quad
    k=1,\ldots,c_4\,.
  \end{aligned}
\end{align} 
They arrange into two strings, see
figure~\ref{fig:bethe-gl2-four-line}. The Bethe vector
\eqref{eq:bethe-gl2-eigenvector} with these roots is constructed from
the reference state \eqref{eq:bethe-gl2-hws-tot}
\begin{align}
  \label{eq:bethe-gl2-sol-four-vac} \bvac =
(\bar\osca_2^1)^{c_3}(\bar\osca_1^3)^{c_3}
(\bar\osca_2^2)^{c_4}(\bar\osca_1^4)^{c_4} |0\rangle\,,
\end{align} 
which is fixed by the representations in
\eqref{eq:bethe-gl2-sol-four-constr}. The manual evaluation of
\eqref{eq:bethe-gl2-eigenvector} for small values of $c_3$ and $c_4$
leads to
\begin{align}
  \label{eq:bethe-gl2-sol-four-inv}
  \begin{aligned} |\Psi\rangle &= B(\brt_1)\cdots
B(\brt_{c_3})B(\brt_{c_3+1})\cdots B(\brt_{c_3+c_4})\bvac\\ &=
(-1)^{c_3+c_4}c_3!c_4!  \!\!  \prod_{l=1}^{\Min(c_3,c_4)} \!\!\!
(\inh_3-\inh_4+c_4-l+1)^{-1} \!\!  \sum_{k=0}^{\Min(c_3,c_4)} \!\!\!
\frac{1} {(c_3-k)!(c_4-k)!k!}  \\ &\quad \cdot\, \!\!
\prod_{l=k+1}^{\Min(c_3,c_4)} \!\!\!  (\inh_3-\inh_4-c_3+l)\;
(1\bullet 3)^{c_3-k} 
(2\bullet 4)^{c_4-k}
(2\bullet 3)^{k}
(1\bullet 4)^{k} |0\rangle\\
&\propto|\Psi_{4,2}(\inh_3-\inh_4)\rangle\,.
  \end{aligned}
\end{align} 
This matches the Yangian invariant $|\Psi_{4,2}(\inhdiff)\rangle$ from
\eqref{eq:osc-psi42} with \eqref{eq:osc-phi}, \eqref{eq:osc-diffinh}
and \eqref{eq:osc-coeff}. Consequently, we can interpret the R-matrix
$R_{\oscrep_{c_3}\oscrep_{c_4}}(\inhdiff)$, which is
equivalent to this Yangian invariant, as a special Bethe vector.

Let us conclude the investigation of sample invariants with a comment
on the general structure of the set of solutions to the functional
relations \eqref{eq:bethe-gl2-ad} and \eqref{eq:bethe-gl2-qaq}. We
observe that the solution of these relations defined by
\eqref{eq:bethe-gl2-sol-four-ad-eval} and
\eqref{eq:bethe-gl2-sol-four-q} is in fact the product of two two-site
solutions, which we derived in section~\ref{sec:bethe-gl2-sol-line}.
There is a general principle behind this simple observation. Provided
two solutions $(\alpha_1(\spec),\delta_1(\spec),Q_1(\spec))$ and
$(\alpha_2(\spec),\delta_2(\spec),Q_2(\spec))$ of the functional
relations, the product
\begin{align}
  \label{eq:bethe-gl2-superpos} (\alpha_1(\spec)\alpha_2(\spec),
\delta_1(\spec)\delta_2(\spec), Q_1(\spec)Q_2(\spec))\,
\end{align} 
forms a new solution of these relations. Therefore we can obtain new
Yangian invariants by ``superposing'' known ones. For instance, it
should also be possible to combine a two-site solution and one of the
three-site solutions of section~\ref{sec:bethe-gl2-sol-three} with
this method.

\subsection{Classification of Solutions}
\label{sec:class-solut}

After studying sample solutions of the functional relations
\eqref{eq:bethe-gl2-ad} and \eqref{eq:bethe-gl2-qaq} in the previous
section, we review a classification of the solutions of
\eqref{eq:bethe-gl2-ad} found in \cite{Kanning:2014maa}.\footnote{The
  publication \cite{Kanning:2014maa} is co-authored by the creator of
  this thesis. However, the results discussed in the present section
  were obtained before he joined that project. We review them in this
  dissertation because they provide important structural insights into
  the Bethe ansatz for Yangian invariants.}  As already mentioned
before and experienced for the sample solutions, this equation is the
crucial part of the functional relations because it constrains the
monodromy matrix. Given a solution, the remaining first order
difference equation \eqref{eq:bethe-gl2-qaq} can typically be solved
uniquely without any problems. Therefore a classification of the
solutions of \eqref{eq:bethe-gl2-ad} should be thought of as a
classification of all compact invariants of the Yangian of
$\mathfrak{gl}(2)$. With mild restrictions on the form of the
monodromy, one finds that each solution of \eqref{eq:bethe-gl2-ad}
corresponds to a permutation and vice versa.

Let us begin by explaining the class of monodromies employed in this
section. We choose the quantum space to be a tensor product of the two
types of oscillator representations of
section~\ref{sec:deta-oscill-repr}. The sites with ``dual''
representations are placed left of those with ``ordinary'' ones,
\begin{align}
  \mathcal{V}_1\otimes\cdots\otimes\mathcal{V}_N
  =
  \bar{\oscrep}_{c_1}\otimes\cdots\otimes\bar{\oscrep}_{c_K}
  \otimes
  \oscrep_{c_{K+1}}\otimes\cdots\otimes\oscrep_{c_N}\,.
\end{align}
We used this setup already for all the sample invariants in
section~\ref{sec:comp-boson-invar}, which we just reexamined in
section~\ref{sec:bethe-gl2-sol}. Here we impose in addition that the
normalization of the Lax operators~\eqref{eq:yangian-def-lax}
contained in the monodromy is trivial, $f_{\mathcal{V}_i}=1$. This
condition is compatible with the sample invariants because we found by
hindsight that the overall normalization of the monodromy is trivial
for all of them. To proceed, we compute the eigenvalues $\alpha(u)$
and $\delta(u)$ in \eqref{eq:bethe-gl2-alphadelta} for this class of
monodromies using the highest
weights~\eqref{eq:comp-osc-weights}. Inserting the result into the
functional relation~\eqref{eq:bethe-gl2-ad} provides us with the
explicit form of the equation we want to investigate,
\begin{align}
  \label{eq:class-functrel}
  \prod_{i=K+1}^N\frac{u-v_i+c_i}{u-v_i}
  \prod_{i=1}^K\frac{u-1-v_i+c_i}{u-1-v_i}
  =1\,.
\end{align}
The solutions of this equation are most easily classified after
changing variables to\footnote{This equation for $v_i'$ differs from
  the corresponding equation (40) in \cite{Kanning:2014maa} by a shift
  of $1$ at the dual sites. This shift originates from a shift of the
  inhomogeneities of the Lax operators at those sites.}
\begin{align}
  \label{eq:class-changevars}
  v_i'=v_i-\frac{c_i}{2}+
  \begin{cases}
    1\quad\text{for}\quad i=1,\ldots,K\,,\\
    0\quad\text{for}\quad i=K+1,\ldots,N\,.
  \end{cases}
\end{align}
Furthermore, we introduce, cf.\ \cite{Beisert:2014qba},
\begin{align}
  \label{eq:class-changevars-pm}
  v_i^\pm=v_i'\pm\frac{c_i}{2}\,.
\end{align}
These definitions transform~\eqref{eq:class-functrel} into
\begin{align}
  \label{eq:class-functrel-newvars}
  \prod_{i=1}^N (u-v_i^+)
  =
  \prod_{i=1}^N (u-v_i^-)\,.
\end{align}
For these two $N$-th order polynomials to be equal, their roots have
to agree. Thus the solutions of
\eqref{eq:class-functrel-newvars} are in one-to-one correspondence
with permutations $\sigma$ of $N$ elements,
\begin{align}
  \label{eq:class-sol-perm}
  v^+_{\sigma(i)}=v^-_i\,
\end{align}
for $i=1,\ldots,N$. Consequently, the solutions of
\eqref{eq:bethe-gl2-ad} and therefore also the associated Yangian
invariants are classified by permutations. Equation
\eqref{eq:class-sol-perm} imposes $N$ constraints on the $2N$
parameters $v_i$ and $c_i$ of the monodromy that enters the Yangian
invariance condition~\eqref{eq:yi}.

We remark that to recover the conditions on the monodromies of the
sample invariants $|\Psi_{N,K}\rangle$ in
section~\ref{sec:bethe-gl2-sol} from \eqref{eq:class-sol-perm}, we
have to choose the permutation of $N$ elements to be the cyclic shift
$\sigma(i)=i+K$. For the invariants $|\Psi_{2,1}\rangle$,
$|\Psi_{3,1}\rangle$, $|\Psi_{3,2}\rangle$ and $|\Psi_{4,2}\rangle$
this choice turns \eqref{eq:class-sol-perm} into
\eqref{eq:bethe-gl2-sol-line-constr},
\eqref{eq:bethe-gl2-sol-three1-constr},
\eqref{eq:bethe-gl2-sol-three2-constr} and
\eqref{eq:bethe-gl2-sol-four-constr}, respectively.

Even though we derived the key condition \eqref{eq:class-sol-perm}
within the context of the Bethe ansatz, it is a property of the
Yangian invariants themselves and not tied to the Bethe ansatz
construction. In fact, we already encountered such a condition in the
introductory section~\ref{sec:deform} on Yangian invariant
deformations of SYM scattering amplitudes, cf.\ \eqref{eq:amp-wperm}.
It will appear again in chapter \ref{cha:grassmann-amp} during the
construction of Yangian invariants with oscillator representations of
$\mathfrak{u}(p,q|m)$ using Graßmannian matrix models. Hence it is not
tied to the compact bosonic algebra $\mathfrak{u}(2)$ either. Probably
the simplicity of \eqref{eq:class-sol-perm} is, at least in part,
related to the restricted class of oscillator representations that we
consider in most parts of this thesis.

\section{Six-Vertex Model}
\label{sec:six-vertex}

Let us shift gears for a moment and discuss the \emph{six-vertex
  model}. It is a prime example of an exactly solvable model in
two-dimensional statistical mechanics, see e.g.\ the classic monograph
\cite{Baxter:2007}. Typically, this model is studied on a square
lattice with periodic boundary conditions. In this setting the exact
expression for the partition function is well-known for lattices of
finite size \cite{Lieb:1967zz,Lieb:1967bg,Sutherland:1967}. However,
it is given in an implicit form requiring the solutions of certain
Bethe equations. The model has also been investigated on more general
planar lattices \cite{Baxter:1978xr}, so-called \emph{Baxter
  lattices}, which are in general non-rectangular.  It is probably
less known that on these lattices the partition function for fixed
boundary conditions can be computed exactly using Baxter's
\emph{perimeter Bethe ansatz} \cite{Baxter:1987}. In this approach the
partition function is identified with a Bethe wave
function. Astonishingly, the solutions of the Bethe equations are
known explicitly, which is in contrast to most other applications of
the Bethe ansatz. Of course, we already encountered another of these
rare situations, the Bethe ansatz for Yangian invariants of
section~\ref{sec:bethe-yangian}. This similarity is not a
coincidence. In section~\ref{sec:vertex-yangian} we will explain that
the Bethe ansatz for Yangian invariants can be understood as a
generalization of the perimeter Bethe ansatz.

Before we are able to establish that connection, we have to review
Baxter's work \cite{Baxter:1987}. We restrict our discussion to the
rational limit of the six-vertex model. In this limit the model
exhibits a $\mathfrak{su}(2)$ spin $\frac{1}{2}$
symmetry. Furthermore, our notation differs considerably from that in
his publication. In section~\ref{sec:pba-model} we define the model on
a Baxter lattice. Section~\ref{sec:pba-solution} contains its solution
in terms of the perimeter Bethe ansatz. We do not include Baxter's
proof of this solution here because the connection with the Bethe
ansatz for Yangian invariants in section~\ref{sec:vertex-yangian}
below will provide an alternative proof.

\subsection{Rational Model on Baxter Lattices}
\label{sec:pba-model}

\begin{figure}[!t]
  \begin{center}  
    \begin{align*}
      \begin{aligned}
        \begin{tikzpicture}
          \draw[thick,densely dotted]
          (0,0) +(-180:2.2cm) arc (-180:180:2.2cm) 
          node[draw,solid,fill=black,inner sep=0.75pt,shape=circle] {}
          node[left] {$B$};
          \draw[thick,
          decoration={
            markings, mark=at position 0.9 with {\arrow{latex reversed}}},
          postaction={decorate}
          ] 
          (-160:2.2cm) 
          node[left] {$\epe_1\!=\!1$}
           -- (50:2.2cm)
          node[above right] {$\epb_1\!=\!9$};
          \draw[thick,
          decoration={
            markings, mark=at position 0.9 with {\arrow{latex reversed}}},
          postaction={decorate}
          ] 
          (-135:2.2cm) 
          node[below=0.1cm,left] {$\epe_2\!=\!2$}
          -- (75:2.2cm)
          node[right,above=0.05cm] {$\epb_2\!=\!10$};
          \draw[thick,
          decoration={
            markings, mark=at position 0.95 with {\arrow{latex reversed}}},
          postaction={decorate}
          ] 
          (-110:2.2cm) 
          node[below left] {$\epe_3\!=\!3$}
          -- (28:2.2cm)
          node[right=0.2cm] {$\epb_3\!=\!8$};
          \draw[thick,
          decoration={
            markings, mark=at position 0.85 with {\arrow{latex reversed}}},
          postaction={decorate}
          ] 
          (-85:2.2cm) 
          node[left=0.2cm,below] {$\epe_4\!=\!4$}
          -- (150:2.2cm)
          node[left] {$\epb_4\!=\!12$};
          \draw[thick,
          decoration={
            markings, mark=at position 0.5 with {\arrow{latex reversed}}},
          postaction={decorate}
          ] 
          (-65:2.2cm) 
          node[right=0.4cm,below=0.1cm] {$\epe_5\!=\!5$}
          -- (-20:2.2cm)
          node[right] {$\epb_5\!=\!6$};
          \draw[thick,
          decoration={
            markings, mark=at position 0.85 with {\arrow{latex reversed}}},
          postaction={decorate}
          ] 
          (3:2.2cm) 
          node[right] {$\epe_6\!=\!7$}
          -- (120:2.2cm)
          node[left=0.3cm,above=0.1cm] {$\epb_6\!=\!11$};
        \end{tikzpicture}
      \end{aligned}
       \end{align*}
       \caption{Example of a Baxter lattice containing $L=6$ lines
         that are specified in terms of their endpoints by
         $\graph=((1,9),(2,10),(3,8),(4,12),(5,6),(7,11))$. The $k$-th
         line has the endpoints $(\epe_k,\epb_k)$. An arrow defines
         its orientation and we assign to it a rapidity $\rap_k$,
         which is not displayed in this figure.}
    \label{fig:pba-lattice}
  \end{center}  
\end{figure}
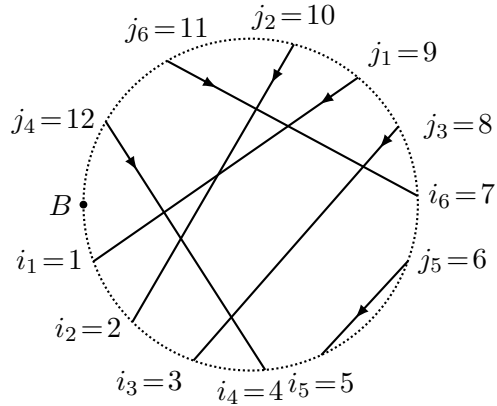

A \emph{Baxter lattice} consists of $L$ straight lines that are placed
arbitrarily in the interior of a circle in such a way that their
endpoints lie on its perimeter. Each line is divided into edges by the
points of intersection with the other lines. At most two lines are
allowed to intersect at a point. An example of such a lattice is
displayed in figure \ref{fig:pba-lattice}, where the perimeter is
represented by a dotted circle. The $\lines$ lines and their $2\lines$
endpoints are labeled counterclockwise starting at a reference point
$B$ on the perimeter. The $k$-th line has the endpoints
$(\epe_k, \epb_k)$ with $1\leq \epe_k < \epb_k\leq2\lines$. Its
orientation is indicated by an arrow pointing from $\epb_k$ to
$\epe_k$. Furthermore, we allocate a complex rapidity $\rap_k$ to the
line. Thus a Baxter lattice is determined by the ordered sets
\begin{align}
  \label{eq:pba-g-theta}
  \graph=((\epe_1,\epb_1),\ldots,(\epe_\lines, \epb_\lines))\,,
  \quad
  \rapset=(\rap_1,\ldots,\rap_\lines)\,.
\end{align}

Each intersection of two lines defines a vertex to which we assign a
Boltzmann weight, that is a matrix element of an R-matrix. Such a
weight is a function of the rapidities of the lines and depends on
states labeled by Gothic letters which are assigned to the adjacent
edges,
\begin{align}
  \label{eq:pba-boltzmann}
  \begin{aligned}
    \langle \mathfrak{a},\mathfrak{c}|
    \tilde{R}_{\square\,\square'}(\rap-\rap')
    |\mathfrak{b},\mathfrak{d}\rangle
    =\,\,\,\\\phantom{}
  \end{aligned}
  \begin{aligned}
    \begin{tikzpicture}
      \draw[thick,
      decoration={
        markings, mark=at position 0.85 with {\arrow{latex reversed}}},
      postaction={decorate}
      ] 
      (0,0) 
      node[left=0.4cm] {$\rap$}
      node[left] {$\mathfrak{a}$} -- 
      (1,0)
      node[right] {$\mathfrak{b}$};
      \draw[thick,
      decoration={
        markings, mark=at position 0.85 with {\arrow{latex reversed}}},
      postaction={decorate}
      ] 
      (0.5,-0.5) node[below=0.5cm] {$\rap'$}
      node[below] {$\mathfrak{c}$} -- 
      (0.5,0.5)
      node[above] {$\mathfrak{d}$};
    \end{tikzpicture}
  \end{aligned}
  \begin{aligned}
    .\\\phantom{}
  \end{aligned}
\end{align}
For the rational limit of the six-vertex model these weights are
elements of the R-matrix
\begin{align}
  \label{eq:pba-r-matrix}
  \tilde{R}_{\square\,\square'}(\rap-\rap')=\frac{1}{\rap-\rap'+1}
    \begin{pmatrix}
      \rap-\rap'+1 & 0 & 0 & 0\\
      0&\rap-\rap'&1&0\\
      0&1&\rap-\rap'&0\\
      0&0&0&\rap-\rap'+1\\
    \end{pmatrix}.
\end{align}
Apart from a change of the normalization, which we indicate by the
tilde, this is the $\mathfrak{gl}(2)$ version of the R-matrix
\eqref{eq:yangian-def-r} used in the definition of the Yangian. The
Gothic indices
$\mathfrak{a}, \mathfrak{b}, \mathfrak{c}, \mathfrak{d}$ take the
values $1$ or $2$ denoting the states states
$|1\rangle = \bigl(\begin{smallmatrix}1\\0\end{smallmatrix}\bigr)$ or
$|2\rangle = \bigl(\begin{smallmatrix}0\\1\end{smallmatrix}\bigr)$,
respectively. The R-matrix acts on the tensor product
$|\mathfrak{b}, \mathfrak{d} \rangle := |\mathfrak{b}\rangle \otimes
|\mathfrak{d}\rangle$
and its matrix elements are built with the bras
$\langle \mathfrak{a}, \mathfrak{c}| := \langle \mathfrak{a}| \otimes
\langle \mathfrak{c}|$.
The six non-zero matrix elements of \eqref{eq:pba-r-matrix} correspond
to vertex configurations with an equal number of incoming states
$|1\rangle$, $|2\rangle$ and outgoing states $\langle 1|$,
$\langle 2|$, respectively. This ``conservation law'' is referred to
as \emph{ice rule}.

\begin{figure}[!t]
  \begin{center}  
    \begin{align*}
      \begin{aligned}
        \begin{tikzpicture}
          \draw[thick,densely dotted]
          (0,0) +(-180:2.2cm) arc (-180:180:2.2cm) 
          node[draw,solid,fill=black,inner sep=0.75pt,shape=circle] {}
          node[left] {$B$};
          \draw[thick,
          decoration={
            markings, mark=at position 0.9 with {\arrow{latex reversed}}},
          postaction={decorate}
          ] 
          (-160:2.2cm) 
          node[left] {$\mathfrak{a}_1\!=\!1$}
           -- (50:2.2cm)
          node[above right] {$\mathfrak{a}_9\!=\!2$};
          \draw[thick,
          decoration={
            markings, mark=at position 0.9 with {\arrow{latex reversed}}},
          postaction={decorate}
          ] 
          (-135:2.2cm) 
          node[below=0.1cm,left] {$\mathfrak{a}_2\!=\!2$}
          -- (75:2.2cm)
          node[right,above=0.05cm] {$\mathfrak{a}_{10}\!=\!1$};
          \draw[thick,
          decoration={
            markings, mark=at position 0.95 with {\arrow{latex reversed}}},
          postaction={decorate}
          ] 
          (-110:2.2cm) 
          node[below left] {$\mathfrak{a}_3\!=\!2$}
          -- (28:2.2cm)
          node[right=0.2cm] {$\mathfrak{a}_8\!=\!1$};
          \draw[thick,
          decoration={
            markings, mark=at position 0.85 with {\arrow{latex reversed}}},
          postaction={decorate}
          ] 
          (-85:2.2cm) 
          node[left=0.2cm,below] {$\mathfrak{a}_4\!=\!1$}
          -- (150:2.2cm)
          node[left] {$\mathfrak{a}_{12}\!=\!2$};
          \draw[thick,
          decoration={
            markings, mark=at position 0.5 with {\arrow{latex reversed}}},
          postaction={decorate}
          ] 
          (-65:2.2cm) 
          node[right=0.4cm,below=0.1cm] {$\mathfrak{a}_5\!=\!2$}
          -- (-20:2.2cm)
          node[right] {$\mathfrak{a}_6\!=\!2$};
          \draw[thick,
          decoration={
            markings, mark=at position 0.85 with {\arrow{latex reversed}}},
          postaction={decorate}
          ] 
          (3:2.2cm) 
          node[right] {$\mathfrak{a}_7\!=\!2$}
          -- (120:2.2cm)
          node[left=0.3cm,above=0.1cm] {$\mathfrak{a}_{11}\!=\!2$};
        \end{tikzpicture}
      \end{aligned}
       \end{align*}
       \caption{The sample Baxter lattice of
         figure~\ref{fig:pba-lattice} including boundary conditions,
         which consist of state labels in $\boldsymbol{\mathfrak{a}}$
         that are assigned to the endpoints. Notice that the ice rule
         \eqref{eq:pba-ice-rule-global} is obeyed because the number
         of endpoints $\epe_k$ with a state label
         $\mathfrak{a}_{\epe_k}=1$ and that of endpoints $\epb_k$ with
         $\mathfrak{a}_{\epb_k}=1$ coincides. Thus
         $\boldsymbol{\mathfrak{a}}$ gives rise to the magnon
         positions $\magnset=(1,4,6,9,11,12)$ via
         \eqref{eq:pba-alpha-position}.  These are instrumental in
         expressing the partition function
         $\mathcal{Z}(\graph,\rapset,\boldsymbol{\mathfrak{a}})$ in
         terms of a Bethe wave function
         $\Phi(\inhtset,\brtset,\magnset)$ in
         \eqref{eq:pba-partition-wave}.}
    \label{fig:pba-lattice-states}
  \end{center}  
\end{figure}
The boundary conditions of the vertex model are specified by fixing
the states at the boundary edges of the Baxter lattice. This is done
by assigning state labels $\mathfrak{a}_{\epe_k}$ and
$\mathfrak{a}_{\epb_k}$ with values $1$ or $2$ to the endpoints
$(\epe_k, \epb_k)$ of the lines, see
figure~\ref{fig:pba-lattice-states}. These labels are collectively
denoted by
\begin{align}
  \label{eq:pba-alpha}
  \boldsymbol{\mathfrak{a}}=(\mathfrak{a}_1, \ldots, \mathfrak{a}_{2 \lines})\,.
\end{align}

The partition function of the vertex model is
\begin{align}
  \label{eq:pba-partition}
  \mathcal{Z}(\graph,\rapset,\boldsymbol{\mathfrak{a}})
  =
  \sum_{\substack{\text{internal}\\\text{state}\\\text{config.}}}\,
  \prod_{\text{vertices}} \text{Boltzmann weight}\,,
\end{align} 
where we sum over all possible state configurations of the internal
edges. An additional prescription is necessary for lines which do not
contain any vertex, see e.g.\ the line with endpoints $(5,6)$ in
figure~\ref{fig:pba-lattice-states}. Such a line $k$ contributes a
factor of $1$ to the partition functions if the state labels at the
endpoints are identical,
$\mathfrak{a}_{\epe_k}=\mathfrak{a}_{\epb_k}$. It contributes a factor
of $0$ if they differ,
$\mathfrak{a}_{\epe_k} \neq \mathfrak{a}_{\epb_k}$. Thus the entire
partition function vanishes in the latter case.

The ice rule for each vertex implies at a global level that for the
partition function to be non-zero, the number of endpoints $\epe_k$ at
outward pointing boundary edges with $\mathfrak{a}_{\epe_k}=1$ must be equal
to that of endpoints $\epb_k$ at inward pointing edges with
$\mathfrak{a}_{\epb_k}=1$,
\begin{align}
  \label{eq:pba-ice-rule-global}
  \big|\{\epe_k\,|\,\mathfrak{a}_{\epe_k}\!=\!1\}\big|
  =
  \big|\{\epb_k\,|\,\mathfrak{a}_{\epb_k}\!=\!1\}\big|\,.
\end{align}
The analogous condition must be fulfilled for endpoints with the state
label $2$.

The R-matrix \eqref{eq:pba-r-matrix}, which contains the Boltzmann
weights of the vertex model, solves a Yang-Baxter equation, cf.\
section~\ref{sec:yangian}. This equation implies that the partition
function does not change if a line is moved through a vertex without
changing the order of endpoints at the perimeter. This property of the
partition function is called Z-invariance.

\subsection{Perimeter Bethe Ansatz Solution}
\label{sec:pba-solution}

The partition function \eqref{eq:pba-partition} was computed exactly
by Baxter in \cite{Baxter:1987} by identifying it with a Bethe wave
function. Such wave functions were introduced by Bethe in his original
solution of the Heisenberg model \cite{Bethe:1931hc}. Pedagogical
accounts on that coordinate Bethe ansatz may be found in
\cite{Karbach:1997,Sutherland:2004}. It presents an alternative to the
algebraic Bethe ansatz, which we recapitulated in
section~\ref{sec:bethe-ansatze}. To reproduce Baxter's result we need
an extension of the coordinate Bethe ansatz to inhomogeneous spin
chains \cite{Yang:1967bm,Gaudin:1967}.

In case of a chain with $N$ sites and $P$ magnon excitations the Bethe
wave function is parametrized by
\begin{align}
  \label{eq:pba-psi-param}
  \inhtset=(\inht_1,\ldots,\inht_{\sites})\,,
  \quad
  \brtset=(\brt_1,\ldots,\brt_\brts)\,,
  \quad
  \magnset=(\magn_1,\ldots,\magn_{\brts})\,,
\end{align}
which denote the inhomogeneities, the Bethe roots and the magnon
positions satisfying $1\leq \magn_1<\ldots<\magn_\brts\leq \sites$,
respectively. The wave function is given by
\begin{align}
  \label{eq:pba-psi}
  \Phi(\inhtset,\brtset,\magnset)
  =
  \sum_\rho
  \mathrm{A}(\brt_{\rho(1)},\ldots,\brt_{\rho(P)})
  \prod_{k=1}^{\brts}\phi_{\magn_k}(\brt_{\rho(k)},\inhtset)\,,
\end{align} 
where the summation runs over all permutations $\rho$ of $\brts$
elements. Furthermore, the factor
\begin{align}
  \label{eq:pba-amp}
  \mathrm{A}(\brt_{\rho(1)},\ldots,\brt_{\rho(P)})=
  \prod_{1\leq k<l\leq \brts}
  \!\!\!
  \frac{\brt_{\rho(k)}-\brt_{\rho(l)}+1}{\brt_{\rho(k)}-\brt_{\rho(l)}}\,
\end{align}
is independent of the inhomogeneities and 
\begin{align}
  \label{eq:pba-phi}
  \phi_{\magn}(\brt,\inhtset)
  =
  \prod_{j=1}^{\magn-1}(\brt-\inht_j+1)
  \prod_{j=\magn+1}^{\sites}(\brt-\inht_j)\,
\end{align}
is a single particle wave function, see also
\cite{Essler:2010}.\footnote{ For a homogeneous spin chain with
  $\inht_j=0$, the Bethe wave function \eqref{eq:pba-psi} takes a more
  common form after dividing \eqref{eq:pba-psi} by
  $\mathrm{A}(\brt_1,\ldots,\brt_P)$ to obtain the S-matrix and
  changing variables to $p_k=-i\log\tfrac{\brt_k+1}{\brt_k}$.}
The Bethe equations 
\begin{align}
  \label{eq:pba-bethe}
  \prod_{i=1}^{\sites}\frac{\brt_k-\inht_i+1}{\brt_k-\inht_i}
  =
  -\prod_{l=1}^{\brts}\frac{\brt_k-\brt_l+1}{\brt_k-\brt_l-1}
\end{align}
with $1\leq k \leq \brts$ are obtained by imposing periodicity of
\eqref{eq:pba-psi} in the magnon positions. These equations ensure
that the wave functions \eqref{eq:pba-psi} for different magnon
configurations $\magnset$ are components of the transfer matrix
eigenvectors of the inhomogeneous Heisenberg spin chain. For generic
Bethe roots $\brtset$, \eqref{eq:pba-psi} is often called
``off-shell'' Bethe wave function. It becomes ``on-shell'' once Bethe
roots obeying \eqref{eq:pba-bethe} are inserted.

Next, we will identify the partition function \eqref{eq:pba-partition}
with the Bethe wave function \eqref{eq:pba-psi}. We restrict to
lattice configurations for which the ice rule
\eqref{eq:pba-ice-rule-global} holds because otherwise the partition
function vanishes. To perform the identification we proceed as
follows, where in particular the parameters $\inhtset$, $\brtset$ and
$\magnset$ of \eqref{eq:pba-psi} are related to the variables
$\graph$, $\rapset$ and $\boldsymbol{\mathfrak{a}}$ of
\eqref{eq:pba-partition}:
\begin{enumerate}
\item For a Baxter lattice consisting of $L$ lines, we choose a wave
  function of a spin chain with $N=2L$ sites and $P=L$
  excitations. Such a configuration is referred to as
  ``half-filling''.
\item The lattice configuration in $\boldsymbol{\mathfrak{a}}$ and
  $\graph$ determines the magnon positions $\magnset$. They are
  defined as the endpoint positions $\epe_k$ at outward pointing edges
  with $\mathfrak{a}_{\epe_k}=1$ and $\epb_k$ at edges directed inwards with
  $\mathfrak{a}_{\epb_k}=2$,
  \begin{align}
    \label{eq:pba-alpha-position}
    \{\magn_k\}
    =
    \{\epe_k|\mathfrak{a}_{\epe_k}\!=\!1\}
    \cup
    \{\epb_k|\mathfrak{a}_{\epb_k}\!=\!2\}\,.
  \end{align}
  These positions are then ordered by imposing
  $1\leq \magn_1<\ldots<\magn_\lines\leq 2\lines$. An example is
  provided in figure~\ref{fig:pba-lattice-states}.
\item $\graph$ and the rapidities $\rapset$ fix the inhomogeneities
  $\inhtset$ and the Bethe roots $\brtset$. For each line $k$ with
  endpoints $(\epe_k,\epb_k)$ we define
  \begin{align}
    \label{eq:pba-inhomo-rap}
    \inht_{\epe_k}=\rap_k+1\,,
    \quad
    \inht_{\epb_k}=\rap_k+2\,,
    \quad
    \brt_k=\rap_k+1\,.
  \end{align}
  Perhaps surprisingly, this constitutes an exact solution of the
  Bethe equations~\eqref{eq:pba-bethe}. This claim is most easily
  verified after writing the Bethe equations in polynomial form to
  circumvent divergencies, see also the comment below
  \eqref{eq:bethe-gl2-betheeq-ordinary}.
\end{enumerate}

After following these steps, the partition function
\eqref{eq:pba-partition} is given by the Bethe wave function
\eqref{eq:pba-psi},
\begin{align}
  \label{eq:pba-partition-wave}
  \mathcal{Z}(\graph,\rapset,\boldsymbol{\mathfrak{a}})
  =
  \mathcal{C}(\graph,\rapset)^{-1}
  (-1)^{\mathcal{K}(\graph,\boldsymbol{\mathfrak{a}})}
  \Phi(\inhtset,\brtset,\magnset)\,.
\end{align}
Here $\mathcal{K}(\graph,\boldsymbol{\mathfrak{a}})$ denotes the number of endpoints
$\epe_k$ with state label $\mathfrak{a}_{\epe_k}=2$,
\begin{align}
  \label{eq:pba-partition-wave-exp}
  \mathcal{K}(\graph,\boldsymbol{\mathfrak{a}})
  =
  \big|\{\epe_k\,|\,\mathfrak{a}_{\epe_k}\!=\!2\}\big|\,.
\end{align}
The normalization is independent of $\boldsymbol{\mathfrak{a}}$ and
reads
\begin{align}
  \label{eq:-partition-wave-norm}
  \mathcal{C}(\graph,\rapset)
  =
  \Phi(\inhtset,\brtset,\magnset_0)\,.
\end{align}
Here $\magnset_0=(\epe_1,\ldots,\epe_\lines)$ is computed from
\eqref{eq:pba-alpha-position} with the boundary conditions
$\boldsymbol{\mathfrak{a}}_0=(1,\ldots,1)$. The explicit expression
\eqref{eq:pba-partition-wave} for the partition function is the
\emph{perimeter Bethe ansatz} solution of the six-vertex model on a
Baxter lattice in the rational limit \cite{Baxter:1987}. As mentioned
earlier, we refrain from presenting the original proof here. Instead,
we will prove this expression in section~\ref{sec:conn-perim-bethe} by
showing it to be a special case of the Bethe ansatz for Yangian
invariants.

\section{From Vertex Models to Yangian Invariance}
\label{sec:vertex-yangian}

We already alluded to a relation between the rational six-vertex model
and Yangian invariance in the introduction of the previous
section~\ref{sec:six-vertex}. Here we expose the details of this
connection. As it turns out, it is not limited to the rational
six-vertex model but extends to more general vertex models on Baxter
lattices, which we define in
section~\ref{sec:vertex-models-baxter}. In these models the symmetry
algebra is generalized from $\mathfrak{gl}(2)$ to
$\mathfrak{gl}(n)$. Furthermore, the lines of the lattice can carry
representations that are different from the defining representation
$\square$.  In particular, we may use the compact oscillator
representations of section~\ref{sec:deta-oscill-repr}. In
section~\ref{sec:part-funct-as} we show that the partition functions
of these vertex models are components of Yangian invariant
vectors. Consequently, the Bethe ansatz for Yangian invariants of
section~\ref{sec:bethe-yangian} is applicable for the construction of
these partition functions in the $\mathfrak{gl}(2)$ case, as detailed
in section~\ref{sec:bethe-ansatz-solut}. Finally, in
section~\ref{sec:conn-perim-bethe} we demonstrate explicitly that this
Bethe ansatz reduces to the perimeter Bethe ansatz reviewed in
section~\ref{sec:pba-solution} if we restrict the representations of
the lines to get back to the rational six-vertex model.

\subsection{Vertex Models on Baxter Lattices}
\label{sec:vertex-models-baxter}

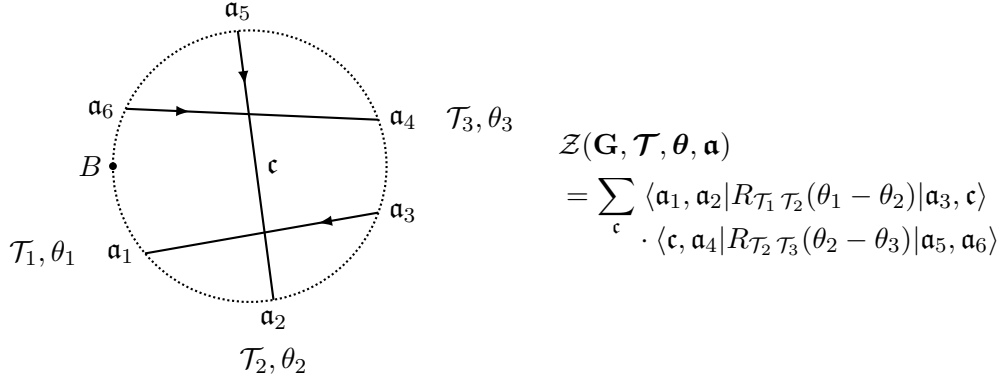
\begin{figure}[!t]
  \begin{center}  
    \begin{align*}
      \begin{aligned}
        \begin{tikzpicture}
          \draw[thick,densely dotted]
          (0,0) +(-180:1.8cm) arc (-180:180:1.8cm) 
          node[draw,solid,fill=black,inner sep=0.75pt,shape=circle] {}
          node[left] {$B$};
          \draw[thick,
          decoration={
            markings, mark=at position 0.8 with {\arrow{latex reversed}}},
          postaction={decorate}
          ] 
          (-140:1.8cm) 
          node[left] {$\mathfrak{a}_1$}
          node[left=0.75cm] {$\mathcal{T}_1, \rap_{1}$}
          -- (-20:1.8cm)
          node[right] {$\mathfrak{a}_3$};
          \draw[thick,
          decoration={
            markings, mark=at position 0.85 with {\arrow{latex reversed}}},
          postaction={decorate}
          ] 
          (-80:1.8cm) 
          node[below] {$\mathfrak{a}_2$}
          node[below=0.5cm] {$\mathcal{T}_2, \rap_{2}$}
          -- (95:1.8cm)
          node[above] {$\mathfrak{a}_5$};
          \draw[thick,
          decoration={
            markings, mark=at position 0.8 with {\arrow{latex reversed}}},
          postaction={decorate}
          ] 
          (20:1.8cm) 
          node[right] {$\mathfrak{a}_4$}
          node[right=0.75cm] {$\mathcal{T}_3, \rap_{3}$}
          -- (155:1.8cm)
          node[left] {$\mathfrak{a}_6$};
          \node at (0.3,0) {$\mathfrak{c}$};
        \end{tikzpicture}
      \end{aligned}
      \quad
      \begin{aligned}
        &\mathcal{Z}
        (\graph,\boldsymbol{\mathcal{T}},
        \rapset,\boldsymbol{\mathfrak{a}})\\
        &=
        \smash[b]{\sum_{\mathfrak{c}}}\;
        \langle\mathfrak{a}_1,\mathfrak{a}_2|
        R_{\mathcal{T}_1\,\mathcal{T}_2}(\rap_1-\rap_2)
        |\mathfrak{a}_3,\mathfrak{c}\rangle\\
        &\hspace{1cm}\cdot\langle\mathfrak{c},\mathfrak{a}_4|
        R_{\mathcal{T}_2\,\mathcal{T}_3}(\rap_2-\rap_3)
        |\mathfrak{a}_5,\mathfrak{a}_6\rangle
      \end{aligned}
    \end{align*}
    \caption{The left side shows a sample (generalized) Baxter lattice
      with $\lines=3$ lines whose endpoints are encoded in
      $\graph=((1,3),(2,5),(4,6))$. The $k$-th line carries a
      $\mathfrak{gl}(n)$ representation $\mathcal{T}_k$, a spectral
      parameter $\rap_k$, two state labels $\mathfrak{a}_{\epe_k}$ and
      $\mathfrak{a}_{\epb_k}$ at the endpoints $(\epe_k,\epb_k)$ and
      an arrow indicating its orientation. The dotted circle and the
      reference point $B$ are not part of the lattice. On the left we
      present the associated partition function
      $\mathcal{Z} (\graph,\boldsymbol{\mathcal{T}},
      \rapset,\boldsymbol{\mathfrak{a}})$.}
    \label{fig:yi-baxter-lattice}
  \end{center}  
\end{figure}

We generalize the vertex model of section~\ref{sec:pba-model}, where
all lines of the Baxter lattice carry the defining representation
$\square$ of $\mathfrak{gl}(2)$, to a class of models for which each
line is associated with a, from the outset, different representation
of $\mathfrak{gl}(n)$. Let us reiterate the definition of the Baxter
lattice for such models. A sample lattice is provided in the left part
of figure ~\ref{fig:yi-baxter-lattice}. The construction of the Baxter
lattice makes use of a dotted circle on which we mark a reference
point $B$. However, both, the circle and the reference point, are not
part of the lattice itself. We choose $L$ straight lines whose
endpoints lie on the dotted circle. Moreover, we demand that at most
two lines intersect at a point in the interior of the circle. The $L$
lines and their $2L$ endpoints are labeled counterclockwise starting
at the reference point. The two endpoints of the $k$-th line are
$\epe_k<\epb_k$. An arrow pointing from $\epb_k$ to $\epe_k$ provides
an orientation of the line. Notice that the orientation clearly
depends on the position of the reference point. Furthermore, the
$k$-th line possesses a spectral parameter $\rap_k$ and it carries a
representation $\mathcal{T}_k$ of $\mathfrak{gl}(n)$. Recall that our
terminology does not distinguish between a representation and the
associated vector space, as mentioned after \eqref{eq:gl-supercomm}.
We assign basis states of $\mathcal{T}_k$ labeled by
$\mathfrak{a}_{\epe_k}$ and $\mathfrak{a}_{\epb_k}$ to the endpoints
of the line. These Gothic state labels may take the values
$1,2,\ldots,\dim(\mathcal{T}_k)$. In summary, a (generalized)
\emph{Baxter lattice} including boundary conditions is defined by the
ordered sets
\begin{align}
  \label{eq:yi-baxter-data}
  \begin{gathered}
    \graph=((\epe_1,\epb_1),\ldots,(\epe_\lines,\epb_\lines))\,,\\
    \boldsymbol{\mathcal{T}}=(\mathcal{T}_{1},\ldots,\mathcal{T}_{\lines})\,,
    \quad
    \rapset=(\rap_{1},\ldots,\rap_{\lines})\,,
    \quad
    \boldsymbol{\mathfrak{a}}=(\mathfrak{a}_1,\ldots,\mathfrak{a}_{2\lines})\,.
  \end{gathered}
\end{align}
Let us remark that we may choose the $\mathcal{T}_k$ out of the
classes of oscillator representations introduced in
section~\ref{sec:osc-rep}, yet this restriction is not necessary
here.\footnote{\label{fn:gothicgreek}The Gothic letters used here as
  state labels coincide with the Greek oscillator indices of
  section~\ref{sec:osc-rep} for the $\mathfrak{u}(n)$ representations
  $\oscrep_{1}$ and $\bar{\oscrep}_{-1}$ because these are spanned by
  the states $\bar{\mathbf{a}}_\alpha|0\rangle$ with a single Greek
  index $\alpha=1,\ldots,n$.}  However, we will have to impose some
conditions on the $\mathcal{T}_k$ later on in
section~\ref{sec:part-funct-as}.

To define a vertex model on this type of Baxter lattice, we generalize
the Boltzmann weights of section~\ref{sec:pba-model} to
\begin{align}
  \label{eq:yi-boltzmann}
  \begin{aligned}
    \langle \mathfrak{a},\mathfrak{c}|
    R_{\mathcal{T}\,\mathcal{T}'}(\rap-\rap')
    |\mathfrak{b},\mathfrak{d}\rangle
    =
    \,\,\,\\\phantom{}
  \end{aligned}
  \begin{aligned}
    \begin{tikzpicture}
      \draw[thick,
      decoration={
        markings, mark=at position 0.85 with {\arrow{latex reversed}}},
      postaction={decorate}
      ] 
      (0,0) 
      node[left=0.4cm] {$\mathcal{T},\rap$}
      node[left] {$\mathfrak{a}$} -- 
      (1,0)
      node[right] {$\mathfrak{b}$};
      \draw[thick,
      decoration={
        markings, mark=at position 0.85 with {\arrow{latex reversed}}},
      postaction={decorate}
      ] 
      (0.5,-0.5) node[below=0.5cm] {$\mathcal{T}',\rap'$}
      node[below] {$\mathfrak{c}$} -- 
      (0.5,0.5)
      node[above] {$\mathfrak{d}$};
    \end{tikzpicture}
  \end{aligned}
  \begin{aligned}
    .\\\phantom{}
  \end{aligned}
\end{align}
These weights are computed with respect to orthonormal bases of the
$\mathfrak{gl}(n)$ representations $\mathcal{T}$ and $\mathcal{T}'$
whose states are labeled by the Gothic indices $\mathfrak{a}$,
$\mathfrak{b}$ and $\mathfrak{c}$, $\mathfrak{d}$, respectively. They
are elements of the R-matrix
\begin{align}
  \label{eq:yi-r-matrix}
  \begin{aligned}
    R_{\mathcal{T}\,\mathcal{T}'}(\rap-\rap')
    =
    \,\,\,\\\phantom{}    
  \end{aligned}
  \begin{aligned}
    \begin{tikzpicture}
      \draw[thick,
      decoration={
        markings, mark=at position 0.8 with {\arrow{latex reversed}}},
      postaction={decorate}
      ] 
      (0,0) 
      node[left] {$\mathcal{T},\rap$} -- 
      (1,0);
      \draw[thick,
      decoration={
        markings, mark=at position 0.8 with {\arrow{latex reversed}}},
      postaction={decorate}
      ] 
      (0.5,-0.5) 
      node[below] {$\mathcal{T}',\rap'$} -- 
      (0.5,0.5);
    \end{tikzpicture}
  \end{aligned}
  \begin{aligned}
    ,\\\phantom{}
  \end{aligned}
\end{align}
which acts on the tensor product $\mathcal{T}\otimes\mathcal{T}'$ and
is a function of the spectral parameters $\rap$ and $\rap'$. We
briefly recall the graphical notation for such R-matrices from
section~\ref{sec:yangian}. Each line is associated with one
representation space. The arrow on the line defines the order in which
multiple R-matrices act on that space. An R-matrix ``earlier'' on the
line acts before one appearing ``later'' on the line. Thus the arrows
are directed from the ``inputs'' of the R-matrix towards the
``outputs''. In the component language \eqref{eq:yi-boltzmann} we may
think of them as pointing from the ket to the bra states. In the
following we will use the operator and the component language
interchangeably. The R-matrix \eqref{eq:yi-r-matrix} is a solution of
the Yang-Baxter equation
\begin{align}
  \label{eq:yi-ybe}
  \begin{aligned}
    &R_{\mathcal{T}\,\mathcal{T}'}(\rap-\rap')
    R_{\mathcal{T}\,\mathcal{T}''}(\rap-\rap'')
    R_{\mathcal{T}'\,\mathcal{T}''}(\rap'-\rap'')\\
    &=
    R_{\mathcal{T}'\,\mathcal{T}''}(\rap'-\rap'')
    R_{\mathcal{T}\,\mathcal{T}''}(\rap-\rap'')
    R_{\mathcal{T}\,\mathcal{T}'}(\rap-\rap')\,
  \end{aligned}
\end{align}
acting in the tensor product
$\mathcal{T}\otimes\mathcal{T}'\otimes\mathcal{T}''$. Employing the
graphical notation this becomes
\begin{align}
  \label{eq:yi-ybe-graphical}
  \begin{aligned}
    \begin{tikzpicture}
      \draw[thick,
      decoration={
        markings, mark=at position 0.9 with {\arrow{latex reversed}}},
      postaction={decorate}
      ]
      (-0.25,0.5) 
      node[left] {$\mathcal{T},\rap$} -- 
      (1.75,0.5);
      \draw[thick,
      decoration={
        markings, mark=at position 0.85 with {\arrow{latex reversed}}},
      postaction={decorate}
      ]
      (0,0) 
      node[below] {$\mathcal{T}',\rap'$} -- 
      (1.5,1.5);
      \draw[thick,
      decoration={
        markings, mark=at position 0.85 with {\arrow{latex reversed}}},
      postaction={decorate}
      ]
      (1.5,0)
      node[below] {$\mathcal{T}'',\rap''$}  -- 
      (0,1.5);
    \end{tikzpicture}
  \end{aligned}
  \begin{aligned}
    \,\,\,=\,\,\,\\\phantom{}
  \end{aligned}
  \begin{aligned}
    \begin{tikzpicture}
      \draw[thick,
      decoration={
        markings, mark=at position 0.9 with {\arrow{latex reversed}}},
      postaction={decorate}
      ]
      (-0.25,1) 
      node[left] {$\mathcal{T},\rap$} -- 
      (1.75,1);
      \draw[thick,
      decoration={
        markings, mark=at position 0.85 with {\arrow{latex reversed}}},
      postaction={decorate}
      ]
      (0,0) 
      node[below] {$\mathcal{T}',\rap'$} -- 
      (1.5,1.5);
      \draw[thick,
      decoration={
        markings, mark=at position 0.85 with {\arrow{latex reversed}}},
      postaction={decorate}
      ]
      (1.5,0) 
      node[below] {$\mathcal{T}'',\rap''$} -- 
      (0,1.5);
    \end{tikzpicture}
  \end{aligned}
  \begin{aligned}
    .\\\phantom{}
  \end{aligned}
\end{align}
These equations are the analogues of \eqref{eq:yangian-ybe-def} and
\eqref{eq:yangian-ybe-def-graphical} for the general class of
$\mathfrak{gl}(n)$ representations considered here.

This setup allows us to define the partition function of a vertex
model on a (generalized) Baxter lattice following
\eqref{eq:pba-partition} as
\begin{align}
  \label{eq:yi-partition-function}  
  \mathcal{Z}
  (\graph,\boldsymbol{\mathcal{T}},
  \rapset,\boldsymbol{\mathfrak{a}})
  =
  \sum_{\substack{\text{internal}\\\text{state}\\\text{config.}}}\,
  \prod_{\text{vertices}}
  \text{Boltzmann weight}\,,
\end{align}
see once more the example in figure~\ref{fig:yi-baxter-lattice}. The
sum in this formula runs over all possible state configurations at the
internal edges of the lattice. For an internal edge belonging to a
line with a representation $\mathcal{T}$ these are the basis states of
that representation. The product is over all vertices of the
lattice. Each vertex is associated with a Boltzmann weight of the form
\eqref{eq:yi-boltzmann}. The states at the boundary edges of the
lattice are fixed by the state labels in
$\boldsymbol{\mathfrak{a}}$. In case a line consists of a single edge
and the two state labels assigned to it by $\boldsymbol{\mathfrak{a}}$
differ, the partition function vanishes.

After defining the partition function in
\eqref{eq:yi-partition-function} in component language using the
Boltzmann weights in \eqref{eq:yi-boltzmann}, we rephrase it using
directly the R-matrix \eqref{eq:yi-r-matrix}. In such an operator
language the partition function is a matrix element of a certain
product of R-matrices. The R-matrices appearing in this product as
well as their order are prescribed by the form of the Baxter
lattice. The matrix multiplications in this product correspond to the
sum and the product in \eqref{eq:yi-partition-function}. A
non-intersecting line in the Baxter lattice becomes an identity
operator on the corresponding representation space. The matrix element
of the product of R-matrices is selected by the boundary conditions in
$\boldsymbol{\mathfrak{a}}$.

\subsection{Partition Function as Yangian Invariant}
\label{sec:part-funct-as}

To understand the relation between the partition function
$\mathcal{Z} (\graph,\boldsymbol{\mathcal{T}},
\rapset,\boldsymbol{\mathfrak{a}})$
on a Baxter lattice and a Yangian invariant $|\Psi\rangle$, we start
out by deriving an identity obeyed by this partition function. In a
second step, we then reformulate this identity in the QISM language to
show that it is identical to the Yangian invariance condition
\eqref{eq:yi}.

\begin{figure}[!t]
  \begin{center}
    \begin{tikzpicture}
      \draw[thick,densely dashed,
      decoration={
        markings, mark=at position 0.9 with {\arrow{latex reversed}}},
      postaction={decorate}
      ]
      (0,0) +(-165:1.8cm) 
      node[right=0.5cm,above] {$\alpha=:\gamma_0$}
      node[left] {$\square, \spec$}
      arc (-165:165:1.8cm)
      node[right=0.5cm,below=-0.1cm] {$\beta=:\gamma_6$};
      \node at (-0.7,-1.95) {$\gamma_1$};
      \node at (1.3,-1.65) {$\gamma_2$};
      \node at (2.1,0) {$\gamma_3$};
      \node at (1.1,1.7) {$\gamma_4$};
      \node at (-1.2,1.7) {$\gamma_5$};
      \node at (-0.5,-0.9) {$\mathfrak{b}_1$};
      \node at (-0.0,-1.45) {$\mathfrak{b}_2$};
      \node at (0.9,-0.65) {$\mathfrak{b}_3$};
      \node at (0.8,0.5) {$\mathfrak{b}_4$};
      \node at (0.25,1.3) {$\mathfrak{b}_5$};
      \node at (-0.7,1.15) {$\mathfrak{b}_6$};
      \draw[thick,
      decoration={
        markings, mark=at position 0.8 with {\arrow{latex reversed}}},
      postaction={decorate}
      ] 
      (-140:2.2cm) 
      node[left] {$\mathfrak{a}_1$}
      node[left=0.75cm] {$\mathcal{T}_{1}, \rap_{1}$}
      -- (-20:2.2cm)
      node[right] {$\mathfrak{a}_3$};
      \draw[thick,
      decoration={
        markings, mark=at position 0.85 with {\arrow{latex reversed}}},
      postaction={decorate}
      ] 
      (-80:2.2cm) 
      node[below] {$\mathfrak{a}_2$}
      node[below=0.5cm] {$\mathcal{T}_{2}, \rap_{2}$}
      -- (95:2.2cm)
      node[above] {$\mathfrak{a}_5$};
      \draw[thick,
      decoration={
        markings, mark=at position 0.8 with {\arrow{latex reversed}}},
      postaction={decorate}
      ] 
      (20:2.2cm) 
      node[right] {$\mathfrak{a}_4$}
      node[right=0.75cm] {$\mathcal{T}_{3}, \rap_{3}$}
      -- (155:2.2cm)
      node[left] {$\mathfrak{a}_6$};
    \end{tikzpicture}
    \caption{The Baxter lattice introduced in the example of
      figure~\ref{fig:yi-baxter-lattice} after the dotted circle has
      been replaced by a dashed auxiliary space line in the defining
      representation $\square$ with a spectral parameter $\spec$ and
      states labeled $\alpha$, $\beta$ at the endpoints. The indices
      $\gamma_i$ are assigned to the edges of this auxiliary
      space. The states at the edges connecting this space with the
      Baxter lattice are labeled $\mathfrak{b}_i$.}
    \label{fig:yi-baxter-lattice-aux}
  \end{center}
\end{figure}
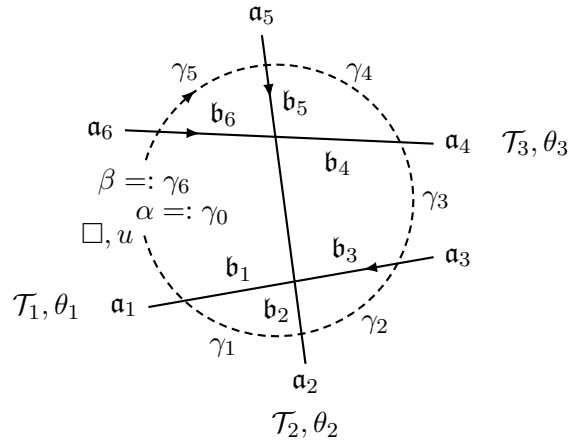

In order to derive said identity, we replace the dotted circle, which
appeared in the construction of the Baxter lattice, by a dashed arc
that is opened at the reference point $B$. Furthermore, we extend the
lines of the Baxter lattice such that they intersect the arc. See the
example lattice in figure~\ref{fig:yi-baxter-lattice-aux}. The dashed
arc now represents an actual space. This auxiliary space carries the
defining representation $\square=\mathbb{C}^n$ of $\mathfrak{gl}(n)$
and a spectral parameter $u$. It is oriented counterclockwise. The bra
and ket states at its endpoints are labeled by $\alpha$ and $\beta$,
respectively, which can take the values $1,\ldots,n$. Each line of the
Baxter lattice intersects the arc twice. This creates an additional
layer of vertices at the boundary of the lattice. The Boltzmann
weights associated with these vertices are elements of R-matrices of
the type $R_{\square\,\mathcal{T}}(\spec-\rap)$ or
$R_{\mathcal{T}\,\square}(\rap-\spec)$. These R-matrices are referred
to as Lax operators, cf.\ section~\ref{sec:yangian}, and they satisfy
Yang-Baxter equations like
\begin{align}
  \label{eq:yi-ybe-boundary}
  \begin{aligned}
    \begin{tikzpicture}
      \draw[thick,
      decoration={
        markings, mark=at position 0.85 with {\arrow{latex reversed}}},
      postaction={decorate}
      ]
      (0,0) 
      node[below] {$\mathcal{T},\rap$} -- 
      (1.5,1.5);
      \draw[thick,
      decoration={
        markings, mark=at position 0.85 with {\arrow{latex reversed}}},
      postaction={decorate}
      ]
      (1.5,0) 
      node[below] {$\mathcal{T}',\rap'$} -- 
      (0,1.5);
      \draw[thick,densely dashed,
      decoration={
        markings, mark=at position 0.9 with {\arrow{latex reversed}}},
      postaction={decorate}
      ]
      (-0.25,0.5)
      node[left] {$\square,\spec$} -- 
      (1.75,0.5);
    \end{tikzpicture}
  \end{aligned}
  \begin{aligned}
    \,\,\,=\,\,\,\\\phantom{}
  \end{aligned}
  \begin{aligned}
    \begin{tikzpicture}
      \draw[thick,
      decoration={
        markings, mark=at position 0.85 with {\arrow{latex reversed}}},
      postaction={decorate}
      ]
      (0,0) 
      node[below] {$\mathcal{T},\rap$} -- 
      (1.5,1.5);
      \draw[thick,
      decoration={
        markings, mark=at position 0.85 with {\arrow{latex reversed}}},
      postaction={decorate}
      ]
      (1.5,0) 
      node[below] {$\mathcal{T}',\rap'$} -- 
      (0,1.5);
      \draw[thick,densely dashed,
      decoration={
        markings, mark=at position 0.9 with {\arrow{latex reversed}}},
      postaction={decorate}
      ]
      (-0.25,1) 
      node[left] {$\square,\spec$} -- 
      (1.75,1);
    \end{tikzpicture}
  \end{aligned}
  \begin{aligned}
    .\\\phantom{}
  \end{aligned}
\end{align}
Furthermore, we impose the unitarity condition
\begin{align}
  \label{eq:yi-unitarity}
  \begin{aligned}
    R_{\square\,\mathcal{T}}(\spec-\rap)
    R_{\mathcal{T}\,\square}(\rap-\spec)
    =
    1\,,
  \end{aligned}
  \quad\text{i.e.}\quad
  \begin{aligned}
    \begin{aligned}
      \begin{tikzpicture}
        \draw[thick,densely dashed,
        decoration={
          markings, mark=at position 0.90 with {\arrow{latex reversed}}},
        postaction={decorate}
        ]
        (0,0) 
        node[left] {$\square,\spec$} -- 
        (2.0,0);
        \draw[thick,
        decoration={
          markings, mark=at position 0.95 with {\arrow{latex reversed}}},
        postaction={decorate}
        ]
        (0.5,-0.5)
        node[below] {$\mathcal{T},\rap$} -- 
        (0.5,0.25) .. controls (0.5,1) and (1.5,1) .. (1.5,0.25) -- 
        (1.5,-0.5);
        \path
        (0,1.25) node[left] {\phantom{$\square,\spec$}} -- 
        (2.0,1.25);
      \end{tikzpicture}
    \end{aligned}\;
    \,\,\,=
    \begin{aligned}
      \begin{tikzpicture}
        \draw[thick,densely dashed,
        decoration={
          markings, mark=at position 0.9 with {\arrow{latex reversed}}},
        postaction={decorate}
        ]
        (0,1.25) 
        node[left] {$\square,\spec$} -- 
        (2.0,1.25);
        \draw[thick,
        decoration={
          markings, mark=at position 0.95 with {\arrow{latex reversed}}},
        postaction={decorate}
        ]
        (0.5,-0.5) 
        node[below] {$\mathcal{T},\rap$} -- 
        (0.5,0.25) .. controls (0.5,1) and (1.5,1) .. (1.5,0.25) -- 
        (1.5,-0.5);
      \end{tikzpicture}
    \end{aligned}
  \end{aligned}
\end{align}
employing the graphical notation. We encountered such a condition
already for the Lax operators of oscillator representations in
section~\ref{sec:deta-oscill-repr}. The Yang-Baxter equation
\eqref{eq:yi-ybe-boundary} and the unitarity
condition~\eqref{eq:yi-unitarity} allow us to disentangle the dashed
auxiliary space line from the $L$ solid lines constituting the Baxter
lattice.
\begin{figure}[!t]
  \begin{center}
    \begin{align*}
      \begin{aligned}
        \begin{tikzpicture}
          \draw[thick,densely dashed,
          decoration={
            markings, mark=at position 0.95 with {\arrow{latex reversed}}},
          postaction={decorate}
          ]
          (0,0) 
          node[left=0.4cm] {$\square,\spec$}
          node[left] {$\alpha$} -- 
          (3.5,0)
          node[right] {$\beta$};
          \draw[thick,
          decoration={
            markings, mark=at position 0.96 with {\arrow{latex reversed}}},
          postaction={decorate}
          ] 
          (0.5,-0.5) 
          node[below] {$\mathfrak{a}_1$}-- 
          (0.5,0.5)
          node[above=0.5cm] {$\mathcal{T}_1,\rap_1$} 
          .. controls (0.5,1.25) and (1.5,1.25) .. (1.5,0.5) -- 
          (1.5,-0.5)
          node[below] {$\mathfrak{a}_3$};
          \draw[thick,
          decoration={
            markings, mark=at position 0.95 with {\arrow{latex reversed}}},
          postaction={decorate}
          ] 
          (2,-0.5)
          node[below] {$\mathfrak{a}_4$} -- 
          (2,0.25) .. controls (2,1) and (3,1) .. 
          (3,0.25) 
          node[above=0.5cm] {$\mathcal{T}_3,\rap_3$} -- 
          (3,-0.5)
          node[below] {$\mathfrak{a}_6$};
          \draw[thick,
          decoration={
            markings, mark=at position 0.965 with {\arrow{latex reversed}}},
          postaction={decorate}
          ] 
          (1,-0.5) 
          node[below] {$\mathfrak{a}_2$} -- 
          (1,0.5) 
          node[above right=0.8cm and 0.2cm] {$\mathcal{T}_2,\rap_2$} 
          .. controls (1,1.5) and (2.5,1.5) .. (2.5,0.5) -- 
          (2.5,-0.5)
          node[below] {$\mathfrak{a}_5$};
          \path
          (0,2.25) node[left] {\phantom{$\square,\spec$}} -- 
          (2,2.25);
        \end{tikzpicture}
      \end{aligned}\;
      \,\,\,=
      \begin{aligned}
        \begin{tikzpicture}
          \draw[thick,densely dashed,
          decoration={
            markings, mark=at position 0.95 with {\arrow{latex reversed}}},
          postaction={decorate}
          ]
          (0,2.25) 
          node[left=0.4cm] {$\square,\spec$}
          node[left] {$\alpha$} -- 
          (3.5,2.25)
          node[right] {$\beta$};
          \draw[thick,
          decoration={
            markings, mark=at position 0.96 with {\arrow{latex reversed}}},
          postaction={decorate}
          ] 
          (0.5,-0.5) 
          node[below] {$\mathfrak{a}_1$} -- 
          (0.5,0.5)
          node[above=0.5cm] {$\mathcal{T}_1,\rap_1$} 
          .. controls (0.5,1.25) and (1.5,1.25) .. (1.5,0.5) -- 
          (1.5,-0.5)
          node[below] {$\mathfrak{a}_3$};
          \draw[thick,
          decoration={
            markings, mark=at position 0.95 with {\arrow{latex reversed}}},
          postaction={decorate}
          ] 
          (2,-0.5)
          node[below] {$\mathfrak{a}_4$} -- 
          (2,0.25) .. controls (2,1) and (3,1) .. 
          (3,0.25) 
          node[above=0.5cm] {$\mathcal{T}_3,\rap_3$} -- 
          (3,-0.5)
          node[below] {$\mathfrak{a}_6$};
          \draw[thick,
          decoration={
            markings, mark=at position 0.965 with {\arrow{latex reversed}}},
          postaction={decorate}
          ] 
          (1,-0.5) 
          node[below] {$\mathfrak{a}_2$} -- 
          (1,0.5) 
          node[above right=0.8cm and 0.2cm] {$\mathcal{T}_2,\rap_2$} 
          .. controls (1,1.5) and (2.5,1.5) .. (2.5,0.5) -- 
          (2.5,-0.5)
          node[below] {$\mathfrak{a}_5$};
        \end{tikzpicture}
      \end{aligned}
    \end{align*}
    \caption{Disentangling the dashed auxiliary line from the solid
      lines by means of \eqref{eq:yi-ybe-boundary} and
      \eqref{eq:yi-unitarity} results in a non-trivial identity for
      $\mathcal{Z}(\graph,\boldsymbol{\mathcal{T}},
      \rapset,\boldsymbol{\mathfrak{a}})$
      of the sample Baxter lattice in
      figure~\ref{fig:yi-baxter-lattice}. We deformed the lattice to
      highlight the horizontal row of vertices involving the auxiliary
      line which will be expressed as a spin chain monodromy shortly,
      cf.\ figure~\ref{fig:yi-mono-element}.}
    \label{fig:yi-disentangle}
  \end{center}
\end{figure}
As shown for an example in figure~\ref{fig:yi-disentangle}, this
yields an identity for the partition function
$\mathcal{Z}(\graph,\boldsymbol{\mathcal{T}},\rapset,\boldsymbol{\mathfrak{a}})$. To
derive this identity for a general Baxter lattice, let us denote the
Boltzmann weights involving the auxiliary space by
\begin{align}
  \label{eq:yi-add-boltzmann}
    \begin{aligned}
      \mathcal{M}_{\alpha\beta}(\spec,\graph,
      \boldsymbol{\mathcal{T}},
      \rapset,
      \boldsymbol{\mathfrak{a}},
      \boldsymbol{\mathfrak{b}})\\
      =
      \smash[b]{\sum_{\gamma_1,\ldots,\gamma_{2\lines-1}=1}^n}\;
      \Bigg(\;
      &
      \smash[b]{\prod_{k=1}^\lines}\;
      \langle \gamma_{\epe_k-1},\mathfrak{a}_{\epe_k}|
      R_{\square\,\mathcal{T}_k}(\spec-\rap_k)
      |\gamma_{\epe_k},\mathfrak{b}_{\epe_k}\rangle\\
      &\quad\cdot
      \langle \mathfrak{b}_{\epb_k},\gamma_{\epb_k-1}|
      R_{\mathcal{T}_k\,\square}(\rap_k-\spec)
      |\mathfrak{a}_{\epb_k},\gamma_{\epb_k}\rangle
      \smash[t]{
        \Bigg)_{
          \begin{subarray}{l}
            \gamma_0:=\alpha\\\gamma_{2\lines}:=\beta
          \end{subarray}
        }
      }\,.
    \end{aligned}
\end{align}
Here each of the $L$ lines of the Baxter lattice leads to two
Boltzmann weights in the product. Recall from the definition of the
lattice around \eqref{eq:yi-baxter-data} that the endpoints of the
$k$-th line are denoted
$i_k<j_k$. Figure~\ref{fig:yi-baxter-lattice-aux} provides an example
for the assignment of the state labels $\gamma_i$ and $\mathfrak{b}_i$
to the edges of the lattice. States at the edges of the auxiliary
space line are labeled by $\gamma_i$ with $i=0,\ldots,2L$. The state
labels
$\boldsymbol{\mathfrak{b}}=(\mathfrak{b}_1,\ldots,\mathfrak{b}_{2\lines})$
are placed at those edges that connect the layer of vertices involving
the auxiliary space with the Baxter lattice on which the partition
function is defined. Eventually, equating the original entangled
situation to the disentangled one results in the sought after identity
\begin{align}
  \label{eq:yi-condition-z-general}
  \sum_{\boldsymbol{\mathfrak{b}}}
  \mathcal{M}_{\alpha\beta}(\spec,\graph,
  \boldsymbol{\mathcal{T}},\rapset,
  \boldsymbol{\mathfrak{a}},\boldsymbol{\mathfrak{b}})\,
  \mathcal{Z}
  (\graph,
  \boldsymbol{\mathcal{T}},\rapset,\boldsymbol{\mathfrak{b}})
  =
  \delta_{\alpha\beta}\,
  \mathcal{Z}
  (\graph,
  \boldsymbol{\mathcal{T}},\rapset,\boldsymbol{\mathfrak{a}})\,.
\end{align}
In this equation the $\delta_{\alpha\beta}$ on the r.h.s.\ corresponds
to the unraveled auxiliary line. For a sample Baxter lattice we know
this identity already from figure~\ref{fig:yi-disentangle}.

For the purpose of translating the identity
\eqref{eq:yi-condition-z-general} in the QISM formalism, we show that
the combination of Boltzmann weights in
$\mathcal{M}_{ab}(\spec,\graph,\boldsymbol{\mathcal{T}},\rapset,\boldsymbol{\mathfrak{a}},\boldsymbol{\mathfrak{b}})$
can be interpreted as a matrix element of a monodromy $M(u)$. The
monodromy of a spin chain with $N$ sites is defined as in
\eqref{eq:yangian-mono-spinchain} by
\begin{align}
  \label{eq:yi-mono}
  \begin{aligned}
    \mon(\spec)
    =
    R_{\square\,\mathcal{V}_1}(\spec-\inh_1)
    \cdots R_{\square\,\mathcal{V}_\sites}(\spec-\inh_\sites)
    =
    \,\,\,\\\phantom{}
  \end{aligned}
  \begin{aligned}
    \begin{tikzpicture}
      \draw[thick,densely dashed,
      decoration={
        markings, mark=at position 0.95 with {\arrow{latex reversed}}},
      postaction={decorate}
      ] 
      (0,0) 
      node[left] {$\square,\spec$} -- 
      (3,0);
      \draw[thick,
      decoration={
        markings, mark=at position 0.85 with {\arrow{latex reversed}}},
      postaction={decorate}
      ] 
      (0.5,-0.5) 
      node[below] {$\mathcal{V}_1,\inh_1$} -- 
      (0.5,0.5);
      \node at (1.5,-0.25) {$\ldots$};
      \draw[thick,
      decoration={
        markings, mark=at position 0.85 with {\arrow{latex reversed}}},
      postaction={decorate}
      ] 
      (2.5,-0.5) 
      node[below] {$\mathcal{V}_\sites,\inh_\sites$} -- 
      (2.5,0.5);
    \end{tikzpicture}
  \end{aligned}
  \begin{aligned}
    .\\\phantom{}
  \end{aligned}
\end{align}
To the $j$-th site we assign an inhomogeneity $\inh_j\in\mathbb{C}$
and a representation $\mathcal{V}_j$ of $\mathfrak{gl}(n)$. Hence the
total quantum space of the monodromy is
$\mathcal{V}_{1}\otimes\cdots\otimes \mathcal{V}_{\sites}$. Its
auxiliary space is in the defining representation $\square$ and the
matrix elements with respect to this space are
\begin{align}
  \label{eq:yi-mono-elements}
  \mon_{\alpha\beta}(\spec):=\langle \alpha|\mon(\spec)|\beta\rangle\,.
\end{align}
They are operators in the total quantum space. To proceed, we assume
that the Lax operators belonging to the Boltzmann weights in
\eqref{eq:yi-add-boltzmann} fulfill the crossing relation
\begin{align}
  \label{eq:yi-crossing}
  R_{\square\,\bar{\mathcal{T}}}(\spec-\rap+\kappa_{\mathcal{T}})
  =
  R_{\mathcal{T}\,\square}(\rap-\spec)^\dagger\,,
\end{align}
where the crossing parameter $\kappa_{\mathcal{T}}$ depends on the
representation and the Hermitian conjugation acts solely on the
representation space $\mathcal{T}$. The generators
$\bar J_{\alpha\beta}$ of the dual representation $\bar{\mathcal{T}}$
appearing in \eqref{eq:yi-crossing} are defined in terms of the
generators $J_{\alpha\beta}$ of $\mathcal{T}$ by
\begin{align}
  \label{eq:yi-conj-rep}
  \bar J_{\alpha\beta}=-J_{\alpha\beta}^\dagger\,.
\end{align}
Recall from section~\ref{sec:deta-oscill-repr} that the crossing
relation~\eqref{eq:yi-crossing} holds in particular in case of the
compact oscillator representations discussed there. In what follows,
we require the matrix elements
$\langle\mathfrak{a}|J_{\alpha\beta}|\mathfrak{b}\rangle$ of the
generators to be real. This leads to the component form of
\eqref{eq:yi-crossing},
\begin{align}
  \label{eq:yi-crossing-coord}
  \begin{aligned}
    &\langle \gamma,\mathfrak{b}|
    R_{\square\,\bar{\mathcal{T}}}(\spec-\rap+\kappa_{\mathcal{T}})
    |\delta,\mathfrak{a}\rangle\\
    &=
    \langle \mathfrak{a},\gamma|
    R_{\mathcal{T}\,\square}(\rap-\spec)
    |\mathfrak{b},\delta\rangle\,,
  \end{aligned}
  \begin{aligned}
    \quad\text{i.e.}\quad
  \end{aligned}
  \begin{aligned}
    \begin{aligned}
      \begin{tikzpicture}
        \draw[thick,densely dashed,
        decoration={
          markings, mark=at position 0.85 with {\arrow{latex reversed}}},
        postaction={decorate}
        ]
        (0,0)
        node[left=0.4cm] {$\square,\spec$}
        node[left] {$\gamma$} -- 
        (1,0)
        node[right] {$\delta$};
        \draw[thick,
        decoration={
          markings, mark=at position 0.85 with {\arrow{latex reversed}}},
        postaction={decorate}
        ] 
        (0.5,-0.5)
        node[below=0.5cm] {$\bar{\mathcal{T}},\rap-\kappa_{\mathcal{T}}$}
        node[below] {$\mathfrak{b}$} -- 
        (0.5,0.5)
        node[above] {$\mathfrak{a}$};
      \end{tikzpicture}
    \end{aligned}
    \begin{aligned}
      \,\,\,=\,\,\,\\\phantom{}  
    \end{aligned}
    \begin{aligned}
      \begin{tikzpicture}
        \draw[thick,densely dashed,
        decoration={
          markings, mark=at position 0.85 with {\arrow{latex reversed}}},
        postaction={decorate}
        ] 
        (0,0) 
        node[left=0.4cm] {$\square,\spec$}
        node[left] {$\gamma$} -- 
        (1,0)
        node[right] {$\delta$};
        \draw[thick,
        decoration={
          markings, mark=at position 0.3 with {\arrow{latex}}},
        postaction={decorate}
        ] 
        (0.5,-0.5) 
        node[below=0.5cm] {$\mathcal{T},\rap$}
        node[below] {$\mathfrak{b}$} -- 
        (0.5,0.5)
        node[above] {$\mathfrak{a}$};
      \end{tikzpicture}
    \end{aligned}
  \end{aligned}
  \begin{aligned}    
  \end{aligned}
\end{align}
in the graphical notation. Applying \eqref{eq:yi-crossing-coord} for
the Boltzmann weights in the second line of
\eqref{eq:yi-add-boltzmann} results in
\begin{align}
  \label{eq:yi-add-boltzmann-cross}
  \begin{aligned}
    \mathcal{M}_{\alpha\beta}(\spec,\graph,
    \boldsymbol{\mathcal{T}},\rapset,
    \boldsymbol{\mathfrak{a}},\boldsymbol{\mathfrak{b}})\\
    =
    \smash[b]{\sum_{\gamma_1,\ldots,\gamma_{2\lines-1}=1}^n}\;
    \Bigg(\;
    &
    \smash[b]{\prod_{k=1}^\lines}\;
    \langle \gamma_{\epe_k-1},\mathfrak{a}_{\epe_k}|
    R_{\square\,\mathcal{T}_k}(\spec-\rap_k)
    |\gamma_{\epe_k},\mathfrak{b}_{\epe_k}\rangle\\
    &\quad\cdot
    \langle \gamma_{\epb_k-1},\mathfrak{a}_{\epb_k}|
    R_{\square\,\bar{\mathcal{T}}_k}(\spec-\rap_k+\kappa_{\mathcal{T}_k})
    |\gamma_{\epb_k},\mathfrak{b}_{\epb_k}\rangle
    \smash[t]{
      \Bigg)_{
        \begin{subarray}{l}
          \gamma_0:=\alpha\\\gamma_{2\lines}:=\beta
        \end{subarray}
      }
    }\,.
  \end{aligned}
\end{align}
From the index structure of this expression we infer that it is the
matrix element of the monodromy \eqref{eq:yi-mono} with
$\sites=2\lines$ sites,
\begin{align}
  \label{eq:yi-mono-element}
  \begin{aligned}
    \mathcal{M}_{\alpha\beta}(\spec,\graph,
    \boldsymbol{\mathcal{T}},\rapset,
    \boldsymbol{\mathfrak{a}},\boldsymbol{\mathfrak{b}})
    =
    \langle\boldsymbol{\mathfrak{a}}|
    \mon_{\alpha\beta}(\spec)
    |\boldsymbol{\mathfrak{b}}\rangle\,.
  \end{aligned}
\end{align}
See the example in figure~\ref{fig:yi-mono-element}. The labels
$\graph$, $\boldsymbol{\mathcal{T}}$, $\rapset$ on the l.h.s.\ of
\eqref{eq:yi-mono-element} encode the total quantum space of the
monodromy. As is usual in the QISM, this information is hidden on the
r.h.s.\ of the equation.
\begin{figure}[!t]
  \begin{center}
    \begin{align*}
      \begin{aligned}
        \begin{tikzpicture}
          \draw[thick,densely dashed,
          decoration={
            markings, mark=at position 0.97 with {\arrow{latex reversed}}},
          postaction={decorate}
          ] 
          (0,0) 
          node[left=0.4cm] {$\square,\spec$}
          node[left] {$\alpha$} -- 
          (6.5,0)
          node[right] {$\beta$};
          \draw[thick,
          decoration={
            markings, mark=at position 0.85 with {\arrow{latex reversed}}},
          postaction={decorate}
          ] 
          (0.5,-0.5) 
          node[below] {$\mathfrak{a}_1$}
          node[below=0.5cm] {$\mathcal{T}_1,\rap_1$} -- 
          (0.5,0.5)
          node[above] {$\mathfrak{b}_1$};
          \draw[thick,
          decoration={
            markings, mark=at position 0.85 with {\arrow{latex reversed}}},
          postaction={decorate}
          ] 
          (1.6,-0.5) 
          node[below] {$\mathfrak{a}_2$}
          node[below=0.5cm] {$\mathcal{T}_2,\rap_2$} -- 
          (1.6,0.5)
          node[above] {$\mathfrak{b}_2$};
          \draw[thick,
          decoration={
            markings, mark=at position 0.3 with {\arrow{latex}}},
          postaction={decorate}
          ] 
          (2.7,-0.5) 
          node[below] {$\mathfrak{a}_3$}
          node[below=0.5cm] {$\mathcal{T}_1,\rap_1$} -- 
          (2.7,0.5)
          node[above] {$\mathfrak{b}_3$};
          \draw[thick,
          decoration={
            markings, mark=at position 0.85 with {\arrow{latex reversed}}},
          postaction={decorate}
          ] 
          (3.8,-0.5) 
          node[below] {$\mathfrak{a}_4$}
          node[below=0.5cm] {$\mathcal{T}_3,\rap_3$} -- 
          (3.8,0.5)
          node[above] {$\mathfrak{b}_4$};
          \draw[thick,
          decoration={
            markings, mark=at position 0.3 with {\arrow{latex}}},
          postaction={decorate}
          ] 
          (4.9,-0.5) 
          node[below] {$\mathfrak{a}_5$}
          node[below=0.5cm] {$\mathcal{T}_2,\rap_2$} -- 
          (4.9,0.5)
          node[above] {$\mathfrak{b}_5$};
          \draw[thick,
          decoration={
            markings, mark=at position 0.3 with {\arrow{latex}}},
          postaction={decorate}
          ] 
          (6.0,-0.5) 
          node[below] {$\mathfrak{a}_6$}
          node[below=0.5cm] {$\mathcal{T}_3,\rap_3$} -- 
          (6.0,0.5)
          node[above] {$\mathfrak{b}_6$};
        \end{tikzpicture}
      \end{aligned}
      \begin{aligned}
        \,\,\,=\,\,\,\\\phantom{}
      \end{aligned}
      \begin{aligned}
        \begin{tikzpicture}
          \draw[thick,densely dashed,
          decoration={
            markings, mark=at position 0.95 with {\arrow{latex reversed}}},
          postaction={decorate}
          ] 
          (0,0) 
          node[left=0.4cm] {$\square,\spec$}
          node[left] {$\alpha$} -- 
          (2.5,0)
          node[right] {$\beta$};
          \draw[thick,
          decoration={
            markings, mark=at position 0.85 with {\arrow{latex reversed}}},
          postaction={decorate}
          ] 
          (0.5,-0.5) 
          node[below] {$\mathfrak{a}_1$}
          node[below=0.5cm] {$\mathcal{V}_1,\inh_1$} -- 
          (0.5,0.5)
          node[above] {$\mathfrak{b}_1$};
          \node at (1.25,-0.25) {$\ldots$};
          \draw[thick,
          decoration={
            markings, mark=at position 0.85 with {\arrow{latex reversed}}},
          postaction={decorate}
          ] 
          (2.0,-0.5) 
          node[below] {$\mathfrak{a}_6$}
          node[below=0.5cm] {$\mathcal{V}_6,\inh_6$} 
          -- 
          (2.0,0.5)
          node[above] {$\mathfrak{b}_6$};
        \end{tikzpicture}
      \end{aligned}
    \end{align*}
    \caption{Rewriting of the summed Boltzmann weights in
      $\mathcal{M}_{\alpha\beta}(\spec,\graph,\boldsymbol{\mathcal{T}},
      \rapset,\boldsymbol{\mathfrak{a}},\boldsymbol{\mathfrak{b}})$
      on the l.h.s.\ as a matrix element
      $\langle\boldsymbol{\mathfrak{a}}|\mon_{\alpha\beta}(\spec)|\boldsymbol{\mathfrak{b}}\rangle$
      of a monodromy on the r.h.s.\ for the example discussed in
      figure~\ref{fig:yi-disentangle}. After applying
      \eqref{eq:yi-crossing-coord} to the l.h.s.\ all vertical lines
      have the same orientation. $\mathcal{V}_i$ and $\inh_i$ of the
      resulting monodromy are given by \eqref{eq:yi-ident-rep-inhomo}
      with $\graph$ specified in the caption of
      figure~\ref{fig:yi-baxter-lattice}.}
    \label{fig:yi-mono-element}
  \end{center}
\end{figure}
We employed the notation
$|\boldsymbol{\mathfrak{b}}\rangle:=|\mathfrak{b}_1\rangle\otimes\cdots\otimes|\mathfrak{b}_{2\lines}\rangle\in
\mathcal{V}_1\otimes\cdots\otimes \mathcal{V}_{2\lines}$.
The $k$-th line of the Baxter lattice whose endpoints $\epe_k<\epb_k$
are defined in $\graph$, cf.\ \eqref{eq:yi-baxter-data}, determines
two sites of the monodromy matrix with representations and
inhomogeneities
\begin{align}
  \label{eq:yi-ident-rep-inhomo}
  \mathcal{V}_{\epe_k}=\mathcal{T}_k\,,
  \quad
  \inh_{\epe_k}=\rap_k
  \quad
  \text{and}
  \quad
  \mathcal{V}_{\epb_k}=\bar{\mathcal{T}}_k\,,
  \quad
  \inh_{\epb_k}=\rap_k-\kappa_{\mathcal{T}_k}\,.
\end{align}
What is more, we can associate a vector $|\Psi\rangle$ in the total
quantum space with the partition function via the relation
\begin{align}
  \label{eq:yi-partition-psi}
  \langle\boldsymbol{\mathfrak{a}}|\Psi\rangle
  :=
  \mathcal{Z}(\graph,
  \boldsymbol{\mathcal{T}},\rapset,\boldsymbol{\mathfrak{a}})\,.
\end{align}
Making use of \eqref{eq:yi-mono-element}, \eqref{eq:yi-partition-psi}
and the orthonormality relation for the states
$|\boldsymbol{\mathfrak{b}}\rangle$, the identity
\eqref{eq:yi-condition-z-general} for the partition function becomes
\begin{align}
  \label{eq:yi-eigenvalue-bra}
  \langle\boldsymbol{\mathfrak{a}}|\mon_{\alpha\beta}(\spec)|\Psi\rangle
  =
  \delta_{\alpha\beta}\langle\boldsymbol{\mathfrak{a}}|\Psi\rangle\,.
\end{align}
After dropping the bra state $\langle\boldsymbol{\mathfrak{a}}|$, this
is precisely the component version \eqref{eq:yi-components} of the
Yangian invariance condition in the QISM language. The Yangian
invariant vector $|\Psi\rangle$ comprises the partition functions of a
fixed Baxter lattice for all possible boundary conditions
$\boldsymbol{\mathfrak{a}}$. 

In this sense vertex models on Baxter lattices give rise to a special
class of Yangian invariants with a monodromy defined by
\eqref{eq:yi-ident-rep-inhomo}. The vertex model origin of these
invariants allows us to interpret the associated intertwiners, cf.\
section~\ref{sec:yang-invar-as}, as products of R-matrices. The
two-site sample invariant of section~\ref{sec:comp-bos-2-site} with
compact oscillator representations falls into this class. Its Baxter
lattice consists of just a single line. The four-site invariant of
section~\ref{sec:comp-bos-4-site} may be understood as originating
from a Baxter lattice with two intersecting lines. In contrast, the
three-site Yangian invariants constructed in
section~\ref{sec:comp-bos-3-site} clearly leave the vertex model
framework because their monodromies are not of the form
\eqref{eq:yi-ident-rep-inhomo}. Notably, the vertex model
interpretation established in this section only applies to Yangian
invariants with an even number of sites.

\subsection{Bethe Ansatz Solution}
\label{sec:bethe-ansatz-solut}

We argued in section~\ref{sec:bethe-yangian} that Yangian invariants
$|\Psi\rangle$ can be constructed using a Bethe ansatz and we showed
this in detail for finite-dimensional $\mathfrak{gl}(2)$
representations. In the previous section we demonstrated that the
partition functions of vertex models on Baxter lattices are encoded in
a certain class of Yangian invariants. Here we present the Bethe
ansatz solution corresponding to these invariants. We restrict to
Baxter lattices with compact oscillator representations of
$\mathfrak{gl}(2)$. This restriction also ensures that the two- and
four-site sample solutions of the Bethe ansatz from
section~\ref{sec:bethe-gl2-sol} are included in the following result
as the special case of Baxter lattices with one and two lines,
respectively.

Let us consider a Baxter lattice with $L$ lines. The $k$-th line with
endpoints $\epe_k<\epb_k$ and spectral parameter $\rap_k$ carries the
compact oscillator representation
$\mathcal{T}_k=\bar{\oscrep}_{c_{i_k}}$ of $\mathfrak{gl}(2)$, see
section~\ref{sec:deta-oscill-repr}. According to
\eqref{eq:yi-ident-rep-inhomo} the monodromy of the associated Yangian
invariant has $N=2L$ sites that are given by
\begin{align}
  \label{eq:bethe-gl2-sol-lattice-constr}
  \begin{gathered}
  \mathcal{V}_{\epe_k}=\bar{\oscrep}_{c_{i_k}}\,,
  \quad
  \mathcal{V}_{\epb_k}=\oscrep_{c_{j_k}}\,,\\
  \inh_{\epe_k}=\rap_k\,,
  \quad
  \inh_{\epb_k}=\rap_k-c_{\epe_k}+1\,,
  \quad
  c_{\epe_k}+c_{\epb_k}=0\,.
  \end{gathered}
\end{align}
With this data the monodromy eigenvalues
\eqref{eq:bethe-gl2-alphadelta} read
\begin{align}
  \label{eq:bethe-gl2-sol-lattice-ad-eval}
  \begin{aligned}
    \alpha(\spec)
    &=
    \prod_{k=1}^\lines
    f_{\bar{\oscrep}_{c_{i_k}}}(\spec-\inh_{\epe_k})f_{\oscrep_{c_{j_k}}}(\spec-\inh_{\epb_k})
    \frac{\spec-\inh_{\epb_k}+c_{j_k}}{\spec-\inh_{\epb_k}}
    =\prod_{k=1}^\lines\frac{\spec-\inh_{\epb_k}+c_{\epb_k}}{\spec-\inh_{\epb_k}}\,,\\
    \delta(\spec)
    &=
    \prod_{k=1}^\lines
    f_{\bar{\oscrep}_{c_{i_k}}}(\spec-\inh_{\epe_k})f_{\oscrep_{c_{j_k}}}(\spec-\inh_{\epb_k})
    \frac{\spec-\inh_{\epe_k}+c_{\epe_k}}{\spec-\inh_{\epe_k}}
    =\prod_{k=1}^\lines\frac{\spec-\inh_{\epb_k}+1}{\spec-\inh_{\epb_k}+1+c_{\epb_k}}\,.
  \end{aligned}
\end{align}
To show the last equality for each eigenvalue, we observe that the
conditions \eqref{eq:bethe-gl2-sol-lattice-constr} for the $k$-th line
of the Baxter lattice are equivalent to those for the two-site
invariant in \eqref{eq:osc-m21-vs}. Therefore the normalization
factors in the eigenvalues trivialize like in case of the two-site
invariant in \eqref{eq:osc-m21-norm}. Obviously,
\eqref{eq:bethe-gl2-sol-lattice-ad-eval} solves the functional
relation \eqref{eq:bethe-gl2-ad}. The other functional relation
\eqref{eq:bethe-gl2-qaq} has the unique solution
\begin{align}
  \label{eq:bethe-gl2-sol-lattice-q}  
  \begin{aligned}
    Q(\spec)
    =
    \prod_{k=1}^\lines
    \frac{\Gamma(\spec-\inh_{\epb_k}+c_{\epb_k}+1)}{\Gamma(\spec-\inh_{\epb_k}+1)}
    =
    \prod_{k=1}^\lines\prod_{l=1}^{c_{\epb_k}}(\spec-\inh_{\epb_k}+l)\,
  \end{aligned}
\end{align}
because we demand the Q-function to be of the form
\eqref{eq:bethe-gl2-qfunct}. The zeros of this function are the Bethe
roots
\begin{align}
  \label{eq:bethe-gl2-sol-lattice-roots}  
  \begin{aligned}
    \brt_l&
    =
    \inh_{\epb_1}-l\quad 
    \text{for}\quad 
    l=1,\ldots,c_{\epb_1}\,,\\
    \brt_{l+c_{\epb_1}}&
    =
    \inh_{\epb_2}-l\quad 
    \text{for}\quad 
    l=1,\ldots,c_{\epb_2}\,,\\
    &\hspace{6pt}\vdots\\
    \brt_{l+c_{\epb_{\lines-1}}}&
    =
    \inh_{\epb_\lines}-l\quad 
    \text{for}\quad 
    l=1,\ldots,c_{\epb_\lines}\,.
  \end{aligned}
\end{align}
These roots form $L$ strings in the complex plane. The $k$-th line of
the Baxter lattice yields a string of $c_{j_k}=-c_{i_k}\in\mathbb{N}$
uniformly spaced Bethe roots lying between the inhomogeneities
$\inh_{\epe_k}$ and $\inh_{\epb_k}$. Thus the representation
$\mathcal{T}_k=\bar{\oscrep}_{c_{i_k}}$ of this line determines the
length of the string and the spectral parameter $\theta_k$ fixes its
position in the complex plane, cf.\
\eqref{eq:bethe-gl2-sol-lattice-constr}. Next, we compute the
reference state \eqref{eq:bethe-gl2-hws-tot} of the associated Bethe
vector using \eqref{eq:bethe-gl2-sol-lattice-constr} and the form of
the highest weight states in \eqref{eq:osc-hws},
\begin{align}
  \label{eq:bethe-gl2-sol-lattice-vac}  
  \bvac
  =
  \prod_{k=1}^\lines
  (\bar\osca_2^{\epe_k})^{c_{\epb_k}}(\bar\osca_1^{\epb_k})^{c_{\epb_k}}|0\rangle\,.
\end{align}
Finally, the Yangian invariant $|\Psi\rangle$ is the Bethe vector
\eqref{eq:bethe-gl2-eigenvector}. From \eqref{eq:yi-partition-psi} we
know that the components of this Bethe vector are the partition
functions of the Baxter lattice for different boundary conditions.

\subsection{Relation to Perimeter Bethe Ansatz}
\label{sec:conn-perim-bethe}

After discussing the solution of the functional relations
\eqref{eq:bethe-gl2-ad} and \eqref{eq:bethe-gl2-qaq} that is
associated with a Baxter lattice of $\lines$ lines in the previous
section~\ref{sec:bethe-ansatz-solut}, we show that in a special case
it reproduces the perimeter Bethe ansatz of
section~\ref{sec:pba-solution}. For this we first have to derive some
special properties of the $\mathfrak{gl}(2)$ Lax operators. Then we
restrict to a Baxter lattice where each line carries the dual of the
defining representation. Finally, the algebraic Bethe vector of the
associated Yangian invariant $|\Psi\rangle$ is expressed in terms of a
coordinate Bethe ansatz wave function. The result is the perimeter
Bethe ansatz formula \eqref{eq:pba-partition-wave} for the partition
function $\mathcal{Z}(\graph,\rapset,\boldsymbol{\mathfrak{a}})$.

Let us first concentrate on the special properties of the Lax
operators. These are based on a relation between the compact
oscillator representations $\oscrep_c$ and
$\bar{\oscrep}_{-c}$, which holds for $\mathfrak{gl}(2)$ but does
not extend to the higher rank $\mathfrak{gl}(n)$ case. The generators
\eqref{eq:osc-gen-s-bs} and the highest weight states
\eqref{eq:osc-hws} of these representations are related by
\begin{align}
  \label{eq:bethe-gl2-special-rep}
  U\mathbf{J}_{\alpha\beta}U^{-1}=\bar{\mathbf{J}}_{\alpha\beta}+c\,\delta_{\alpha\beta}\,,
  \quad
  U|\sigma\rangle=(-1)^c|\bar\sigma\rangle\,,
\end{align}
with the unitary operator
\begin{align}
  \label{eq:bethe-gl2-special-v}
  U=e^{\frac{\pi}{2}(\bar\osca_1\osca_2-\bar\osca_2\osca_1)}
  \quad
  \text{obeying}
  \quad
  U|0\rangle=|0\rangle\,,
  \quad
  \bar\osca_1 U=U\bar\osca_2\,,
  \quad
  \bar\osca_2 U=-U\bar\osca_1\,.  
\end{align}
In what follows, it proves to be convenient to employ Lax operators
with a certain fixed normalization to avoid spurious divergencies,
\begin{align}
  \label{eq:bethe-gl2-special-lax}
  \tilde{R}_{\square\,\oscrep_c}(\spec-\inht)
  =
  (\spec-\inht)1+\sum_{\alpha,\beta=1}^2\elemm_{\alpha\beta}\bar\osca_\beta\osca_\alpha\,.
\end{align}
From these we construct a monodromy matrix with inhomogeneities
$\inht_i$,
\begin{align}
  \label{eq:bethe-gl2-special-mono}
  \tilde{M}(\spec)
  =
  \tilde{R}_{\square\,\oscrep_{c_1}}(\spec-\inht_1)
  \cdots
  \tilde{R}_{\square\,\oscrep_{c_N}}(\spec-\inht_\sites)\,.
\end{align}
The standard Lax operators \eqref{eq:osc-lax-fund-s} for the
representation $\oscrep_{c}$ and \eqref{eq:osc-lax-fund-bs} for
$\bar{\oscrep}_{-c}$ can be reformulated in terms of
\eqref{eq:bethe-gl2-special-lax} with the help of
\eqref{eq:bethe-gl2-special-rep},
\begin{align}
  \label{eq:bethe-gl2-special-lax-symm}
  R_{\square\,\oscrep_{c}}(\spec)
  =
  \frac{f_{\oscrep_{c}}(\spec)}{\spec}
  \tilde{R}_{\square\,\oscrep_{c}}(\spec)\,,
  \quad
  R_{\square\,\bar{\oscrep}_{-c}}(\spec)
  =
  \frac{f_{\bar{\oscrep}_{-c}}(\spec)}{\spec}U
  \tilde{R}_{\square\,\oscrep_{c}}(\spec-c)U^{-1}\,.
\end{align}
Employing these relations, any $\mathfrak{gl}(2)$ monodromy $M(\spec)$
built from $R_{\square\,\oscrep_{c}}(\spec)$ and
$R_{\square\,\bar{\oscrep}_{-c}}(\spec)$ can be expressed via
$\tilde{\mon}(\spec)$, which only contains Lax operators of the type
$\tilde{R}_{\square\,\oscrep_{c}}(\spec)$.

We will use this reformulation for the monodromy corresponding to a
Baxter lattice with $L$ lines defined in
\eqref{eq:bethe-gl2-sol-lattice-constr}. To end up with the perimeter
Bethe ansatz, we restrict to a lattice consisting solely of lines
carrying the dual of the defining representation. Notice that
according to \eqref{eq:bethe-gl2-special-rep}, the defining
representation of $\mathfrak{gl}(2)$ and its dual are essentially
equivalent. However, choosing lines with the dual representation is
more natural in our conventions. Employing the notation of
\eqref{eq:yi-baxter-data} this choice means
\begin{align}
  \label{eq:bethe-gl2-special-reps}
  \boldsymbol{\mathcal{T}}=(\bar{\square}_1,\ldots,\bar{\square}_L)\,.
\end{align}
We realize these representations in terms of oscillators,
$\bar{\square}_i=\bar{\oscrep}_{c_{i_k}}$ with $c_{i_k}=-1$. This
choice of representations together with
\eqref{eq:bethe-gl2-sol-lattice-constr} implies $c_{j_k}=-c_{i_k}=1$.
Consequently, the strings of Bethe roots in
\eqref{eq:bethe-gl2-sol-lattice-roots} reduce to single points,
\begin{align}
  \label{eq:bethe-gl2-special-roots}
  \brt_k=\rap_k+1
  \quad
  \text{for}
  \quad
  k=1,\ldots,\lines\,.
\end{align}
These Bethe roots already match those of the perimeter Bethe ansatz in
\eqref{eq:pba-inhomo-rap}. Next, the monodromy matrix defined by
\eqref{eq:bethe-gl2-sol-lattice-constr} and
\eqref{eq:bethe-gl2-special-reps} is rewritten using
\eqref{eq:bethe-gl2-special-lax-symm} as
\begin{align}
  \label{eq:bethe-gl2-special-trafo-mono}
  \mon(\spec)
  =
  \prod_{i=1}^{2\lines}\frac{1}{\spec-\inh_i}
  W\tilde{\mon}(\spec)W^{-1}
  \quad
  \text{with}
  \quad
  W=\prod_{k=1}^\lines U^{\epe_k}\,.
\end{align}
In this monodromy the normalizations of the Lax operators cancel, as
we observed already after \eqref{eq:bethe-gl2-sol-lattice-ad-eval}.
All sites carrying a dual representation are transformed by $W$
because the unitary operator $U^{\epe_k}$ acts on site $\epe_k$. The
representation labels and inhomogeneities of $\tilde{\mon}(\spec)$ in
\eqref{eq:bethe-gl2-special-mono} are
\begin{align}
  \label{eq:bethe-gl2-special-rep-inhomo}
  c_i=1\,,
  \quad
  \inht_{\epe_k}=\rap_k+1\,,
  \quad
  \inht_{\epb_k}=\rap_k+2\,.
\end{align}
The inhomogeneities $\inht_{\epe_k}$, which stem from the dual sites
of $\mon(\spec)$, are shifted by $1$ with respect to the
$\inh_{\epe_k}$ in \eqref{eq:bethe-gl2-sol-lattice-constr}. The
inhomogeneities in \eqref{eq:bethe-gl2-special-rep-inhomo} match those
of the perimeter Bethe ansatz in \eqref{eq:pba-inhomo-rap}.  The
highest weight state $|\tilde{\Omega}\rangle$ in the total quantum
space of $\tilde{\mon}(\spec)$ is derived from $\bvac$ in
\eqref{eq:bethe-gl2-sol-lattice-vac} with the help of
\eqref{eq:bethe-gl2-special-rep},
\begin{align}
  \label{eq:bethe-gl2-special-refstate}
  |\tilde{\Omega}\rangle
  =
  (-1)^\lines W^{-1}\bvac
  =
  \bar\osca_1^1\cdots\bar\osca_1^{\sites}|0\rangle\,.
\end{align}
Employing \eqref{eq:bethe-gl2-special-trafo-mono} and
\eqref{eq:bethe-gl2-special-refstate} we can express the Bethe vector
\eqref{eq:bethe-gl2-eigenvector}, which is constructed from the
monodromy element $M_{12}(u)=B(u)$, as a Bethe vector built up from
the element $\tilde{\mon}_{12}(\spec)=\tilde{B}(\spec)$ of the new
monodromy,
\begin{align}
  \label{eq:bethe-gl2-special-trafo-vector}
  |\Psi\rangle
  =
  (-1)^\lines\prod_{k=1}^\lines\prod_{i=1}^{2\lines}\frac{1}{\brt_k-\inh_i}
  W|\tilde{\Psi}\rangle
  \quad
  \text{with}
  \quad
  |\tilde{\Psi}\rangle
  =
  \tilde{B}(\brt_1)\cdots \tilde{B}(\brt_\lines)
  |\tilde{\Omega}\rangle\,.
\end{align} 
Therefore, also the Yangian invariant $|\Psi\rangle$ of the Baxter
lattice can be expressed in terms of $|\tilde{\Psi}\rangle$.

We continue by representing the algebraic Bethe vector
$|\tilde{\Psi}\rangle$ in \eqref{eq:bethe-gl2-special-trafo-vector}
using coordinate Bethe ansatz wave functions. In case of a monodromy
$\tilde{\mon}(\spec)$ of the type \eqref{eq:bethe-gl2-special-mono}
with representation labels $c_i=1$ at all sites, the vector reads, see
e.g.\ \cite{Ovchinnikov:2010vb} and appendix~3.E of
\cite{Essler:2010},\footnote{A proof of the analogous relation for
  more general compact $\mathfrak{gl}(2)$ representations, that are
  equivalent to $\oscrep_c$ with $c\in\mathbb{N}$, albeit without
  inhomogeneities, $\inht_i=0$, may be found in \cite{Avdeev:1985cx}.}
\begin{align}
  \label{eq:bethe-gl2-special-vector}
  \begin{aligned}
  |\tilde{\Psi}\rangle
  =
  \tilde{B}(\brt_1)\cdots \tilde{B}(\brt_\brts)|\tilde{\Omega}\rangle
  =
  \sum_{1\leq \magn_1<\cdots <\magn_\brts\leq \sites}
  \!\!\!\!\!\!
  \Phi(\inhtset,\brtset,\magnset)\,
  \mathbf{J}_{21}^{\magn_1}\cdots \mathbf{J}_{21}^{\magn_\brts}|\tilde{\Omega}\rangle\,,
  \end{aligned}
\end{align}
with $\mathfrak{gl}(2)$ generators
$\mathbf{J}_{\alpha\beta}^i=\bar\osca_\alpha^i\osca_\beta^i$ and the
Bethe wave function $\Phi(\inhtset,\brtset,\magnset)$ from
\eqref{eq:pba-psi}. The arguments $\inhtset$, $\brtset$ and $\magnset$
defined in \eqref{eq:pba-psi-param} encode respectively the
inhomogeneities $\inht_i$, Bethe roots $\brt_k$ and magnon positions
$\magn_k$. To apply \eqref{eq:bethe-gl2-special-vector} in
\eqref{eq:bethe-gl2-special-trafo-vector} for the case of Yangian
invariants we need $\sites=2\lines$ sites and $\brts=\lines$ Bethe
roots.

Recall the connection between the partition function and the Yangian
invariant vector $|\Psi\rangle$ in \eqref{eq:yi-partition-psi},
\begin{align}
  \label{eq:bethe-gl2-special-z-psi}
  \mathcal{Z}(\graph,
  \boldsymbol{\mathcal{T}},\rapset,\boldsymbol{\mathfrak{a}})
  \propto
  \langle\boldsymbol{\mathfrak{a}}|\Psi\rangle\,.
\end{align}
For the representations specified in \eqref{eq:bethe-gl2-special-reps}
the possible boundary states of the Baxter lattice are
$|\boldsymbol{\mathfrak{a}}\rangle =
|\mathfrak{a}_1\rangle\otimes\cdots\otimes|\mathfrak{a}_{2\lines}\rangle$
with $\mathfrak{a}_i=1,2$. Here the Gothic labels $\mathfrak{a}_i$ are
identical to the Greek indices $\alpha_i=1,2$ of the oscillators which
build up the states
$|\mathfrak{a}_i\rangle\equiv|\alpha_i\rangle=\bar{\mathbf{a}}^i_{\alpha_i}|0\rangle$
at each site, cf.\ footnote~\ref{fn:gothicgreek}. Inserting
\eqref{eq:bethe-gl2-special-trafo-vector} and
\eqref{eq:bethe-gl2-special-vector} into
\eqref{eq:bethe-gl2-special-z-psi}, the scalar product in the latter
equation reduces to that for each term in
\eqref{eq:bethe-gl2-special-vector}.  This is non-vanishing only if
the state labels $\boldsymbol{\mathfrak{a}}$ satisfy the ice rule
\eqref{eq:pba-ice-rule-global}, and if $\magnset$ is given in terms of
$\graph$ and $\boldsymbol{\mathfrak{a}}$ by
\eqref{eq:pba-alpha-position}. In this case
\begin{align}
  \label{eq:bethe-gl2-special-scalarprod}
  \langle\boldsymbol{\mathfrak{a}}|
  W
  \mathbf{J}_{21}^{\magn_1}\cdots \mathbf{J}_{21}^{\magn_\lines}|\tilde{\Omega}\rangle
  =
  (-1)^{\mathcal{K}(\graph,\boldsymbol{\mathfrak{a}})}\,,
\end{align}
where $\mathcal{K}(\graph,\boldsymbol{\mathfrak{a}})$ is specified in
\eqref{eq:pba-partition-wave-exp} and the factor of $-1$ originates
from sites transformed by $W$.

Taken together, \eqref{eq:bethe-gl2-special-trafo-vector},
\eqref{eq:bethe-gl2-special-vector} and
\eqref{eq:bethe-gl2-special-scalarprod} yield the final formula for
the partition function \eqref{eq:bethe-gl2-special-z-psi}. For it to
be non-zero the state labels $\boldsymbol{\mathfrak{a}}$ have to obey
\eqref{eq:pba-ice-rule-global}. In this case
\begin{align}
  \label{eq:bethe-gl2-special-partitionfct}
  \mathcal{Z}(\graph,
  \boldsymbol{\mathcal{T}},
  \rapset,
  \boldsymbol{\mathfrak{a}})
  \propto
  \langle\boldsymbol{\mathfrak{a}}|\Psi\rangle
  =
  (-1)^\lines
  \prod_{k=1}^\lines\prod_{i=1}^{2\lines}\frac{1}{\brt_k-\inh_i}\,
  (-1)^{\mathcal{K}(\graph,\boldsymbol{\mathfrak{a}})}
  \Phi(\inhtset,\brtset,\magnset)\,,
\end{align}
where the representations in $\boldsymbol{\mathcal{T}}$ are specified
by \eqref{eq:bethe-gl2-special-reps}. Furthermore, the wave function
arguments $\inhtset$, $\brtset$, $\magnset$ are fixed in terms of the
variables $\graph$, $\rapset$, $\boldsymbol{\mathfrak{a}}$ of the
partition function using \eqref{eq:pba-alpha-position} and
\eqref{eq:pba-inhomo-rap}. The l.h.s.\ of
\eqref{eq:bethe-gl2-special-partitionfct} agrees with the perimeter
Bethe ansatz expression \eqref{eq:pba-partition-wave} up to an
$\boldsymbol{\mathfrak{a}}$-independent normalization.

Yet it is not possible to fix this normalization factor from the Bethe
ansatz. Let us now argue why the choice in
\eqref{eq:pba-partition-wave} gives the correct partition function
\eqref{eq:pba-partition}. Obviously,
$\mathcal{Z}(\graph,\rapset, \boldsymbol{\mathfrak{a}}_0)=1$ for the
particular state labels $\boldsymbol{\mathfrak{a}}_0=(1,\ldots,1)$
because the Boltzmann weight in the upper left entry of the R-matrix
\eqref{eq:pba-r-matrix} is equal to $1$. The
$\boldsymbol{\mathfrak{a}}$-independent normalization chosen in
\eqref{eq:pba-partition-wave} clearly reproduces this value of the
partition function for
$\boldsymbol{\mathfrak{a}}=\boldsymbol{\mathfrak{a}}_0$. From the
Bethe ansatz derivation of \eqref{eq:bethe-gl2-special-partitionfct}
we already know that the $\boldsymbol{\mathfrak{a}}$-dependence of
\eqref{eq:pba-partition-wave} agrees with that of the partition
function \eqref{eq:pba-partition}. This concludes the derivation of
\eqref{eq:pba-partition-wave}. What is more, we showed that Baxter's
perimeter Bethe ansatz reviewed in section~\ref{sec:pba-solution} is a
very particular case of the Bethe ansatz for Yangian invariants that
we established in section~\ref{sec:bethe-yangian}.

\chapter{Graßmannian Integrals and Scattering Amplitudes}
\label{cha:grassmann-amp}

After utilizing the Bethe ansatz in the preceding chapter, we develop
a further method for the construction of Yangian invariants: the
unitary Graßmannian integral. It is a refinement of the Graßmannian
integral in the introductory
section~\ref{sec:grassmannian-integral}. In particular, we integrate
over the unitary group manifold, whereas the integration contour has
to be imposed ``by hand'' in the original proposal. Our approach is
applicable for oscillator representations of the non-compact
superalgebra $\mathfrak{u}(p,q|m)$. If $p=q$, we are able to change
the basis from oscillators to spinor~helicity-like variables. This
allows us to examine the relation between the Yangian invariants
obtained from our unitary Graßmannian integral and tree-level
superamplitudes of $\mathcal{N}=4$ SYM.

We begin in section~\ref{sec:grassmann-osc} by introducing a
Graßmannian integral formula for oscillator representations of
$\mathfrak{u}(p,q|m)$, though at first without specifying a contour.
For special representation labels and inhomogeneities, and upon
enforcing a unitary contour, the integral reduces to the
Brezin-Gross-Witten matrix model. This observation motivates the use
of the unitary contour for general values of the parameters. We prove
that this contour guarantees the Yangian invariance of what we then
call unitary Graßmannian integral or matrix model. We employ this
method to recover several sample invariants from
section~\ref{sec:sample-inv}. The unitary Graßmannian integral
approach is complementary to the Bethe ansatz for Yangian invariants
of chapter \ref{cha:bethe-vertex}. On the one hand, it is limited to
Yangian invariants corresponding to specific permutations in the
classification of section~\ref{sec:class-solut}, which was derived
from the Bethe ansatz. On the other hand, it allows for the
construction of invariants for $\mathfrak{u}(p,q|m)$, whereas the
Bethe ansatz is currently limited to $\mathfrak{u}(2)$.

In section~\ref{sec:osc-spinor} we change the basis in the integrand
of the unitary Graßmannian formula for $\mathfrak{u}(p,p|m)$ from
oscillators to spinor helicity-like variables. For the bosonic
oscillators this amounts to a Bargmann transformation. Such a
transformation is known e.g.\ from the one-dimensional harmonic
oscillator in quantum mechanics, where it implements the transition
from Fock space to position space.

Next, in section~\ref{sec:grassmann-spinor} we apply the change of
basis to the entire unitary Graßmannian integral. The resulting
formula in spinor helicity-like variables is then compared to the
original Graßmannian integral from
section~\ref{sec:grassmannian-integral}. Besides the presence of a
unitary contour in our approach, we also work in the physical
Minkowski signature, which is not the case in the original
framework. In addition, our integral inherently contains deformation
parameters in the form of inhomogeneities and representation
labels. The branch cuts of the integrand, which caused problems with
such parameters for the deformed amplitudes in
section~\ref{sec:deform}, disappear because of the unitary contour. To
put our proposal to the test, we compare sample Yangian invariants
computed with the unitary Graßmannian integral to known expressions
for superamplitudes of $\mathcal{N}=4$ SYM and deformations thereof.

Lastly, some additional material on the unitary Graßmannian integral
is deferred to appendix~\ref{cha:add-grass-int}.

\section{Graßmannian Integral in Oscillator Variables}
\label{sec:grassmann-osc}

\subsection{Graßmannian Formula}
\label{sec:formula}

We delve into this chapter by directly presenting one of the main
results, a Graßmannian integral formula for Yangian invariants with
oscillator representations of the non-compact superalgebra
$\mathfrak{u}(p,q|r+s)$. Recall the notation $p+q=n$ and $r+s=m$ from
section~\ref{sec:osc-rep}. We motivate our formula by combining our
knowledge of the Graßmannian integral for deformed scattering
amplitudes \eqref{eq:grassint-amp-def} with that of the simple
two-site oscillator sample invariant \eqref{eq:nc21-int}. Here we
merely state the resulting formula. Its implications, some refinements
and examples will be explored in detail in the subsequent sections. In
particular, the proof of its Yangian invariance is deferred to
section~\ref{sec:proof-yang-invar}.

We consider the monodromy $M_{N,K}(u)$ in \eqref{eq:yi-mono-nc} with
$N=2K$ sites, out of which the first $K$ carry a ``dual'' oscillator
representation $\bar{\oscrep}_{c_i}$ and the remaining $K=N-K$ sites
carry an ``ordinary'' one $\oscrep_{c_i}$. The normalization factors
of the Lax operators \eqref{eq:yangian-def-lax} are chosen to be
trivial, i.e.\ $f_{\oscrep_{c_i}}=f_{\bar{\oscrep}_{c_i}}=1$. A
Yangian invariant for this monodromy is given by the \emph{Graßmannian
  integral formula}
\begin{align}
  \label{eq:grass-int}
  |\Psi_{N,K}\rangle
  =\int\D\mathcal{C}
  \frac{e^{\tr(\mathcal{C}\mathbf{I}_\bullet^t+\mathbf{I}_\circ \mathcal{C}^{-1})}
    |0\rangle}
  {(\det\mathcal{C})^{q-s}(1,\ldots,K)^{1+v_{K}^+-v_1^-}
    \cdots (N,\ldots,K-1)^{1+v_{K-1}^+-v_N^-}}\,.
\end{align}
Here the numerator can be understood as a matrix generalization of
that of the two-site sample invariant \eqref{eq:nc21-int}. The single
contractions of oscillators in the exponent in that formula are
replaced by the $K\times K$ matrices
\begin{align}
  \label{eq:osc-matrix}
  \mathbf{I}_{\mathrel{\ooalign{\raisebox{0.4ex}{$\scriptstyle\bullet$}\cr\raisebox{-0.4ex}{$\scriptstyle\circ$}}}}=
  \begin{pmatrix}
    (1\mathrel{\ooalign{\raisebox{0.7ex}{$\bullet$}\cr\raisebox{-0.3ex}{$\circ$}}} K+1)&
    \cdots&
    (1\mathrel{\ooalign{\raisebox{0.7ex}{$\bullet$}\cr\raisebox{-0.3ex}{$\circ$}}} N)\\
    \vdots&&\vdots\\
    (K\mathrel{\ooalign{\raisebox{0.7ex}{$\bullet$}\cr\raisebox{-0.3ex}{$\circ$}}} K+1)&
    \cdots&
    (K\mathrel{\ooalign{\raisebox{0.7ex}{$\bullet$}\cr\raisebox{-0.3ex}{$\circ$}}} N)\\
  \end{pmatrix}.
\end{align}
These matrices contain all possible contractions of oscillators
between dual and ordinary sites that we defined in
\eqref{eq:bullets-nc},
\begin{align}
  \label{eq:bullets-nc-recall}
    (k\bullet l)=
    \sum_{\indssub{A}}\bar{\mathbf{A}}^l_{\indssub{A}}\bar{\mathbf{A}}^k_{\indssub{A}}\,,\quad
    (k\circ l)=
    \sum_{\dot{\indssub{A}}}\bar{\mathbf{A}}^l_{\dot{\indssub{A}}}\bar{\mathbf{A}}^k_{\dot{\indssub{A}}}\,.
\end{align}
These entries of the matrices $\mathbf{I}_{\bullet}$ and
$\mathbf{I}_{\circ}$ are respectively $\mathfrak{u}(p|r)$ and
$\mathfrak{u}(q|s)$ invariant, cf.\ \eqref{eq:bullets-nc-symm}.  We
may think of these invariants of compact subalgebras of
$\mathfrak{u}(p,q|r+s)$ as ``elementary building blocks'' of the
Yangian invariant. The denominator of \eqref{eq:grass-int} is
analogous to that of the Graßmannian integral for deformed scattering
amplitudes \eqref{eq:grassint-amp-def}. It contains the minors
$(i,\ldots,i+K-1)$ of the $K\times N$ matrix $C$ defined in
\eqref{eq:grassint-matrix}. However, notice the extra factor
$(\det\mathcal{C})^{q-s}=(N-K+1,\ldots,N)^{q-s}$, which depends on the
symmetry algebra. The gauge fixing of the matrix $C$ in
\eqref{eq:grassint-matrix} corresponds to the order of dual and
ordinary sites. Furthermore, the integral in \eqref{eq:grass-int} is
over the holomorphic $K^2$-form
$\D\mathcal{C}=\bigwedge_{k,l}\D C_{k l}$, which we already
encountered in \eqref{eq:grassint-amp-def}. The $2N$ parameters
$v_i^+,v_i^-$ appearing as exponents of the minors are related to the
$2N$ parameters $v_i,c_i$ of the monodromy \eqref{eq:yi-mono-nc} by,
cf.\ \cite{Beisert:2014qba},
\begin{align}
  \label{eq:spec-redef}
  v^\pm_i=v_i'\pm\frac{c_i}{2}\,,\quad
  v_i'=v_i-\frac{c_i}{2}+
  \begin{cases}
    n-m-1&\text{for}\quad i=1,\ldots, K\,,\\
    0&\text{for}\quad i=K+1,\ldots, N\,.
  \end{cases}
\end{align}
Finally, for $|\Psi_{N,K}\rangle$ in
\eqref{eq:grass-int} to be Yangian invariant, the parameters
$v_i^+,v_i^-$ have to satisfy the $N$ relations
\begin{align}
  \label{eq:spec-perm}
  v^+_{i+K}=v^-_i\,
\end{align}
for $i=1,\ldots,N$. These relations are analogous to
\eqref{eq:amp-wperm} for the deformed amplitudes.\footnote{Because
  $N=2K$, the relations \eqref{eq:spec-perm} are equivalent to
  $v^-_{i+K}=v^+_i$, which is exactly the form of
  \eqref{eq:amp-wperm}.} Recall that while the inhomogeneities $v_i$
are complex numbers, the labels $c_i$ of the oscillator
representations $\oscrep_{c_i}$ and $\bar{\oscrep}_{c_i}$ have to be
integers, see section~\ref{sec:osc-rep}. This yields further
constraints on the parameters $v_i^+,v_i^-$. For now, this
completes the specification of the Graßmannian integral
formula~\eqref{eq:grass-int}.

One obvious omission in this specification is the choice of a
multi-dimensional contour of integration in \eqref{eq:grass-int}. The
proof of Yangian invariance in the following section only assumes that
certain boundary terms vanish upon integration by parts, which is
satisfied in particular for closed contours. The choice of a suitable
integration contour will be of paramount importance in the sections
thereafter.

We add some further remarks. The condition $N=2K$ guarantees
$\mathcal{C}$ to be a square matrix. Thus it is sensible to use its
inverse in \eqref{eq:grass-int}. In the compact special case
$\mathfrak{u}(p,0|r)$ we have $\mathbf{I}_\circ=0$, thus
$\mathcal{C}^{-1}$ is absent from \eqref{eq:grass-int} and the
Graßmannian integral yields Yangian invariants also for $N\neq 2K$.
However, we do not elaborate on the compact case in this work.  We
note that because of $\mathbf{I}_\circ=0$, the \emph{compact} case of
\eqref{eq:grass-int} is reminiscent of the link representation of
scattering amplitudes, cf.\ \cite{ArkaniHamed:2009dn}. It is different
though, as the amplitudes transform under the \emph{non-compact}
superconformal algebra. Let us also point out a relation between the
Graßmannian integral \eqref{eq:grass-int} and the Bethe ansatz for
Yangian invariants of section~\ref{sec:bethe-yangian}. In the
$\mathfrak{u}(2,0|0)$ case the redefinition of variables in
\eqref{eq:spec-redef} reduces to \eqref{eq:class-changevars}. This
equation was important to obtain the classification
\eqref{eq:class-sol-perm} of the solutions to the functional relation
\eqref{eq:bethe-gl2-ad} in terms of permutations. We can identify
\eqref{eq:spec-perm} with \eqref{eq:class-sol-perm} for the special
permutation $\sigma(i)=i+K$. Thus we expect that for $\mathfrak{u}(2)$
the Yangian invariants produced by the Graßmannian
integral~\eqref{eq:grass-int} are precisely those which we already
know from the Bethe ansatz. We will verify this explicitly for some
examples in section~\ref{sec:sample-invariants}.

\subsection{Proof of Yangian Invariance}
\label{sec:proof-yang-invar}

Here we prove the Yangian invariance of the Graßmannian integral
\eqref{eq:grass-int} for the invariant $|\Psi_{N, K}\rangle$ with
$N=2K$ sites and oscillator representations of the non-compact
superalgebra $\mathfrak{u}(p,q|r+s)$. We will verify the expanded form
\eqref{eq:yi-exp-1} of the Yangian invariance condition. As argued
there, it is sufficient to check this equation for the expansion
coefficients $M_{\indnm{AB}}^{(1)}$ and $M_{\indnm{AB}}^{(2)}$ of the
monodromy elements $M_{\indnm{AB}}(u)$. With straightforward
modifications the following proof also applies to the compact case
with $q=s=0$ where $\mathbf{I}_\circ=0$ and $N\neq 2K$ is possible.

Let us start with the ansatz
\begin{align}
  \label{eq:proof-ansatz}
  |\Phi\rangle=e^{\tr(\mathcal{C}\mathbf{I}_\bullet^t+\mathbf{I}_\circ \mathcal{C}^{-1})}
  |0\rangle\,,
\end{align}
which we recognize as the exponential function in
\eqref{eq:grass-int}. We want to show that this ansatz satisfies
\eqref{eq:yi-exp-1} with the Yangian generators
$M_{\indnm{AB}}^{(1)}$, that is to say $\mathfrak{gl}(n|m)$
invariance. Using the expression \eqref{eq:yangian-mono-coeff-glnm} of
these Yangian generators in terms of $\mathfrak{gl}(n|m)$ generators
for the monodromy \eqref{eq:yi-mono-nc} yields
\begin{align}
  \label{eq:level1-dual-normal}
  M^{(1)}_{\indnm{AB}}=\sum_{k=1}^K \bar{\mathbf{J}}^k_{\indnm{BA}}+\sum_{l=K+1}^N \mathbf{J}^l_{\indnm{BA}}\,.
\end{align}
The generators $\bar{\mathbf{J}}^k_{\indnm{BA}}$ of the ``dual''
oscillator representation $\bar{\oscrep}_{c_k}$ are given in
\eqref{eq:gen-dual}. Likewise, the $\mathbf{J}^l_{\indnm{BA}}$ defined
in \eqref{eq:gen-ordinary} generate the ``ordinary'' representation
$\oscrep_{c_l}$. To evaluate the action of the operator
\eqref{eq:level1-dual-normal} on the ansatz \eqref{eq:proof-ansatz} we
compute
\begin{align}
  \label{eq:action-level-one}
  \begin{aligned}
    (\bar{\mathbf{J}}^k_{\indnm{AB}})\,|\Phi\rangle
    &=
    \left(
      \begin{array}{c:c}
        -\sum_w\bar{\mathbf{A}}^w_{\indssub{A}} \,\bar{\mathbf{A}}^k_{\indssub{B}} \,C_{k w}&
        -\sum_{w,w'}\bar{\mathbf{A}}^w_{\indssub{A}}\,\bar{\mathbf{A}}^{w'}_{\dot{\indssub{B}}} D_{w' k}C_{k w}\\[0.3em]
        \hdashline\\[-1.0em]
        (-1)^{|\dot{\indssub{A}}|}\bar{\mathbf{A}}^k_{\dot{\indssub{A}}}\,\bar{\mathbf{A}}^k_{\indssub{B}}&
        (-1)^{|\dot{\indssub{A}}|}(\sum_w\bar{\mathbf{A}}^k_{\dot{\indssub{A}}}\,\bar{\mathbf{A}}^w_{\dot{\indssub{B}}}\,D_{w k}+\delta_{\dot{\indssub{A}}\dot{\indssub{B}}})\\
      \end{array}
    \right)
    |\Phi\rangle\,,\\ 
    (\mathbf{J}^l_{\indnm{AB}})\,|\Phi\rangle
    &=
    \left(
      \begin{array}{c:c}
        \sum_v\bar{\mathbf{A}}^l_{\indssub{A}}\,\bar{\mathbf{A}}^v_{\indssub{B}} \,C_{v l}&
        \bar{\mathbf{A}}^l_{\indssub{A}}\,\bar{\mathbf{A}}^l_{\dot{\indssub{B}}}\\[0.3em]
        \hdashline\\[-1.0em]
        -(-1)^{|\dot{\indssub{A}}|}\sum_{v,v'}\bar{\mathbf{A}}^v_{\dot{\indssub{A}}}\,\bar{\mathbf{A}}^{v'}_{\indssub{B}} C_{v' l}\,D_{l v}&
        -(-1)^{|\dot{\indssub{A}}|}(\sum_v\bar{\mathbf{A}}^v_{\dot{\indssub{A}}}\,\bar{\mathbf{A}}^l_{\dot{\indssub{B}}}\,D_{l v}+\delta_{\dot{\indssub{A}}\dot{\indssub{B}}})\\
      \end{array}
    \right)
    |\Phi\rangle\,,
  \end{aligned}
\end{align}
where components of the matrix $\mathcal{C}^{-1}$ are denoted by
$D_{lk}$. Here and in the remainder of this proof the indices $k,v,v'$
always take the values $1,\ldots,K$ while $l,w,w'$ are in the range
$K+1,\ldots,N$. Now one immediately obtains
\begin{align}
  \label{eq:level-one}
  M^{(1)}_{\indnm{AB}}\,|\Phi\rangle=0\,.
\end{align}
Hence \eqref{eq:yi-exp-1} with the Yangian generators
$M_{\indnm{AB}}^{(1)}$ holds for the ansatz~\eqref{eq:proof-ansatz}.

However, each site of the ansatz \eqref{eq:proof-ansatz} does not yet
transform in an irreducible representation of the superalgebra
$\mathfrak{u}(p,q|r+s)$. In fact, \eqref{eq:proof-ansatz} is not an
eigenstate of the central elements
$\mathbf{C}^l=\sum_{\indnm{A}=1}^{n+m}\mathbf{J}_{\indnm{AA}}^l$ and
$\bar{\mathbf{C}}^k=\sum_{\indnm{A}=1}^{n+m}\bar{\mathbf{J}}_{\indnm{AA}}^k$
that were defined in \eqref{eq:central-ord} and
\eqref{eq:central-dual}, respectively. To obtain eigenstates we have
to pick special linear combinations of the ansatz
\eqref{eq:proof-ansatz},
\begin{align}
  \label{eq:ansatz-general}
  |\Psi_{N, K}\rangle=\int\D\mathcal{C}\,\mathscr{G}(\mathcal{C})\,
  |\Phi\rangle\,.
\end{align}
It turns out to be suitable to choose an integrand that contains only
consecutive minors of the matrix $C$ defined in \eqref{eq:grassint-matrix},
\begin{align}
  \label{eq:ansatz-minors}
  \mathscr{G}(\mathcal{C})=\frac{1}{(1,\ldots,K)^{1+\alpha_1}\cdots(N,\ldots,K-1)^{1+\alpha_N}}\,
\end{align}
with arbitrary complex constants $\alpha_i$. With this integrand the
ansatz \eqref{eq:ansatz-general} is an eigenstate of the central
elements,
\begin{align}
  \label{eq:ccbar-eval}
  \bar{\mathbf{C}}^k\,|\Psi_{N, K}\rangle
  =\biggl(\,\,q-s-\sum_{\mathclap{i=k+1}}^{\mathclap{k+N-K}}\alpha_i\,\,\biggr)\,|\Psi_{N, K}\rangle\,,\quad
  \mathbf{C}^l\,|\Psi_{N, K}\rangle
  =\biggl(-q+s+\sum_{\mathclap{i=l-K+1}}^l\alpha_i\,\,\biggr)\,|\Psi_{N, K}\rangle\,.
\end{align}
To show this property we assumed that upon integration by parts the
boundary terms vanish. Furthermore, we employed the identity
\begin{align}
  \label{eq:diff-ansatz}
  \frac{\D}{\D C_{k l}}e^{\tr(\mathcal{C}\mathbf{I}_\bullet^t+\mathbf{I}_\circ \mathcal{C}^{-1})}|0\rangle=
  \biggl((k\bullet l)-\sum_{v,w}D_{w k}D_{l v}(v\circ w)\biggr)
  e^{\tr(\mathcal{C}\mathbf{I}_\bullet^t+\mathbf{I}_\circ \mathcal{C}^{-1})}|0\rangle\,,
\end{align}
which is easily verified taking into account
$\frac{\D}{\text{d} C_{k l}}D_{w v}=-D_{w k}D_{l v}$. In addition, in
evaluating derivatives of the minors in $\mathscr{G}(\mathcal{C})$ we
used, cf.\ \cite{Drummond:2010qh,Drummond:2010uq},
\begin{align}
  \label{eq:deriv-minor}
  \sum_{w} C_{k w} \frac{\D}{\D C_{k w}} (i, \ldots, i+K-1)^{1+\alpha_i} 
  = (1+\alpha_i )\,(i, \ldots, i+K-1)^{1+\alpha_i}\,
\end{align}
for $i = k+1, \ldots, k+N-K$. For other values of $i$ the left hand
side in \eqref{eq:deriv-minor} vanishes due to the gauge fixing of $C$
in \eqref{eq:grassint-matrix}.

Next, we turn our attention to the invariance condition
\eqref{eq:yi-exp-1} with the Yangian generators
$M_{\indnm{AB}}^{(2)}$. From the commutation relations
\eqref{eq:yangian-def-gen} with $r=2$ and $s=1$ one sees that if a
state $|\Psi\rangle$ is annihilated by all $M_{\indnm{AB}}^{(1)}$ and
by one of the generators $M_{\indnm{AB}}^{(2)}$, e.g.\ by
$M_{11}^{(2)}$, then it is annihilated by all
$M_{\indnm{AB}}^{(2)}$. Thus in our case it is sufficient to verify
\eqref{eq:yi-exp-1} for one of the four blocks of generators, say for
$M^{(2)}_{\indssub{AB}}$. Expressions for these generators can be
found in \eqref{eq:yangian-mono-coeff-glnm}. We compute the action of
all terms appearing therein on our ansatz \eqref{eq:proof-ansatz},
\begin{align}
  \label{eq:jj-action}
  \begin{aligned}
    &\sum_{\indnm{I}} \,(-1)^{|\indnm{I}|}\mathbf{J}_{\indssub{B} \indnm{I}}^l\bar{\mathbf{J}}^k_{\indnm{I}\indssub{A}}\,
    |\Phi\rangle
    =-\bar{\mathbf{A}}^l_{\indssub{B}}\,\bar{\mathbf{A}}^k_{\indssub{A}}
    \biggl(\,\sum_{v,w}\,C_{v l}C_{k w}\,\frac{\D}{\D C_{v w}}+(p-r)\,C_{k l}\biggr)
    |\Phi\rangle\,,\\
    &\sum_{\indnm{I}}\,(-1)^{|\indnm{I}|}\bar{\mathbf{J}}_{\indssub{B} \indnm{I}}^{k'}\bar{\mathbf{J}}^k_{\indnm{I}\indssub{A}}\,
    |\Phi\rangle
    =
    \sum_{w,w'}\bar{\mathbf{A}}^w_{\indssub{B}}\,\bar{\mathbf{A}}^k_{\indssub{A}}\, C_{k' w}\,C_{k w'}\frac{\D}{\D C_{k' w'}}
    |\Phi\rangle\,,\\
    &\sum_{\indnm{I}}\,(-1)^{|\indnm{I}|}\mathbf{J}_{\indssub{B} \indnm{I}}^{l'}\mathbf{J}^l_{\indnm{I}\indssub{A}}\,
    |\Phi\rangle
    =
    \sum_{v,v'}\bar{\mathbf{A}}^{l'}_{\indssub{B}}\,\bar{\mathbf{A}}^v_{\indssub{A}}\, C_{v l}\,C_{v' l'}\,\frac{\D}{\D C_{v' l}}\,
    |\Phi\rangle\,
  \end{aligned}
\end{align}
for $k\neq k'$ and $l\neq l'$, and furthermore
\begin{align}
  \label{eq:vj-action}
  \biggl(\sum_kv_k\,\bar{\mathbf{J}}_{\indssub{BA}}^k+\sum_lv_l\,\mathbf{J}_{\indssub{BA}}^l\biggr)
  |\Phi\rangle
  =\sum_{k,l}\bar{\mathbf{A}}^l_{\indssub{B}}\,\bar{\mathbf{A}}^k_{\indssub{A}} \,C_{k l}\,(v_l-v_k)\,
  |\Phi\rangle\,.
\end{align}
Making use of these formulas we can evaluate the action on
\eqref{eq:ansatz-general},
\begin{align}
  \label{eq:m2-eval}
  \!\!\! M_{\indnm{AB}}^{(2)}\,|\Psi_{N, K}\rangle
  =\sum_{k,l}\,\biggl(v_l-v_k-p+r+1-\,\,\,\sum_{\mathclap{i=l-K+1}}^{\mathclap{k+N-K}}\,\,\alpha_i\,\biggr)
  \,\bar{\mathbf{A}}^l_{\indssub{B}}\,\bar{\mathbf{A}}^k_{\indssub{A}}
  \int\D\mathcal{C}\mathscr{G}(\mathcal{C})\,C_{k l}\,
  |\Phi\rangle\,.
\end{align}
Here we assumed once more that the boundary terms of the integration
by parts vanish. Furthermore, we used \eqref{eq:diff-ansatz} and
properties of the minors in $\mathscr{G}(\mathcal{C})$ similar to
\eqref{eq:deriv-minor}. To ensure Yangian invariance of the ansatz,
the parameters $\alpha_i$ have to be chosen such that the bracket in
\eqref{eq:m2-eval} vanishes.

In conclusion, for the ansatz \eqref{eq:ansatz-general} to be Yangian
invariant, the parameters $v_i,c_i$ of the monodromy and the
$\alpha_i$ appearing in this ansatz have to obey the equations
obtained from \eqref{eq:ccbar-eval} and \eqref{eq:m2-eval},
\begin{gather}
  \label{eq:proof-final-ccbar}
  c_k=q-s\,-\,\,\,\sum_{\mathclap{i=k+1}}^{\mathclap{k+N-K}}\,\,\alpha_i\,,\quad
  c_l=-q+s\,+\,\,\,\sum_{\mathclap{i=l-K+1}}^l\,\,\alpha_i\,,\quad
  v_k-v_l=-p+r+1-\,\,\,\sum_{\mathclap{i=l-K+1}}^{\mathclap{k+N-K}}\,\,\alpha_i\,
\end{gather}
for $k=1,\ldots K$ and $l=K+1,\ldots,N$. These equations are
conveniently addressed after changing from the variables $v_i,c_i$ to
$v^+_i,v^-_i$ with \eqref{eq:spec-redef}. In these variables they are
solved by
\begin{align}
  \label{eq:alpha}
  \alpha_i=v^+_{i+K-1}-\,v^-_i+(q-s)\,\delta_{i,N-K+1}\,
\end{align}
and imposing the $N$ constraints in
\eqref{eq:spec-perm}. Equation~\eqref{eq:alpha} turns the ansatz
\eqref{eq:ansatz-general} into the Graßmannian integral formula
\eqref{eq:grass-int}. This concludes the proof of its Yangian
invariance.

\subsection{Unitary Matrix Models}
\label{sec:unit-matr-models}

In the introductory section~\ref{sec:amplitudes} we saw that the
Graßmannian integral for $\mathcal{N}=4$ SYM scattering amplitudes
\eqref{eq:grassint-amp} is that special case of the deformed integral
\eqref{eq:grassint-amp-def} where the exponents of all minors are
equal to $1$. Here we investigate a special case of our Graßmannian
integral formula~\eqref{eq:grass-int} in oscillator variables. We
choose deformations parameters $v_i^\pm$ such that the exponents of
almost all of the minors are identical to $0$. Although this choice
seems trivial from an amplitudes perspective, it reveals an
interesting connection between the Graßmannian integral
\eqref{eq:grass-int} and certain unitary matrix models. The
\emph{integrand} of \eqref{eq:grass-int} reduces to that of the
Brezin-Gross-Witten matrix model or even a slight generalization
thereof, the Leutwyler-Smilga model. To identify the entire
\emph{integral} in \eqref{eq:grass-int} with these unitary matrix
models, we choose the contour of integration to be the unitary group
manifold. This ``unitary contour'' will be of pivotal importance in
the rest of this chapter. Furthermore, in the special case explored
here the Graßmannian integral \eqref{eq:grass-int} can be computed
easily by applying well established matrix model techniques. In this
way, we obtain a representation of these Yangian invariants in terms
of Bessel functions.

In order to reduce \eqref{eq:grass-int} with $N=2K$ to the
Leutwyler-Smilga integral, we restrict to a special solution of the
constraints in \eqref{eq:spec-perm} on the deformation parameters
$v_i^\pm$. The solution has to be such that all minors in
\eqref{eq:grass-int}, except for $(1,\ldots,K)=1$ and
$(N-K+1,\ldots,N)=\det\mathcal{C}$, have a vanishing exponent. A short
calculation shows that this solution depends only on two parameters
$v\in\mathbb{C}, c\in\mathbb{Z}$. It is given by
\begin{align}
  \label{eq:ls-int-para}
  \begin{aligned}
    v_i&=v-c-n+m+1+(i-1)\,,\quad&c_i&=-c &\quad&\text{for}\quad i=1,\ldots,K\,,\\
    v_i&=v+(i-K-1)\,,\quad&c_i&=\phantom{-}c &\quad&\text{for}\quad i=K+1,\ldots,2K\,.\\
  \end{aligned}
\end{align}
Here we used \eqref{eq:spec-redef} to change from the variables
$v_i^+,v_i^-$ employed in \eqref{eq:grass-int} to the variables
$v_i,c_i$. Let us now focus on the measure
$\D\mathcal{C}=\bigwedge_{k,l}\D C_{k,l}$ in \eqref{eq:grass-int}. One
readily verifies that
\begin{align}
  \label{eq:measure-haar}
  [\D\mathcal{C}]=\chi_K\,\frac{\D\mathcal{C}}{(\det\mathcal{C})^K}\,,
\end{align}
with a constant number $\chi_K\in\mathbb{C}$, is invariant under
$\mathcal{C}\mapsto\mathcal{V}\mathcal{C}$ and
$\mathcal{C}\mapsto\mathcal{C}\mathcal{V}$ for any constant matrix
$\mathcal{V}\in GL(\mathbb{C}^K)$. Because of these properties, for unitary
$\mathcal{C}$ the differential form $[\D\mathcal{C}]$ defined in
\eqref{eq:measure-haar} gives rise to the Haar measure on the unitary
group $U(K)$, cf.\ \cite{Knapp:2002,Fuchs:2003}. The normalization
$\chi_K$ is chosen such that $\int_{U(K)}[\D\mathcal{C}]= 1$. We
select a ``unitary contour'' in the Graßmannian integral
\eqref{eq:grass-int} by demanding
$\mathcal{C}^\dagger=\mathcal{C}^{-1}$. This allows us to express the
Yangian invariant with the special choice of deformation parameters
\eqref{eq:ls-int-para} as
\begin{align}
  \label{eq:red-matrix-fin-inv}
  |\Psi_{2K,K}\rangle
  =\chi_K^{-1}\int_{U(K)}[\D\mathcal{C}]
  \frac{e^{\tr(\mathcal{C}\mathbf{I}_\bullet^t+\mathbf{I}_\circ \mathcal{C}^\dagger)}|0\rangle}
  {(\det\mathcal{C})^{c+q-s}}\,,
\end{align}
where $c\in\mathbb{Z}$ is a free
parameter. Equation~\eqref{eq:red-matrix-fin-inv} is known as
\emph{Leutwyler-Smilga model} \cite{Leutwyler:1992yt}, where the
matrices $\mathbf{I}_\bullet^t$ and $\mathbf{I}_\circ$ are considered
as sources. For $c+q-s=0$ it becomes the \emph{Brezin-Gross-Witten
  model} \cite{Gross:1980he,Brezin:1980rk,Bars:1979xb}. Remarkably,
the integral \eqref{eq:red-matrix-fin-inv} can be computed
exactly. For two \emph{independent} source matrices
$\mathbf{I}_\bullet^t$ and $\mathbf{I}_\circ$ this was achieved in
\cite{Schlittgen:2002tj} using the character expansion methods of
\cite{Balantekin:2000vn},
\begin{align}
  \label{eq:ls-integral-bessel}
  |\Psi_{2K,K}\rangle=
  \chi_K^{-1}\prod_{j=0}^{K-1}j!
  \frac{(\det\mathbf{I}_\bullet^t)^{c+q-s}}
  {\Delta(\mathbf{I}_\circ\mathbf{I}_\bullet^t)}
  \det\left(\frac{I_{k+c+q-s-K}\big(2\sqrt{(\mathbf{I}_\circ\mathbf{I}_\bullet^t)_l}\big)}
    {\sqrt{(\mathbf{I}_\circ\mathbf{I}_\bullet^t)_l}^{\,k+c+q-s-K}}\right)_{k,l}
  |0\rangle\,.
\end{align}
The entries of the matrices $\mathbf{I}_\bullet^t$ and
$\mathbf{I}_\circ$ are bosonic even in the supersymmetric setting, see
the discussion after \eqref{eq:bullets-nc}. Assuming the matrix
$\mathbf{I}_\circ\mathbf{I}_\bullet^t$ to be diagonalizable, we denote
its $l$-th eigenvalue by
$(\mathbf{I}_\circ\mathbf{I}_\bullet^t)_l$. Furthermore,
$\Delta(\mathbf{I}_\circ\mathbf{I}_\bullet^t)=\det
({(\mathbf{I}_\circ\mathbf{I}_\bullet^t)_l}^{k-1})_{k,l}$
is the Vandermonde determinant. The formula
\eqref{eq:ls-integral-bessel} involving a determinant of Bessel
functions $I_\nu(x)$ generalizes the single Bessel function that we
found for the sample Yangian invariant $|\Psi_{2,1}\rangle$ in
\eqref{eq:sample-inv21}.

We conclude this section by adding some background material on the
unitary matrix integral \eqref{eq:red-matrix-fin-inv}. It is of
relevance in multiple physical contexts. The Brezin-Gross-Witten model
appears in two-dimensional massless lattice QCD, see
\cite{Gross:1980he}. The partition function of this gauge theory can
be reduced to \eqref{eq:red-matrix-fin-inv} with a vanishing exponent
of $\det{\mathcal{C}}$. Both source matrices
$\mathbf{I}_\bullet^t=\mathbf{I}_\circ\propto 1_K$ are multiples of
the theory's coupling constant and the description is valid at any
value of this coupling. The unitary group $U(K)$ corresponds to the
gauge group of the Yang-Mills field. Let us move on to a different
context in which the integral \eqref{eq:red-matrix-fin-inv}
occurs. The Leutwyler-Smilga model describes four-dimensional
continuum QCD with non-vanishing quark masses in a certain low energy
regime. In this theory the partition function in a sector with a fixed
topological charge of the gauge field is given by
\eqref{eq:red-matrix-fin-inv}, where the exponent of
$\det{\mathcal{C}}$ corresponds to that charge. Furthermore, the group
$U(K)$ is associated with the flavor symmetry of the $K$ quarks. The
source matrices $\mathbf{I}_\bullet^t$ and $\mathbf{I}_\circ$ are
parametrized by the quark masses. A more detailed account on this
interpretation of the integral \eqref{eq:red-matrix-fin-inv} is
provided in the lecture notes \cite{Verbaarschot:2005rj}. Our interest
in this integral is mostly of mathematical nature. Basics of the group
theoretical character expansion method, which yields the determinant
formula \eqref{eq:ls-integral-bessel}, are discussed e.g.\ in the
concise review \cite{Morozov:2009jv}.

Numerous matrix models are long known to be related to classically
integrable hierarchies of partial differential equations, see the
reviews \cite{DiFrancesco:1993cyw,Morozov:1994hh} and references
therein. This notion of integrability is closely linked to the
Korteweg-de\ Vries equation, which we encountered as a
$1+1$-dimensional model for waves in shallow water right at the
beginning of this thesis in section~\ref{sec:int-mod}. There exists an
integrable generalization of this model to $2+1$ dimensions, the
Kadomtsev-Petviashvili (KP) equation, see e.g.\
\cite{Ablowitz:1991,Biondini:2008}. Besides its interpretation as a
model for water waves, it appears in multiple further physical
contexts. In addition, it is of importance because many other
integrable differential equations can be obtained from this equation
by means of a symmetry reduction. Notably, the KP equation was found
to belong to an infinite set of compatible integrable partial
differential equations, the KP hierarchy, see the substantial review
in \cite{Miwa:2000}. Solutions of this hierarchy are given in the form
of so-called $\tau$-functions. After this digression, we return to the
matrix models encountered in this section. The partition function of
the Brezin-Gross-Witten model is known to be a $\tau$-function of the
KP hierarchy \cite{Mironov:1994mv}. This also applies to the partition
function of the slightly more general Leutwyler-Smilga model, cf.\
\cite{Orlov:2002ka}. The determinant representation
\eqref{eq:ls-integral-bessel} of the partition function is the key to
establish these relations. The connection between this integrable
structure and the Yangian invariance of these models, that is
investigated in this thesis, seems to be far from obvious. It would be
interesting to clarify this connection.

\subsection{Unitary Graßmannian Matrix Models}
\label{sec:unitary-contour}

We just established that the choice of a ``unitary contour'' together
with \emph{special} deformation parameters $v_i^\pm$ reduces the
Graßmannian integral \eqref{eq:grass-int} to a well-known unitary
matrix model. In what follows, we show that this contour remains
appropriate for \emph{general} deformation parameters. This leads to
a, to the best of our knowledge, novel class of unitary Graßmannian
matrix models.

\subsubsection{Single-Valuedness of Integrand}
\label{sec:single-valu-integr}

The multi-dimensional contour for the Graßmannian integral
\eqref{eq:grass-int} should be closed. This ensures that the boundary
terms in the proof of its Yangian invariance in
section~\ref{sec:proof-yang-invar} vanish. Choosing the contour to be
the unitary group manifold $U(K)$ seems to assure this because it is
compact. However, we also have to verify that the integrand of
\eqref{eq:grass-int} is a single-valued function on this
contour. Otherwise, the compactness of the contour does not imply that
the boundary terms vanish. Due to the complex exponents $v_i^\pm$ of
the minors and the resulting branch cuts, the single-valuedness of the
integrand in \eqref{eq:grass-int} is far from obvious. In fact, we
have to modify the integrand in a minute way to be able to prove that
it is single-valued.

We transcribe the Graßmannian integral for $N=2K$ and the symmetry
algebra $\mathfrak{u}(p,q|r+s)$ from \eqref{eq:grass-int},
\begin{align}
  \label{eq:grass-int-unitary}
  |\Psi_{2K,K}\rangle
  =\chi_K^{-1}\int_{U(K)}[\D\mathcal{C}]\,
  \mathscr{F}(\mathcal{C})\,
  e^{\tr(\mathcal{C}\mathbf{I}_\bullet^t+\mathbf{I}_\circ \mathcal{C}^{^\dagger})}
    |0\rangle\,.
\end{align}
Here we imposed the unitary contour
$\mathcal{C}^{-1}=\mathcal{C}^{\dagger}$. In addition, we expressed
$\D\mathcal{C}$ in terms of the Haar measure $[\D\mathcal{C}]$ via
\eqref{eq:measure-haar}. Most importantly, the integrand containing
the minors of the Graßmannian matrix $C$ reads
\begin{align}
  \label{eq:grass-int-unitary-integrand}
  \begin{aligned}
    \mathscr{F}(\mathcal{C})^{-1}=
    \,(\det{\mathcal{C}})^{q-s-K}
    \cdot(1,\ldots,K)^{1+v_K^+-v_1^-}\cdots(2K,\ldots,K-1)^{1+v_{K-1}^+-v_{2K}^-}\,.
  \end{aligned}
\end{align}
This integrand is still somewhat formal because we have not specified
its analytic structure yet. In the following we will do this
implicitly by manipulating it into a form that is manifestly
single-valued.

We start by expressing the minors of the $K\times 2K$ matrix $C$
defined in \eqref{eq:grassint-matrix} in terms of those of the
$K\times K$ matrix $\mathcal{C}$,
\begin{align}
  (i,\ldots,i+K-1)=(-1)^{(K-i+1)(i-1)}
  \begin{cases}
    [1,\ldots,i-1]&\text{for}\quad i=1,\ldots,K\,,\\
    [i-K,\ldots,K]&\text{for}\quad i=K+1,\ldots,2K\,.
  \end{cases}
\end{align}
In this formula the principal minor of $\mathcal{C}$ corresponding to
the rows and columns $i$ to $j$ is denoted $[i,\ldots,j]$, e.g.\
$[\hphantom{1}]=1$, $[1]=C_{1\,K+1}$ and
$[1,\ldots,K]=\det{\mathcal{C}}$. Furthermore, using the unitarity of
$\mathcal{C}$ we obtain
\begin{align}
  \label{eq:prop-unitary-minors}
  [i+1,\ldots,K]=\overline{[1,\ldots,i]}\det{\mathcal{C}}\,,
\end{align}
where the bar denotes complex conjugation, see e.g.\
\cite{Salaff:1967} and the reference mentioned therein. This identity
can be proven using a block decomposition of $\mathcal{C}$. It turns
out to be of great utility here and in the remainder of the
chapter. We will use it frequently without explicit reference. Next,
using \eqref{eq:spec-redef} and \eqref{eq:spec-perm} the exponents of
the minors in \eqref{eq:grass-int-unitary-integrand} are expressed in
terms of the variables $v_i\in\mathbb{C}$ and $c_i\in\mathbb{Z}$,
\begin{align}
  \begin{aligned}
  1+v^+_K-v^-_1&=1+v_K-v_1+c_1\,,\\
  1+v^+_{K+1}-v^-_2&=1+v_1-v_2-c_1+c_2\,,\\
  &\;\;\vdots\\  
  1+v^+_{2K-1}-v^-_K&=1+v_{K-1}-v_K-c_{K-1}+c_K\,,\\
  1+v^+_{2K}-v^-_{K+1}&=1+v_K-v_1-c_K\,,\\
  1+v^+_1-v^-_{K+2}&=1+v_1-v_2\,,\\
  &\;\;\vdots\\
  1+v^+_{K-1}-v^-_{2K}&=1+v_{K-1}-v_K\,.
  \end{aligned}
\end{align}
These variables obey
\begin{align}
  \label{eq:grass-int-parameters}
  \begin{aligned}
    v_{K+1}&=v_1+n-m-1-c_1\,,&\quad\ldots\,,\quad v_{2K}&=v_K+n-m-1-c_K\,,\\
    c_{K+1}&=-c_1\,,&\quad\ldots\,,\quad c_{2K}&=-c_K\,.
  \end{aligned}
\end{align}
Using the relations obtained here and disregarding the analytic
structure momentarily by combining products of minors with common
\emph{complex} exponents, we rewrite the integrand
\eqref{eq:grass-int-unitary-integrand} as
\begin{align}
  \label{eq:eq:grass-int-unitary-integrand-final}
  \begin{aligned}
  \mathscr{F}(\mathcal{C})^{-1}=&(-1)^{(c_1+\cdots +c_K)(K+1)}(\det\mathcal{C})^{q-s-c_K}\\
  &\cdot
  |[1]|^{2(1+v_1-v_2)}[1]^{c_2-c_1}\cdots
  |[1,\ldots,K-1]|^{2(1+v_{K-1}-v_K)}[1,\ldots,K-1]^{c_K-c_{K-1}}\,.
  \end{aligned}
\end{align}
This function of $\mathcal{C}=(C_{kl})$ is manifestly single-valued as
only non-negative numbers are exponentiated to non-integer
powers. Together with the formal proof of
section~\ref{sec:proof-yang-invar} this shows the Yangian invariance
of \eqref{eq:grass-int-unitary} with the integrand
\eqref{eq:eq:grass-int-unitary-integrand-final}. The integral
\eqref{eq:grass-int-unitary} with
\eqref{eq:eq:grass-int-unitary-integrand-final} is a novel matrix
model that we refer to as \emph{unitary Graßmannian matrix model}. It
generalizes the well established Brezin-Gross-Witten and
Leutwyler-Smilga model in \eqref{eq:red-matrix-fin-inv} by including
principal minors of the unitary matrix $\mathcal{C}$ other than
$\det{\mathcal{C}}$.

We append some remarks. Note the crucial importance of \emph{integer}
representation labels $c_i$ in
\eqref{eq:eq:grass-int-unitary-integrand-final}. Hence we stay within
the class of oscillator representations introduced in
section~\ref{sec:osc-rep}. This is in contrast to previous attempts
\cite{Ferro:2014gca} to find a suitable contour for the Graßmannian
integral \eqref{eq:grassint-amp-def} for deformed $\mathcal{N}=4$ SYM
scattering amplitudes, where the representation labels are typically
\emph{complex} numbers. Let us also remark that the convergence of the
integral \eqref{eq:grass-int-unitary} is not guaranteed because some
minors might vanish. Furthermore, a function of the form of
$\mathscr{F}(\mathcal{C})$ in
\eqref{eq:eq:grass-int-unitary-integrand-final} appears in the
classification of $U(K)$ representations in \S~49 of
\cite{Zhelobenko:1973}. We also observe that the function
\eqref{eq:eq:grass-int-unitary-integrand-final} is not a class
function, i.e.\ it is not invariant under conjugation of $\mathcal{C}$
with an arbitrary unitary matrix. This is arguably the most important
difference to the Leutwyler-Smilga model
\eqref{eq:red-matrix-fin-inv}, where the corresponding function is
just a power of $\det\mathcal{C}$. In particular, it hinders the
direct application of character expansion methods
\cite{Morozov:2009jv,Balantekin:2000vn,Schlittgen:2002tj} for the
evaluation of \eqref{eq:grass-int-unitary}.

For later use we list the explicit form of the integrand
\eqref{eq:eq:grass-int-unitary-integrand-final} for the simplest
invariants. For $|\Psi_{2,1}\rangle$ we have
\begin{align}
  \label{eq:integrand21}
  \mathscr{F}(\mathcal{C})^{-1}=(\det\mathcal{C})^{q-s-c_1}\,.
\end{align}
The integrand of $|\Psi_{4,2}\rangle$ is
\begin{align}
  \label{eq:integrand42}
  \mathscr{F}(\mathcal{C})^{-1}=(-1)^{c_1+c_2}
    (\det{\mathcal{C}})^{q-s-c_2}
    |[1]|^{2(1+v_1-v_2)}[1]^{c_2-c_1}\,.
\end{align}
In case of the invariant $|\Psi_{6,3}\rangle$ we obtain
\begin{align}
  \label{eq:integrand63}
  \begin{aligned}
    \mathscr{F}(\mathcal{C})^{-1}&=(\det\mathcal{C})^{q-s-c_3}
    |[1]|^{2(1+v_1-v_2)}[1]^{c_2-c_1}
    |[1,2]|^{2(1+v_2-v_3)}[1,2]^{c_3-c_2}\\
    &=
    (\det\mathcal{C})^{q-s-c_2}
    |[1]|^{2(1+v_1-v_2)}[1]^{c_2-c_1}
    |\overline{[3]}|^{2(1+v_2-v_3)}\overline{[3]}^{c_3-c_2}\,.
  \end{aligned}
\end{align}

\subsubsection{Parameterization of Unitary Contour}
\label{sec:parameterization}

So far we analyzed the matrix integral \eqref{eq:grass-int-unitary} by
making use of the unitarity of the integration variable
$\mathcal{C}$. In order to eventually evaluate this integral, we
resort to an explicit parameterization of $\mathcal{C}$. In this
section we first briefly discuss a parameterization of the group
$U(K)$ and the corresponding Haar measure. Then we work out the
examples $U(1)$, $U(2)$ and $U(3)$ associated with the simplest
Yangian invariants in some detail.

We formulate the unitary group as the semidirect product
$U(K)=SU(K)\rtimes U(1)$, see e.g.\ \cite{Lee:2003}. Hence,
\begin{align}
  \mathcal{C}=\tilde{\mathcal{C}}
  \left(
  \begin{array}{c:c}
    e^{i\gamma}&0\\[0.3em]
    \hdashline\\[-1.0em]
    0&1_{K-1}\\
  \end{array}
  \right)\,
\end{align}
with $\tilde{\mathcal{C}}\in SU(K)$ and $\gamma\in[0,2\pi]$. We work
with a parameterization of the $SU(K)$ factor in terms of products of
$SU(2)$ matrices, which is already known since the 19th century
\cite{Hurwitz:1897}, see also e.g.\
\cite{Zyczkowski:1994,Guralnik:1985ue,Murnaghan:1962}. The Haar
measure, cf.\ \cite{Knapp:2002,Fuchs:2003}, of the $K^2$-dimensional
group $U(K)$ is obtained from the left- and right-invariant
top-dimensional form
\begin{align}
  \label{eq:haar-general}
  [\D\mathcal{C}]=\chi_K\frac{\D \mathcal{C}}{(\det\mathcal{C})^K}\,,
\end{align}
which already appeared in \eqref{eq:measure-haar}. In slight abuse of
notation we use the symbol $[\D\mathcal{C}]$ for the Haar measure as
well as for the form. The normalization $\chi_K$ is fixed by demanding
$\int_{U(K)}[\D \mathcal{C}]=1$. To evaluate \eqref{eq:haar-general}
below for examples, it is helpful to state the transformation law
\begin{align}
  \begin{aligned}
    \D\mathcal{C}&=
    \D C_{1\,K+1}\wedge\D C_{1\,K+2}\wedge\cdots\wedge\D C_{K\,2K}
    =
    \det\left(\frac{\partial \boldsymbol{C}}{\partial \boldsymbol{\phi}}\right)
    \D\phi_1\wedge\cdots\wedge\D\phi_{K^2}
  \end{aligned}
\end{align}
for a parameterization $\mathcal{C}=\mathcal{C}(\boldsymbol{\phi})$ of
$U(K)$ in terms of variables
$\boldsymbol{\phi}=(\phi_1,\ldots,\phi_{K^2})$. Here we denote
$\boldsymbol{C}=(C_{1\,K+1}, C_{1\,K+2},\ldots,C_{K\,2K})$. Recall the
notation for the components of $\mathcal{C}$ from
\eqref{eq:grassint-matrix}. In order to obtain the Haar measure, we
assume $\D\phi_1\wedge\cdots\wedge\D\phi_{K^2}$ to be positively
oriented and thus replace it by the measure
$\D\phi_1\cdots\D\phi_{K^2}$.

Let us continue by discussing some examples. The formulas stated here
will be employed in subsequent sections for the computation of sample
Yangian invariants. We parameterize $U(1)$ as
\begin{align}
  \label{eq:para-u1}
  \mathcal{C}=C_{12}=e^{i\gamma}\quad\text{with}\quad \gamma\in[0,2\pi]\,.
\end{align}
The Haar measure is given by
\begin{align}
  \label{eq:haar-u1}
  [\D\mathcal{C}]=\chi_1\,i\D\gamma\,
\end{align}
with $\chi_1=-i(2\pi)^{-1}$. The group $U(2)=SU(2)\rtimes U(1)$ is
parameterized as
\begin{align}
  \label{eq:para-u2}
  \begin{aligned}
    \mathcal{C}=
    \begin{pmatrix}
      C_{13}&C_{14}\\
      C_{23}&C_{24}\\
    \end{pmatrix}
    &=
    \begin{pmatrix}
      e^{i\alpha} \cos{\theta}&-e^{i\beta}\sin{\theta}\\
      e^{-i\beta}\sin{\theta}&e^{-i\alpha}\cos{\theta}\\
    \end{pmatrix}
    \begin{pmatrix}
      e^{i\gamma}&0\\
      0&1\\
    \end{pmatrix}\\
    &=  
    \begin{pmatrix}
      e^{i(\gamma+\alpha)} \cos{\theta}&-e^{i\beta}\sin{\theta}\\
      e^{i(\gamma-\beta)}\sin{\theta}&e^{-i\alpha}\cos{\theta}\\
    \end{pmatrix}
  \end{aligned}
\end{align}
with
\begin{align}
  \alpha,\beta,\gamma\in[0,2\pi],\quad
  \theta\in[0,\tfrac{\pi}{2}]\,.
\end{align}
The Haar measure is 
\begin{align}
  \label{eq:haar-u2}
  [\D\mathcal{C}]
  =\chi_2\,i\sin{(2\theta)}
  \D\theta\D\alpha\D\beta\D\gamma\,
\end{align}
with $\chi_2=-i(2\pi)^{-3}$. A parameterization of
$U(3)=SU(3)\rtimes U(1)$ is
\begin{align}
  \label{eq:para-u3}
  \begin{aligned}
    \mathcal{C}=
    \begin{pmatrix}
      C_{14}&C_{15}&C_{16}\\
      C_{24}&C_{25}&C_{26}\\
      C_{34}&C_{35}&C_{36}\\
    \end{pmatrix}
    =
    \phantom{\cdot}&\begin{pmatrix}
      e^{i\alpha_1} \cos{\theta_1}&-e^{i\beta_1}\sin{\theta_1}&0\\
      e^{-i\beta_1}\sin{\theta_1}&e^{-i\alpha_1}\cos{\theta_1}&0\\
      0&0&1\\
    \end{pmatrix}\\
    \cdot&\begin{pmatrix}
      e^{i\alpha_2}\cos{\theta_2}&0&-e^{i\beta_2}\sin{\theta_2}\\
      0&1&0\\
      e^{-i\beta_2}\sin{\theta_2}&0&e^{-i\alpha_2}\cos{\theta_2}\\
    \end{pmatrix}\\
    \cdot&\begin{pmatrix}
      1&0&0\\
      0&e^{i\alpha_3} \cos{\theta_3}&-e^{i\beta_3}\sin{\theta_3}\\
      0&e^{-i\beta_3}\sin{\theta_3}&e^{-i\alpha_3}\cos{\theta_3}\\
    \end{pmatrix}
    \begin{pmatrix}
      e^{i\gamma}&0&0\\
      0&1&0\\
      0&0&1\\
    \end{pmatrix}
    \,.
  \end{aligned}
\end{align}
Here the $SU(3)$ part is given in terms of $SU(2)$ matrices and
\begin{align}
  \alpha_1,\alpha_3,\beta_1,\beta_2,\beta_3,\gamma\in[0,2\pi],\quad
  \alpha_2=0,\quad
  \theta_1,\theta_2,\theta_3\in[0,\tfrac{\pi}{2}]\,.
\end{align}
More explicitly the parameterization \eqref{eq:para-u3} reads
\begin{align}
  \mathcal{C}=
  \begin{pmatrix}
    e^{i(\gamma+\alpha_1)}\cos\theta_1\cos\theta_2&C_{15}&C_{16}\\
    e^{i(\gamma-\beta_1)}\sin\theta_1\cos\theta_2&C_{25}&C_{26}\\
    e^{i(\gamma-\beta_2)}\sin\theta_2&e^{-i\beta_3}\cos\theta_2\sin\theta_3&e^{-i\alpha_3}\cos\theta_2\cos\theta_3\\
  \end{pmatrix}
\end{align}
with
\begin{align}
  \begin{aligned}
    C_{15}&=-e^{i(\beta_1+\alpha_3)}\sin\theta_1\cos\theta_3-e^{i(\alpha_1+\beta_2-\beta_3)}\cos\theta_1\sin\theta_2\sin\theta_3\,,\\
    C_{16}&=e^{i(\beta_1+\beta_3)}\sin\theta_1\sin\theta_3-e^{i(\alpha_1+\beta_2-\alpha_3)}\cos\theta_1\sin\theta_2\cos\theta_3\,,\\
    C_{25}&=e^{i(-\alpha_1+\alpha_3)}\cos\theta_1\cos\theta_3-e^{i(-\beta_1+\beta_2-\beta_3)}\sin\theta_1\sin\theta_2\sin\theta_3\,,\\
    C_{26}&=-e^{i(-\alpha_1+\beta_3)}\cos\theta_1\sin\theta_3-e^{i(-\beta_1+\beta_2-\alpha_3)}\sin\theta_1\sin\theta_2\cos\theta_3\,.
  \end{aligned}
\end{align}
The Haar measure is given by
\begin{align}
  \label{eq:haar-u3}
  \begin{aligned}
    [\D\mathcal{C}]
    =\chi_3\, (\cos\theta_2)^2\sin(2\theta_1)\sin(2\theta_2)\sin(2\theta_3)
    \D\theta_1\D\alpha_1\D\beta_1\D\theta_2\D\beta_2\D\theta_3
    \D\alpha_3\D\beta_3\D\gamma\,
  \end{aligned}
\end{align}
with $\chi_3=2(2\pi)^{-6}$. Let us at this point refer the reader to
appendix~\ref{sec:glue-contours}. There the parameterization of the
$U(3)$ contour in \eqref{eq:para-u3}, which is associated with the
invariant $|\Psi_{6,3}\rangle$, emerges naturally by gluing three
invariants of the type $|\Psi_{4,2}\rangle$, each of which is obtained
from a $U(2)$ integral.

\subsection{Sample Invariants}
\label{sec:sample-invariants}

Currently, there are no efficient techniques available to evaluate our
unitary Graßmannian matrix model~\eqref{eq:grass-int-unitary} with the
integrand \eqref{eq:eq:grass-int-unitary-integrand-final} for the
Yangian invariant $|\Psi_{2K,K}\rangle$ in full generality. In
particular, there is no analogue of the formula
\eqref{eq:ls-integral-bessel}, which applies to the special case where
our model reduces to the Leutwyler-Smilga integral. Thus we study
\eqref{eq:grass-int-unitary} by evaluating it ``by hand'' for the
simplest sample invariants $|\Psi_{2,1}\rangle$, $|\Psi_{4,2}\rangle$
and $|\Psi_{6,3}\rangle$. For these computations we make use of the
parameterizations of the unitary contour and the formulas for the Haar
measure from section~\ref{sec:parameterization}.

\subsubsection{Two-Site Invariant}
\label{sec:sample-inv21}

We evaluate the Graßmannian integral \eqref{eq:grass-int-unitary} for
the two-site invariant $|\Psi_{2,1}\rangle$ with representations of
the non-compact superalgebra $\mathfrak{u}(p+q|r+s)$. With the
integrand \eqref{eq:integrand21}, the parameterization
\eqref{eq:para-u1} of $U(1)$ and the Haar measure \eqref{eq:haar-u1},
we obtain
\begin{align}
  \label{eq:gr-sample-inv21}
  \begin{aligned}
    |\Psi_{2,1}\rangle
    &=2\pi i\;\sum_{\mathclap{\substack{g_{12},h_{12}=0\\g_{12}-h_{12}=q-s-c_1}}}^\infty\;
    \frac{(1\bullet 2)^{g_{12}}}{g_{12}!}\frac{(1\circ 2)^{h_{12}}}{h_{12}!}|0\rangle\\
    &=2\pi i\;
    \frac{I_{q-s-c_1}\big(2\sqrt{(1\bullet 2)(1\circ 2)}\big)}
    {\sqrt{(1\bullet 2)(1\circ 2)}^{\,q-s-c_1}}
    (1\bullet 2)^{q-s-c_1}|0\rangle\,.
  \end{aligned}
\end{align}
To derive this result we treated the $U(1)$ integral in the variable
$\gamma$, cf.\ \eqref{eq:para-u1}, as a complex contour integral in
$e^{i\gamma}$ and applied the residue theorem. Note that the integrand
\eqref{eq:integrand21} of \eqref{eq:grass-int-unitary} for
$|\Psi_{2,1}\rangle$ only contains a factor $\det{\mathcal{C}}$. Thus
\eqref{eq:grass-int-unitary} for $|\Psi_{2,1}\rangle$ is a
Leutwyler-Smilga integral~\eqref{eq:red-matrix-fin-inv} even for
general deformation parameters $v_i^\pm$. We presented the
formula~\eqref{eq:gr-sample-inv21} for $|\Psi_{2,1}\rangle$, up to the
choice of the normalization, already in
section~\ref{sec:ncomp-susy-2-site}. Here we saw how this most simple
non-compact Yangian invariant originates from the general unitary
Graßmannian integral \eqref{eq:grass-int-unitary}.

\subsubsection{Four-Site Invariant}
\label{sec:sample-inv42}

We continue with the evaluation of the integral
\eqref{eq:grass-int-unitary} for the invariant $|\Psi_{4,2}\rangle$ in
case of the algebra $\mathfrak{u}(p,q|r+s)$. Its integrand can be
found in \eqref{eq:integrand42}. We use the parameterization
\eqref{eq:para-u2} of $U(2)$ with the Haar measure
\eqref{eq:haar-u2}. The integrals in the variables
$e^{i\alpha},e^{i\beta}$ and $e^{i\gamma}$ are performed using the
residue theorem. The remaining integral in $\theta$ then reduces to
the Euler beta function,
\begin{align}
  \label{eq:beta-int}
  \text{B}(x,y)
  =2\int_0^{\frac{\pi}{2}}\D\theta(\sin\theta)^{2x-1}(\cos\theta)^{2y-1}
  =\frac{\Gamma(x)\Gamma(y)}{\Gamma(x+y)}
\end{align}
for $\Real x,\Real y>0$, cf.\ \cite{AbramowitzStegun:1964}. This
leads to the invariant
\begin{align}
  \label{eq:gr-sample-inv42}
  \begin{aligned}
    |\Psi_{4,2}\rangle=
    -(-1)^{c_1+c_2}(2\pi i)^3\;
    \,\smash{\sum_{\mathclap{\substack{g_{13},\ldots,g_{24}=0\\h_{13},\ldots,h_{24}=0\\\text{with \eqref{eq:gr-sample-inv42-constraints}}}}}^\infty}\,\quad\quad
    &\frac{(1\bullet 3)^{g_{13}}}{g_{13}!}\frac{(1\bullet 4)^{g_{14}}}{g_{14}!}
    \frac{(2\bullet 3)^{g_{23}}}{g_{23}!}\frac{(2\bullet 4)^{g_{24}}}{g_{24}!}\\
    \cdot\,&\frac{(1\circ 3)^{h_{13}}}{h_{13}!}\frac{(1\circ 4)^{h_{14}}}{h_{14}!}
    \frac{(2\circ 3)^{h_{23}}}{h_{23}!}\frac{(2\circ 4)^{h_{24}}}{h_{24}!}
    |0\rangle\\
    \cdot\,(-1)^{g_{14}+h_{14}}&\text{B}(g_{14}+h_{23}+1,h_{13}+g_{24}-v_1+v_2)\,.
  \end{aligned}
\end{align}
In this formula the summation range is constrained by
\begin{align}
  \label{eq:gr-sample-inv42-constraints}
  \begin{aligned}
    g_{13}-h_{13}+g_{14}-h_{14}&=-c_1+q-s\,,&\quad
    g_{23}-h_{23}+g_{24}-h_{24}&=-c_2+q-s\,,\\
    g_{13}-h_{13}+g_{23}-h_{23}&=\phantom{-}c_3+q-s\,,&\quad
    g_{14}-h_{14}+g_{24}-h_{24}&=\phantom{-}c_4+q-s\,.\\
  \end{aligned}
\end{align}
Furthermore, we have to assume $\Real(v_2-v_1)>0$ in order for the
beta function integral to converge. We displayed the expression
\eqref{eq:gr-sample-inv42} for $|\Psi_{4,2}\rangle$ with a different
normalization already above in section~\ref{sec:ncomp-susy-4-site}
without giving a derivation nor showing its Yangian invariance. These
gaps are filled now. Recall also from this section that the invariant
$|\Psi_{4,2}\rangle$ is of special importance because it is equivalent
to an R-matrix. Moreover, this invariant is the first case where the
unitary Graßmannian integral \eqref{eq:grass-int-unitary} goes beyond
the Leutwyler-Smilga model \eqref{eq:red-matrix-fin-inv} because the
integrand \eqref{eq:integrand42} contains more than just a factor of
$\det\mathcal{C}$.

\subsubsection{Six-Site Invariant}
\label{sec:63-invariant}

For the computation of the invariant $|\Psi_{6,3}\rangle$ from the
Graßmannian integral \eqref{eq:grass-int-unitary} we restrict for
simplicity to the compact algebra $\mathfrak{u}(p)$, i.e.\
$\mathbf{I}_\circ=0$. The integrand is given in
\eqref{eq:integrand63}. We use the parameterization \eqref{eq:para-u3}
of $U(3)$ with the Haar measure \eqref{eq:haar-u3}.  In the compact
case the representation labels satisfy
$c_4=-c_1,c_5=-c_2,c_6=-c_3\geq 0$. The integral is addressed using
the residue theorem for the integration variables $e^{i\alpha_1}$,
$e^{i\alpha_3}$, $e^{i\beta_1}$, $e^{i\beta_2}$, $e^{i\beta_3}$ and
$e^{i\gamma}$. The remaining integrals in $\theta_1$, $\theta_2$ and
$\theta_3$ then reduce to Euler beta functions, cf.\
\eqref{eq:beta-int}. In this way we obtain
\begin{align}
  \label{eq:inv63-comp}
  \begin{aligned}
    |\Psi_{6,3}\rangle=&
    -(2\pi i)^6
    \sum_{\substack{k_{14},k_{15},k_{24},k_{25}=0\\l_{15},l_{16},l_{25}=0}}^{\infty}
    \frac{(1\bullet 4)^{k_{14}}}{k_{14}!}
    \frac{(1\bullet 5)^{k_{15}}}{k_{15}!}
    \frac{(1\bullet 6)^{c_4-k_{14}-k_{15}}}{(c_4-k_{14}-k_{15})!}\\
    &\cdot\frac{(2\bullet 4)^{k_{24}}}{k_{24}!}
    \frac{(2\bullet 5)^{k_{25}}}{k_{25}!}
    \frac{(2\bullet 6)^{c_5-k_{24}-k_{25}}}{(c_5-k_{24}-k_{25})!}\\
    &\cdot\frac{(3\bullet 4)^{c_4-k_{14}-k_{24}}}{(c_4-k_{14}-k_{24})!}
    \frac{(3\bullet 5)^{c_5-k_{15}-k_{25}}}{(c_5-k_{15}-k_{25})!}
    \frac{(3\bullet 6)^{-c_4-c_5+c_6+k_{14}+k_{15}+k_{24}+k_{25}}}{(-c_4-c_5+c_6+k_{14}+k_{15}+k_{24}+k_{25})!}|0\rangle\\
    &\cdot
    \binom{k_{15}}{l_{15}}
    \binom{c_4-k_{14}-k_{15}}{l_{16}}
    \binom{k_{25}}{l_{25}}
    \binom{c_5-k_{24}-k_{25}}{-c_4+c_5+k_{14}+k_{15}-l_{15}+l_{16}-l_{25}}\\
    &\cdot(-1)^{c_5+k_{15}+k_{24}+l_{16}+l_{25}}\text{B}(1+c_4-k_{14}-k_{24},-c_4+c_6+k_{14}+k_{24}-v_1+v_3)\\
    &\cdot\text{B}(1+c_4-k_{14}-k_{15}+l_{15}-l_{16},-c_4+c_5+k_{14}+k_{15}-l_{15}+l_{16}-v_1+v_2)\\
    &\cdot\text{B}(1+c_5-l_{15}-l_{25},-c_5+c_6+l_{15}+l_{25}-v_2+v_3)\,.
  \end{aligned}
\end{align}
Here we expressed some combinatorial factors as binomial
coefficients. Furthermore, we have to assume
$-c_5+c_6>\Real{(v_2-v_3)}$ and $-c_4+c_6>\Real{(v_1-v_3)}$ for the
beta function integrals to converge. Note that the infinite sums
truncate to finite ones due to the factorials. Hence the invariant is
a polynomial in the oscillator contractions $(k\bullet l)$. We checked
the Yangian invariance of \eqref{eq:inv63-comp} also independently of
the proof in section~\ref{sec:proof-yang-invar} using computer algebra
for small values of the representation labels. The complicated
structure of the formula \eqref{eq:inv63-comp} emphasizes the need for
a more efficient method to evaluate the unitary Graßmannian matrix
model \eqref{eq:grass-int-unitary}.  However, this route is not
pursued further in this thesis.

\section{From Oscillators to Spinor Helicity Variables}
\label{sec:osc-spinor}

So far we investigated the unitary Graßmannian matrix model
\eqref{eq:grass-int-unitary} for Yangian invariants with oscillator
representations of $\mathfrak{u}(p,q|m)$. In particular, this includes
representations of the superconformal algebra
$\mathfrak{psu}(2,2|4)$. As we reviewed in the introductory
section~\ref{sec:amplitudes}, tree-level scattering amplitudes of
$\mathcal{N}=4$ SYM are Yangian invariants with certain
representations of this algebra. This raises the question how the
invariants computed by \eqref{eq:grass-int-unitary} are related to
these amplitudes. We address it in the following by applying a change
of basis to the oscillators of $\mathfrak{u}(2,2|4)$ that turns them
into the spinor helicity variables of section~\ref{sec:amplitudes}. In
fact, we implement this basis transformation to spinor helicity-like
variables more generally for $\mathfrak{u}(p,p|m)$.

We proceed in several steps. In section~\ref{sec:bargm-transf} we
introduce the Bargmann transformation. This integral transformation is
known from the one-dimensional quantum mechanical harmonic oscillator.
There it essentially relates the Fock states to the wave functions in
position space. We apply this transformation in
section~\ref{sec:transf-boson-non} to express the generators of the
bosonic $\mathfrak{u}(p,p)$ oscillator representations in terms of
spinor helicity-like variables. In section~\ref{sec:change-basis-lax}
we comment on the resulting form of the Lax operators, that contain
these generators. The Bargmann transformation is applied to the
integrand of the Graßmannian matrix model \eqref{eq:grass-int-unitary}
for $\mathfrak{u}(p,p)$ in
section~\ref{sec:bargm-transf-integrand}. Finally, we work out the
extension of these calculations to the superalgebra
$\mathfrak{u}(p,p|m)$ in section~\ref{sec:transf-ferm-oscill}. This
material provides the necessary groundwork for addressing the
computation of $\mathcal{N}=4$ SYM amplitudes by means of a
Graßmannian integral with a unitary contour in
section~\ref{sec:grassmann-spinor}.

\subsection{Bargmann Transformation}
\label{sec:bargm-transf}

We introduce the Bargmann transformation along the lines of the
original publication \cite{Bargmann:1961gm}. From the outset, we work
in a multi-dimensional setting because it is needed for our
application of the transformation later on. All formulas
straightforwardly reduce to the one-dimensional case, where they
describe the simple harmonic oscillator in quantum mechanics. At times
we employ this example to provide some intuition for key equations.

We start out with a family of bosonic oscillators on a Fock space
obeying
\begin{align}
  \label{eq:barg-barg}
  [\mathbf{A}_{\indnm{A}},\bar{\mathbf{A}}_{\indnm{B}}]=\delta_{\indnm{AB}}\,,\quad 
  {\mathbf{A}_{\indnm{A}}}^\dagger=\bar{\mathbf{A}}_{\indnm{A}}\,,\quad
  \mathbf{A}_{\indnm{A}}|0\rangle=0
\end{align}
with $\indnm{A},\indnm{B}=1,\ldots,n$. Let
$\mathbf{A}=(\mathbf{A}_{\indnm{A}})$ etc.\ denote an $n$-component
column vector. The relations in \eqref{eq:barg-barg} are realized by
the \emph{Bargmann representation}
\begin{align}
  \label{eq:barg-map-b}
  \bar{\mathbf{A}}\mapsto z\,,\quad
  \mathbf{A}\mapsto\partial_z\,,\quad
  |0\rangle\mapsto \Psi_0(z)=1
\end{align}
on the Bargmann space $\mathcal{H}_\text{B}$. This is the Hilbert
space of holomorphic functions of $z\in\mathbb{C}^n$ with the inner
product
\begin{align}
  \label{eq:barg-inner-barg}
  \langle\Psi(z),\Phi(z)\rangle_{\text{B}}=
  \int_{\mathbb{C}^n}
  \frac{\D^{\,n}\!\overline{z}\D^{\,n}\!z}{(2\pi i)^n}
  e^{-\overline{z}^tz}\overline{\Psi(z)}\Phi(z)\,,
\end{align}
where
$(2i)^{-n}\D^{\,n}\!\overline{z}\D^{\,n}\!z=\D^{\,n}\!\Real{z}\D^{\,n}\!\Imag{z}$
is understood as the measure on $\mathbb{R}^{2n}$. In particular, this
inner product implements the reality condition in
\eqref{eq:barg-barg}, i.e.\
${\partial_{z_{\indnm{A}}}}^\dagger=z_{\indnm{A}}$. The Bargmann
representation can be thought of as a concrete realization of the
formal Fock space operators. For recent expositions of this
representation see also e.g.\ \cite{Takhtajan:2008,ZinnJustin:2005},
where it is, however, called ``holomorphic representation''.

In addition, we introduce another family of canonical variables
obeying different reality conditions,
\begin{align}
  \label{eq:barg-sch}
  [\partial_{x_{\indnm{A}}},x_{\indnm{B}}]=\delta_{\indnm{AB}}\,,\quad 
  {\partial_{x_{\indnm{A}}}}^\dagger=-\partial_{x_{\indnm{A}}}\,,\quad
  {x_{\indnm{A}}}^\dagger=x_{\indnm{A}}\,.
\end{align}
These are considered as operators on the Hilbert space
$\mathcal{H}_{\text{Sch}}$ of square integrable functions of the
variable $x\in\mathbb{R}^n$ with the inner product
\begin{align}
  \label{eq:barg-inner-sch}
  \langle\Psi(x),\Phi(x)\rangle_{\text{Sch}}=
  \int_{\mathbb{R}^n}\D^{\,n}\!x\, \overline{\Psi(x)}\Phi(x)\,.
\end{align}
This realization of \eqref{eq:barg-sch} is referred to as
\emph{Schrödinger representation}. For the example of the
one-dimensional harmonic oscillator, this may be interpreted as the
realization in position space.

We observe that by a naive counting the degrees of freedom in
$\mathcal{H}_{\text{B}}$ and $\mathcal{H}_{\text{Sch}}$ do match. A
function $\Psi(z)$ in $\mathcal{H}_{\text{B}}$ depends on $n$ complex
coordinates $z_{\indnm{A}}$ but not on their conjugates
$\overline{z}_{\indnm{A}}$. Similarly, $\Psi(x)$ in
$\mathcal{H}_{\text{Sch}}$ is a function of $n$ real coordinates
$x_{\indnm{A}}$. Thus we want to identify the canonical variables in
$\mathcal{H}_{\text{B}}$ and $\mathcal{H}_{\text{Sch}}$. For this
purpose we make the ansatz
\begin{align}
  \label{eq:barg-rel}
  \partial_z\leftrightarrow A x+B\partial_x\,,\quad
  z\leftrightarrow \overline{A}x-\overline{B}\partial_x\,,
\end{align}
where we allow for $n\times n$ matrices $A,B$ and their complex
conjugates $\overline{A},\overline{B}$. The latter relation is
obtained from the first one by taking the Hilbert space adjoint. For
\eqref{eq:barg-rel} to be compatible with the commutation relations
and reality conditions in \eqref{eq:barg-barg} and \eqref{eq:barg-sch}
we have to impose
\begin{align}
  \label{eq:barg-rel-cond}
  A B^\dagger+B A^\dagger=1_n\,,\quad B A^t=AB^t\,,
\end{align}
where ${}^\dagger$ stands for Hermitian conjugation and ${}^t$ for
transposition of matrices. From now on we concentrate for simplicity
on the special class of solutions of \eqref{eq:barg-rel-cond} where
\begin{align}
  \label{eq:barg-rel-cond-simp}
  2\gamma\, AA^\dagger=1_n\,,\quad
  B=\gamma A\quad
  \text{with}\quad
  \gamma\in\mathbb{R}\,.
\end{align}
Note that this condition can be solved trivially by taking
$A\propto 1_n$, in which case the components of the relations in
\eqref{eq:barg-rel} decouple. The identification \eqref{eq:barg-rel}
of the Hilbert spaces $\mathcal{H}_{\text{B}}$ and
$\mathcal{H}_{\text{Sch}}$ is implemented by the \emph{Bargmann
  transformation}
\begin{align}
  \label{eq:barg-trafo}
  \Psi(z)=\langle\overline{\mathcal{K}(z,x)},\Psi(x)\rangle_{\text{Sch}}\,,\quad
  \Psi(x)=\langle\mathcal{K}(z,x),\Psi(z)\rangle_{\text{B}}\,
\end{align}
with the kernel
\begin{align}
  \label{eq:barg-kernel}
  \mathcal{K}(z,x)=(\pi\gamma)^{-\frac{n}{4}}e^{-\gamma z^tAA^tz-\frac{1}{2\gamma}x^tx+2z^tAx}\,.
\end{align}
This kernel solves the differential equations obtained by imposing
\eqref{eq:barg-rel} on \eqref{eq:barg-trafo},
\begin{align}
  \partial_z\mathcal{K}(z,x)=A(x-\gamma\partial_x)\mathcal{K}(z,x)\,,\quad
  z \mathcal{K}(z,x)=\overline{A}(x+\gamma\partial_x)\mathcal{K}(z,x)\,.
\end{align}
The prefactor in \eqref{eq:barg-kernel} is fixed by demanding that the
transformation \eqref{eq:barg-trafo} preserves the norm of the vacuum
state, $\|\Psi_0(z)\|_{\text{B}}=\|\Psi_0(x)\|_{\text{Sch}}=1$, where
\begin{align}
  \label{eq:barg-vac}
  \Psi_0(x)=(\pi \gamma)^{-\frac{n}{4}}e^{-\frac{1}{2\gamma}x^tx}\,.
\end{align}
The Bargmann transformation is unitary, i.e.\ 
\begin{align}
  \label{eq:barg-unitary}
  \langle\Psi(z),\Phi(z)\rangle_{\text{B}}=
  \langle\Psi(x),\Phi(x)\rangle_{\text{Sch}}\,,
\end{align}
because an orthonormal basis of both Hilbert spaces can be built by
acting with the canonical variables on the vacuum. In the example of
the one-dimensional quantum mechanical oscillator, a Bargmann
transformation like \eqref{eq:barg-trafo} implements the change of
basis between the Fock (or rather Bargmann) and the position space
(Schrödinger) representation.

\subsection{Transformation of Bosonic Non-Compact Generators}
\label{sec:transf-boson-non}

Here we express the generators of the $\mathfrak{u}(p,p)$ oscillator
representations $\oscrep_c$ and $\bar{\oscrep}_c$ from
section~\ref{sec:osc-rep} in terms of spinor helicity-like
variables. An analogous calculation in the special case of the
conformal algebra $\mathfrak{su}(2,2)$ can be found in
\cite{Stoyanov:1968tn}, see also \cite{Mack:1969dg}.\footnote{The
  so-called ladder representations of the conformal algebra
  $\mathfrak{su}(2,2)$ in terms of spinor helicity variables from
  \cite{Mack:1969dg} can be exponentiated to representations of the
  Lie group $SU(2,2)$. This gives rise to transformations laws under
  \emph{finite} conformal transformations \cite{Post:1976}. The
  resulting formulas are also known in the mathematical literature,
  see e.g.\ \cite{Gross:1972,Jakobsen:1977}. Our discussion here is
  confined to the \emph{infinitesimal} Lie algebra representations.}

We identify the oscillators $\bar{\mathbf{A}}_{\indnm{A}}$ of the
previous section, and therefore the Bargmann variables
$z_{\indnm{A}}$, with those in the generators of the
$\mathfrak{u}(p,p)$ representations from
section~\ref{sec:osc-rep}. Thus we choose the number of oscillators in
\eqref{eq:barg-barg} to be $n=2p$. Likewise, we want to relate the
Schrödinger variables $x_{\indnm{A}}$ to analogues of the spinor
helicity variables that we employed for the description of
$\mathcal{N}=4$ SYM amplitudes in section~\ref{sec:amplitudes}. Recall
that those spinor helicity variables are complex. Thus to establish
the relation we have to introduce complex coordinates in
$\mathbb{R}^{2p}$ by
\begin{align}
  \left(
  \begin{array}{c}
    \sigma_\alpha\\[0.3em]
    \hdashline\\[-1.0em]
    \overline{\sigma}_\alpha
  \end{array}
  \right)
  =  E
  \left(
  \begin{array}{c}
    x_{\indssub{A}}\\[0.3em]
    \hdashline\\[-1.0em]
    x_{\dot{\indssub{A}}}
  \end{array}
  \right)
  \,,\quad
  \left(
  \begin{array}{c}
    \partial_{\sigma_\alpha}\\[0.3em]
    \hdashline\\[-1.0em]
    \partial_{\overline{\sigma}_\alpha}
  \end{array}
  \right)
  =\frac{1}{2}\overline{E}
  \left(
  \begin{array}{c}
    \partial_{x_{\indssub{A}}}\\[0.3em]
    \hdashline\\[-1.0em]
    \partial_{x_{\dot{\indssub{A}}}}
  \end{array}
  \right)
  \,,\quad\text{with}\quad
  E=\left(
  \begin{array}{c:c}
    1_p&i 1_p\\[0.3em]
    \hdashline\\[-1.0em]
    1_p&-i 1_p
  \end{array}
  \right)
  \,.
\end{align}
Here we split $x=(x_{\indnm{A}})\in\mathbb{R}^{2p}$ into its
components $x_{\indssub{A}}$ with $\indssub{A}=1,\ldots,p$ and
$x_{\dot{\indssub{A}}}$ with $\dot{\indssub{A}}=p+1,\ldots,2p$. The
new coordinates are $\sigma=(\sigma_\alpha)\in\mathbb{C}^p$, i.e.\
$\alpha=1,\ldots,p$, and its complex conjugate
$\overline{\sigma}$. See also \eqref{eq:osc-split} and the text before
and after that equation for explanations on the index
ranges. Employing these variables, the properties \eqref{eq:barg-sch}
of the operators in $\mathcal{H}_{\text{Sch}}$ read
\begin{align}
  [\partial_{\sigma_\alpha},\sigma_\beta]=\delta_{\alpha\beta}\,,\quad
  {\sigma_\alpha}^\dagger=\overline{\sigma}_\alpha\,,\quad
  {\partial_{\sigma_\alpha}}^\dagger=-\partial_{\overline{\sigma}_\alpha}\,.
\end{align}
The inner product \eqref{eq:barg-inner-sch} becomes
\begin{align}
  \label{eq:barg-inner-sch-comp}
  \langle\Psi(\sigma,\overline{\sigma}),\Phi(\sigma,\overline{\sigma})\rangle_{\text{Sch}}=
  \int_{\mathbb{C}^p}
  \frac{\D^{\,p}\!\overline{\sigma}\D^{\,p}\!\sigma}{(2i)^p}\,
  \overline{\Psi(\sigma,\overline{\sigma})}\Phi(\sigma,\overline{\sigma})\,,
\end{align}
where the measure on $\mathbb{C}^p$ is defined as in the context of
\eqref{eq:barg-inner-barg}. Furthermore, we select the particular
solution $A=\frac{1}{\sqrt{2}}\overline{E}$ and $\gamma=\frac{1}{2}$
of \eqref{eq:barg-rel-cond-simp}. This transforms the relation between
the operators in $\mathcal{H}_{\text{B}}$ and
$\mathcal{H}_{\text{Sch}}$ from \eqref{eq:barg-rel} and the vacuum
state in \eqref{eq:barg-vac} into
\begin{align}
  \label{eq:barg-rel-spin-hel}
  \begin{gathered}
    \left(
      \begin{array}{c}
        z_{\indssub{A}}\\[0.3em]
        \hdashline\\[-1.0em]
        z_{\dot{\indssub{A}}}
      \end{array}
    \right)
    \leftrightarrow\frac{1}{\sqrt{2}}
    \left(
      \begin{array}{c}
        \sigma_\alpha-\partial_{\overline{\sigma}_\alpha}\\[0.3em]
        \hdashline\\[-1.0em]
        \overline{\sigma}_\alpha-\partial_{\sigma_\alpha}
      \end{array}
    \right)\,,
    \quad
    \left(
      \begin{array}{c}
        \partial_{z_{\indssub{A}}}\\[0.3em]
        \hdashline\\[-1.0em]
        \partial_{z_{\dot{\indssub{A}}}}
      \end{array}
    \right)
    \leftrightarrow\frac{1}{\sqrt{2}}
    \left(
      \begin{array}{c}
        \overline{\sigma}_\alpha+\partial_{\sigma_\alpha}\\[0.3em]
        \hdashline\\[-1.0em]
        \sigma_\alpha+\partial_{\overline{\sigma}_\alpha}
      \end{array}
    \right)\,,\\
    \Psi_0(\sigma,\overline{\sigma})=\sqrt{\frac{2}{\pi}}^{\,p}e^{-\overline{\sigma}^t\sigma}\,.
  \end{gathered}
\end{align}
The Bargmann transformation \eqref{eq:barg-trafo} mapping
$\mathcal{H}_{\text{B}}\to\mathcal{H}_{\text{Sch}}$ becomes explicitly
\begin{align}
  \label{eq:barg-trafo-ordinary}
  \Psi(\sigma,\overline{\sigma})=
  \sqrt{\frac{2}{\pi}}^{\,p}
  e^{-\overline{\sigma}^t\sigma}
  \int_{\mathbb{C}^{2p}}
  \frac{\D^{\,2p}\!\overline{z}\D^{\,2p}\!z}{(2\pi i)^{2p}}
  e^{-\overline{\dt{z}}^t \dt{z}
    -\overline{\mathring{z}}^t \mathring{z}
    -\overline{\mathring{z}}^t\overline{\dt{z}}
    +\sqrt{2}(\overline{\dt{z}}^t\sigma+\overline{\mathring{z}}^t\overline{\sigma})}
  \Psi(z)\,,
\end{align}
where we defined
$\dt{z}=(z_\indssub{A}),\mathring{z}=(z_{\dot{\indssub{A}}})\in\mathbb{C}^p$.
The bullet and the circle in this notation are in analogy to those of
the oscillator contractions in \eqref{eq:bullets-nc-recall}. The
integral transformation \eqref{eq:barg-trafo-ordinary} will we be
crucial in section~\ref{sec:bargm-transf-integrand} because it allows
us to express the unitary Graßmannian matrix model
\eqref{eq:grass-int-unitary} in terms of spinor helicity-like
variables.

We turn our attention to the generators of the ``ordinary'' oscillator
representation $\oscrep_c$ and the ``dual'' one $\bar{\oscrep}_c$,
whose Fock space realizations can be found in \eqref{eq:gen-ordinary}
and \eqref{eq:gen-dual}, respectively.  For their realization in the
Bargmann space $\mathcal{H}_{\text{B}}$, we use the same symbol as in
the Fock space and denote them by $\mathbf{J}_{\indnm{AB}}$ at
ordinary and by $\bar{\mathbf{J}}_{\indnm{AB}}$ at dual sites. In
$\mathcal{H}_{\text{Sch}}$ we write $\mathfrak{J}_{\indnm{AB}}$ and
$\bar{\mathfrak{J}}_{\indnm{AB}}$, respectively. To match with the
spinor helicity variables later on, we have to redefine the variables
for $\mathcal{H}_{\text{Sch}}$ once more depending on the type of site
by introducing
\begin{align}
  \label{eq:barg-rename}
  \lambda=
  \begin{cases}
    \sigma\quad\text{for ordinary sites}\,,\\
    -\overline{\sigma}\quad\text{for dual sites}\,.
  \end{cases}
\end{align}
From \eqref{eq:barg-rel-spin-hel} we then obtain
\begin{align}
  \begin{aligned}
    \label{eq:barg-gen}
    (\mathbf{J}_{\indnm{AB}})=
    \left(
      \begin{array}{c:c}
        z_{\indssub{A}}\partial_{z_{\indssub{B}}}&z_{\indssub{A}} z_{\dot{\indssub{B}}}\\[0.3em]
        \hdashline\\[-1.0em]
        -\partial_{z_{\dot{\indssub{A}}}}\partial_{z_{\indssub{B}}}&-\partial_{z_{\dot{\indssub{A}}}}z_{\dot{\indssub{B}}}
      \end{array}
    \right)
    \quad
    &\leftrightarrow
    \quad
    (\mathfrak{J}_{\indnm{AB}})
    =
    D
    \left(
      \begin{array}{c:c}
      \lambda_\alpha\partial_{\lambda_\beta}&\lambda_\alpha\overline{\lambda}_\beta\\[0.3em]
        \hdashline\\[-1.0em]
      -\partial_{\overline{\lambda}_\alpha}\partial_{\lambda_\beta}&-\partial_{\overline{\lambda}_\alpha}\overline{\lambda}_\beta
      \end{array}
    \right)\,
    D^{-1}\,,\\
    (\bar{\mathbf{J}}_{\indnm{AB}})=
    \left(
      \begin{array}{c:c}
      -z_{\indssub{B}}\partial_{z_{\indssub{A}}}&-\partial_{z_{\dot{\indssub{B}}}}\partial_{z_{\indssub{A}}}\\[0.3em]
        \hdashline\\[-1.0em]
      z_{\indssub{B}} z_{\dot{\indssub{A}}}&\partial_{z_{\dot{\indssub{B}}}}z_{\dot{\indssub{A}}}
      \end{array}
    \right)
    \quad
    &\leftrightarrow
    \quad
    (\bar{\mathfrak{J}}_{\indnm{AB}})
    =
    D
    \left(
      \begin{array}{c:c}
      \partial_{\lambda_\beta}\lambda_\alpha&-\overline{\lambda}_\beta\lambda_\alpha\\[0.3em]
        \hdashline\\[-1.0em]
      \partial_{\lambda_\beta}\partial_{\overline{\lambda}_\alpha}&-\overline{\lambda}_\beta\partial_{\overline{\lambda}_\alpha}
      \end{array}
    \right)
    D^{-1}
  \end{aligned}
\end{align}
with
\begin{align}
  D=
  \left(
  \begin{array}{c:c}
    1_p&1_p\\[0.3em]
    \hdashline\\[-1.0em]
    -1_p&1_p
  \end{array}
          \right)\,.
\end{align}
Notice that in $\mathcal{H}_{\text{B}}$ the form of the
$\mathfrak{u}(p,p)$ generators $\mathbf{J}_{\indnm{AB}}$ at ordinary
sites and $\bar{\mathbf{J}}_{\indnm{AB}}$ at dual sites differs
``considerably''. For example the upper right block contains two
coordinates for the former generators and two derivatives for the
latter. In contrast, the generators look ``almost alike'' in
$\mathcal{H}_{\text{Sch}}$,
\begin{align}
  \label{eq:barg-gen-replace}
  \left.\bar{\mathfrak{J}}_{\indnm{AB}}\right|_{(\lambda,\overline{\lambda})\mapsto(\lambda,-\overline{\lambda})}
  =
  \mathfrak{J}_{\indnm{AB}}+\delta_{\indnm{AB}}\,.
\end{align}
The central elements \eqref{eq:central-ord} and
\eqref{eq:central-dual} that function as representation labels become,
respectively,
\begin{align}
  \label{eq:barg-replabel}
  \begin{aligned}
    \mathbf{C}=\tr(\mathbf{J}_{\indnm{AB}})
    \leftrightarrow
    \mathfrak{C}=\sum_{\alpha=1}^p
    (\lambda_\alpha\partial_{\lambda_\alpha}-\overline{\lambda}_\alpha\partial_{\overline{\lambda}_\alpha})-p\,,\\
    \bar{\mathbf{C}}=\tr(\bar{\mathbf{J}}_{\indnm{AB}})
    \leftrightarrow
    \bar{\mathfrak{C}}=\sum_{\alpha=1}^p
    (\lambda_\alpha\partial_{\lambda_\alpha}-\overline{\lambda}_\alpha\partial_{\overline{\lambda}_\alpha})+p\,.
  \end{aligned}
\end{align}

At this point we can compare the $\mathfrak{u}(2,2)$ case of the
generators in \eqref{eq:barg-gen} with the bosonic part of the
generators \eqref{eq:sym-gen-gl44} for scattering amplitudes, which
are expressed in terms of spinor helicity variables. We disregard the
similarity transformation with the matrix $D$ in \eqref{eq:barg-gen}
for this comparison, see, however, section~\ref{sec:change-basis-lax}
below. Under these premises the generators $\mathfrak{J}_{\indnm{AB}}$
of the representation $\oscrep_c$ in \eqref{eq:barg-gen} agree with
those in \eqref{eq:sym-gen-gl44} after setting
$(\tilde{\lambda}_{\dot{\alpha}})=+(\overline{\lambda}_\alpha)$.  The
generators $\bar{\mathfrak{J}}_{\indnm{AB}}$ of $\bar{\oscrep}_c$ in
\eqref{eq:barg-gen} match those in \eqref{eq:sym-gen-gl44} with
$(\tilde{\lambda}_{\dot{\alpha}})=-(\overline{\lambda}_\alpha)$ up to
a shift as in \eqref{eq:barg-gen-replace}. We recall from the
definition of the spinor helicity variables around
\eqref{eq:spinors-real} that the sign in the relation
$\tilde{\lambda}=\pm\overline{\lambda}$ determines the sign of the
energy. Therefore the ``ordinary'' oscillator representations
$\oscrep_c$ correspond to positive energies.  Analogously
representations of the ``dual'' class $\bar{\oscrep}_c$ are associated
with negative energies. This explains why in
section~\ref{sec:amplitudes} seemingly only the \emph{one} type of
generators \eqref{eq:sym-gen-gl44} appears at all legs of the
scattering amplitudes, whereas the \emph{two} types of representations
$\oscrep_c$ and $\bar{\oscrep}_c$ are omnipresent in the main part of
this thesis.

Let us add a comment about the just mentioned shift of the generators
at the dual sites. It is the reason why we were only able to show the
invariance of the amplitudes under a \emph{special} linear Lie
superalgebra, and the associated Yangian, in the introductory
section~\ref{sec:amplitudes}. In the main part of this thesis we can
always work with \emph{general} linear Lie (super)algebras because our
definition of the dual representation properly incorporates this
shift.  On a different note, in \eqref{eq:spinors-hel-op} we defined
the operator
$\mathfrak{h}=\frac{1}{2}\sum_{\alpha=1}^2(\overline{\lambda}_\alpha\partial_{\overline{\lambda}_\alpha}-\lambda_\alpha\partial_{\lambda_\alpha})$
that measures the helicity $h$ of a particle. For the gluon amplitudes
discussed in section~\ref{sec:gluon-amp} it may take the values
$h=\pm 1$. Using \eqref{eq:barg-replabel} we translate $h$ into the
$\mathfrak{u}(2,2)$ oscillator representation label $c$, which is the
eigenvalue of the central elements in that equation,
\begin{align}
  \label{eq:hel-rep-label}
  \begin{aligned}
    &h=+1\Leftrightarrow c=-4\,,\quad&
    &h=-1\Leftrightarrow c=\phantom{-}0\quad&
    &\text{for}\quad \oscrep_c\,,\\
    &h=+1\Leftrightarrow c=\phantom{+}0\,,\quad&
    &h=-1\Leftrightarrow c=+4\quad&
    &\text{for}\quad \bar{\oscrep}_c\,.\\
  \end{aligned}
\end{align}

\subsection{Change of Basis in Lax Operators}
\label{sec:change-basis-lax}

A quick calculation shows that the map
$J_{\indnm{AB}}\mapsto\tilde{J}_{\indnm{AB}}$ defined by
\begin{align}
  \label{eq:auto-sim}
  (J_{\indnm{AB}})=D (\tilde{J}_{\indnm{AB}})D^{-1}\,,\quad
  \text{i.e.}\quad
  J_{\indnm{AB}}=\sum_{\indnm{C},\indnm{D}}D_{\indnm{AC}}\tilde{J}_{\indnm{CD}}D^{-1}_{\indnm{DB}}\,
\end{align}
with an even $n|m\times n|m$ supermatrix $D$ is an automorphism of the
$\mathfrak{gl}(n|m)$ superalgebra \eqref{eq:gl-superalg}. This allows
us to reformulate the Lax operator \eqref{eq:yangian-def-lax} as
\begin{align}
  \begin{aligned}
    R_{\square\,\mathcal{V}}(u-v)&=f_{\mathcal{V}}(u-v)\left(1+(u-v)^{-1}\sum_{\indnm{A},\indnm{B}}\elemm_{\indnm{AB}}J_{\indnm{BA}}(-1)^{|\indnm{B}|}\right)\\
    &=f_{\mathcal{V}}(u-v)\left(1+(u-v)^{-1}\sum_{\indnm{C},\indnm{D}}\tilde{\elemm}_{\indnm{DC}}\tilde{J}_{\indnm{CD}}(-1)^{|\indnm{C}|}\right)\,,
  \end{aligned}
\end{align}
where we introduced
\begin{align}
  \tilde{\elemm}_{\indnm{DC}}=\sum_{\indnm{A},\indnm{B}}D^{-1}_{\indnm{DA}}\elemm_{\indnm{AB}}D_{\indnm{BC}}=(D^{-1})^te_{\indnm{DC}}D^{t}\,.
\end{align}
We can apply this observation to express the matrix elements the
monodromy \eqref{eq:yangian-mono-spinchain} in the basis
$\tilde{\elemm}_{\indnm{AB}}$ instead of
$\elemm_{\indnm{AB}}$. Therefore the similarity transformation
\eqref{eq:auto-sim} of the $\mathfrak{gl}(n|m)$ generators can be
absorbed in a redefinition of the Yangian generators
\eqref{eq:yangian-mono}. This justifies the negligence of such a
transformation towards the end of the previous section, where we
compared the generators $\mathfrak{J}_{\indnm{AB}}$ and
$\bar{\mathfrak{J}}_{\indnm{AB}}$ in \eqref{eq:barg-gen} for the
$\mathfrak{u}(2,2)$ case with the bosonic part of those for amplitudes
in \eqref{eq:sym-gen-gl44}. We will refer to the superalgebra case of
the observation presented here later in
section~\ref{sec:transf-ferm-oscill}.

\subsection{Transformation of Bosonic Graßmannian Integrand}
\label{sec:bargm-transf-integrand}

The map in \eqref{eq:barg-rel-spin-hel} transforms the generators of
the $\mathfrak{u}(p,p)$ oscillator representations into spinor
helicity-like variables. Let us now apply the appendant Bargmann
transformation \eqref{eq:barg-trafo-ordinary} to the unitary
Graßmannian matrix model~\eqref{eq:grass-int-unitary} in order to
transform it into those variables. We can focus on the transformation
of the exponential function in the integrand of
\eqref{eq:grass-int-unitary} because it is the only part containing
oscillators. In essence, the Bargmann transformation of this
exponential function reduces to a multi-dimensional Gaußian
integral. The evaluation of this integral yields a delta function of
the spinor helicity-like variables. In the following paragraph we
state this result in detail. Its proof occupies the rest of this
section.

We concentrate on the oscillator-dependent part of the integrand in
the Graßmannian matrix model \eqref{eq:grass-int-unitary} for
$\mathfrak{u}(p,p)$, which we denote by $|\Phi\rangle$. To be able to
apply the Bargmann transformation, we first have to realize it in the
space $\mathcal{H}_{\text{B}}$ using the replacement
\eqref{eq:barg-map-b},
\begin{align}
  \label{eq:barg-inv-matrix}
  \begin{aligned}
    |\Phi\rangle=e^{\tr(\mathcal{C}\mathbf{I}_\bullet^t+\mathbf{I}_\circ \mathcal{C}^{\dagger})}|0\rangle
    \mapsto
    \Phi(\boldsymbol{z})\,,
  \end{aligned}
\end{align}
where $\boldsymbol{z}=(z_{\indnm{A}}^i)$ with $i=1,\ldots,N=2K$ and
$\indnm{A}=1,\ldots,2p$ collectively denotes all complex Bargmann
variables. The functional forms of $|\Phi\rangle$ and
$\Phi(\boldsymbol{z})$ are identical. Only the oscillator-valued
entries of the matrices $\mathbf{I}_\bullet$ and $\mathbf{I}_\circ$
defined in \eqref{eq:osc-matrix} and \eqref{eq:bullets-nc-recall} get
replaced by
\begin{align}
  (k\bullet l)
  \mapsto
  \sum_{\indssub{A}=1}^pz^l_{\indssub{A}}z^k_{\indssub{A}}\,,\quad
  (k\circ l)
  \mapsto
  \sum_{\dot{\indssub{A}}=p+1}^{2p}z^l_{\dot{\indssub{A}}}z^k_{\dot{\indssub{A}}}\,,
\end{align}
and, furthermore, $|0\rangle\mapsto 1$. Here $k=1,\ldots,K$ refers to
a dual site and $l=K+1,\ldots,2K$ to an ordinary one. The Bargmann
transformation \eqref{eq:barg-trafo-ordinary} can now be applied to
$\Phi(\boldsymbol{z})$ in \eqref{eq:barg-inv-matrix}. After relabeling
the variables according to \eqref{eq:barg-rename}, this yields the
expression of the Graßmannian integrand in $\mathcal{H}_{\text{Sch}}$,
\begin{align}
  \label{eq:barg-trafo-integrand}
  \Phi(\boldsymbol{z})
  \leftrightarrow
  \Phi(\boldsymbol{\lambda},\overline{\boldsymbol{\lambda}})
  =\delta_{\mathbb{C}}^{pK}(\boldsymbol{\lambda}^{\text{d}}+\mathcal{C}\boldsymbol{\lambda}^{\text{o}})
  =\delta_{\mathbb{C}}^{pK}(C\boldsymbol{\lambda})\,.
\end{align}
Here the spinor helicity-like variables at the dual and ordinary sites
are arranged in the $K\times p$ matrices
$\boldsymbol{\lambda}^{\text{d}}$ and
$\boldsymbol{\lambda}^{\text{o}}$, respectively. The $2K\times p$
matrix $\boldsymbol{\lambda}$ is built from these two matrices,
\begin{align}
\label{eq:barg-trafo-mult-sch}
  \boldsymbol{\lambda}=
  \left(
  \begin{array}{c}
    \boldsymbol{\lambda}^{\text{d}}\\[0.3em]
    \hdashline\\[-1.0em]
    \boldsymbol{\lambda}^{\text{o}}
  \end{array}
  \right)
  \,,\quad
  \boldsymbol{\lambda}^{\text{d}}=
  \begin{pmatrix}
    \lambda^1_1&\cdots&\lambda_p^1\\
    \vdots&&\vdots\\
    \lambda^K_1&\cdots&\lambda_p^K\\
  \end{pmatrix}
  \,,\quad
  \boldsymbol{\lambda}^{\text{o}}=
  \begin{pmatrix}
    \lambda^{K+1}_1&\cdots&\lambda_p^{K+1}\\
    \vdots&&\vdots\\
    \lambda^{2K}_1&\cdots&\lambda_p^{2K}\\
  \end{pmatrix}\,.
\end{align}
The $K\times 2K$ matrix $C$ is an element of the Graßmannian
$\text{Gr}(2K,K)$ and it contains the unitary $K\times K$ block
$\mathcal{C}$, recall \eqref{eq:grassint-matrix}. Notice that the
arguments of the delta functions in \eqref{eq:barg-trafo-integrand}
are complex as this applies to the entries of $\boldsymbol{\lambda}$
and $C$. A complex delta function is defined as the product of the
delta function for the real part of the argument times that for the
imaginary part, see also \eqref{eq:proof-complex-delta} below.
Equation~\eqref{eq:barg-trafo-integrand} contains the form of the
Graßmannian integrand in spinor helicity-like variables that we
proclaimed already in the introductory paragraph.

Let us set out to prove the result of the Bargmann transformation
presented in \eqref{eq:barg-trafo-integrand}. To begin with, we
reformulate the r.h.s.\ of \eqref{eq:barg-inv-matrix} as
\begin{align}
  \label{eq:phi-z-new}
  \Phi(\boldsymbol{z})
  =
  \prod_{\alpha=1}^p e^{\frac{1}{2}\boldsymbol{z}_\alpha^t\mathtt{C}\, \boldsymbol{z}_\alpha}\,,
\end{align}
where we introduced $\boldsymbol{z}_\alpha\in\mathbb{C}^{4K}$ and a $4K\times 4K$
matrix $\mathtt{C}$ presented in terms of its $K\times K$ blocks,
\begin{align}
  \boldsymbol{z}_\alpha=
  \left(
  \begin{array}{c}
    \dt{\boldsymbol{z}}_\alpha^{\text{d}}\\[0.3em]
    \hdashline\\[-1.0em]
    \dt{\boldsymbol{z}}_\alpha^{\text{o}}\\[0.3em]
    \hdashline\\[-1.0em]
    \mathring{\boldsymbol{z}}_\alpha^{\text{d}}\\[0.3em]
    \hdashline\\[-1.0em]
    \mathring{\boldsymbol{z}}_\alpha^{\text{o}}
  \end{array}
  \right)
  =
  \left(
  \begin{array}{c}
    (z_\alpha^k)\\[0.3em]
    \hdashline\\[-1.0em]
    (z_\alpha^{l})\\[0.3em]
    \hdashline\\[-1.0em]
    (z_{\alpha+p}^k)\\[0.3em]
    \hdashline\\[-1.0em]
    (z_{\alpha+p}^{l})
  \end{array}
  \right)\,,
  \quad
  \mathtt{C}=
  \left(
  \begin{array}{c:c:c:c}
    0&\mathcal{C}&0&0\\[0.3em]
    \hdashline\\[-1.0em]
    \mathcal{C}^t&0&0&0\\[0.3em]
    \hdashline\\[-1.0em]
    0&0&0&(\mathcal{C}^{\dagger})^t\\[0.3em]
    \hdashline\\[-1.0em]
    0&0&\mathcal{C}^{\dagger}&0
  \end{array}
  \right)\,.
\end{align}
Let us explain our notation. Here
$\dt{\boldsymbol{z}}_\alpha^{\text{d}}\in\mathbb{C}^K$ contains the
variables $z_\alpha^k$ at dual sites with $k=1,\ldots,K$ and
$\dt{\boldsymbol{z}}_\alpha^{\text{o}}\in\mathbb{C}^K$ is built from
$z_\alpha^l$ at ordinary sites with $l=K+1,\ldots,2K$ etc. We apply
the Bargmann transformation \eqref{eq:barg-trafo-ordinary} to all
sites of $\Phi(\boldsymbol{z})$ in \eqref{eq:phi-z-new}. This yields
\newenvironment{smallarray}[1] {\null\,\vcenter\bgroup\scriptsize
  \renewcommand{\arraystretch}{0.7}%
  \arraycolsep=.13885em
  \hbox\bgroup$\array{@{}#1@{}}} {\endarray$\egroup\egroup\,\null}
\begin{align}
  \label{eq:barg-trafo-fact}
  \begin{aligned}
    \Phi(\boldsymbol{\sigma},\overline{\boldsymbol{\sigma}})
    &=\prod_{\alpha=1}^p\Phi_\alpha(\boldsymbol{\sigma}_\alpha,\overline{\boldsymbol{\sigma}}_\alpha)\\
    &=\prod_{\alpha=1}^p
    \Bigg(\frac{2}{\pi}\Bigg)^K
    e^{-\overline{\boldsymbol{\sigma}}_\alpha^t\boldsymbol{\sigma}_\alpha}
    \int_{\mathbb{C}^{4K}}
    \frac{\D^{\,4K}\!\overline{\boldsymbol{z}}_\alpha\D^{\,4K}\! \boldsymbol{z}_\alpha}{(2\pi i)^{4K}}\,
    e^{\frac{1}{2}
      \left(
      \begin{smallarray}{c}
        \boldsymbol{z}_\alpha\\[0.1em]
        \hdashline\\[-0.6em]
        \overline{\boldsymbol{z}}_\alpha
      \end{smallarray}
      \right)^t
      \mathtt{H}
      \left(
      \begin{smallarray}{c}
        \boldsymbol{z}_\alpha\\[0.1em]
        \hdashline\\[-0.6em]
        \overline{\boldsymbol{z}}_\alpha
      \end{smallarray}
      \right)
      +\sqrt{2}\;
      \overline{\boldsymbol{z}}_\alpha^t
      \smash{\left(\!
      \begin{smallarray}{c}
        \boldsymbol{\sigma}_\alpha\\[0.1em]
        \hdashline\\[-0.6em]
        \overline{\boldsymbol{\sigma}}_\alpha
      \end{smallarray}
      \!\right)}
    }\,,
  \end{aligned}
\end{align}
where
\begin{align}
  \label{eq:barg-trafo-sigma-h}
  \boldsymbol{\sigma}_\alpha=
  \left(
  \begin{array}{c}
    \boldsymbol{\sigma}^{\text{d}}_\alpha\\[0.3em]
    \hdashline\\[-1.0em]
    \boldsymbol{\sigma}^{\text{o}}_\alpha
  \end{array}
  \right)
  =
  \left(
  \begin{array}{c}
    (\sigma_\alpha^k)\\[0.3em]
    \hdashline\\[-1.0em]
    (\sigma_\alpha^l)
  \end{array}
  \right)\,,\quad
  \mathtt{H}=
  \left(
  \begin{array}{c:c:c:c:c:c:c:c}
    0&\mathcal{C}&0&0&-1&0&0&0\\[0.3em]
    \hdashline\\[-1.0em]  
    \mathcal{C}^t&0&0&0&0&-1&0&0\\[0.3em]
    \hdashline\\[-1.0em]
    0&0&0&(\mathcal{C}^{\dagger})^t&0&0&-1&0\\[0.3em]
    \hdashline\\[-1.0em]
    0&0&\mathcal{C}^{\dagger}&0&0&0&0&-1\\[0.3em]
    \hdashline\\[-1.0em]
    -1&0&0&0&0&0&-1&0\\[0.3em]
    \hdashline\\[-1.0em]
    0&-1&0&0&0&0&0&-1\\[0.3em]
    \hdashline\\[-1.0em]
    0&0&-1&0&-1&0&0&0\\[0.3em]
    \hdashline\\[-1.0em]
    0&0&0&-1&0&-1&0&0
  \end{array}
  \right)\,.
\end{align}
The variables $\boldsymbol{\sigma}^{\text{d}}_\alpha$ and
$\boldsymbol{\sigma}^{\text{o}}_\alpha$ are in $\mathbb{C}^K$. In
block matrices we abbreviate the unit matrix $1_K$ by
$1$. Furthermore, $0$ can represent a $K\times K$ block of zeros or a
$K$-dimensional null vector. We observe a factorization the Bargmann
transformation of $\Phi(\boldsymbol{z})$ into $p$ Gaußian integrals of
$4K$ complex dimensions in \eqref{eq:barg-trafo-fact}. These integrals
can be evaluated after bringing $\mathtt{H}$ into block-diagonal form,
\begin{align}
  \label{eq:proof-bdiag-h}
  \mathtt{H}=\mathtt{V}^{t}
  \left(
  \begin{array}{c:c:c c c}
    0&0&0&\cdots&0\\[0.3em]
    \hdashline\\[-1.0em]
    0&0&0&\cdots&0\\[0.3em]
    \hdashline\\[-1.0em]
    0&0\\
    \vdots&\vdots&\multicolumn{3}{c}{\check{\mathtt{H}}}\\
    0&0
  \end{array}
  \right)\mathtt{V}\,,\quad
  \check{\mathtt{H}}=
  \left(
  \begin{array}{c:c:c:c:c:c}
    0&0&0&-1&0&0\\[0.3em]
    \hdashline\\[-1.0em]
    0&0&(\mathcal{C}^{\dagger})^t&0&-1&0\\[0.3em]
    \hdashline\\[-1.0em]
    0&\mathcal{C}^{\dagger}&0&0&0&-1\\[0.3em]
    \hdashline\\[-1.0em]
    -1&0&0&0&0&-1\\[0.3em]
    \hdashline\\[-1.0em]
    0&-1&0&0&0&0\\[0.3em]
    \hdashline\\[-1.0em]
    0&0&-1&-1&0&0
  \end{array}\right)\,,
\end{align}
where the transformation matrix is given by
\begin{align}
  \mathtt{V}=
  \left(
  \begin{array}{c:c:c:c:c:c:c:c}
    \sqrt{2}&0&0&0&\sqrt{2}&0&0&0\\[0.3em]
    \hdashline\\[-1.0em]
    -i\sqrt{2}&0&0&0&i\sqrt{2}&0&0&0\\[0.3em]
    \hdashline\\[-1.0em]
    0&1&0&0&-\mathcal{C}^{\dagger}&0&0&0\\[0.3em]
    \hdashline\\[-1.0em]
    0&0&1&0&1&0&0&0\\[0.3em]
    \hdashline\\[-1.0em]
    \mathcal{C}^t&0&0&1&0&0&0&0\\[0.3em]
    \hdashline\\[-1.0em]
    -\mathcal{C}^t&0&0&0&0&1&0&0\\[0.3em]
    \hdashline\\[-1.0em]
    1&0&0&0&0&0&1&0\\[0.3em]
    \hdashline\\[-1.0em]
    0&0&0&0&\mathcal{C}^{\dagger}&0&0&1
  \end{array}
  \right)
  \,.
\end{align}
To change the integration variables in \eqref{eq:barg-trafo-fact} to
ones which are adapted to this block-diagonal form of $\mathtt{H}$, we
compute
\begin{align}
  \mathtt{V}
  \left(
  \begin{array}{c}
    \vcenter{\vspace{2.8cm}}\boldsymbol{z}_\alpha\\[0.3em]
    \hdashline\\[-1.0em]
    \vcenter{\vspace{2.8cm}}\overline{\boldsymbol{z}}_\alpha
  \end{array}
  \right)
  =
  \left(
  \begin{array}{c}
    2\sqrt{2}\Real \dt{\boldsymbol{z}}_\alpha^{\text{d}}\\[0.3em]
    \hdashline\\[-1.0em]
    2\sqrt{2}\Imag \dt{\boldsymbol{z}}_\alpha^{\text{d}}\\[0.3em]
    \hdashline\\[-1.0em]
    \dt{\boldsymbol{z}}_\alpha^{\text{o}}-\mathcal{C}^{\dagger}\overline{\dt{\boldsymbol{z}}}_\alpha^{\text{d}}\\[0.3em]
    \hdashline\\[-1.0em]
    \mathring{\boldsymbol{z}}_\alpha^{\text{d}}+\overline{\dt{\boldsymbol{z}}}_\alpha^{\text{d}}\\[0.3em]
    \hdashline\\[-1.0em]
    \mathring{\boldsymbol{z}}_\alpha^{\text{o}}+\mathcal{C}^t\dt{\boldsymbol{z}}_\alpha^{\text{d}}\\[0.3em]
    \hdashline\\[-1.0em]
    \overline{\dt{\boldsymbol{z}}}_\alpha^{\text{o}}-\mathcal{C}^t\dt{\boldsymbol{z}}_{\alpha}^{\text{d}}\\[0.3em]
    \hdashline\\[-1.0em]
    \overline{\mathring{\boldsymbol{z}}}_\alpha^{\text{d}}+\dt{\boldsymbol{z}}_\alpha^{\text{d}}\\[0.3em]
    \hdashline\\[-1.0em]
    \overline{\mathring{\boldsymbol{z}}}_\alpha^{\text{o}}+\mathcal{C}^{\dagger}\overline{\dt{\boldsymbol{z}}}_\alpha^{\text{d}}
  \end{array}
  \right)
  =
  \left(
  \begin{array}{c}
    \boldsymbol{x}_\alpha\\[0.3em]
    \hdashline\\[-1.0em]
    \boldsymbol{y}_\alpha\\[0.3em]
    \hdashline\\[-1.0em]
    \vcenter{\vspace{2.2cm}}\boldsymbol{w}_\alpha\\[0.3em]
    \hdashline\\[-1.0em]
    \vcenter{\vspace{2.1cm}}\overline{\boldsymbol{w}}_\alpha\vphantom{}
  \end{array}
  \right)
  \,,\quad
  (\mathtt{V}^{-1})^t
  \left(
  \begin{array}{c}
    0\\[0.3em]
    \hdashline\\[-1.0em]
    0\\[0.3em]
    \hdashline\\[-1.0em]
    0\\[0.3em]
    \hdashline\\[-1.0em]
    0\\[0.3em]
    \hdashline\\[-1.0em]
    \vcenter{\vspace{1.25cm}}\boldsymbol{\sigma}_\alpha\\[0.3em]
    \hdashline\\[-1.0em]
    \vcenter{\vspace{1.25cm}}\overline{\boldsymbol{\sigma}}_\alpha
  \end{array}
  \right)
  =
  \left(
  \begin{array}{c}
   \frac{-i}{\sqrt{2}}\Imag(\overline{\boldsymbol{\sigma}}_\alpha^{\text{d}}-\mathcal{C}\boldsymbol{\sigma}_\alpha^{\text{o}})\\[0.3em]
    \hdashline\\[-1.0em]
   \frac{-i}{\sqrt{2}}\Real(\overline{\boldsymbol{\sigma}}_\alpha^{\text{d}}-\mathcal{C}\boldsymbol{\sigma}_\alpha^{\text{o}})\\[0.3em]
    \hdashline\\[-1.0em]
    0\\[0.3em]
    \hdashline\\[-1.0em]
    0\\[0.3em]
    \hdashline\\[-1.0em]
    0\\[0.3em]
    \hdashline\\[-1.0em]
    \boldsymbol{\sigma}_\alpha^{\text{o}}\\[0.3em]
    \hdashline\\[-1.0em]
    \overline{\boldsymbol{\sigma}}_\alpha^{\text{d}}\\[0.3em]
    \hdashline\\[-1.0em]
    \overline{\boldsymbol{\sigma}}_\alpha^{\text{o}}
  \end{array}
  \right)\,.
\end{align}
Expressing \eqref{eq:barg-trafo-fact} in terms of the new variables
$\boldsymbol{x}_\alpha,\boldsymbol{y}_\alpha\in\mathbb{R}^K$ and
$\boldsymbol{w}_\alpha\in\mathbb{C}^{3K}$ leads to
\begin{align}
  \label{eq:barg-trafo-reduced}
  \begin{aligned}
    \Phi_\alpha(\boldsymbol{\sigma}_\alpha,\overline{\boldsymbol{\sigma}}_\alpha)
    &=    
    \frac{\delta_{\mathbb{C}^K}(\overline{\boldsymbol{\sigma}}_\alpha^{\text{d}}-\mathcal{C}\boldsymbol{\sigma}_\alpha^{\text{o}})}{e^{\overline{\boldsymbol{\sigma}}_\alpha^t\boldsymbol{\sigma}_\alpha}}
    \int_{\mathbb{C}^{3K}}
    \frac{\D^{\,3K}\!\overline{\boldsymbol{w}}_\alpha\D^{\,3K}\! \boldsymbol{w}_\alpha}{(2\pi i)^{3K}}\;
    e^{\frac{1}{2}
      \left(
        \begin{smallarray}{c}
          \boldsymbol{w}_\alpha\\[0.1em]
          \hdashline\\[-0.6em]
          \overline{\boldsymbol{w}}_\alpha
        \end{smallarray}
      \right)^t
      \check{\mathtt{H}}
      \left(
        \begin{smallarray}{c}
          \boldsymbol{w}_\alpha\\[0.1em]
          \hdashline\\[-0.6em]
          \overline{\boldsymbol{w}}_\alpha
        \end{smallarray}
      \right)
      +\sqrt{2}\,
      \overline{\boldsymbol{w}}_\alpha^t
      \smash{\left(
        \begin{smallarray}{c}
          \boldsymbol{\sigma}^{\text{o}}_\alpha\\[0.1em]
          \hdashline\\[-0.6em]
          \overline{\boldsymbol{\sigma}}^{\text{d}}_\alpha\\[0.1em]
          \hdashline\\[-0.6em]
          \overline{\boldsymbol{\sigma}}^{\text{o}}_\alpha\\
        \end{smallarray}
      \right)}
    }\,.
  \end{aligned}
\end{align}
Here the integrals over $\boldsymbol{x}_\alpha$ and
$\boldsymbol{y}_\alpha$, which are associated with the vanishing
diagonal blocks of $\mathtt{H}$ in \eqref{eq:proof-bdiag-h}, reduce to
Fourier representations of delta functions. These are defined by
\begin{align}  
  \label{eq:proof-complex-delta}
  \int_{\mathbb{R}^K}\D^{\,K}\!\boldsymbol{x}_\alpha\, e^{-i\boldsymbol{x}_\alpha^t \boldsymbol{\theta}}=(2\pi)^K\delta^K(\boldsymbol{\theta})\,,\quad
  \delta_{\mathbb{C}}^K(\boldsymbol{\theta}+i\boldsymbol{\varphi})=\delta^{K}(\boldsymbol{\theta})\delta^{K}(\boldsymbol{\varphi})\,
\end{align}
for $\boldsymbol{\theta},\boldsymbol{\varphi}\in\mathbb{R}^K$. Recall
that the symbol $\delta$ denotes an ordinary delta function of a real
argument. Next, we focus on the evaluation of the remaining integral
in \eqref{eq:barg-trafo-reduced}. Recall the standard result on
multi-dimensional Gaußian integrals,
\begin{align}
  \label{eq:barg-gaussian-general}
  \int_{\mathbb{R}^n}\D^{\,n}\!\mathtt{u}\, e^{-\frac{1}{2}\mathtt{u}^t\mathcal{A}\mathtt{u}+\mathtt{b}^t\mathtt{u}}=
  (2\pi)^\frac{n}{2}\sqrt{\det{\mathcal{A}}}^{\,-1}e^{\frac{1}{2}\mathtt{b}^t\mathcal{A}^{-1}\mathtt{b}}\,,
\end{align}
where $\mathcal{A}$ is a symmetric complex $n\times n$ matrix whose
eigenvalues have a strictly positive real part and
$\mathtt{b}\in\mathbb{C}^n$, see e.g.\ \cite{ZinnJustin:2005}. The
$3K$-complex-dimensional Gaußian integral in
\eqref{eq:barg-trafo-reduced} is brought into this form by defining
\begin{align}
  \mathtt{u}=
  \left(
  \begin{array}{c}
    \Real \boldsymbol{w}_\alpha\\[0.3em]
    \hdashline\\[-1.0em]
    \Imag \boldsymbol{w}_\alpha
  \end{array}
  \right)\,,\quad
  \mathcal{A}=-
  {\mathcal{E}}^t
  \check{\mathtt{H}}
  \mathcal{E}\,,\quad
  \mathcal{E}=
  \left(
  \begin{array}{c:c}
    1_{3K}&i 1_{3K}\\[0.3em]
    \hdashline\\[-1.0em]
    1_{3K}&-i 1_{3K}
  \end{array}\right)\,,\quad
  \mathtt{b}=\sqrt{2}
  \left(
  \begin{array}{c}
    \boldsymbol{\sigma}^{\text{o}}_\alpha\\[0.3em]
    \hdashline\\[-1.0em]
    \overline{\boldsymbol{\sigma}}^{\text{d}}_\alpha\\[0.3em]
    \hdashline\\[-1.0em]
    \overline{\boldsymbol{\sigma}}^{\text{o}}_\alpha\\[0.3em]
    \hdashline\\[-1.0em]
    -i\boldsymbol{\sigma}^{\text{o}}_\alpha\\[0.3em]
    \hdashline\\[-1.0em]
    -i\overline{\boldsymbol{\sigma}}^{\text{d}}_\alpha\\[0.3em]
    \hdashline\\[-1.0em]
    -i\overline{\boldsymbol{\sigma}}^{\text{o}}_\alpha
  \end{array}\right)\,.
\end{align}
One easily verifies that this matrix $\mathcal{A}$ is symmetric and
all its eigenvalues are equal to $2$. Consequently
\eqref{eq:barg-gaussian-general} can be applied and we obtain
\begin{align}
  \Phi_\alpha(\boldsymbol{\sigma}_\alpha,\overline{\boldsymbol{\sigma}}_\alpha)=
  e^{-\overline{\boldsymbol{\sigma}}_\alpha^t\boldsymbol{\sigma}_\alpha}
  \delta_{\mathbb{C}^K}(\overline{\boldsymbol{\sigma}}_\alpha^{\text{d}}-\mathcal{C}\boldsymbol{\sigma}^{\text{o}}_\alpha)
  e^{(\mathcal{C}^{\dagger}\overline{\boldsymbol{\sigma}}_\alpha^{\text{d}})^t\overline{\boldsymbol{\sigma}}_\alpha^{\text{o}}+(\overline{\boldsymbol{\sigma}}_\alpha^{\text{o}})^t\mathcal{C}^{\dagger}\overline{\boldsymbol{\sigma}}^{\text{d}}_\alpha}
  =\delta_{\mathbb{C}^K}(\overline{\boldsymbol{\sigma}}_\alpha^{\text{d}}-\mathcal{C}\boldsymbol{\sigma}^{\text{o}}_\alpha)\,,
\end{align}
where we made use of the delta function for the last equality.
Finally, to arrive at the desired result
\eqref{eq:barg-trafo-integrand}, we rename the variables
$\sigma_\alpha^i$ contained in $\boldsymbol{\sigma}_\alpha$ according
to \eqref{eq:barg-rename} into $\lambda_\alpha^i$. Q.E.D.

\subsection{Extension to Superalgebras}
\label{sec:transf-ferm-oscill}

In the previous sections we detailed the change of basis from
oscillator to spinor helicity-like variables for the bosonic algebra
$\mathfrak{u}(p,p)$. In particular, we explained how the symmetry
generators and the integrand of the unitary Graßmannian matrix
model~\eqref{eq:grass-int-unitary} are transformed. To cover the case
relevant for tree-level $\mathcal{N}=4$ SYM scattering amplitudes, we
have to extend these calculations to superalgebras. In the present
section we show how our results generalize from $\mathfrak{u}(p,p)$ to
the superalgebra $\mathfrak{u}(p,p|r+s)$.

Let us start by discussing the realization of fermionic oscillators is
terms of a Graßmann algebra, cf.\
\cite{Kirillov:2013,ZinnJustin:2005,Takhtajan:2008}. This can be
viewed as the fermionic analogue of the Bargmann representation from
section~\ref{sec:bargm-transf}. Consider a family of fermionic
oscillators on a Fock space,
\begin{align}
  \label{eq:fermi-osc}
  \{\mathbf{c}_{a},\bar{\mathbf{c}}_b\}=\delta_{ab}\,,\quad
  {\mathbf{c}_a}^\dagger=\bar{\mathbf{c}}_a\,,\quad
  \mathbf{c}_a|0\rangle=0\,,
\end{align}
where $a,b=1\,\ldots,r$ and the bracket denotes the
anticommutator. These oscillators can be thought of as being
associated with the $\mathfrak{u}(0|r)$ subalgebra of
$\mathfrak{u}(p,p|r+s)$ according to \eqref{eq:osc-split}. However,
this interpretations is not yet of importance in this paragraph. The
commutation relations and the action on the vacuum in
\eqref{eq:fermi-osc} are realized using a Graßmann algebra with
anticommuting generators $\chi_a$,
\begin{align}
  \label{eq:fermi-map-gr}
  \bar{\mathbf{c}}\mapsto \chi\,,\quad
  \mathbf{c}\mapsto\partial_{\chi}\,,\quad
  |0\rangle\mapsto 1\,,
\end{align}
where $\bar{\mathbf{c}}=(\bar{\mathbf{c}}_a)$ and $\chi=(\chi_a)$ are
$r$-component column vectors. In order to implement the adjoint in
\eqref{eq:fermi-osc} we need additional structure. We define the
one-dimensional Berezin integral as
$\int\D\chi_a(\alpha+\beta\chi_a)=\int(\alpha-\beta\chi_a)\D\chi_a=\beta$
for complex numbers $\alpha,\beta$. Multi-dimensional integrals are
obtained by iteration. Note that $\chi_a$ and $\D\chi_b$
anticommute. Furthermore, we append the $r$ anticommuting generators
$\overline{\chi}_a$ to the original Graßmann algebra and demand
$\{\overline{\chi}_a,\chi_b\}=0$. This extended Graßmann algebra is
equipped with an antilinear antiinvolution that we denote, with some
abuse of notation, also by $\overline{\,\cdot\,}$. It is
defined by mapping $\chi_a$ to the generator $\overline{\chi}_a$ of
the extended algebra. Furthermore, we have
$\overline{\chi_a\chi_b}=\overline{\chi}_b\overline{\chi}_a$ and
$\overline{\overline{\chi}}_a=\chi_a$. These structures allow us to
define the inner product
\begin{align}
  \label{eq:fermi-inner}
  \langle \Psi(\chi),\Phi(\chi)\rangle=
  \int\D^{\,r}\!\chi\D^{\,r}\!\overline{\chi}
  e^{-\chi^t\overline{\chi}}\,
  \overline{\Psi(\chi)}\Phi(\chi)\,,
\end{align}
where $\D^{\,r}\!\chi=\D\chi_1\cdots\D\chi_r$. Here $\Psi(\chi)$ and
$\Phi(\chi)$ are ``holomorphic'' in the sense that they do not depend
on the generators $\overline{\chi}_a$. One then verifies that with
respect to this inner product ${\chi_a}^\dagger=\partial_{\chi_a}$,
i.e.\ \eqref{eq:fermi-inner} implements the adjoint in
\eqref{eq:fermi-osc}. The realization of the fermionic oscillators in
\eqref{eq:fermi-map-gr} and the inner product \eqref{eq:fermi-inner}
are very much reminiscent, respectively, of \eqref{eq:barg-map-b} and
\eqref{eq:barg-inner-barg} from the Bargmann representation of bosonic
oscillators.

We turn to the superoscillators that are used to build the
representations $\oscrep_c$ and $\bar{\oscrep}_c$ of
$\mathfrak{u}(p,p|r+s)$ in \eqref{eq:gen-ordinary} and
\eqref{eq:gen-dual}, respectively. We want to reformulate the
generators of these representations in such a way that they can be
identified with those appearing for $\mathcal{N}=4$ SYM amplitudes in
\eqref{eq:sym-gen-gl44}. To this end, the bosonic oscillators among
the superoscillators are realized using the Bargmann representation of
section~\ref{sec:bargm-transf}. For the fermionic ones we employ the
realization in terms of a Graßmann algebra as just explained.  Let us
state our naming of the variables and indices based on the creation
operators, cf.\ \eqref{eq:osc-split},
\begin{align}
  \label{eq:fermi-real}
  (\bar{\mathbf{A}}_{\indnm{A}})=
  \left(
  \begin{array}{c}
    \bar{\mathbf{a}}_{\alpha}\\[0.3em]
    \hdashline\\[-1.0em]
    \bar{\mathbf{c}}_{a}\\[0.3em]
    \hdashline\\[-1.0em]
    \bar{\mathbf{b}}_{\dot{\alpha}}\\[0.3em]
    \hdashline\\[-1.0em]
    \bar{\mathbf{d}}_{\dot{a}}
  \end{array}
  \right)
  \mapsto
  \left(
  \begin{array}{c}
    z_\alpha\\[0.3em]
    \hdashline\\[-1.0em]
    \chi_a\\[0.3em]
    \hdashline\\[-1.0em]
    z_{\alpha+p}\\[0.3em]
    \hdashline\\[-1.0em]
    \psi_{\dot{a}}
  \end{array}
  \right)\,
\end{align}
with complex commuting variables $z_\alpha$ and $z_{\alpha+p}$ as well
as anticommuting Graßmann variables $\chi_a$ and $\psi_{\dot{a}}$. The
index ranges of the bosonic variables are
$\alpha,\dot{\alpha}=1,\ldots,p$ and those of the fermionic ones read
$a=1,\ldots,r$ and $\dot{a}=1,\ldots,s$. Thus
$\indnm{A}=1,\ldots,2p+r+s$. We already dealt with the bosonic
variables in the previous sections, where we expressed them in terms
of the spinor helicity-like variables $\lambda_\alpha$. Therefore we
can concentrate on the fermionic ones from now on. As in the bosonic
case, cf.\ \eqref{eq:barg-rename}, there is a slight distinction
between ordinary sites with representations $\oscrep_c$ and dual ones
with $\bar{\oscrep}_c$. At the ordinary sites we just rename
\begin{align}
  \label{eq:fermi-trafo-ord}
  \chi\mapsto\theta\,,\quad
  \partial_{\chi}\mapsto\partial_{\theta}\,,\quad
  \psi\mapsto\eta\,,\quad
  \partial_\psi\mapsto\partial_\eta\,,\quad
  1\mapsto 1\,.
\end{align}
At the dual sites we apply a fermionic analogue of the Fourier
transformation,
\begin{align}
\label{eq:fermi-ft}
  \Phi(\chi,\psi)\mapsto
  \Phi(\theta,\eta)=
  \int e^{\theta^t\chi-\eta^t\psi}
  \Phi(\chi,\psi)\D^{\,r}\!\chi\D^{\,s}\!\psi\,.
\end{align}
This transformation amounts to
\begin{align}
  \label{eq:fermi-trafo-dual}
  \chi\mapsto\partial_\theta\,,\quad
  \partial_\chi\mapsto\theta\,,\quad
  \psi\mapsto-\partial_\eta\,,\quad
  \partial_\psi\mapsto-\eta\,,\quad
  1\mapsto(-1)^r\theta_r\cdots\theta_1\eta_1\cdots\eta_s\,.
\end{align}
The ``measure'' in \eqref{eq:fermi-ft} is on the right hand side of
the integrand in order to avoid additional signs in
\eqref{eq:fermi-trafo-dual}. Let us apply these transformations to the
fermionic variables of the generators
$\mathbf{J}_{\indnm{AB}}$ of $\oscrep_c$ and
$\bar{\mathbf{J}}_{\indnm{AB}}$ of $\bar{\oscrep}_c$ in the $\mathfrak{u}(p,p|r+s)$ case, which are given
in terms of oscillators in \eqref{eq:gen-ordinary} and
\eqref{eq:gen-dual}, respectively. Together with the result
\eqref{eq:barg-gen} for the bosonic variables this yields
\begin{align}
  \label{eq:fermi-gen}
  \begin{aligned}
    (\mathbf{J}_{\indnm{AB}})\leftrightarrow
    (\mathfrak{J}_{\indnm{AB}})&=
    D
    \left(
    \begin{array}{c:c:c:c}
      \lambda_\alpha\partial_{\lambda_\beta}&
      \lambda_\alpha\partial_{\theta_b}&
      \lambda_\alpha\overline{\lambda}_\beta&
      \lambda_\alpha\eta_{\dot{b}}\\[0.3em]
      \hdashline\\[-1.0em]
      \theta_a\partial_{\lambda_\beta}&
      \theta_a\partial_{\theta_b}&
      \theta_a\overline{\lambda}_\beta&
      \theta_a\eta_{\dot{b}}\\[0.3em]
      \hdashline\\[-1.0em]
      -\partial_{\overline{\lambda}_\alpha}\partial_{\lambda_\beta}&
      -\partial_{\overline{\lambda}_\alpha}\partial_{\theta_b}&
      -\partial_{\overline{\lambda}_\alpha}\overline{\lambda}_\beta&
      -\partial_{\overline{\lambda}_\alpha}\eta_{\dot{b}}\\[0.3em]
      \hdashline\\[-1.0em]
      \partial_{\eta_{\dot{a}}}\partial_{\lambda_\beta}&
      \partial_{\eta_{\dot{a}}}\partial_{\theta_b}&
      \partial_{\eta_{\dot{a}}}\overline{\lambda}_\beta&
      \partial_{\eta_{\dot{a}}}\eta_{\dot{b}}
    \end{array}
  \right)
    D^{-1}\,,\\
    (\bar{\mathbf{J}}_{\indnm{AB}})\leftrightarrow
    (\bar{\mathfrak{J}}_{\indnm{AB}})&=
    D
    \left(
    \begin{array}{c:c:c:c}
      \partial_{\lambda_\beta}\lambda_\alpha&
      \partial_{\theta_b}\lambda_\alpha&
      -\overline{\lambda}_\beta\lambda_\alpha&
      \eta_{\dot{b}}\lambda_\alpha\\[0.3em]
      \hdashline\\[-1.0em]
      \partial_{\lambda_\beta}\theta_a&
      -\partial_{\theta_b}\theta_a&
      -\overline{\lambda}_\beta\theta_a&
      -\eta_{\dot{b}}\theta_a\\[0.3em]
      \hdashline\\[-1.0em]
      \partial_{\lambda_\beta}\partial_{\overline{\lambda}_\alpha}&
      \partial_{\theta_b}\partial_{\overline{\lambda}_\alpha}&
      -\overline{\lambda}_\beta\partial_{\overline{\lambda}_\alpha}&
      \eta_{\dot{b}}\partial_{\overline{\lambda}_\alpha}\\[0.3em]
      \hdashline\\[-1.0em]
      \partial_{\lambda_\beta}\partial_{\eta_{\dot{a}}}&
      -\partial_{\theta_b}\partial_{\eta_{\dot{a}}}&
      -\overline{\lambda}_\beta\partial_{\eta_{\dot{a}}}&
      -\eta_{\dot{b}}\partial_{\eta_{\dot{a}}}
    \end{array}
    \right)
    D^{-1}\,\\
  \end{aligned}
\end{align}
with the $(p+p|r+s)\times (p+p|r+s)$ supermatrix
\begin{align}
  D=
  \left(
  \begin{array}{c:c:c:c}
    1_p&0&1_p&0\\[0.3em]
      \hdashline\\[-1.0em]
    0&\sqrt{2}\,1_r&0&0\\[0.3em]
      \hdashline\\[-1.0em]
    -1_p&0&1_p&0\\[0.3em]
      \hdashline\\[-1.0em]
    0&0&0&\sqrt{2}\,1_s
  \end{array}\right)\,.
\end{align}
This matrix is even with respect to the grading \eqref{eq:grading} and
thus satisfies the criteria of
section~\ref{sec:change-basis-lax}. Therefore the similarity
transformation in \eqref{eq:fermi-gen} can be absorbed in a
redefinition of the Yangian generators. Furthermore, we remark that
the ordinary and dual generators in \eqref{eq:fermi-gen} are identical
up to some signs and a shift,
\begin{align}
  \label{eq:fermi-shift}
  \left.\bar{\mathfrak{J}}_{\indnm{AB}}\right|_{(\lambda,\overline{\lambda})\mapsto(\lambda,-\overline{\lambda})}
  =
  \mathfrak{J}_{\indnm{AB}}+\delta_{\indnm{AB}}(-1)^{|\indnm{A}|}\,.
\end{align}
This generalizes the bosonic relation \eqref{eq:barg-gen-replace}. We
also translate the central elements \eqref{eq:central-ord} and
\eqref{eq:central-dual}, which function as representation labels, into
the new basis,
\begin{align}
  \label{eq:fermi-replabel}
  \begin{aligned}
    \mathbf{C}=\tr(\mathbf{J}_{\indnm{AB}})
    \leftrightarrow
    \mathfrak{C}=\sum_{\alpha=1}^p
    (\lambda_\alpha\partial_{\lambda_\alpha}-\overline{\lambda}_\alpha\partial_{\overline{\lambda}_\alpha})
    +\sum_{a=1}^r\theta_a\partial_{\theta_a}
    -\sum_{\dot{a}=1}^s\eta_{\dot{a}}\partial_{\eta_{\dot{a}}}
    -p+s\,,\\
    \bar{\mathbf{C}}=\tr(\bar{\mathbf{J}}_{\indnm{AB}})
    \leftrightarrow
    \bar{\mathfrak{C}}=\sum_{\alpha=1}^p
    (\lambda_\alpha\partial_{\lambda_\alpha}-\overline{\lambda}_\alpha\partial_{\overline{\lambda}_\alpha})
    +\sum_{a=1}^r\theta_a\partial_{\theta_a}
    -\sum_{\dot{a}=1}^s\eta_{\dot{a}}\partial_{\eta_{\dot{a}}}
    +p-r\,.
  \end{aligned}
\end{align}
Let us for the moment concentrate on the algebra
$\mathfrak{u}(2,2|r+s=0+4)$, which appears in our discussion of the
superconformal symmetry of $\mathcal{N}=4$ SYM amplitudes in
section~\ref{sec:symmetries}. In this case we can identify the
generators $\mathfrak{J}_{\indnm{AB}}$ of $\oscrep_c$ in
\eqref{eq:fermi-gen} with those in \eqref{eq:sym-gen-gl44} for
$\tilde\lambda=+\overline{\lambda}$, where we neglect the similarity
transformation with the matrix $D$. The generators
$\bar{\mathfrak{J}}_{\indnm{AB}}$ of $\bar{\oscrep}_c$ in
\eqref{eq:fermi-gen} match those in \eqref{eq:sym-gen-gl44} for
$\tilde\lambda=-\overline{\lambda}$ up to the same similarity
transformation and a shift as in \eqref{eq:fermi-shift}. Furthermore,
for this algebra the expressions for $\mathfrak{C}$ and
$\bar{\mathfrak{C}}$ in \eqref{eq:fermi-replabel} coincide because
$-p+s=p-r=2$. The condition that the eigenvalue of these central
elements equals $c=0$ is identical to the constraint on the
``superhelicity'' of the amplitudes in \eqref{eq:amp-super-hel}. Thus
in our language the tree-level amplitudes of $\mathcal{N}=4$ SYM
transform in the representations $\oscrep_0$ and $\bar{\oscrep}_0$ of
$\mathfrak{u}(2,2|0+4)$. These belong, respectively, to particles with
positive and negative energy as we know from the last paragraphs of
section~\ref{sec:transf-boson-non}.

Finally, we want to apply the change of basis to the integrand of the
Graßmannian matrix model \eqref{eq:grass-int-unitary}. We observe that
this integrand factorizes into one part containing the bosonic
oscillators and one with the fermionic ones,
\begin{align}
  |\Phi\rangle=e^{\tr(\mathcal{C}\mathbf{I}_\bullet^t+\mathbf{I}_\circ \mathcal{C}^{\dagger})}|0\rangle
  =e^{\tr(\mathcal{C}{\mathbf{I}_\bullet^t}_{\text{b}}+{\mathbf{I}_{\circ}}_{\text{b}} \, \mathcal{C}^{\dagger})}|0\rangle_{\text{b}}\,
  e^{\tr(\mathcal{C}{\mathbf{I}_{\bullet}^t}_{\text{f}}+{\mathbf{I}_{\circ}}_{\text{f}} \, \mathcal{C}^{\dagger})}|0\rangle_{\text{f}}
  =|\Phi\rangle_{\text{b}}|\Phi\rangle_{\text{f}}\,.
\end{align}
This factorization is based on the structure of the entries
\eqref{eq:bullets-nc-recall} of
$\mathbf{I}_{\mathrel{\ooalign{\raisebox{0.4ex}{$\scriptstyle\bullet$}\cr\raisebox{-0.4ex}{$\scriptstyle\circ$}}}}$,
that can be written as
$(k\mathrel{\ooalign{\raisebox{0.7ex}{$\bullet$}\cr\raisebox{-0.3ex}{$\circ$}}}
l)=(k\mathrel{\ooalign{\raisebox{0.7ex}{$\bullet$}\cr\raisebox{-0.3ex}{$\circ$}}}
l)_{\text{b}}+(k\mathrel{\ooalign{\raisebox{0.7ex}{$\bullet$}\cr\raisebox{-0.3ex}{$\circ$}}}
l)_{\text{f}}$.
The subscripts $\text{b}$ and $\text{f}$ refer to the parts with
bosonic and fermionic oscillators, respectively. We already studied
the transformation of $|\Phi\rangle_{\text{b}}$ to spinor
helicity-like variables in
section~\ref{sec:bargm-transf-integrand}. Thus we can concentrate on
$|\Phi\rangle_{\text{f}}$ here. We realize the fermionic oscillators
as in \eqref{eq:fermi-real}. Then we apply the replacement
\eqref{eq:fermi-trafo-ord} at the ordinary sites and the Fourier
transformation \eqref{eq:fermi-ft} at the dual ones. In our
conventions the ``measure'' in the Fourier transformation from site
$k$ is left of that from site $k+1$. This yields
\begin{align}
  \label{eq:fermi-int}
  |\Phi\rangle_{\text{f}}\mapsto\Phi(\boldsymbol{\theta},\boldsymbol{\eta})_{\text{f}}=
  \epsilon\,
  \delta^{0|rK}(\boldsymbol{\theta}^{\text{d}}+\mathcal{C}\boldsymbol{\theta}^{\text{o}})
  \delta^{0|sK}(\boldsymbol{\eta}^{\text{d}}-\overline{\mathcal{C}}\boldsymbol{\eta}^{\text{o}})\,
\end{align}
with the sign $\epsilon=(-1)^{\frac{1}{2}rs(K-1)K+rK}$. Here we
arranged the Graßmann variables into the matrices
\begin{align}
  \label{eq:fermi-theta-eta}
  \begin{aligned}
    \boldsymbol{\eta}=
    \left(
    \begin{array}{c}
      \boldsymbol{\eta}^{\text{d}}\\[0.3em]
      \hdashline\\[-1.0em]
      \boldsymbol{\eta}^{\text{o}}
    \end{array}
    \right)
    \,,\quad \boldsymbol{\eta}^{\text{d}}=
    \begin{pmatrix}
      \eta_1^1&\cdots&\eta_s^1\\
      \vdots&&\vdots\\
      \eta_1^K&\cdots&\eta_s^K\\
    \end{pmatrix}
    \,,\quad \boldsymbol{\eta}^{\text{o}}=
    \begin{pmatrix}
      \eta_1^{K+1}&\cdots&\eta_s^{K+1}\\
      \vdots&&\vdots\\
      \eta_1^{2K}&\cdots&\eta_s^{2K}\\
    \end{pmatrix}
    \,,\quad \\
    \boldsymbol{\theta}=
    \left(
    \begin{array}{c}
      \boldsymbol{\theta}^{\text{d}}\\[0.3em]
      \hdashline\\[-1.0em]
      \boldsymbol{\theta}^{\text{o}}
    \end{array}
    \right)
    \,,\quad \boldsymbol{\theta}^{\text{d}}=
    \begin{pmatrix}
      \theta_1^1&\cdots&\theta_r^1\\
      \vdots&&\vdots\\
      \theta_1^K&\cdots&\theta_r^K\\
    \end{pmatrix}\,,\quad \boldsymbol{\theta}^{\text{o}}=
    \begin{pmatrix}
      \theta_1^{K+1}&\cdots&\theta_r^{K+1}\\
      \vdots&&\vdots\\
      \theta_1^{2K}&\cdots&\theta_r^{2K}\\
    \end{pmatrix}\,.\quad
  \end{aligned}
\end{align}
The fermionic delta functions occurring in \eqref{eq:fermi-int}
are defined by
\begin{align}
  \label{eq:fermi-delta}
 \delta^{0|uv}(\mathtt{A})=\prod_{i=1}^u\prod_{j=1}^v\mathtt{A}_{ij}\,
\end{align}
for a Graßmann-valued $u\times v$ matrix
$\mathtt{A}=(\mathtt{A}_{ij})$. Factors with smaller values of the
indices appear left in the products. Notice that with a bosonic
$u\times u$ matrix $\mathtt{B}$ we have
$\delta^{0|uv}(\mathtt{BA})=\det(\mathtt{B})^v\delta^{0|uv}(\mathtt{A})$. We
conclude by combining the transformation of $|\Phi\rangle_{\text{f}}$
from \eqref{eq:fermi-int} with that of the bosonic part
$|\Phi\rangle_{\text{b}}$ in \eqref{eq:barg-trafo-integrand},
\begin{align}
  \label{eq:trafo-complete-integrand}
  |\Phi\rangle \mapsto
  \Phi(\boldsymbol{\lambda},\overline{\boldsymbol{\lambda}},\boldsymbol{\theta},\boldsymbol{\eta})
  =
  \epsilon\,
  \delta_{\mathbb{C}}^{pK|0}(\boldsymbol{\lambda}^{\text{d}}+\mathcal{C}\boldsymbol{\lambda}^{\text{o}})
  \delta^{0|rK}(\boldsymbol{\theta}^{\text{d}}+\mathcal{C}\boldsymbol{\theta}^{\text{o}})
  \delta^{0|sK}(\boldsymbol{\eta}^{\text{d}}-\overline{\mathcal{C}}\boldsymbol{\eta}^{\text{o}})\,.
\end{align}
This is the integrand of the unitary Graßmannian matrix
model~\eqref{eq:grass-int-unitary} for oscillator representations of
$\mathfrak{u}(p,p|m)$ expressed in analogues of the spinor helicity
variables for $\mathcal{N}=4$ SYM scattering amplitudes.

\section{Graßmannian Integral in Spinor Helicity Variables}
\label{sec:grassmann-spinor}

Here we utilize the change of basis from oscillators to spinor
helicity-like variables, which we just derived, to obtain a unitary
Graßmannian integral formula in terms of the latter variables. This
formula is explained in section~\ref{sec:grassmannian-formula}. There
we also point out its differences to the original Graßmannian integral
proposal, which we reviewed in
section~\ref{sec:grassmannian-integral}. What is more, we find a tight
relation between the unitary of the integration contour and momentum
conservation, which is the topic of
section~\ref{sec:mom-supermom}. Section~\ref{sec:sample-invar-ampl} is
devoted to the evaluation of the integral for sample Yangian
invariants. In particular, we study examples that are related to
certain tree-level gluon amplitudes, superamplitudes of
$\mathcal{N}=4$ SYM and to integrable deformations thereof.

\subsection{Unitary Graßmannian Integral}
\label{sec:grassmannian-formula}

After we transformed the integrand of the unitary Graßmannian matrix
model \eqref{eq:grass-int-unitary} in the foregoing
section~\ref{sec:osc-spinor} and summarized the result in
\eqref{eq:trafo-complete-integrand}, we are able to state the
transformation of the entire integral. This results in a refined
\emph{Graßmannian integral in spinor helicity-like variables}. It
computes Yangian invariants for representations of
$\mathfrak{u}(p,p|r+s)$ with $N=2K$ sites, out of which the first $K$
are dual, and reads
\begin{align}
  \label{eq:grass-int-barg}
  \Psi_{2K,K}(\boldsymbol{\lambda},\overline{\boldsymbol{\lambda}},\boldsymbol{\theta},\boldsymbol{\eta})
  =\epsilon\chi_K^{-1}\int_{U(K)}[\D\mathcal{C}]\mathscr{F}(\mathcal{C})\,
  \delta_{\mathbb{C}}^{pK|0}(C\boldsymbol{\lambda})  
  \delta^{0|rK}(C\boldsymbol{\theta})
  \delta^{0|sK}(C^\perp\boldsymbol{\eta})\,.
\end{align}
In this formula the prefactor involves the sign $\epsilon$ introduced
after \eqref{eq:fermi-int}. The $U(K)$ invariant Haar measure
$[\D\mathcal{C}]$ is defined in \eqref{eq:haar-general}. Its
normalization is denoted by $\chi_K$. The integrand
$\mathscr{F}(\mathcal{C})$ is specified in
\eqref{eq:eq:grass-int-unitary-integrand-final}. Recall also its
explicit form for $N=2,4,6$ in \eqref{eq:integrand21},
\eqref{eq:integrand42} and \eqref{eq:integrand63}, respectively. The
unitary $K\times K$ matrix $\mathcal{C}$ is embedded into the
$K\times 2K$ matrix
$C=\big(\begin{array}{c:c}1_{K}&\mathcal{C}\end{array}\big)$, which is
an element of the Graßmannian $\text{Gr}(2K,K)$, cf.\
\eqref{eq:grassint-matrix}. We also introduced
$C^\perp=\big(\begin{array}{c:c}1_K&-\overline{\mathcal{C}}\end{array}\big)$
satisfying $C(C^\perp)^t=0$. The external bosonic and fermionic
variables are encoded in the matrices
\begin{align}
  \label{eq:matrices-split}
  \boldsymbol{\lambda}=
  \left(
  \begin{array}{c}
    \boldsymbol{\lambda}^{\text{d}}\\[0.3em]
    \hdashline\\[-1.0em]
    \boldsymbol{\lambda}^{\text{o}}
  \end{array}
  \right)
  \,,\quad
  \boldsymbol{\eta}=
  \left(
  \begin{array}{c}
    \boldsymbol{\eta}^{\text{d}}\\[0.3em]
    \hdashline\\[-1.0em]
    \boldsymbol{\eta}^{\text{o}}
  \end{array}
  \right)
  \,,\quad 
  \boldsymbol{\theta}=
  \left(
  \begin{array}{c}
    \boldsymbol{\theta}^{\text{d}}\\[0.3em]
    \hdashline\\[-1.0em]
    \boldsymbol{\theta}^{\text{o}}
  \end{array}
  \right)\,.
\end{align}
Here the $K\times p$ blocks
$\boldsymbol{\lambda}^{\text{d}}=(\lambda^k_\alpha)$ and
$\boldsymbol{\lambda}^{\text{o}}=(\lambda^l_\alpha)$ contain,
respectively, the spinor helicity-like variables at the dual sites
$k=1,\ldots,K$ and ordinary sites $l=K+1,\ldots,2K$, see
\eqref{eq:barg-trafo-mult-sch}. Analogously, the fermionic variables
are arranged into the $K\times s$ blocks
$\boldsymbol{\eta}^{\text{d}}=(\eta^k_{\dot{a}})$,
$\boldsymbol{\eta}^{\text{o}}=(\eta^l_{\dot{a}})$ and into the
$K\times r$ blocks $\boldsymbol{\theta}^{\text{d}}=(\theta^k_{a})$,
$\boldsymbol{\theta}^{\text{o}}=(\theta^l_{a})$, see
\eqref{eq:fermi-theta-eta}. Lastly, the complex bosonic delta function
in \eqref{eq:grass-int-barg} is defined as the delta function of the
real part of its the argument times that of the imaginary part, see
\eqref{eq:proof-complex-delta}. The definition of the fermionic delta
function is provided in \eqref{eq:fermi-delta}.

Let us compare our refined formula \eqref{eq:grass-int-barg} to the
original proposal of the Graßmannian integral for tree-level
$\mathcal{N}=4$ SYM amplitudes, as reviewed in the introductory
section~\ref{sec:grassmannian-integral}. We know from the
identification of the generators belonging to $\mathfrak{u}(p,p|r+s)$
representations with those of the amplitudes in
section~\ref{sec:transf-ferm-oscill} that we have to restrict to the
algebra $\mathfrak{u}(2,2|0+4)$ for this comparison. This is
essentially the superconformal algebra discussed in
section~\ref{sec:symmetries}. First, we observe some notational
differences. The matrix $C^\perp$ is determined by $C(C^\perp)^t=0$
only up to a $GL(\mathbb{C}^K)$ transformation. Our choice of this
matrix after \eqref{eq:grass-int-barg} differs from that in
\eqref{eq:grassint-matrix-perp} by such a transformation. Moreover,
the roles of the matrices $C$ and $C^\perp$ in
\eqref{eq:grass-int-barg} and the original formula
\eqref{eq:grassint-amp} are exchanged. Both matrices are elements of
the same Graßmannian $\text{Gr}(2K,K)$ in the case $N=2K$ under
consideration. Therefore this exchange may be viewed as an alternative
choice of parameterization.

Of considerably more importance are conceptual differences. These
address the issues of the original Graßmannian integral that we
identified in section~\ref{sec:grassmannian-integral}. In
\eqref{eq:grass-int-barg} we work at all times in real Minkowski space
with $(1,3)$ signature. That is, the complex spinor helicity variables
$\lambda^i=(\lambda^i_\alpha)$ and
$\tilde{\lambda}^i=(\tilde{\lambda}^i_{\dot{\alpha}})$, which
according to \eqref{eq:nullvec-spinors} make up the particle momenta
$p^i$, obey the reality condition~\eqref{eq:spinors-real}, see the
discussion in section~\ref{sec:transf-ferm-oscill}. We have
\begin{align}
  \label{eq:def-lambda-tilde}
  \tilde{\lambda}^i=
  \begin{cases}
    -\overline{\lambda}^i\quad\text{for dual sites}\quad& i=1,\ldots,K\,,\\
    \phantom{-}\overline{\lambda}^i\quad\text{for ordinary sites}\quad& i=K+1,\ldots,2K\,.\\
  \end{cases}
\end{align}
These reality conditions are imperative considering the oscillator
representations we started out with in
section~\ref{sec:grassmann-osc}. In contrast, in the original proposal
\eqref{eq:grassint-amp} one works in $(2,2)$ signature or a
complexified momentum space, which entails, respectively, real or
complex \emph{independent} variables $\lambda^i, \tilde{\lambda}^i$.
Furthermore, our formula \eqref{eq:grass-int-barg} does not feature a
delta function involving the matrix
$\boldsymbol{\tilde{\lambda}}=(\tilde{\lambda}^i_{\dot{\alpha}})$ as
in \eqref{eq:grassint-amp}. In our setting the reality conditions
\eqref{eq:def-lambda-tilde} yield
\begin{align}
  \label{eq:def-lambda-tilde-mat}
  \boldsymbol{\tilde{\lambda}}=
  \left(
  \begin{array}{c}
    \boldsymbol{\tilde{\lambda}}^{\text{d}}\\[0.3em]
    \hdashline\\[-1.0em]
    \boldsymbol{\tilde{\lambda}}^{\text{o}}
  \end{array}
  \right)
  =
  \left(
  \begin{array}{c}
    -\overline{\boldsymbol{\lambda}}^{\text{d}}\\[0.3em]
    \hdashline\\[-1.0em]
    \phantom{-}\overline{\boldsymbol{\lambda}}^{\text{o}}
  \end{array}
  \right)\,.
\end{align}
Thus $\boldsymbol{\tilde{\lambda}}$ is determined by
$\boldsymbol{\lambda}$ and does not have to be constrained by a
separate delta function. Let us also stress that the complex bosonic
delta function $\delta_{\mathbb{C}}$ in \eqref{eq:grass-int-barg} is
defined by \eqref{eq:proof-complex-delta} in terms of ordinary real
delta functions and thus differs from the somewhat formal
$\delta_\ast$ in \eqref{eq:grassint-amp}. This brings us to another
issue raised in the context of the original Graßmannian integral
\eqref{eq:grassint-amp}. In that approach a number of delta functions
is usually eliminated in a purely algebraic fashion without the
specification of a corresponding contour of integration. Then a
contour is enforced ``by hand'' onto the remaining integral. The
unitary contour in \eqref{eq:grass-int-barg} is supposed to unify both
steps in a natural way. As we shall see in the next section, the
unitary of the integration variable $\mathcal{C}$ and the reality
conditions on the spinor helicity variables are in fact tightly
interlocked.

On a different note, we reviewed the extension of the Graßmannian
integral \eqref{eq:grassint-amp} to deformed amplitudes in
section~\ref{sec:deform}. In the resulting formula
\eqref{eq:grassint-amp-def} the challenge of selecting an appropriate
contour of integration seems to be quite involved due to the branch
cuts of the integrand. In particular, a satisfactory solution is not
known even for the six-particle $\text{NMHV}$ amplitude. The situation
changes drastically for our refined Graßmannian integral in
\eqref{eq:grass-int-barg}. Its integrand $\mathscr{F}(\mathcal{C})$
defined in \eqref{eq:eq:grass-int-unitary-integrand-final} also
incorporates deformation parameters. Nevertheless, it is manifestly
free from any branch cuts as discussed in
section~\ref{sec:single-valu-integr}. Recall that the way we obtained
\eqref{eq:eq:grass-int-unitary-integrand-final} from the standard form
of the integrand in \eqref{eq:grass-int-unitary-integrand} is heavily
based on the unitarity of $\mathcal{C}$. Furthermore, it makes
decisive use of integer representation labels $c_i$, as opposed to
complex ones employed in the original deformed Graßmannian integral
\eqref{eq:grassint-amp-def}.

Despite these numerous advantages of the Graßmannian integral
\eqref{eq:grass-int-barg} with a unitary contour, it remains to
be shown whether it reproduces the well-known expressions for the
$\mathcal{N}=4$ SYM tree-level amplitudes. This will be addressed in
section~\ref{sec:sample-invar-ampl} by evaluating
\eqref{eq:grass-int-barg} for sample invariants.

\subsection{Momentum Conservation and Unitarity of Contour}
\label{sec:mom-supermom}

Before we investigate sample invariants, it is instructive to study
the relation between momentum conservation and the unitary contour in
the Graßmannian integral \eqref{eq:grass-int-barg} on a general
level. As this analysis applies to the $\mathfrak{u}(p,p|r+s)$ case of
the integral, we work with an appropriate generalization of
four-dimensional Minkowski momenta for this algebra.

To begin with, we have to introduce some notation. We define the
``momentum'' and two notions of ``supermomenta'' by
\begin{align}
  \label{eq:def-mon-supermon}
  \begin{aligned}
    P_{\alpha\dot{\alpha}}=\sum_{i=1}^N\lambda^i_\alpha\tilde{\lambda}^i_{\dot{\alpha}}\,,\quad
    Q_{\alpha \dot{a}}=\sum_{i=1}^N\lambda^i_{\alpha}\eta^i_{\dot{a}}\,,\quad
    \hat{Q}_{\dot{\alpha} a}=\sum_{i=1}^N\tilde{\lambda}_{\dot{\alpha}}^i\theta^i_a\,
  \end{aligned}
\end{align}
with the index ranges $\alpha,\dot{\alpha}=1,\ldots,p$, $a=1,\ldots,r$
and $\dot{a}=1,\ldots,s$. For the algebra $\mathfrak{u}(2,2|0+4)$
these reduce to the four-dimensional Minkowski momentum
$P_{\alpha\dot{\alpha}}$ and the supermomentum $Q_{\alpha \dot{a}}$
defined in \eqref{eq:amp-bos-delta} and \eqref{eq:amp-superdelta},
respectively. The $p\times p$ matrix $P=(P_{\alpha\dot{\alpha}})$ is
Hermitian because of \eqref{eq:def-lambda-tilde}. This observation
allows us to specify the $p^2$-dimensional real bosonic delta function
\begin{align}
  \label{eq:def-delta-bosonic}
  \delta^{p^2|0}(P)=\prod_{\alpha=\dot{\alpha}=1}^p\delta(P_{\alpha\dot{\alpha}})
  \prod_{\substack{\alpha,\dot{\alpha}=1\\\alpha>\dot{\alpha}}}^p\delta(\Real P_{\alpha\dot{\alpha}})\delta(\Imag P_{\alpha\dot{\alpha}})\,.
\end{align}
It will be referred to as ``momentum conserving'' delta function from
now on. Furthermore, let us introduce the $p\times s$ matrix
$Q=(Q_{\alpha \dot{a}})$ and the $p\times r$ matrix
$\hat{Q}=(\hat{Q}_{\dot{\alpha} a})$ with Graßmann-valued entries. The
``supermomentum conserving'' delta functions are
\begin{align}
  \label{eq:def-delta-fermionic}
  \delta^{0|ps}(Q)=\prod_{\alpha=1}^p\prod_{\dot{a}=1}^r Q_{\alpha\dot{a}}\,,\quad 
  \delta^{0|pr}(\hat{Q})=\prod_{\dot{\alpha}=1}^p\prod_{a=1}^s \hat{Q}_{\dot{\alpha} a}\,.
\end{align}
Recall \eqref{eq:fermi-delta} for the definition of a fermionic delta
function.

Next, we show that the Graßmannian integral~\eqref{eq:grass-int-barg}
with the unitary contour implies momentum conservation. From the delta
functions in the Graßmannian integral \eqref{eq:grass-int-barg} we
obtain
\begin{align}
  \label{eq:on-support}
  C\boldsymbol{\lambda}=0\,,\quad
  C^\perp\boldsymbol{\tilde{\lambda}}=0\,,\quad
  C\boldsymbol{\theta}=0\,,\quad
  C^\perp \boldsymbol{\eta}=0\,.
\end{align}
The second equation is minus the complex conjugate of the first
one. Nevertheless, it is instructional to display it here
explicitly. Of course, the first two equations are only valid on the
support of the bosonic delta function in
\eqref{eq:grass-int-barg}. Analogously, the remaining equations hold
in the presence of the fermionic delta functions in
\eqref{eq:grass-int-barg}.  The equations in \eqref{eq:on-support}
together with, importantly, the unitarity of the integration variable
$\mathcal{C}$ imply momentum and supermomentum conservation,
\begin{align}
  \label{eq:mon-consv}
  \begin{aligned}
    \boldsymbol{\lambda}^t\boldsymbol{\tilde{\lambda}}=0
    &\quad\Leftrightarrow\quad P_{\alpha\dot{\alpha}}=0\,,\\
    \boldsymbol{\lambda}^t\boldsymbol{\eta}=0
    &\quad\Leftrightarrow\quad Q_{\alpha \dot{a}}=0\,,\\
    \boldsymbol{\tilde{\lambda}}^t\boldsymbol{\theta}=0
    &\quad\Leftrightarrow\quad \hat{Q}_{\dot{\alpha} a}=0\,.
  \end{aligned}
\end{align}
This is easily verified after splitting the matrices
$\boldsymbol{\lambda},\boldsymbol{\tilde{\lambda}},\boldsymbol{\theta},\boldsymbol{\eta}$
into ``dual'' and ``ordinary'' blocks as in \eqref{eq:matrices-split}
and \eqref{eq:def-lambda-tilde-mat}. Therefore the Yangian invariant
$\Psi_{2K,K}(\boldsymbol{\lambda},\overline{\boldsymbol{\lambda}},\boldsymbol{\theta},\boldsymbol{\eta})$
computed by the Graßmannian integral \eqref{eq:grass-int-barg} is
proportional to the momentum and supermomentum conserving delta
functions \eqref{eq:def-delta-bosonic} and
\eqref{eq:def-delta-fermionic}, respectively.

We move on to show, in a certain sense, the converse statement, i.e.\
that demanding momentum conservation implies the unitary of the
integration variable $\mathcal{C}$ in \eqref{eq:grass-int-barg}. Let
us first state our assumptions. We do not specify a contour of
integration in \eqref{eq:grass-int-barg} and hence at the outset
$\mathcal{C}\in GL(\mathbb{C}^K)$. The spinor helicity-like variables
$\boldsymbol{\lambda}$ and $\boldsymbol{\tilde{\lambda}}$ satisfy the
reality conditions \eqref{eq:def-lambda-tilde-mat}. The bosonic delta
function in the integral \eqref{eq:grass-int-barg} enforces
$C\boldsymbol{\lambda}=0$. Lastly, we assume that the integration
contour, and therefore $\mathcal{C}$, does not depend on the external
data $\boldsymbol{\lambda}$ and $\boldsymbol{\tilde{\lambda}}$. Under
these premises, momentum conservation becomes
\begin{align}
  0=\boldsymbol{\lambda}^t\boldsymbol{\tilde{\lambda}}=
  (\boldsymbol{\lambda}^{\text{o}})^t
  \big(1_K-\mathcal{C}^{t}\overline{\mathcal{C}}\big)
  \overline{\boldsymbol{\lambda}}^{\text{o}}\,,
\end{align}
which has to hold for any complex $K\times p$ matrix
$\boldsymbol{\lambda}^{\text{o}}$. We rephrase this equation by
introducing $\mathtt{h}=\overline{\boldsymbol{\lambda}}^{\text{o}}$
and
$\mathtt{A}=\mathtt{A}^\dagger=1_K-\mathcal{C}^{t}\overline{\mathcal{C}}$. This yields
\begin{align}
  0=\mathtt{h}^\dagger\mathtt{A}\mathtt{h}\,,
\end{align}
which has to be satisfied for all $\mathtt{h}$. Diagonalizing the
Hermitian matrix $\mathtt{A}$ by means of a unitary transformation, we
see that this equation implies $\mathtt{A}=0$. Thus $\mathcal{C}$ has
to be unitary. In this spirit momentum conservation implies a unitary
contour.

We should add a comment regarding one of our assumptions in this
argument. In the usual Graßmannian integral approach to
$\mathcal{N}=4$ SYM amplitudes, which we reviewed in
section~\ref{sec:grassmannian-integral}, the integration contour does
depend on the external data $\boldsymbol{\lambda}$ and
$\boldsymbol{\tilde{\lambda}}$. It encircles certain poles of the
integrand in \eqref{eq:grassint-amp}. The positions of these poles
depend on $\boldsymbol{\lambda}$ and $\boldsymbol{\tilde{\lambda}}$,
and therefore so does the contour.  This violates our assumption.
Nevertheless, in this thesis we retain the assumption of a contour
that is independent of the external data. In fact, it is very natural
from the point of view put forward in this thesis. We not only have
the Graßmannian integral \eqref{eq:grass-int-barg} in spinor helicity
variables but also know how to transform it into the oscillator form
\eqref{eq:grass-int-unitary}. In this basis a contour which depends on
the ``values'' of the oscillators does not seem to be well-defined.
In contrast, the unitary contour is known to yield correct oscillator
sample Yangian invariants, see
section~\ref{sec:sample-invariants}. Moreover, we proved the Yangian
invariance of the Graßmannian integral \eqref{eq:grass-int-unitary}
for this contour. Let us emphasize that this proof also applies to
the spinor helicity version \eqref{eq:grass-int-barg} of the
integral.

\subsection{Sample Invariants and Amplitudes}
\label{sec:sample-invar-ampl}

Now we are in a position to actually evaluate the unitary Graßmannian
integral \eqref{eq:grass-int-barg} in order to obtain sample Yangian
invariants in spinor helicity-like variables. We focus on invariants
with representations of the algebra $\mathfrak{u}(2,2)$ and the
superalgebra $\mathfrak{u}(2,2|0+4)$ because of their relevance for
gluon amplitudes and superamplitudes of $\mathcal{N}=4$ SYM,
respectively. We identify the four-site invariant with the
four-particle $\text{MHV}$ amplitude. Furthermore, the six-site
invariant is computed and its relation to the six-particle
$\text{NMHV}$ amplitude is discussed. Recall that this is the first
case where one has to impose a contour ``by hand'' in the usual
Graßmannian integral approach of
section~\ref{sec:grassmannian-integral}. More generally, our sample
invariants contain deformation parameters, which allow us to relate
them to the deformed amplitudes of section~\ref{sec:deform}.  Before
we turn our attention to the above-mentioned algebras, it is
instructive to compute some sample invariants for
$\mathfrak{u}(1,1)$. These share key features with the higher rank
examples but are technically easier to compute. Even before that, we
discuss some tools which will be of great utility for the evaluation
of the unitary Graßmannian integral \eqref{eq:grass-int-barg} in all
the examples considered. Let us also mention that some additional
sample invariants are computed in
appendix~\ref{sec:add-sample-inv-gr}.

\subsubsection{Tools for Evaluation of Integrals}
\label{sec:tools}

The unitary Graßmannian integral \eqref{eq:grass-int-barg} contains a
complex bosonic delta function that we have to manipulate in the
course of evaluating the integral. Thus we recall some of its
properties. We begin with the definition in terms of real delta
functions. For those we have
\begin{align}
  \int_{\mathbb{R}^2}\D x\D y\,\delta(x)\delta(y)f(x,y)=f(0,0)\,
\end{align}
for a suitable test function $f(x,y)$. Let us introduce the complex
coordinate $z=x+iy$. Then the measure reads
$(2i)^{-1}\D\overline{z}\D z=\D x\D y$. Defining
$\delta_{\mathbb{C}}(z)=\delta(x)\delta(y)$
and denoting the test function by $g(z,\overline{z})=f(x,y)$,
the above equation turns into
\begin{align}
  \int_{\mathbb{C}}\frac{\D\overline{z}\D z}{2i}\delta_{\mathbb{C}}(z)
  g(z,\overline{z})=g(0,0)\,.
\end{align}
Therefore $\delta_{\mathbb{C}}(z)$ is a complex delta function, cf.\
\eqref{eq:proof-complex-delta}. Using a linear change of variables one
readily verifies
\begin{align}
  \label{eq:delta-matrix}
  \begin{aligned}
    \delta^K(Ax)&=\frac{\delta^K(x)}{|\det A|}\quad&
    &\text{for}\quad
    x\in\mathbb{R}^K\,,\quad A\in GL(\mathbb{R}^K)\,,\\
    \delta_{\mathbb{C}}^K(Az)&=\frac{\delta_{\mathbb{C}}^K(z)}{\det AA^\dagger}\quad&
    &\text{for}\quad
    z\in\mathbb{C}^K\,,\quad A\in GL(\mathbb{C}^K)\,.
  \end{aligned}
\end{align}
Especially the second line will be used frequently. At times we will
also need non-linear coordinate transformations. If we change
variables from $x\in\mathbb{R}^K$ to $y(x)\in\mathbb{R}^K$, the
measure transforms as
$\D^{\,K}\!x=\D^{\,K}\!y|\det\frac{\partial x}{\partial y}|$ and the
real delta function in the new variables becomes
\begin{align}
  \label{eq:delta-change-var}
  \delta^K(y-y_0)
  =\left|\det\frac{\partial x}{\partial y}\right|\,\delta^K(x-x_0)\,.
\end{align}

There is another transformation that we will apply regularly to
simplify the argument of the bosonic delta function in the Graßmannian
integral \eqref{eq:grass-int-barg}. We want to express a unit vector
$\frac{\mathtt{v}}{\|\mathtt{v}\|}\in\mathbb{C}^K$ with
$\|\mathtt{v}\|=\sqrt{\mathtt{v}^\dagger \mathtt{v}}$ as a matrix
$\mathtt{L}\in U(K)$ acting on a reference vector,
\begin{align}
  \label{eq:vec-as-mat}
  \frac{\mathtt{v}}{\|\mathtt{v}\|}
  =
  \mathtt{L}
  \begin{pmatrix}
    1\\
    0\\
    \vdots\\
    0\\
  \end{pmatrix}\,.
\end{align}
A solution to this equation is
\begin{align}
  \label{eq:para-mat}
  \mathtt{L}=\frac{1}{\|\mathtt{v}\|}
  \left(
  \begin{array}{c:c}
    \mathtt{v}^1&-\tilde{\mathtt{v}}^\dagger\\[0.3em]
    \hdashline\\[-1.0em]
    \tilde{\mathtt{v}}&
    \frac{\overline{\mathtt{v}}^1}{|\mathtt{v}^1|}
    \Big(\|\mathtt{v}\|1_{n-1}-\frac{\tilde{\mathtt{v}}\tilde{\mathtt{v}}^\dagger}{\|\mathtt{v}\|+|\mathtt{v}^1|}\Big)
  \end{array}
  \right)\,,
  \quad\text{where}\quad
  \mathtt{v}=
  \left(
  \begin{array}{c}
    \mathtt{v}^1\\[0.3em]
    \hdashline\\[-1.0em]
    \tilde{\mathtt{v}}
  \end{array}
  \right)
\end{align}
with $\tilde{\mathtt{v}}\in\mathbb{C}^{K-1}$. We note that
$\det\mathtt{L}=\left(\frac{\overline{\mathtt{v}}^1}{|\mathtt{v}^1|}\right)^{K-2}$. A
different solution of \eqref{eq:vec-as-mat} is provided by replacing
$\mathtt{L}\mapsto\mathtt{L}\,\text{diag}(1,\mathtt{W})$ with any
$U(K-1)$ matrix $\mathtt{W}$. See e.g.\ the discussion of coset spaces
of the unitary group in \cite{Gilmore:2012}. Notice that for $K=2$ the
matrix in \eqref{eq:para-mat} becomes
\begin{align}
  \label{eq:mat-2}
  \mathtt{L}=
  \frac{1}{\|\mathtt{v}\|}
  \begin{pmatrix}
    \mathtt{v}^1&-\overline{\mathtt{v}}^2\\
    \mathtt{v}^2&\overline{\mathtt{v}}^1\\
  \end{pmatrix}\,.
\end{align}

\subsubsection{Two-Site Invariant for
  \texorpdfstring{$\mathfrak{u}(1,1)$}{u(1,1)}}
\label{sec:inv21-u11}

To begin with, we evaluate the unitary Graßmannian integral
\eqref{eq:grass-int-barg} in the simplest case possible. That is for
$(N,K)=(2,1)$ and the bosonic algebra $\mathfrak{u}(1,1)$. In this
case \eqref{eq:grass-int-barg} becomes a $U(1)$ integral. We employ
the parameterization \eqref{eq:para-u1} of $U(1)$, the Haar measure
\eqref{eq:haar-u1} and the integrand $\mathscr{F}(\mathcal{C})$ given
in \eqref{eq:integrand21}. Then the integral \eqref{eq:grass-int-barg}
evaluates to
\begin{align}
  \label{eq:inv21-u11}
  \Psi_{2,1}(\boldsymbol{\lambda},\overline{\boldsymbol{\lambda}})
  =2i\,
  \delta^{1}(P)\left(-\frac{\lambda^1_1}{\lambda^2_1}\right)^{c_1-1}\,
\end{align}
with the momentum conserving delta function given in
\eqref{eq:def-delta-bosonic}. To obtain this result we trade the one
complex delta function in \eqref{eq:grass-int-barg} for two real
ones. Using \eqref{eq:delta-change-var} we change variables such that
one real delta function constrains the phase of the original complex
argument and the other one the square of its absolute value. The delta
function constraining the phase disappears because of the $U(1)$
integral. The other one remains in \eqref{eq:inv21-u11} implementing
momentum conservation.

\subsubsection{Four-Site Invariant for
  \texorpdfstring{$\mathfrak{u}(1,1)$}{u(1,1)}}
\label{sec:inv42-u11}

The $\mathfrak{u}(1,1)$ Yangian invariant with $(N,K)=(4,2)$ is of
importance primarily for two reasons. First, we will learn how to
utilize the tool introduced at the end of section~\ref{sec:tools} for
the evaluation of the Graßmannian integral
\eqref{eq:grass-int-barg}. Second, the resulting invariant shares a
characteristic feature with the six-site invariant of
$\mathfrak{u}(2,2)$, which is related to the simplest tree-level
$\text{NMHV}$ gluon amplitude.

For the case under consideration the Graßmannian integral
\eqref{eq:grass-int-barg} is equipped with the $U(2)$ contour given in
\eqref{eq:para-u2}, the corresponding Haar measure \eqref{eq:haar-u2}
and the function $\mathscr{F}(\mathcal{C})$ from
\eqref{eq:integrand42}. Let us denote the first columns of the
matrices $\boldsymbol{\lambda}^{\text{d}}$ and
$\boldsymbol{\lambda}^{\text{o}}$ by
$\boldsymbol{\lambda}^{\text{d}}_1$ and
$\boldsymbol{\lambda}^{\text{o}}_1$, respectively. This notation is
superfluous at this point because for $\mathfrak{u}(1,1)$ these
matrices consist only of one column. However, it will become necessary
in subsequent examples. For the evaluation of
\eqref{eq:grass-int-barg} we express the vectors
$\boldsymbol{\lambda}^{\text{d}}_1$ and
$\boldsymbol{\lambda}^{\text{o}}_1$ in terms of $U(2)$ matrices
$\mathtt{L}^{\text{d}}_1$ and $\mathtt{L}^{\text{o}}_1$ that obey
\begin{align}
  \label{eq:int-42-u11-matrices-prop}
  \mathtt{L}^{\text{d}}_1
  \begin{pmatrix}
    1\\
    0\\
  \end{pmatrix}  
  =
  \frac{\boldsymbol{\lambda}^{\text{d}}_1}{\|\boldsymbol{\lambda}^{\text{d}}_1\|}\,,
  \quad  
  \mathtt{L}^{\text{o}}_1
  \begin{pmatrix}
    1\\
    0\\
  \end{pmatrix}
  =
  \frac{\boldsymbol{\lambda}^{\text{o}}_1}{\|\boldsymbol{\lambda}^{\text{o}}_1\|}\,.
\end{align}
These matrices are parameterized in terms of the vectors as defined in
\eqref{eq:para-mat} or more explicitly in \eqref{eq:mat-2}. The $U(2)$
integral \eqref{eq:grass-int-barg} is conveniently addressed after
introducing the new integration variable
$(\mathtt{L}_1^{\text{d}})^\dagger\mathcal{C}\mathtt{L}_1^{\text{o}}$
and using the left and right invariance of the Haar measure. Employing
\eqref{eq:delta-matrix}, the arguments of the delta functions then
become independent of $\boldsymbol{\lambda}^{\text{d}}$ and
$\boldsymbol{\lambda}^{\text{o}}$. After parameterizing the new $U(2)$
integration variable as in \eqref{eq:para-u2}, we obtain
\begin{align}
  \label{eq:int-42-u11-momentum}
  \Psi_{4,2}(\boldsymbol{\lambda},\overline{\boldsymbol{\lambda}})=
  4 i\,  
  \frac{\delta^{1}(P)}
  {\lambda_1^1\overline{\lambda}_1^1+\lambda_1^2\overline{\lambda}_1^2}\,
  \mathcal{I}(v_1-v_2,c_1,c_2)\,.
\end{align}
Here we assumed
$\lambda_1^1\overline{\lambda}_1^1+\lambda_1^2\overline{\lambda}_1^2\neq
0$
and the momentum conserving delta function is defined in
\eqref{eq:def-delta-bosonic}. The remaining integral is
\begin{align}
  \label{eq:int-42-u11-before-lastint}
    \mathcal{I}(v_1-v_2,c_1,c_2)&=\int_0^{2\pi}\D \gamma\, \mathscr{F}(\mathcal{C}(\gamma))\,,
\end{align}
where the function $\mathscr{F}$ depends on the parameters
$v_1-v_2,c_1,c_2$ and
\begin{align}
  \begin{aligned}
    \label{eq:int-42-u11-lastint-general}
    \mathcal{C}(\gamma)
    &=
    \mathtt{L}^{\text{d}}_1\,\text{diag}(-1,-e^{i\gamma})(\mathtt{L}^{\text{o}}_1)^\dagger\\
    &=
    \frac{1}{\lambda_1^1\overline{\lambda}_1^1+\lambda_1^2\overline{\lambda}_1^2}
    \begin{pmatrix}    
      -\lambda^1_1\overline{\lambda}^3_1-\lambda^4_1\overline{\lambda}^2_1e^{i\gamma}&
      -\lambda^1_1\overline{\lambda}^4_1+\lambda^3_1\overline{\lambda}^2_1e^{i\gamma}\\
      -\lambda^2_1\overline{\lambda}^3_1+\lambda^4_1\overline{\lambda}^1_1e^{i\gamma}&
      -\lambda^2_1\overline{\lambda}^4_1-\lambda^3_1\overline{\lambda}^1_1e^{i\gamma}\\
    \end{pmatrix}\,.
  \end{aligned}
\end{align}
This is basically the original unitary integration variable of the
Graßmannian integral \eqref{eq:grass-int-barg} after most of its
degrees of freedom have been fixed in terms of the external data
$\boldsymbol{\lambda}$ by the delta functions in its integrand. Using
the form of $\mathscr{F}$ specified in \eqref{eq:integrand42}, the
integral \eqref{eq:int-42-u11-before-lastint} becomes
\begin{align}
  \label{eq:int-42-u11-lastint}
    \mathcal{I}(v_1-v_2,c_1,c_2)= \int_0^{2\pi}\D \gamma
    \frac{1}{(e^{i\gamma})^{1-c_2}
      |\mathtt{A}-e^{i\gamma}\mathtt{B}|^{2(1+v_1-v_2)}
      (\mathtt{A}-e^{i\gamma}\mathtt{B})^{c_2-c_1}}
\end{align}
with
\begin{align}
  \label{eq:int-42-u11-ab}  
  \mathtt{A}=\frac{\lambda^1_1\overline{\lambda}^3_1}
  {\lambda_1^1\overline{\lambda}_1^1+\lambda_1^2\overline{\lambda}_1^2}\,,\quad
  \mathtt{B}=\frac{-\lambda^4_1\overline{\lambda}^2_1}
  {\lambda_1^1\overline{\lambda}_1^1+\lambda_1^2\overline{\lambda}_1^2}\,.
\end{align}

It remains to evaluate the one-dimensional integral
\eqref{eq:int-42-u11-lastint}. Focusing on equal representation labels
$c_1=c_2$, we rewrite this integral as
\begin{align}
  \label{eq:int-42-u11-lastint-different}
  \mathcal{I}(z,c_1,c_1)=
  \big||\mathtt{A}|^2-|\mathtt{B}|^2\big|^{-1-z}
  \int_0^{2\pi}\D\gamma\,\big(w+\sqrt{w^2-1}\cos(\gamma+\text{arg}(-\overline{\mathtt{A}}\mathtt{B}))\big)^{-1-z}e^{i\gamma(c_1-1)}
\end{align}
with the variable
$w=(|\mathtt{A}|^2+|\mathtt{B}|^2)(||\mathtt{A}|^2-|\mathtt{B}|^2|)^{-1}$
in the range $[1,\infty)$. This formula is independent of the branch
of the $\text{arg}$ function. We disregard the case
$|\mathtt{A}|=|\mathtt{B}|$ from now on. To identify the integral
\eqref{eq:int-42-u11-lastint-different} with a known function, let us
recall some results about the Legendre function $P_\nu^\mu(u)$, where
we follow the conventions of \cite{Bateman:2007}.\footnote{These
  conventions do not comply with the implementation of those functions
  in the computer algebra program Mathematica
  \cite{Wolfram:2014}. Mathematica can be used after expressing the
  Legendre functions in terms of hypergeometric ones with
  \eqref{eq:legendre-hyper} below.} It has the integral representation
\begin{align}
  \label{eq:legendre-int}
  P_\nu^m(u)=
  \frac{\Gamma(\nu+m+1)}{2\pi\,\Gamma(\nu+1)}
  \int_{-\pi}^\pi\D t\,\big(u+\sqrt{u^2-1}\cos(t)\big)^\nu e^{imt}
\end{align}
for a non-negative integer $m$, $\nu\in\mathbb{C}$ and $\Real u>0$.
Furthermore, it obeys
\begin{align}
  \label{eq:legendre-prop}
  P^m_\nu(u)=\frac{\Gamma(\nu+m+1)}{\Gamma(\nu-m+1)}P^{-m}_\nu(u)\,,
  \quad
  P^\mu_\nu(u)=P^\mu_{-\nu-1}(u)  
\end{align}
with $\mu\in\mathbb{C}$. This function can also be expressed in terms
of a hypergeometric function,
\begin{align}
  \label{eq:legendre-hyper}
  P^\mu_\nu(u)=2^{-\nu}(u+1)^{\frac{\mu}{2}+\nu}(u-1)^{-\frac{\mu}{2}}\frac{{}_2F_1(-\nu,-\nu-\mu;1-\mu;\frac{u-1}{u+1})}{\Gamma(1-\mu)}
\end{align}
for $|u-1|<|u+1|$. Using \eqref{eq:legendre-int} and
\eqref{eq:legendre-prop}, we identify
\eqref{eq:int-42-u11-lastint-different} with
\begin{align}
  \label{eq:int-42-u11-lastint-eval}
  \mathcal{I}(z,c_1,c_1)=
  \left(\frac{-\overline{\mathtt{A}}\mathtt{B}}
    {|\mathtt{A}||\mathtt{B}|}\right)^{1-c_1}
  \frac{2\pi}{\big||\mathtt{A}|^2-|\mathtt{B}|^2\big|^{1+z}}  
  \frac{\Gamma(-z)}{\Gamma(-z+c_1-1)}P_z^{-1+c_1}
  \left(\frac{|\mathtt{A}|^2+|\mathtt{B}|^2}
    {\big||\mathtt{A}|^2-|\mathtt{B}|^2\big|}\right)\,.
\end{align}
This expression inserted into \eqref{eq:int-42-u11-momentum} is our
final form of the four-site Yangian invariant
$\Psi_{4,2}(\boldsymbol{\lambda},\overline{\boldsymbol{\lambda}})$ for
the algebra $\mathfrak{u}(1,1)$.

The expression \eqref{eq:int-42-u11-lastint-eval} contains the complex
deformation parameter $z$. In the undeformed limit $z\to 0$ it
simplifies using \eqref{eq:legendre-hyper} together with
${}_2F_1(0,\,\cdot\,;\,\cdot\,;\,\cdot)=1$,
\begin{align}
  \label{eq:int-42-u11-lastint-def}
  \mathcal{I}(0,c_1,c_1)=
  \left(\frac{\overline{\mathtt{A}}\mathtt{B}}{|\mathtt{A}||\mathtt{B}|}\right)^{1-c_1}
  \frac{2\pi}{\big||\mathtt{A}|^2-|\mathtt{B}|^2\big|}
  \left(
    \frac{|\mathtt{A}|^2+|\mathtt{B}|^2-\big||\mathtt{A}|^2-|\mathtt{B}|^2\big|}
    {|\mathtt{A}|^2+|\mathtt{B}|^2+\big||\mathtt{A}|^2-|\mathtt{B}|^2\big|}
  \right)^{\frac{|1-c_1|}{2}}\,.
\end{align}
Note that in case of $1-c_1<0$, the fraction of gamma functions in
\eqref{eq:int-42-u11-lastint-eval} diverges for $z\to 0$ whereas the
Legendre function vanishes. Thus in this case we have to apply the
first identity in \eqref{eq:legendre-prop} before taking $z\to 0$.
Lastly, we can write \eqref{eq:int-42-u11-lastint-def} more explicitly
as
\begin{align}
  \label{eq:int-42-u11-lastint-def-expl}
  \mathcal{I}(0,c_1,c_1)=
  \frac{2\pi}{\big||\mathtt{A}|^2-|\mathtt{B}|^2\big|}
  \begin{cases}
    \left(\frac{\mathtt{B}}{\mathtt{A}}\right)^{1-c_1}
    &\text{for}\quad
    \begin{aligned}
      |\mathtt{A}|&>|\mathtt{B}|,\,\\ c_1 &\leq 1
    \end{aligned}    
    \quad\text{or}\quad
    \begin{aligned}
    |\mathtt{A}|&<|\mathtt{B}|,\,\\ c_1 &\geq 1\,,  
    \end{aligned}
    \\[0.6cm]
    \left(\frac{\overline{\mathtt{A}}}{\overline{\mathtt{B}}}\right)^{1-c_1}
    &\text{for}\quad
    \begin{aligned}
      |\mathtt{A}|&<|\mathtt{B}|,\,\\ c_1 &\leq 1
    \end{aligned}    
    \quad\text{or}\quad
    \begin{aligned}
    |\mathtt{A}|&>|\mathtt{B}|,\,\\ c_1 &\geq 1\,.  
    \end{aligned}
    \\
  \end{cases}
\end{align}
Notice that for $c_1=1$ the expressions for the different cases
coincide. In appendix~\ref{sec:discrete-symmetry} we study a discrete
parity symmetry of the Graßmannian integral
\eqref{eq:grass-int-barg}. In case of the invariant
$\Psi_{4,2}(\boldsymbol{\lambda},\overline{\boldsymbol{\lambda}})$ for
the algebra $\mathfrak{u}(1,1)$ it exchanges the two regions
$|\mathtt{A}|>|\mathtt{B}|$ and $|\mathtt{B}|>|\mathtt{A}|$ in
\eqref{eq:int-42-u11-lastint-def-expl}.

\begin{figure}[!t]
  \begin{center}
    \begin{tikzpicture}
      \draw[thick,densely dotted,
      decoration={markings, mark=at position 0.0 with {\arrow{latex reversed}}},
      postaction={decorate}]
      (3,0) node[right=.2] {$\Real{u}$} -- (-3,0);
      \draw[thick,densely dotted,
      decoration={markings, mark=at position 0.0 with {\arrow{latex reversed}}},
      postaction={decorate}]
      (0,3) node[above=.2] {$\Imag{u}$} -- (0,-3);
      \draw[thick,decoration={markings, mark=at position 0.125 with {\arrow{latex}}},
      postaction={decorate}] (0,0) circle [radius=2];
      \filldraw[thick] (0,0) circle (1pt) node[below left]{$0$};
      \filldraw[thick] (2.5,2.5) circle (1pt) node[below right]{$\infty$};
      \filldraw[thick] (55:1.5) circle (1pt) node[below right]
      {$\frac{\mathtt{A}}{\mathtt{B}}$};
      \filldraw[thick] (55:2.66) circle (1pt) node[below right]
      {$\frac{\overline{\mathtt{B}}}{\overline{\mathtt{A}}}$};
      \node at (2,0) [below right] {$1$};
    \end{tikzpicture}
    \caption{The computation of the undeformed Yangian invariant
      $\Psi_{4,2}(\boldsymbol{\lambda},\overline{\boldsymbol{\lambda}})$
      with $v_1=v_2$ and $c_1=c_2$ for $\mathfrak{u}(1,1)$ can be
      reduced to the one-dimensional complex contour integral
      \eqref{eq:inv42-u11-contourint} in the variable $u$. Its contour
      is depicted by a solid circle. The integrand has poles whose
      positions we indicate by a dot. Here we display these for sample
      external data $\boldsymbol{\lambda}$ satisfying
      $|\mathtt{A}|<|\mathtt{B}|$. The relation between
      $\boldsymbol{\lambda}$ and $\mathtt{A},\mathtt{B}$ is given in
      \eqref{eq:int-42-u11-ab}. In case of
      $|\mathtt{A}|>|\mathtt{B}|$, the pole at
      $\frac{\overline{\mathtt{B}}}{\overline{\mathtt{A}}}$ moves
      inside of the contour and that at
      $\frac{\mathtt{A}}{\mathtt{B}}$ is outside.}
    \label{fig:grint-inv42-u11-contour}
  \end{center}
\end{figure}
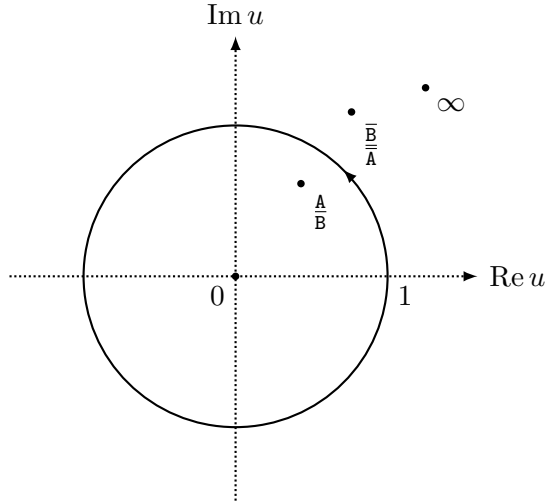

We may obtain the result \eqref{eq:int-42-u11-lastint-def-expl} for
the undeformed case alternatively by rewriting
\eqref{eq:int-42-u11-lastint} as a contour integral in the variable
$u=e^{i\gamma}$,
\begin{align}
  \label{eq:inv42-u11-contourint}
  \mathcal{I}(0,c_1,c_1)=
  \frac{i}{\overline{\mathtt{A}}\mathtt{B}}\oint\D u
  \frac{1}{u^{1-c_1}\Big(u-\frac{\mathtt{A}}{\mathtt{B}}\Big)
  \Big(u-\frac{\overline{\mathtt{B}}}{\overline{\mathtt{A}}}\Big)}\,,
\end{align}
where the contour of integration is the counterclockwise unit circle,
see figure \ref{fig:grint-inv42-u11-contour}. The integrand can have
poles at four points. There can be a pole at $u=0$, which is inside of
the contour, and one at $u=\infty$, which is outside. Furthermore,
there is a pair of poles at
$u=\frac{\mathtt{A}}{\mathtt{B}},\frac{\overline{\mathtt{B}}}{\overline{\mathtt{A}}}$.
For all values of $\mathtt{A}$ and $\mathtt{B}$, one of these poles is
inside the contour, whereas the other one is outside. Which pole is
inside depends on whether $|\mathtt{A}|$ or $|\mathtt{B}|$ is larger,
and therefore on the external data $\boldsymbol{\lambda}$, cf.\
\eqref{eq:int-42-u11-ab}. In combination, there are always two poles
inside the unit circle. The integral \eqref{eq:inv42-u11-contourint}
is then computed easily employing the residue theorem. To obtain
\eqref{eq:int-42-u11-lastint-def-expl} in this way, we observe that
the residues at $u=0$ and $u=\infty$ vanish for certain values of
$c_1$ and we use that the sum of all residues is equal to zero. We
want to emphasize that the varying position of the pair of poles at
$u=\frac{\mathtt{A}}{\mathtt{B}},\frac{\overline{\mathtt{B}}}{\overline{\mathtt{A}}}$
is the reason for the case distinction in
\eqref{eq:int-42-u11-lastint-def-expl} for fixed $c_1$.

Let us comment on the relevance of this calculation for the six-site
invariant of $\mathfrak{u}(2,2)$, and thus for the six-particle
tree-level $\text{NMHV}$ gluon amplitude. In the discussion of that
invariant we will encounter an integral which is very similar to
\eqref{eq:inv42-u11-contourint}. The main difference being that there
will be two pairs of poles instead of only one here.

\subsubsection{Four-Site Invariant for
  \texorpdfstring{$\mathfrak{u}(2,2)$}{u(2,2)} and MHV Gluon
  Amplitude}
\label{sec:inv42-u22}

Finally, we are ready compute the first amplitude by means of the
unitary Graßmannian integral \eqref{eq:grass-int-barg}. We will
identify this integral for four sites, the algebra $\mathfrak{u}(2,2)$
and a certain choice of the deformation parameters $v_i$ and the
representation labels $c_i$ with the four-particle $\text{MHV}$ gluon
amplitude $A^{(\text{tree})}_{4,2}$ from \eqref{eq:amp-bos-mhv}. For
general values of $v_i$ and $c_i$ it will be matched with a gluonic
component of the deformed superamplitude
$\mathcal{A}^{(\text{def.})}_{4,2}$ in \eqref{eq:amp-4-2-def}.

For this purpose, we evaluate the integral \eqref{eq:grass-int-barg}
with the $U(2)$ parameterization \eqref{eq:para-u2}, the Haar measure
\eqref{eq:haar-u2} and the function $\mathscr{F}(\mathcal{C})$ given
in \eqref{eq:integrand42}. We express the first column
$\boldsymbol{\lambda}^{\text{d}}_1$ of the matrix
$\boldsymbol{\lambda}^{\text{d}}$ and
$\boldsymbol{\lambda}^{\text{o}}_1$ of
$\boldsymbol{\lambda}^{\text{o}}$ in terms of the $U(2)$ matrices
defined already in \eqref{eq:int-42-u11-matrices-prop}. The delta
function in \eqref{eq:grass-int-barg} that involves
$\boldsymbol{\lambda}_1^{\text{d}}$ and
$\boldsymbol{\lambda}_1^{\text{o}}$ is then addressed analogously to
the discussion of the $\mathfrak{u}(1,1)$ invariant in the previous
section~\ref{sec:inv42-u11}. This leads to
\begin{align}
  \label{eq:inv42-u22-step1}
  \begin{aligned}
    \Psi_{4,2}(\boldsymbol{\lambda},\overline{\boldsymbol{\lambda}})=
    4
    i\,\frac{\delta(\|\boldsymbol{\lambda}^{\text{d}}_1\|^2-\|\boldsymbol{\lambda}_1^{\text{o}}\|^2)}{\|\boldsymbol{\lambda}^{\text{d}}_1\|^2}
    \int_0^{2\pi}&\D\gamma\,
    \mathscr{F}\left(\mathtt{L}^{\text{d}}_1\,\text{diag}(-1,-e^{i\gamma})
      (\mathtt{L}_1^{\text{o}})^\dagger
    \right)\\
    &\cdot\delta_{\mathbb{C}}^2\left((\mathtt{L}^{\text{d}}_1)^\dagger\boldsymbol{\lambda}^{\text{d}}_2+
      \text{diag}(-1,-e^{i\gamma}) (\mathtt{L}^{\text{o}}_1)^\dagger\boldsymbol{\lambda}^{\text{o}}_2
    \right)\,,
  \end{aligned}
\end{align}
where we assumed $\|\boldsymbol{\lambda}^{\text{d}}_1\|\neq 0$. The evaluation of the
remaining integral yields
\begin{align}  
  \label{eq:inv42-u22-step2}
  \Psi_{4,2}(\boldsymbol{\lambda},\overline{\boldsymbol{\lambda}})=
  8i\,
  \delta^{4}(P)
  \mathscr{F}(\mathcal{C})
\end{align}
with 
\begin{align}
  \label{eq:inv42-u22-e}
  \mathcal{C}=
  \mathtt{L}^{\text{d}}_1\,
  \text{diag}\left(-1,\frac{\langle 21 \rangle}{\langle 3 4 \rangle}\right)
  (\mathtt{L}_1^{\text{o}})^\dagger
\end{align}
and the momentum conserving delta function
\eqref{eq:def-delta-bosonic} in the form
\begin{align}
  \label{eq:inv42-u22-mom-delta}
  \delta^{4}(P)=
  \delta(\|\boldsymbol{\lambda}_1^{\text{d}}\|^2-\|\boldsymbol{\lambda}_1^{\text{o}}\|^2)
  \delta(\|\boldsymbol{\lambda}_2^{\text{d}}\|^2-\|\boldsymbol{\lambda}_2^{\text{o}}\|^2)
  \delta_{\mathbb{C}}((\boldsymbol{\lambda}_1^{\text{d}})^\dagger\boldsymbol{\lambda}_2^{\text{d}}
  -(\boldsymbol{\lambda}_1^{\text{o}})^\dagger\boldsymbol{\lambda}_2^{\text{o}})\,.
\end{align}
Here we used \eqref{eq:delta-change-var} to manipulate the delta
function in \eqref{eq:inv42-u22-step1}. In particular, the absolute
value and the first component of the $\mathbb{C}^2$-vector in its
argument directly lead to the delta functions in
\eqref{eq:inv42-u22-mom-delta}. Moreover, the phase $e^{i\gamma}$ is
fixed by the remaining integral in \eqref{eq:inv42-u22-step1} and
expressed in terms of the angle bracket defined in
\eqref{eq:spinors-brackets}.

As a brief interlude, we recall for convenience the definition of said
bracket from \eqref{eq:spinors-brackets},
\begin{align}
  \label{eq:bracket}
  \langle ij\rangle=\lambda^i_1\lambda^j_2-\lambda^i_2\lambda^j_1\,.
\end{align}
Let us also recapitulate the definition of the square bracket from
that equation,
\begin{align}
  [ij]=-\tilde{\lambda}^i_1\tilde{\lambda}^j_2+\tilde{\lambda}^i_2\tilde{\lambda}^j_1\,.
\end{align}
Using the reality conditions for the spinor helicity variables in
\eqref{eq:def-lambda-tilde} we obtain, cf.\
\eqref{eq:spinors-brackets-conj},
\begin{align}
  \label{eq:brackets-cc-angle}
  [k k']=-\overline{\langle k k'\rangle}\,,\quad
  [l l']=-\overline{\langle l l'\rangle}\,,\quad
  [k l]=\overline{\langle k l\rangle}
\end{align}
for dual sites $k,k'=1,\ldots,K$ and ordinary sites
$l,l'=K+1,\ldots,N$. The brackets obey the Schouten identity, cf.\
\eqref{eq:spinors-schouten},
\begin{align}
  \label{eq:bracket-schouten}
  \langle ij\rangle\langle kl\rangle-\langle ik\rangle\langle jl\rangle=\langle il\rangle\langle kj\rangle\,.
\end{align}
Furthermore, due to momentum conservation
\eqref{eq:inv42-u22-mom-delta} we have
\begin{align}
  \label{eq:bracket-mom-4}
  \sum_{k=1}^{K}\langle ik\rangle\overline{\langle kj\rangle}=\sum_{l=K+1}^{N}\langle il\rangle\overline{\langle lj\rangle}\,
  \quad\text{or equivalently}\quad
  \sum_{h=1}^N\langle ih\rangle[hj]=0\,,
\end{align}
where $i$ and $j$ are can denote any site. The latter from of this
condition was presented already in \eqref{eq:spinors-momcons}. As
indicated by the notation, the formulas in this paragraph are valid
for general $N$ and $K$. We will utilize them also for further sample
Yangian invariants below.

Let us return to the evaluation of
$\Psi_{4,2}(\boldsymbol{\lambda},\overline{\boldsymbol{\lambda}})$.
Note that using \eqref{eq:bracket-mom-4} the combination of brackets
in \eqref{eq:inv42-u22-e}, which originated from $e^{i\gamma}$, is
indeed a phase. Moreover, with \eqref{eq:bracket-schouten} and
\eqref{eq:bracket-mom-4} we obtain for the entire matrix in
\eqref{eq:inv42-u22-e}
\begin{align}
  \label{eq:inv42-u22-e-eval}
  \mathcal{C}
  =
  \frac{1}{\langle 3 4 \rangle}
  \begin{pmatrix}
    \langle 4 1 \rangle&\langle 1 3 \rangle\\
    \langle 4 2 \rangle&\langle 2 3 \rangle\\
  \end{pmatrix}
  =
  \frac{1}{\,\overline{\langle 12\rangle}\,}
  \begin{pmatrix}
  \overline{\langle 23\rangle}&\overline{\langle 24\rangle}\\
  \overline{\langle 31\rangle}&\overline{\langle 41\rangle}\\
  \end{pmatrix}
\end{align}
and verify the unitarity of this matrix. Finally, we insert the
function $\mathscr{F}$ from \eqref{eq:integrand42} into
\eqref{eq:inv42-u22-step2} to obtain the Yangian invariant with
$(N,K)=(4,2)$ for the algebra $\mathfrak{u}(2,2)$,
\begin{align}
  \label{eq:inv42-u22-final}
  \Psi_{4,2}(\boldsymbol{\lambda},\overline{\boldsymbol{\lambda}})=8i\,
  \delta^{4}(P)
  \frac{\langle 34 \rangle^4}{\langle 12 \rangle\langle 23 \rangle\langle 34 \rangle\langle 41 \rangle}
  \left(\frac{\langle 14 \rangle}{\langle 34 \rangle}\right)^{c_1}
  \left(\frac{\langle 12 \rangle}{\langle 14 \rangle}\right)^{c_2}
  \left(\frac{\langle 34 \rangle\overline{\langle 34\rangle}}{\langle 14 \rangle\overline{\langle 14\rangle}}\right)^{v_1-v_2}\,.
\end{align}
We conclude by identifying this invariant with known results. We start
by focusing on the representation labels $c_1=c_2=0$. According to
\eqref{eq:hel-rep-label} the four sites then carry the helicities
$(+1,+1,-1,-1)$. If in addition $v_1=v_2$, the invariant reduces, up
to a prefactor, to the well-known Parke-Taylor formula
\eqref{eq:amp-bos-mhv} for the tree-level $\text{MHV}$ gluon amplitude
$A^{(\text{tree})}_{4,2}$. We move on the discuss the case of general
deformation parameters $c_1,c_2\in\mathbb{Z}$ and
$v_1,v_2\in\mathbb{C}$. Observe that the invariant
\eqref{eq:inv42-u22-final} is a single-valued function in
$\lambda^i_\alpha$ because the complex deformation parameters appear
only as the exponent of a non-negative real number. This property can
be traced back to the single-valuedness of the integrand of the
unitary Graßmannian formula discussed in
section~\ref{sec:single-valu-integr}. To identify
\eqref{eq:inv42-u22-final} with a deformed amplitude from the
introductory section~\ref{sec:deform}, we have to give up the manifest
single-valuedness. Employing the momentum conservation in
\eqref{eq:bracket-mom-4}, we write
\begin{align}
  \label{eq:inv42-u22-real-brackets}
  \frac{\langle 34 \rangle\overline{\langle 34\rangle}}{\langle 14 \rangle\overline{\langle 14\rangle}}=
  \frac{\langle 12\rangle\langle 34\rangle}{\langle 23\rangle\langle 41\rangle}\,.
\end{align}
This equality shows that \eqref{eq:inv42-u22-final} is, up to a
numerical constant, a gluonic component $A^{(\text{def.})}_{4,2}$ of
the deformed superamplitude $\mathcal{A}^{(\text{def.})}_{4,2}$ given
in \eqref{eq:amp-4-2-def}. Merely the deformation parameters $v_1,v_2$
are parameterized in a slightly different way in that formula.

\subsubsection{Six-Site Invariant for
  \texorpdfstring{$\mathfrak{u}(2,2)$}{u(2,2)} and NMHV Gluon
  Amplitude}
\label{sec:inv63-u22}

After the successful identification of the first amplitude, we move on
to the evaluation of the unitary Graßmannian integral
\eqref{eq:grass-int-barg} for six sites and the algebra
$\mathfrak{u}(2,2)$. From the usual Graßmannian integral approach to
amplitudes reviewed in section~\ref{sec:grassmannian-integral} we
expect our invariant to be related to the six-particle $\text{NMHV}$
gluon amplitude $A^{(\text{tree})}_{6,3}$. This is a crucial test of
our method because it is the first instance where one has to fix a
integration contour ``by hand'' in the usual approach. We do not have
this freedom with our unitary contour. Therefore it has to produce the
correct result automatically. Furthermore, the unitary Graßmannian
integral \eqref{eq:grass-int-barg} naturally includes deformation
parameters. Thus we may also hope to relate it to the sought-after
deformed gluon amplitude $A^{(\text{def.})}_{6,3}$. Recall from
section~\ref{sec:deform} that the construction of the deformed
superamplitude $\mathcal{A}^{(\text{def.})}_{6,3}$, and thus also of
$A^{(\text{def.})}_{6,3}$, is still an open problem.

We want to compute the Graßmannian integral \eqref{eq:grass-int-barg}
for $\Psi_{6,3}(\boldsymbol{\lambda},\overline{\boldsymbol{\lambda}})$
with the $U(3)$ contour \eqref{eq:para-u3}, the associated Haar
measure \eqref{eq:haar-u3} and the function $\mathscr{F}(\mathcal{C})$
in \eqref{eq:integrand63}. To address those delta functions in the
Graßmannian integral involving the first column
$\boldsymbol{\lambda}^{\text{d}}_1$ of the matrix
$\boldsymbol{\lambda}^{\text{d}}$ and
$\boldsymbol{\lambda}^{\text{o}}_1$ of
$\boldsymbol{\lambda}^{\text{o}}$, we introduce $U(3)$ matrices
$\mathtt{L}^{\text{d}}_1$ and $\mathtt{L}^{\text{o}}_1$ that satisfy
\begin{align}
  \mathtt{L}^{\text{d}}_1
  \begin{pmatrix}
    1\\
    0\\
    0\\
  \end{pmatrix}
  =
  \frac{\boldsymbol{\lambda}^{\text{d}}_1}{\|\boldsymbol{\lambda}^{\text{d}}_1\|}\,,\quad
  \mathtt{L}^{\text{o}}_1
  \begin{pmatrix}
    1\\
    0\\
    0\\
  \end{pmatrix}
  =
  \frac{\boldsymbol{\lambda}^{\text{o}}_1}{\|\boldsymbol{\lambda}^{\text{o}}_1\|}\,.
\end{align}
These matrices are parameterized in terms of the column vectors as
defined in \eqref{eq:para-mat}. Then we change the integration
variable of the $U(3)$ Graßmannian integral from $\mathcal{C}$ to
$(\mathtt{L}^{\text{d}}_1)^\dagger\mathcal{C}\,\mathtt{L}^{\text{o}}_1$
and parameterize this new variable as in \eqref{eq:para-u3}. The delta
functions containing $\boldsymbol{\lambda}^{\text{d}}_1$ and
$\boldsymbol{\lambda}^{\text{o}}_1$ then fix parts of the integration
variable such that we are left with the $U(2)$ integral
\begin{align}
  \label{eq:inv63-u22-step1}
  \begin{aligned}
    \Psi_{6,3}(\boldsymbol{\lambda},\overline{\boldsymbol{\lambda}})
    =\;&
    8\frac{\delta(\|\boldsymbol{\lambda}^{\text{d}}_1\|^2-\|\boldsymbol{\lambda}^{\text{o}}_1\|^2)}{\|\boldsymbol{\lambda}^{\text{d}}_1\|^4}
    \delta_{\mathbb{C}}([(\mathtt{L}_1^{\text{d}})^\dagger\boldsymbol{\lambda}_2^{\text{d}}]^1-[(\mathtt{L}_1^{\text{o}})^\dagger\boldsymbol{\lambda}_2^{\text{o}}]^1)(\chi_2 i)^{-1}\\
    &\cdot    
    \int_{U(2)}[\D \mathcal{D}]\,\mathscr{F}\left(\mathtt{L}^{\text{d}}_1\,\text{diag}(-1,\mathcal{D})(\mathtt{L}^{\text{o}}_1)^\dagger\right)
    \delta_{\mathbb{C}}^2([(\mathtt{L}_1^{\text{d}})^\dagger\boldsymbol{\lambda}_2^{\text{d}}]\tilde{\phantom{l}}+\mathcal{D}[(\mathtt{L}_1^{\text{o}})^\dagger\boldsymbol{\lambda}_2^{\text{o}}]\tilde{\phantom{l}})\,,
  \end{aligned}
\end{align}
where we assumed $\|\boldsymbol{\lambda}^{\text{d}}_1\|\neq 0$. Here,
as in \eqref{eq:para-mat},
$[(\mathtt{L}_1^{\text{d}})^\dagger\boldsymbol{\lambda}_2^{\text{d}}]^1$
denotes the first component of the vector
$(\mathtt{L}_1^{\text{d}})^\dagger\boldsymbol{\lambda}_2^{\text{d}}$
and
$[(\mathtt{L}_1^{\text{d}})^\dagger\boldsymbol{\lambda}_2^{\text{d}}]\tilde{\phantom{l}}$
refers to its remaining two components. To proceed with the $U(2)$
integral, we introduce the $U(2)$ matrices $\mathtt{L}^{\text{d}}_2$
and $\mathtt{L}^{\text{o}}_2$ obeying
\begin{align}
  \mathtt{L}^{\text{d}}_2
  \begin{pmatrix}
    1\\
    0\\
  \end{pmatrix}
  =
  \frac{[(\mathtt{L}_1^{\text{d}})^\dagger\boldsymbol{\lambda}_2^{\text{d}}]\tilde{\phantom{l}}}
  {\|[(\mathtt{L}_1^{\text{d}})^\dagger\boldsymbol{\lambda}_2^{\text{d}}]\tilde{\phantom{l}}\|}\,,\quad
  \mathtt{L}^{\text{o}}_2
  \begin{pmatrix}
    1\\
    0\\
  \end{pmatrix}
  =
  \frac{[(\mathtt{L}_1^{\text{o}})^\dagger\boldsymbol{\lambda}_2^{\text{o}}]\tilde{\phantom{l}}}
  {\|[(\mathtt{L}_1^{\text{o}})^\dagger\boldsymbol{\lambda}_2^{\text{o}}]\tilde{\phantom{l}}\|}\,.
\end{align}
These matrices are parametrized as shown in \eqref{eq:mat-2}. Then the
$U(2)$ integral in \eqref{eq:inv63-u22-step1} is evaluated after
changing variables from $\mathcal{D}$ to
$(\mathtt{L}^{\text{d}}_2)^\dagger\mathcal{D}\,\mathtt{L}^{\text{o}}_2$. This
yields
\begin{align}
  \label{eq:inv63-u22-final}
  \Psi_{6,3}(\boldsymbol{\lambda},\overline{\boldsymbol{\lambda}})
  =32\frac{\delta^{4}(P)}{s_{123}}\mathcal{I}(v_1-v_2,v_2-v_3,c_1,c_2,c_3)\,
\end{align}
with the momentum conserving delta function defined in
\eqref{eq:def-delta-bosonic}. Furthermore, we introduced
\begin{align}
  \label{eq:inv63-u22-kininv}
  \begin{aligned}
      s_{123}
      &=\|\boldsymbol{\lambda}^{\text{d}}_1\|^2\|[(\mathtt{L}_1^{\text{d}})^\dagger\boldsymbol{\lambda}_2^{\text{d}}]\tilde{\phantom{l}}\|^2
      =\|\boldsymbol{\lambda}^{\text{d}}_1\|^2\|\boldsymbol{\lambda}^{\text{d}}_2\|^2
      -(\boldsymbol{\lambda}^{\text{d}}_1)^\dagger\boldsymbol{\lambda}^{\text{d}}_2\,(\boldsymbol{\lambda}^{\text{d}}_2)^\dagger\boldsymbol{\lambda}^{\text{d}}_1\\
      &=\langle 12\rangle\overline{\langle 12\rangle}
      +\langle 13\rangle\overline{\langle 13\rangle}
      +\langle 23\rangle\overline{\langle 23\rangle}\,.
  \end{aligned}
\end{align}
Analogously one defines $s_{456}$ in terms of
$\boldsymbol{\lambda}^{\text{o}}$. Due to momentum conservation
\eqref{eq:bracket-mom-4} we have $s_{123}=s_{456}$. In
\eqref{eq:inv63-u22-final} there remains the $U(1)$ integral
\begin{align}
  \label{eq:inv63-u22-lastint}
  \begin{aligned}
    \mathcal{I}(v_1-v_2,v_2-v_3,c_1,c_2,c_3)
    =
    \int_0^{2\pi}\D\gamma\,
    \mathscr{F}\left(\mathcal{C}(\gamma)\right)\,,
  \end{aligned}
\end{align}
where the integral depends on the parameters $v_i$ and $c_i$ through
the function $\mathscr{F}$ and we introduced the unitary matrix
\begin{align}
  \mathcal{C}(\gamma)=
  \mathtt{L}^{\text{d}}_1\,\text{diag}(1,\mathtt{L}^{\text{d}}_2)\,
  \text{diag}(-1_2,-e^{i\gamma})
  \big(\mathtt{L}^{\text{o}}_1\,\text{diag}(1,\mathtt{L}^{\text{o}}_2)\big)^\dagger\,.
\end{align}
In essence, this is the integration variable of the original
Graßmannian integral \eqref{eq:grass-int-barg} after the $U(3)$
degrees of freedom, except for one $U(1)$ phase, have been fixed by
the delta functions in the integrand.

In order to obtain an explicit representation of the integral
\eqref{eq:inv63-u22-lastint} we compute
\begin{align}
  \begin{aligned}
    \mathtt{L}^{\text{d}}_1\,\text{diag}(1,\mathtt{L}^{\text{d}}_2)&=
    \begin{pmatrix}
      \frac{\lambda^1_1}{\|\boldsymbol{\lambda}^{\text{d}}_1\|}&
      \frac{-\overline{\lambda}^2_1\langle 12 \rangle-\overline{\lambda}^3_1\langle 13\rangle}{\sqrt{s_{123}}\|\boldsymbol{\lambda}^{\text{d}}_1\|}&
      \frac{\overline{\lambda}^1_1\overline{\langle 23\rangle}}{\sqrt{s_{123}}|\lambda^1_1|}\\
      \frac{\lambda^2_1}{\|\boldsymbol{\lambda}^{\text{d}}_1\|}&
      \frac{\overline{\lambda}^1_1\langle 12 \rangle-\overline{\lambda}^3_1\langle 23\rangle}{\sqrt{s_{123}}\|\boldsymbol{\lambda}^{\text{d}}_1\|}&
      \frac{\overline{\lambda}^1_1\overline{\langle 31\rangle}}{\sqrt{s_{123}}|\lambda^1_1|}\\
      \frac{\lambda^3_1}{\|\boldsymbol{\lambda}^{\text{d}}_1\|}&
      \frac{\overline{\lambda}^1_1\langle 13 \rangle+\overline{\lambda}^2_1\langle 23\rangle}{\sqrt{s_{123}}\|\boldsymbol{\lambda}^{\text{d}}_1\|}&
      \frac{\overline{\lambda}^1_1\overline{\langle 12\rangle}}{\sqrt{s_{123}}|\lambda^1_1|}
    \end{pmatrix}\,.
  \end{aligned}
\end{align}
The matrix
$\mathtt{L}^{\text{o}}_1\,\text{diag}(1,\mathtt{L}^{\text{o}}_2)$ is
of the same form with $\boldsymbol{\lambda}^{\text{d}}_1$ replaced by
$\boldsymbol{\lambda}^{\text{o}}_1$ and the site indices $1,2,3$ by
$4,5,6$, respectively. Using the Schouten identity
\eqref{eq:bracket-schouten} and momentum conservation
\eqref{eq:bracket-mom-4}, we obtain
\begin{align}
  \label{eq:inv63-u22-e-explicit}
  \begin{aligned}
    \mathcal{C}(\gamma)&=
    \frac{1}{s_{123}}
    \begin{pmatrix}
      \langle 1|2+3\overline{|4\rangle}&
      \langle 1|2+3\overline{|5\rangle}&
      \langle 1|2+3\overline{|6\rangle}\\
      \langle 2|1+3\overline{|4\rangle}&
      \langle 2|1+3\overline{|5\rangle}&
      \langle 2|1+3\overline{|6\rangle}\\
      \langle 3|1+2\overline{|4\rangle}&
      \langle 3|1+2\overline{|5\rangle}&
      \langle 3|1+2\overline{|6\rangle}\\
    \end{pmatrix}\\
    &\quad-\frac{1}{s_{123}}e^{i\gamma}\frac{\overline{\lambda}^1_1}{|\lambda^1_1|}\frac{\lambda^4_1}{|\lambda^4_1|}
    \begin{pmatrix}
      \langle 56\rangle \overline{\langle 23\rangle}&
      \langle 64\rangle \overline{\langle 23\rangle}&
      \langle 45\rangle \overline{\langle 23\rangle}\\
      \langle 56\rangle \overline{\langle 31\rangle}&
      \langle 64\rangle \overline{\langle 31\rangle}&
      \langle 45\rangle \overline{\langle 31\rangle}\\
      \langle 56\rangle \overline{\langle 12\rangle}&
      \langle 64\rangle \overline{\langle 12\rangle}&
      \langle 45\rangle \overline{\langle 12\rangle}\\
    \end{pmatrix}\,,
  \end{aligned}
\end{align}
where we employed the shorthand notation
$\langle 1|2+3\overline{|4\rangle}=\langle 12\rangle\overline{\langle
  24\rangle}+\langle 13\rangle\overline{\langle 34\rangle}$
etc. We work with the complex conjugate $\overline{\langle ij\rangle}$
of the angle brackets instead of the square brackets $[ij]$ to make
the analytic structure of the expressions more transparent. One can
easily translate between the two with the help of
\eqref{eq:brackets-cc-angle}. According to \eqref{eq:integrand63} the
function $\mathscr{F}$ in \eqref{eq:inv63-u22-lastint} involves the
minors
\begin{align}
  \label{eq:inv63-u22-minors-abcd}
  \begin{aligned}
    \det\mathcal{C}(\gamma)
    &=-e^{i\gamma}\frac{\overline{\lambda}^1_1}{|\lambda^1_1|}\frac{\lambda^4_1}{|\lambda^4_1|}\,,\\
    [1]_{\mathcal{C}(\gamma)}
    &
    =
    \mathtt{A}-e^{i\gamma}\frac{\overline{\lambda}^1_1}{|\lambda^1_1|}\frac{\lambda^4_1}{|\lambda^4_1|}\mathtt{B}
    \quad\text{with}\quad
    \mathtt{A}=\frac{\langle 1|2+3\overline{|4\rangle}}{s_{123}}\,,\quad
    \mathtt{B}=\frac{\langle 56\rangle\overline{\langle 23\rangle}}{s_{123}}\,,\\
    [3]_{\mathcal{C}(\gamma)}
    &
    =
    \overline{\mathtt{C}}-e^{i\gamma}\frac{\overline{\lambda}^1_1}{|\lambda^1_1|}\frac{\lambda^4_1}{|\lambda^4_1|}\overline{\mathtt{D}}
    \quad\text{with}\quad
    \mathtt{C}=\frac{\overline{\langle 3|}1+2|6\rangle}{s_{123}}\,,\quad
    \mathtt{D}=\frac{\overline{\langle 45\rangle}\langle 12\rangle}{s_{123}}\,.\\
  \end{aligned}
\end{align}
With these we can provide the desired explicit expression for the
$U(1)$ integral \eqref{eq:inv63-u22-lastint},
\begin{align}
  \label{eq:inv63-u22-lastint-explicit}
  \begin{aligned}
    \mathcal{I}(z_1,z_2,c_1,c_2,c_3)=
    \int_0^{2\pi}\D\gamma\,&\frac{1}{(-e^{i\gamma})^{2-c_2}
      |\mathtt{A}-e^{i\gamma}\mathtt{B}|^{2(1+z_1)}(\mathtt{A}-e^{i\gamma}\mathtt{B})^{c_2-c_1}}\\
    &\cdot\frac{1}{|\mathtt{C}-e^{-i\gamma}\mathtt{D}|^{2(1+z_2)}(\mathtt{C}-e^{-i\gamma}\mathtt{D})^{c_3-c_2}}\,,
  \end{aligned}
\end{align}
where we shifted the integration variable by the phase appearing e.g.\
in the first line of \eqref{eq:inv63-u22-minors-abcd}. Let us comment
that we were able to identify a similar integral in
\eqref{eq:int-42-u11-lastint}, which occurred for the invariant
$\Psi_{4,2}(\boldsymbol{\lambda},\overline{\boldsymbol{\lambda}})$ of
$\mathfrak{u}(1,1)$, with a Legendre function in
\eqref{eq:int-42-u11-lastint-eval}. It would be desirable to also
understand \eqref{eq:inv63-u22-lastint-explicit} in terms of a known
special function. Returning to our main discussion, equation
\eqref{eq:inv63-u22-final} together with
\eqref{eq:inv63-u22-lastint-explicit} is our final form of the
invariant
$\Psi_{6,3}(\boldsymbol{\lambda},\overline{\boldsymbol{\lambda}})$ for
$\mathfrak{u}(2,2)$. This result immediately raises a pressing
question. Is
$\Psi_{6,3}(\boldsymbol{\lambda},\overline{\boldsymbol{\lambda}})$ the
sought-after deformed gluon amplitude $A^{(\text{def.})}_{6,3}$? It
certainly contains the deformation parameters
$v_1,v_2,v_3\in\mathbb{C}$ and $c_1,c_2,c_3\in\mathbb{Z}$. What is
more, it is Yangian invariant because of the unitary contour in the
Graßmannian integral \eqref{eq:grass-int-barg}. From the perspective
of the introductory section~\ref{sec:deform}, the combination of these
two properties is already highly non-trivial. Recall that the
integrand of the deformed Graßmannian formula
\eqref{eq:grassint-amp-def} has branch cuts, which make the choice of
a closed contour guaranteeing Yangian invariance difficult.
\begin{figure}[!t]
  \begin{center}
    \begin{tikzpicture}
      \draw[thick,densely dotted,
      decoration={markings, mark=at position 0.0 with {\arrow{latex reversed}}},
      postaction={decorate}]
      (3,0) node[right=.2] {$\Real{u}$} -- (-3,0);
      \draw[thick,densely dotted,
      decoration={markings, mark=at position 0.0 with {\arrow{latex reversed}}},
      postaction={decorate}]
      (0,3) node[above=.2] {$\Imag{u}$} -- (0,-3);
      \draw[thick,decoration={markings, mark=at position 0.125 with {\arrow{latex}}},
      postaction={decorate}] (0,0) circle [radius=2];
      \filldraw[thick] (0,0) circle (1pt) node[below left]{$0$};
      \filldraw[thick] (2.5,2.5) circle (1pt) node[below right]{$\infty$};
      \filldraw[thick] (55:1.5) circle (1pt) node[below right]
      {$\frac{\mathtt{A}}{\mathtt{B}}$};
      \filldraw[thick] (55:2.66) circle (1pt) node[below right]
      {$\frac{\overline{\mathtt{B}}}{\overline{\mathtt{A}}}$};
      \filldraw[thick] (-170:1.3) circle (1pt) node[below right]
      {$\frac{\overline{\mathtt{C}}}{\overline{\mathtt{D}}}$};
      \filldraw[thick] (-170:3.01) circle (1pt) node[below right]
      {$\frac{\mathtt{D}}{\mathtt{C}}$};
      \node at (2,0) [below right] {$1$};
    \end{tikzpicture}
    \caption{The evaluation of the invariant
      $\Psi_{6,3}(\boldsymbol{\lambda},\overline{\boldsymbol{\lambda}})$
      for $\mathfrak{u}(2,2)$ with the parameters
      $v_1=v_2=v_3\in\mathbb{C}$, $c_1=c_2=c_3\in\mathbb{Z}$ yields
      the complex contour integral \eqref{eq:inv63-u22-complex-int} in
      the variable $u$. The positions of the pairs of poles at
      $u=\frac{\mathtt{A}}{\mathtt{B}},\frac{\overline{\mathtt{B}}}{\overline{\mathtt{A}}}$
      and
      $u=\frac{\mathtt{D}}{\mathtt{C}},\frac{\overline{\mathtt{C}}}{\overline{\mathtt{D}}}$
      depend on the external data $\boldsymbol{\lambda}$, cf.\
      \eqref{eq:inv63-u22-minors-abcd}. Exactly one pole of each pair
      is inside the contour. This divides the external data into four
      regions controlled by $s_{234},s_{126}\gtrless 0$, see
      \eqref{eq:inv63-u22-four-cases}. The pole configuration for the
      region with $s_{234},s_{126}>0$ is depicted here. For this
      region
      $\Psi_{6,3}(\boldsymbol{\lambda},\overline{\boldsymbol{\lambda}})$
      matches the amplitude $A^{(\text{tree})}_{6,3}$ as shown in
      \eqref{eq:inv63-u22-compare}.}
    \label{fig:grint-inv63-u22-contour}
  \end{center}
\end{figure}

To approach the above question we have to investigate the undeformed
limit of
$\Psi_{6,3}(\boldsymbol{\lambda},\overline{\boldsymbol{\lambda}})$ and
relate it to the gluon amplitude $A^{(\text{tree})}_{6,3}$. To this
end, we study the integral \eqref{eq:inv63-u22-lastint-explicit} for
the special case with $z_1=z_2=0$ and $c_1=c_2=c_3$. We rewrite it as
the complex contour integral
\begin{align}
  \label{eq:inv63-u22-complex-int}
  \begin{aligned}
    \mathcal{I}(0,0,c_1,c_1,c_1)&=
    \frac{i}{\overline{\mathtt{A}}\mathtt{B}\mathtt{C}\overline{\mathtt{D}}}
    \oint\D u\,
    \frac{1}{(-u)^{1-c_1}
      \Big(u-\frac{\mathtt{A}}{\mathtt{B}}\Big)
      \Big(u-\frac{\overline{\mathtt{B}}}{\overline{\mathtt{A}}}\Big)
      \Big(u-\frac{\mathtt{D}}{\mathtt{C}}\Big)
      \Big(u-\frac{\overline{\mathtt{C}}}{\overline{\mathtt{D}}}\Big)
    }\\
    &=
    \oint\D u\,\mathscr{I}(u)\,,
  \end{aligned}
\end{align}
where we integrate counterclockwise along the unit circle, see
figure~\ref{fig:grint-inv63-u22-contour}. 
For later use, we introduced the symbol $\mathscr{I}(u)$ for the
integrand. Notice the striking similarity to the integral
\eqref{eq:inv42-u11-contourint}, which appears for the invariant
$\Psi_{4,2}(\boldsymbol{\lambda},\overline{\boldsymbol{\lambda}})$ of
$\mathfrak{u}(1,1)$. Consequently these integrals share some key
features. The integrand of \eqref{eq:inv63-u22-complex-int} can have a
pole at $u=0$ inside of the contour and at $u=\infty$ outside of
it. In addition, there are two pairs of poles at
$u=\frac{\mathtt{A}}{\mathtt{B}},\frac{\overline{\mathtt{B}}}{\overline{\mathtt{A}}}$
and at
$u=\frac{\mathtt{D}}{\mathtt{C}},\frac{\overline{\mathtt{C}}}{\overline{\mathtt{D}}}$. The
positions of these poles depend on the external data
$\boldsymbol{\lambda}$, cf.\ \eqref{eq:inv63-u22-minors-abcd}.
Interestingly, for generic external data with
$|\mathtt{A}|\neq |\mathtt{B}|$ and $|\mathtt{C}|\neq |\mathtt{D}|$,
one pole of each pair is always inside of the contour and the other
one is outside. Hence the contour in \eqref{eq:inv63-u22-complex-int}
encircles three poles. We want to evaluate this integral using the
residue theorem. The residues are conveniently expressed in terms of
\begin{align}
  \label{eq:inv63-u22-var-res}
  \begin{aligned}
    |\mathtt{B}|^2-|\mathtt{A}|^2&=\frac{s_{234}}{s_{123}}\,,\quad&
    |\mathtt{D}|^2-|\mathtt{C}|^2&=\frac{s_{126}}{s_{123}}\,,\\
    \mathtt{A}\overline{\mathtt{D}}-\mathtt{B}\overline{\mathtt{C}}&=\frac{\langle 5|1-6\overline{|2\rangle}}{s_{123}}\,,\quad&
    \mathtt{B}\mathtt{D}-\mathtt{A}\mathtt{C}&=\frac{\langle 16\rangle \overline{\langle 34\rangle}}{s_{123}}\,,
  \end{aligned}
\end{align}
where
$s_{234}=\langle 23\rangle\overline{\langle 23\rangle}-\langle
24\rangle\overline{\langle 24\rangle}-\langle
34\rangle\overline{\langle 34\rangle}=\langle 23\rangle[32]+\langle
24\rangle[42]+\langle 34\rangle[43]$
and $s_{126}$ is defined analogously with $2,3,4$ in $s_{234}$
replaced by $1,2,6$, respectively. We used the Schouten identity
\eqref{eq:bracket-schouten} and momentum conservation
\eqref{eq:bracket-mom-4} to derive \eqref{eq:inv63-u22-var-res} from
the expressions for $\mathtt{A},\mathtt{B},\mathtt{C},\mathtt{D}$ in
\eqref{eq:inv63-u22-minors-abcd}. The residues of the integrand in
\eqref{eq:inv63-u22-complex-int} then read
\begin{align}
  \label{eq:inv63-u22-residues}
  \begin{aligned}
    \text{res}_{0}\mathscr{I}(u)&=
    \begin{cases}
      0&\quad\text{for}\quad c_1 \geq 1\,,\\
      \begin{aligned}
        -i\frac{s_{123}^4}{\langle
          1|2+3\overline{|4\rangle}\,\overline{\langle 56\rangle}
          \langle 23\rangle\langle
          3|1+2\overline{|6\rangle}\,\overline{\langle
            45\rangle}\langle 12\rangle}
      \end{aligned}
      &\quad\text{for}\quad c_1=0\,,\\
    \end{cases}\\
    \text{res}_{\infty}\mathscr{I}(u)&=0\quad\text{for}\quad c_1 \leq 3\,,\\
    \text{res}_{\frac{\mathtt{A}}{\mathtt{B}}}\mathscr{I}(u)
    &=-i
    \frac{s_{123}}{\langle 5|1-6\overline{|2\rangle}}
    \frac{\langle 1|2+3\overline{|4\rangle}\langle 56\rangle\overline{\langle 23\rangle}}
    {s_{234}\langle 16\rangle \overline{\langle 34\rangle}}
    \left(-\frac{\langle 1|2+3\overline{|4\rangle}}{\langle 56\rangle \overline{\langle 23\rangle}}\right)^{c_1-2}\,,
    \\
    \text{res}_{\frac{\overline{\mathtt{B}}}{\overline{\mathtt{A}}}}\mathscr{I}(u)
    &=\phantom{-}i
    \frac{s_{123}}{\overline{\langle 5|}1-6|2\rangle}
    \frac{\overline{\langle 1|}2+3|4\rangle\overline{\langle 56\rangle}\langle 23\rangle}
    {s_{234}\overline{\langle 16\rangle}\langle 34\rangle}
    \left(-\frac{\overline{\langle 1|}2+3|4\rangle}{\overline{\langle 56\rangle}\langle 23\rangle}\right)^{2-c_1}\,,
    \\
    \text{res}_{\frac{\mathtt{D}}{\mathtt{C}}}\mathscr{I}(u)
    &=-i
    \frac{s_{123}}{\overline{\langle 5|}1-6|2\rangle}
    \frac{\overline{\langle 3|}1+2|6\rangle\overline{\langle 45\rangle}\langle 12\rangle}
    {s_{126}\overline{\langle 34\rangle}\langle 16\rangle}
    \left(-\frac{\overline{\langle 3|}1+2|6\rangle}{\overline{\langle 45\rangle}\langle 12\rangle}\right)^{2-c_1}\,,
    \\
    \text{res}_{\frac{\overline{\mathtt{C}}}{\overline{\mathtt{D}}}}\mathscr{I}(u)
    &=\phantom{-}i
    \frac{s_{123}}{\langle 5|1-6\overline{|2\rangle}}
    \frac{\langle 3|1+2\overline{|6\rangle}\langle 45\rangle\overline{\langle 12\rangle}}
    {s_{126}\langle 34\rangle\overline{\langle 16\rangle}}
    \left(-\frac{\langle 3|1+2\overline{|6\rangle}}{\langle 45\rangle\overline{\langle 12\rangle}}\right)^{c_1-2}\,.
    \\
  \end{aligned}
\end{align}
Employing the residue theorem, the integral
\eqref{eq:inv63-u22-complex-int} becomes
\begin{align}
  \label{eq:inv63-u22-four-cases}
  \mathcal{I}(0,0,c_1,c_1,c_1)=
  2\pi i
  \begin{cases}
    \text{res}_{0}\mathscr{I}(u)+
    \text{res}_{\frac{\mathtt{A}}{\mathtt{B}}}\mathscr{I}(u)+
    \text{res}_{\frac{\mathtt{D}}{\mathtt{C}}}\mathscr{I}(u)
    &\text{for}
    \quad
    \begin{aligned}
      s_{234}>0\,,\\
      s_{126}<0\,,\\
    \end{aligned}
    \\
    \text{res}_{0}\mathscr{I}(u)+
    \text{res}_{\frac{\overline{\mathtt{B}}}{\overline{\mathtt{A}}}}\mathscr{I}(u)+
    \text{res}_{\frac{\mathtt{D}}{\mathtt{C}}}\mathscr{I}(u)
    &\text{for}
    \quad
    \begin{aligned}
      s_{234}<0\,,\\
      s_{126}<0\,,\\
    \end{aligned}
    \\
    \text{res}_{0}\mathscr{I}(u)+
    \text{res}_{\frac{\mathtt{A}}{\mathtt{B}}}\mathscr{I}(u)+
    \text{res}_{\frac{\overline{\mathtt{C}}}{\overline{\mathtt{D}}}}\mathscr{I}(u)
    &\text{for}
    \quad
    \begin{aligned}
      s_{234}>0\,,\\
      s_{126}>0\,,\\
    \end{aligned}
    \\
    \text{res}_{0}\mathscr{I}(u)+
    \text{res}_{\frac{\overline{\mathtt{B}}}{\overline{\mathtt{A}}}}\mathscr{I}(u)+
    \text{res}_{\frac{\overline{\mathtt{C}}}{\overline{\mathtt{D}}}}\mathscr{I}(u)
    &\text{for}
    \quad
    \begin{aligned}
      s_{234}<0\,,\\
      s_{126}>0\,.\\
    \end{aligned}
    \\
  \end{cases}
\end{align}
The four kinematic regions result from the two pairs of poles
explained after the contour integral formula
\eqref{eq:inv63-u22-complex-int}. Which pole of the pair at
$u=\frac{\mathtt{A}}{\mathtt{B}},\frac{\overline{\mathtt{B}}}{\overline{\mathtt{A}}}$
is inside of the contour is determined by
$\frac{|\mathtt{A}|}{|\mathtt{B}|}\lessgtr 1$. Here we used
\eqref{eq:inv63-u22-var-res} to translate this into the condition
$s_{234}\gtrless 0$. Notice from \eqref{eq:inv63-u22-kininv} that
$s_{123}>0$ cannot change the sign because of the reality conditions
on the spinor helicity variables. Analogously, for the pair of poles
at
$u=\frac{\mathtt{D}}{\mathtt{C}},\frac{\overline{\mathtt{C}}}{\overline{\mathtt{D}}}$,
we write the condition $\frac{|\mathtt{C}|}{|\mathtt{D}|}\gtrless 1$
as $s_{126}\lessgtr 0$. The formula \eqref{eq:inv63-u22-final}
combined with \eqref{eq:inv63-u22-four-cases} is the final result for
the invariant
$\Psi_{6,3}(\boldsymbol{\lambda},\overline{\boldsymbol{\lambda}})$
with trivial complex deformation parameters $v_1=v_2=v_3$.

Let us add an aside. As the reader might have already noted, we did
not present the residues $\text{res}_{0}\mathscr{I}(u)$ and
$\text{res}_{\infty}\mathscr{I}(u)$ in \eqref{eq:inv63-u22-residues}
for the entire range of the representation label
$c_1\in\mathbb{Z}$. However, using that the sum of all residues
vanishes, we can still express the result
\eqref{eq:inv63-u22-four-cases} for all $c_1\in\mathbb{Z}$ in terms of
those residue that we computed explicitly. This yields
\begin{align}
  \label{eq:inv63-u22-Icases}
  \begin{aligned}
    &\mathcal{I}(0,0,c_1,c_1,c_1)=\\
    &\cdot2\pi i \frac{s_{234}}{|s_{234}|}
    \begin{cases}
      \text{res}_{\frac{\mathtt{A}}{\mathtt{B}}}\mathscr{I}(u)
      +\text{res}_{\frac{\overline{\mathtt{C}}}{\overline{\mathtt{D}}}}\mathscr{I}(u)\quad
      &\text{for}\quad
      \begin{aligned}
        s_{234}&<0\,,\\s_{126}&<0\,,\\c_1&\leq 3
      \end{aligned}
      \quad
      \text{or}\quad 
      \begin{aligned}
        s_{234}&>0\,,\\s_{126}&>0\,,\\c_1&\geq 1\,,        
      \end{aligned}
      \\[0.6cm]
      -\text{res}_{\frac{\overline{\mathtt{B}}}{\overline{\mathtt{A}}}}\mathscr{I}(u)
      -\text{res}_{\frac{\overline{\mathtt{C}}}{\overline{\mathtt{D}}}}\mathscr{I}(u)\quad
      &\text{for}\quad
      \begin{aligned}
        s_{234}&>0\,,\\s_{126}&<0\,,\\c_1&\leq 3
      \end{aligned}
      \quad
      \text{or}\quad 
      \begin{aligned}
        s_{234}&<0\,,\\s_{126}&>0\,,\\c_1&\geq 1\,,  
      \end{aligned}
      \\[0.6cm]
      \text{res}_{\frac{\mathtt{A}}{\mathtt{B}}}\mathscr{I}(u)
      +\text{res}_{\frac{\mathtt{D}}{\mathtt{C}}}\mathscr{I}(u)\quad
      &\text{for}\quad
      \begin{aligned}
        s_{234}&<0\,,\\s_{126}&>0\,,\\c_1&\leq 3
      \end{aligned}
      \quad
      \text{or}\quad 
      \begin{aligned}
        s_{234}&>0\,,\\s_{126}&<0\,,\\c_1&\geq 1\,,  
      \end{aligned}
      \\[0.6cm]
      -\text{res}_{\frac{\overline{\mathtt{B}}}{\overline{\mathtt{A}}}}\mathscr{I}(u)
      -\text{res}_{\frac{\mathtt{D}}{\mathtt{C}}}\mathscr{I}(u)\quad
      &\text{for}\quad
      \begin{aligned}
        s_{234}&>0\,,\\s_{126}&>0\,,\\c_1&\leq 3
      \end{aligned}
      \quad
      \text{or}\quad 
      \begin{aligned}
        s_{234}&<0\,,\\s_{126}&<0\,,\\c_1&\geq 1\,.
      \end{aligned}
      \\
     \end{cases}
  \end{aligned}
\end{align}
We remark that the number of cases reduces for $c_1=1,2,3$.

Returning to our objective of relating the undeformed invariant
$\Psi_{6,3}(\boldsymbol{\lambda},\overline{\boldsymbol{\lambda}})$ to
the gluon amplitude $A_{6,3}^{(\text{tree})}$, we choose the
representation labels $c_1=c_2=c_3=0$ in addition to $v_1=v_2=v_3$.
The six sites of the invariant then carry the helicities
$(+1,+1,+1,-1,-1,-1)$, cf.\ \eqref{eq:hel-rep-label}. The expressions
for the invariant in \eqref{eq:inv63-u22-four-cases} and
\eqref{eq:inv63-u22-Icases} reduce in the kinematic region
$s_{234},s_{126}>0$ to
\begin{align}
  \label{eq:inv63-u22-compare}
  \begin{aligned}
    \Psi_{6,3}(\boldsymbol{\lambda},\overline{\boldsymbol{\lambda}})&=
    64\pi i\delta^{4}(P)\frac{
      \text{res}_{0}\mathscr{I}(u)+
      \text{res}_{\frac{\mathtt{A}}{\mathtt{B}}}\mathscr{I}(u)+
      \text{res}_{\frac{\overline{\mathtt{C}}}{\overline{\mathtt{D}}}}\mathscr{I}(u)
    }{s_{123}}\\
    &=64\pi i\delta^{4}(P)\frac{
      -\text{res}_{\frac{\overline{\mathtt{B}}}{\overline{\mathtt{A}}}}\mathscr{I}(u)
      -\text{res}_{\frac{\mathtt{D}}{\mathtt{C}}}\mathscr{I}(u)
    }{s_{123}}\\
    &=64\pi\delta^{4}(P)
    \frac{1}{\overline{\langle 5|}1-6|2\rangle}
    \Bigg(
    \frac{\langle 6|1+2\overline{|3\rangle}^3}
    {\langle 61\rangle\langle 12\rangle\overline{\langle 34\rangle}\,\overline{\langle 45\rangle}s_{126}}
    -
    \frac{\langle 4|5+6\overline{|1\rangle}^3}
    {\langle 23\rangle\langle 34\rangle\overline{\langle 16\rangle}\,\overline{\langle 65\rangle}s_{156}}
    \Bigg)\\
    &
    =64\pi\delta^{4}(P)
    \frac{1}{[5|1+6|2\rangle}
    \Bigg(
    \frac{\langle 6|1+2|3]^3}
    {\langle 61\rangle\langle 12\rangle[34][45]s_{126}}
    +
    \frac{\langle 4|5+6|1]^3}
    {\langle 23\rangle\langle 34\rangle[16][65]s_{156}}
    \Bigg)\,.
  \end{aligned}
\end{align}
Here
$s_{156}=-\langle15\rangle\overline{\langle 15\rangle}-\langle
16\rangle\overline{\langle 16\rangle}+\langle
56\rangle\overline{\langle 56\rangle}=\langle 15\rangle[51]+\langle
16\rangle[61]+\langle 56\rangle[65]=s_{234}$.
This formula is proportional to the six-particle $\text{NMHV}$ gluon
amplitude $A^{(\text{tree})}_{6,3}$ from
\eqref{eq:amp-bos-nmhv}. Hence in this one kinematic region the $U(3)$
contour we started out with in the Graßmannian integral
\eqref{eq:grass-int-barg} automatically selects the desired three
residues out of six. Curiously, our formula seems to differ from
$A^{(\text{tree})}_{6,3}$ for external data in the other three
regions. We will elaborate on this result in the conclusions presented
in chapter~\ref{cha:concl}. See also
appendix~\ref{sec:discrete-symmetry} where we investigate a parity
symmetry of
$\Psi_{6,3}(\boldsymbol{\lambda},\overline{\boldsymbol{\lambda}})$
that relates two of the four regions in
\eqref{eq:inv63-u22-four-cases}.

\subsubsection{Four-Site Invariant for
  \texorpdfstring{$\mathfrak{u}(2,2|0+4)$}{u(2,2|0+4)} and MHV
  Superamplitude}
\label{sec:inv42-u2204}

Let us continue with an example that yields the deformation
$\mathcal{A}^{(\text{def.})}_{4,2}$ of the $\mathcal{N}=4$ SYM
$\text{MHV}$ superamplitude $\mathcal{A}^{(\text{tree})}_{4,2}$. To
this end, we evaluate the unitary Graßmannian integral
\eqref{eq:grass-int-barg} for the invariant with $(N,K)=(4,2)$ of the
superalgebra $\mathfrak{u}(2,2|0+4)$. We proceed analogously to the
bosonic case presented in section~\ref{sec:inv42-u22}. Instead of
\eqref{eq:inv42-u22-step2}, we end up with
\begin{align}  
  \label{eq:inv42-u2204-step2}
  \Psi_{4,2}(\boldsymbol{\lambda},\overline{\boldsymbol{\lambda}},\boldsymbol{\eta})=
  8i\,
  \delta^{4|0}(P)
  \mathscr{F}(\mathcal{C})
  \delta^{0|8}(\boldsymbol{\eta}^{\text{d}}-\overline{\mathcal{C}} \boldsymbol{\eta}^{\text{o}})\,.
\end{align}
Here the momentum conserving delta function can be found in
\eqref{eq:def-delta-bosonic} and the matrix $\mathcal{C}$ is given in
\eqref{eq:inv42-u22-e-eval} in terms of the external data
$\boldsymbol{\lambda}$. The function $\mathscr{F}$ is specified in
\eqref{eq:integrand42}. Note that the exponents in this function
depend on the algebra and thus differ slightly from the bosonic
case. We obtain from \eqref{eq:inv42-u2204-step2} the final expression
\begin{align}
  \label{eq:inv42-u2204-final}
  \Psi_{4,2}(\boldsymbol{\lambda},\overline{\boldsymbol{\lambda}},\boldsymbol{\eta})=
  8i\,
  \frac{\delta^{4|0}(P)\delta^{0|8}(Q)}
  {\langle 12 \rangle\langle 23 \rangle\langle 34 \rangle\langle 41 \rangle}
  \left(\frac{\langle 14 \rangle}{\langle 34 \rangle}\right)^{c_1}
  \left(\frac{\langle 12 \rangle}{\langle 14 \rangle}\right)^{c_2}
  \left(\frac{\langle 34 \rangle\overline{\langle 34\rangle}}{\langle 14 \rangle\overline{\langle 14\rangle}}\right)^{v_1-v_2}\,
\end{align}
with the supermomentum conserving delta function
\eqref{eq:def-delta-fermionic}. Employing the relation
\eqref{eq:inv42-u22-real-brackets} for the fraction in the last
bracket, we realize that
$\Psi_{4,2}(\boldsymbol{\lambda},\overline{\boldsymbol{\lambda}},\boldsymbol{\eta})$
is proportional to the deformed $\text{MHV}$ superamplitude
$\mathcal{A}^{(\text{def.})}_{4,2}$ from \eqref{eq:amp-4-2-def}.

At this point we pause for a moment to bring to mind a structural
connection with results discussed earlier in this thesis. In
section~\ref{sec:sample-inv42} we obtained the Yangian invariant
$|\Psi_{4,2}\rangle$ for oscillator representations of
$\mathfrak{u}(p,q|r+s)$ from a unitary Graßmannian integral. Already
prior to this, in section~\ref{sec:ncomp-susy-4-site} we pointed out
that $|\Psi_{4,2}\rangle$ for $\mathfrak{u}(2,2|4)$ is essentially the
R-matrix of the planar $\mathcal{N}=4$ SYM one-loop spectral
problem. The function
$\Psi_{4,2}(\boldsymbol{\lambda},\overline{\boldsymbol{\lambda}},\boldsymbol{\eta})$
computed in this section, and thus the deformed amplitude
$\mathcal{A}^{(\text{def.})}_{4,2}$, is the very same Yangian
invariant as $|\Psi_{4,2}\rangle$. It is merely expressed in a
different basis, i.e.\ spinor helicity variables instead of
oscillators. The idea to identify the R-matrix of the spectral problem
with a deformation of the amplitude
$\mathcal{A}_{4,2}^{(\text{tree})}$ goes back to \cite{Ferro:2012xw}
and was in part inspired by \cite{Zwiebel:2011bx}. However, the
necessary change of basis has never been worked out explicitly. We
filled this gap with the Bargmann transformation of
section~\ref{sec:osc-spinor}.\footnote{In the last paragraph of
  section~\ref{sec:ncomp-susy-4-site} we brought to attention that
  slightly non-standard ``oscillator'' representations are frequently
  used in the $\mathcal{N}=4$ SYM spectral problem. This subtlety is
  not taken into account by our transformation.}

\subsubsection{Six-Site Invariant for
  \texorpdfstring{$\mathfrak{u}(2,2|0+4)$}{u(2,2|0+4)} and NMHV
  Superamplitude}
\label{sec:inv63-u2204}

Here we extend the calculation of the six-site Yangian invariant for
$\mathfrak{u}(2,2)$ from section~\ref{sec:inv63-u22} to the
superalgebra $\mathfrak{u}(2,2|0+4)$. We compare our result to the
NMHV superamplitude $\mathcal{A}^{(\text{tree})}_{6,3}$ of
$\mathcal{N}=4$ SYM. Furthermore, we comment on its relation to the
deformed superamplitude $\mathcal{A}^{(\text{def}.)}_{6,3}$ whose
existence has not been settled yet.

Computing the unitary Graßmannian integral \eqref{eq:grass-int-barg}
for $(N,K)=(6,3)$ and the superalgebra $\mathfrak{u}(2,2|0+4)$, the
bosonic equation \eqref{eq:inv63-u22-final} gets replaced by
\begin{align}
  \Psi_{6,3}(\boldsymbol{\lambda},\overline{\boldsymbol{\lambda}},\boldsymbol{\eta})
  =32\frac{\delta^{4|0}(P)}{s_{123}}
  \int_0^{2\pi}\D\gamma\,\mathscr{F}(\mathcal{C}(\gamma))\delta^{0|12}(\boldsymbol{\eta}^{\text{d}}-\overline{\mathcal{C}(\gamma)}\boldsymbol{\eta}^{\text{o}})
\end{align}
with $\mathcal{C}(\gamma)$ given in terms of the external data
$\boldsymbol{\lambda}$ by
\eqref{eq:inv63-u22-e-explicit}. Manipulating the fermionic delta
function and using the form of the function $\mathscr{F}$ in
\eqref{eq:integrand63} leads to
\begin{align}
  \label{eq:inv63-u2204-formula}
  \Psi_{6,3}(\boldsymbol{\lambda},\overline{\boldsymbol{\lambda}},\boldsymbol{\eta})
  =32\frac{\delta^{4|0}(P)\delta^{0|8}(Q)}{s_{123}^5}\mathcal{I}(v_1-v_2,v_2-v_3,c_1,c_2,c_3)\,,
\end{align}
where
\begin{align}
  \label{eq:inv63-u2204-lastint-general}
  \begin{aligned}
    \mathcal{I}(z_1,z_2,c_1,c_2,c_3)=
    \int_0^{2\pi}\D\gamma\,&\frac{
      \delta^{0|4}(\mathtt{a}+e^{-i \gamma}\mathtt{b})
    }{
      (-e^{i\gamma})^{-2-c_2}
      |\mathtt{A}-e^{i\gamma}\mathtt{B}|^{2(1+z_1)}(\mathtt{A}-e^{i\gamma}\mathtt{B})^{c_2-c_1}      
    }\\
    &\cdot\frac{
      1
    }{
      |\mathtt{C}-e^{-i\gamma}\mathtt{D}|^{2(1+z_2)}(\mathtt{C}-e^{-i\gamma}\mathtt{D})^{c_3-c_2}
    }
    \,.
  \end{aligned}
\end{align}
Here we shifted the integration variable $\gamma$ as in the bosonic
case and the quantities $\mathtt{A},\mathtt{B},\mathtt{C},\mathtt{D}$
are defined in \eqref{eq:inv63-u22-minors-abcd}. Furthermore, we
introduced the Graßmann variables
\begin{align}
  \mathtt{a}=\overline{\langle 23\rangle}\eta^1+\overline{\langle 31\rangle}\eta^2+\overline{\langle 12\rangle}\eta^3\,,\quad
  \mathtt{b}=\overline{\langle 56\rangle}\eta^4+\overline{\langle 64\rangle}\eta^5+\overline{\langle 45\rangle}\eta^6\,
\end{align}
with the vectors $\eta^i=(\eta^i_{\dot{a}})$. The expression in
\eqref{eq:inv63-u2204-formula} with the integral
\eqref{eq:inv63-u2204-lastint-general} is our ultimate result for the
Yangian invariant
$\Psi_{6,3}(\boldsymbol{\lambda},\overline{\boldsymbol{\lambda}},\boldsymbol{\eta})$
with generic deformation parameters $v_1,v_2,v_3\in\mathbb{C}$ and
$c_1,c_2,c_3\in\mathbb{Z}$. It should be thought of as the deformed
superamplitude $\mathcal{A}^{(\text{def.})}_{6,3}$ whose existence has
been questioned, see section~\ref{sec:deform}.

However, this interpretation of
$\Psi_{6,3}(\boldsymbol{\lambda},\overline{\boldsymbol{\lambda}},\boldsymbol{\eta})$
comes with a caveat. As for the bosonic version, its undeformed limit
reduces only in a certain kinematic region to the superamplitude
$\mathcal{A}^{(\text{tree})}_{6,3}$. Let us show this explicitly. For
this purpose we study the integral \eqref{eq:inv63-u2204-lastint-general} in the
special case $z_1=z_2=0$ and $c_1=c_2=c_3$. Introducing
$u=e^{i\gamma}$, it becomes
\begin{align}
  \begin{aligned}
    \mathcal{I}(0,0,c_1,c_1,c_1)&=
    \frac{i}{\overline{\mathtt{A}}\mathtt{B}\mathtt{C}\overline{\mathtt{D}}}
    \oint\D u\,  
    \frac{
      \delta^{0|4}(u\mathtt{a}+\mathtt{b})
    }
    {
      (-u)^{1-c_1}
      \Big(u-\frac{\mathtt{A}}{\mathtt{B}}\Big)
      \Big(u-\frac{\overline{\mathtt{B}}}{\overline{\mathtt{A}}}\Big)
      \Big(u-\frac{\mathtt{D}}{\mathtt{C}}\Big)
      \Big(u-\frac{\overline{\mathtt{C}}}{\overline{\mathtt{D}}}\Big)
    }\\
    &=\oint\D u\,\mathscr{I}(u)\,,
  \end{aligned}
\end{align}
where we integrate counterclockwise along the unit circle. We denote
the integrand by $\mathscr{I}(u)$. Let us compute its residues,
\begingroup
\allowdisplaybreaks
\begin{align}
  \label{eq:inv63-u2204-residues}
  \notag
  \text{res}_{0}\mathscr{I}(u)&=
    \begin{cases}
      \begin{aligned}
        -i
        \frac{\delta^{0|4}(\mathtt{b})}{\mathtt{A}\overline{\mathtt{B}}\overline{\mathtt{C}}\mathtt{D}}
        &=
        -i \frac{s_{123}^4\delta^{0|4}(\overline{\langle 56\rangle}\eta^4+\overline{\langle 64\rangle}\eta^5+\overline{\langle 45\rangle}\eta^6)}
        {\langle 1|2+3\overline{|4\rangle}\, \overline{\langle 56\rangle}\langle 23\rangle\langle 3|1+2\overline{|6\rangle}\,\overline{\langle 45\rangle}\langle 12\rangle}\\[1ex]
        &=
        -i \frac{s_{123}^4\delta^{0|4}([65]\eta^4+[46]\eta^5+[54]\eta^6)}
        {\langle 1|2+3|4] [65]\langle 23\rangle\langle 3|1+2|6][54]\langle 12\rangle}      
      \end{aligned}
      &\text{for}\quad c_1=0\,,\\
      0&\text{for}\quad c_1>0\,,
    \end{cases}\\
    \notag
    \text{res}_{\infty}\mathscr{I}(u)&=
    \begin{cases}
      \begin{aligned}
        \phantom{-}i
        \frac{\delta^{0|4}(\mathtt{a})}{\overline{\mathtt{A}}\mathtt{B}\mathtt{C}\overline{\mathtt{D}}}
        &=
        \phantom{-}i \frac{s_{123}^4\delta^{0|4}(\overline{\langle 23\rangle}\eta^1+\overline{\langle 31\rangle}\eta^2+\overline{\langle 12\rangle}\eta^3)}
        {\overline{\langle 1|}2+3|4\rangle \langle 56\rangle\overline{\langle 23\rangle}\,\overline{\langle 3|}1+2|6\rangle\langle 45\rangle\overline{\langle 12\rangle}}\\[1ex]
        &=
        \phantom{-}i \frac{s_{123}^4\delta^{0|4}([32]\eta^1+[13]\eta^2+[21]\eta^3)}
        {[1|2+3|4\rangle \langle 56\rangle[32][3|1+2|6\rangle\langle 45\rangle[21]}
      \end{aligned}
      &\text{for}\quad c_1=0\,,\\
      0&\text{for}\quad c_1<0\,,
    \end{cases}\\
  \notag
    \text{res}_{\frac{\mathtt{A}}{\mathtt{B}}}\mathscr{I}(u)&=
    -i
    \frac{\left(-\frac{\mathtt{A}}{\mathtt{B}}\right)^{c_1}
      \delta^{0|4}(\mathtt{A}\mathtt{a}+\mathtt{B}\mathtt{b})
    }{
      \mathtt{A}\mathtt{B}
      (|\mathtt{A}|^2-|\mathtt{B}|^2)
      (\mathtt{A}\mathtt{C}-\mathtt{B}\mathtt{D})
      (\mathtt{A}\overline{\mathtt{D}}-\mathtt{B}\overline{\mathtt{C}})
    }\\
  \notag
    &=
    -i \frac{s_{123}^5}{\langle 5|1-6\overline{|2\rangle}\langle 16\rangle\overline{\langle 34\rangle}}
    \frac{\delta^{0|4}(\overline{\langle 43\rangle}\eta^2+\overline{\langle 24\rangle}\eta^3+\overline{\langle 23\rangle}\eta^4)}
    {s_{234}\langle 1|2+3\overline{|4\rangle}\langle 56\rangle \overline{\langle 23\rangle}}
    \left(-\frac{\langle 1|2+3\overline{|4\rangle}}{\langle 56\rangle \overline{\langle 23\rangle}}\right)^{c_1}
    \\
  \notag
    &=
    \phantom{-}i \frac{s_{123}^5}{\langle 5|1+6|2]\langle 16\rangle[34]}
    \frac{\delta^{0|4}([43]\eta^2+[24]\eta^3+[32]\eta^4)}
    {s_{234}\langle 1|2+3|4]\langle 56\rangle [32]}\left(-\frac{\langle 1|2+3|4]}{\langle 56\rangle [32]}\right)^{c_1}
    \,,\\
  \notag
    \text{res}_{\frac{\overline{\mathtt{B}}}{\overline{\mathtt{A}}}}\mathscr{I}(u)&=
    \phantom{-}i
    \frac{\left(-\frac{\overline{\mathtt{B}}}{\overline{\mathtt{A}}}\right)^{c_1}
      \delta^{0|4}(\overline{\mathtt{B}}\mathtt{a}+\overline{\mathtt{A}}\mathtt{b})
    }{
      \overline{\mathtt{A}}\overline{\mathtt{B}}
      (|\mathtt{A}|^2-|\mathtt{B}|^2)
      (\overline{\mathtt{A}}\overline{\mathtt{C}}-\overline{\mathtt{B}}\overline{\mathtt{D}})
      (\overline{\mathtt{A}}\mathtt{D}-\overline{\mathtt{B}}\mathtt{C})
    }\\
  \notag
    &=
    \phantom{-}i
    \frac{s_{123}^5}{\overline{\langle 5|}1-6|2\rangle \overline{\langle 16\rangle}\langle 34\rangle}
    \frac{\delta^{0|4}(\overline{\langle 56\rangle}\eta^1+\overline{\langle 16\rangle}\eta^5+\overline{\langle 51\rangle}\eta^6)}
    {s_{234}\overline{\langle 1|}2+3|4\rangle \overline{\langle 56\rangle}\langle 23\rangle}
    \left(-\frac{\overline{\langle 56\rangle}\langle 23\rangle}{\overline{\langle 1|}2+3|4\rangle}\right)^{c_1}
    \\
  \notag
    &=
    -i
    \frac{s_{123}^5}{[5|1+6|2\rangle [16]\langle 34\rangle}
    \frac{\delta^{0|4}([65]\eta^1+[16]\eta^5+[51]\eta^6)}{s_{234}[1|2+3|4\rangle [65]\langle 23\rangle}
    \left(\frac{[65]\langle 23\rangle}{[1|2+3|4\rangle}\right)^{c_1}
    \,,\\
  \notag
    \text{res}_{\frac{\mathtt{D}}{\mathtt{C}}}\mathscr{I}(u)&=
    -i
    \frac{\left(-\frac{\mathtt{D}}{\mathtt{C}}\right)^{c_1}
      \delta^{0|4}(\mathtt{D}\mathtt{a}+\mathtt{C}\mathtt{b})
    }{
      \mathtt{C}\mathtt{D}
      (|\mathtt{C}|^2-|\mathtt{D}|^2)
      (\mathtt{A}\mathtt{C}-\mathtt{B}\mathtt{D})
      (\overline{\mathtt{A}}\mathtt{D}-\overline{\mathtt{B}}\mathtt{C})
    }\\
  \notag
    &=
    -i
    \frac{s_{123}^5}{\overline{\langle 5|}1-6|2\rangle\langle 16\rangle \overline{\langle 34\rangle}}
    \frac{\delta^{0|4}(\overline{\langle 45\rangle}\eta^3+\overline{\langle 35\rangle}\eta^4+\overline{\langle 43\rangle}\eta^5)}
    {s_{126}\overline{\langle 3|}1+2|6\rangle\overline{\langle 45\rangle}\langle 12\rangle}
    \left(-\frac{\overline{\langle 45\rangle}\langle 12\rangle}{\overline{\langle 3|}1+2|6\rangle}\right)^{c_1}
    \\
  \notag
    &=
    \phantom{-}i
    \frac{s_{123}^5}{[5|1+6|2\rangle\langle 16\rangle [34]}
    \frac{\delta^{0|4}([54]\eta^3+[35]\eta^4+[43]\eta^5)}{s_{126}[3|1+2|6\rangle[54]\langle 12\rangle}
    \left(\frac{[54]\langle 12\rangle}{[3|1+2|6\rangle}\right)^{c_1}
    \,,\\
  \notag
    \text{res}_{\frac{\overline{\mathtt{C}}}{\overline{\mathtt{D}}}}\mathscr{I}(u)&=
    \phantom{-}i
    \frac{\left(-\frac{\overline{\mathtt{C}}}{\overline{\mathtt{D}}}\right)^{c_1}
      \delta^{0|4}(\overline{\mathtt{C}}\mathtt{a}+\overline{\mathtt{D}}\mathtt{b})
    }{
      \overline{\mathtt{C}}\overline{\mathtt{D}}
      (|\mathtt{C}|^2-|\mathtt{D}|^2)
      (\overline{\mathtt{A}}\overline{\mathtt{C}}-\overline{\mathtt{B}}\overline{\mathtt{D}})
      (\mathtt{A}\overline{\mathtt{D}}-\mathtt{B}\overline{\mathtt{C}})
    }\\
  \notag
    &=
    \phantom{-}i
    \frac{s_{123}^5}{\langle 5|1-6\overline{|2\rangle}\,\overline{\langle 16\rangle}\langle 34\rangle}
    \frac{\delta^{0|4}(\overline{\langle 62\rangle}\eta^1+\overline{\langle 16\rangle}\eta^2+\overline{\langle 12\rangle}\eta^6)}
    {s_{126}\langle 3|1+2\overline{|6\rangle}\langle 45\rangle \overline{\langle 12\rangle}}
    \left(-\frac{\langle 3|1+2\overline{|6\langle }}{\langle 45\rangle \overline{\langle 12\rangle}}\right)^{c_1}
    \\
    &=
    -i
    \frac{s_{123}^5}{\langle 5|1+6|2][16]\langle 34\rangle}
    \frac{\delta^{0|4}([62]\eta^1+[16]\eta^2+[21]\eta^6)}{s_{126}\langle 3|1+2|6]\langle 45\rangle [21]}
    \left(-\frac{\langle 3|1+2|6]}{\langle 45\rangle [21]}\right)^{c_1}
    \,.
\end{align}
\endgroup
We used the momentum and supermomentum conservation in
\eqref{eq:inv63-u2204-formula} to obtain these formulas. What is more,
we presented three expressions for each residue. The first two make
the analytic structure most transparent. The third version involving
the angle brackets is obtained using \eqref{eq:brackets-cc-angle}. It
is appropriate to identify the residues with known expressions later
on. By means of the residue theorem, the integral
\eqref{eq:inv63-u2204-lastint-general} then becomes
\begin{align}
  \label{eq:inv63-u2204-lastint-eval}
  \mathcal{I}(0,0,c_1,c_1,c_1)=
  2\pi i
  \begin{cases}
    \text{res}_{0}\mathscr{I}(u)+
    \text{res}_{\frac{\mathtt{A}}{\mathtt{B}}}\mathscr{I}(u)+
    \text{res}_{\frac{\mathtt{D}}{\mathtt{C}}}\mathscr{I}(u)
    &\text{for}
    \quad
    \begin{aligned}
      s_{234}>0\,,\\
      s_{126}<0\,,\\
    \end{aligned}
    \\
    \text{res}_{0}\mathscr{I}(u)+
    \text{res}_{\frac{\overline{\mathtt{B}}}{\overline{\mathtt{A}}}}\mathscr{I}(u)+
    \text{res}_{\frac{\mathtt{D}}{\mathtt{C}}}\mathscr{I}(u)
    &\text{for}
    \quad
    \begin{aligned}
      s_{234}<0\,,\\
      s_{126}<0\,,\\
    \end{aligned}
    \\
    \text{res}_{0}\mathscr{I}(u)+
    \text{res}_{\frac{\mathtt{A}}{\mathtt{B}}}\mathscr{I}(u)+
    \text{res}_{\frac{\overline{\mathtt{C}}}{\overline{\mathtt{D}}}}\mathscr{I}(u)
    &\text{for}
    \quad
    \begin{aligned}
      s_{234}>0\,,\\
      s_{126}>0\,,\\
    \end{aligned}
    \\
    \text{res}_{0}\mathscr{I}(u)+
    \text{res}_{\frac{\overline{\mathtt{B}}}{\overline{\mathtt{A}}}}\mathscr{I}(u)+
    \text{res}_{\frac{\overline{\mathtt{C}}}{\overline{\mathtt{D}}}}\mathscr{I}(u)
    &\text{for}
    \quad
    \begin{aligned}
      s_{234}<0\,,\\
      s_{126}>0\,.\\
    \end{aligned}
    \\
  \end{cases}
\end{align}
The four cases, which are distinguished by kinematic regions of the
external data $\boldsymbol{\lambda}$, appear in precisely the same
manner as for the bosonic version in \eqref{eq:inv63-u22-four-cases}.
At this point we insert a remark. Because the sum of all residues
vanishes, we can express \eqref{eq:inv63-u2204-lastint-eval} for all
$c_1\in\mathbb{Z}$ in terms of the residues computed in
\eqref{eq:inv63-u2204-residues}. That is even though we evaluated
$\text{res}_{0}\mathscr{I}(u)$ and $\text{res}_{\infty}\mathscr{I}(u)$
only for restricted ranges of $c_1$.

Finally, we want to compare our findings with the superamplitude
$\mathcal{A}^{(\text{tree})}_{6,3}$ of $\mathcal{N}=4$ SYM presented
in \eqref{eq:amp-super-nmhv6}. Thus we have to specialize to the
representation label $c_1=0$, see the discussion after
\eqref{eq:fermi-replabel}. In this case the residues can be expressed
in terms of the quantities $\mathcal{R}^{r;st}$ from
\eqref{eq:amp-super-r},
\begin{align}
  \begin{aligned}
    \text{res}_{0}\mathscr{I}(u)\Big|_{c_1=0}
    &=
    -i\frac{s_{123}^{5} \mathcal{R}^{1;46}}
    {\langle 12\rangle\langle 23\rangle\langle 34\rangle\langle 45\rangle\langle 56\rangle\langle 61\rangle}\,,\\
    \text{res}_{\infty}\mathscr{I}(u)\Big|_{c_1=0}
    &=
    \phantom{-}i\frac{s_{123}^{5} \mathcal{R}^{6;24}}
    {\langle 12\rangle\langle 23\rangle\langle 34\rangle\langle 45\rangle\langle 56\rangle\langle 61\rangle}\,,\\
    \text{res}_{\frac{\mathtt{A}}{\mathtt{B}}}\mathscr{I}(u)\Big|_{c_1=0}
    &=
    -i\frac{s_{123}^{5} \mathcal{R}^{1;35}}
    {\langle 12\rangle\langle 23\rangle\langle 34\rangle\langle 45\rangle\langle 56\rangle\langle 61\rangle}\,,\\
    \text{res}_{\frac{\overline{\mathtt{B}}}{\overline{\mathtt{A}}}}\mathscr{I}(u)\Big|_{c_1=0}
    &=
    \phantom{-}i\frac{s_{123}^{5} \mathcal{R}^{6;25}}
    {\langle 12\rangle\langle 23\rangle\langle 34\rangle\langle 45\rangle\langle 56\rangle\langle 61\rangle}\,,\\
    \text{res}_{\frac{\mathtt{D}}{\mathtt{C}}}\mathscr{I}(u)\Big|_{c_1=0}
    &=
    \phantom{-}i\frac{s_{123}^{5} \mathcal{R}^{6;35}}
    {\langle 12\rangle\langle 23\rangle\langle 34\rangle\langle 45\rangle\langle 56\rangle\langle 61\rangle}\,,\\
    \text{res}_{\frac{\overline{\mathtt{C}}}{\overline{\mathtt{D}}}}\mathscr{I}(u)\Big|_{c_1=0}
    &=
    -i\frac{s_{123}^{5} \mathcal{R}^{1;36}}
    {\langle 12\rangle\langle 23\rangle\langle 34\rangle\langle 45\rangle\langle 56\rangle\langle 61\rangle}\,,\\
  \end{aligned}
\end{align}
where we used once more momentum and supermomentum conservation. The
statement that the sum over all residues vanishes translates into the
identity
\begin{align}
  \mathcal{R}^{6;24}+\mathcal{R}^{6;25}+\mathcal{R}^{6;35}=\mathcal{R}^{1;35}+\mathcal{R}^{1;36}+\mathcal{R}^{1;46}\,,
\end{align}
cf.\ equation (4.20) in \cite{Drummond:2008bq}. The invariant
\eqref{eq:inv63-u2204-formula} with
\eqref{eq:inv63-u2204-lastint-eval} in the region $s_{234},s_{126}>0$
and for $c_1=0$ becomes
\begin{align}
  \begin{aligned}
    \Psi_{6,3}(\boldsymbol{\lambda},\overline{\boldsymbol{\lambda}},\boldsymbol{\eta})
    &=
    64\pi i\delta^{4|0}(P)\delta^{0|8}(Q)
    \frac{\text{res}_{0}\mathscr{I}(u)+\text{res}_{\frac{\mathtt{A}}{\mathtt{B}}}\mathscr{I}(u)+\text{res}_{\frac{\overline{\mathtt{C}}}{\overline{\mathtt{D}}}}\mathscr{I}(u)}{s_{123}^5}\\
    &=
    64\pi\delta^{4|0}(P)\delta^{0|8}(Q)
    \frac{\mathcal{R}^{1;46}+\mathcal{R}^{1;35}+\mathcal{R}^{1;36}}{\langle 12\rangle\langle 23\rangle\langle 34\rangle\langle 45\rangle\langle 56\rangle\langle 61\rangle}\,.
  \end{aligned}
\end{align}
Up to a numerical prefactor, this is
$\mathcal{A}^{(\text{tree})}_{6,3}$ from
\eqref{eq:amp-super-nmhv6}. Thus the situation for the
$\mathfrak{u}(2,2|0+4)$ Yangian invariant
$\Psi_{6,3}(\boldsymbol{\lambda},\overline{\boldsymbol{\lambda}},\boldsymbol{\eta})$,
which we obtained from the unitary Graßmannian integral
\eqref{eq:grass-int-barg}, is completely analogous to the bosonic case
of section~\ref{sec:inv63-u22}. In one of the four kinematic regions
it agrees with the amplitude and in the other three it does not seem
to match. Once again we refer the reader to the conclusions in
chapter~\ref{cha:concl} for further thoughts on this issue.

\chapter{Conclusions and Outlook}
\label{cha:concl}

In this dissertation we investigated the integrable structure of
tree-level $\mathcal{N}=4$ SYM scattering amplitudes. It manifests
itself in the invariance of these amplitudes under the Yangian of the
superconformal algebra $\mathfrak{psu}(2,2|4)$. We adopted a
mathematical perspective on this topic by studying Yangian invariants
of the large class of algebras $\mathfrak{u}(p,q|m)$. This broadened
scope revealed fascinating connections to various concepts in the
field of integrable models. Notably, in chapter~\ref{cha:yang-rep} we
derived an elegant characterization of Yangian invariants within the
QISM, which is an extensive toolbox to study integrable spin
chains. It allowed us to show that Yangian invariants for
$\mathfrak{u}(2)$ can be constructed by means of a Bethe ansatz in
chapter~\ref{cha:bethe-vertex}. Therein we also established a link to
vertex models of statistical physics. A complementary method for the
construction of a class of Yangian invariants for
$\mathfrak{u}(p,q|m)$ was developed in
chapter~\ref{cha:grassmann-amp}. It has its origin in the Graßmannian
integral formulation of scattering amplitudes and generalizes certain
unitary matrix models. In the present chapter we recapitulate these
main results and highlight further findings chapter by chapter. Along
the way we provide some additional perspectives and mention future
directions. In particular, we emphasize the relevance of our work for
the ``physical'' problem we started out with, i.e.\ understanding the
integrable structure of planar $\mathcal{N}=4$ SYM
amplitudes. Finally, we conclude with some general remarks.

The goal of \textbf{chapter~\ref{cha:yang-rep}} was to establish a
common algebraic and representation theoretic language that covers
integrable spin chains and amplitudes alike, and can be applied for
our study of Yangian invariants throughout this thesis. The use of the
QISM form of the Yangian of $\mathfrak{gl}(n|m)$ was instrumental for
achieving this. In our review of this formulation in
section~\ref{sec:yangian} we emphasized that the Yangian generators
can be obtained from an expansion of a spin chain monodromy in its
spectral parameter. Building on this, we translated the Yangian
invariance condition into the QISM language in
section~\ref{sec:yangian-inv}. We identified Yangian invariants with
spin chain states that are specific eigenstates of the monodromy
matrix elements. This allows to interpret the deformation parameters
of the amplitudes in section~\ref{sec:deform} as inhomogeneities and
representation labels of spin chains. Our formulation of the Yangian
invariance condition makes the tools of the QISM applicable for the
construction of such invariants. Some of these tools were put to use
in chapter~\ref{cha:bethe-vertex}. The second cornerstone for
achieving the aforementioned goal was the choice of certain oscillator
representations of $\mathfrak{u}(p,q|m)\subset\mathfrak{gl}(n|m)$,
which we reviewed in section \ref{sec:osc-rep}, at the spin chain
sites. These include a wide range of spin chain models. They allow to
interpolate between the spin $\frac{1}{2}$ representation of
$\mathfrak{su}(2)$, which appears in the Heisenberg model, and an
infinite-dimensional representation of $\mathfrak{psu}(2,2|4)$ that
features in $\mathcal{N}=4$ SYM. We found that for the construction of
Yangian invariants we actually need two series of oscillator
representations, $\oscrep_c$ and the dual family
$\bar{\oscrep}_{c}$. Importantly, the representation label $c$ has to
be an integer in order to stay inside the Fock space. Such constraints
on the representation labels are usually neglected in the context of
deformed amplitudes, cf.\ section~\ref{sec:deform}. However, they
proved to be crucial for the results we obtained in
chapter~\ref{cha:grassmann-amp}. Furthermore, we paid attention to the
conjugation properties of the oscillators building up the
representations. These are of importance because they determine the
generators \eqref{eq:gen-dual} at the dual sites. In addition, they
influence the real form $\mathfrak{u}(p,q|m)$ of
$\mathfrak{gl}(p+q|m)$. The impact of this real form is nicely
illustrated by the two-site sample invariant $|\Psi_{2,1}\rangle$ from
section~\ref{sec:ncomp-susy-2-site}. In the compact case this
invariant is just a monomial, while in the non-compact setting it is a
Bessel function, which has an infinite power series expansion. Let us
also recapitulate the four-site sample invariant $|\Psi_{4,2}\rangle$
from sections~\ref{sec:comp-bos-4-site} and
\ref{sec:ncomp-susy-4-site}. Its Yangian invariance condition is
equivalent to a Yang-Baxter equation. Hence $|\Psi_{4,2}\rangle$ can
be interpreted as an R-matrix whose spectral parameter $z$ is given by
the difference of two inhomogeneities of the associated monodromy
$M_{4,2}(u)$. The two ``spectral parameters'' $z$ and $u$ have to be
distinguished. We also pointed out that for $\mathfrak{u}(2,2|4)$ this
R-matrix is essentially that of the one-loop $\mathcal{N}=4$ SYM spin
chain. This interesting link to the spectral problem was first
observed in \cite{Ferro:2012xw,Ferro:2013dga}.

\textbf{Chapter~\ref{cha:yang-rep}} led to some open questions which
deserve further attention. In section~\ref{sec:ncomp-susy-3-site} we
observed that non-compact three-site invariants do not seem to exist
for all algebras $\mathfrak{u}(p,q|m)$. We constructed a sample
invariant for $\mathfrak{u}(p,1)$ but we were not able to do so for
$\mathfrak{u}(2,2)$, which is in agreement with the non-existence of
three-particle scattering amplitudes for real momenta. It would be
desirable to study systematically for which algebras there can be a
Yangian invariant $|\Psi_{N,K}\rangle$. Because this invariant is an
element of the tensor product of $N-K$ ordinary representations
$\oscrep_c$ and $K$ dual ones $\bar{\oscrep}_c$, this can be
investigated by comparing the decomposition of the $N-K$-fold tensor
product of $\oscrep_c$ with that of the $K$-fold one. These
decompositions were studied in \cite{Kashiwara:1978} for bosonic
algebras and in \cite{Cheng:2004} for superalgebras. Another starting
point for inquiries is formula \eqref{eq:inv42-bos-hyper} that
expresses the compact Yangian invariant $|\Psi_{4,2}\rangle$ in terms
of the Gauß hypergeometric function ${}_2F_1$. There is a class of
multivariate hypergeometric functions defined on the complex
Graßmannian $\text{Gr}(N,K)$ which reduces to Gauß' function for
$\text{Gr}(4,2)$. This class is discussed in \cite{Gelfand:1990}, see
also the substantial review \cite{Gelfand:1992} and the lightning
introduction \cite{Beukers:2014} to similar functions. It remains to
be seen whether the invariant $|\Psi_{N,K}\rangle$ can be related to
said functions on $\text{Gr}(N,K)$.\footnote{Relations between
  multivariate hypergeometric functions and Yangian invariants were
  observed independently in the context of deformed amplitudes
  \cite{Ferro:2014gca}.} On a different note, it would be very
illuminating to obtain a closed expression for the non-compact
invariant $|\Psi_{4,2}\rangle$ in \eqref{eq:sample-inv42}, which is
presently only available in form of a multiple sum.

The major observation of \textbf{chapter~\ref{cha:bethe-vertex}} was
that Yangian invariants are specific eigenstates of spin chain
transfer matrices. This follows directly from our QISM
characterization of Yangian invariants put forward in
chapter~\ref{cha:yang-rep}. It applies to invariants with
representations of the superalgebra\footnote{Note that we spelled out
  the argument explicitly only in the bosonic case in
  section~\ref{sec:bethe-yangian}.} $\mathfrak{u}(p,q|m)$ and thus in
particular to the deformed amplitudes of section~\ref{sec:deform} that
are associated with the superconformal algebra
$\mathfrak{psu}(2,2|4)$. The diagonalization of spin chain transfer
matrices is the central problem addressed by the toolbox of the
QISM. Therefore these tools should be applicable for the construction
of Yangian invariants. In section~\ref{sec:bethe-ansatze} we reviewed
the most basic method of the toolbox, i.e.\ the algebraic Bethe ansatz
for the $\mathfrak{su}(2)$ Heisenberg spin chain. Thereby we filled in
some details on this model, which already served as an example in the
introductory section~\ref{sec:int-mod}. In
section~\ref{sec:bethe-yangian} we employed this Bethe ansatz for the
construction of $\mathfrak{u}(2)$ Yangian invariants, which can be
considered as toy models for amplitudes. We showed that those
invariants are special Bethe vectors. The Bethe roots parameterizing
these vectors are obtained from functional relations that are a
special case of the usual Baxter equation. Remarkably, unlike the full
equation, this special case can be solved with ease. The Bethe roots
form real strings in the complex plane and also the admissible
inhomogeneities and representation labels of the monodromy are
constrained. What is more, we found a simple superposition principle
for the solutions of the functional relations. While the simplest
two-site Yangian invariant $|\Psi_{2,1}\rangle$ corresponds to one
string of Bethe roots, the four-site invariant $|\Psi_{4,2}\rangle$ is
obtained by placing two of those strings in the complex
plane. Furthermore, our work on the Bethe ansatz led to a
classification of $\mathfrak{u}(2)$ Yangian invariants in terms of
permutations in \cite{Kanning:2014maa}. Astonishingly, these
permutations also appear in the study of deformed amplitudes, cf.\
section~\ref{sec:deform}. This illustrates the importance of our Bethe
ansatz for the investigation of the structure of Yangian invariants
even far beyond the $\mathfrak{u}(2)$ case. Finally, in
section~\ref{sec:vertex-yangian} we identified Yangian invariants with
partition functions of certain vertex models on in general
non-rectangular lattices. In particular, we demonstrated that our
Bethe ansatz for $\mathfrak{u}(2)$ Yangian invariants generalizes the
rational limit of Baxter's perimeter Bethe ansatz, which only covers
the spin $\frac{1}{2}$ representation.

Our work on the Bethe ansatz for Yangian invariants in
\textbf{chapter~\ref{cha:bethe-vertex}} suggests several natural
generalizations. The most urgent issue is to find an efficient
technique to evaluate the Yangian invariant Bethe vectors. For the
sample invariants in section~\ref{sec:bethe-gl2-sol} we computed the
Bethe vectors explicitly for small representation labels to arrive at
the formulas involving the oscillator contractions $(k\bullet l)$. We
were able to surpass this method for the invariant
$|\Psi_{2,1}\rangle$ by the elegant derivation presented in
appendix~\ref{sec:derivation-two-site}. Clearly it would be desirable
to extend this method to further invariants, say $|\Psi_{4,2}\rangle$
to begin with. A different direction concerns the extension of our
Bethe ansatz for $\mathfrak{u}(2)$ to more general algebras, i.e.\ the
exploration of a larger part of the ``landscape'' of invariants in
figure~\ref{fig:map-inv}. There are no conceptual difficulties to
cover the $\mathfrak{u}(n)$ case. In fact, we already provided the
relevant functional relations in appendix~\ref{sec:bethe-gln}. On a
technical level, however, the evaluation of the Bethe vectors appears
to be quite intricate due to nesting. Yet our results in
appendix~\ref{sec:high-rank-invar} show that at least for certain
sample invariants the nesting completely disappears and thereby the
complexity is reduced to that of the $\mathfrak{u}(2)$ case. These
simplifications deserve continuing attention. Furthermore, the
extension of the Bethe ansatz for Yangian invariants to compact
superalgebras $\mathfrak{u}(n|m)$, e.g.\ along the lines of
\cite{Kulish:1985bj}, should impose no obstacles. Interesting
conceptual questions are to be expected in the non-compact
$\mathfrak{u}(p,q|m)$ setting. The algebraic Bethe ansatz for
$\mathfrak{u}(2)$ is based on a reference state $|\Omega\rangle$,
which is given in \eqref{eq:bethe-gl2-hws-tot} by the tensor product
of the highest weight states at the spin chain sites. However, out of
the non-compact oscillator representations $\oscrep_c$ and
$\bar{\oscrep}_c$ only the latter has a highest weight, cf.\
section~\ref{sec:osc-rep}. Thus there is no such reference state in
the non-compact case. A way out of this dilemma might be to replace
the algebraic Bethe ansatz with another method from the QISM
toolbox. A suitable tool might be Sklyanin's separation of variables
\cite{Sklyanin:1991ss,Sklyanin:1995bm} as it does not require a
reference state. Some recent developments of this method are discussed
e.g.\ in \cite{Niccoli:2013bw}. Let us remark that it was also used to
solve the $\mathfrak{sl}(\mathbb{C}^2)$ spin chain appearing in QCD
\cite{Faddeev:1994zg}, which we encountered as an example in
section~\ref{sec:int-mod}. Lastly, we want to mention work that
potentially has close ties with our Bethe ansatz for Yangian
invariants. In \cite{Babujian:2006km} a set of equations
characterizing form factors of $1+1$-dimensional integrable quantum
field theories was solved in terms of Bethe vectors. One of these
equations appears to generalize our QISM version \eqref{eq:yi} of the
Yangian invariance condition. This equation is also known to be
related to a special case of the so-called quantum
Knizhnik-Zamolodchikov (qKZ) equation \cite{Frenkel:1991gx}, which is
solved by means of Bethe ansatz techniques in
\cite{Reshetikhin:1992}.\footnote{Similarities between the Yangian
  invariance condition and the qKZ equation were pointed out
  independently in \cite{Ferro:2014gca}.}

In \textbf{chapter~\ref{cha:grassmann-amp}} we advocated to equip the
Graßmannian integral with a unitary contour. We were led to that idea
by applying this method from the field of $\mathcal{N}=4$ SYM
scattering amplitudes, cf.\ sections~\ref{sec:grassmannian-integral}
and~\ref{sec:deform}, to the construction of Yangian invariants of a
wide range of algebras. We started out in
section~\ref{sec:grassmann-osc} by utilizing the Graßmannian integral
to build Yangian invariants $|\Psi_{N=2K,K}\rangle$ with oscillator
presentations of $\mathfrak{u}(p,q|m)$. In the resulting formula the
usual formal delta functions of spinor helicity variables of the SYM
case in section~\ref{sec:grassmannian-integral} are replaced by an
exponential function of oscillators. We were able to choose the
integration variable $\mathcal{C}$ to be the unitary group manifold
$U(K)$. We found that this contour eliminates branch cuts of the
integrand which plagued the Graßmannian integral for deformed
amplitudes in section~\ref{sec:deform}. This observation is tightly
interlocked with the restriction to integer representation labels
$c_i$. Let us emphasize that the choice of the contour is independent
of the algebra. Notably, it works for compact and non-compact algebras
alike. We termed our construction unitary Graßmannian matrix model
because for special values of the deformation parameters $v_i,c_i$ it
reduces to a well-known unitary matrix model, the Brezin-Gross-Witten
model. Our reasoning implies that this matrix model is Yangian
invariant in the external source fields. We evaluated our unitary
Graßmannian matrix model for several sample invariants, some which
were obtained by other means in chapters~\ref{cha:yang-rep}
and~\ref{cha:bethe-vertex}. In particular, we evaluated the invariant
$|\Psi_{4,2}\rangle$, i.e.\ the R-matrix, which takes the form of a
$U(2)$ matrix integral in our approach. This formula for the R-matrix
is an example of ideas from scattering amplitudes contributing to our
understanding of integrable models as such. In
section~\ref{sec:osc-spinor} we established a direct connection
between our results for oscillator representations of
$\mathfrak{u}(p,q=p|m)$ and the study of deformed scattering
amplitudes by applying a change of basis to spinor helicity-like
variables.  Technically, this was implemented by a Bargmann
transformation, which is an integral transformation known from the
harmonic oscillator in quantum mechanics. We identified those
oscillator representations that are relevant for tree-level amplitudes
by matching the $\mathfrak{u}(2,2|4)$ symmetry generators. The
``ordinary'' representation $\oscrep_{0}$ corresponds to a particle
with positive energy, which is enforced by
$(\tilde{\lambda}_{\dot{\alpha}})=+(\overline{\lambda}_\alpha)$. Likewise,
the dual representation $\bar{\oscrep}_{0}$ is associated with a
negative energy particle, i.e.\
$(\tilde{\lambda}_{\dot{\alpha}})=-(\overline{\lambda}_\alpha)$. Computing
the Bargmann transformation of the unitary Graßmannian matrix model
for $\mathfrak{u}(p,p|m)$ yielded a formula for the Yangian invariant
$\Psi_{2K,K}$ in spinor helicity-like variables that we investigated
in section~\ref{sec:grassmann-spinor}. It refines the Graßmannian
formula for deformed amplitudes of section~\ref{sec:deform} in several
aspects. Importantly, it is defined for the physical Minkowski
signature instead of split signature or complexified momentum
space. We showed that for this signature and our order of positive and
negative energy particles the choice of the unitary contour is, in a
way, dictated by momentum conservation. Lastly, we put our formula to
the test by evaluating sample invariants in the $\mathfrak{u}(2,2|4)$
case. We identified $\Psi_{4,2}$ with the deformed amplitude
$\mathcal{A}_{4,2}^{(\text{def.})}$. Thereby we constructed is essence
an explicit change of basis from the oscillator R-matrix of the planar
$\mathcal{N}=4$ SYM one-loop spectral problem to this deformed
amplitude. Furthermore, we evaluated the invariant $\Psi_{6,3}$, which
is a natural candidate for the presently unknown deformed amplitude
$\mathcal{A}_{6,3}^{(\text{def.})}$. Vexingly, however, its undeformed
limit $v_i,c_i=0$ coincides with the amplitude
$\mathcal{A}_{6,3}^{(\text{tree})}$ merely in the kinematic region
$s_{234}, s_{126}>0$.

Obviously, the most pressing open problem in
\textbf{chapter~\ref{cha:grassmann-amp}} is to clarify the precise
relation between the unitary Graßmannian matrix integral for the
undeformed $\Psi_{6,3}$ in the $\mathfrak{u}(2,2|4)$ case and the NMHV
superamplitude $\mathcal{A}_{6,3}^{(\text{tree})}$. Let us elaborate
on some of the logical possibilities for the apparent mismatch between
these two quantities. First, one might question the correctness of our
evaluation of the unitary Graßmannian integral
\eqref{eq:grass-int-barg} for the $\mathfrak{u}(2,2)$ version of
$\Psi_{6,3}$ in section~\ref{sec:inv63-u22}. Here we restrict to the
bosonic setting because supersymmetry does not seem to affect the
essential features. We certainly did make systematic mistakes because
we only focused on the contribution with maximal kinematic support,
which is proportional to a momentum conserving delta function. For
example we explicitly assumed
$\|\boldsymbol{\lambda}^{\text{d}}_1\|\neq 0$ to obtain the
intermediate step \eqref{eq:inv63-u22-step1}. Thus there might be
additional terms in case this norm vanishes. However, such terms
cannot make up for the mismatch between the undeformed Yangian
invariant $\Psi_{6,3}$ and the gluon amplitude
$A_{6,3}^{(\text{tree})}$ because of their restricted kinematic
support. The most puzzling feature of the undeformed invariant
$\Psi_{6,3}$ is the appearance of four kinematic regions in
\eqref{eq:inv63-u22-four-cases}, in only one of which the invariant
matches $A_{6,3}^{(\text{tree})}$. It would be highly desirable to
have a principle argument for the necessity of these regions rather
than merely the direct computation that led to
\eqref{eq:inv63-u22-four-cases}. We addressed this point in
appendix~\ref{sec:discrete-symmetry}, where we exposed a discrete
parity symmetry of the unitary Graßmannian integral. We showed that
the two kinematic regions of the simpler invariant $\Psi_{4,2}$ for
$\mathfrak{u}(1,1)$ in \eqref{eq:int-42-u11-lastint-def-expl} are
inevitable because of this symmetry. In case of the
$\mathfrak{u}(2,2)$ version of $\Psi_{6,3}$ the parity symmetry
interrelates only two of the four regions in
\eqref{eq:inv63-u22-four-cases}. In particular, the amplitude region
$s_{234}, s_{126}>0$ is parity invariant by itself. Therefore this
symmetry of not sufficient to explain the need for all four regions.
It would be interesting to search for a larger discrete symmetry group
of $\Psi_{6,3}$ that does relate all regions. For this purpose it
might be helpful to study in detail the kinematics of six massless
relativistic particles e.g.\ by means of the techniques in
\cite{Byckling:1973}. Let us mention in this regard that we verified
numerically the existence of physical momentum configurations in each
of the four regions using \cite{Maitre:2007jq}. The second logical
possibility we want to comment on is that the unitary contour of the
Graßmannian integral might be conceptually wrong for the construction
of the sought after Yangian invariants. However, we provided a formal
proof of the Yangian invariance of the Graßmannian integral with
oscillator variables in
section~\ref{sec:proof-yang-invar}. Assumptions in this proof about
the then unspecified contour were precisely satisfied by the selection
of the unitary contour in section~\ref{sec:single-valu-integr}.  What
is more, we explicitly verified the Yangian symmetry of sample
invariants computed using the unitary contour in
section~\ref{sec:sample-invariants}. Therefore we do not doubt the
Yangian invariance of the unitary Graßmannian integral
\eqref{eq:grass-int-unitary} in oscillator variables. We also have no
reason to assume that this is affected by the Bargmann transformation
to the spinor helicity version \eqref{eq:grass-int-barg} of that
integral. Still, one might argue that some amplitudes belong to a
different class of Yangian invariants which is not captured by the
unitary contour. We tried to address this concern in
appendix~\ref{sec:glue-contours}, where we showed in the bosonic case
that once the $U(2)$ contour of $|\Psi_{4,2}\rangle$ is fixed, the
$U(3)$ contour of $|\Psi_{6,3}\rangle$ follows from gluing. Recall in
this context from section~\ref{sec:inv42-u22} the agreement between
the Bargmann transformation of the undeformed $|\Psi_{4,2}\rangle$ for
$\mathfrak{u}(2,2)$ and the gluon amplitude
$A_{4,2}^{(\text{tree})}$. As a third logical possibility for the
mismatch between the undeformed Yangian invariant $\Psi_{6,3}$ for
$\mathfrak{u}(2,2|4)$ and the superamplitude
$\mathcal{A}_{6,3}^{(\text{tree})}$, let us speculate that the
symmetries of this amplitude might not quite be as commonly
expected. The breakdown of its superconformal and Yangian invariance
at certain singularities is well established, cf.\
\cite{Bargheer:2009qu,Sever:2009aa,Bargheer:2011mm}. Possibly these
symmetries are in addition broken in a subtle way by the lack of the
different kinematic regions, which are separated by such
singularities. This speculation is motivated by our analysis of
$\Psi_{4,2}$ for $\mathfrak{u}(1,1)$ in
appendix~\ref{sec:parity-four-site-spin-hel}, where we discussed for
this sample invariant the possibility of extending the expression from
one of the two kinematic regions to the entire domain. Even though
this function would satisfy the Yangian invariance condition
\eqref{eq:yi-exp-1} as a differential equation for generic
$\lambda^i_\alpha$, it would violate the parity symmetry. This
symmetry is present in the oscillator invariant $|\Psi_{4,2}\rangle$
for $\mathfrak{u}(1,1)$ and therefore also in the associated R-matrix,
see appendix~\ref{sec:parity-osc}. The lesson learned from this
example might apply to Yangian invariants with infinite-dimensional
representations of $\mathfrak{u}(p,p|m)$ in general. In the spinor
helicity-like basis it seems to be insufficient to verify the Yangian
invariance condition \eqref{eq:yi-exp-1} only for generic external
data. In principle, one would probably also have to do the intricate
analysis for external data at the singularities. In practice, however,
it might often be possible to avoid this by exploiting discrete
symmetries. This brings us back to the search for the discrete
symmetry group of $\Psi_{6,3}$, which we already proposed above. On a
slightly different note, we would like to mention the study of NMHV
superamplitudes for $(2,2)$ signature in
\cite{Korchemsky:2009jv}. There the authors modified the usual
expressions for these amplitudes by introducing sign factors that
depend on certain kinematic regions. Only after this modification they
were able to apply the conformal inversion, which is not an
infinitesimal but a finite conformal transformation, to those
expressions. Their kinematic regions are somewhat reminiscent of our
regions for the $\mathfrak{u}(2,2)$ version of $\Psi_{6,3}$ in
Minkowski signature in \eqref{eq:inv63-u22-four-cases}. Thus it might
be instructive to compare the behavior of $\Psi_{6,3}$ and
$A_{6,3}^{(\text{tree})}$ under finite conformal transformations,
whose action is given e.g.\ in
\cite{Gross:1972,Post:1976,Jakobsen:1977}. In this context it could
also be necessary to think about ``finite Yangian transformations''.
Ultimately, we could not clarify the relation between our function
$\Psi_{6,3}$ for $\mathfrak{u}(2,2|4)$ and the NMHV superamplitude
$\mathcal{A}_{6,3}^{(\text{tree})}$ in this thesis. We believe,
however, that this issue is of crucial importance to understand the
role of integrability for amplitudes.

Once this conceptual problem is resolved,
\textbf{chapter~\ref{cha:grassmann-amp}} offers a host of interesting
further directions to be explored. Clearly, our aim is to relate the
Yangian invariant $\Psi_{2K,K}$ computed by the unitary Graßmannian
integral \eqref{eq:grass-int-barg} to the tree-level amplitude
$\mathcal{A}_{2K,K}^{(\text{tree})}$ and deformations
thereof,\footnote{Let us mention a subtlety concerning the parameter
  $K$. It is defined as the number of negative energy representations
  in $\Psi_{2K,K}$. After selecting the representation labels $c_i=0$,
  it agrees with the degree of helicity violation of the amplitude
  $\mathcal{A}_{2K,K}^{(\text{tree})}$, see
  section~\ref{sec:superamplitudes}. In the context of amplitudes one
  usually does not specify the number of negative energy particles.}
at least in one kinematic region. This would demonstrate the relevance
of the very natural unitary contour for a large class of tree-level
amplitudes. Fortunately, this class contains amplitudes of all
$\text{MHV}$ degrees and thus ranges from amplitudes whose explicit
expressions are very simple to those of immense complexity, cf.\
section~\ref{sec:gluon-amp}. For the invariant $\Psi_{6,3}$ of
$\mathfrak{u}(2,2|4)$ we reduced the defining $U(3)$ integral in
\eqref{eq:grass-int-barg} to a $U(1)$ integral that was then solved in
the undeformed case by means of the residue theorem. The $U(K)$
integral for the invariant $\Psi_{2K,K}$ of $\mathfrak{u}(p,p|m)$ is
expected to reduce to a $U(K-p)$ integral after exploiting the bosonic
delta functions. In general, this is still a multi-dimensional
integral, albeit with a fully specified contour. A technique to
evaluate its undeformed version might be the global residue theorem,
which has already been employed in the context of amplitudes
\cite{ArkaniHamed:2009dn}, cf.\
section~\ref{sec:grassmannian-integral}. Another puzzling question is
to understand the geometric role of the $U(K)$ contour for
$\mathcal{C}$ within the Graßmannian $\text{Gr}(2K,K)$, that we
parameterized in \eqref{eq:grassint-matrix} by
$C=\big(\begin{array}{c:c}1_{K}&\mathcal{C}\end{array}\big)$. It would
be desirable to specify the contour in a way that does not rely on
this particular gauge fixing of $C$. Throughout
chapter~\ref{cha:grassmann-amp} we concentrated on invariants with
$N=2K$ because only in this case $\mathcal{C}$ is a \emph{square}
matrix. Naturally, we would like to extend the unitary Graßmannian
integral \eqref{eq:grass-int-barg} to $N\neq 2K$ and thereby access
all tree-level amplitudes. Here the issue is to use an appropriate
measure on the complex Stiefel manifold of \emph{rectangular}
$K\times(N-K)$ matrices $\mathcal{C}$ with
$\mathcal{C}\mathcal{C}^\dagger=1_{K\times K}$, see e.g.\
\cite{Mathai:1997}. This generalizes the unitary group manifold to the
case of rectangular matrices. The extension to $N\neq 2K$ is also of
interest for the Graßmannian matrix model \eqref{eq:grass-int-unitary}
for $|\Psi_{N,K}\rangle$ in the oscillator basis. In
section~\ref{sec:ncomp-susy-3-site} we observed that the
$\mathfrak{gl}(p|r)$ invariant oscillator contractions $(k\bullet l)$
and the $\mathfrak{gl}(q|s)$ invariants $(k\circ l)$ are not
sufficient to build up the non-compact Yangian invariant
$|\Psi_{3,1}\rangle$. In fact, classical invariant theory, see e.g.\
\cite{Kraft:1996}, suggests to supplement these elementary building
blocks by determinants of oscillators. Consequently, the oscillator
dependence of the exponential function in the Graßmannian matrix model
\eqref{eq:grass-int-unitary} will have to be modified for $N\neq 2K$.
Another open question concerns formula \eqref{eq:grass-int-unitary} in
the $N=2K$ case. We pointed out the lack of efficient technology for
its evaluation during the computation of sample invariants in
section~\ref{sec:sample-invariants}. We want to overcome this issue by
applying matrix model methods for the evaluation of the Graßmannian
integral \eqref{eq:grass-int-unitary} beyond the Leutwyler-Smilga case
\eqref{eq:red-matrix-fin-inv}. One might wonder whether the Bessel
function formula \eqref{eq:ls-integral-bessel} generalizes to Yangian
invariants with general deformation parameters $v_i,c_i$. One
technique for this endeavor could be a character expansion, which was
successfully employed for the Leutwyler-Smilga model in
\cite{Schlittgen:2002tj,Balantekin:2000vn}. Another auspicious method
may be the use of Gelfand-Tzetlin coordinates, which has been applied
to compute correlation functions of the Itzykson-Zuber model
\cite{Shatashvili:1992fw}. In our setting these coordinates might be
well adapted to the minors appearing in the Graßmannian integral
\eqref{eq:grass-int}. Yet another approach could be to employ
Weingarten functions \cite{Collins:2003}, which provide explicit
formulas for integrals of products of matrix elements over the unitary
group. A different point to be addressed in the future is the
transformation of the unitary Graßmannian matrix model
\eqref{eq:grass-int-unitary} to twistor variables because these are
used in a large part of the amplitudes literature. While we
implemented the relation between oscillators and spinor helicity-like
variables via a Bargmann transformation in this thesis, we completely
left aside twistors. As the unitary contour is intrinsically related
to Minkowski signature, we cannot simply employ the half Fourier
transform of \cite{Witten:2003nn} for the transition from spinor
helicity-like variables to twistors. However, a point of departure
could be a twistorial description of the $\mathfrak{u}(p,q)$
oscillator representations, a.k.a.\ ``ladder representations'',
discussed e.g.\ in \cite{Dunne:1990}. Let us move on to another
promising topic. At the end of section~\ref{sec:unit-matr-models} we
mentioned that the partition functions of the Brezin-Gross-Witten and
of the Leutwyler-Smilga model are solutions, so-called
$\tau$-functions, of the KP hierarchy, which is an infinite set of
classically integrable equations. This immediately leads to the
question whether also the more general unitary Graßmannian matrix
model introduced in section~\ref{sec:unitary-contour} is a KP
$\tau$-function. Moreover, one might hope for a direct connection
between the KP hierarchy and integrability in the sense of Yangian
invariance because the latter was our guiding principle for the
construction of the unitary matrix models in
sections~\ref{sec:unit-matr-models} and
\ref{sec:unitary-contour}. Such a structural insight would make the
large body of work on the KP hierarchy applicable for the construction
of Yangian invariants and possibly also tree-level $\mathcal{N}=4$ SYM
scattering amplitudes. There is a further independent indication for a
connection between Yangian invariants and the KP hierarchy. The
partition function of the six-vertex model with domain wall boundary
conditions is known to be a special KP $\tau$-function
\cite{Foda:2009}. In the rational limit this setup belongs to the
class of vertex models considered in
section~\ref{sec:vertex-yangian}. There we established that the
partition functions of these models are components of Yangian
invariant vectors. Additional clues might be provided by a link
between transfer matrices of quantum integrable spin chains and
$\tau$-functions \cite{Alexandrov:2011aa,Zabrodin:2012gx}. Finally, we
would like to discuss the relevance of our results for loop
amplitudes. In section~\ref{sec:deform} we mentioned that a deformed
amplitude $\mathcal{A}_{6,3}^{(\text{def.})}$ might yield a
regularization of the one-loop amplitude $\mathcal{A}_{4,2}^{(1)}$. It
remains to be seen if the deformed Yangian invariant $\Psi_{6,3}$ in
\eqref{eq:inv63-u2204-formula} with
\eqref{eq:inv63-u2204-lastint-general} serves this purpose. Let us
instead speculate about a conceptually clear route to
loop-amplitudes. We already emphasized an important link between the
unitary Graßmannian matrix model \eqref{eq:grass-int-unitary} for
$\mathfrak{u}(2,2|4)$ and the one-loop spectral problem of planar
$\mathcal{N}=4$ SYM. Both can be formulated in terms of essentially
the same oscillator representations. We may use a grading for which
each of the building blocks $(k\bullet l)$ and $(k\circ l)$ of
\eqref{eq:grass-int-unitary} is invariant under one of the two compact
subalgebras
$\mathfrak{su}(2|2)\oplus\mathfrak{su}(2|2)\subset\mathfrak{u}(2,2|2+2)$,
see also appendix~\ref{sec:inv42-u2222}. In the spectral problem a
central extension of such subalgebras is the key to all-loop results,
cf.\ section~\ref{sec:spectrum}. Appealing to a common integrable
structure of the entire $\mathcal{N}=4$ model, we suspect that in this
way a coupling constant can also be introduced in our unitary
Graßmannian integral formula. Such an approach would also shed light
on an infinite-dimensional symmetry algebra that possibly governs the
all-loop amplitudes.

After specifying our objectives in
section~\ref{sec:objectives-outline}, we set out in this dissertation
to build a robust bridge between integrable models and planar
$\mathcal{N}=4$ SYM scattering amplitudes. Even though this
construction is most definitely not completed, we established
promising ties between these two fields in both directions. The Bethe
ansatz construction of simple Yangian invariants, which can be
considered as toy models for amplitudes, is one example. The unitary
Graßmannian integral with its origin in the field of amplitudes and
its potential connection to integrable hierarchies is another one for
the reverse direction. Our representation theoretic setup, that in
particular respects the reality conditions of the variables involved,
allows us to ask very detailed questions. Arguably the most urgent one
at present is to clarify the relation of the Yangian invariants
computed by our unitary Graßmannian integral and tree-level
amplitudes. We hope to extend the framework established in this thesis
in future work because we believe that it bears the potential to
contribute to both areas of research, integrable models and scattering
amplitudes.

\chapter*{Acknowledgments}
\phantomsection
\addcontentsline{toc}{chapter}{Acknowledgments}
\markboth{Acknowledgments}{Acknowledgments}
\label{cha:ackn}

I would like to express my gratitude to my advisor Matthias
Staudacher. In particular, I thank him for sharing his insights in
frequent inspiring discussions and for his encouragement throughout
the course of this thesis. Next, I would like to thank Rouven Frassek,
Yumi Ko, Tomasz Łukowski, Gregor Richter and Matthias Staudacher for
instructive and enjoyable collaboration. I am indebted to Rouven
Frassek, Gregor Richter and Matthias Staudacher for valuable comments
on the manuscript. Moreover, I am much obliged to Jacob Bourjaily, Jan
Plefka and Matthias Staudacher for preparing referee reports on this
dissertation.

I thank Lorenzo Bianchi, Marco Bianchi, Johannes Brödel, Ita Brunke,
Harald Dorn, Burkhard Eden, Livia Ferro, Jan Fokken, Valentina Forini,
Rouven Frassek, Sergey Frolov, Ilmar Gahramanov, Raquel
Gomez-Ambrosio, Philipp Hähnel, Martin Heinze, Ben Hoare, Chrysostomos
Kalousios, Stephan Kirsten, Thomas Klose, Yumi Ko, Laura Koster, Pedro
Liendo, Florian Loebbert, Tomasz Łukowski, Christian Marboe, David
Meidinger, Carlo Meneghelli, Dennis Müller, Hagen Münkler, Dhritiman
Nandan, Vladimir Mitev, Hans-Peter Pavel, Brenda Penante, Jan Plefka,
Jonas Pollok, Elli Pomoni, Israel Ramirez, Gregor Richter, Sylvia
Richter, Sourav Sarkar, Annegret Schalke, Jan Schlenker, Alessandro
Sfondrini, Christoph Sieg, Vladimir Smirnov, Matthias Staudacher,
Stijn van Tongeren, Zengo Tsuboi, Vitaly Velizhanin, Edoardo Vescovi,
Matthias Wilhelm, Sebastian Wuttke, Yingying Xu, Gang Yang, Leonard
Zippelius and the entire group at Humboldt University Berlin for
discussions, company and creating a pleasant atmosphere. For these
reasons I am also grateful to the mathematical physics group at the
Ludwig Maximilians University of Munich and in particular to Livia
Ferro's group, where I had the possibility to complete this
manuscript. Furthermore, I thank Changrim Ahn, Ludvig Faddeev, Frank
Göhmann, Andrew Hodges, Shota Komatsu, Ivan Kostov, Marius de Leeuw,
Lionel Mason and Didina Serban for helpful conversations and answers
to some burning questions.

I also greatly appreciated the opportunity to take part in numerous
``class trips'' with Matthias Staudacher's research group to physics
institutes around the globe. I am thankful to the Perimeter Institute
for Theoretical Physics, the Institut Henri Poincaré, the Israel
Institute for Advanced Studies, the Kavli Institute for the Physics
and Mathematics of the Universe, the CERN Theory Division, the
C.~N.~Yang Institute for Theoretical Physics at Stony Brook University
and the Simons Center for Geometry and Physics for kind hospitality
and for providing stimulating working environments.

I am grateful to the Studienstiftung des Deutschen Volkes and the
Albert Einstein Institute for Ph.D. scholarships. In addition, my work
was supported by the GK\ 1504 ``Masse, Spektrum, Symmetrie'', the SFB\
647 ``Raum, Zeit, Materie'', the Marie Curie Network GATIS of the EU's
FP7-2007-2013 under grant agreement No.\ 317089, the Marie Curie
International Research Staff Exchange Network UNIFY of
FP7-People-2010-IRSES under grant agreement No.\ 269217 and the DFG
grant FE 1529/1-1.

Schließlich bedanke ich mich ganz herzlich bei meiner Familie und
insbesondere bei meinen Eltern für die fortwährende Unterstützung.

\appendix

\addtocontents{toc}{\protect\setcounter{tocdepth}{1}}

\chapter{Loose Ends of Bethe Ansatz}
\label{cha:app-bethe}

In this appendix we take initial steps in addressing some questions
that remained unresolved in the discussion of the Bethe ansatz for
Yangian invariants in section~\ref{sec:bethe-yangian}. We begin in
section~\ref{sec:derivation-two-site} by developing a technique to
evaluate the Bethe vector leading to the compact two-site Yangian
invariant for $\mathfrak{gl}(2)$. Section~\ref{sec:bethe-gln} contains
a generalization of the functional relations which characterize
compact $\mathfrak{gl}(2)$ Yangian invariants to the
$\mathfrak{gl}(n)$ case. Lastly, in section~\ref{sec:high-rank-invar}
we demonstrate that for certain $\mathfrak{gl}(n)$ Yangian invariants
the complicated nesting, which is inherent to higher rank Bethe
ansätze, disappears.

\section{Derivation of Two-Site Invariant}
\label{sec:derivation-two-site}

We constructed sample solutions of the functional relation
\eqref{eq:bethe-gl2-ad} for finite-dimensional $\mathfrak{gl}(2)$
Yangian invariants in section \ref{sec:bethe-gl2-sol}. What is more,
we even explained a classification of its solutions, in
section~\ref{sec:class-solut}. However, we did not present a
satisfactory method to evaluate the corresponding Bethe vectors,
although we proved them to be Yangian invariant. Instead, we relied on
explicit case-by-case calculations for small values of the
representation labels. Here we fill this gap for the simplest example
of the two-site invariant $|\Psi_{2,1}\rangle$, which we discussed in
section~\ref{sec:bethe-gl2-sol-line}. The explicit form
\eqref{eq:osc-psi21} of $|\Psi_{2,1}\rangle$ is a product of $c_2$
factors $(1\bullet 2)$. Similarly, the associated algebraic Bethe
vector \eqref{eq:bethe-gl2-eigenvector} is a $c_2$-fold product of
operators $B(u_k)$ with the Bethe roots $u_k$ from
\eqref{eq:bethe-gl2-sol-line-roots}. The idea of the following
derivation is to keep this product structure manifest and to show how
each factor $(1\bullet 2)$ corresponds to one operator $B(u_k)$.

We reformulate the $\mathfrak{gl}(2)$ two-site Yangian invariant
\eqref{eq:osc-psi21} as
\begin{align}
\label{eq:derivation-aba-2site-rewrite-explcit}
  \begin{aligned}
    |\Psi_{2,1}\rangle 
    &= 
    (1\bullet 2)^{c_2}
    |0\rangle 
    =
    (\bar{\mathbf{a}}_1^1\bar{\mathbf{a}}_1^2+\bar{\mathbf{a}}_2^1\bar{\mathbf{a}}_2^2)^{c_2}
    |0\rangle\\
    &=
    \left(\bar{\mathbf{a}}_1^1\mathbf{a}_2^1\frac{1}{\bar{\mathbf{a}}_2^1\mathbf{a}_2^1}
      +\bar{\mathbf{a}}_2^2\mathbf{a}_1^2\frac{1}{\bar{\mathbf{a}}_1^2\mathbf{a}_1^2}\right)^{c_2}
    (\bar{\mathbf{a}}_2^1\bar{\mathbf{a}}_1^2)^{c_2}
    |0\rangle 
    =
    \left(
      \bar{\mathbf{J}}_{21}^1\frac{1}{\bar{\mathbf{J}}_{22}^1}
      +\mathbf{J}_{21}^2\frac{1}{\mathbf{J}_{11}^2}
    \right)^{c_2}
    |\Omega\rangle\,,
  \end{aligned}
\end{align}
where the reference state $|\Omega\rangle$ is that of
\eqref{eq:bethe-gl2-sol-line-vac}. The inverse powers of the number
operators $\mathbf{J}_{11}^2$ and $\bar{\mathbf{J}}_{22}^1$ are
well-defined because they act on states with at least one of the
respective oscillators. Next, we move the inverse powers of
$\mathbf{J}_{11}^2$ to the right using
$\mathbf{J}_{11}^2\mathbf{J}_{21}^2=\mathbf{J}_{21}^2(\mathbf{J}_{11}^2-1)$,
which implies
$f(\mathbf{J}_{11}^2)\mathbf{J}_{21}^2=\mathbf{J}_{21}^2f(\mathbf{J}_{11}^2-1)$
for a function $f(\mathbf{J}_{11}^2)$. Then we evaluate their action
on the highest weight state with
$\mathbf{J}_{11}^2|\Omega\rangle=c_2|\Omega\rangle$,
\begin{align}
  \label{eq:derivation-aba-2site-initial-order}
  \begin{aligned}
    |\Psi_{2,1}\rangle
    &=
    \frac{1}{c_2!}
    \left(\bar{\mathbf{J}}_{21}^1\frac{1}{\bar{\mathbf{J}}_{22}^1}\mathbf{J}_{11}^2+\mathbf{J}_{21}^2\right)
    \left(\bar{\mathbf{J}}_{21}^1\frac{1}{\bar{\mathbf{J}}_{22}^1}(\mathbf{J}_{11}^2-1)+\mathbf{J}_{21}^2\right)\\
    &\quad\quad\quad\cdots
    \left(\bar{\mathbf{J}}_{21}^1\frac{1}{\bar{\mathbf{J}}_{22}^1}(\mathbf{J}_{11}^2-c_2+1)+\mathbf{J}_{21}^2\right)
    |\Omega\rangle\,.
  \end{aligned}
\end{align}
As the factors in this product commute, we reverse their order. Next,
the inverse powers of $\bar{\mathbf{J}}_{22}^1$ are moved to the right
with the help of
$\bar{\mathbf{J}}_{22}^1\bar{\mathbf{J}}_{21}^1=\bar{\mathbf{J}}_{21}^1(\bar{\mathbf{J}}_{22}^1+1)$
and $\bar{\mathbf{J}}_{22}^1|\Omega\rangle=-c_{2}|\Omega\rangle$,
\begin{align}
  \label{eq:derivation-aba-2site-reversed-order}
  \begin{aligned}
    |\Psi_{2,1}\rangle
    =
    \frac{(-1)^{c_2}}{c_2!^2}
    &\left(\bar{\mathbf{J}}_{21}^1(\mathbf{J}_{11}^2-c_2+1)+\mathbf{J}_{21}^2\bar{\mathbf{J}}_{22}^1\right)\\
    &\cdots
    \left(\bar{\mathbf{J}}_{21}^1(\mathbf{J}_{11}^2-1)+\mathbf{J}_{21}^2(\bar{\mathbf{J}}_{22}^1+c_2-2)\right)
    \left(\bar{\mathbf{J}}_{21}^1\mathbf{J}_{11}^2+J_{21}^2(\bar{\mathbf{J}}_{22}^1+c_2-1)\right)
    |\Omega\rangle\,.
  \end{aligned}
\end{align}
Taking into account $\mathbf{J}_{11}^2+\mathbf{J}_{22}^2=c_2$ and
$\bar{\mathbf{J}}_{11}^1+\bar{\mathbf{J}}_{22}^1=-c_2$, which is valid
for the representations at hand, yields
\begin{align}
  \label{eq:derivation-aba-2site-reformulated}
  \begin{aligned}
    |\Psi_{2,1}\rangle
    =
    \frac{1}{c_2!^2}
    &\left(\bar{\mathbf{J}}_{21}^1(\mathbf{J}_{22}^2-1)+\mathbf{J}_{21}^2(\bar{\mathbf{J}}_{11}^1+c_2)\right)\\
    &\cdots
    \left(\bar{\mathbf{J}}_{21}^1(\mathbf{J}_{22}^2-c_2+1)+\mathbf{J}_{21}^2(\bar{\mathbf{J}}_{11}^1+2)\right)
    \left(\bar{\mathbf{J}}_{21}^1(\mathbf{J}_{22}^2-c_2)+\mathbf{J}_{21}^2(\bar{\mathbf{J}}_{11}^1+1)\right)
    |\Omega\rangle\,.
  \end{aligned}
\end{align}
The Bethe vector \eqref{eq:bethe-gl2-eigenvector} is built from the
operator
\begin{align}
  B(u)
  =
  \frac{\bar{\mathbf{J}}_{21}^1(\mathbf{J}_{22}^2+u-v_2)+\mathbf{J}_{21}^2(\bar{\mathbf{J}}_{11}^1+u-v_1)}
  {(u-v_1)(u-v_2)}\,.
\end{align}
This is a matrix element of the monodromy \eqref{eq:osc-m21} and we
inserted the trivial normalization \eqref{eq:osc-m21-norm}. Next, we
use the two-site solution \eqref{eq:bethe-gl2-sol-line-constr} of the
functional relation \eqref{eq:bethe-gl2-ad}, in particular
$v_1=v_2-1-c_2$, and the Bethe roots
\eqref{eq:bethe-gl2-sol-line-roots}, i.e.\ $u_k=v_2-k$. This shows
that each factor in \eqref{eq:derivation-aba-2site-reformulated}
matches the numerator of one operator $B(u_k)$. The denominators
account for the prefactor in
\eqref{eq:derivation-aba-2site-reformulated}. Consequently, we proved
that
\begin{align}
  \label{eq:derivation-aba-2site-identification}
  |\Psi_{2,1}\rangle
  =
  (1\bullet 2)^{c_2}
  |0\rangle
  =
  (-1)^{c_2}
  B(u_1)\cdots B(u_{c_2-1})B(u_{c_2})
  |\Omega\rangle\,
\end{align}
for any $c_2\in\mathbb{N}$. It would certainly be interesting to
extend this derivation to the other sample invariants of
section~\ref{sec:bethe-gl2-sol} and even more generally to the
solutions of the functional relations classified in
section~\ref{sec:class-solut}.

\section{Functional Equations for Higher Rank}
\label{sec:bethe-gln}

In section~\ref{sec:bethe-gl2} we showed that the algebraic Bethe
ansatz for an inhomogeneous spin chain with finite-dimensional
$\mathfrak{gl}(2)$ representations can be specialized to the case of
Yangian invariant Bethe vectors. The Yangian invariants are then
characterized by the functional relations
\eqref{eq:bethe-gl2-specbaxter}. They restrict the admissible
inhomogeneities and representations of the monodromy and furthermore
determine the Bethe roots. Our method is based on the key observation
that a Yangian invariant is a special eigenstate of a transfer matrix,
cf.\ \eqref{eq:bethe-inv-eigen}. This observation is clearly valid
beyond the $\mathfrak{gl}(2)$ case, in particular it remains true for
the higher rank $\mathfrak{gl}(n)$ algebra. Transfer matrices with
finite-dimensional $\mathfrak{gl}(n)$ representations can be
diagonalized using the \emph{nested algebraic Bethe ansatz} in
\cite{Kulish:1983rd}. In the present section we specialize this
construction to the case of Yangian invariant Bethe vectors. This
leads to a $\mathfrak{gl}(n)$ generalization of the functional
relations \eqref{eq:bethe-gl2-specbaxter}. We skip cumbersome
technical details and do not present the derivation of these relations
here, nor do we work out the intricate form of the associated Yangian
invariant nested Bethe vectors.

The functional relations for $\mathfrak{gl}(n)$ read
\begin{align}
  \label{eq:bethe-gln-specbaxter}
  \begin{aligned} 1&=\mu_1(\spec) \frac{Q_1(\spec-1)}{Q_1(\spec)}\,,\\
1&=\mu_2(\spec) \frac{Q_1(\spec+1)}{Q_1(\spec)}\,
\frac{Q_2(\spec-1)}{Q_2(\spec)}\,,\\ 1&=\mu_3(\spec)
\frac{Q_2(\spec+1)}{Q_2(\spec)}\, \frac{Q_3(\spec-1)}{Q_3(\spec)}\,,\\
&\;\;\vdots\\ 1&=\mu_{n-1}(\spec)
\frac{Q_{n-2}(\spec+1)}{Q_{n-2}(\spec)}\,
\frac{Q_{n-1}(\spec-1)}{Q_{n-1}(\spec)}\,,\\ 1&=\mu_{n}(\spec)
\frac{Q_{n-1}(\spec+1)}{Q_{n-1}(\spec)}\,.
  \end{aligned}
\end{align} 
These relations restrict the inhomogeneities and the representation
labels of $\mathfrak{gl}(n)$ monodromies $M(u)$ that admit Yangian
invariants. Furthermore, they determine the corresponding Bethe roots.
The monodromy matrix $M(u)$ is defined in
\eqref{eq:yangian-mono-spinchain} in terms of the Lax operators
\eqref{eq:yangian-def-lax}. The eigenvalues of its elements
$\mon_{11}(\spec),\ldots,\mon_{nn}(\spec)$ on the reference state of
the Bethe ansatz are denoted $\mu_1(\spec),\ldots,\mu_n(\spec)$, cf.\
\eqref{eq:bethe-gl2-vacuum} for the $\mathfrak{gl}(2)$ case. We work
with a finite-dimensional $\mathfrak{gl}(n)$ representation
$\mathcal{V}_i$ with a highest weight
$\Xi_i=(\xi_{i}^{(1)},\ldots,\xi_{i}^{(n)})$ at the $i$-th spin chain
site. In analogy to the $\mathfrak{gl}(2)$ equation
\eqref{eq:bethe-gl2-alphadelta}, the monodromy eigenvalues are
parametrized as
\begin{align}
  \label{eq:bethe-gln-mu} \mu_{a}(\spec) =
\prod_{i=1}^{\sites}f_{\mathcal{V}_i}(\spec-\inh_i)
\frac{\spec-\inh_{i}+\xi_{i}^{(a)}}{\spec-\inh_{i}}\,.
\end{align} 
The Q-function
\begin{align}
  \label{eq:bethe-gln-q}
  Q_k(\spec)=\prod_{i=1}^{\brts_k}(\spec-\brt_{i}^{(k)})\,,
\end{align} 
encodes $\brts_k$ Bethe roots $\brt_i^{(k)}$, where $k$ is the nesting
level taking the values $1,\ldots,n-1$. We observe that for $n=2$ the
functional relations \eqref{eq:bethe-gln-specbaxter} reduce to the
$\mathfrak{gl}(2)$ case \eqref{eq:bethe-gl2-specbaxter}.

The relations \eqref{eq:bethe-gln-specbaxter} decouple into
\begin{align}
  \label{eq:bethe-gln-separated-mu}
  1&=\;\prod_{\mathclap{a=1}}^n\;\mu_a(\spec-a+1)\,,\\
  \label{eq:bethe-gln-separated-qk} 
  \frac{Q_k(\spec)}{Q_k(\spec+1)}
   &=\;\prod_{\mathclap{a=k+1}}^n\;\mu_a(\spec-a+k+1)
\end{align} 
with $k=1,\ldots,n-1$. Here \eqref{eq:bethe-gln-separated-mu} is free
of Bethe roots and solely constraints the inhomogeneities and
representations of the monodromy. Equation
\eqref{eq:bethe-gln-separated-qk} contains only Bethe roots of the
nesting level $k$. The equations \eqref{eq:bethe-gln-separated-mu} and
\eqref{eq:bethe-gln-separated-qk} generalize the respective
$\mathfrak{gl}(2)$ versions \eqref{eq:bethe-gl2-ad} and
\eqref{eq:bethe-gl2-qaq}. We may think of
\eqref{eq:bethe-gln-separated-mu} as the most important result of this
section because its constraints on $\mathfrak{gl}(n)$ monodromies
admitting Yangian invariants are valid independent of the Bethe ansatz
construction.

\section{Higher Rank Invariants from Bethe Vectors}
\label{sec:high-rank-invar}

The algebraic Bethe ansatz for $\mathfrak{gl}(2)$ spin chains uses a
monodromy $M(u)$ that is a $2\times 2$ matrix in the auxiliary space,
see section~\ref{sec:bethe-ansatze}. The Bethe vectors $|\Psi\rangle$
are products of the monodromy element $M_{12}(u)\equiv B(u)$ acting on
a reference state $|\Omega\rangle$. Consequently, also the Yangian
invariant Bethe vectors in section~\ref{sec:bethe-yangian} are of this
form. In case of the algebra $\mathfrak{gl}(n)$ the monodromy matrix
$M(u)$ in \eqref{eq:yangian-mono-spinchain} is a $n\times n$ matrix in
the auxiliary space. The nested Bethe ansatz construction
\cite{Kulish:1983rd} of a general Bethe vector $|\Psi\rangle$ involves
contributions from all monodromy elements $M_{\alpha\beta}(u)$ with
$\alpha<\beta$. This leads to rather involved formulas for these
vectors, which is why we did not include them in
section~\ref{sec:bethe-gln}. Here we show, by way of example, that
certain $\mathfrak{gl}(n)$ Yangian invariants $|\Psi\rangle$ can be
expressed as the action of only operators $M_{1n}(u)$ on a reference
state $|\Omega\rangle$. Thereby the usual nesting procedure is
bypassed completely.

Once again, we employ the compact $\mathfrak{gl}(n)$ oscillators
representations $\oscrep_c$ and $\bar{\oscrep}_{-c}$ with
$c\in\mathbb{N}$ of section~\ref{sec:deta-oscill-repr} at the sites of
the monodromy $M(u)$. In contrast to the sample invariants considered
in section~\ref{sec:comp-boson-invar}, we place the sites with dual
representations $\bar{\oscrep}_{-c}$ right of those with ordinary
representations $\oscrep_c$. This order seems to be necessary for the
simple structure of the Yangian invariants $|\Psi\rangle$ that we are
aiming at. In addition, to find solutions of the Yangian invariance
condition \eqref{eq:yi} for such monodromies, we work with the
non-trivial normalization
\begin{align}
  f_{\mathcal{V}_i}(u-v_i)=\frac{u-v_i}{u-v_i-1}\,
\end{align}
of the Lax operators \eqref{eq:yangian-def-lax}.

We remark that the simple expressions for the sample invariants, which
we will present shortly, are at present purely based on explicit
calculations. It would be desirable to understand how they come about
as a reduction of the nested Bethe vectors of \cite{Kulish:1983rd} and
thereby also establish a connection with
section~\ref{sec:bethe-gln}. Alternatively, one could attempt to show
the Yangian invariance of our expressions directly using the
commutation relations \eqref{eq:yangian-def} of the monodromy
elements. Either approach should lead to a better understanding of the
hidden simplicity of Yangian invariant Bethe vectors. Let us now
present our list of sample invariants.

\subsection{Two-Site Invariant and Identity Operator}
\label{sec:two-site-invariant}

We introduce the $\mathfrak{gl}(n)$ monodromy matrix
\begin{align}
  \begin{aligned}
    M(u)
    =
    R_{\square\,\oscrep_{c_1}}(u-v_1)
    R_{\square\,\bar{\oscrep}_{c_2}}(u-v_2)
  \end{aligned}
\end{align}
with
\begin{align}
  v_1=v_2+c_1-1\,,\quad c_1+c_2=0\,.
\end{align}
A computation along the lines of that for the $\mathfrak{gl}(2)$
invariant in section~\ref{sec:derivation-two-site} shows
\begin{align}
  \label{eq:bvec-gln-21-inv}
  \begin{aligned}
    |\Psi\rangle
    =
    M_{1n}(u_1)\cdots M_{1n}(u_P)|\Omega\rangle
    \propto(1\bullet 2)^{c_1}|0\rangle\,,
  \end{aligned}
\end{align}
where
$|\Omega\rangle=(\bar{\mathbf{a}}^1_1)^{c_1}(\bar{\mathbf{a}}^2_n)^{c_1}|0\rangle$
and the $P=c_1$ ``Bethe roots'' are given by
\begin{align}
  u_k=v_2+k-1\,.
\end{align}
We use quotation marks to remind the reader that we did not derive the
form of the vector \eqref{eq:bvec-gln-21-inv} from a Bethe
ansatz. Anyhow, \eqref{eq:bvec-gln-21-inv} satisfies the Yangian
invariance condition \eqref{eq:yi}, as one easily checks by a direct
computation. Notice that \eqref{eq:bvec-gln-21-inv} is an element of
$\oscrep_{c_1}\otimes\bar{\oscrep}_{c_2}$ and therefore
differs from $|\Psi_{2,1}\rangle$ of section~\ref{sec:comp-bos-2-site}
which is in
$\bar{\oscrep}_{c_1}\otimes\oscrep_{c_2}$. Nevertheless, we
may interpret also the intertwiner version of
\eqref{eq:bvec-gln-21-inv} as an identity operator.

\subsection{Four-Site Invariant and R-Matrix}

We consider the monodromy matrix
\begin{align}
  \label{eq:bvec-gln-42-mono}
  M(u)=
  R_{\square\,\oscrep_{c_1}}(u-v_1)R_{\square\,\oscrep_{c_2}}(u-v_2)
  R_{\square\,\bar{\oscrep}_{c_3}}(u-v_3)R_{\square\,\bar{\oscrep}_{c_4}}(u-v_4)
\end{align}
with
\begin{align}
  v_2=v_4+c_2-1\,,\quad c_2+c_4=0\,,\quad
  v_1=v_3+c_1-1\,,\quad c_1+c_3=0\,.
\end{align}
A case-by-case computation for small values of the representation
labels $c_1,c_2$ yields
\begin{align}
  \label{eq:bvec-gln-42-inv}
  \begin{aligned}
    |\Psi\rangle
    &=M_{1n}(u_1)\cdots M_{1n}(u_P)|\Omega\rangle\\
    &\propto
    \sum_{k=0}^{\text{min}(c_1,c_2)}
    \frac{k!}{\Gamma(v_3-v_4-c_2+k+1)}
    \frac{(1\bullet 3)^{c_1-k}}{(c_1-k)!}
    \frac{(2\bullet 4)^{c_2-k}}{(c_2-k)!}
    \frac{(1\bullet 4)^{k}}{k!}
    \frac{(2\bullet 3)^{k}}{k!}
    |0\rangle\,,
  \end{aligned}
\end{align}
where
$|\Omega\rangle=(\bar{\mathbf{a}}^1_1)^{c_1}(\bar{\mathbf{a}}^2_1)^{c_2}(\bar{\mathbf{a}}^3_n)^{c_1}(\bar{\mathbf{a}}^4_n)^{c_2}|0\rangle$
and we have $P=c_1+c_2$ ``Bethe roots'' given by
\begin{align}
  \begin{aligned}
    u_k=v_3+k-1\quad\text{for}\quad k=1,\ldots,c_1\,,\\
    u_{k+c_1}=v_4+k-1\quad\text{for}\quad k=1,\ldots,c_2\,.\\
  \end{aligned}
\end{align}
By means of a direct calculation we verify that
\eqref{eq:bvec-gln-42-inv} solves the Yangian invariance
condition~\eqref{eq:yi}. The intertwiner version of this condition is
a Yang-Baxter equation. Thus the invariant \eqref{eq:bvec-gln-42-inv}
corresponds to a $\mathfrak{gl}(n)$ R-matrix. Furthermore, it is akin
to the invariant $|\Psi_{4,2}(z)\rangle$ investigated in
section~\ref{sec:comp-bos-4-site}.

\subsection{Another Four-Site Invariant and Identity Operators}

We use a monodromy of the same form as in \eqref{eq:bvec-gln-42-mono},
\begin{align}
  M(u)=
  R_{\square\,\oscrep_{c_1}}(u-v_1)R_{\square\,\oscrep_{c_2}}(u-v_2)
  R_{\square\,\bar{\oscrep}_{c_3}}(u-v_3)R_{\square\,\bar{\oscrep}_{c_4}}(u-v_4)\,.
\end{align}
This time, however, the parameters obey the constraints
\begin{align}
  v_1=v_4+c_1-1\,,\quad c_1+c_4=0\,,\quad
  v_2=v_3+c_2-1\,,\quad c_2+c_3=0\,.
\end{align}
For small values of the representation labels we compute
\begin{align}
  \label{eq:bvec-gln-42-id-inv}
  \begin{aligned}
    |\Psi\rangle
    =M_{1n}(u_1)\cdots M_{1n}(u_P)|\Omega\rangle
    \propto
    (1\bullet 4)^{c_1}
    (2\bullet 3)^{c_2}
    |0\rangle\,,
  \end{aligned}
\end{align}
where
$|\Omega\rangle=(\bar{\mathbf{a}}^1_1)^{c_1}(\bar{\mathbf{a}}^2_1)^{c_2}(\bar{\mathbf{a}}^3_n)^{c_2}(\bar{\mathbf{a}}^4_n)^{c_1}|0\rangle$
and there are $P=c_1+c_2$ ``Bethe roots''
\begin{align}
  \begin{aligned}
    u_k=v_4+k-1\quad\text{for}\quad k=1,\ldots,c_1\,,\\
    u_{k+c_1}=v_3+k-1\quad\text{for}\quad k=1,\ldots,c_2\,.\\
  \end{aligned}
\end{align}
The Yangian invariance condition \eqref{eq:yi} for
\eqref{eq:bvec-gln-42-id-inv} is checked by a direct
calculation. Comparing with the two-site invariant in
section~\ref{sec:two-site-invariant}, we can interpret the intertwiner
version of \eqref{eq:bvec-gln-42-id-inv} as the product of two
identity operators.

\subsection{Yet Another Four-Site Invariant}

Let us choose the monodromy
\begin{align}
  \label{eq:bvec-gln-42-yet-mono}
  M(u)=
  R_{\square\,\oscrep_{c_1}}(u-v_1)R_{\square\,\bar{\oscrep}_{c_2}}(u-v_2)
  R_{\square\,\oscrep_{c_3}}(u-v_3)R_{\square\,\bar{\oscrep}_{c_4}}(u-v_4)
\end{align}
with
\begin{align}
  \label{eq:bvec-gln-42-yet-constr}
  v_1=v_2+c_1-1\,,\quad c_1+c_2=0\,,\quad
  v_3=v_4+c_3-1\,,\quad c_3+c_4=0\,.
\end{align}
Again we do a computation for small representation labels to obtain
\begin{align}
  \label{eq:bvec-gln-42-yet-inv}
  \begin{aligned}
    |\Psi\rangle
    =M_{1n}(u_1)\cdots M_{1n}(u_P)|\Omega\rangle
    \propto
    (1\bullet 2)^{c_1}
    (3\bullet 4)^{c_3}
    |0\rangle\,,
  \end{aligned}
\end{align}
where
$|\Omega\rangle=(\bar{\mathbf{a}}^1_1)^{c_1}(\bar{\mathbf{a}}^2_n)^{c_1}(\bar{\mathbf{a}}^3_1)^{c_3}(\bar{\mathbf{a}}^4_n)^{c_3}|0\rangle$
and we have the $P=c_1+c_3$ ``Bethe roots''
\begin{align}
  \begin{aligned}
    u_k=v_2+k-1\quad\text{for}\quad k=1,\ldots,c_1\,,\\
    u_{k+c_1}=v_4+k-1\quad\text{for}\quad k=1,\ldots,c_3\,.\\
  \end{aligned}
\end{align}
Also here we fall back to an explicit calculation to show the Yangian
invariance \eqref{eq:yi} of the vector \eqref{eq:bvec-gln-42-yet-inv}.
We included this example because in the monodromy
\eqref{eq:bvec-gln-42-yet-mono} not all dual sites are right of the
ordinary ones. However, this constraint on the order of sites still
holds within those that are linked by
\eqref{eq:bvec-gln-42-yet-constr}, i.e.\ $\bar{\oscrep}_{c_2}$ is
right of $\oscrep_{c_1}$ and $\bar{\oscrep}_{c_4}$ is right of
$\oscrep_{c_3}$.

\chapter{Additional Material on Graßmannian Integral}
\label{cha:add-grass-int}

This appendix contains supplementary results on the unitary
Graßmannian integral introduced in chapter~\ref{cha:grassmann-amp}. In
section~\ref{sec:glue-contours} we show that the $U(3)$ contour of the
bosonic oscillator Yangian invariant $|\Psi_{6,3}\rangle$ emerges from
``gluing'' the $U(2)$ contours of three invariants of the type
$|\Psi_{4,2}\rangle$. Section~\ref{sec:discrete-symmetry} deals with a
discrete parity symmetry of the unitary Graßmannian integral. It
interrelates some of the kinematic regions which we encountered for
the sample invariants in terms of spinor helicity-like variables in
section~\ref{sec:sample-invar-ampl}. Lastly, we present some
additional instructive sample invariants in terms of these variables
in section~\ref{sec:add-sample-inv-gr}.

\section{Gluing of Contours}
\label{sec:glue-contours}

The invariant $|\Psi_{6,3}\rangle$ can be obtained by combining, let's
say ``gluing together'', three invariants of the type
$|\Psi_{4,2}\rangle$. Another way of phrasing this is that the
intertwiner version of $|\Psi_{6,3}\rangle$ is the product of three
R-matrices, cf.\ sections~\ref{sec:yang-invar-as}
and~\ref{sec:comp-bos-4-site}. In the vertex model language of
section~\ref{sec:vertex-yangian} such an invariant is associated with
a Baxter lattice consisting of three lines with three intersections,
each of which represents one R-matrix. We discussed intertwiners as
well as Baxter lattices in detail only for compact bosonic
algebras. Yet the assertion in the first sentence of this paragraph is
valid more generally. Here we construct the invariant
$|\Psi_{6,3}\rangle$ from three invariants of the type
$|\Psi_{4,2}\rangle$ for the non-compact algebra $\mathfrak{u}(p,q|0)$
using the framework of chapter~\ref{cha:grassmann-amp}. We restrict to
the bosonic case for clarity.  Importantly, we show that this
construction is compatible with the unitary contour of the Graßmannian
matrix model \eqref{eq:grass-int-unitary} with the integrand
\eqref{eq:eq:grass-int-unitary-integrand-final}. We demonstrate that
the three $U(2)$ contours of the invariants $|\Psi_{4,2}\rangle$
combine into one $U(3)$ contour of $|\Psi_{6,3}\rangle$. The procedure
described here should be thought of as analogue of the gluing of
on-shell diagrams for $\mathcal{N}=4$ SYM scattering amplitudes, cf.\
sections~\ref{sec:grassmannian-integral} and~\ref{sec:deform}. In
contrast to our approach, the gluing of the contour is usually
neglected for amplitudes.

Let us show that the $U(3)$ integral \eqref{eq:grass-int-unitary} with
the integrand \eqref{eq:integrand63} for the
invariant $|\Psi_{6,3}\rangle$ can be obtained from the vector
\begin{align}
  \label{eq:glue-3inv42}
  \Big(|\Psi_{4,2}^{(3)}\rangle\Big)^{\dagger_8\dagger_9}
  \Big(|\Psi_{4,2}^{(2)}\rangle\Big)^{\dagger_7}
  |\Psi_{4,2}^{(1)}\rangle\,.
\end{align}
Here the three invariants of type $|\Psi_{4,2}\rangle$ are given
by the integral \eqref{eq:grass-int-unitary} with the $U(2)$ matrices
\begin{align}
  \mathcal{C}^{(1)}=
  \begin{pmatrix}
    C_{17}&C_{18}\\
    C_{27}&C_{28}\\
  \end{pmatrix}\,,\quad
  \mathcal{C}^{(2)}=
  \begin{pmatrix}
    C_{74}&C_{79}\\
    C_{34}&C_{39}\\
  \end{pmatrix}\,,\quad
  \mathcal{C}^{(3)}=
  \begin{pmatrix}
    C_{85}&C_{86}\\
    C_{95}&C_{96}\\
  \end{pmatrix}\,,
\end{align}
the oscillator contractions
\begin{align}
  \mathbf{I}_{\mathrel{\ooalign{\raisebox{0.4ex}{$\scriptstyle\bullet$}\cr\raisebox{-0.4ex}{$\scriptstyle\circ$}}}}^{(1)}=
  \begin{pmatrix}
    (1\mathrel{\ooalign{\raisebox{0.7ex}{$\bullet$}\cr\raisebox{-0.3ex}{$\circ$}}} 7)&
    (1\mathrel{\ooalign{\raisebox{0.7ex}{$\bullet$}\cr\raisebox{-0.3ex}{$\circ$}}} 8)\\
    (2\mathrel{\ooalign{\raisebox{0.7ex}{$\bullet$}\cr\raisebox{-0.3ex}{$\circ$}}} 7)&
    (2\mathrel{\ooalign{\raisebox{0.7ex}{$\bullet$}\cr\raisebox{-0.3ex}{$\circ$}}} 8)\\
  \end{pmatrix}\,,\quad
  \mathbf{I}_{\mathrel{\ooalign{\raisebox{0.4ex}{$\scriptstyle\bullet$}\cr\raisebox{-0.4ex}{$\scriptstyle\circ$}}}}^{(2)}=
  \begin{pmatrix}
    (7\mathrel{\ooalign{\raisebox{0.7ex}{$\bullet$}\cr\raisebox{-0.3ex}{$\circ$}}} 4)&
    (7\mathrel{\ooalign{\raisebox{0.7ex}{$\bullet$}\cr\raisebox{-0.3ex}{$\circ$}}} 9)\\
    (3\mathrel{\ooalign{\raisebox{0.7ex}{$\bullet$}\cr\raisebox{-0.3ex}{$\circ$}}} 4)&
    (3\mathrel{\ooalign{\raisebox{0.7ex}{$\bullet$}\cr\raisebox{-0.3ex}{$\circ$}}} 9)\\
  \end{pmatrix}\,,\quad
  \mathbf{I}_{\mathrel{\ooalign{\raisebox{0.4ex}{$\scriptstyle\bullet$}\cr\raisebox{-0.4ex}{$\scriptstyle\circ$}}}}^{(3)}=
  \begin{pmatrix}
    (8\mathrel{\ooalign{\raisebox{0.7ex}{$\bullet$}\cr\raisebox{-0.3ex}{$\circ$}}} 5)&
    (8\mathrel{\ooalign{\raisebox{0.7ex}{$\bullet$}\cr\raisebox{-0.3ex}{$\circ$}}} 6)\\
    (9\mathrel{\ooalign{\raisebox{0.7ex}{$\bullet$}\cr\raisebox{-0.3ex}{$\circ$}}} 5)&
    (9\mathrel{\ooalign{\raisebox{0.7ex}{$\bullet$}\cr\raisebox{-0.3ex}{$\circ$}}} 6)\\
  \end{pmatrix}\,\quad
\end{align}
and the integrands \eqref{eq:integrand42}
\begin{align}
  \label{eq:glue-inv42-integrands}
  \begin{aligned}
    \mathscr{F}^{(1)}(\mathcal{C}^{(1)})^{-1}&=
    (-1)^{c_1+c_2}(\det\mathcal{C}^{(1)})^{q-c_2}|[1]_{\mathcal{C}^{(1)}}|^{2(1+v_1-v_2)}([1]_{\mathcal{C}^{(1)}})^{c_2-c_1}\,,\\
    \mathscr{F}^{(2)}(\mathcal{C}^{(2)})^{-1}&=
    (-1)^{c_4+c_3}(\det\mathcal{C}^{(2)})^{q-c_3}|[1]_{\mathcal{C}^{(2)}}|^{2(1+v_4-n+1-c_4-v_3)}([1]_{\mathcal{C}^{(2)}})^{c_3+c_4}\,,\\
    \mathscr{F}^{(3)}(\mathcal{C}^{(3)})^{-1}&=
    (-1)^{c_5+c_6}(\det\mathcal{C}^{(3)})^{q+c_6}|[1]_{\mathcal{C}^{(3)}}|^{2(1+v_5-v_6-c_5+c_6)}([1]_{\mathcal{C}^{(3)}})^{c_5-c_6}\,.\\
  \end{aligned}
\end{align}
The expression \eqref{eq:glue-3inv42} contains oscillators at sites
$7,8,9$ only in inner products. Thus it is a vector in the tensor
product of the spaces associated with the sites $1,\ldots,6$. In
\eqref{eq:glue-inv42-integrands} we used
\eqref{eq:grass-int-parameters} for $|\Psi_{4,2}\rangle$ to express
the integrands in terms of the parameters $v_i$ and $c_i$ of only
those spaces. We can write \eqref{eq:glue-3inv42} as
\begin{align}
  \label{eq:glue-3inv42-step}
  \begin{aligned}
    &\Big(|\Psi_{4,2}^{(3)}\rangle\Big)^{\dagger_8\dagger_9}
    \Big(|\Psi_{4,2}^{(2)}\rangle\Big)^{\dagger_7}
    |\Psi_{4,2}^{(1)}\rangle\\
    &\quad=
    \chi_{2}^{-3}\int_{U(2)}[\D\mathcal{C}^{(1)}]\int_{U(2)}[\D\mathcal{C}^{(2)}]\int_{U(2)}[\D\mathcal{C}^{(3)}]
    |[1]_{\mathcal{C}^{(2)}}|^2\mathscr{F}(\mathcal{C})
    e^{\tr(\mathcal{C}\mathbf{I}_\bullet^t+\mathbf{I}_\circ \mathcal{C}^{^\dagger})}
    |0\rangle\,,
  \end{aligned}
\end{align}
where $\mathscr{F}(\mathcal{C})$ is already the integrand of
$|\Psi_{6,3}\rangle$ given in \eqref{eq:integrand63}. Furthermore,
\begin{align}
  \label{eq:gluing-u3-mat}
  \mathcal{C}=
  \begin{pmatrix}
    C_{17}&C_{18}&0\\
    C_{27}&C_{28}&0\\
    0&0&1\\
  \end{pmatrix}
  \begin{pmatrix}
    C_{74}&0&C_{79}\\
    0&1&0\\
    C_{34}&0&C_{39}\\
  \end{pmatrix}
  \begin{pmatrix}
    1&0&0\\
    0&C_{85}&C_{86}\\
    0&C_{95}&C_{96}\\
  \end{pmatrix}
\end{align}
and
\begin{align}
  \mathbf{I}_{\mathrel{\ooalign{\raisebox{0.4ex}{$\scriptstyle\bullet$}\cr\raisebox{-0.4ex}{$\scriptstyle\circ$}}}}=
  \begin{pmatrix}
    (1\mathrel{\ooalign{\raisebox{0.7ex}{$\bullet$}\cr\raisebox{-0.3ex}{$\circ$}}} 4)&
    (1\mathrel{\ooalign{\raisebox{0.7ex}{$\bullet$}\cr\raisebox{-0.3ex}{$\circ$}}} 5)&
    (1\mathrel{\ooalign{\raisebox{0.7ex}{$\bullet$}\cr\raisebox{-0.3ex}{$\circ$}}} 6)\\
    (2\mathrel{\ooalign{\raisebox{0.7ex}{$\bullet$}\cr\raisebox{-0.3ex}{$\circ$}}} 4)&
    (2\mathrel{\ooalign{\raisebox{0.7ex}{$\bullet$}\cr\raisebox{-0.3ex}{$\circ$}}} 5)&
    (2\mathrel{\ooalign{\raisebox{0.7ex}{$\bullet$}\cr\raisebox{-0.3ex}{$\circ$}}} 6)\\
    (3\mathrel{\ooalign{\raisebox{0.7ex}{$\bullet$}\cr\raisebox{-0.3ex}{$\circ$}}} 4)&
    (3\mathrel{\ooalign{\raisebox{0.7ex}{$\bullet$}\cr\raisebox{-0.3ex}{$\circ$}}} 5)&
    (3\mathrel{\ooalign{\raisebox{0.7ex}{$\bullet$}\cr\raisebox{-0.3ex}{$\circ$}}} 6)\\
  \end{pmatrix}\,.\quad
\end{align}
Here we eliminated the oscillators $\mathbf{A}^i_{\indnm{A}}$ and
$\bar{\mathbf{A}}^i_{\indnm{A}}$ at sites $i=7,8,9$ from
\eqref{eq:glue-3inv42} using
$\langle
0|e^{\mathbf{B}\mathbf{A}^i_{\indnm{A}}}e^{\mathbf{C}\bar{\mathbf{A}}^i_{\indnm{A}}}|0\rangle=e^{\mathbf{B}\mathbf{C}}$
for commuting operators $\mathbf{B}$ and $\mathbf{C}$. In addition, we
used \eqref{eq:prop-unitary-minors} to relate the minors of the $U(2)$
matrices $\mathcal{C}^{(1)}$, $\mathcal{C}^{(2)}$ and
$\mathcal{C}^{(3)}$ to those of the $U(3)$ matrix $\mathcal{C}$. We
also made use of the relation \eqref{eq:grass-int-parameters} for the
parameters $v_i$ and $c_i$ of $|\Psi_{6,3}\rangle$. Finally, we can
identify \eqref{eq:glue-3inv42-step} with the $U(3)$ integral for
$|\Psi_{6.3}\rangle$ in \eqref{eq:grass-int-unitary},
\begin{align}
  i(2\pi)^{-3}
  \Big(|\Psi_{4,2}^{(3)}\rangle\Big)^{\dagger_8\dagger_9}
  \Big(|\Psi_{4,2}^{(2)}\rangle\Big)^{\dagger_7}
  |\Psi_{4,2}^{(1)}\rangle
  =\chi_3^{-1}\int_{U(3)}[\D\mathcal{C}]\mathscr{F}(\mathcal{C})e^{\tr(\mathcal{C}\mathbf{I}_\bullet^t+\mathbf{I}_\circ \mathcal{C}^{^\dagger})}
  |0\rangle=
  |\Psi_{6,3}\rangle\,.
\end{align}
For this step we parameterized each of the $U(2)$ matrices $\mathcal{C}^{(1)}$,
$\mathcal{C}^{(2)}$ and $\mathcal{C}^{(3)}$ as in \eqref{eq:para-u2}
with variables $\theta^{(1)},\alpha^{(1)},\beta^{(1)},\gamma^{(1)}$
etc.\ Then we introduced the new variables
\begin{align}
  \begin{gathered}
    \alpha_1=\alpha^{(1)}+\alpha^{(2)}-\gamma^{(3)}\,,\quad
    \alpha_3=\alpha^{(3)}+\alpha^{(2)}\,,\quad
    \gamma=\gamma^{(1)}+\gamma^{(2)}+\gamma^{(3)}\,,\\
    \beta_1=\beta^{(1)}-\alpha^{(2)}+\gamma^{(3)}\,,\quad
    \beta_2=\beta^{(2)}+\gamma^{(1)}+\gamma^{(3)}\,,\quad
    \beta_3=\beta^{(3)}+\alpha^{(2)}-\gamma^{(3)}\,,\\
    \theta_1=\theta^{(1)}\,,\quad\theta_2=\theta^{(2)}\,,\quad\theta_3=\theta^{(3)}\,,
  \end{gathered}
\end{align}
which bring the $U(3)$ matrix $\mathcal{C}$ of
\eqref{eq:gluing-u3-mat} into the form given in \eqref{eq:para-u3}. We
reparameterized the integral \eqref{eq:glue-3inv42-step} in terms of
these new variables together with $\alpha^{(2)}$, $\gamma^{(2)}$ and
$\gamma^{(3)}$. The integrals in the latter three variables only
contribute a factor of $2\pi$ each because these variables do not
appear in the integrand of \eqref{eq:glue-3inv42-step} anymore. The
remaining integrals of the three $U(2)$ Haar measures and the one
minor of $\mathcal{C}^{(2)}$ in \eqref{eq:glue-3inv42-step} combine
into the $U(3)$ Haar measure \eqref{eq:haar-u3}. This concludes our
calculation.

We add some comments. The parameterization of $U(3)$ by $U(2)$
matrices in \eqref{eq:para-u3} is well adapted to the gluing procedure
because we just saw that each invariant of the type
$|\Psi_{4,2}\rangle$ is associated with one of the $U(2)$
matrices. Furthermore, we can argue that once the contour of the
Graßmannian integral \eqref{eq:grass-int} is fixed to be $U(2)$ for
the invariant $|\Psi_{4,2}\rangle$, the gluing implies that we have to
choose $U(3)$ for $|\Psi_{6,3}\rangle$.  Moreover, it should be
possible to construct the general invariant $|\Psi_{2K,K}\rangle$ from
gluing. In that case we should end up with a parameterization of
$U(K)$ in terms of $U(2)$ matrices as in \cite{Hurwitz:1897}. We
expect our calculation to generalize straightforwardly to
superalgebras $\mathfrak{u}(p,q|r+s)$, except for the usual tedious
subtleties with signs.

\section{Discrete Parity Symmetry}
\label{sec:discrete-symmetry}

In this section we investigate a discrete symmetry transformation of
the unitary Graßmannian integral formulas \eqref{eq:grass-int-unitary}
in oscillator variables and \eqref{eq:grass-int-barg} in terms of
spinor helicity-like variables. What is more, we study this symmetry
explicitly for some sample invariants. The investigation of the
oscillator invariant $|\Psi_{4,2}\rangle$ shows that for this example
it is identical to the ``parity'' symmetry of the corresponding
R-matrix. Interestingly, when applied to the invariants
$\Psi_{4,2}(\boldsymbol{\lambda},\overline{\boldsymbol{\lambda}})$ for
$\mathfrak{u}(1,1)$ and
$\Psi_{6,3}(\boldsymbol{\lambda},\overline{\boldsymbol{\lambda}})$ for
$\mathfrak{u}(2,2)$, this symmetry transformation relates some of the
kinematic regions that we encountered for these invariants in
section~\ref{sec:sample-invar-ampl}.

\subsection{On Level of Graßmannian Integral}
\label{sec:level-grass-int}

We define the parity transformation $\mathcal{P}$ of the Graßmannian integral
\eqref{eq:grass-int-unitary} for $\mathfrak{u}(p,q|r+s)$ oscillator
Yangian invariants by reversing the order of the oscillators at the
dual sites as well as at the ordinary sites,
\begin{align}
  \label{eq:parity-osc}
  \bar{\mathbf{A}}_{\indnm{A}}^i
  \stackrel{\mathcal{P}}{\mapsto}
  \begin{cases}
    \bar{\mathbf{A}}_{\indnm{A}}^{K-i+1}&\text{for}\quad i=1,\ldots,K\,,\\
    \bar{\mathbf{A}}_{\indnm{A}}^{2K+K-i+1}&\text{for}\quad i=K+1,\ldots,2K\,.
  \end{cases}
\end{align}
For the matrices of oscillator contractions entering
\eqref{eq:grass-int-unitary} this translates into
\begin{align}
  \label{eq:parity-contr}
  \mathbf{I}_{\mathrel{\ooalign{\raisebox{0.4ex}{$\scriptstyle\bullet$}\cr\raisebox{-0.4ex}{$\scriptstyle\circ$}}}}
  \stackrel{\mathcal{P}}{\mapsto}
  \mathscr{E}\,\mathbf{I}_{\mathrel{\ooalign{\raisebox{0.4ex}{$\scriptstyle\bullet$}\cr\raisebox{-0.4ex}{$\scriptstyle\circ$}}}}\,\mathscr{E}
  \quad\text{with}\quad
  \mathscr{E}=\mathscr{E}^{-1}=
  \begin{pmatrix}
    0&0&\cdots&0&1\\
    0&&&1&0\\
    \vdots&&\iddots&&\vdots\\
    0&1&&&0\\
    1&0&\cdots&0&0\\
  \end{pmatrix}
  \in U(K)\,.
\end{align}
Using the left- and right-invariance of the Haar measure we find that
the Graßmannian integral \eqref{eq:grass-int-unitary} is invariant
under the transformation \eqref{eq:parity-osc},
\begin{align}
  |\Psi_{2K,K}\rangle
  \stackrel{\mathcal{P}}{\mapsto}
  \chi_K^{-1}\int_{U(K)}[\D\mathcal{C}]\,
  \mathscr{F}(\mathscr{E}\mathcal{C}\mathscr{E})\,
  e^{\tr(\mathcal{C}\mathbf{I}_\bullet^t+\mathbf{I}_\circ \mathcal{C}^{^\dagger})}
  |0\rangle
  =|\Psi_{2K,K}\rangle\,,
\end{align}
provided that
$\mathscr{F}(\mathscr{E}\mathcal{C}\mathscr{E})=\mathscr{F}(\mathcal{C})$. This
constraint is satisfied if the parameters in the function
$\mathscr{F}(\mathcal{C})$ defined in
\eqref{eq:eq:grass-int-unitary-integrand-final} obey
\begin{align}
  \label{eq:parity-constr}
  \begin{aligned}
    v_1-v_2&=v_{K-1}-v_K+c_K-c_{K-1}\,,&
    \quad\ldots\,,\quad&&
    v_{K-1}-v_K&=v_1-v_2+c_2-c_1\,,\\
    c_2&=c_{K-1}\,,&
    \quad\ldots\,,\quad&&
    c_K&=c_1\,.
  \end{aligned}
\end{align}
Here we used
$[1,2,\ldots,j]_{\mathscr{E}\mathcal{C}\mathscr{E}}=\overline{[1,2,\ldots,K-j]}_{\mathcal{C}}\det\mathcal{C}$
for the minors appearing in
$\mathscr{F}(\mathscr{E}\mathcal{C}\mathscr{E})$.  Notice that a
particular solution of the constraints in \eqref{eq:parity-constr} is
given by
\begin{align}
  \label{eq:parity-constr-special}
  v_1-v_2=v_2-v_3=\cdots=v_{K-1}-v_{K}\,,\quad
  c_1=c_2=\cdots= c_K\,.
\end{align}

This discussion of the parity symmetry easily translates to the
Graßmannian integral \eqref{eq:grass-int-barg} for $\mathfrak{u}(p,p)$
in spinor helicity-like variables, where we restrict to bosonic
algebras to avoid nasty sign factors. With the change of variables
from oscillators to these variables given in \eqref{eq:barg-map-b},
\eqref{eq:barg-rel-spin-hel} and \eqref{eq:barg-rename}, the
transformation \eqref{eq:parity-osc} reads
\begin{align}
  \label{eq:parity-spinh}
  \boldsymbol{\lambda}=
  \left(
  \begin{array}{c}
    \boldsymbol{\lambda}^{\text{d}}\\[0.3em]
    \hdashline\\[-1.0em]
    \boldsymbol{\lambda}^{\text{o}}
  \end{array}
  \right)
  \stackrel{\mathcal{P}}{\mapsto}
  \left(
  \begin{array}{c}
    \mathscr{E}\boldsymbol{\lambda}^{\text{d}}\\[0.3em]
    \hdashline\\[-1.0em]
    \mathscr{E}\boldsymbol{\lambda}^{\text{o}}
  \end{array}
  \right)\,.
\end{align}
Similar to the oscillator case one verifies that the Graßmannian
integral \eqref{eq:grass-int-barg} is invariant under this
transformation,
\begin{align}
  \Psi_{2K,K}(\boldsymbol{\lambda},\overline{\boldsymbol{\lambda}})
  \stackrel{\mathcal{P}}{\mapsto}
  \Psi_{2K,K}(\boldsymbol{\lambda},\overline{\boldsymbol{\lambda}})\,,
\end{align}
if the parameters obey \eqref{eq:parity-constr}.

\subsection{On Level of Sample Invariants}
\label{sec:level-sample-invar}

\subsubsection{Four-Site Invariant in Oscillator Variables}
\label{sec:parity-osc}

Let us study the parity transformation $\mathcal{P}$ more explicitly
for the oscillator invariant $|\Psi_{4,2}\rangle$ defined by the
Graßmannian integral \eqref{eq:grass-int-unitary}. In this case the
transformation \eqref{eq:parity-contr} becomes
\begin{align}
  \label{eq:parity-42-osc}
  \mathbf{I}_{\mathrel{\ooalign{\raisebox{0.4ex}{$\scriptstyle\bullet$}\cr\raisebox{-0.4ex}{$\scriptstyle\circ$}}}}
  =
  \begin{pmatrix}
    (1\mathrel{\ooalign{\raisebox{0.7ex}{$\bullet$}\cr\raisebox{-0.3ex}{$\circ$}}} 3)&
    (1\mathrel{\ooalign{\raisebox{0.7ex}{$\bullet$}\cr\raisebox{-0.3ex}{$\circ$}}} 4)\\
    (2\mathrel{\ooalign{\raisebox{0.7ex}{$\bullet$}\cr\raisebox{-0.3ex}{$\circ$}}} 3)&
    (2\mathrel{\ooalign{\raisebox{0.7ex}{$\bullet$}\cr\raisebox{-0.3ex}{$\circ$}}} 4)\\
  \end{pmatrix}
  \stackrel{\mathcal{P}}{\mapsto}
  \begin{pmatrix}
    (2\mathrel{\ooalign{\raisebox{0.7ex}{$\bullet$}\cr\raisebox{-0.3ex}{$\circ$}}} 4)&
    (2\mathrel{\ooalign{\raisebox{0.7ex}{$\bullet$}\cr\raisebox{-0.3ex}{$\circ$}}} 3)\\
    (1\mathrel{\ooalign{\raisebox{0.7ex}{$\bullet$}\cr\raisebox{-0.3ex}{$\circ$}}} 4)&
    (1\mathrel{\ooalign{\raisebox{0.7ex}{$\bullet$}\cr\raisebox{-0.3ex}{$\circ$}}} 3)\\
  \end{pmatrix}\,.
\end{align}
According to \eqref{eq:parity-constr-special} the Yangian invariant
$|\Psi_{4,2}\rangle$ is parity invariant for $c_1=c_2\in\mathbb{Z}$
and unconstrained $v_1,v_2\in\mathbb{C}$. Of course, this property can
also be verified on the level of the explicit expression
\eqref{eq:gr-sample-inv42} for the $\mathfrak{u}(p,q|r+s)$ version of
$|\Psi_{4,2}\rangle$. We do not detail this computation here and just
mention that it is important to pay close attention to the constraints
of the summation range in
\eqref{eq:gr-sample-inv42-constraints}. However, it is helpful to take
a look at the compact $\mathfrak{u}(n)$ special case of
\eqref{eq:gr-sample-inv42} that we discussed in detail in
section~\ref{sec:comp-bos-4-site}, albeit with a different
normalization. In this case $|\Psi_{4,2}\rangle$ is given in
\eqref{eq:osc-psi42} as a sum over terms specified
in~\eqref{eq:osc-phi}. We immediately see that each of these terms is
invariant under the transformation \eqref{eq:parity-42-osc}. In the
aforementioned section we also reformulated $|\Psi_{4,2}\rangle$ as an
R-matrix acting on the tensor product
$\oscrep_{-{c_1}}\otimes\oscrep_{-{c_2}}$. For this R-matrix the
invariance under the transformation \eqref{eq:parity-42-osc}
translates into the invariance under the permutation of tensor
factors. In the literature on integrable models this property is known
as ``parity invariance'', see e.g.\ \cite{Gomez:2005}. Therefore we
also refer to the general transformation $\mathcal{P}$ defined in
section~\ref{sec:level-grass-int} as parity transformation.

\subsubsection{Four-Site Invariant for
  \texorpdfstring{$\mathfrak{u}(1,1)$}{u(1,1)} in Spinor Helicity
  Variables}
\label{sec:parity-four-site-spin-hel}

We move on to discuss the Yangian invariant
$\Psi_{4,2}(\boldsymbol{\lambda},\overline{\boldsymbol{\lambda}})$ for
$\mathfrak{u}(1,1)$, which we constructed in
section~\ref{sec:inv42-u11} using the unitary Graßmannian integral
\eqref{eq:grass-int-barg}. According to section
\ref{sec:level-grass-int} it is invariant under the parity
transformation \eqref{eq:parity-spinh} if the representation labels
satisfy $c_1=c_2$. Let us in addition restrict the deformation
parameters to $v_1=v_2$. In this case
$\Psi_{4,2}(\boldsymbol{\lambda},\overline{\boldsymbol{\lambda}})$ is
given by \eqref{eq:int-42-u11-momentum} with
\eqref{eq:int-42-u11-lastint-def-expl}. The latter equation shows that
for a fixed value of $c_1=c_2$ there are two kinematic regions in
which the expression for the invariant differs. These regions are
characterized by the absolute values of the variables $\mathtt{A}$ and
$\mathtt{B}$ defined in \eqref{eq:int-42-u11-ab}. It is instructive to
study the action of the parity transformation \eqref{eq:parity-spinh}
on these regions. For the invariant under consideration this
transformation reads
\begin{align}
  \begin{pmatrix}
    \lambda^1_1\\
    \lambda^2_1\\
  \end{pmatrix}
  \stackrel{\mathcal{P}}{\mapsto}
  \begin{pmatrix}
    \lambda^2_1\\
    \lambda^1_1\\
  \end{pmatrix}\,,\quad
  \begin{pmatrix}
    \lambda^3_1\\
    \lambda^4_1\\
  \end{pmatrix}
  \stackrel{\mathcal{P}}{\mapsto}
  \begin{pmatrix}
    \lambda^4_1\\
    \lambda^3_1\\
  \end{pmatrix}\,.
\end{align}
It implies
\begin{align}
  \mathtt{A}\stackrel{\mathcal{P}}{\mapsto} -\overline{\mathtt{B}}\,,\quad
  \mathtt{B}\stackrel{\mathcal{P}}{\mapsto} -\overline{\mathtt{A}}\,.
\end{align}
Consequently, the two kinematic regions and, respectively, the
associated expressions for the invariant in
\eqref{eq:int-42-u11-lastint-def-expl} are interchanged,
\begin{align}
  |\mathtt{A}|>|\mathtt{B}|
  \;\;
  \stackrel{\;\mathcal{P}}{\leftrightarrow}
  \;\;
  |\mathtt{A}|<|\mathtt{B}|\,.
\end{align}
Thus the existence of these two regions in
\eqref{eq:int-42-u11-lastint-def-expl} is of crucial importance for
the parity symmetry of
$\Psi_{4,2}(\boldsymbol{\lambda},\overline{\boldsymbol{\lambda}})$ for
$\mathfrak{u}(1,1)$.

Let us discuss what happens if we use the expression of either region
in \eqref{eq:int-42-u11-lastint-def-expl} and declare it to be valid
for all values of $\mathtt{A}$ and $\mathtt{B}$. The resulting
function still satisfies the Yangian invariance condition
\eqref{eq:yi}, which takes the form of a differential equation in
$\lambda^i_1$, for generic values of $\lambda^i_1$. However, it
violates parity symmetry. Hence this function cannot be the correct
evaluation of the unitary Graßmannian integral
\eqref{eq:grass-int-barg}. Moreover, it cannot be related to the
oscillator Yangian invariant $|\Psi_{4,2}\rangle$ for
$\mathfrak{u}(1,1)$ via the change of basis introduced in
section~\ref{sec:osc-spinor} because $|\Psi_{4,2}\rangle$ is parity
symmetric.

\subsubsection{Six-Site Invariant for
  \texorpdfstring{$\mathfrak{u}(2,2)$}{u(2,2)} in Spinor Helicity
  Variables}
\label{sec:parity-six-site-spin-hel}

In section~\ref{sec:inv63-u22} we constructed the Yangian invariant
$\Psi_{6,3}(\boldsymbol{\lambda},\overline{\boldsymbol{\lambda}})$ for
$\mathfrak{u}(2,2)$ from the unitary Graßmannian integral
\eqref{eq:grass-int-barg}. Here we focus on the case with
representation labels $c_1=c_2=c_3$ and deformation parameters
$v_1=v_2=v_3$. Section~\ref{sec:level-grass-int} shows that in this
case the Yangian invariant
$\Psi_{6,3}(\boldsymbol{\lambda},\overline{\boldsymbol{\lambda}})$ is
also invariant under the parity transformation
\eqref{eq:parity-spinh}.
$\Psi_{6,3}(\boldsymbol{\lambda},\overline{\boldsymbol{\lambda}})$ is
given explicitly by \eqref{eq:inv63-u22-final} with
\eqref{eq:inv63-u22-four-cases} as a sum of residues. The selection of
residues differs for four kinematic regions, which are characterized
by the absolute values of the variables
$\mathtt{A},\mathtt{D},\mathtt{C},\mathtt{D}$ defined in
\eqref{eq:inv63-u22-minors-abcd}. After
\eqref{eq:inv63-u22-four-cases} we explained that this
characterization can be translated into conditions on the generalized
Mandelstam variables $s_{126}$ and $s_{234}$. We want to study the
action of the parity transformation~\eqref{eq:parity-spinh},
\begin{align}
  \label{eq:parity-inv63-u22-trafo}
  \begin{pmatrix}
    \lambda^1_\alpha\\
    \lambda^2_\alpha\\
    \lambda^3_\alpha\\
  \end{pmatrix}
  \stackrel{\mathcal{P}}{\mapsto}
  \begin{pmatrix}
    \lambda^3_\alpha\\
    \lambda^2_\alpha\\
    \lambda^1_\alpha\\
  \end{pmatrix}\,,\quad
  \begin{pmatrix}
    \lambda^4_\alpha\\
    \lambda^5_\alpha\\
    \lambda^6_\alpha\\
  \end{pmatrix}
  \stackrel{\mathcal{P}}{\mapsto}
  \begin{pmatrix}
    \lambda^6_\alpha\\
    \lambda^5_\alpha\\
    \lambda^4_\alpha\\
  \end{pmatrix}\,,
\end{align}
on this explicit form of
$\Psi_{6,3}(\boldsymbol{\lambda},\overline{\boldsymbol{\lambda}})$.
First, we observe
\begin{align}
  \mathtt{A}\stackrel{\mathcal{P}}{\mapsto}\overline{\mathtt{C}}\,,\quad
  \mathtt{B}\stackrel{\mathcal{P}}{\mapsto}\overline{\mathtt{D}}\,,\quad
  \mathtt{C}\stackrel{\mathcal{P}}{\mapsto}\overline{\mathtt{A}}\,,\quad
  \mathtt{D}\stackrel{\mathcal{P}}{\mapsto}\overline{\mathtt{B}}\,.
\end{align}
From this we compute the transformation of the individual residues
appearing in \eqref{eq:inv63-u22-four-cases},
\begin{align}
  \begin{aligned}
    &\text{res}_{0}\mathscr{I}(u)\stackrel{\mathcal{P}}{\mapsto}\text{res}_{0}\mathscr{I}(u)\,,&&
    \text{res}_{\infty}\mathscr{I}(u)\stackrel{\mathcal{P}}{\mapsto}\text{res}_{\infty}\mathscr{I}(u)\,,\\
    &\text{res}_{\frac{\mathtt{A}}{\mathtt{B}}}\mathscr{I}(u)\stackrel{\;\mathcal{P}}{\leftrightarrow}\text{res}_{\frac{\overline{\mathtt{C}}}{\overline{\mathtt{D}}}}\mathscr{I}(u)\,,&&
    \text{res}_{\frac{\overline{\mathtt{B}}}{\overline{\mathtt{A}}}}\mathscr{I}(u)\stackrel{\;\mathcal{P}}{\leftrightarrow}\text{res}_{\frac{\mathtt{D}}{\mathtt{C}}}\mathscr{I}(u)\,.
  \end{aligned}
\end{align}
It follows the transformation of the four kinematic regions and the
corresponding expressions for the invariant in
\eqref{eq:inv63-u22-four-cases},
\begin{align}
  \label{eq:parity-inv63-u22}
  \begin{aligned}
    s_{234}>0\,, s_{126}>0\;\;\stackrel{\mathcal{P}}{\mapsto}\;\;s_{234}>0\,, s_{126}>0\,,\\
    s_{234}<0\,, s_{126}<0\;\;\stackrel{\mathcal{P}}{\mapsto}\;\;s_{234}<0\,, s_{126}<0\,,\\
    s_{234}>0\,, s_{126}<0\;\;\stackrel{\;\mathcal{P}}{\leftrightarrow}\;\;s_{234}<0\,, s_{126}>0\,.\\
  \end{aligned}
\end{align}
Thus two of the kinematic regions of
$\Psi_{6,3}(\boldsymbol{\lambda},\overline{\boldsymbol{\lambda}})$ for
$\mathfrak{u}(2,2)$ are exchanged by the parity transformation. Each
of the remaining two regions is parity invariant by itself.

As we argued in section~\ref{sec:inv63-u22}, the Yangian invariant
$\Psi_{6,3}(\boldsymbol{\lambda},\overline{\boldsymbol{\lambda}})$
with $c_1=c_2=c_3=0$ and $v_1=v_2=v_3$ agrees with the gluon amplitude
$A^{(\text{tree})}_{6,3}$ merely in the kinematic region
$s_{234},s_{126}>0$. We would like to show on principle grounds that
the evaluation of the unitary Graßmannian integral
\eqref{eq:grass-int-barg} necessarily leads to the four kinematic
regions of
$\Psi_{6,3}(\boldsymbol{\lambda},\overline{\boldsymbol{\lambda}})$ in
\eqref{eq:inv63-u22-four-cases} despite the mismatch with
$A^{(\text{tree})}_{6,3}$ in three of those. At the end of
section~\ref{sec:parity-four-site-spin-hel} we successfully employed
the parity invariance to demonstrate the necessity of the two
kinematic region of the Yangian invariant considered there. Proceeding
analogously for
$\Psi_{6,3}(\boldsymbol{\lambda},\overline{\boldsymbol{\lambda}})$, we
find that we cannot extend the expression in
\eqref{eq:inv63-u22-four-cases} for either of the regions
$s_{234}>0, s_{126}<0$ and $s_{234}<0, s_{126}>0$ to the entire
kinematic regime. The result would violate parity invariance according
to \eqref{eq:parity-inv63-u22}. However, the expressions for the
regions $s_{234}>0, s_{126}>0$ and $s_{234}<0, s_{126}<0$ could be
extended to the entire domain without violating parity
invariance. Consequently, the discrete parity symmetry is not
sufficient to establish the necessity of the four kinematic regions of
$\Psi_{6,3}(\boldsymbol{\lambda},\overline{\boldsymbol{\lambda}})$. For
this we would need a discrete symmetry group of
$\Psi_{6,3}(\boldsymbol{\lambda},\overline{\boldsymbol{\lambda}})$
which does not leave any of the kinematic regions invariant. It would
be very interesting to search for such a symmetry.

Let us remark that the parity transformation in
\eqref{eq:parity-inv63-u22-trafo} is part of a known discrete symmetry
group of the amplitude $A^{(\text{tree})}_{6,3}$. This group includes
a cyclic shift of the particle index $i\mapsto i+1$ and a reversal of
the order of all particles, see e.g.\ \cite{Bern:1990ux} and recall
the brief mention towards the end of
section~\ref{sec:symmetries}. These transformations act on the
particle momenta and helicities. The transformation
\eqref{eq:parity-inv63-u22-trafo} is obtained from three shifts and
one reversal. Most of the other transformations of this group are not
straightforwardly implemented on the level of the unitary Graßmannian
integral \eqref{eq:grass-int-barg} because in our setting they would
mix dual and ordinary sites.

\section{Further Sample Invariants}
\label{sec:add-sample-inv-gr}

In section~\ref{sec:sample-invar-ampl} we evaluated the unitary
Graßmannian integral \eqref{eq:grass-int-barg} in spinor helicity-like
variables for a number of sample Yangian invariants. We concentrated
on those examples that are of help to understand how the scattering
amplitudes of the introductory section~\ref{sec:amplitudes} are
related to our integral. Here we supplement the list of invariants by
some further examples that do not directly serve this purpose but
display other noteworthy features.

\subsection{Two-Site Invariant for
  \texorpdfstring{$\mathfrak{u}(2,2)$}{u(2,2)}}
\label{sec:inv21-u22}

We evaluated the unitary Graßmannian integral
\eqref{eq:grass-int-barg} with $(N,K)=(2,1)$ for the algebra
$\mathfrak{u}(1,1)$ in section~\ref{sec:inv21-u11}. The computation of
this two-site invariant for $\mathfrak{u}(2,2)$ shows a
peculiarity. From \eqref{eq:grass-int-barg} with the parameterization
of $U(1)$ in \eqref{eq:para-u1}, the associated Haar measure
\eqref{eq:haar-u1} and the integrand $\mathscr{F}(\mathcal{C})$ from
\eqref{eq:integrand21}, we obtain
\begin{align}
  \label{eq:inv21-u22-spin-hel}
  \Psi_{2,1}(\boldsymbol{\lambda},\overline{\boldsymbol{\lambda}})=
  2i\,
  \delta(P_{11})\delta_{\mathbb{C}}(P_{21})
  \lambda^1_1\overline{\lambda}^1_1\left(-\frac{\lambda^1_1}{\lambda^2_1}\right)^{c_1-2}\,,
\end{align}
where we assumed $\lambda_1^2\neq 0$ and the total momentum
$P_{\alpha\dot{\alpha}}$ is defined in
\eqref{eq:def-mon-supermon}. Notice that \eqref{eq:inv21-u22-spin-hel}
does not contain explicitly the momentum conserving delta function
$\delta^{4|0}(P)$ from \eqref{eq:def-delta-bosonic}. However, the
delta functions in \eqref{eq:inv21-u22-spin-hel} impose three real
equations that imply the fourth equation $P_{22}=0$ and thus implement
the momentum conservation constraint \eqref{eq:mon-consv}. This is a
special feature of two-particle kinematics in four-dimensional
Minkowski space.

\subsection{Four-Site Invariant for
  \texorpdfstring{$\mathfrak{u}(2,2|4+0)$}{u(2,2|4+0)}}
\label{sec:inv42-u2240}

In section \ref{sec:inv42-u2204} we computed the unitary Graßmannian
integral \eqref{eq:grass-int-barg} with four sites for the
superalgebra $\mathfrak{u}(2,2|0+4)$. The notation $0+4=r+s$ signifies
the choice of the grading, cf.\ \eqref{eq:grading} and
\eqref{eq:osc-split} where it affects the grouping of bosonic and
fermionic oscillators. We made this particular choice in order to make
contact with the superamplitudes reviewed in the introductory
section~\ref{sec:superamplitudes}. Here we investigate how the grading
affects the form of the invariant.  Computing the integral
\eqref{eq:grass-int-barg} for the four-site invariant with the
superalgebra $\mathfrak{u}(2,2|4+0)$ leads to
\begin{align}
  \label{eq:inv42-u2240}
  \Psi_{4,2}(\boldsymbol{\lambda},\overline{\boldsymbol{\lambda}},\boldsymbol{\theta})=
  8i\,
  \frac{\delta^{4|0}(P)\delta^{0|8}(\hat{Q})}
  {\,\overline{\langle 12\rangle}\,\overline{\langle 23\rangle}\,\overline{\langle 34\rangle}\,\overline{\langle 41\rangle}\,}
  \left(\frac{\langle 14 \rangle}{\langle 34 \rangle}\right)^{c_1}
  \left(\frac{\langle 12 \rangle}{\langle 14 \rangle}\right)^{c_2}
  \left(\frac{\langle 34 \rangle\overline{\langle 34\rangle}}{\langle 14 \rangle\overline{\langle 14\rangle}}\right)^{v_1-v_2}\,,
\end{align}
where the fermionic delta function is defined in
\eqref{eq:def-delta-fermionic}. Interestingly, only the spinor
brackets that remain in the undeformed case get complex conjugated
compared to the $\mathfrak{u}(2,2|0+4)$ version in
\eqref{eq:inv42-u2204-final}. Those factors involving the deformation
parameters are identical for both gradings.

\subsection{Four-Site Invariant for
  \texorpdfstring{$\mathfrak{u}(2,2|2+2)$}{u(2,2|2+2)}}
\label{sec:inv42-u2222}

Let us investigate the four-site invariant for yet another
grading. The evaluation of the Graßmannian integral
\eqref{eq:grass-int-barg} with $(N,K)=(4,2)$ for the superalgebra
$\mathfrak{u}(2,2|2+2)$ yields
\begin{align}
  \label{eq:inv42-u2222}
  \Psi_{4,2}(\boldsymbol{\lambda},\overline{\boldsymbol{\lambda}},\boldsymbol{\theta},\boldsymbol{\eta})=
  8i\,
  \frac{\delta^{4|0}(P)\delta^{0|4}(\hat{Q})\delta^{0|4}(Q)}
  {\langle 12\rangle \overline{\langle 23\rangle}\langle 34\rangle \overline{\langle 41\rangle}}
  \left(\frac{\langle 14 \rangle}{\langle 34 \rangle}\right)^{c_1}
  \left(\frac{\langle 12 \rangle}{\langle 14 \rangle}\right)^{c_2}
  \left(\frac{\langle 34 \rangle\overline{\langle 34\rangle}}{\langle 14 \rangle\overline{\langle 14\rangle}}\right)^{v_1-v_2}\,.
\end{align}
Note that
$\langle 12\rangle \overline{\langle 23\rangle}\langle 34\rangle
\overline{\langle 41\rangle}= \langle 12\rangle \overline{\langle
  12\rangle}\langle 23\rangle \overline{\langle 23\rangle}$.
Hence the combination of spinor brackets appearing in the undeformed
Yangian invariant of the superalgebra $\mathfrak{u}(2,2|2+2)$ is
real. Once more, those factors involving the deformation parameters
are not affected by the choice of the grading, cf.\ the invariants
\eqref{eq:inv42-u2204-final} and \eqref{eq:inv42-u2240}.

We remark that the Yangian invariant corresponding to
\eqref{eq:inv42-u2222} in the oscillator basis is $|\Psi_{4,2}\rangle$
from \eqref{eq:gr-sample-inv42}. Each of its two building blocks
$(k\bullet l)$ and $(k\circ l)$ is invariant under one of the two
compact subalgebras in
$\mathfrak{su}(2|2)\oplus\mathfrak{su}(2|2)\subset\mathfrak{u}(2,2|2+2)$,
as shown in \eqref{eq:bullets-nc-symm}. Similar $\mathfrak{su}(2|2)$
subalgebras play an important role to obtain asymptotic all-loop
results in the planar $\mathcal{N}=4$ SYM spectral problem, cf.\
\cite{Beisert:2005tm} and recall section~\ref{sec:spectrum}. This
motivates our study of the Yangian invariant
$\Psi_{4,2}(\boldsymbol{\lambda},\overline{\boldsymbol{\lambda}},\boldsymbol{\theta},\boldsymbol{\eta})$
in spinor helicity variables with $2+2$ grading.

\addtocontents{toc}{\protect\setcounter{tocdepth}{2}}

\cleardoublepage
\phantomsection
\addcontentsline{toc}{chapter}{Bibliography}
\bibliographystyle{utphys}
\bibliography{literature}




\end{document}